\documentclass{aa} %
\usepackage{graphicx}
\usepackage{txfonts}
\usepackage{hyperref} %
\usepackage{textcomp} %
\usepackage{rotating}
\usepackage{threeparttable} %
\usepackage{multicol} %
\usepackage{multirow}
\usepackage{xcolor} %

\newcommand{\mic}{\textmu m}
\begin{document}

\title{Infrared and sub-mm observations of outbursting young stars with Herschel and Spitzer\thanks{{\it \textit{Herschel}} is an ESA space observatory with science instruments provided by European-led Principal Investigator consortia and with important participation from NASA.}}

\subtitle{}

\author{A. Postel\inst{1}, M. Audard\inst{1}, E. Vorobyov\inst{2,3}, O. Dionatos\inst{2}, C. Rab\inst{4}
        \and M. Güdel\inst{2}}

\institute{University of Geneva, Department of Astronomy, Chemin d'Ecogia 16, 1290 Versoix, Switzerland\\ %
                   \email{andreas.postel@unige.ch}
                \and
               University of Vienna, Department of Astrophysics, Türkenschanzstrasse 17, Vienna, Austria 
                \and
                   Research Institute of Physics, Southern Federal University, Stachki 194, Rostov-on-Don, 344090, Russia
                \and
                   Kapteyn Astronomical Institute, University of Groningen, P.O. Box 800, 9700 AV Groningen, The
                   Netherlands
}

\date{Received date /
Accepted date }

\abstract
    {Episodic accretion plays an important role in the evolution of young stars. Although it has been under investigation for a
    long time, the origin of such episodic accretion events is not yet understood.
        }
    {We   investigate the dust and gas emission of a sample of young outbursting sources in the infrared
    to get a better understanding of their properties and circumstellar material, and we use the results in a further work to model the objects.}
    {We used \textit{Herschel} data, from our PI program of 12 objects and complemented with archival observations to obtain the spectral energy distributions (SEDs) and spectra of our targets. 
    We report here the main characteristics of our sample, focussing  on the SED properties and on the gas emission lines detected in the PACS and SPIRE spectra.}
    {The SEDs of our sample show the diversity of the outbursting sources, with several targets showing strong emission in the far-infrared from the
    embedded objects. Most of our targets reside in a complex environment, which we discuss in detail. We   detected several atomic and molecular lines,
in particular rotational CO emission from several transitions from 
    $J=38-37$ to $J=4-3$. We   constructed rotational diagrams for the CO lines, and derived in three domains of assumed local thermodynamic equilibrium (LTE) temperatures and column densities, ranging
    mainly between $0-100$ K and $400-500$K. We confirm correlation in our
    sample between intense CO $J=16-15$ emission and the column density of the warm domain of CO, N(warm). We
     notice a strong increase in luminosity of HH 381 IRS and a weaker increase for PP 13 S, which shows the beginning of an outburst.}
    {}

\keywords{
    stars: formation -- stars: protostars -- stars: pre-main sequence -- protoplanetary disks -- accretion, accretion disks -- infrared: stars}
\authorrunning{Postel et al.}
\maketitle{}

\section{Introduction}

It has been recognized that episodic accretion   plays a central role in the accretion history of young stars. The vivid manifestation of episodic accretion are FU Orionis-type objects (FUors), which are  young low-mass stars displaying  strong optical outbursts that last several decades. FUors are named after the archetype, FU Orionis, which had its outburst in 1936, when it experienced an increased B-band magnitude of 6 mag at peak, and has been slowly fading since then \citep{2000ApJ...531.1028K}. 
After this outburst, a few tens of objects with similar  behavior were detected (\citealt{1996ARA&A..34..207H}; \citealt{2010vaoa.conf...19R}), as well as possible candidates for FUors which had not been observed in the pre-outburst quiescent state, but possess characteristics typical to FUors  (\citealt{2007ApJ...658..487Q}; \citealt{2010ATel.2801....1S};
\citealt{2012ApJ...748L...5R}; \citealt{2012ApJ...756...99F}; \citealt{2013AN....334...53F}).  The majority of FUors appear to be young protostars either still accreting from their parental envelopes or  in the early T Tauri phase with tenuous envelopes.

In addition to FUors, whose accretion rate rapidly increases from typically
$10^{-7}~\textnormal{M}_{\odot}~\textnormal{yr}^{-1}$ to
\begin{math} 10^{-4}~\textnormal{M}_{\odot}~\textnormal{yr}^{-1} \end{math} and stays high over decades, another type of young low-mass star showing weaker and shorter but multiple outburst was detected \citep{1989ESOC...33..233H}. The archetype of this class is EX Lupi and the corresponding class is called EXors. These objects appear to be more evolved than FUors, but the distinction may be somewhat blurred \citep{Audard2014}. Moreover, episodic accretion appears to be inherent to massive star formation  \citep[e.g.,][]{2017sfcc.confE..15D,2017arXiv171002320M}, which implies a universal character of this phenomenon.
        
In  recent years, observations in the IR and sub-millimeter regime were carried out, finally allowing investigations of the dust emission and lines of FUors and EXors in this wavelength range, which show the circumstellar disk and possible envelopes of these objects
(\citealt{Lorenzetti2005}, \citealt{Green2013}). Activity in the form of significant optical and IR variability has been found to be extremely common for young stars. The inner regions of FUors and EXors, highly interesting to reveal the origin of outbursts, are a challenge for single-dish
observations as the phenomena cannot be proven unambiguously
to occur in the inner disk alone.  Atacama Large Millimeter/submillimeter Array (ALMA) and  Karl G. Jansky Very Large Array (JVLA)  studies in recent  years have opened a new window for investigating the outburst objects with very high spatial resolution, tracing a shift of the snow line during outburst
\citep{2016Natur.535..258C}, outflows of the objects (\citealt{2017MNRAS.468.3266R,2017MNRAS.466.3519R}),
density waves in the disks \citep{2016Sci...353.1519P}, and hot inner disks \citep{2017A&A...602A..19L}.

The origin of accretion bursts remains unclear and could be due to the viscous-thermal and/or magnetorotational instabilities 
in the inner disk,  tidal effects from close companions or close flybys of external stars, or accretion of gaseous clumps in a gravitationally unstable disk \citep{Audard2014}. The authors of this paper aim to investigate the properties of FUors and EXors by analyzing observational data, along with numerical hydrodynamics simulations of protostellar disks \citep{2015ApJ...805..115V}, stellar evolution models of outburst stars \citep{2019MNRAS.484..146E}, and thermo-chemical models of star--disk systems in the outburst state \citep{Rab2017}. Our purpose is to obtain a better understanding of the outburst processes and their origins. In this work, we present the spectral energy distributions (SEDs) of 12 outbursting sources, that have been observed with \textit{Herschel} and \textit{Spitzer}. We analyze  the SEDs and the line maps, and provide the first results of the environmental conditions, namely CO rotational temperatures in three different domains of local thermodynamic equilibrium (LTE). We also compare the line strengths of different molecules
and isotopologs.

\section{Observations}
        We observed with the ESA \textit{Herschel} Space Observatory \citep{Herschel2010} a sample of 11 FUors and 1 EXor,
        namely Haro 5a IRS, HH 354 IRS, HH 381 IRS, Parsamian 21, PP 13 S,
        Re 50 N IRS (=HBC 494), V346 Nor, V733 Cep, V883 Ori, EX Lup, and V1647 Ori (PI: M. Audard). The aim of the program
        was to obtain the photometry and spectroscopy of these targets with   the Photodetector Array Camera and
        Spectrometer (PACS; \citealt{PACS2010}) and  Spectral and Photometric Imaging REceiver (SPIRE; \citealt{SPIRE2010}), complementing them
        with data from the   Heterodyne Instrument for the Far-Infrared (HIFI;
        \citealt{HIFI2010}). In this paper we present the PACS and SPIRE photometry and spectra from our program,
        complemented with archival \textit{Herschel} data (e.g., from key or guaranteed time programs). HIFI data will be
        presented elsewhere. The spectra cover the 60-671 \textmu m range of PACS and SPIRE.
        Furthermore, we present \textit{Spitzer} data \citep{2004ApJS..154....1W}, together with photometry at different wavelengths published in the
        literature (shown later for each target individually), which allow us in many cases to show the SED down to the optical. 

\section{Data reduction}
        The \textit{Herschel} data were retrieved from the \textit{Herschel} Science Archive (HSA). To ensure that we
        used the latest improvements in the calibration or data reduction scripts and task, we re-processed the data
        in the Herschel Interactive Processing Environment (HIPE) \citep{HIPE2010}. The data were processed with HIPE (version 15) and its corresponding calibration files.
        
        \subsection{PACS photometry}\label{pacs_photometry}
    PACS photometry consists of three color channels, centered at 70, 100, and 160 \textmu m. The data consist 
    of two mini-scan maps, one using the 70 and 160 \textmu m channels at the same time, the other one using
    the 100 and 160 \textmu m channels. For processing, we used the \texttt{scanmap pointsources PhotProject} task.
    This is designed for point-like and small extended sources, using an iterative high-pass filter of the
    timeline of the data to remove the 1/f noise by masking the source.
    
    For the final photometry values of 70 and 100 \textmu m, the two PACS observations were combined. For the
    160 \textmu m channel, we combined the four observations. In cases where no photometry was included in our observation,
    we used existing archival PACS data, which originated from surveys that covered large areas. Those data were already
    processed to a high level, and so  we extracted the photometry without further re-processing.
    We finally performed aperture correction on the
    obtained photometric fluxes (task \texttt{photApertureCorrectionPointSource}), where we used source aperture
    radii of 12", 12", and 22" for 70, 100, and 160 \textmu m, respectively. The photometry is dominated by its central
    source. In many cases, there is some diffuse emission, which we  tried to take into account for the
    background subtraction.

    \subsection{PACS spectroscopy}
    The spectra of the PACS instrument are split into three color channels,
    internally called ``blue'' (55-72 \textmu m), ``green'' (70-95 \textmu m), and ``red'' (103-190 \textmu m) with resolving power 
    \begin{math} R = \lambda/\Delta \lambda \end{math} between 1000 and 4000, depending on wavelength. The beam size is ~9" for
    the blue and green  channels and 10-13" for the red channel. The detector array consists of a grid of 5 $\times$ 5 spatial pixels (spaxels) 9.4" $\times$ 9.4" each,  assembled into an approximately 47" $\times$ 47" area.
        The PACS spectra were re-reduced and background-subtracted using standard procedures in the HIPE software. The data cover
        the full range with the best sensitivity to continuum.
        We extracted the spectra of the source from the resulting data using the \texttt{9to1} task in the pipeline, which  performs
        a background subtraction to the spectrum, using the outermost nine detectors of the array, to avoid any contamination.
        In addition, a correction for the part of the PSF which hits the outer eight detectors is applied. The task combines the
        advantages of providing the best S/N and high robustness.
        We  discarded spectra (or subsets of them) for some sources due to low S/N, which are discussed in detail
        in Sect. \ref{subsection:SEDs_and_photometry}.
        Due to overlap of the different orders of the spectra we had some redundancy and could still obtain a full SED of the respective
        wavelength range. We made use of the PACS photometry to adjust the fluxes of the PACS spectra in order to avoid jumps in the
        final SED. The PACS spectra usually matched relatively well the available photometry (with maximum multiplication factors of 
         1.22, compared to 1.57 when using the sum of the nine detectors instead of the 9to1 task),
        indicating that the central source dominated the PACS spectral emission.

    \subsection{SPIRE photometry}\label{spire_photometry}
    Similar to the PACS data, SPIRE provides three color channels for photometry, internally called ``long'' (500 \textmu m),
    ``medium'' (350 \textmu m), and ``short'' (250 \textmu m). The exact relative spectral response curve can be found in the
    instrument documentation.\footnote{\url{http://herschel.esac.esa.int/Docs/SPIRE/html/spire_om.html\#x1-780005.2.1}}
    For SPIRE, we  used observations in the ``Small Map'' mode. In the cases when archival data of SPIRE were available, we used those
    observations without further processing and directly obtained the aperture photometry from the respective target. Most of them
    originate from a survey that covers a large area where we found the results from the automatic pipeline sufficient for our purposes.
    We also found it beneficial to check the
    wider vicinity of a target for diffuse emission.
    The aperture photometry was corrected with a correction
    factor, typically around 1.27, depending on the  wavelength of the observation and the slope of the SED (task
    \texttt{photApertureCorrectionPointSource}).
    For the wavelength
    windows centered on 250, 350, and 500 \textmu m the  aperture radius is 22", 30", and 42", respectively.\footnote{\url{https://nhscsci.ipac.caltech.edu/workshop/Workshop_Oct2014/Photometry/SPIRE/SPIRE_DP_Oct2014_PhotPhotometry.pptx.pdf}}
        
        \subsection{SPIRE spectroscopy}
        The spectra are separated in two channels,
        SSW covering 191-318 \textmu m and SLW covering 294-671 \textmu m. The spectral resolution ranges from ~40 to 1000 at
        250 \textmu m and is wavelength dependent: ($\Delta f = 25$ GHz ) for low-resolution spectra and ( $\Delta f = 1.2$ GHz) for high-resolution
        spectra.
    The observing mode was single pointing with an intermediate spatial sampling (one beam spacing) to better determine  the
    spatial extent for our sources due to large envelopes (up to the field of view of 20"). We obtained 
        both low-resolution (LR) and high-resolution (HR) spectra for our targets (``High+Low'' spectral mode), the LR being
        eventually used for faint sources, while due to insufficient S/N the HR were discarded in those cases. The SPIRE spectroscopic data also allowed us to obtain a coarse 
    image of the direct neighborhood  of the main source, allowing us to determine whether the line emission was extended, for example.
        The beam size  full width at half maximum (FWHM) for the SPIRE spectrometer is 16.5-20.5" for the SSW and 31.0-42.8" for the
        SLW channel. For the background subtraction, we used the annular fitting of the background around the source. In an
        ideal case this is able to reduce all background emission. However, problems in the archival data cubes caused gaps in the spectra
        for several wavelength ranges which complicated the correction in some cases and caused strong fluctuations for the upper
        limit of the spectra, due to suboptimal background fitting.
        The     specific parameters vary between the different sources because they were individually adapted to avoid any
        contamination of the background fitting area with the signal of another source in the field of view.
        Similar to the PACS data, the SPIRE spectra were finally  adjusted with a multiplicative factor to match the photometric values. 
        We note that for the figures we show the apodized version of the spectra (which eliminates the fringes in the spectra
        introduced by the Fourier
        transform spectrometer technique, but which does not conserve the line fluxes), while the line analysis used the unapodized spectra.

    \subsection{\textit{Spitzer} spectra}
We complemented the data analysis with \textit{Spitzer} spectroscopic data from program 50654 (PI: M. Audard, 2010). The data were processed with the software S18.7.0.  
    We started from the post-BCD files, using  the Spitzer IRS Custom Extraction software (SPICE) for the data extraction.  Dedicated astronomical observation requests (AORs) in the program were used to remove the background emission
    for the short-high (SH) and long-high (LH) spectra, while the background was directly determined near the source in the short-low (SL) and
    long-low (LL) data. 
    We also corrected for defringing in the spectra using the IDL tool SMART available for IRS data (\citealt{2010PASP..122..231L}, \citealt{2004PASP..116..975H}).
    The observations usually were done with one single cycle, and ramp durations of 6~s, with a few exceptions where deeper exposures or more numerous cycles 
    were used to enhance the signal-to-noise ratio. Furthermore, not all targets could be observed with the low-resolution mode due to their brightness. 
        
        \subsection{SEDs and lines}
        The different instruments provide  different angular resolutions over their wavelength ranges. Thus, the spectra require
        calibration to remove jumps in the continuum levels of the SEDs. For each of the \textit{Herschel} instruments we used the corresponding
        photometry to determine the absolute flux at the respective wavelength range. While leaving the photometry after applying
        all correction as is, we shifted the absolute value of the spectra to obtain a continuous SED, taking into account the error
        bars of the photometry. For the \textit{Spitzer} spectra, we applied multiplicative factors (typically on the order of 5-20\%, depending on the module)  to stitch the spectra together, using the high-resolution long (LH) module as the basis.
        We also checked for agreement between published photometry and the overall SED based on the \textit{Spitzer} and \textit{Herschel} data. However, since the photometric data are not simultaneous with our \textit{Spitzer} or \textit{Herschel} data, and since FUors and EXors are variable sources, we discarded discrepant published photometric data for clarity (except for HH 381 IRS, see below).

\begin{figure*}
    \includegraphics[width=0.5\textwidth, trim={0cm 0cm 0cm 0}]{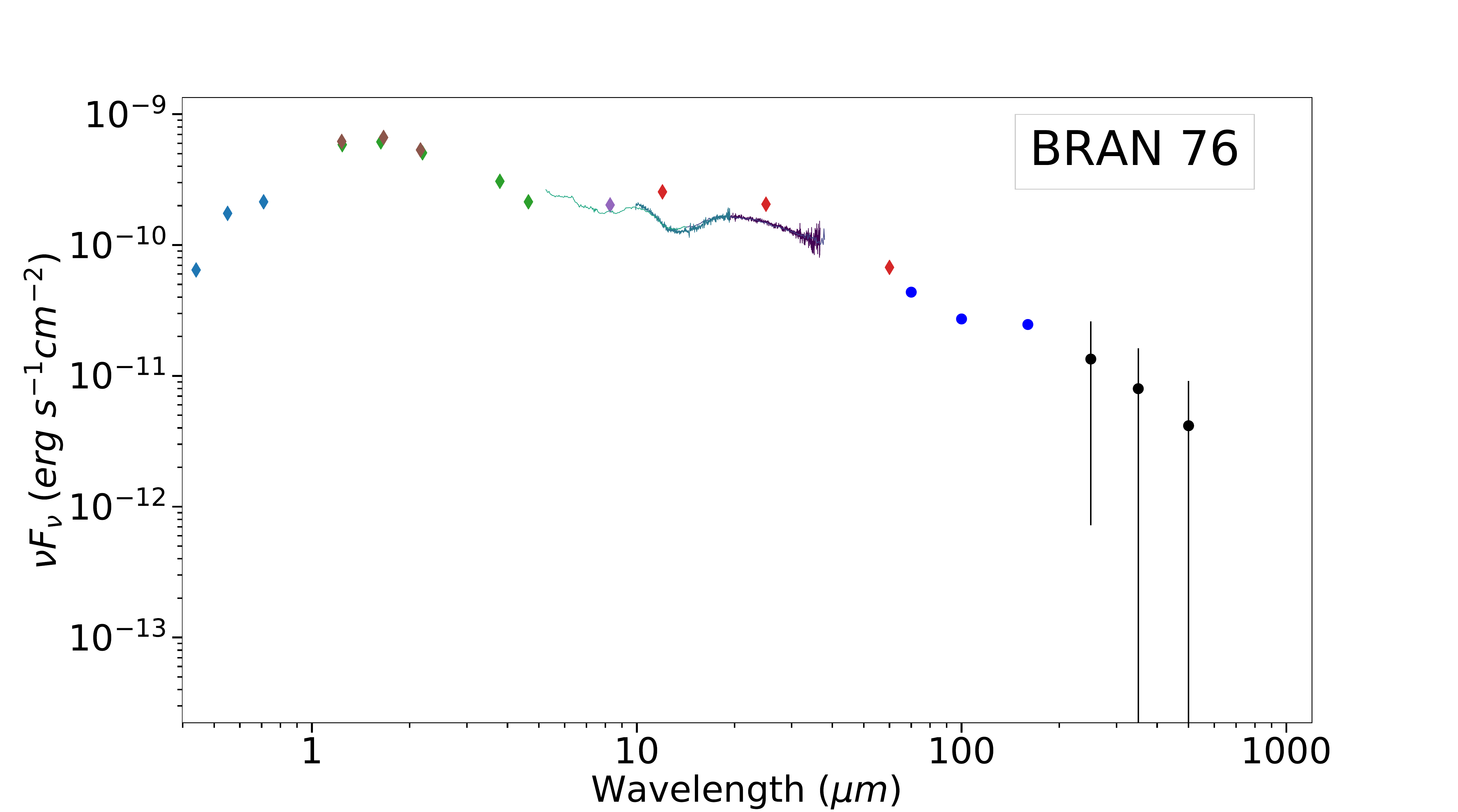}
    \includegraphics[width=0.5\textwidth, trim={0cm 0cm 0cm 0}]{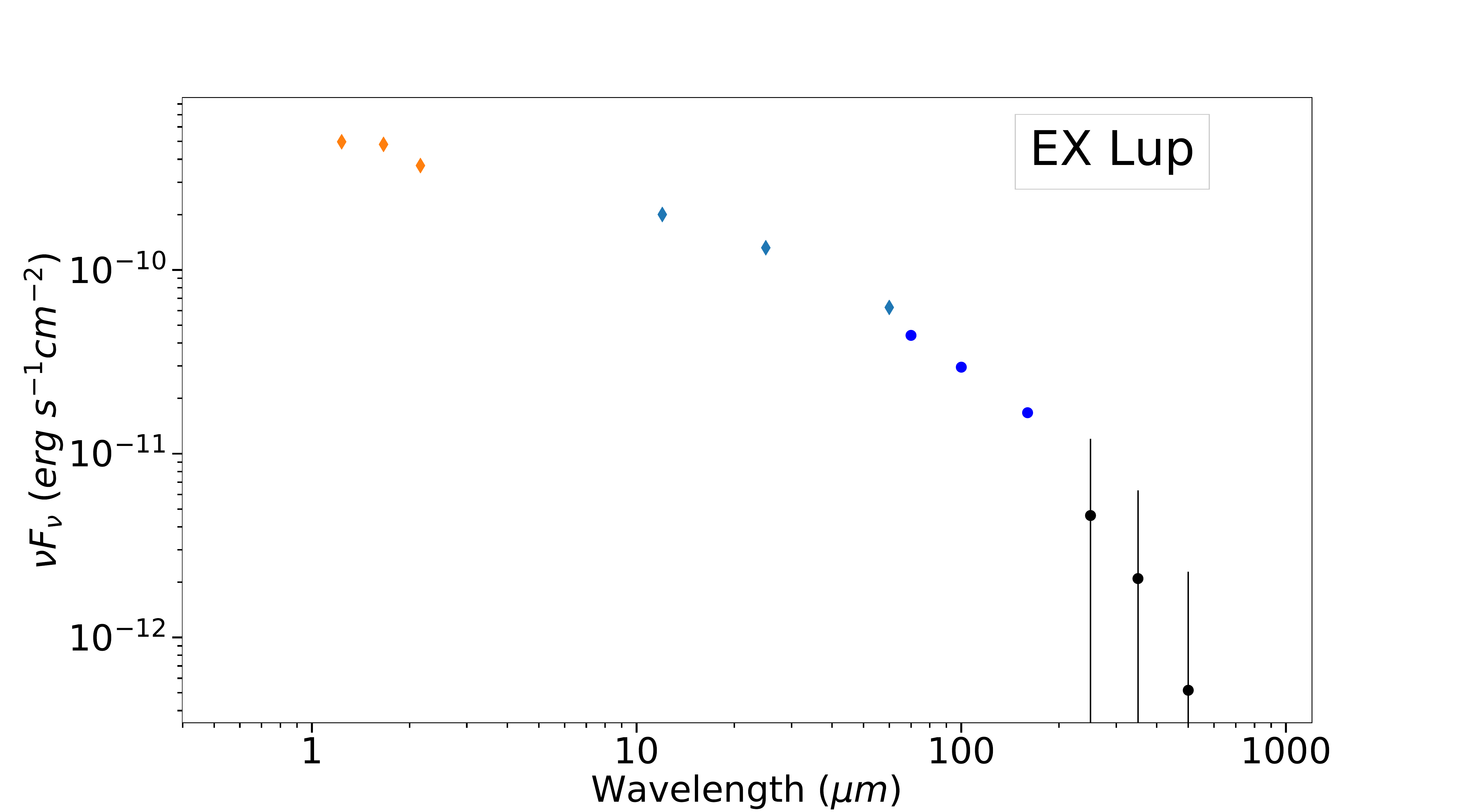} \\
    \includegraphics[width=0.5\textwidth, trim={0cm 0cm 0cm 0}]{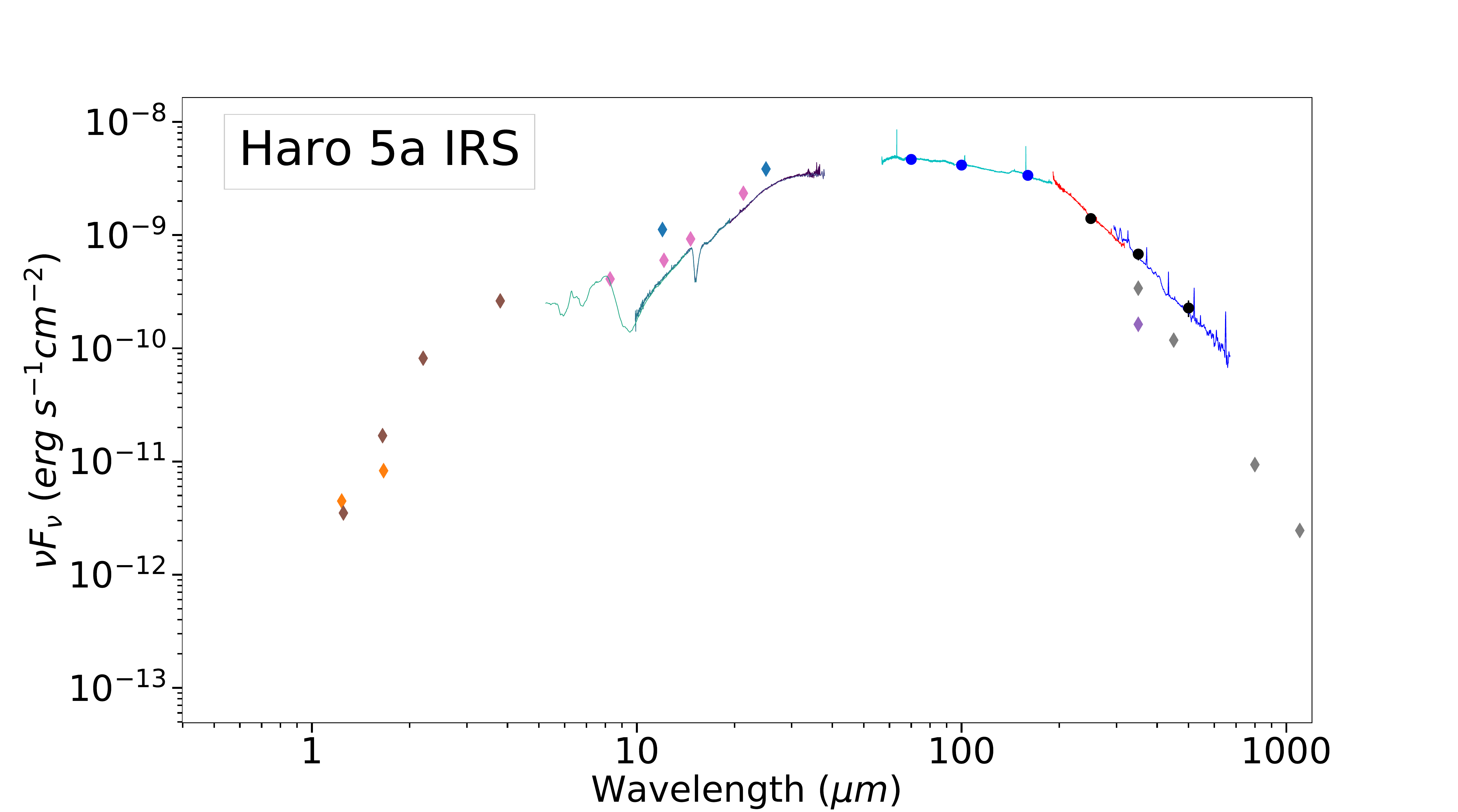}
    \includegraphics[width=0.5\textwidth, trim={0cm 0cm 0cm 0}]{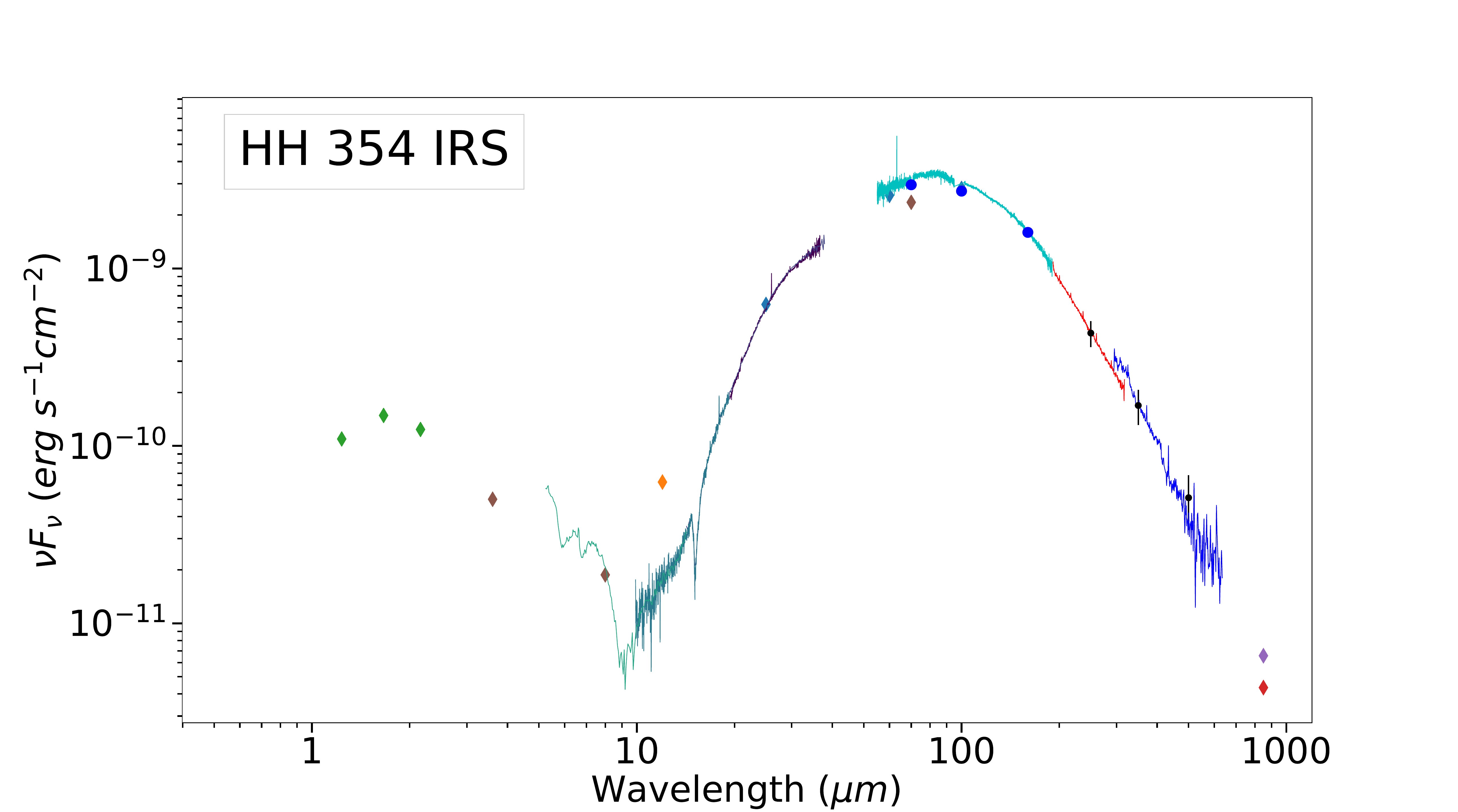} \\
    \includegraphics[width=0.5\textwidth, trim={0cm 0cm 0cm 0}]{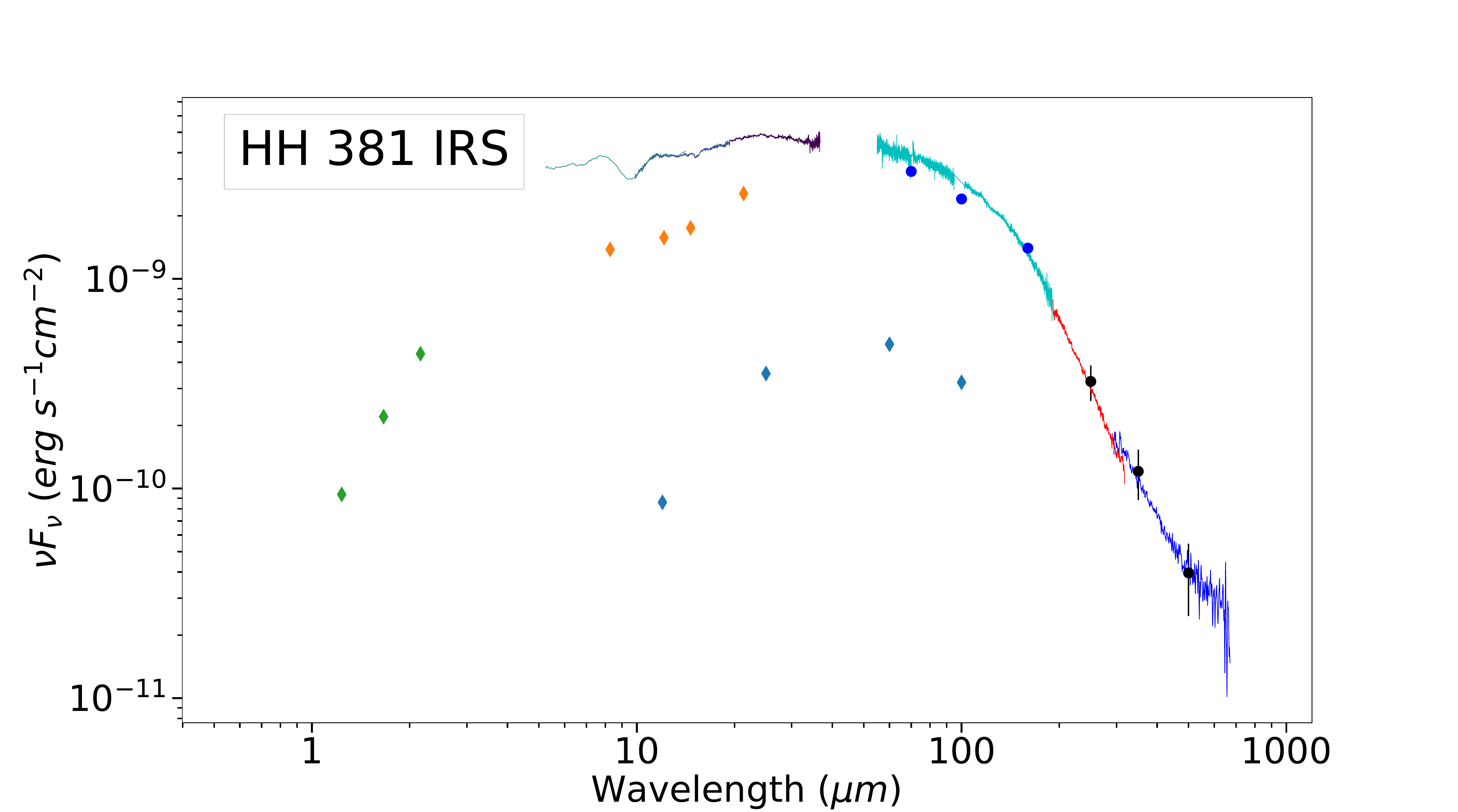}
    \includegraphics[width=0.5\textwidth, trim={0cm 0cm 0cm 0}]{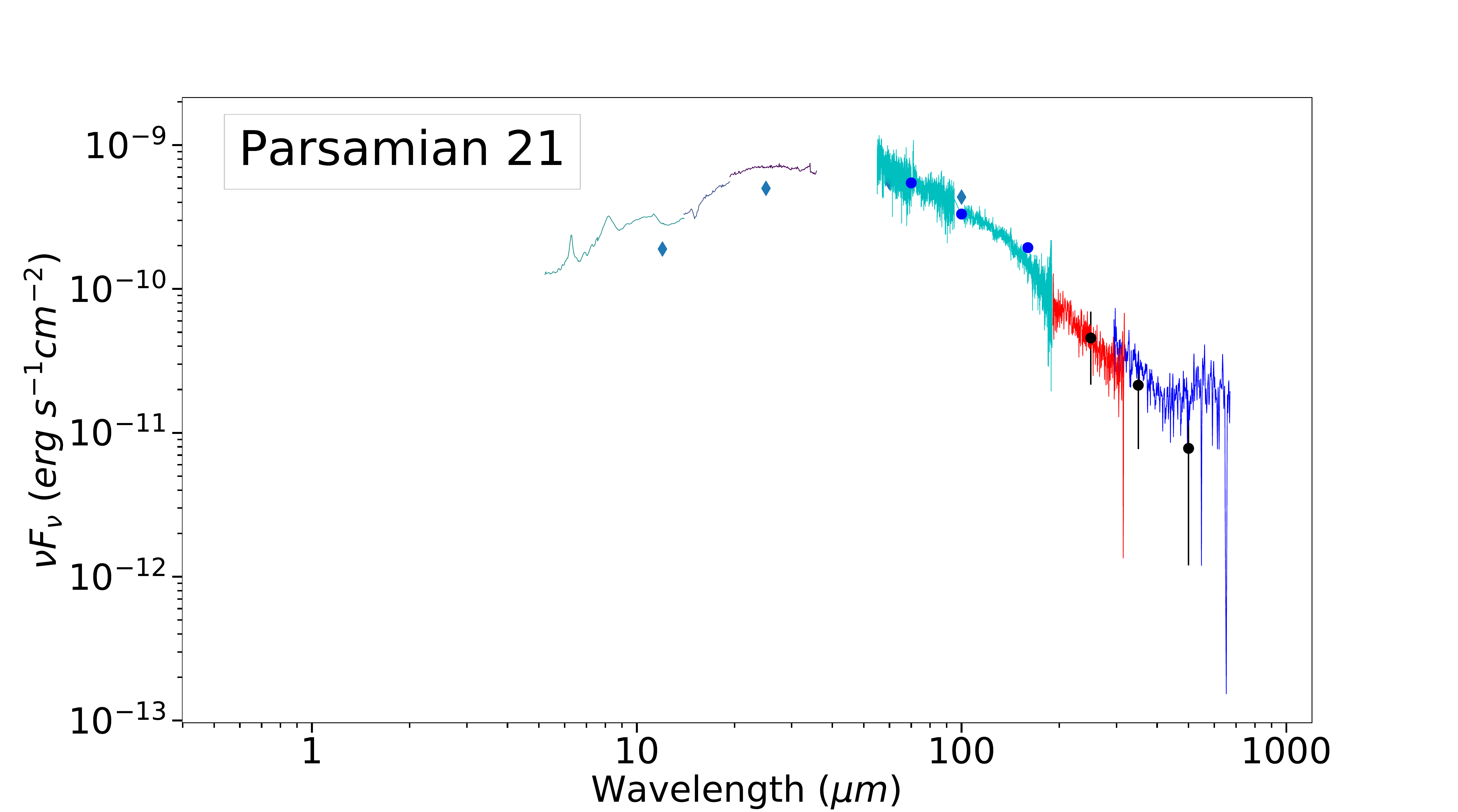}
        \caption{
                \footnotesize
                Spectral energy distributions of the objects BRAN 76 (top left), EX Lupi (top right), Haro 5a IRS (middle left),
                HH 354 IRS (middle right), HH 381 IRS (bottom left), and Parsamian 21 (bottom right) including PACS spectra and
                photometry data (turquoise line + blue dots at 70, 100, and 160 \textmu m), SPIRE high-resolution spectra (dark blue and red lines),
                and SPIRE photometry (black dots at 250, 350, and 500 \textmu m) combined with spectra from \textit{Spitzer} (lines in other colors below 50
                \textmu m) and photometry 
from other sources (diamonds; color-coded by  source).
        }
        \label{SED_multiplot}
\end{figure*}

\begin{figure*}
    \includegraphics[width=0.5\textwidth]{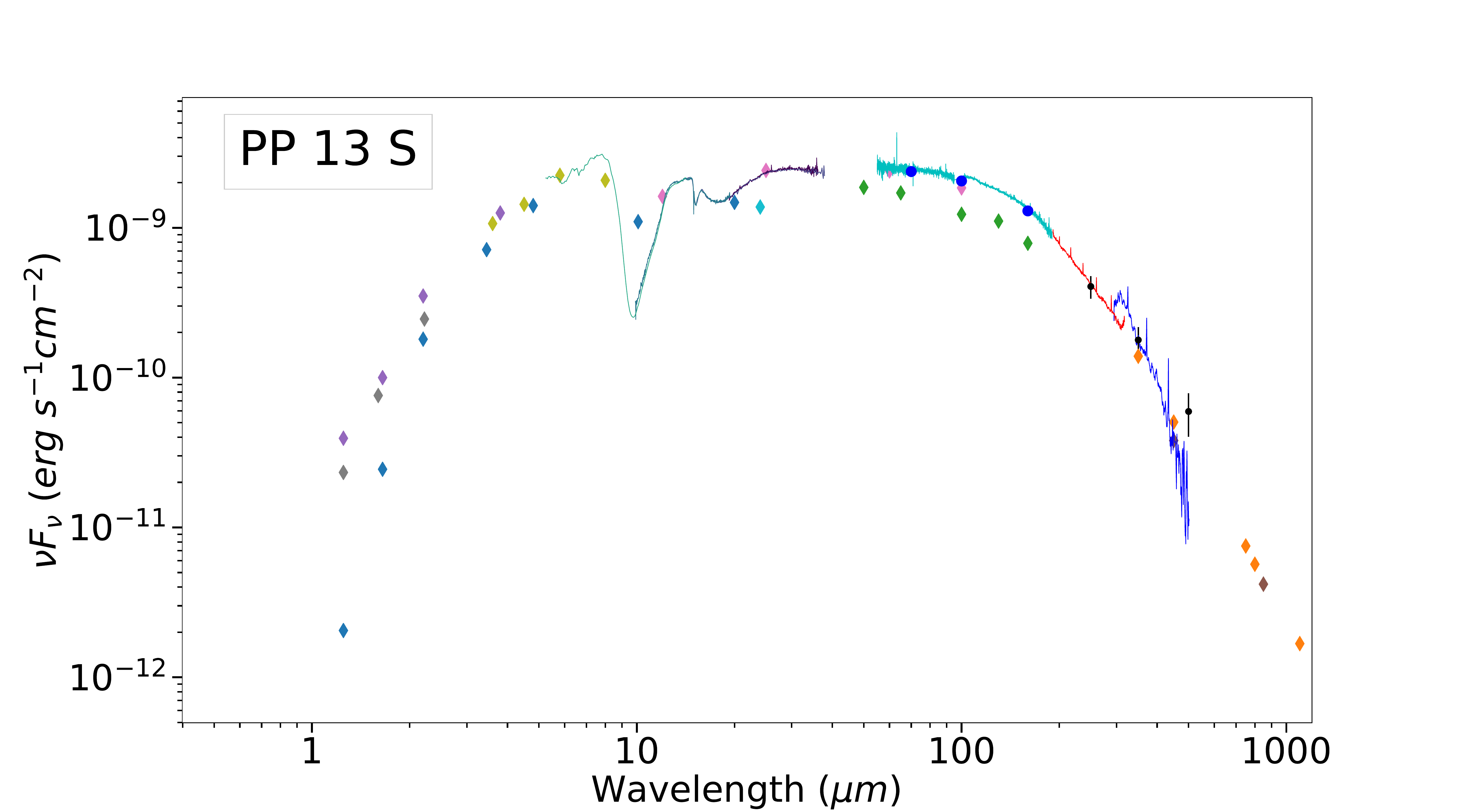}
    \includegraphics[width=0.5\textwidth]{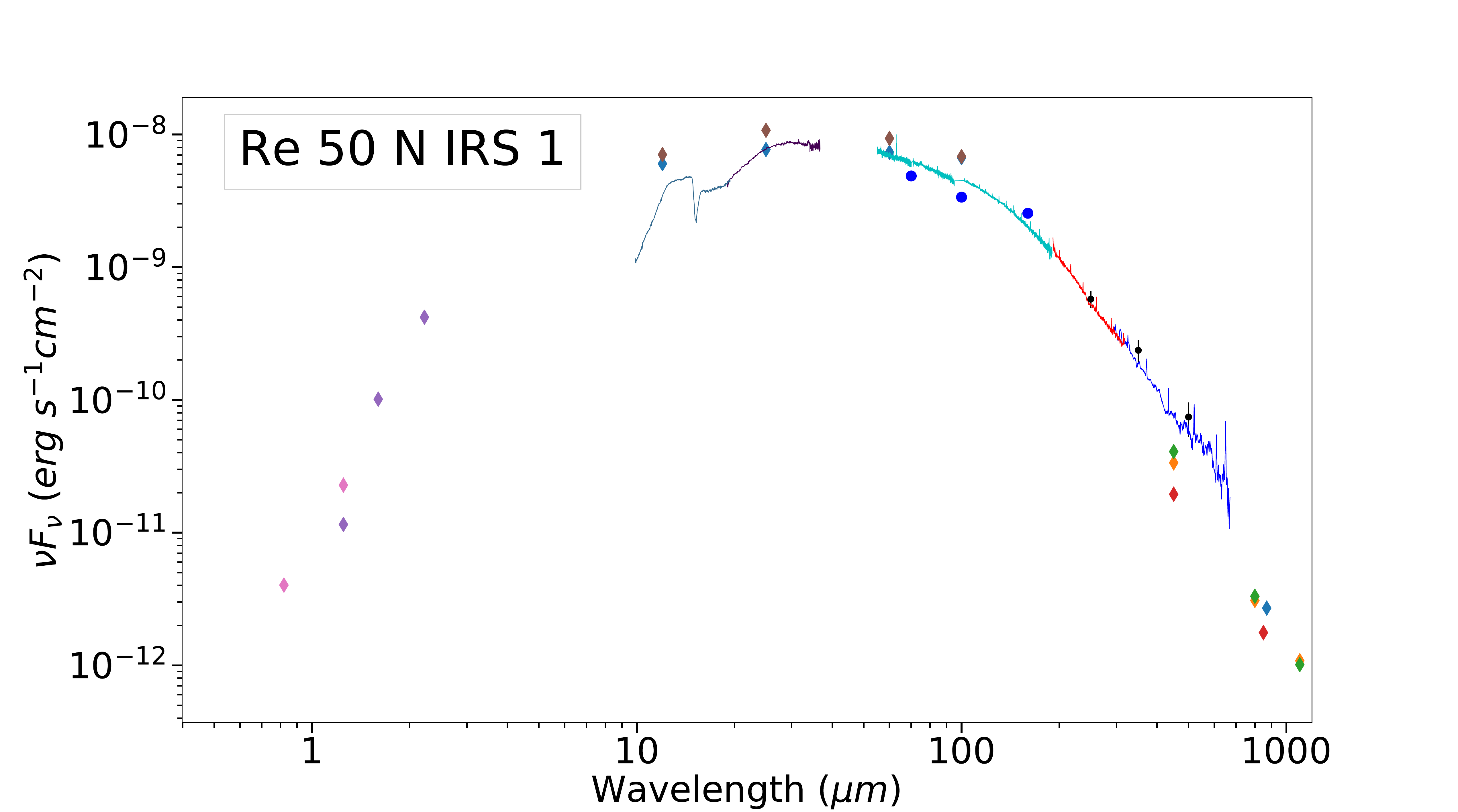} \\
    \includegraphics[width=0.5\textwidth]{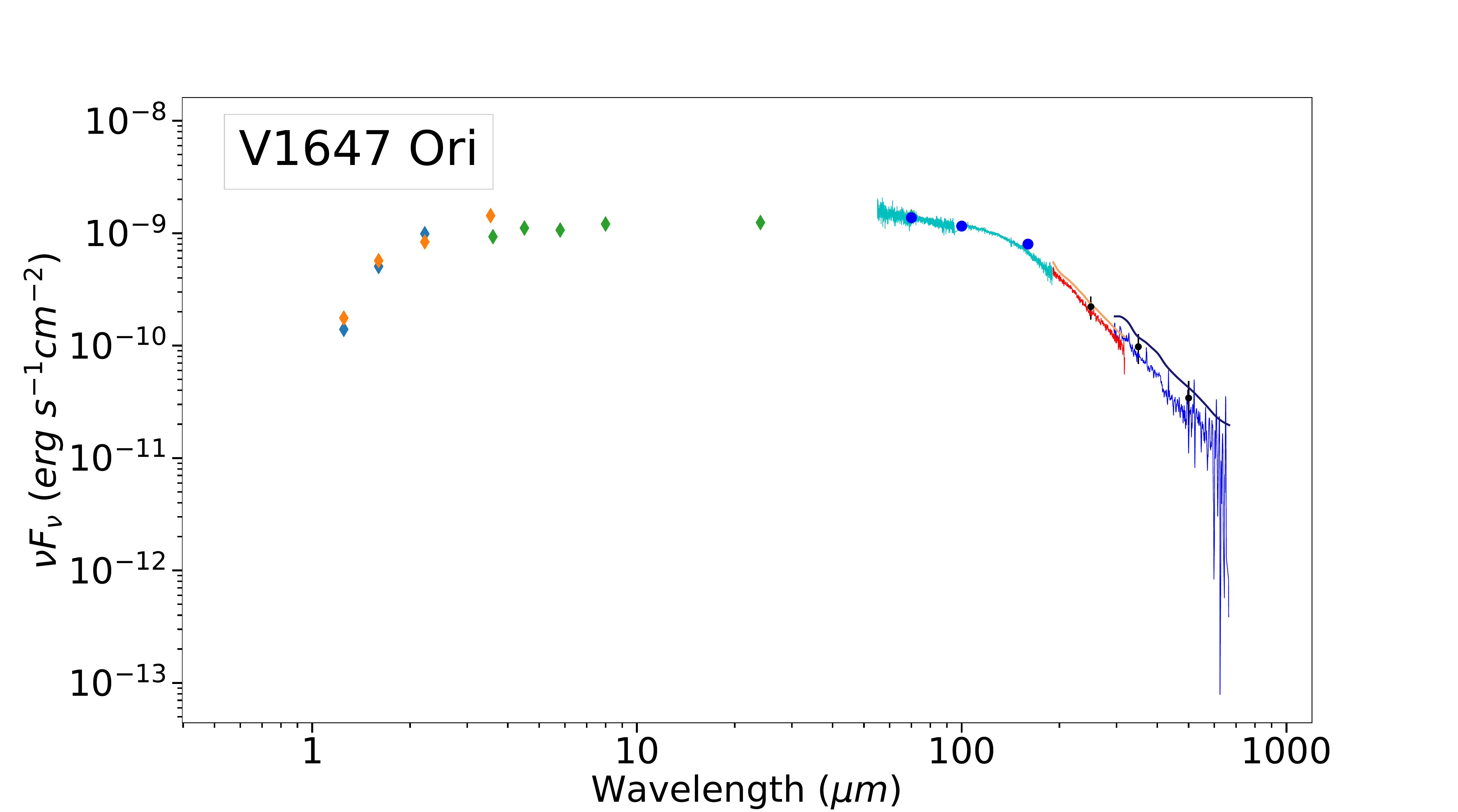}
    \includegraphics[width=0.5\textwidth]{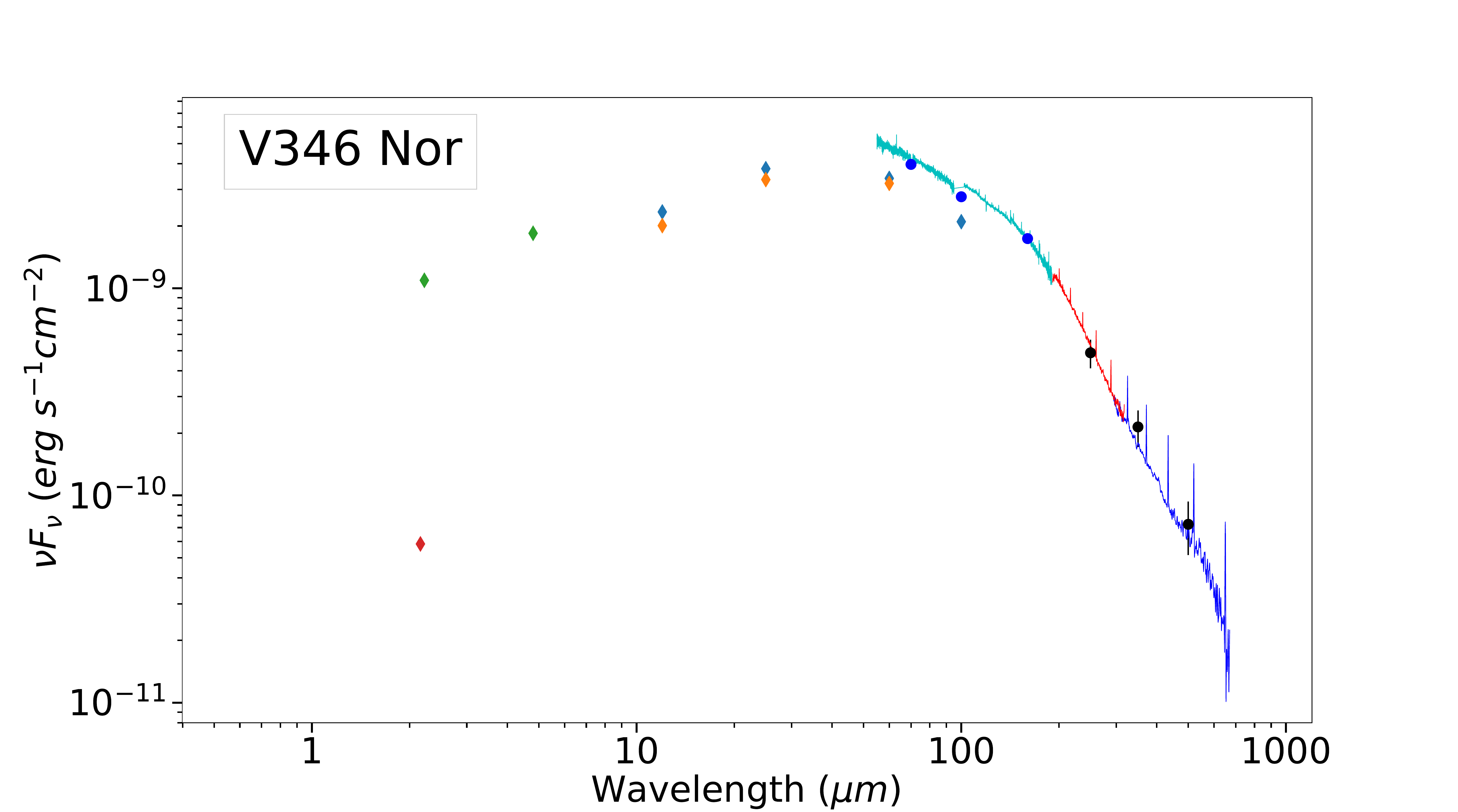} \\
    \includegraphics[width=0.5\textwidth]{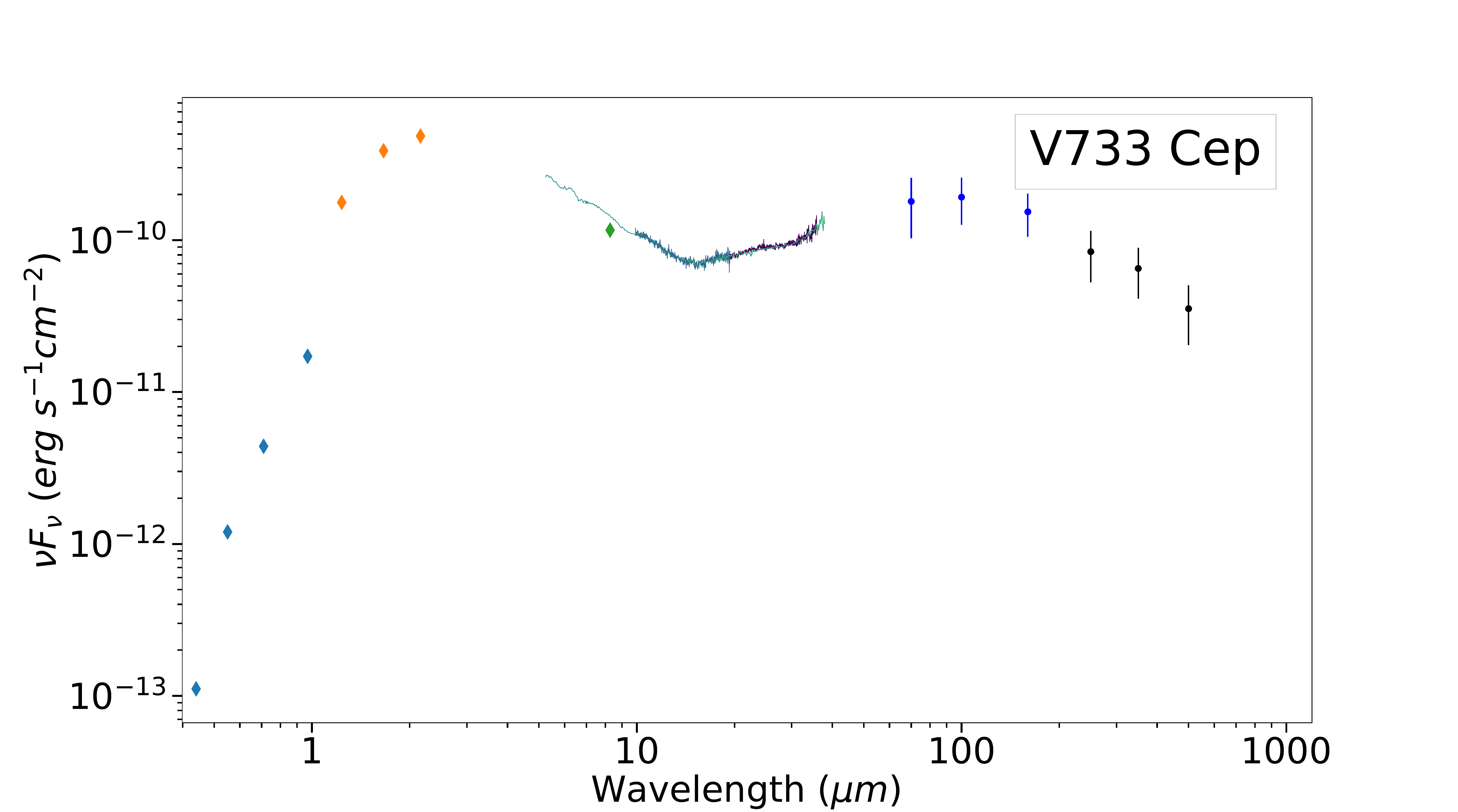}
    \includegraphics[width=0.5\textwidth]{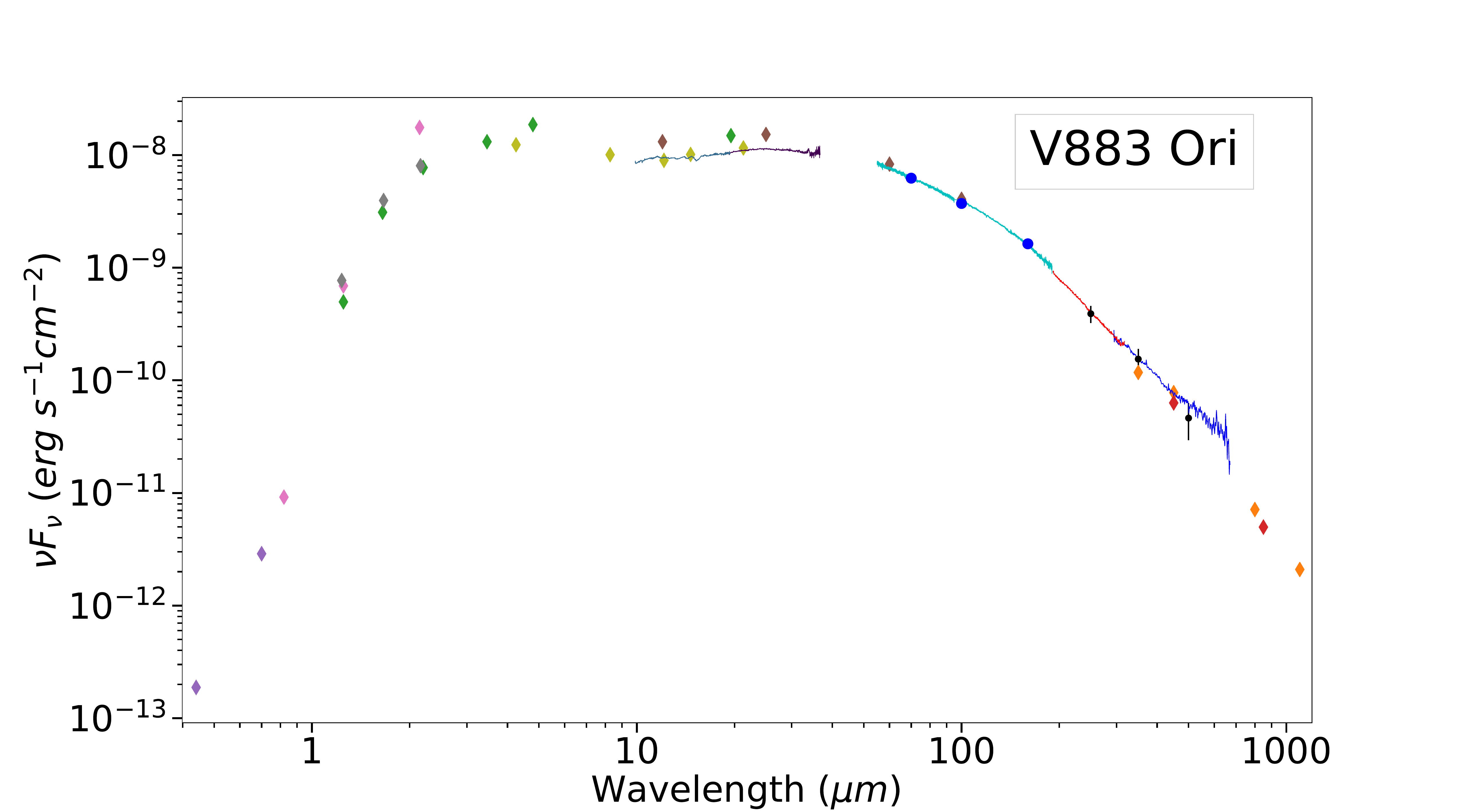}
        \caption{
                \footnotesize
                Spectral energy distributions of the objects PP 13 S (top left), Re 50 N IRS 1 (also known as HBC 494) (top right), V1647 Ori
        (middle left), V346 Nor (middle right), V733 Cep (bottom left), and V883 Ori (bottom right). Data 
        shown  as in Figure \ref{SED_multiplot}. The SED of faint objects (here V1647 Ori and V733 Cep) is complemented
                with low-resolution
                spectra from SPIRE (thick orange and purple lines at the same wavelength window as the high-resolution
                spectra).
        }
        \label{SED_multiplot2}
\end{figure*}

\section{Results}
    \subsection{Spectral properties and surrounding emission} \label{subsection:SEDs_and_photometry}
\begin{figure*}
    \includegraphics[width=0.5\textwidth, trim={0cm -0.5cm 0cm 0}]{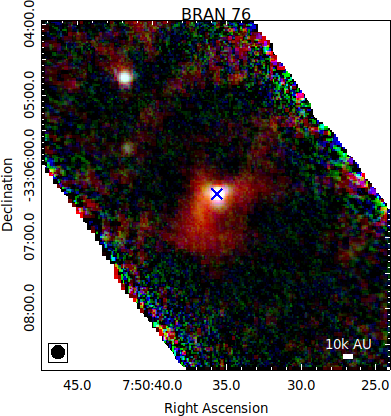}
    \includegraphics[width=0.5\textwidth, trim={0cm -0.5cm 0cm 0}]{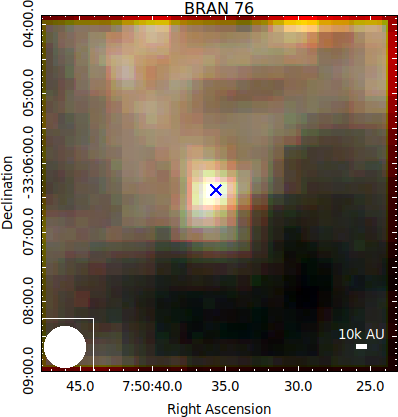}
    \includegraphics[width=0.5\textwidth]{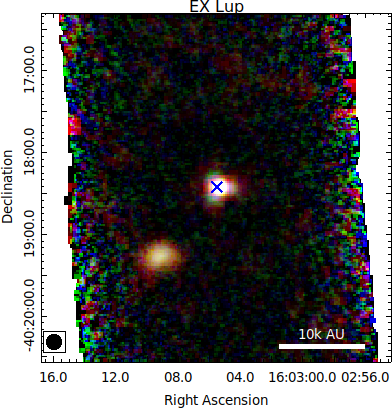}
    \includegraphics[width=0.5\textwidth]{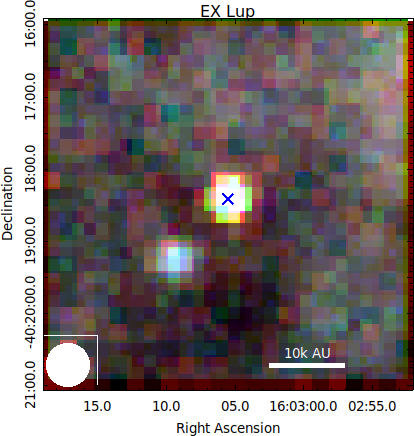}
        \caption{
                \footnotesize
                Three-color composed image of the 70 \textmu m (blue), 100 \textmu m (green), and 160 \textmu m (red) photometry of PACS (left) and 
                250 \textmu m (blue), 350 \textmu m (green), and 500 \textmu m (red) photometry of SPIRE (right) for Bran 76 (top) and EX Lup (bottom). The target coordinates
                of the FUor/EXor are shown with a blue cross. The circle in the lower left corner represents the beam
                for the longest used wavelength of the respective image (11.5" for PACS, 35.95" for SPIRE). For the PACS
                images we used an asinh scaling with 99.5 \%  cutoff but removed values below zero. For SPIRE, we used a linear
                scale with 99.5 \% cutoff.
        } 
        \label{Imaging1_PACS}
\end{figure*}

\begin{figure*}
    \includegraphics[width=0.5\textwidth, trim={0cm -0.5cm 0cm 0}]{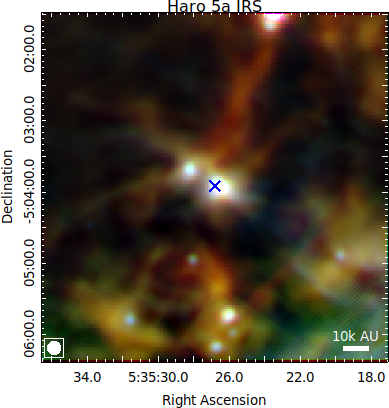}
    \includegraphics[width=0.5\textwidth, trim={0cm -0.5cm 0cm 0}]{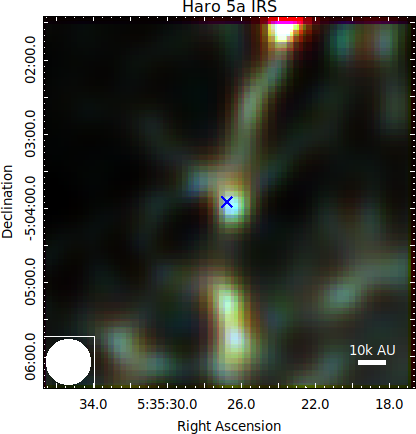}
    \includegraphics[width=0.5\textwidth]{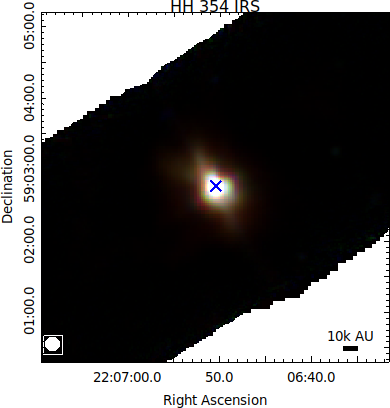}
    \includegraphics[width=0.5\textwidth]{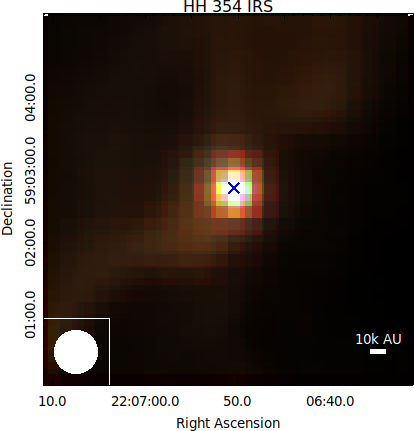}
        \caption{
                \footnotesize
                As in Fig. \ref{Imaging1_PACS}, but for Haro 5a IRS (top) and HH 354 IRS (bottom).
        }
        \label{Imaging2_PACS}
\end{figure*}

\begin{figure*}
    \includegraphics[width=0.5\textwidth, trim={0cm -0.5cm 0cm 0}]{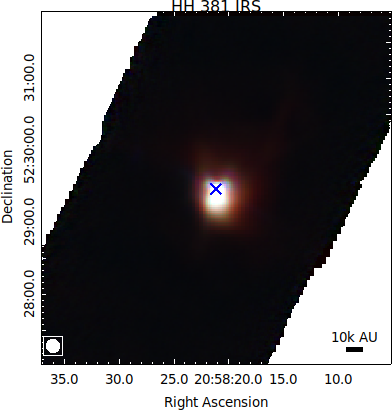}
    \includegraphics[width=0.5\textwidth, trim={0cm -0.5cm 0cm 0}]{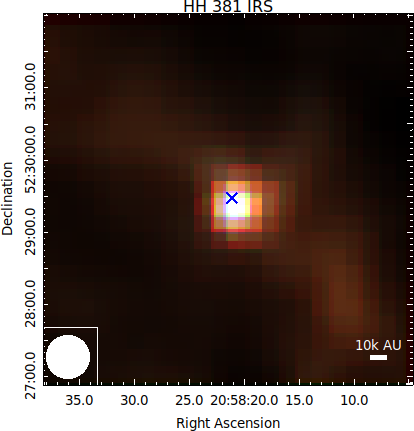}
    \includegraphics[width=0.5\textwidth]{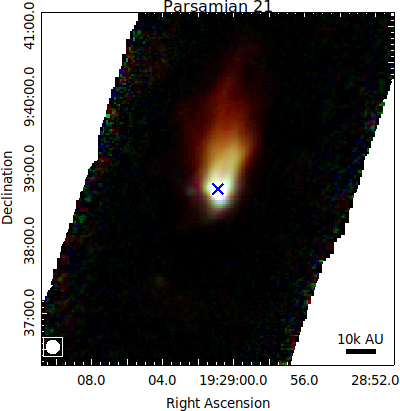}
    \includegraphics[width=0.5\textwidth]{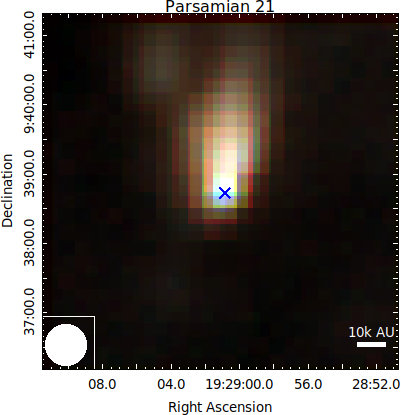}
        \caption{
                \footnotesize
                As in Fig. \ref{Imaging1_PACS}, but for HH 381 IRS (top) and Parsamian 21 (bottom).
        }
        \label{Imaging3_PACS}
\end{figure*}

\begin{figure*}
    \includegraphics[width=0.5\textwidth, trim={0cm -0.5cm 0cm 0}]{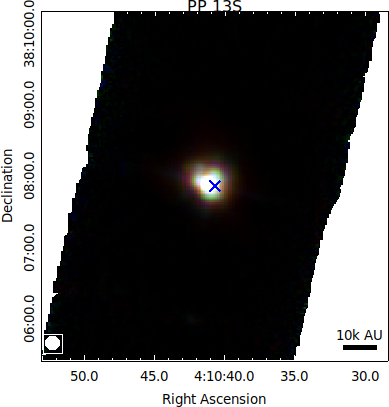}
    \includegraphics[width=0.5\textwidth, trim={0cm -0.5cm 0cm 0}]{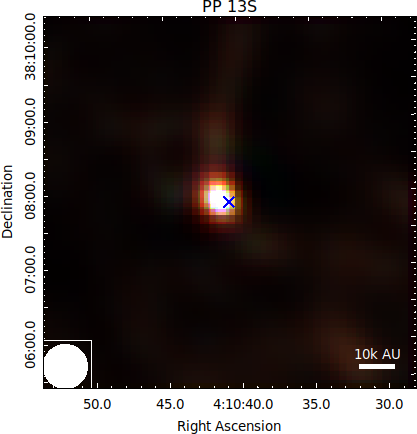}
    \includegraphics[width=0.5\textwidth]{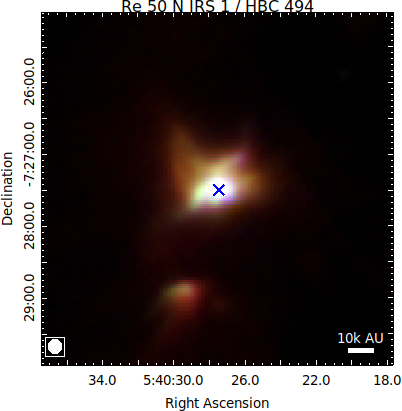}
    \includegraphics[width=0.5\textwidth]{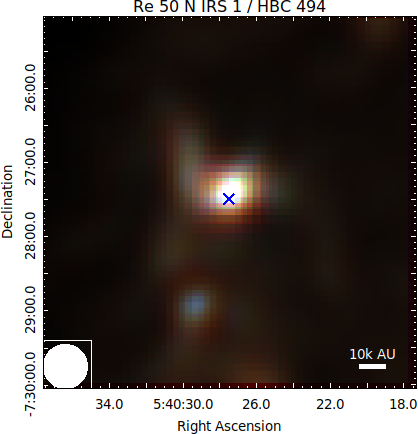}
        \caption{
                \footnotesize
                As in Fig. \ref{Imaging1_PACS}, but for PP 13 S (top) and Re 50 N IRS 1 (bottom).
        }
        \label{Imaging4_PACS}
\end{figure*}

\begin{figure*}
    \includegraphics[width=0.5\textwidth, trim={0cm -0.5cm 0cm 0}]{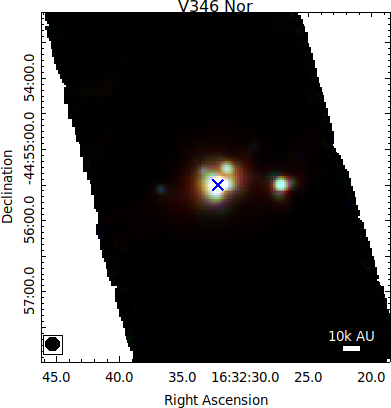}
    \includegraphics[width=0.5\textwidth, trim={0cm -0.5cm 0cm 0}]{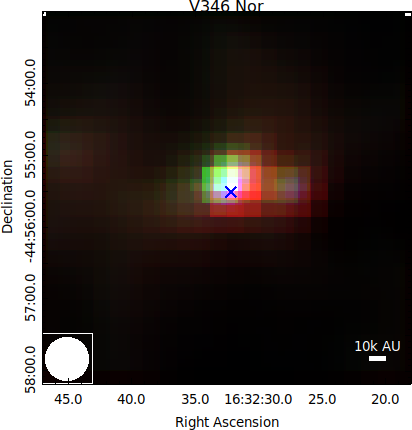}
    \includegraphics[width=0.5\textwidth]{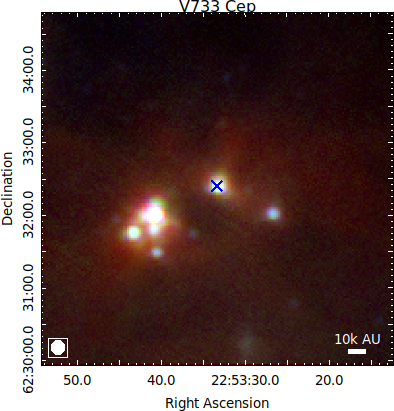}
    \includegraphics[width=0.5\textwidth]{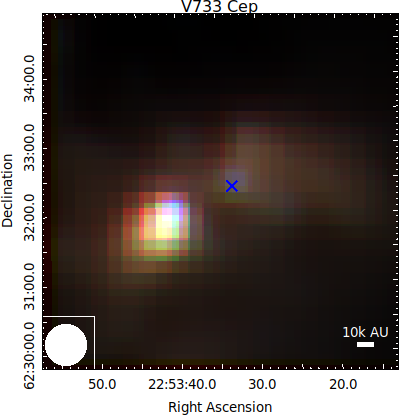}
        \caption{
                \footnotesize
                As in Fig. \ref{Imaging1_PACS}, but for V346 Nor (top) and V733 Cep (bottom).
        }
        \label{Imaging5_PACS}
\end{figure*}

\begin{figure*}
    \includegraphics[width=0.5\textwidth, trim={0cm -1cm 0cm 0}]{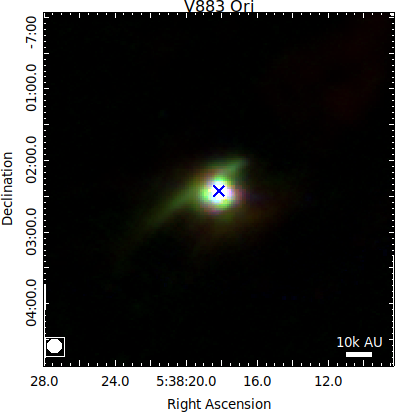}
    \includegraphics[width=0.5\textwidth, trim={0cm -1cm 0cm 0}]{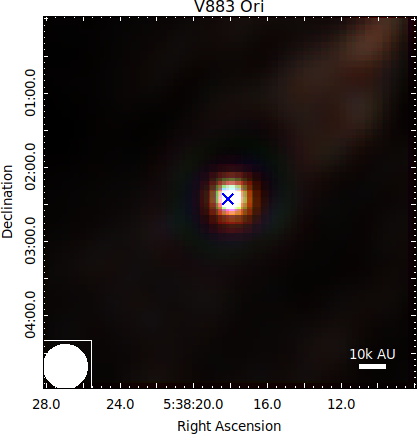}
    \includegraphics[width=0.5\textwidth]{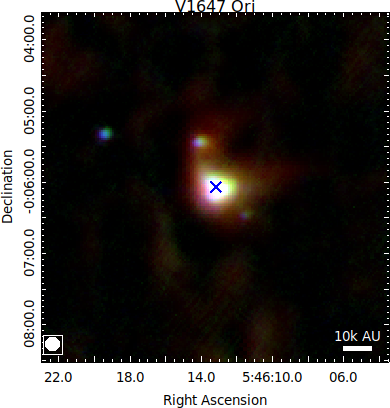}
    \includegraphics[width=0.5\textwidth]{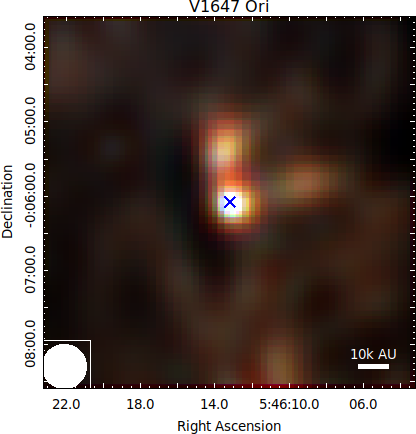}
        \caption{
                \footnotesize
                As in Fig. \ref{Imaging1_PACS}, but for V883 Ori (top) and V1647 Ori (bottom).
        }
        \label{Imaging6_PACS}
\end{figure*}

        The \textit{Herschel} flux densities based on photometry of all twelve sources can be found in Table \ref{photometry_flux_table}.
        The images we used to obtain the photometry are shown in Fig. \ref{Imaging1_PACS} - Fig. \ref{Imaging6_PACS} as combined three-color
        images of PACS and SPIRE with a size of $4.8' \times 4.8'$,
        centered on the target location. The measured values are provided for
        all six color channels of PACS and SPIRE separately. The extraction procedure and chosen parameters are explained in Sects. \ref{pacs_photometry}
        and \ref{spire_photometry}. Although we corrected the values
        for diffuse emission, some   minor contamination might remain depending on the individual shape of the target.
                
\begin{figure*}
        \includegraphics[width=1\textwidth]{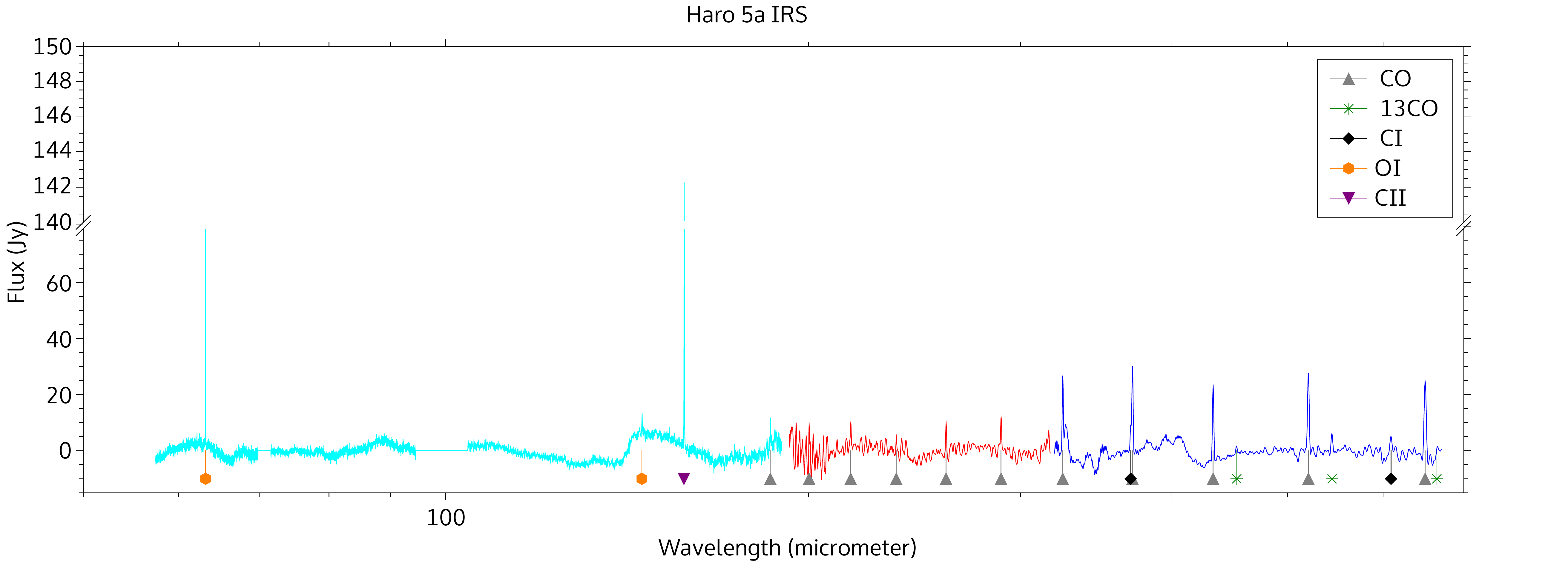}
        \includegraphics[width=1\textwidth]{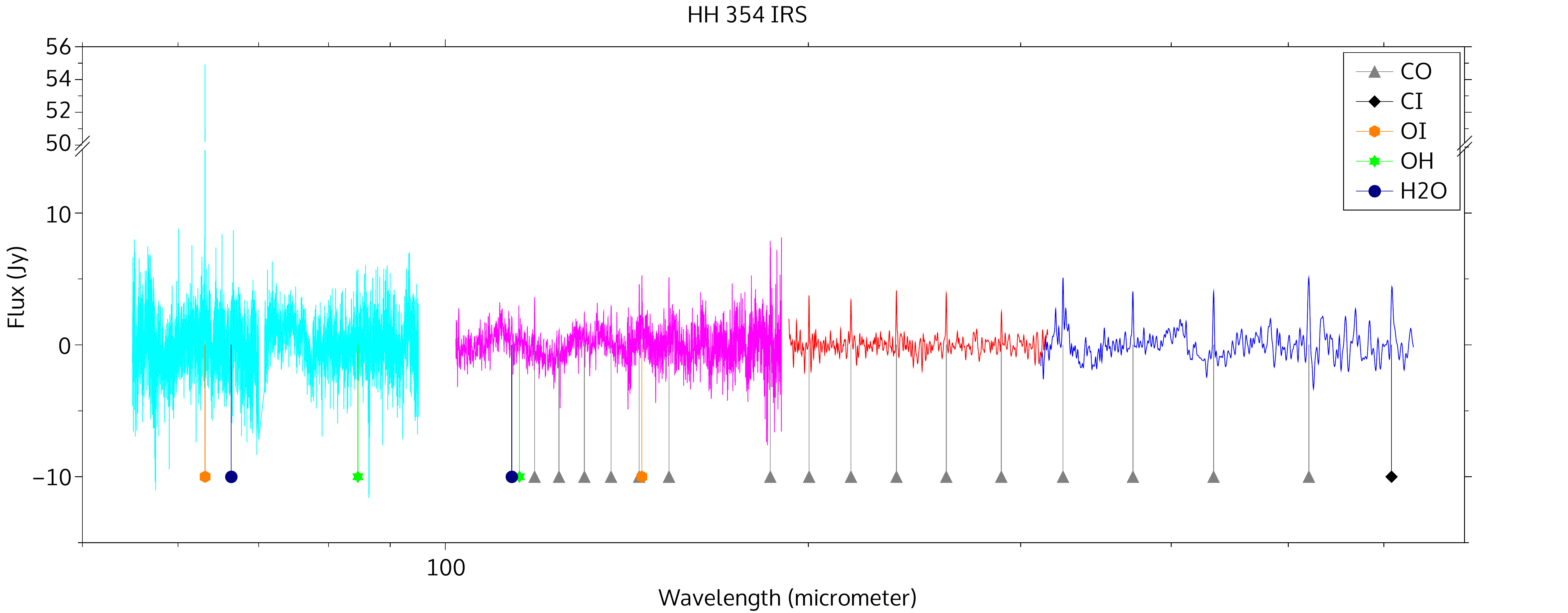}
        \includegraphics[width=1\textwidth]{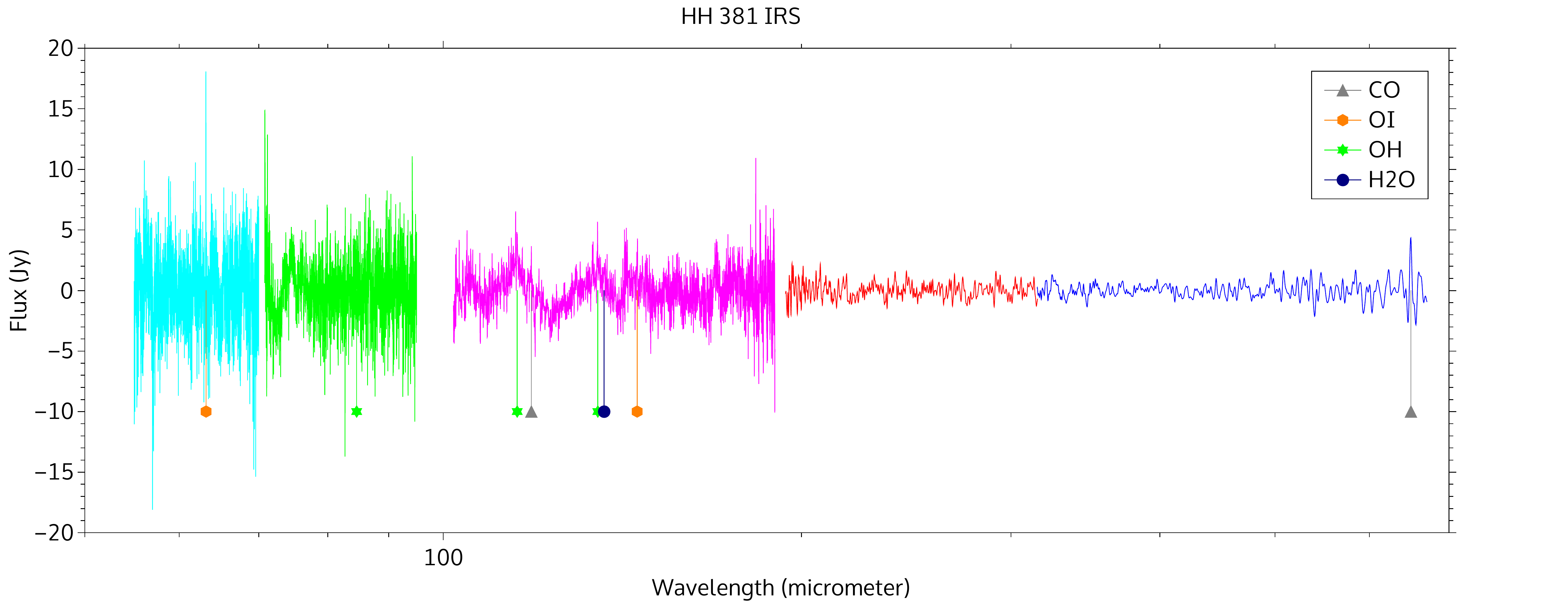}
        \caption{
                \footnotesize
                Continuum-subtracted spectra of PACS and SPIRE for Haro 5a IRS, HH 354 IRS, and HH 381 IRS. For the indicators of the lines, we
                use a $1.5-\sigma$ lower limit.
        \label{PACS_multiplot}
        }
\end{figure*}

\begin{figure*}
        \includegraphics[width=1\textwidth]{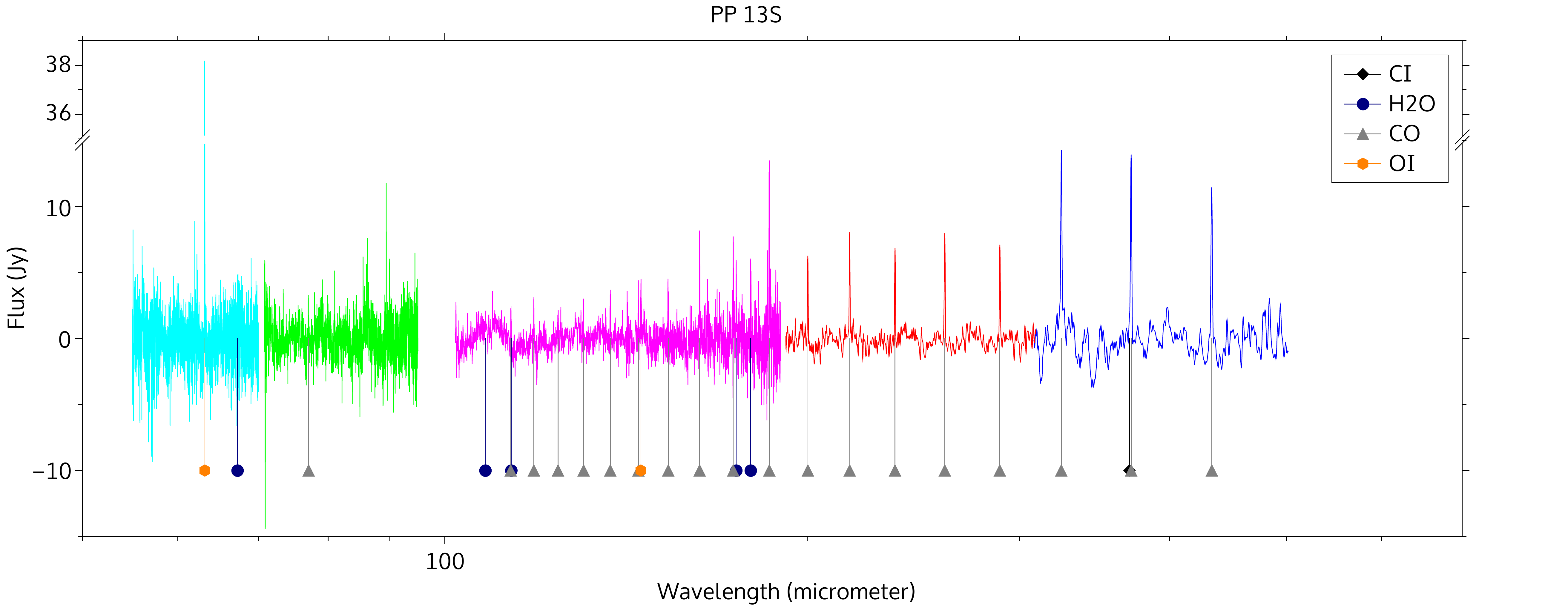}
        \includegraphics[width=1\textwidth]{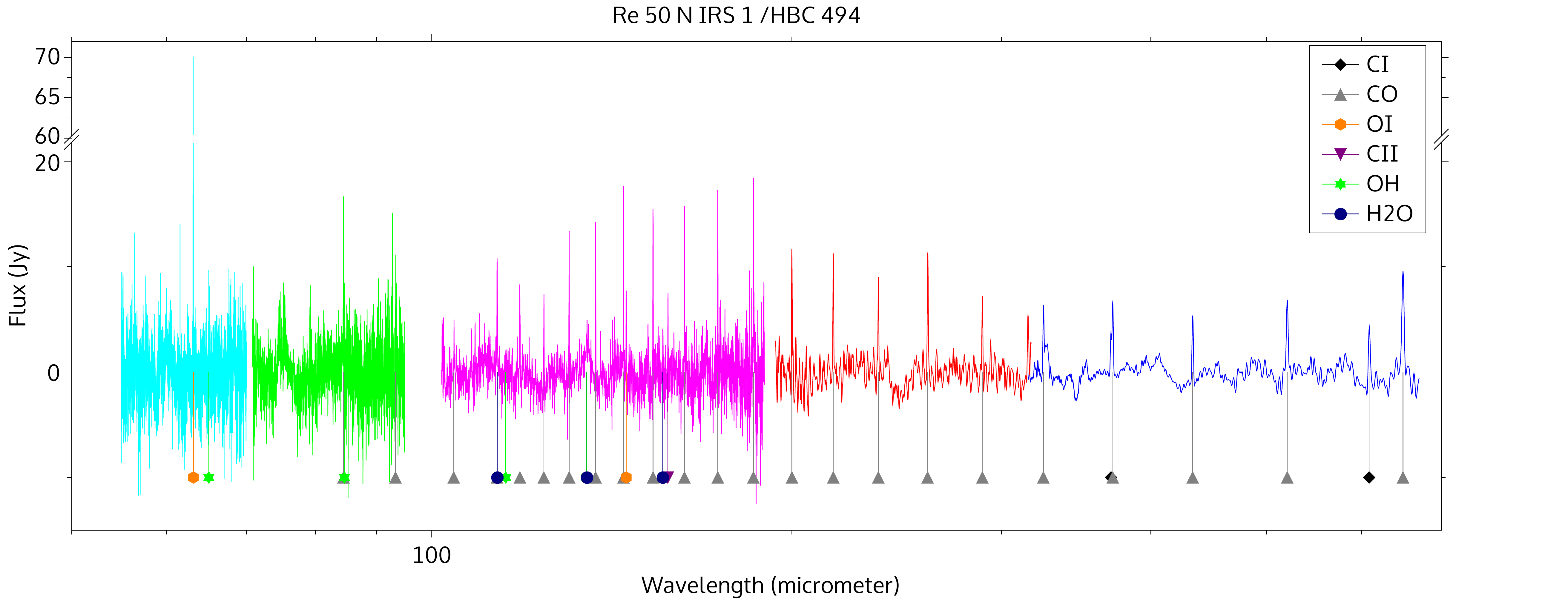}
        \includegraphics[width=1\textwidth]{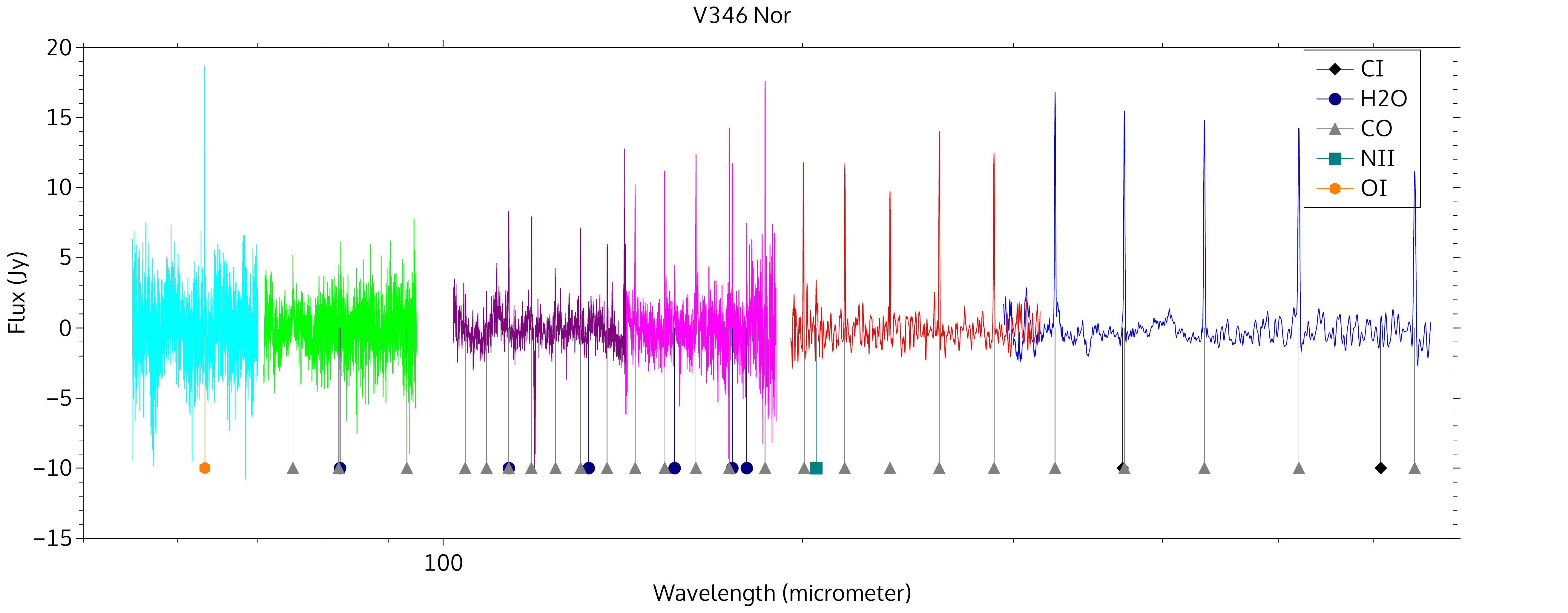}
        \caption{
                \footnotesize
                Continuum-subtracted spectra of PACS and SPIRE for PP 13S, Re 50 N IRS / HBC 494, and V346 Nor.
        }
        \label{PACS_multiplot2}
\end{figure*}

\begin{figure*}
        \includegraphics[width=1\textwidth]{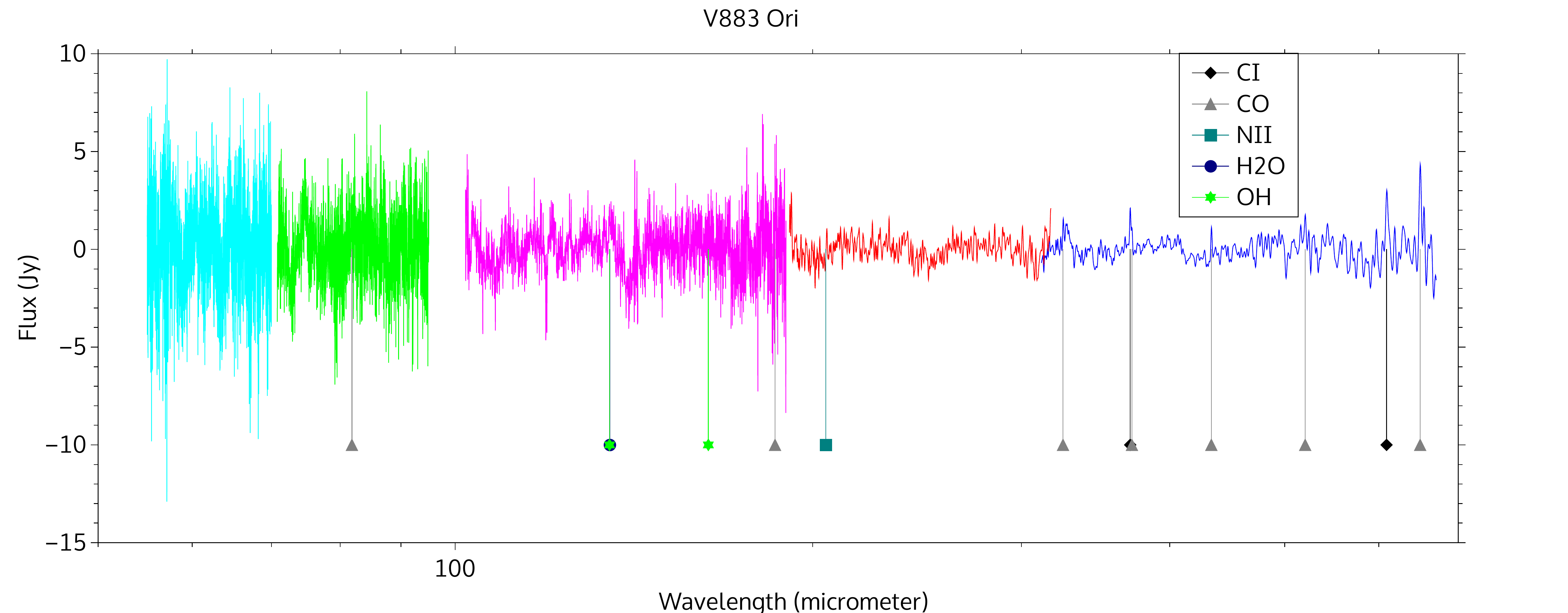}
        \includegraphics[width=1\textwidth]{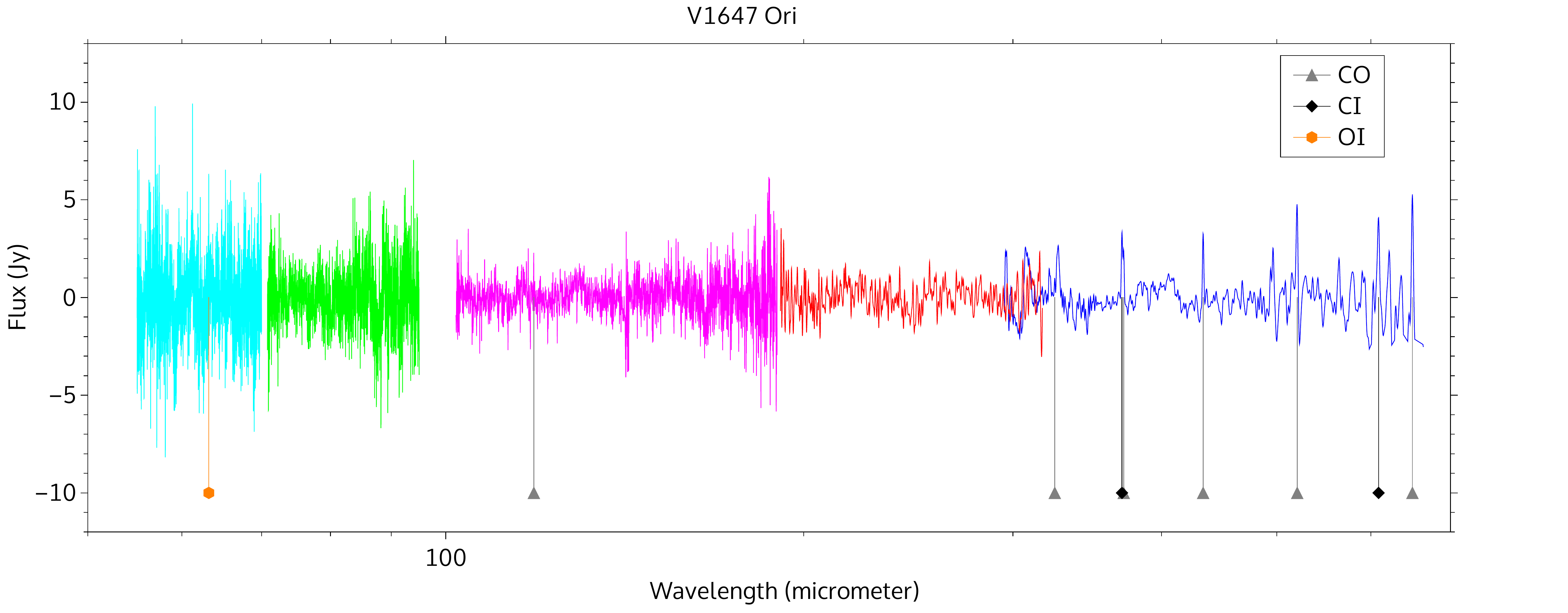}
        \caption{
                \footnotesize
                Continuum-subtracted Spectra of PACS and SPIRE for V883 Ori and V1647 Ori.
        }
        \label{PACS_multiplot3}
\end{figure*}

            Analyzing the spectra, we identified for some of the targets emission lines longward of 100 \textmu m. These lines mainly originate from CO and
                $^{13}$CO emission, but we also detected lines of \mbox{[C I]} ($^{3}P_{2} - ^{3}P_{1},^{3}P_{1} - ^{3}P_{0}$) and \mbox{[C II]} ($^{2}P_{3/2}-^{2}P_{1/2}$),
                both mainly coming from the diffuse emission from the star-forming region.
        More detailed information about the individual line detection and strength
            for the respective targets can be found in Appendix \ref{line_flux_tables}. For the PACS lines we combined the central $3\times3$ grid
            to cover an area comparable to that  for the SPIRE lines. For SPIRE only the spaxel with the coordinates of the object is used. We note here that
            the line fitting was not done on
            the spectra that are shown in the SEDs and the continuum-subtracted spectra (Figs.~\ref{PACS_multiplot}-\ref{PACS_multiplot3}), but directly on
            the data cubes that are provided by the integral field spectroscopy,
            and are therefore more robust to noise.
            The line maps (created by fitting Gaussians to the spectrum of each spaxel after removing a first-order polynomial baseline)
            that we refer to in the discussion of each source are shown in Appendix \ref{line_maps}. The maps are only shown if there is a $3\sigma$
            detection at the source coordinates, which is approximately what other groups used to avoid false detections (S/N ~3-4, see \citealt{Yang_2018}). For the respective pixels in the maps, we follow the nomenclature by \textit{Herschel}, and call them spaxels.\newline
            Due to the limited  resolution of our spectra, we cannot eliminate the possibility of foreground emission.
            However, for most lines (CO, $^{13}$CO, [C~I], [C~II], OH, O~I) we find the peak emission on the coordinates
        of our target, dropping with increasing (angular) distance. This suggests that foreground emission is unlikely. An
        exception is the emission of [N~II], which is in general very diffuse, thus  foreground emission
        is possible.
            In the following paragraphs we discuss the SED (Fig. \ref{SED_multiplot}-\ref{SED_multiplot2}) and imaging for each source as well as lines found in the data cubes
        and different features like silicates and ices from the Spitzer spectra.

\begin{figure*}
    \includegraphics[width=\textwidth]{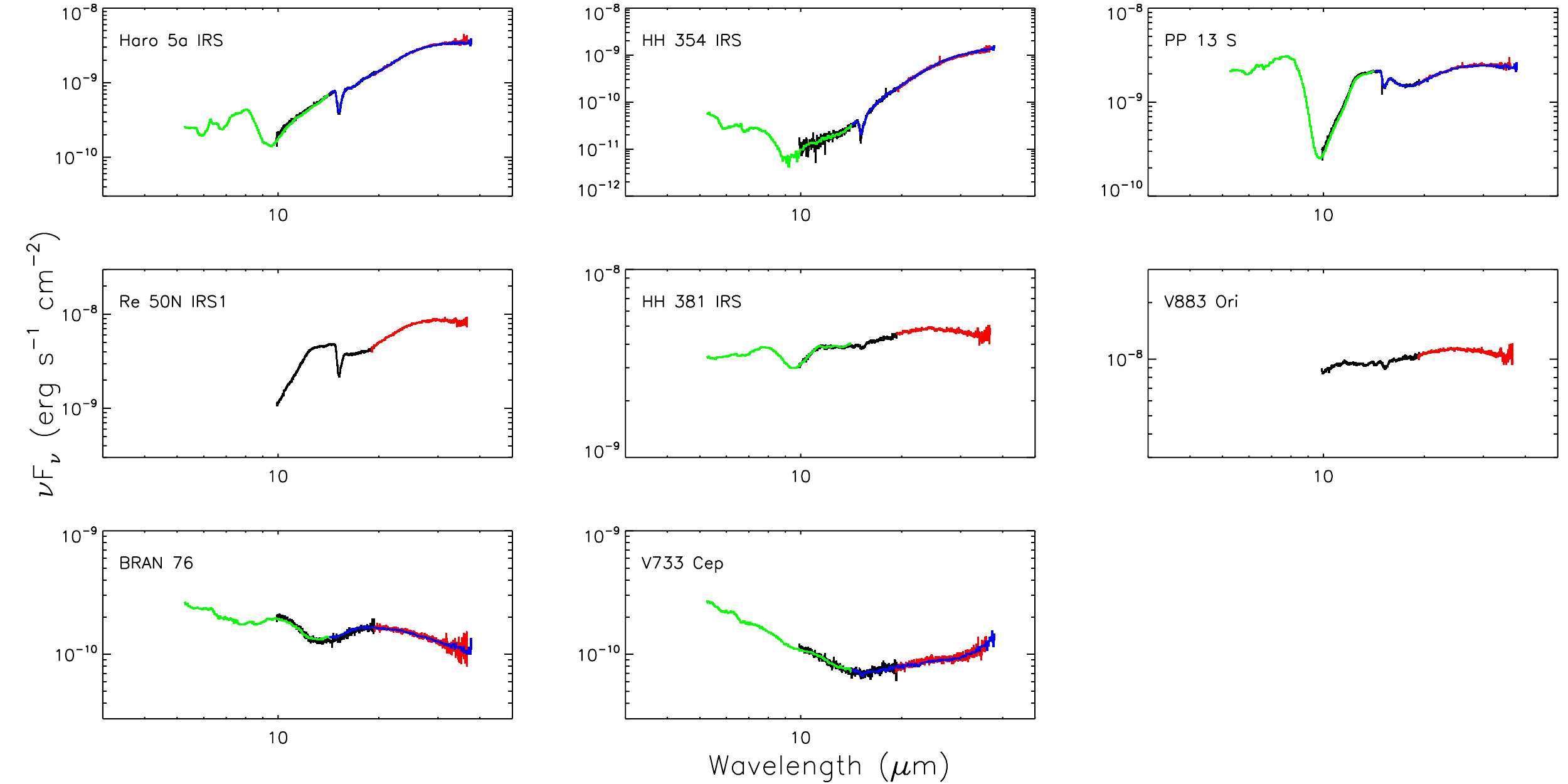}
    \caption{
        \footnotesize
        Zoom-in of the \textit{Spitzer} spectra for our targets (from the PI: Audard program). The colors of the different modules are black (SH), red (LH), green (SL), and blue (LL). The top three targets (and possibly HH 381 IRS) show evidence of H$_2$O and HCOOH ice at $\approx 6$ \textmu m, CH$_3$OH and NH$_4^+$  at $\approx 6.8$ \textmu m, CH$_4$ at $\approx 7.7$ \textmu m, and NH$_3$ at $8.5-9$ \textmu m. In addition, all targets except BRAN 76 and V733 Cep show absorption by CO$_2$ ice at 15 \textmu m.   
    \label{irs_spitzer}
    }
\end{figure*}

        \paragraph{BRAN 76}
        The PACS and SPIRE spectra have poor S/N, and are therefore not presented in the SED.
        The photometry in the SED closely matches   the
        previous observations (\citealt{Sandell2001}, \citealt{Green2006}, \citealt{Reipurth2002}, \citealt{2MASS}), closing the gap above \textit{Spitzer} up
        to the sub-millimeter. The source resembles FU Ori, with silicate in emission (Fig.~\ref{irs_spitzer}) and generally a weak far-infrared component, indicating that
        the source is not as embedded as the other FUors. Due to the higher robustness of the line fitting, as discussed above, we were able to  measure a few
    CO transitions above a $3\sigma$ detection and obtained upper limits up to $J=38-37$.
        We also find tentative evidence of water and OH lines (see Table~\ref{lineFes}).
            Figure \ref{Imaging1_PACS} shows diffuse emission around the object, mainly in the
            160 \textmu m channel, which traces cold dust, in the west and southeast.
            The SPIRE photometry shows mainly diffuse emission in the north, which does not appear             to originate from the target.
            
\begin{figure*}
    \includegraphics[width=0.33\textwidth]{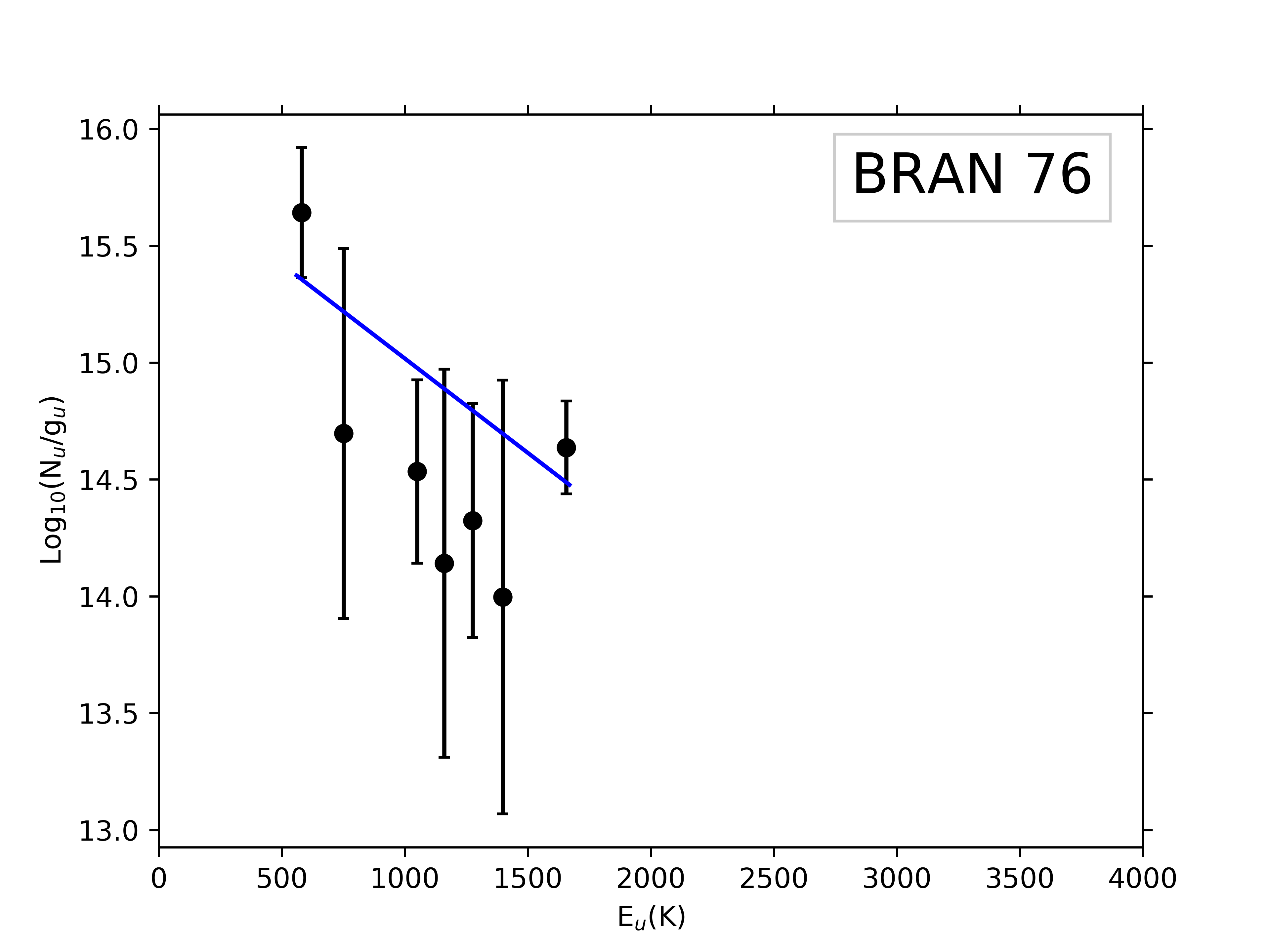}
    \includegraphics[width=0.33\textwidth]{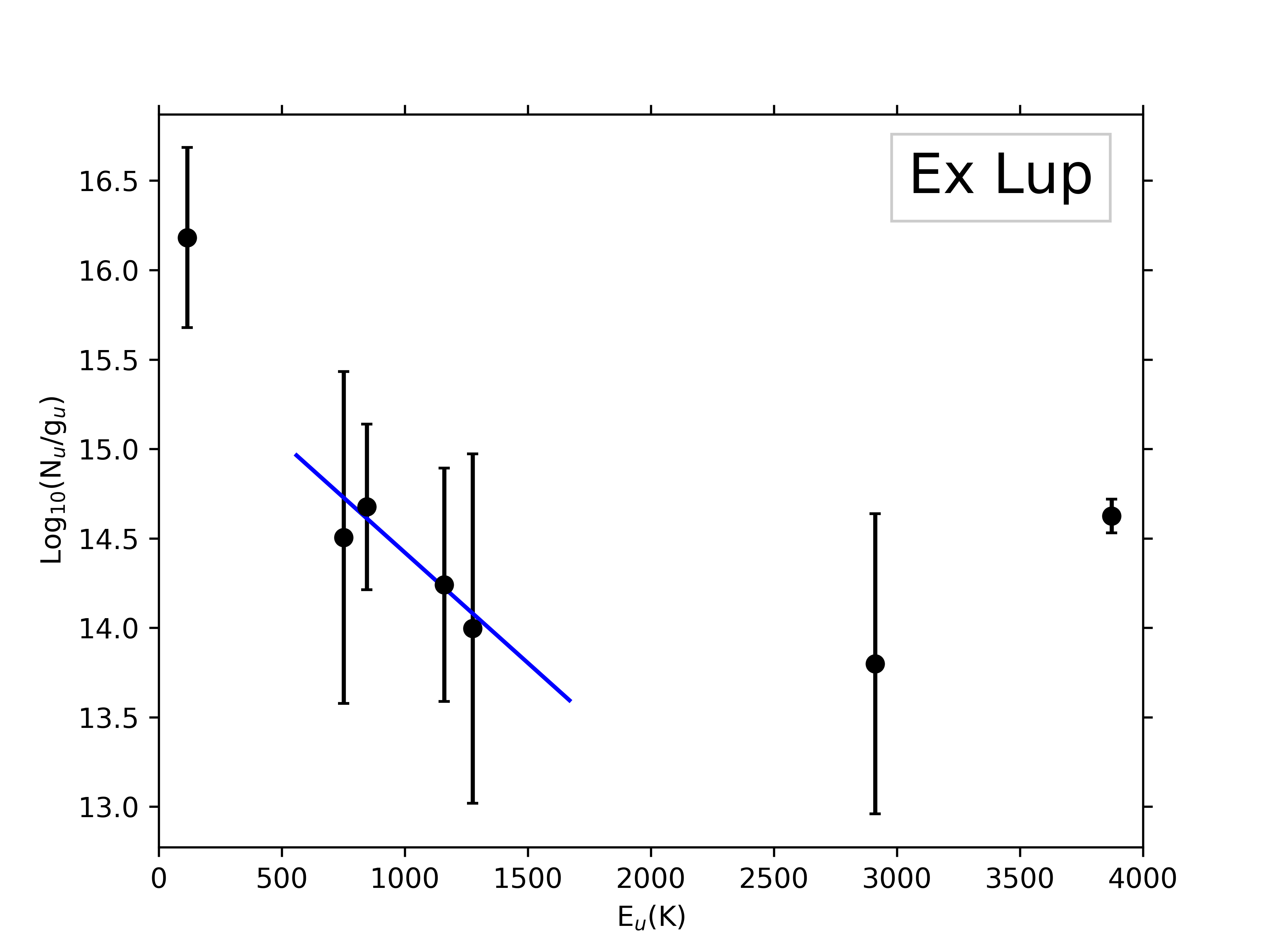}
    \includegraphics[width=0.33\textwidth]{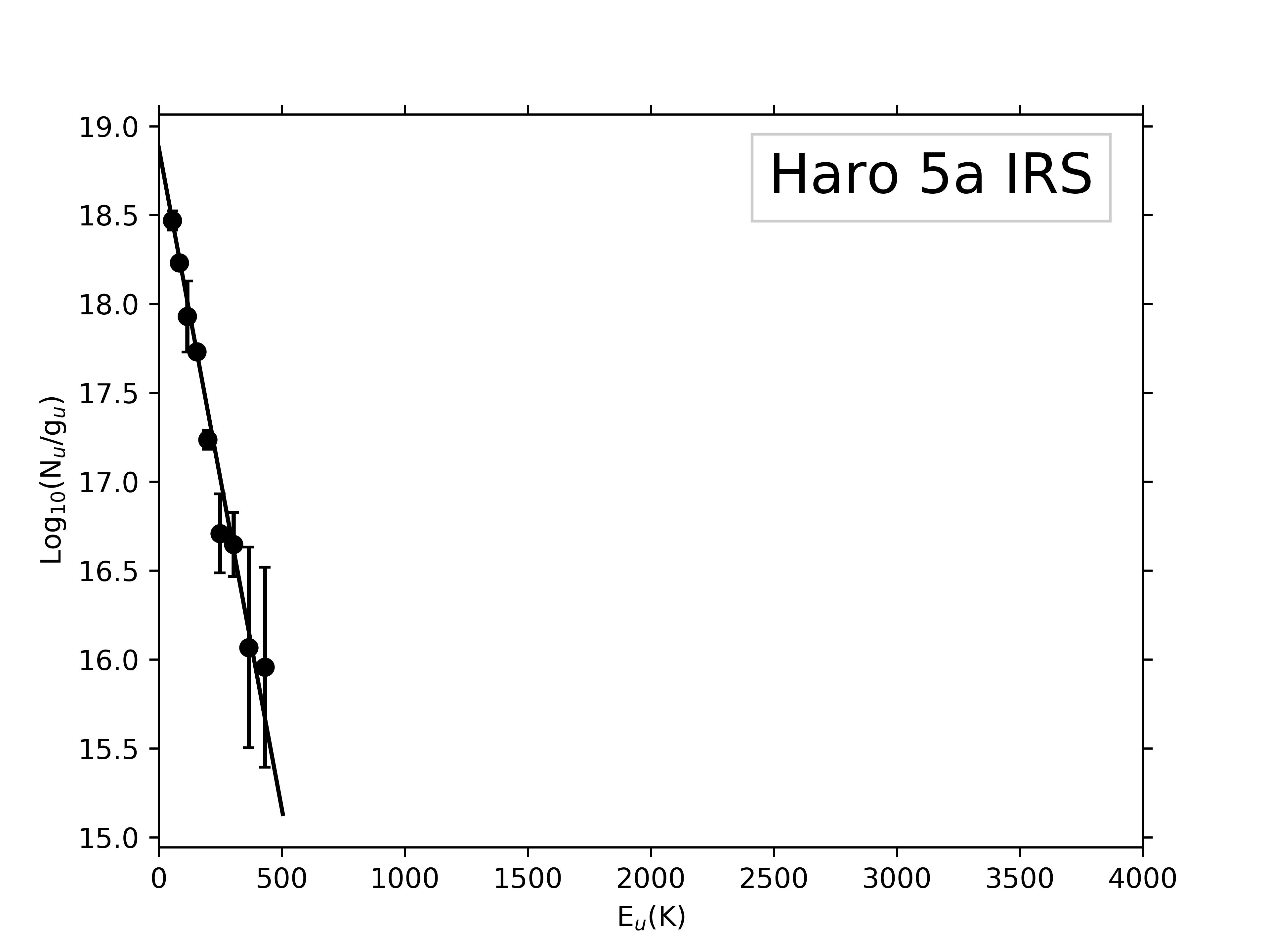}\\
    \includegraphics[width=0.33\textwidth]{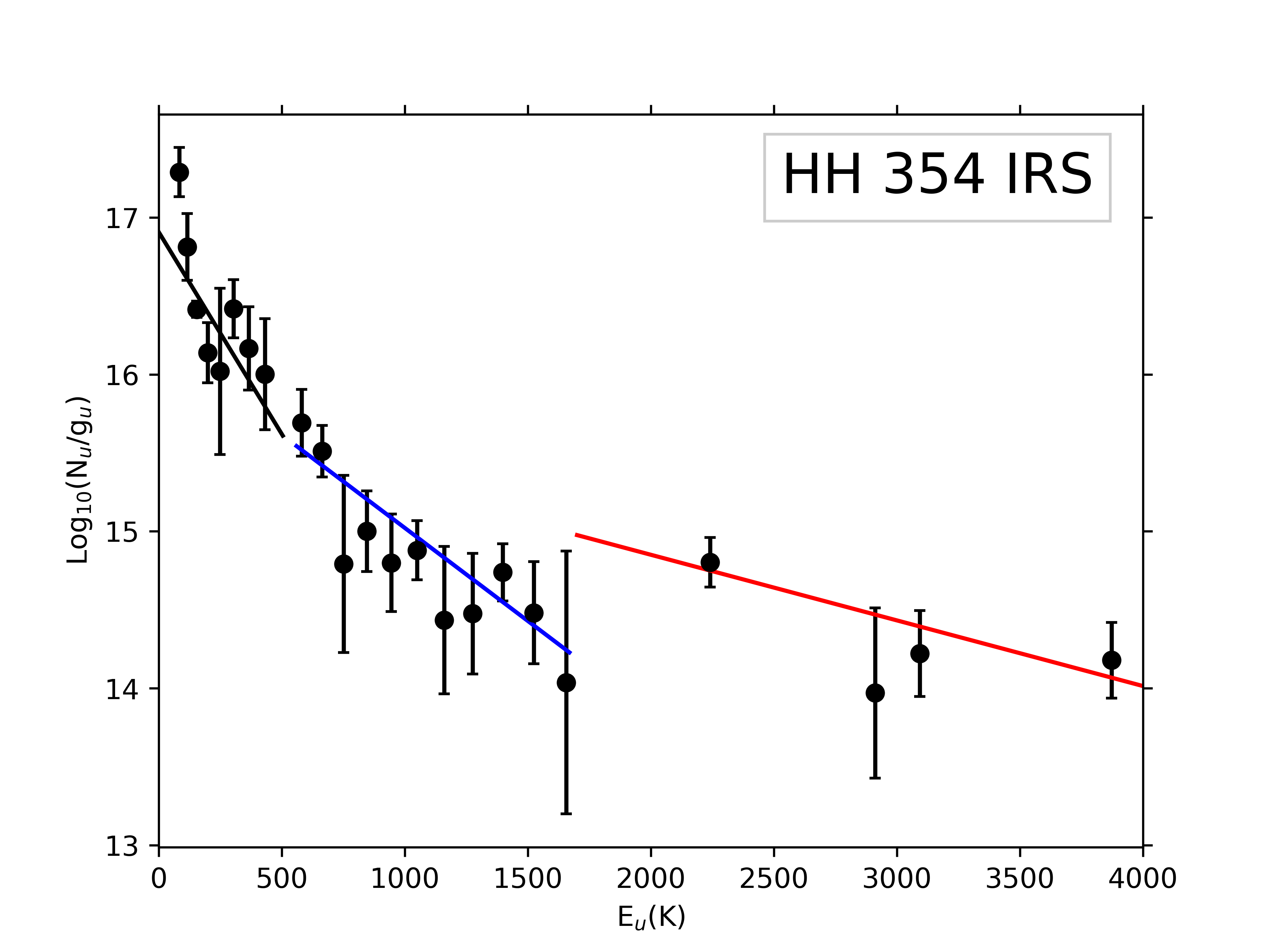}
    \includegraphics[width=0.33\textwidth]{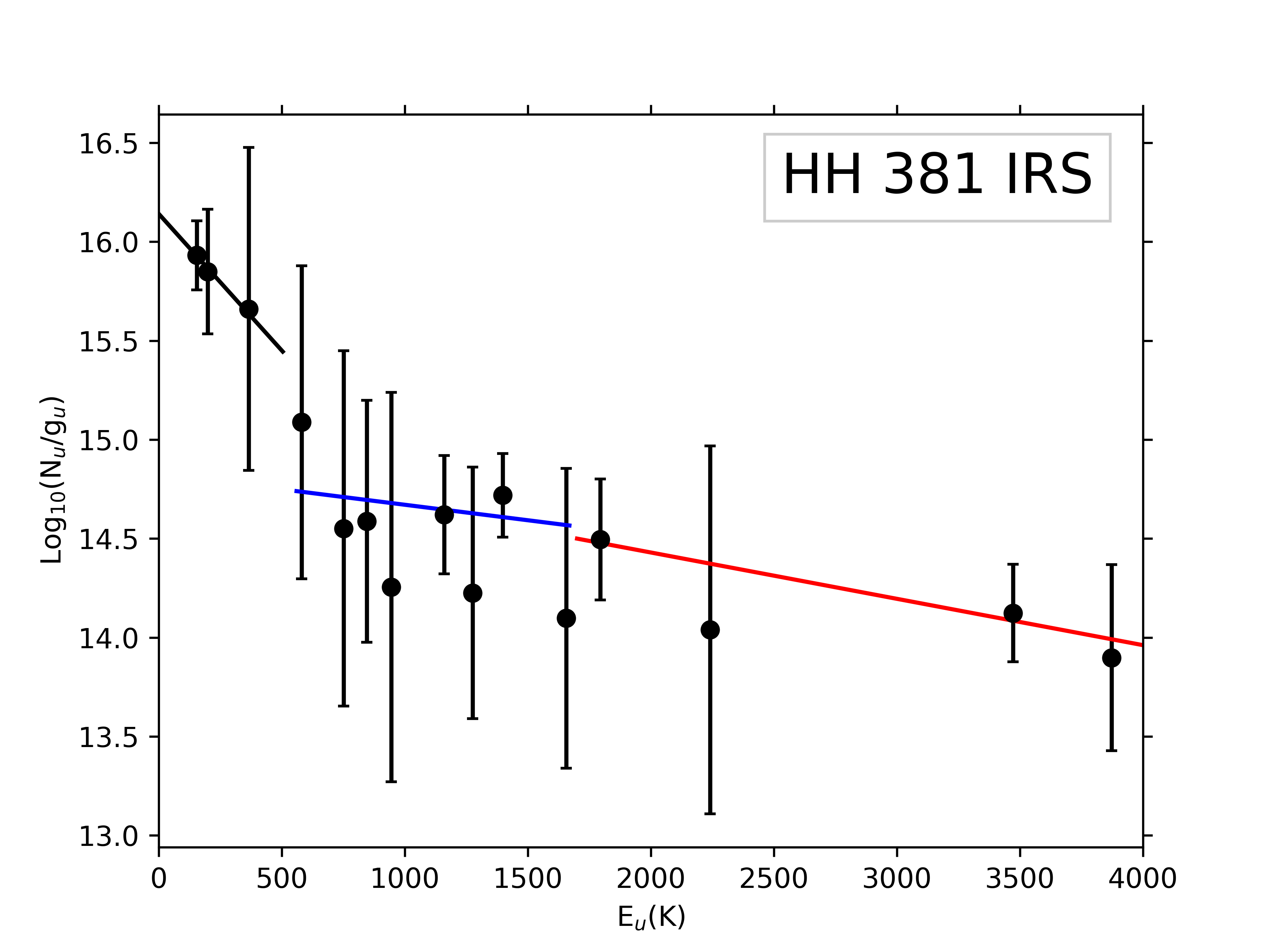}
    \includegraphics[width=0.33\textwidth]{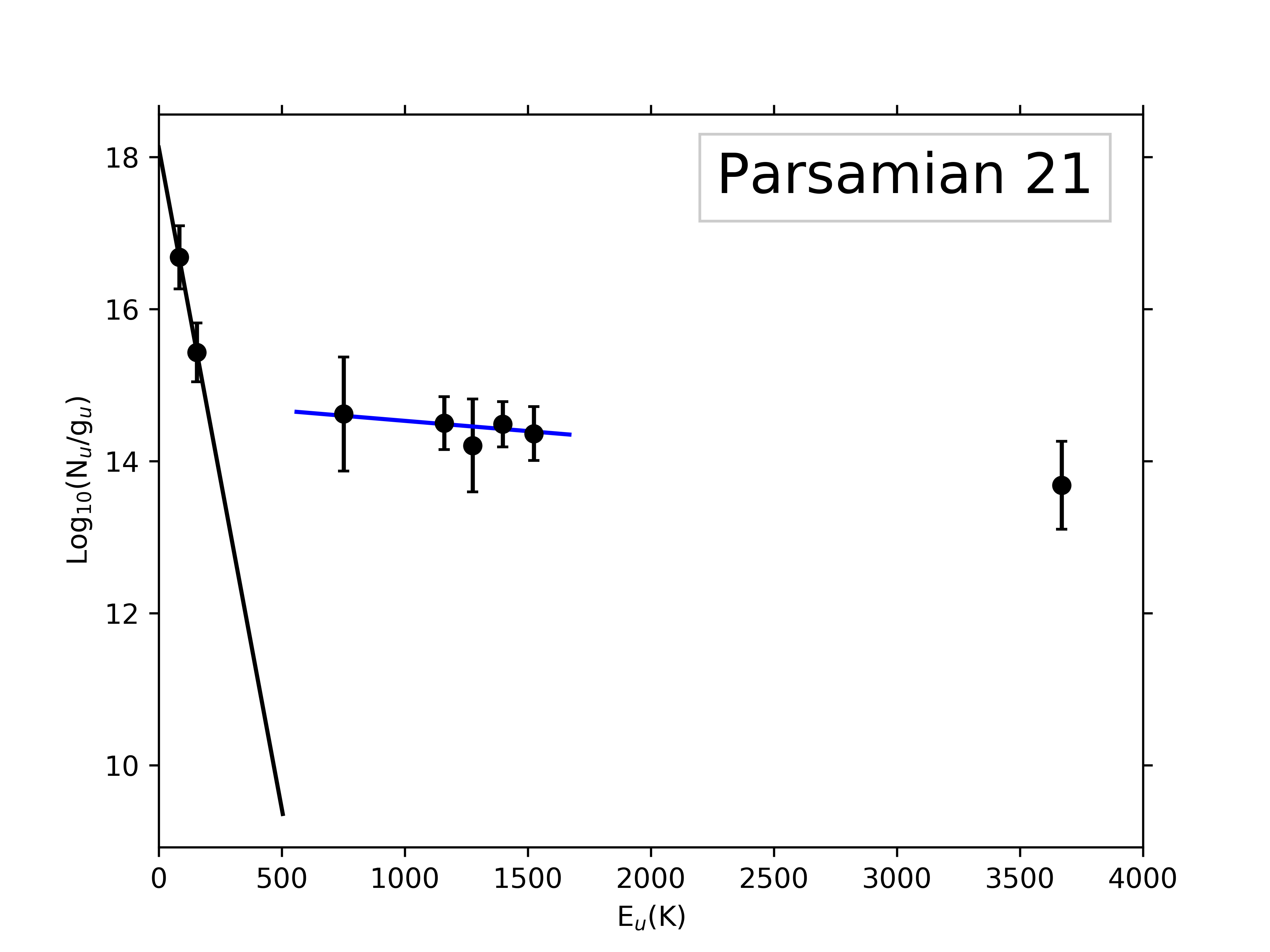}\\
    \includegraphics[width=0.33\textwidth]{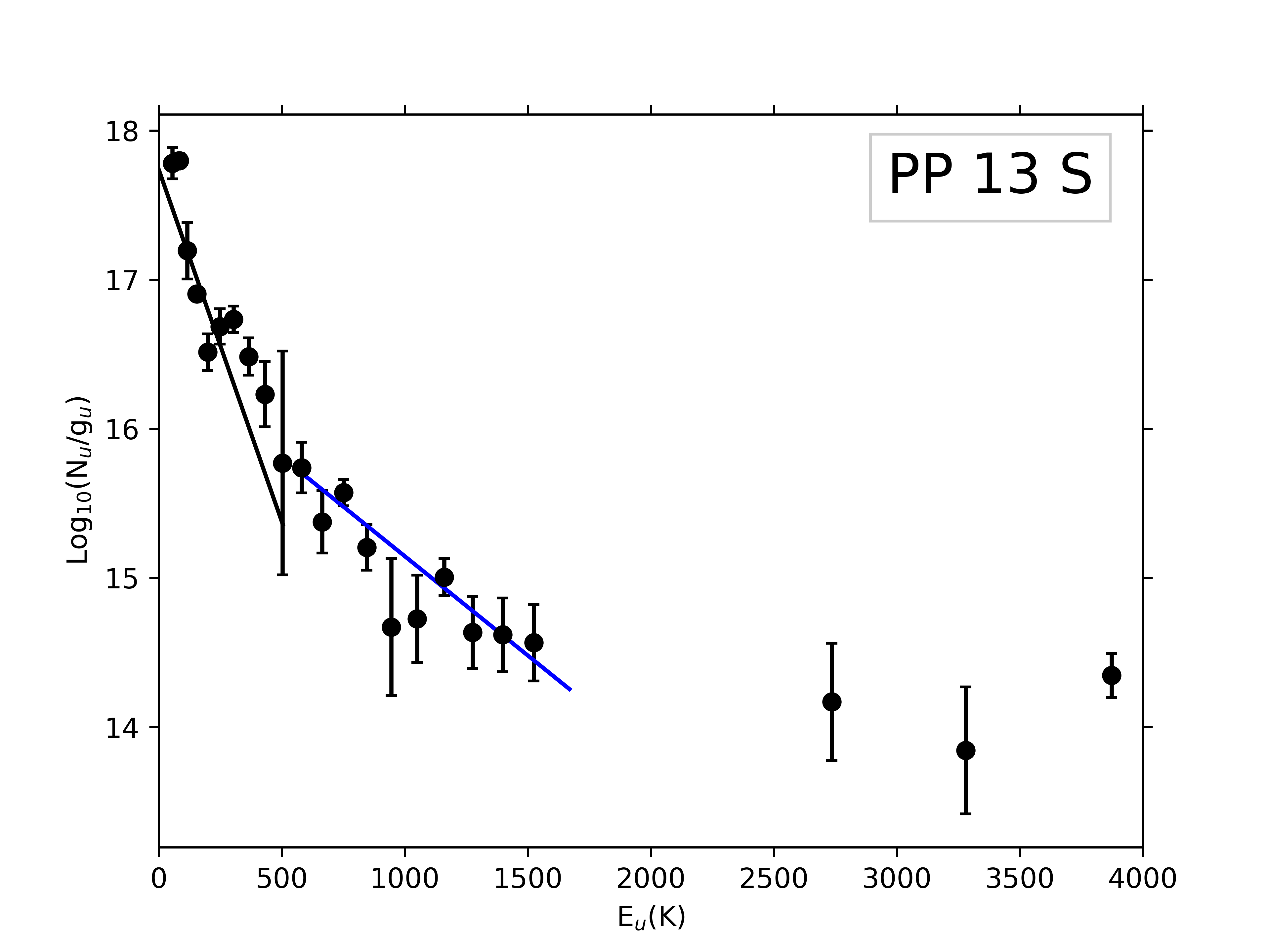}
    \includegraphics[width=0.33\textwidth]{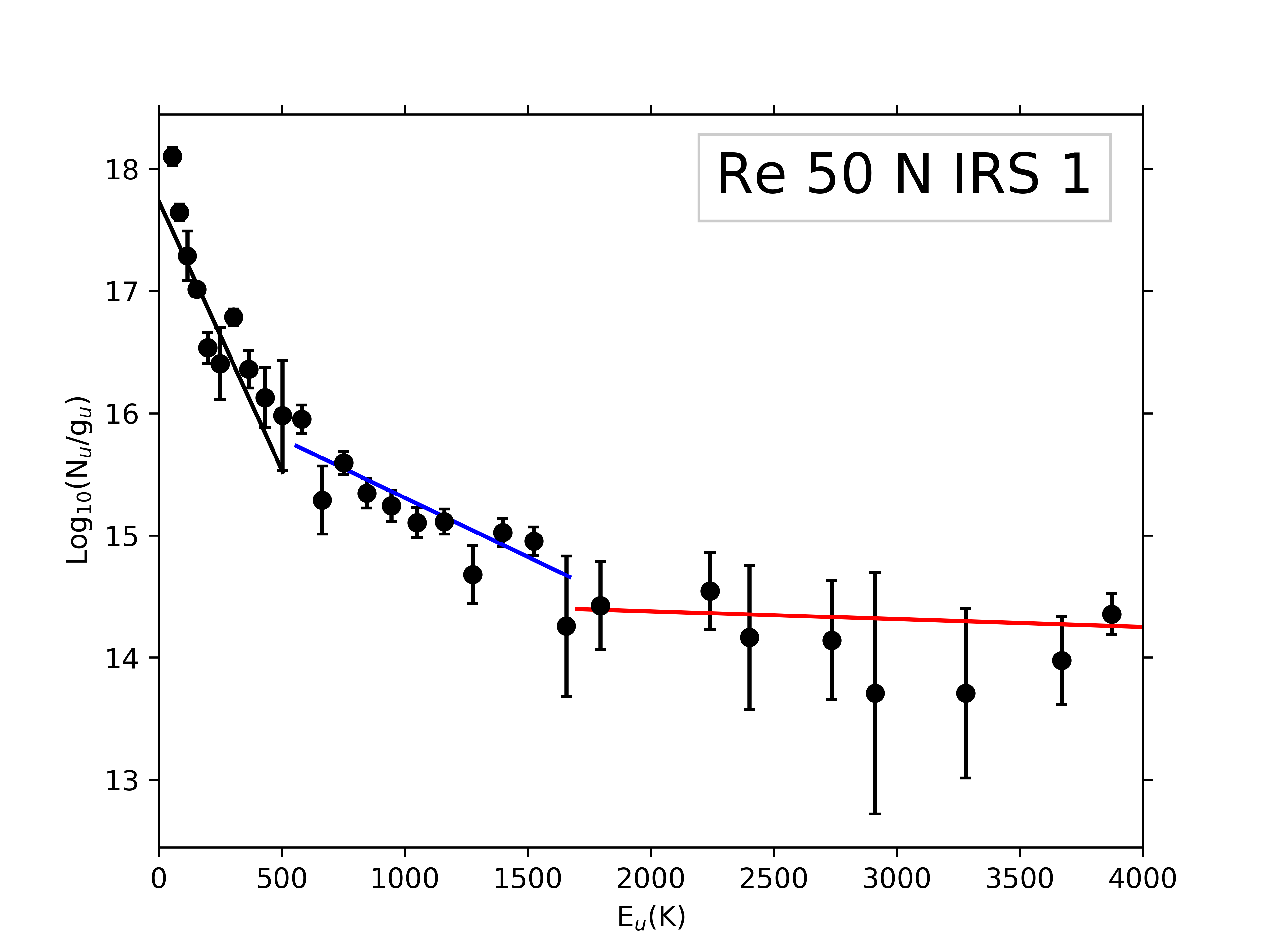}
    \includegraphics[width=0.33\textwidth]{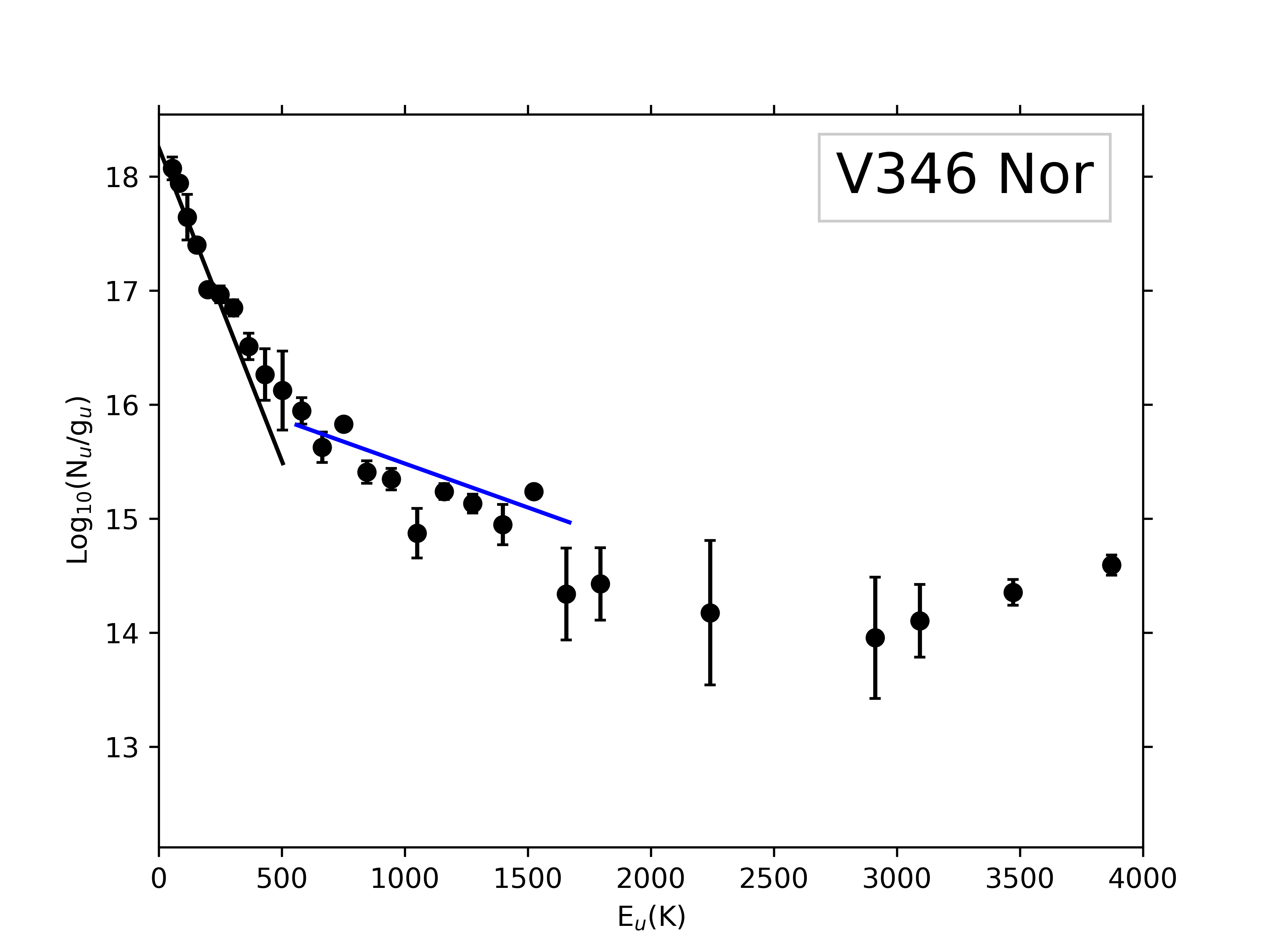}\\
    \includegraphics[width=0.33\textwidth]{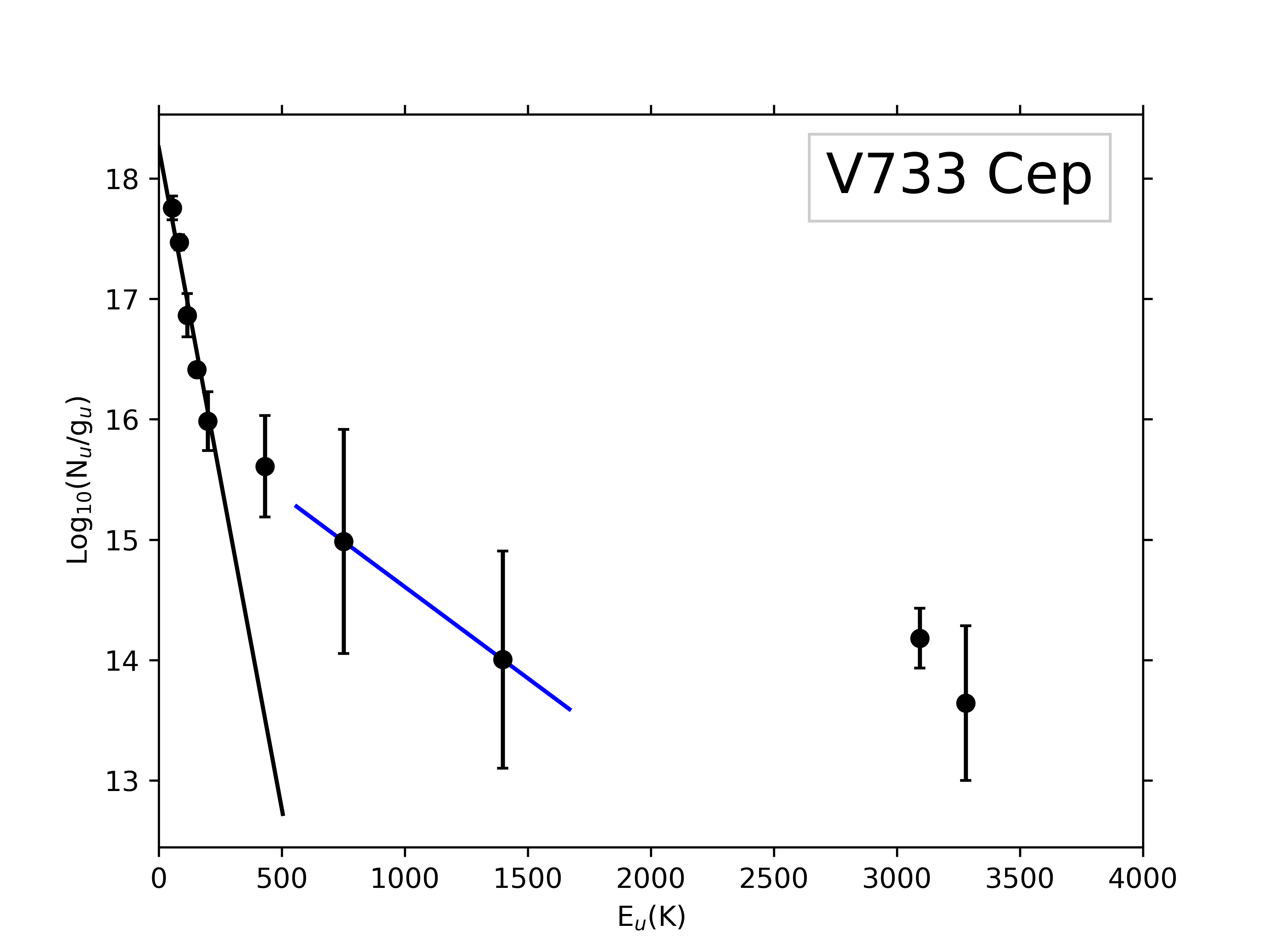}
    \includegraphics[width=0.33\textwidth]{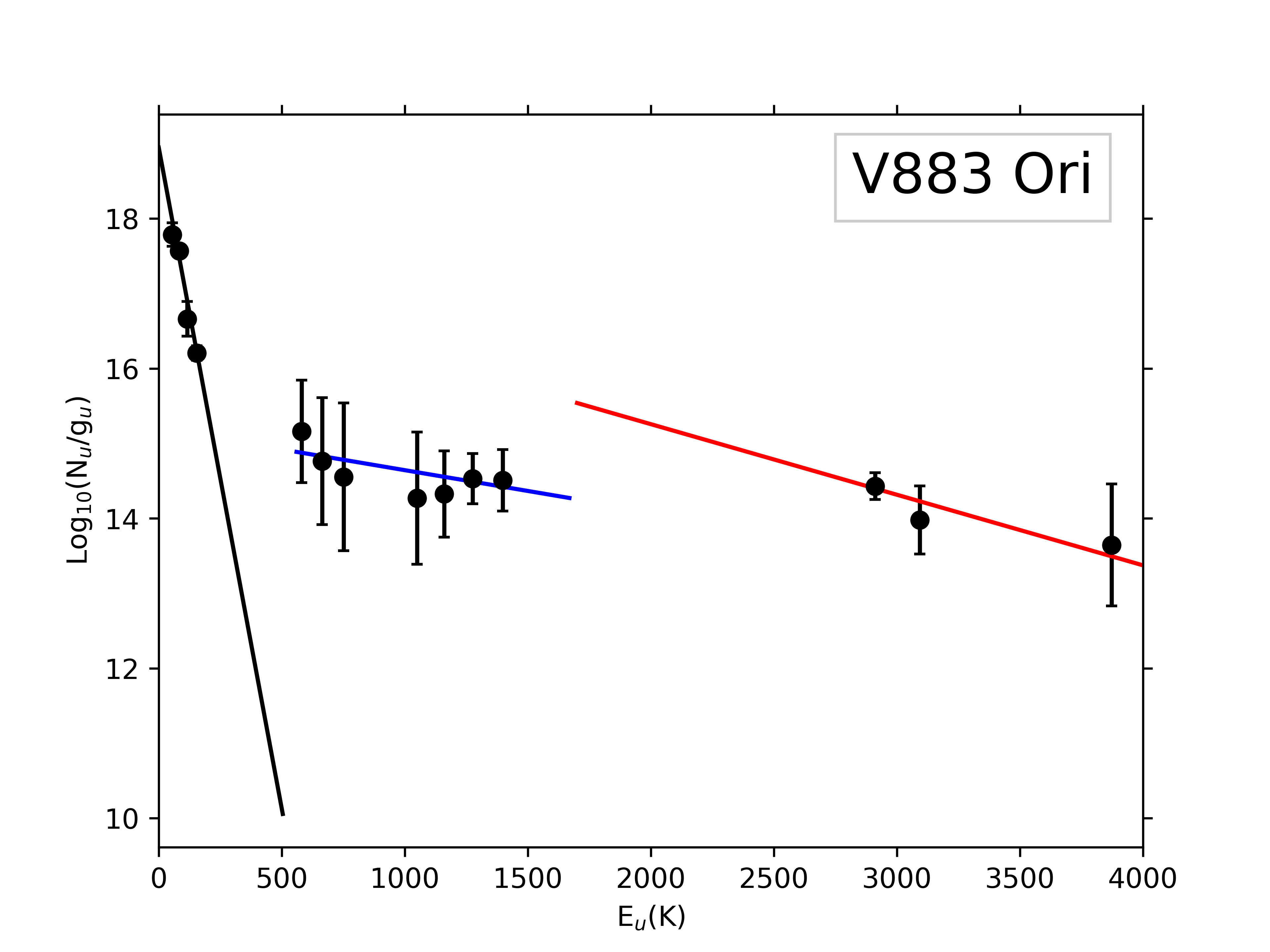}
    \includegraphics[width=0.33\textwidth]{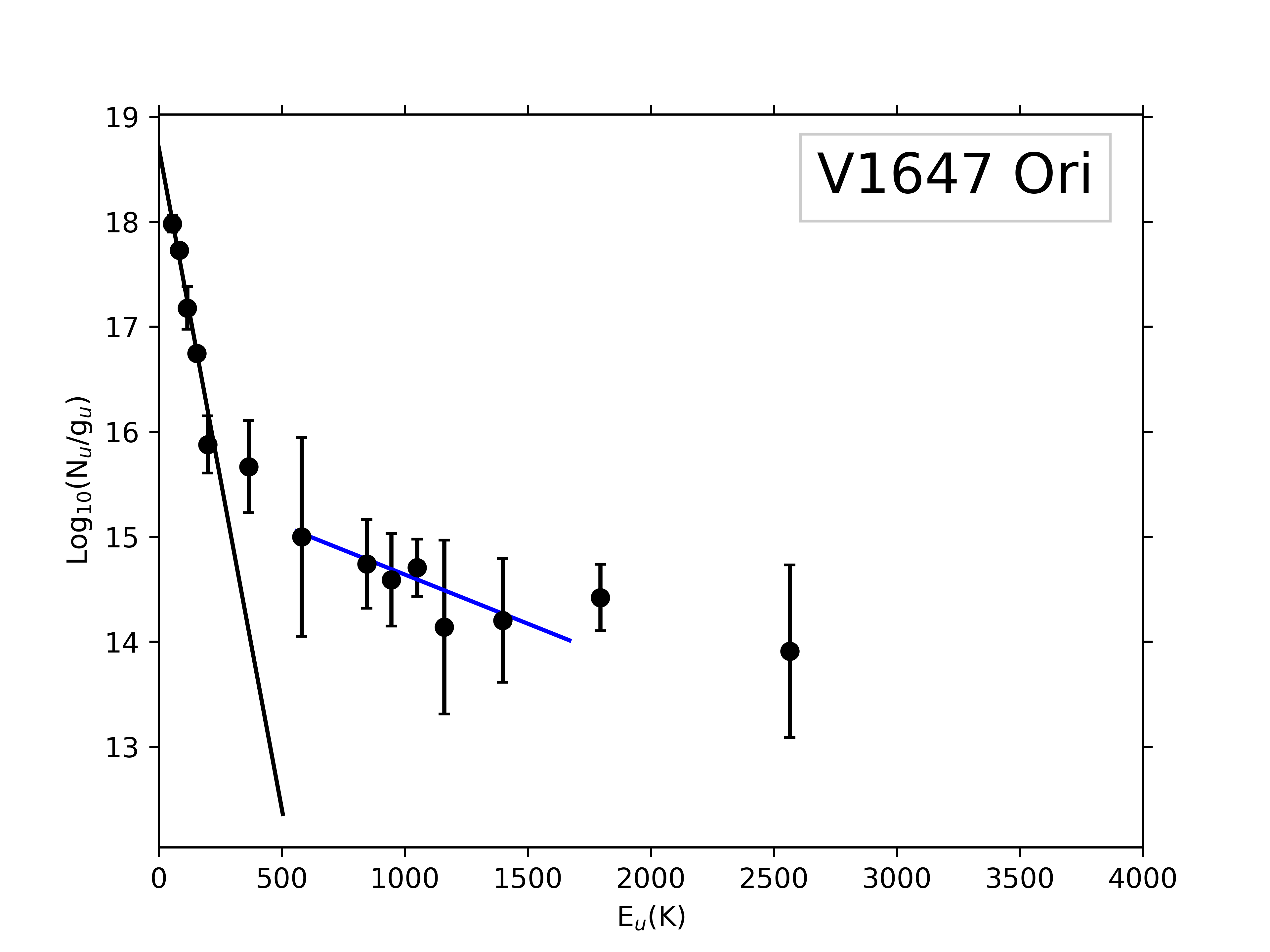}
    \caption{
                \footnotesize
                CO rotational diagrams of all objects. We plot the decadic logarithm of the column densities $N_u$ divided by the statistical weight $g_u$,
                as function of the upper level energy of the transition $E_u$ in Kelvin. SPIRE data are fitted with black, PACS data are fitted
                with blue and red. As the plots contain a logarithmic scale, the errors are actually not symmetric, but we use it here as an approximation.
        }
    \label{rotation_diagrams}
\end{figure*}

            \paragraph{EX Lup}\label{EXLup_results}
        EX Lup is the only EXor in our sample and the faintest object in our analysis (see Table~\ref{OILum} and Fig.~\ref{lbol}). It is the prototype of its class and has shown several outbursts, in particular in 2008 during which its
        silicate, in emission, changed between quiescence and outburst, with evidence of thermal annealing in the surface layer of the inner disk
        by heat from the outburst \citep{2009Natur.459..224A}.
        The SED is shown in Fig. \ref{SED_multiplot} with SPIRE and PACS photometry. We complement
            the SED with additional photometry (\citealt{IRAS88}, \citealt{Cutri2003} 
        down to 12 micrometers, which   match  the PACS photometry in the intersecting
            area at 60 \textmu m. There is no evidence of envelope emission, as previously inferred (e.g., \citealt{Sipos2009}). The \textit{Herschel} photometry of EX Lup is point-like.
            We confirm the presence of LEDA 165681 in the southeast, a background galaxy. The PACS line maps   show centered
            emission on the target coordinates without any sort of outflows. We found only two CO lines with a $3\sigma$ detection at the target coordinates. In addition, these two lines show no significant emission in adjacent spaxels. We found emission above $3\sigma$
        for other CO lines in the central $3 \times 3$ spaxels (see Appendix \ref{line_flux_tables}), which provides further evidence that the CO detection is true. Additional data with higher sensitivity
    for CO are required for a deeper analysis of these lines, as we did for other objects with CO emission in this paper (see Fig. \ref{rotation_diagrams}).
            
            \paragraph{Haro 5a IRS}
            Published photometry (\citealt{2MASS}, \citealt{2004AJ....127.1736R}, \citealt{1998ApJ...509..299L}, \citealt{1986MNRAS.218P...1W}, \citealt{1997ApJ...474L.135C}) complements the emission to the near-IR and
            millimeter range. PACS and SPIRE spectra in the SEDs show strong emission lines of atomic (O~I) and molecular (CO) species, indicating high column
            densities. The wavelength of the SED peak emission lies slightly below 100 \textmu m.
            In the images of PACS and SPIRE in Fig. \ref{Imaging2_PACS},
            we can recognize more significant diffuse emission around the target with another object close by in the northeast (HOPS 85).
            There is also a filament, mainly in the 160 \textmu m channel (red) in the north, which we can also find in the SPIRE data.
            The SPIRE data show also some stronger filament-like emission in the south, so that the target appears to be within a single
            filament. We detected CO lines from $J=4-3$ to $J=10-9$ in the SPIRE spectra, as well
    as $^{13}$CO $J=6-5$ and $J=5-4$. We detected two [C~I] lines (370 and 609 \textmu m) and provide an upper limit for [C~I] at 230 \textmu m.
        The emission in the SPIRE maps is for [C~I] at 370 \textmu m and the CO transmission from $J=9-8$ to $J=5-4$ separated in two
        areas. The first area is localized around the target coordinates, with a slight offset for [C~I] to the  northeast. The CO emission is
        more localized at the target coordinates, with higher emission in the  southwest for CO $J=8-7$. The second area is the southwest itself,
        which is emitting even more strongly than the source. Both areas   correlate well with the RGB images of PACS and SPIRE, and are probably 
    tracing the filament. 
        
\begin{figure*}
  \includegraphics[width=\textwidth]{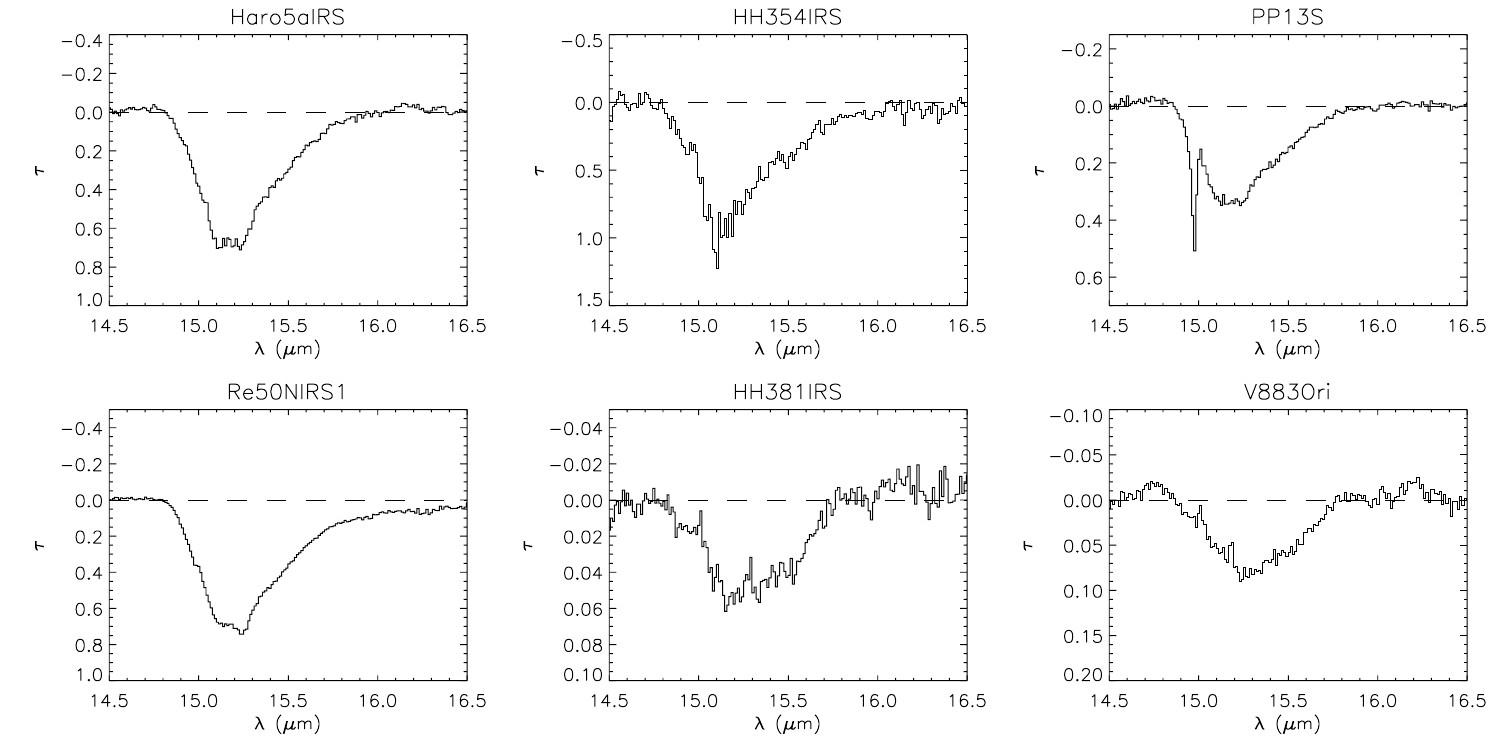}
    \caption{
        \footnotesize
        Zoom-in of the \textit{Spitzer} spectra for targets with CO$_2$ ice absorption. Absorption optical depths $\tau$ near 15 \textmu m were derived from $F=F_0e^{-\tau}$, where $F$ is the observed flux density, and $F_0$ is the continuum derived from a continuum fit, obtained with a polynomial of degree $N=3$ (red curve; HH 381 IRS, V883 Ori, BRAN 76, and V733 Cep needed $N=2$) to fit the ranges 13.5-14.7 and 18.2-19.5 \textmu m (except for HH 381 IRS and V883 Ori for which 13.5-14.7 and 16.0-17.0 \textmu m were used). In addition, in some cases, a Gaussian function centered on 608 cm$^{-1}$ (16.45 \textmu m) and of FWHM=73 cm$^{-1}$ was added to fit the 16.5-18.2 \textmu m range to account for absorption due to silicates. This method is similar to that used by \cite{2008ApJ...678.1005P}.
         Interestingly, we detect \emph{\emph{gas}} absorption by CO$_2$ at 15.0 \textmu m, but only in PP13 S.
    \label{irs_co2}}
\end{figure*}

        The \textit{Spitzer} data show deep silicate absorption at 10 \textmu m (stretching of the Si-O bonds), with a weaker component around 18 \textmu m (O-Si-O bending),
        together with several ice features between 5 and 9 \textmu m, due to H$_2$O and HCOOH ice, CH$_3$OH and NH$_4^+$, CH$_4$, and NH$_3$ at 8.5-9 \textmu m (Fig.~\ref{irs_spitzer};
        see also \cite{Oeberg2010} for identification of ice features). Furthermore, CO$_2$ ice is detected at 15 \textmu m (Fig.\ref{irs_co2}).   
        
        \paragraph{HH 354 IRS}
        The \textit{Herschel} spectra completes the peak emission of the envelope in the SED, which was already partially
        detected in the \textit{Spitzer} data. The SED, complemented by photometry below 10 \textmu m (\citealt{2MASS}, \citealt{Visser02}, \citealt{difran08}, \citealt{IRAS86},
        \citealt{IRAS88}), shows very deep absorption near the silicate feature (Fig. \ref{irs_spitzer}).  The deep narrow absorption peak in the silicate feature can be identified
        as methanol ice (fundamental CO-stretch mode). CO$_2$ ice is also detected at 15 \textmu m (Fig. \ref{irs_co2}). In addition, the 5-9 \textmu m also show ice features as detected in Haro 5a IRS.
        
        While a companion is currently not known, there is a reflection nebula \citep{Reipurth1997} surrounding the object, which likely contributes to some additional emission at lower
        wavelengths.
        HH 354 IRS appears
            to be isolated in the PACS images, with some slight diffuse emission in the northeast. This diffuse emission cannot be confirmed
            in the SPIRE data; however, there is some more emission in the north and the southeast.
        We see a very strong emission of [O~I] at 63 \textmu m and can detect [O~I] at 145.5 \textmu m. Both lines peak at the target coordinates, but the emission seems to  also be affected by diffuse emission.
        We find OH, [N~II] which is slightly enhanced at the source coordinates, several water lines, and CO transitions from $J=5-4$ to $J=37-36$, with an upper limit for the $J=38-37$ transition.
        The SPIRE maps do not show a clear correlation with the RGB images. Some of the CO lines ($J=11-10$, $J=10-9$) show emission distributed
        over a few spaxels, which are not always linked together, while other lines ([C~I] at 370 \textmu m, CO $J=7-6$, [N~II]) have much smoother
        emission, which is also extended over the entire field of view and which might indicate extended emission.
        
        \paragraph{HH 381 IRS}
        Together
        with the \textit{Spitzer} spectra and photometry between 1 and 3 \textmu m, HH 381 IRS shows the SED of an embedded young stellar object,
        similar to Class I with strong continuum emission between 5 and 100 \textmu m.
        Additional photometry (\citealt{2MASS}, \citealt{MSX}) from previous observations do not  match   the \textit{Herschel} and
        \textit{Spitzer} spectra closely, possibly due to significant variations in the mid- and far-infrared emission during the possible FUor outburst. This is further discussed in \ref{subsection:visible_and_missing_features}. 
        The \textit{Spitzer} data show weak silicate absorption at 10 \mic\   (Fig.\ref{irs_spitzer}), much weaker than in other embedded FUors. Nevertheless, there
        is also a weak CO$_2$ absorption detected in the IRS spectra (Fig. \ref{irs_co2}). Other ices $<10$ \textmu m   are possibly detected as well.
        
        Figure \ref{Imaging3_PACS} shows HH 381 IRS as isolated with some weak emission in the south in the PACS data. SPIRE shows
            some more extended emission and a diffuse emission in the northeast and southwest around the target. 2MASS photometry from the $J$, $H$, and $K$ bands
            shows bipolar outflows of the target,  one in the north probably directed away from us, the other  in the south  probably directed toward
            the observer and  with stronger intensity. The emission of OI is centered for the transition at 145.5 \textmu m, but for 63 \textmu m the maximum is
            in the neighboring spaxel southeast of the source.
            In addition to  OI, the spectrum shows water lines: OH and
        CO from $J=7-6$ to $J=37-36$. $^{13}$CO is not detected, so we only provide upper limits.
        The emission of both [C~I] lines and CO $J=7-6$ is extended over a large area, while CO $J=8-7$ is restricted mainly to the target coordinates and
        a small emission region in the southwest. That region matches the bump in the SPIRE RGB image in the southwest, where   the OI emission
        at 145.5 \textmu m also  shows a signal.
        
        \paragraph{Parsamian 21}
        The available spectra of PACS and SPIRE are joined by photometry measurements \citep{IRAS88}, which show the SED down to 10 \textmu m. 
        Parsamian 21 is affected by moderate noise in its spectrum, which makes it difficult to confirm lines above the $3\sigma$ threshold.
        The SED shows that the object is embedded, similar to YSO Class I, with a strong continuum from 10 to 100 \textmu m. The \textit{Spitzer} spectrum was
        shown by \cite{Kospal2008} and shows strong CO$_2$ absorption. Inversely to other FUors, it shows PAH features at 6.3, 8.2, and 11.3 \textmu m  and silicate
        emission at 10 \textmu m. \cite{Kospal2008} argued that Parsamian 21 was an intermediate-age FUor in the evolutionary sequence.
        
        In our \textit{Herschel} data, we find a few CO and OH lines in the data, but mostly provide upper limits. In the RGB images, Parsamian 21 appears
            isolated with some strong diffuse emission in the north in both PACS and SPIRE photometry. This outflow can also be seen in the optical
            (e.g., by DSS2 in the red (F+R) and blue (XJ+S) channel) and was first described by \cite{Parsamyan1978}.
            In SPIRE, we can also recognize a second bump in the east of the northern end of the emission. We could not identify another source in that area, but as this emission is not visible in the PACS data,
            it simply might be material of lower temperature.
            The H$_2$O emission in the PACS line maps shows an additional spaxel, which matches the outflow from the RGB images in the north.
            
            \paragraph{PP 13 S}
        Together with external photometry (\citealt{Sandell1998}, \citealt{Cohen1983}, \citealt{1995ApJ...439..288O},
        \citealt{Sandell2001}, \citealt{Tapia1997}) and the \textit{Spitzer} spectra, the \textit{Herschel} data completes the SED, ranging from 1 \textmu m up to
        above 1 mm. The spectrum is characterized by strong continuum emission ($\approx 10^{-10}$ erg s$^{-1}$ cm$^{-2}$) between 30 and 200 \textmu m, indicative of strong envelope emission. The extinction is very high since the photometry drops significantly below 2 \mic. Figure \ref{irs_spitzer} shows very deep silicate absorption in  the Si-O stretching feature at 10 \textmu m  and in the O-Si-O bending at 18 \mic. In addition, CO$_2$ ice absorption is detected, and  interestingly a CO$_2$ \emph{\emph{gas}} line is detected at 15.0 \textmu m (Fig. \ref{irs_co2}). This is the only target showing such gas absorption feature in our sample. At shorter wavelengths ($5-9$ \mic), the other ice features are detected as well, though NH$_3$ ice appears  weaker than in other targets.

        PP 13S in Fig.
        \ref{Imaging4_PACS} shows slight diffuse emission around the target in PACS data. However, we see some
            filamentary structures in the SPIRE data in the north of the target. There is also a stronger emission in the
            neighborhood around the target in the SPIRE data. We find emission lines in the spectra, namely the two OI lines at 63 and
    145.5 \textmu m, water, OH, [C~I] and [C~II], and CO from $J=4-3$ to $J=37-36,$ and we provide upper limits for $^{13}$CO
    from $J=5-4$ to $J=10-9$. The PACS line maps show the emission of CO, H$_2$O, and [O~I] mostly centered around the object. 
            Some CO lines are extended around the object, while in particular the H$_2$O emission at 108 \textmu m   affects adjacent spaxels.
            The emission of [O~I] at 63 \textmu m shows an outflow in the north, possibly correlated with the filamentary structure in the SPIRE data.
            In the SPIRE maps, the emission
            is more extended, while there appears to be no clear link to the RGB images. 

            \paragraph{Re 50 N IRS 1 / HBC 494}
            The SED fills the area from below 1 \textmu m  to above 1 mm, using published photometry 
            (\citealt{Reipurth1993}, \citealt{Dent1998}, \citealt{Zavagno1997}, \citealt{Sandell2001}, \citealt{Strom1989}, \citealt{IRAS88}) in combination with
            \textit{Spitzer} spectra. We find  again   strong continuum emission in the mid- and far-infrared, indicative of envelope emission. The \textit{Spitzer}
            spectrum also shows   deep silicate absorption (in both bending and stretching bands), although there is no spectrum available below 10 \textmu m 
            (Fig. \ref{irs_spitzer}). CO$_2$ absorption is also detected (Fig. \ref{irs_co2}), although  without a gas absorption line, in spite of the relatively
            similar SED as PP 13 S. The source appears to have rebrightened between 2006 and 2014
            (\cite{2015ApJ...805...54C}; the \textit{Spitzer} data were taken 7 Nov 2008, while the \textit{Herschel} date from Sep 2013). The source is known
            to harbor a very wide outflow, detected with ALMA \citep{2017MNRAS.466.3519R}.

            In the RGB images,
            Re 50 N IRS 1 shows strong diffuse emission in an extended area in both the PACS
            and the SPIRE data. The PACS data extend mostly in the northeast and also in the west of the object, while SPIRE shows some
            weaker diffuse emission in the west, but confirms the emission in the northeast. We can also see a second emission bump
            in the south of the object with both instruments. Unfortunately, the \textit{Spitzer}
        data for Re 50 N IRS 1 stop short of 10 \textmu m. In the \textit{Herschel} spectra, we see water lines, both O~I lines,
            several OH lines, [C~I] and [C~II], and CO lines from  $J=4-3$   to $J=37-36$ transition. We can also detect [N~II] at 205 \textmu m.
        The line maps of PACS are centered around the target coordinates, while the SPIRE maps show a bigger emission of the surrounding area.
            There appears to be no correlation to the shape of the object in the RGB images. However, especially the [C~I] line at 370 \textmu m and
            the CO $J=7-6$ emission peaks clearly at the center, which indicates that the emission is not in the foreground but originates from the object.

        \paragraph{V346 Nor}
            The corresponding spectrum of \textit{Herschel} has been complemented by photometric measurements (\citealt{IRAS88}, \citealt{Kenyon91}, \citealt{2017A&A...597L..10K} (observed 2013.5)) down to 2 \textmu m,
            covering the peak emission of the object.  \textit{Spitzer} data (\cite{Green2006}; not shown in this paper) reveal deep silicate absorption at 10 \mic,  a CO$_2$ ice absorption feature,
            together with the other ice features in the $5-9$ \textmu m  range. Following the light curve of \citealt{2017A&A...597L..10K}, our Herschel data (February--March 2013) were observed during
            the re-brightening phase of V346 Nor, which followed a fade with minimum in 2010.

            The SPIRE data of V346 Nor in Fig. \ref{Imaging5_PACS} shows extended diffuse emission around the
            target in the west, east, and north. The PACS data reveal  multiple individual sources: a strong one in the west, which we identify
            as 2MASS 16322723-4455303 and 2MASS 16322723-4455314; a strong (2MASS 16323031-4455189) and a faint source (2MASS 16323337-4455187)
            in the north; and  a very faint source in the east (2MASS     16323553-4455317) of V346 Nor. In the spectrum of PACS and SPIRE we
            can easily recognize several emission lines. Our fitting found water lines, both OI lines, [N~II], [C~I] and [C~II], and CO lines from $J=4-3$
             to $J=37-36$, in addition to  $^{13}$CO lines between $J=5-4$ and $J=8-7$. Emission in the linemaps of PACS is mainly concentrated around the target
            with some spaxels, mostly of lower transitions, showing emission in the vicinity. The continuum-subtracted  maps of SPIRE show a slightly extended emission,
            similar to the extended emission of the photometry in the RGB image of SPIRE. The [C~I] emission at 370 \textmu m in contrast to the emission
            at 609 \textmu m shows at least an emission peak at the coordinates of the target, so foreground or background emission can be excluded here.
            The CO $J=11-10$ and $J=9-8$
            transition  represents very well what can be seen in the RGB image, while others ($J=4-3$ to $J=7-6$) show  more extended emission over a large area.
        Even though the diffuse emission of [N~II] is extended over the entire field of view, the emission peak at the coordinates of the source, which
        is up to an order of magnitude higher, makes it likely that V346 Nor is the origin of the emission here. The contribution of other nearby objects cannot be eliminated, however.
            
            \paragraph{V733 Cep}
            Due to the poor quality, we do not show the spectrum of PACS, and only provide the low-resolution spectrum of SPIRE. The data are
            combined with spectra from \textit{Spitzer} in the mid-IR and with photometry (\citealt{semkov08}, \citealt{2MASS}, \citealt{MSX}) down to slightly above 400 nm.
        The shape of the SED is peculiar compared to the other targets, as it shows significant emission below 10 \textmu m likely due to the central source and
        disk emission, while there is an additional component in the far-infrared regime, likely a remainder of an envelope, possibly indicating that V733 Cep is
        of intermediate age in the evolutionary sequence proposed for FUors, i.e., between Class I and II YSO. The \textit{Spitzer} data do not show  a clear silicate
        feature, or possibly a very weak absorption feature, with also very weak ice features below 10 \mic. There is, however, no evidence of CO$_2$ ice absorption at 15 \textmu m (Fig. \ref{irs_spitzer}).

            V733 Cep shows in
            PACS data bright emission around the source coordinates, with another object in the southwest and a cluster of objects in the
            southeast of the target. In the SPIRE data, we cannot resolve individual objects of this cluster; however, the emission of V733
            is dominated by the blue (250 \textmu m) emission, while the other bands do not outshine the background. We identified in the
        data O~I at 145.5 \textmu m, [N~II], [C~I], and [C~II], and some CO lines ranging from $J=4-3$ to $J=33-32$. The SPIRE line maps show the strongest
        emission for most lines in the southeast of the target, where we find the already mentioned cluster. We therefore do not assume that
        the (main) emission originates from V733 Cep, but from the objects nearby. For the [N~II] emission, again we cannot tell if this is
        foreground or originating from the observed area.

            \paragraph{V883 Ori}
            The SED of V883 Ori is shown with  \textit{Spitzer} and \textit{Herschel} spectra, and complemented by published photometry
            (\citealt{Reipurth1993}, \citealt{Dent1998}, \citealt{Molinari1993}, \citealt{Sandell2001}, \citealt{Cutri2003}, \citealt{MSX})
            ranging from the visible to above 1 mm. The SED is flat from 3 \mic\  to about 60 \mic. The \textit{Spitzer} spectrum shows weak silicate absorption (though again
            cut below 10 \mic) and weak CO$_2$ ice absorption (Figs. \ref{irs_spitzer} and \ref{irs_co2}).

            For V883 Ori in Fig. \ref{Imaging6_PACS},
            we find diffuse emission in the PACS data in the north of the target, forming a bar from the east to the northwest  of the target.
            The emission overlaps in the east with the position of several 2MASS objects (2MASS 05382028-0702353, 2MASS 05381975-0702261, 2MASS
            05381936-0702241, 2MASS 05381922-0702287, and 2MASS 05381868-0702241). It is uncertain how much of the diffuse emission is contributed by the 2MASS objects and how much originates from V883 Ori.
            We do not find another object in the emission
            in the north, so at least that component likely originates from our target. The SPIRE data appear dominated by V883 Ori, which causes some artifacts in all three bands, with some strong
            diffuse emission in the northwest, for which we also do not find a known object. The object shows [O~I] and some OH lines, as well
    as [N~II], [C~II], and [C~I]. We  report the detection of CO from $J=4-3$  to $J=33-32$ transition.
            The [C~I] lines in the SPIRE maps smoothly peak at the target coordinates, with diffuse emission around it. The CO and [N~II] lines do not
            show such a clear picture.
            
        \paragraph{V1647 Ori}
            V1647 Ori, an object intermediate between FUors and EXors, is shown together with photometry (\citealt{2004ApJ...606L.119R},
            \citealt{2011AJ....142..135A}, \citealt{2013A&A...552A..62M}) in the IR. The object has shown significant flux variations since its initial outburst in 2003
            and was observed on many occasions from the optical to the millimeter ranges, decaying in 2006 and outbursting again in 2008  (see \citealt{Aspin2009} and \citealt{Audard2014} and references therein).

            The SPIRE spectra are quite noisy; therefore, we also provide  the low-resolution spectra, even though they  do not resolve any lines. In the images, we find diffuse emission around the target,
            especially in the PACS data. We identify emission knots in the north (HH 22 MIR), which also shows significant diffuse emission
            in the SPIRE data, and the southwest (2MASS J05461162-0006279), which does not separate from the background emission in SPIRE.
            There is also some faint extension of diffuse emission from V1647 Ori in PACS, going westward, which increases at some
            distance in the SPIRE data. The spectra show both [O~I] transitions, OH, [N~II], and [C~I] and CO transitions from $J=4-3$
            to $J=25-24$. For the higher lines up to $J=37-36$ and for most $^{13}$CO lines we provide upper limits.
            The emission of [C~I] at 370 \textmu m, and the CO $J=7-6$ and $J=6-5$ lines
            show the strongest emission off-source in the northeast where the object HH 22 MIR can be seen in the RGB images. It is not clear
            how much of the emission is contributed by V1647 Ori. The [N~II] emission behaves differently, showing less diffuse emission than most
            other objects, with a peak at the target coordinates, indicating that that emission originates from the source.

        \subsection{SED variability} \label{subsection:visible_and_missing_features}
        We found differences in the SEDs (Fig. \ref{SED_multiplot} and \ref{SED_multiplot2}) between our data and earlier observations for HH 381 IRS (8.28-100 \textmu m) and PP 13 S (50-160 \textmu m). The numerical factors were obtained by measuring the
        difference in the photometry to the spectrum of \textit{Herschel} and \textit{Spitzer} at the same wavelength. In the case of PP 13 S, the offset is up to factor 1.7 \citep{Cohen1983} (observed 1981), while the measurements of the  InfraRed Astronomical Satellite (IRAS, observed 1983.5) and
        Herschel (PACS photometry: March 2013) in the PACS range match. For HH 381 IRS,
        the offset ranges between a factor of  8.5 and 43.8 for the   Midcourse Space Experiment (MSX; data observed in 1996--1997) and 1.8-2.7 for
        IRAS (1983.5), compared to the \textit{Herschel} photometry (PACS photometry: November 2012).
        
        As there are already strong differences in the published data (at least for HH 381 IRS), and since the observations of \textit{Spitzer} (2010) and
        Herschel match quite well, it is unlikely that there is an issue with the observation. We therefore assume that the objects
        increased their luminosities over time, which took place in the IR.

\subsection{Rotational diagrams} \label{subsection:rotational_diagrams}
\begin{figure}
    \includegraphics[width=0.5\textwidth]{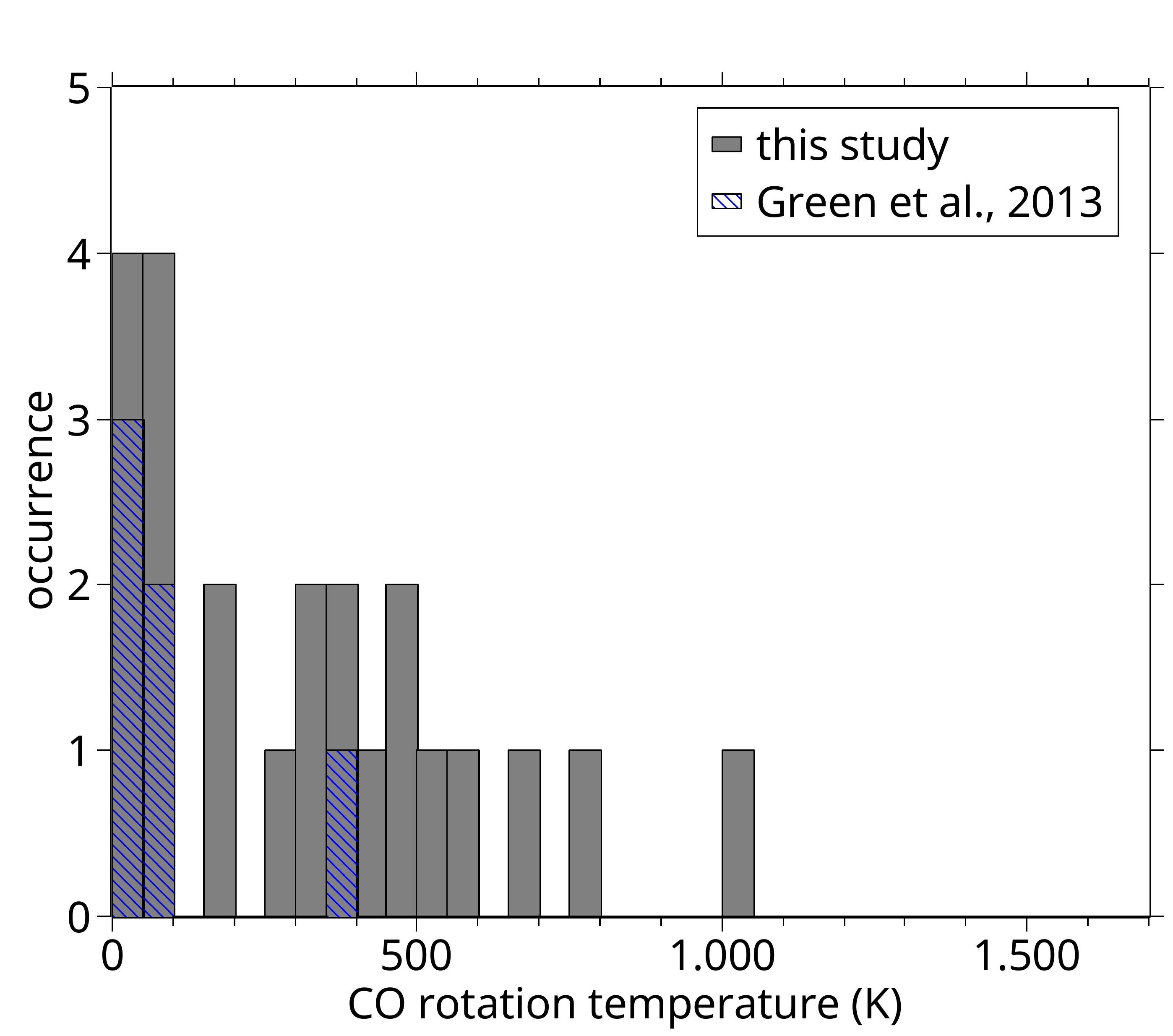}
    \caption{
        \footnotesize
        Occurrence of the CO rotation temperatures found in our targets and for V1057 Cyg,
        V1331 Cyg, and V1515 Cyg from \cite{Green2013}.
        Temperatures above 1700 K are not shown here because of the high uncertainty of the measurements.
    }
    \label{CO_histogram}
\end{figure}

\begin{figure}
    \includegraphics[width=0.5\textwidth]{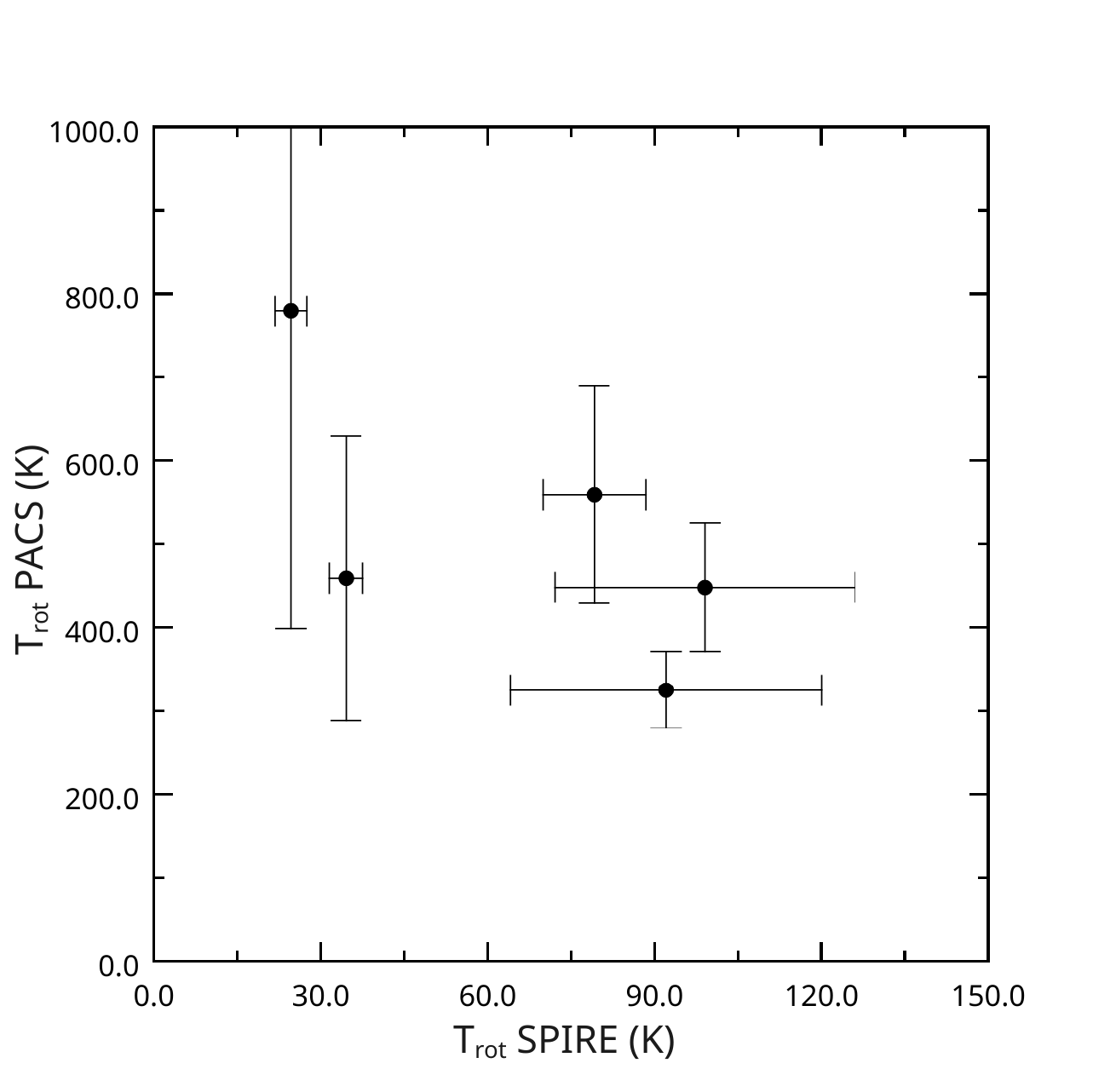}
    \caption{
        \footnotesize
        Temperature measurements of PACS long as a function of SPIRE measurements. Only temperatures derived by at least three CO lines are shown
        (see Table \ref{CO_rotational_temperature}) in order to have more robust error bars.
    }
    \label{CO_cold-mid}
\end{figure}

        We  derived CO rotational temperature for our targets; they  can be found in Table
        \ref{CO_rotational_temperature}, where  the column density is also listed. The method that we used is described in \cite{GoldsmithLanger} in detail and was applied on
    Herschel data in \citealt{GreenRotDiag} and \citealt{Dionatos2013}. We used the same approach as \citealt{GreenRotDiag}, separating the data into three domains of LTE,
    one for SPIRE ($J_{up} \leq 13$ or $E_u \leq 505$ K) and two
    for the different PACS orders (PACS R1, in this paper called PACS long for $14 \leq J_{up} \leq 24$ or $560$ K $\leq E_u \leq 1670$ K, and  PACS B2A/B2B, in this paper called
    PACS short for $J_{up}\geq 25$ or $E_u>1700$ K). However, we use the luminosity $L$ of the CO lines instead of the fluxes (same approach as in \citealt{Dionatos2013}) to eventually obtain the column densities
    $log_{10}(N)$ of the molecules. Data for the upper level energy $E_{u}$, Einstein coefficients $A_{ul}$,  and partition function $Q$ were obtained by the Cologne Database for Molecular Spectroscope (CDMS) \citep{Mueller2005}.
    For the plots, $N_u$ is given by $L \times 4 \pi / (h \times \nu_{ul} \times A_{ul})$, which is divided by the statistical weights of the respective line. As a function of the upper
    energy level $E_u$, a line can be fitted in which the slope $m$ measures the rotation temperature via $T_{rot} = -1/(m \times log(10)) $.
    The height of the line $h$ can be further used to calculate the column density of the data by following $log_{10}(N) = h + log_{10}(Q)$.
        We mention here that the rotation temperature $T_{rot}$ of the CO does not necessarily represent the kinetic temperature $T_{kin}$ of the gas.
        Figure \ref{rotation_diagrams} shows the individual fits to the CO rotation data to derive the CO rotational
        temperatures and column densities. Due to the small number of lines in some ranges, we also used CO lines below the $3\sigma$ limit. However, the fitting routine  takes into account the error and uses a weighting so
        that lines with a large error do not have a too strong impact on the fitting result. Lines below the $3\sigma$ limit are accordingly marked in the tables (see Table~\ref{lineFes}).
        We tried to find a correlation between the different rotational temperatures. The cold temperature from the SPIRE data and the warmer
        temperature from the PACS long data, shown in Figure \ref{CO_cold-mid}, provided enough data points for an analysis. Based on the available data, we could
        not find a correlation or conspicuousness among the objects.
        Figure \ref{CO_histogram} shows a
        histogram of the CO rotational temperature of our targets, and complements it with values derived for other classical FUors, V1057 Cyg, V1331 Cyg, V1515 Cyg, and
    V1735 Cyg, as derived by \citet{Green2013}. We also found  in our sample  several low-temperature components on the  order of $0-100$ K; however,  many targets in our sample
        have temperatures around $400-500$ K, and possibly very hot temperature components of $1000-1500$ K.
        This will be discussed in more  detail in the next section.

\section{Discussion}
 \subsection{CO temperature components}
   Several objects show low-$J$ transitions of CO in the SPIRE range, revealing a cool component ($<100$ K). The rotational diagram indicates temperatures  on the order of $40-50$ K, with a few objects showing lower temperatures, such as V883 Ori, and V1647 Ori, and high column densities.
    No evidence of low temperatures is found in Parsamian 21, with no detected lines in the SPIRE range above $3\sigma$.
    
    For the CO rotational lines in the PACS long range, the uncertainties on the temperature do not allow us  to distinguish differences in the targets. This range consistently provides values on the order of 400 K (Parsamian 21, V883 Ori, and HH 381 IRS show higher temperatures, but large uncertainties due to the low number of lines available).
    Interestingly, even when low-$J$ transitions were not detected (i.e., no low rotational temperature) CO emission lines showing a ``medium'' temperature was almost always found, suggesting that this component is ubiquitous in outbursting young sources.
    
    The PACS short range, where high-$J$ transitions lie, has revealed very few detected lines, making the detection of a hot component ($>1000$ K) difficult to assess. There is a
    suggested presence of such a component in HH 381 IRS and HH 354 IRS, albeit with large  uncertainties. V883 Ori, on the other hand, shows a  temperature similar to that in the PACS long
    range, suggesting that the hot component does not exist in this target; it is interesting to note that this same target also shows the coolest temperature in SPIRE, suggesting that V883 Ori generally has a lower temperature than other targets.
    V346 Nor may display a hot component, again with a rather large uncertainty.
    
    The similarity of the line fluxes in those objects compared to HH 381 IRS, but the strong difference in the high-temperature domain, might be related to the luminosity increase
    of HH 381 IRS (further discussed in Sect. \ref{young_objects}), although we caution that few lines were detected, probably because of the higher noise and stronger continuum. 
    Optical depth effects (e.g., due to the optical thickness of the dust continuum) may play a role, although the SEDs
    (dominated by dust emission) are  similar to other targets with weak or no CO emission in the PACS range, as discussed above. 
     
\subsection{Evolved objects}
    For the more evolved objects in our sample,  EX Lup (with no envelope) or BRAN 76 (a possibly evolved FUor with less envelope and a silicate feature in emission,
    like FU Ori), our spectra are quite noisy. Our analysis has tentatively detected a few CO lines in EX Lup (see Sect. \ref{EXLup_results}), despite the formal statistical uncertainties.
    BRAN 76, on the other hand, displayed CO emission only in the PACS short range. V1647 Ori, whose classification is unclear and with its flat SED, shows enough signal for the detection of several CO lines. 
    For the moment there is not enough evidence to make a link between the detectability of molecular or atomic lines and the properties of the targets. In a subsequent
    analysis, we will investigate this issue by modeling the objects with the radiation thermo-chemical code ProDiMo \citep{2009A&A...501..383W}, and thus derive geometrical information on the emission
    lines, in particular thanks to the enhanced version of the code by \cite{Rab2017} that takes into account the envelope around outbursting sources.
    
\subsection{Young embedded objects} \label{young_objects}
    Our sample of outbursting sources shows a large variety of SEDs. \cite{2007ApJ...658..487Q} suggested that this could be due to the evolutionary stage (Class I versus Class II YSO),
    but some features cannot be explained  by the evolutionary stage alone. PP 13 S, V883 Ori, HH 381 IRS, and V346 Nor, which are  all FUors or FUor-like objects, show similar shapes of their SEDs
    and should in the Quanz scheme be in a similar evolutionary stage. Therefore, the question arises whether there is any correlation between properties among those objects.
    For the emission lines in these targets, they are strongest in PP 13 S and V346 Nor, while V883 Ori and HH 381 IRS show lines on a much lower level in the SEDs,
    despite similar continuum levels in the far-infrared.
    The lines themselves are very different between PP 13 S and V346 Nor. While PP 13 S shows strong atomic oxygen emission and only moderate CO,
    V346 Nor shows strong CO emission but weak oxygen. The CO$_2$ gas absorption at $15$ \textmu m is unique among our targets, and even compared to other YSOs
    it appears very rarely, both in absorption and emission (\citealt{Lahuis_2005}, \citealt{Pontoppidan_2008}).
    For HH 381 IRS we note the increased luminosity over the last years in the IR. A strong increase in luminosity
    on a long timescale indicates that it might be at the beginning phase of a long outburst, with a strong mass accretion onto the embedded object.
    However, we note that, although detected, the [O~I] line flux at 63 \textmu m, a common tracer of high accretion rates \citep{Green2013}, is particularly
    weak in HH 381 IRS, like the other detection of OH, H$_2$O, and CO lines. Taking into account the distance and looking at the luminosity of the line (see Fig. \ref{lbol}), the [O~I] emission of HH381 appears to be among the more luminous objects  in our sample, still below V346 Nor and about one order of magnitude above PP 13 S.
    For HH 381 IRS, the controversy between the rising bolometric luminosity 
    and the absence of strong accretion tracers is very interesting as it shows that the bolometric luminosity is not dominated by accretion luminosity,
    but instead is dominated by photospheric luminosity of the central source. The mass accretion and accretion luminosity may quickly drop after the onset of the burst 
    because of the exhausted mass reservoir in the innermost disk, but the star bloats in response to the accreted energy which leads to a dramatic increase in the photospheric luminosity.
    As a result the total luminosity may keep rising. Similar situations in the evolution of bursting stars were shown recently in \cite{2019MNRAS.484..146E}.
    As we did for the evolved objects, we will investigate the characteristics of the young objects in the future with ProDiMo since
    for the moment there is no correlation visible.
    
\subsection{Correlation between lines}
\begin{figure}
    \includegraphics[width=0.5\textwidth]{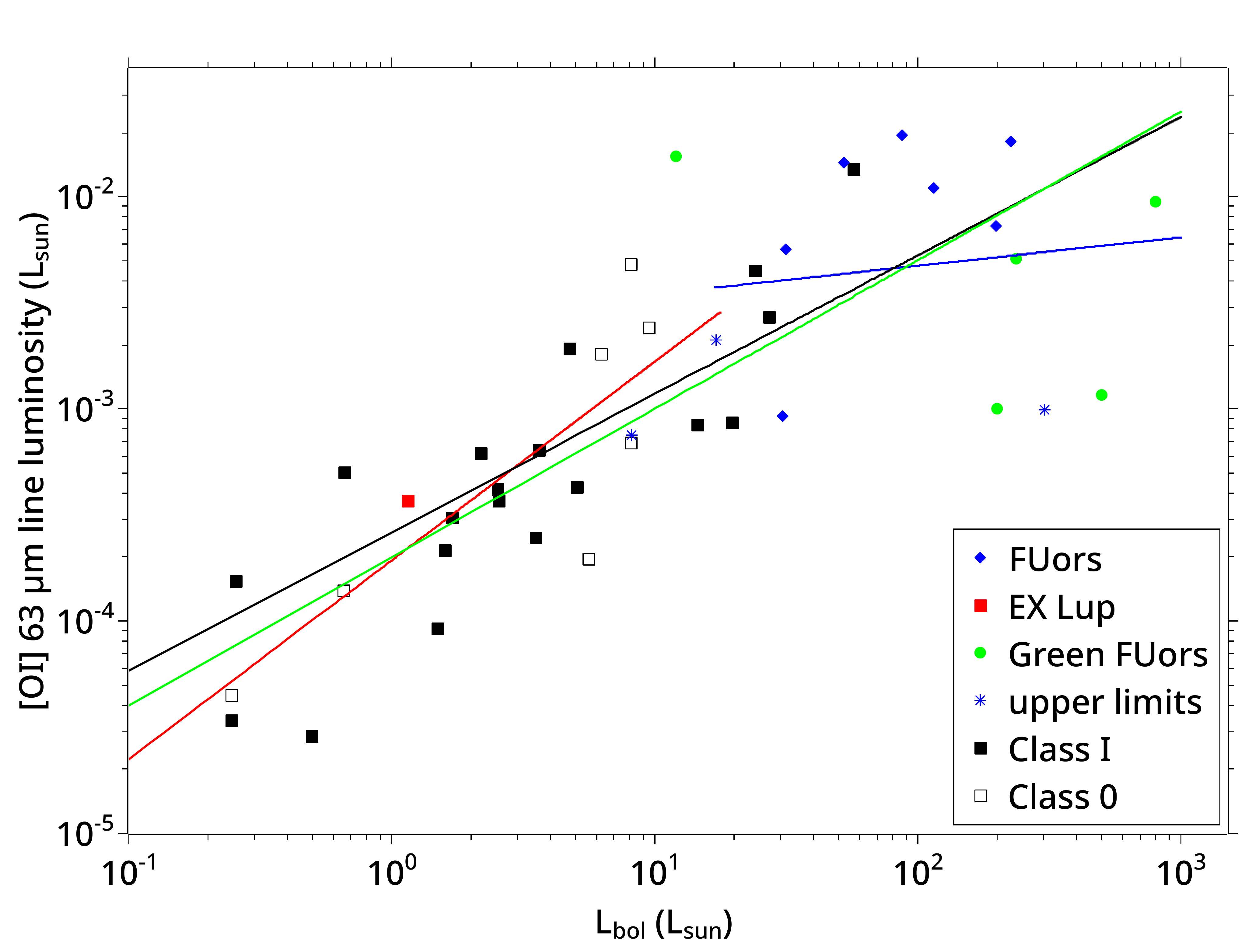}
    \caption{
        \footnotesize
    [O~I] luminosity at 63 \textmu m as a function of the bolometric luminosity of the object. The black curve represents the best-fit result of the data with $f(x)=0.65 \cdot x-3.6$ and $\chi^2 = 7.8$. We show two additional lines which partially fit the data (in red) with $f(x)=0.94 \cdot x-3.7$, $L_{bol}<17 L_{\sun}$, and $\chi^2 = 3.5$, and (in blue) with $f(x)=0.13 \cdot x-2.6$, $L_{bol}>17 L_{\sun}$, and $\chi^2 = 3.2$. We also show  the best fit of \cite{Green2013} with $f(x)=0.70 \cdot x-3.7$ (in green). The data points that are indicated as upper limits ($3\sigma$) and EX Lup are not used to create the fits. 
    }
    \label{lbol}
\end{figure}

    In the following, we discuss the correlation between lines that have been found either in our data or by other groups, in which case we check their consistency within our sample of YSOs.
    In Fig. \ref{lbol} we show the luminosity of the [O~I] emission at 63 \textmu m as a function of the bolometric luminosity of our targets in a log-log plot. Table \ref{OILum} shows the 63 \textmu m [O~I] luminosity of our targets. Haro 5a IRS is missing here.
    For Parsamian 21, V733 Cep, and V883 Ori we only show upper limits. In addition to our sample, we used HBC 722, V1735 Cyg, V1057 Cyg, V1515 Cyg, and FU Ori from \citealt{Green2013}.
    Along with the targets, we show our best-fit result as black line in the plot (not including the upper limits and EX Lup). We find a clear correlation of the bolometric luminosity and
    the luminosity of the 63 \textmu m emission line with $f(x)=0.65 \cdot x-3.6$. However, as mentioned in Sect.  \ref{young_objects}, some objects do not follow this principle.
    Due to the ``knee'' at $\approx$ 17L$_{\sun}$, we created two other fits, one for $L >$ 17 $\cdot L_{\sun}$ and one for $L <$ 17 $\cdot L_{\sun}$. For comparison, the  analysis of \citealt{Green2013} resulted in $f(x)=0.70 \cdot x-3.7$. Our sample of FUors is still small, but it seems that FUors experience a saturation of [O~I] emission compared to Class 0/I objects.
    However, this behavior is  mainly caused by four of the targets from \citealt{Green2013} and needs further investigation.
    
    Remarkably, our analysis suggests the tentative detection of multiple water and OH emission lines in the PACS data for PP 13 S, Re 50 N IRS 1, and V346 Nor,
    and possibly other targets, where we see the water emission in a slightly extended area around each target. We did not find a correlation between the presence of water and OH, but we will investigate this
    possibility during our upcoming analysis when taking into account spatial properties.
    
    Among the DIGIT and HOPS objects, \cite{Green2013} find that a strong CO $J=16-15$ line correlates well with N(PACS long). We find the strongest
    emission of this particular CO line in PP 13 S, Re 50 N IRS 1, and V346 Nor, each   with very strong emission lines in general. These
    objects show the highest column densities in the PACS data at warm temperature (PACS long) among our sample, clearly supporting this result.
    
    Furthermore, Green also discussed the possibility that Class 0/I objects (FUors and FUor-like) have hotter and more excited CO rotation lines than Class II objects.
    The most Class II-like objects in our sample are EX Lupi (EXor) and V1647 Ori (likely EXor).
    Both objects show, if our data allows an analysis, relatively low temperatures compared to the rest of our sample, which   supports this assumption.

    \cite{Green2013} argued that weak silicate features, in emission or absorption, do not correlate well with [O~I] emission.
    In our sample, the \textit{Spitzer} data show strong silicate features
    for Haro 5a IRS, PP 13 S, and Re 50 N IRS 1. HH 354 IRS also shows strong absorption near the silicate feature, although in that case the strong absorption is also
    due to water ice. The targets show several other ice features, such as water, methanol and ammonium, methane, and ammonia and carbon dioxide at 15 \textmu m
    (Figs. \ref{irs_spitzer} and \ref{irs_co2}). We note that a CO$_2$ \emph{\emph{gas}} line was
    detected in absorption in PP13 S, the only target in our sample showing this feature. Each of these targets shows strong emission of [O~I] at 63.1 \textmu m and emission
    of the 145.5 \textmu m line above the $3\sigma$ limit. In addition, V883 Ori and HH 381 IRS, weak [O~I] emitters, show weak silicate absorption. HH 381 IRS, the only
    target for which \textit{Spitzer} covered the 5-10 \textmu m range as well, also did not detect evidence of the other ice components. It provides further evidence that
    the correlation suggested by \cite{Green2013} is indeed strong.
    
\begin{table}
    \caption{Bolometric and 63 \textmu m [OI] luminosities}
    \label{OILum}
    \centering
    \begin{tabular}{l|rrr}
    \hline\hline
        Target      & \multicolumn{1}{c}{L$_{bol}$} & \multicolumn{1}{c}{[OI]}      & \multicolumn{1}{c}{[OI] Error}\\
                    &\multicolumn{3}{c}{($L_{\sun}$)}           \\
        \hline
        BRAN 76     & 52.3    & 1.445e-2          & 9.2e-4    \\
        EX Lup      & 1.15    & 3.69e-4           & 1.3e-5    \\
        HH354 IRS   & 87.2    & 1.936e-2          & 2.3e-4    \\
        HH381 IRS   & 225     & 1.805e-2          & 2.2e-4    \\
        \textit{Parsamian 21}& 8.15    & \textit{7.5e-4}   &           \\
        PP 13S      & 31.4    & 5.668e-3          & 4.7e-5    \\
        Re 50 N IRS 1& 115    & 1.090e-2          & 8.8e-5    \\
        V346 Nor    & 198     & 7.25e-3           & 2.0e-4    \\
        \textit{V733 Cep}    & 17.2    & \textit{2.1e-3}   &           \\
        \textit{V883 Ori}    & 302.5   & \textit{9.9e-4}   &           \\
        V1647 Ori   & 30.5    & 9.25e-4           & 6.3e-5    \\

    \hline
    \end{tabular}
        \tablefoot{
                        We show the bolometric luminosity and the luminosity of the 63 \textmu m [O~I] emission and its error
                        on our targets that was used for the analysis. Targets in italics show upper limits for the [O~I] luminosity.
                }
\end{table}
    
\section{Summary} 
    We provided the SEDs, including photometry and integral field spectroscopy of 12 FUors and EXors using mainly \textit{Herschel} and
    \textit{Spitzer} data, and additional published photometry. In most cases we covered the SED from the optical or near-IR up to the sub-millimeter range, tracing the 
    emission of the envelope and disk of the targets. Our observations revealed the complex environment in which most of our targets reside, which we discuss in detail. The targets show a wide variety of
    spectral features, from silicate in absorption to emission, and 
    often ice features from CO$_2$ and other molecules from $5-9$ \mic, when embedded. We detected for one of our targets a rare CO$_2$ gas absorption at $15$ \textmu m.
    We derived CO rotation temperatures in three temperature domains, finding different temperature regimes: $<100$ K and $400-500$K. A third high-temperature component may be present in some targets as well.
    We provided three-color images of the targets in the PACS and SPIRE  wavelength ranges. Finally, we
    provided line maps and line fluxes. We tentatively detected water emission for PP 13 S, Re 50 N IRS 1,
    and V346 Nor. 
    We found a significant increase of luminosity in HH 381 IRS and PP 13 S, possibly due to early phases of an outburst. With our data set, we aim to use
    the data of all the objects in our sample for further analysis with ProDiMo to refine the models of outbursting sources and derive
    properties of the targets, namely the circumstellar disk and the surrounding envelope. 
 
\begin{acknowledgements}
    We thank the Swiss National Science Foundation (SNSF) (project number 200021L\_163172) and the Austrian
    Science Fund (FWF) (project number I2549-N27) for their funding of this project. M. Audard thanks C. Pearson for his early reduction of the SPIRE data. We thank A. Kospal for providing the SPITZER IRS spectrum for Parsamian 21. O. Dionatos acknowledges support from the
    Austrian Research Promotion Agency (FFG) under the framework of the Austrian Space Applications Program (ASAP) projects JetPro* and PROTEUS (FFG-854025, FFG-866005). This publication makes use of
    data products from the Two Micron All Sky Survey, which is a joint project of the University of Massachusetts and
    the Infrared Processing and Analysis Center/California Institute of Technology, funded by the National Aeronautics
    and Space Administration and the National Science Foundation.
\end{acknowledgements}

\bibliographystyle{aa}
\bibliography{./publication_1}

\begin{sidewaystable*}
        \caption{Photometry flux densities of PACS and SPIRE for each wavelength window
    }
    \label{photometry_flux_table}
    \centering
    \begin{tabular}{l|rlrlrlrlrlrl}
    \hline\hline
    Object& \multicolumn{2}{c}{70 \textmu m}& \multicolumn{2}{c}{100 \textmu m}& \multicolumn{2}{c}{160 \textmu m}& \multicolumn{2}{c}{250 \textmu m}& \multicolumn{2}{c}{350 \textmu m}& \multicolumn{2}{c}{500 \textmu m} \\
   & \multicolumn{2}{c}{(Jy)}& \multicolumn{2}{c}{(Jy)}& \multicolumn{2}{c}{(Jy)}& \multicolumn{2}{c}{(Jy)}& \multicolumn{2}{c}{(Jy)}& \multicolumn{2}{c}{(Jy)} \\
    \hline
    BRAN 76&1.0182&$\pm$0.0071&0.908&$\pm$0.014&1.317&$\pm$0.083&1.1&$\pm$1.1&0.93&$\pm$0.97&0.69&$\pm$0.84 \\
    EX Lup&1.027&$\pm$0.010&0.985&$\pm$0.011&0.891&$\pm$0.026&0.38&$\pm$0.62&0.24&$\pm$0.49&0.09&$\pm$0.29 \\
    Haro 5a IRS&113.5&$\pm$4.2&156.5&$\pm$9.5&230&$\pm$11&117&$\pm$11&79.3&$\pm$9.3&37.8&$\pm$6.4 \\
    HH 354 IRS&69.05&$\pm$0.44&90.8&$\pm$1.1&85.1e&$\pm$0.67&36.0&$\pm$6.0&19.7&$\pm$4.4&8.5&$\pm$2.9 \\
    HH 381 IRS&75.97&$\pm$0.53&80.2&$\pm$1.2&74.76&$\pm$0.81&27.0&$\pm$5.2&14.1&$\pm$3.8&6.6&$\pm$2.5 \\
    Parsamian 21&12.74&$\pm$0.13&11.05&$\pm$0.17&10.33&$\pm$0.12&3.8&$\pm$2.0&2.5&$\pm$1.6&1.3&$\pm$1.1 \\
    PP 13 S&55.34&$\pm$0.21&68.42&$\pm$0.32&69.09&$\pm$0.37&33.8&$\pm$5.8&20.8&$\pm$4.6&9.9&$\pm$3.2 \\
    Re 50 N IRS 1&113.6&$\pm$2.8&112.4&$\pm$9.4&136.0&$\pm$3.9&47.9&$\pm$6.9&27.6&$\pm$5.3&12.4&$\pm$3.6 \\
    V346 Nor&92.50&$\pm$0.48&92.24&$\pm$0.70&92.7&$\pm$1.2&40.7&$\pm$6.4&25.0&$\pm$5.0&12.1&$\pm$3.5 \\
    V733 Cep&4.2&$\pm$1.8&6.4&$\pm$2.2&8.2&$\pm$2.6&7.0&$\pm$2.6&7.6&$\pm$2.8&5.9&$\pm$2.5 \\
    V883 Ori&145.4&$\pm$1.2&124&$\pm$1.9&86.9&$\pm$2.0&32.5&$\pm$5.7&18.0&$\pm$4.2&7.7&$\pm$2.8 \\
    V1647 Ori&32.05&$\pm$0.30&38.54&$\pm$0.92&42.6&$\pm$1.1&18.5&$\pm$4.3&11.4e&$\pm$3.4&5.7&$\pm$2.4 \\
    \hline
    \end{tabular}
    
    \vspace{1 cm}
    
    \caption{CO rotational temperatures and column densities}
    \label{CO_rotational_temperature}
    \centering
    \begin{tabular}{l|rrr|rrr|rrr}
    \hline\hline
    Object          & $T_{rot}$ SPIRE   &$T_{rot}$ PACS long
                                                        &$T_{rot}$ PACS short   & $N$ SPIRE         &$N$ PACS long      &$N$ PACS short &log$_{10}$(Q)$_{SPIRE}$
                                                                                                                                                &log$_{10}$(Q)$_{PACS long}$
                                                                                                                                                        &log$_{10}$(Q)$_{PACS short}$    \\
                    & \multicolumn{3}{c|}{(K)}                                  & \multicolumn{3}{c|}{(column density (log$_{10}$($N$) cm$^{-2}$)}
                                                                                                                                        &                           \\
    \hline

    BRAN 76         &                   & 540$\pm$220   &                       &                   & 18.10$\pm$0.19     &              &       & 2.27  &     \\
    Ex Lup          &                   & 350$\pm$110   & <70000 (4400)         &                   & 17.75$\pm$0.16     & 18.1$\pm$2.6 &       & 2.10  & 3.14 \\
    Haro 5a IRS     & 58.4   $\pm$3.2   &               &                       & 20.234 $\pm$0.027 &                    &              & 1.35  &       &     \\
    HH 354 IRS      & 170    $\pm$90    & 366$\pm$59    & 1040$\pm$380          & 18.70$\pm$0.11    & 18.323$\pm$0.086   &18.23$\pm$0.19& 1.79  & 2.11  & 2.55 \\
    HH 381 IRS      & 314 $\pm$55       & <5700 (2800)  & 1850$\pm$520          & 18.193$\pm$0.018  & 17.78$\pm$0.18     &17.685$\pm$0.086& 2.05& 2.95  & 2.79 \\
    Parsamian 21    & \textit{25}       & <15000 (1600) &                       & \textit{19.13}    & 17.53$\pm$0.14     &              & 1.00  & 2.73  &     \\
    PP 13 S         & 92$\pm$28         & 326 $\pm$46   &                       & 19.28$\pm$0.10    & 18.544$\pm$0.078   &              & 1.54  & 2.07  &     \\
    Re 50 N IRS 1   & 99 $\pm$27            & 449$\pm$77    & <11000 (6800)         & 19.309$\pm$0.086  & 18.473$\pm$0.077   &17.83$\pm$0.15& 1.57  & 2.20  & 3.32 \\
    V346 Nor        & 79.1$\pm$9.2      & 560$\pm$130   &                       & 19.733$\pm$0.043  & 18.546$\pm$0.087   &              & 1.48  & 2.29  &     \\
    V733 Cep        & 39.6$\pm$8.1      & \textit{290}  & \textit{150}          & 19.45$\pm$0.13    & \textit{18.14}     & 24.85$\pm$   & 1.19  & 2.01  & 1.75 \\
    V883 Ori        & 24.6$\pm$2.8      & 780$\pm$380   & 460$\pm$250           & 19.944$\pm$0.097  & 17.63$\pm$0.14     &19.35$\pm$0.66& 1.00  & 2.43  & 2.21\\
    V1647 Ori       & 34.5$\pm$3.0      & 460$\pm$170   & \textit{650}          & 19.843$\pm$0.069       & 17.79$\pm$0.16     &\textit{17.97}& 1.14  & 2.21  & 2.35\\
    \hline
    \end{tabular}
        \tablefoot{
        CO rotational temperature in K, total column density and partition function Q for PACS spectra and SPIRE. In some cases it was
        not possible to derive a temperature because not enough
        lines (<2) were detected. The errors were derived from the uncertainties of the fits (see Fig. \ref{rotation_diagrams}). For values in italics, we found only two lines
        in the spectrum of the respective instrument and consequently do not provide an error on the obtained value. For EX Lup, HH 381 IRS, and Re 50 N IRS 1 we provide an upper limit for the CO rotation
        temperature and the measured value from the rotation diagrams in parentheses.
    }

\end{sidewaystable*}

\begin{appendix}
\section{PACS/SPIRE line maps}
\label{line_maps}
\begin{figure*}
\includegraphics[width=0.33\textwidth, trim={0cm 0 0cm 0}, clip]{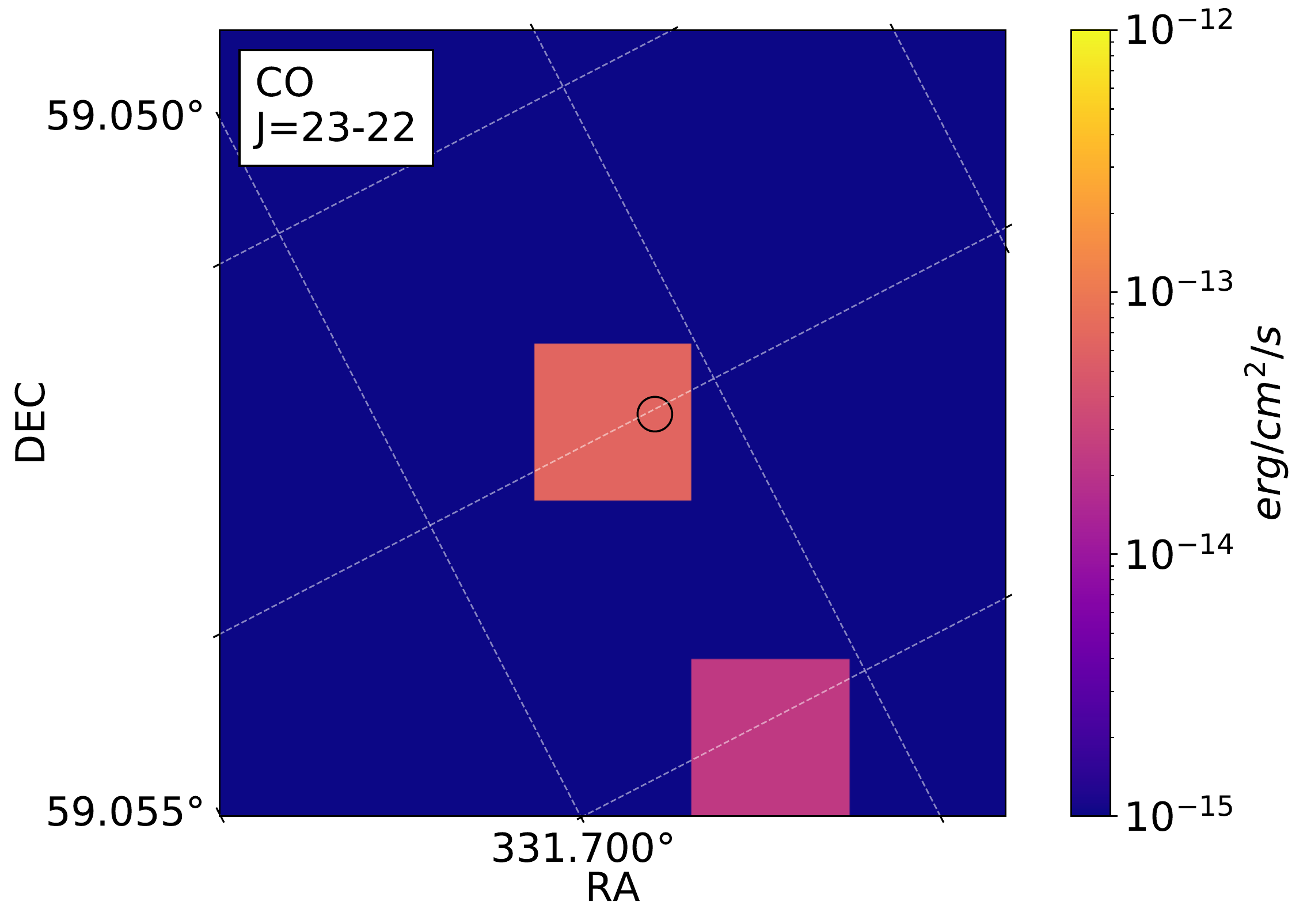}\hspace{-0.1cm}
\includegraphics[width=0.33\textwidth, trim={0cm 0 0cm 0}, clip]{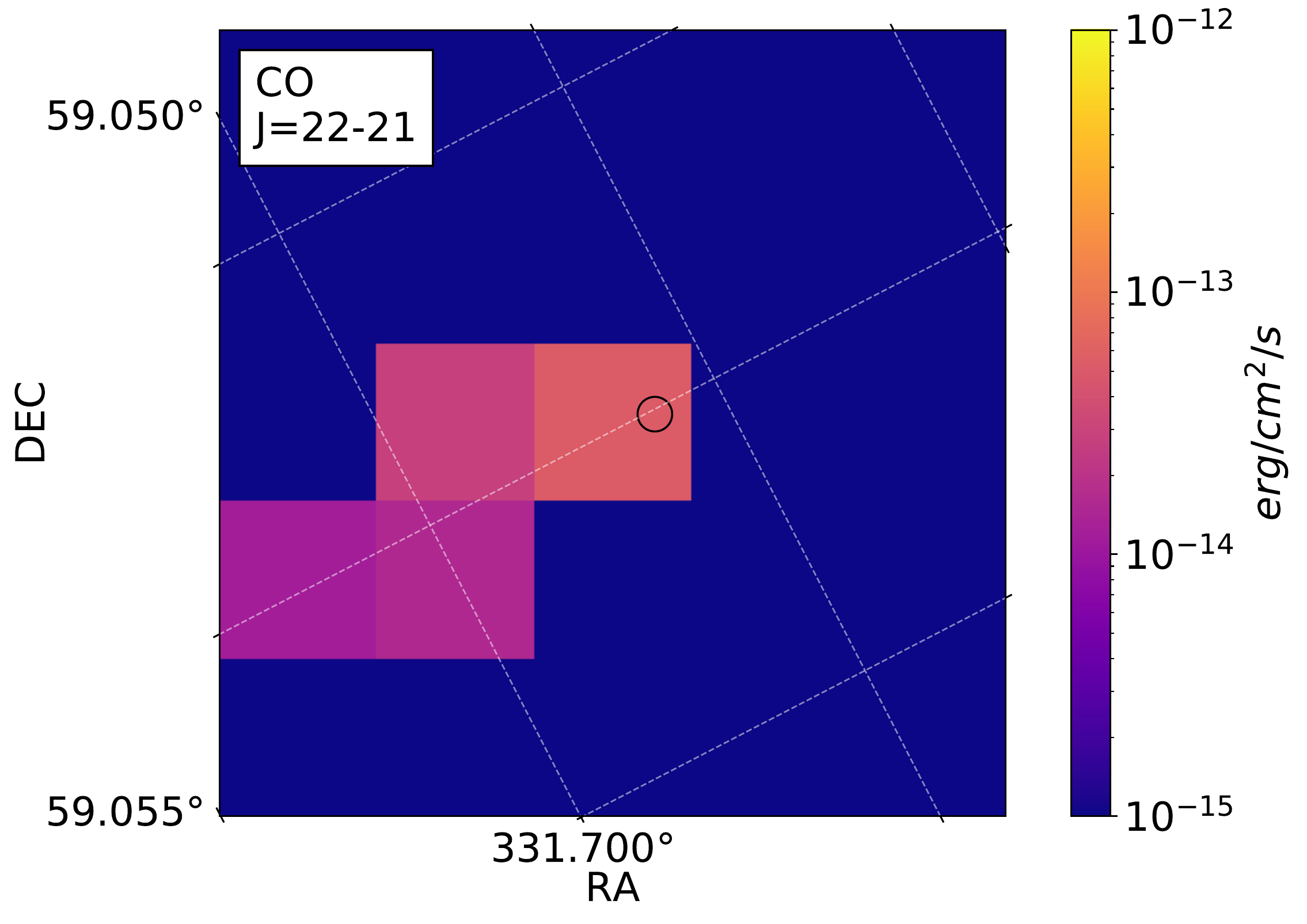}\hspace{-0.1cm}
\includegraphics[width=0.33\textwidth, trim={0cm 0 0cm 0}, clip]{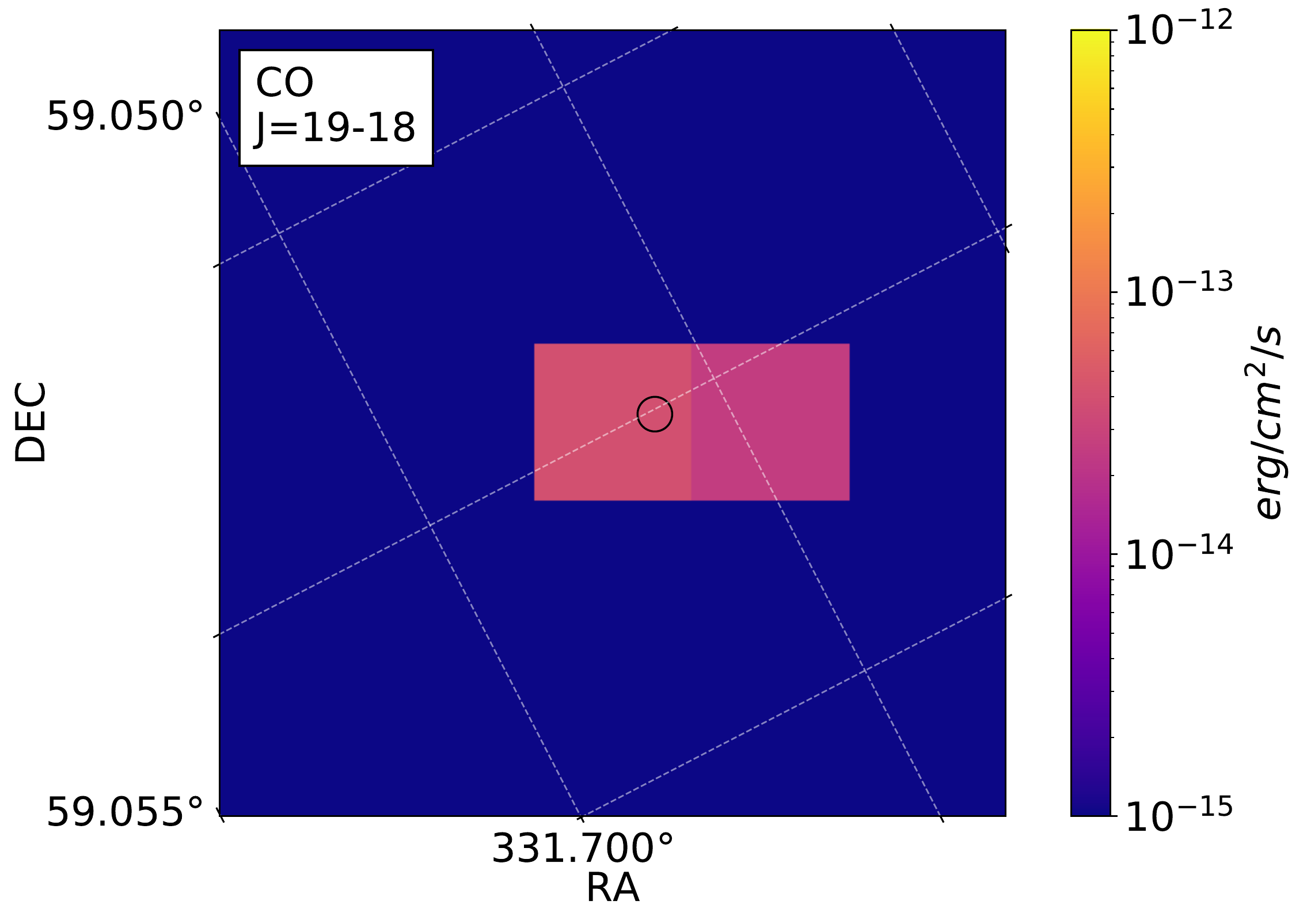}\hspace{-0.1cm}\\
\includegraphics[width=0.33\textwidth, trim={0cm 0 0cm 0}, clip]{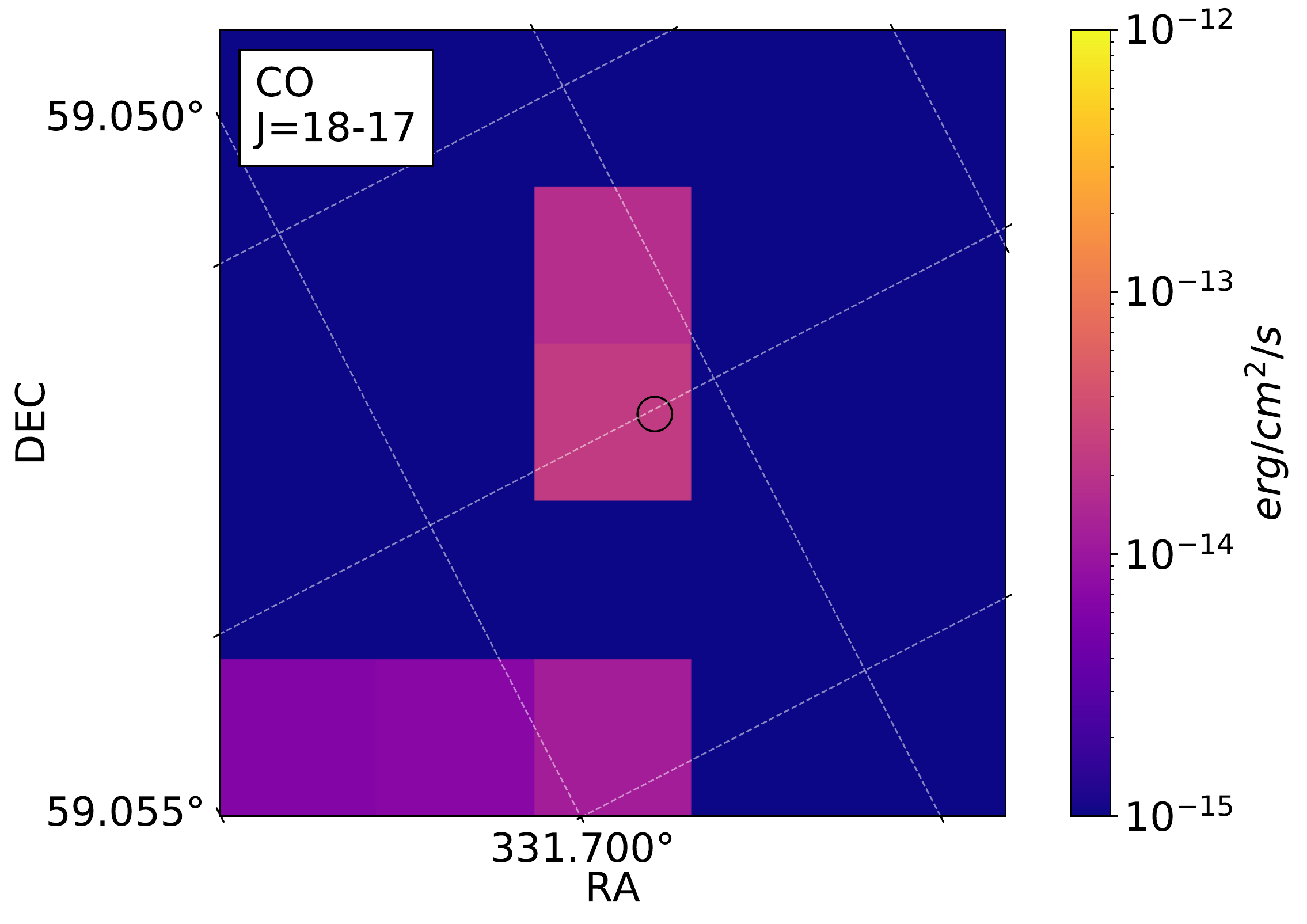}\hspace{-0.1cm}
\includegraphics[width=0.33\textwidth, trim={0cm 0 0cm 0}, clip]{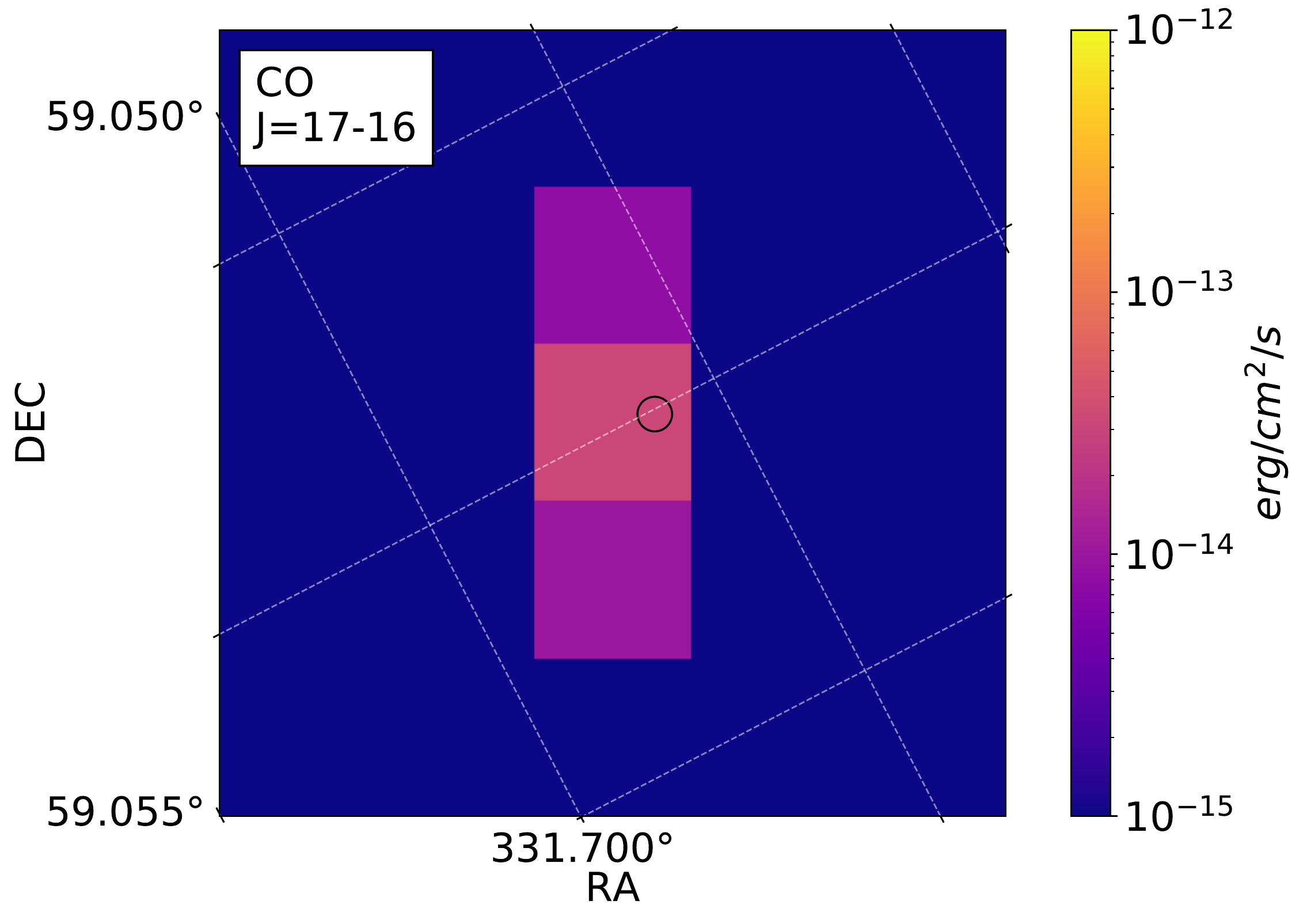}\hspace{-0.1cm}
\includegraphics[width=0.33\textwidth, trim={0cm 0 0cm 0}, clip]{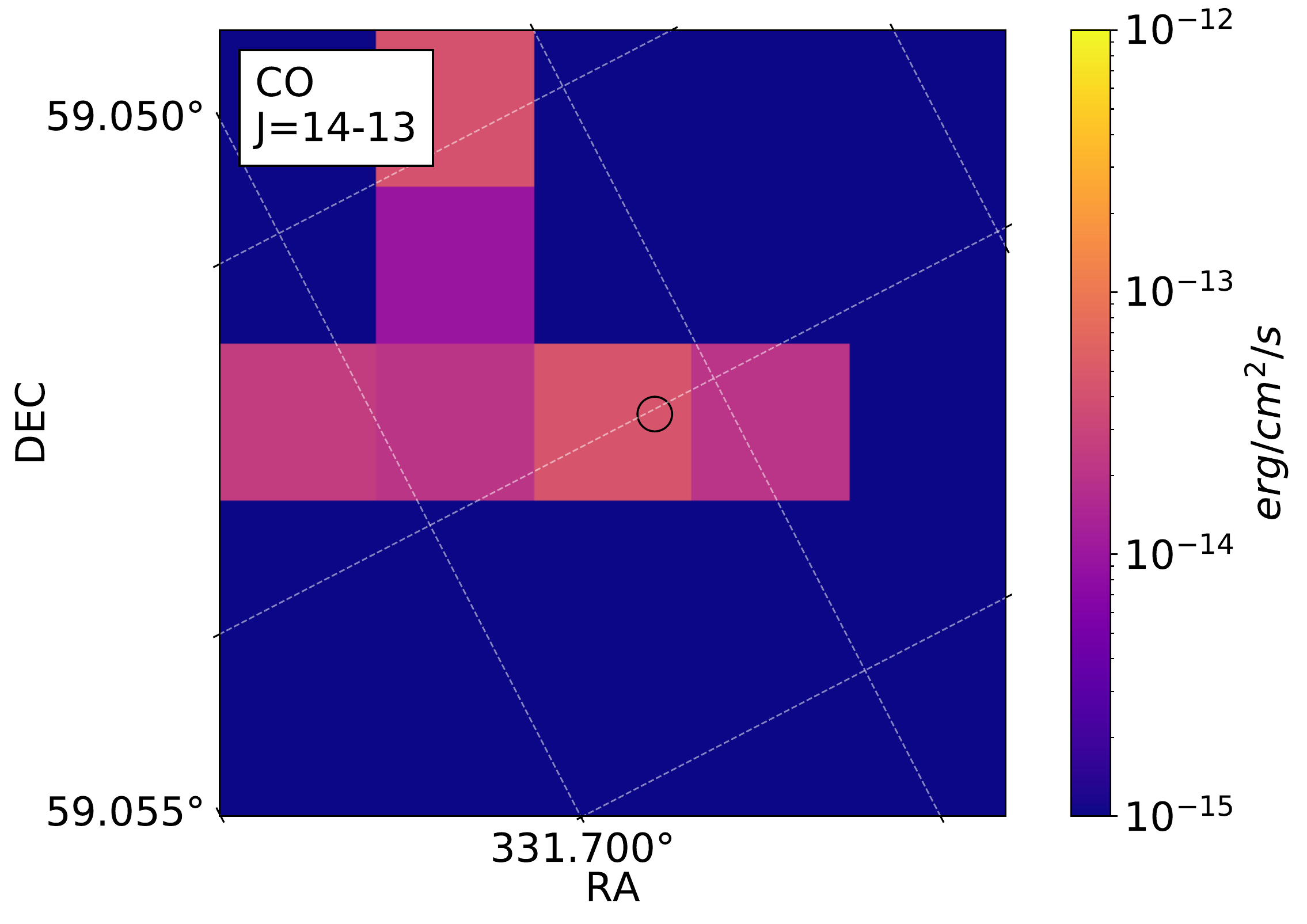}\hspace{-0.1cm}\\
\includegraphics[width=0.33\textwidth, trim={0cm 0 0cm 0}, clip]{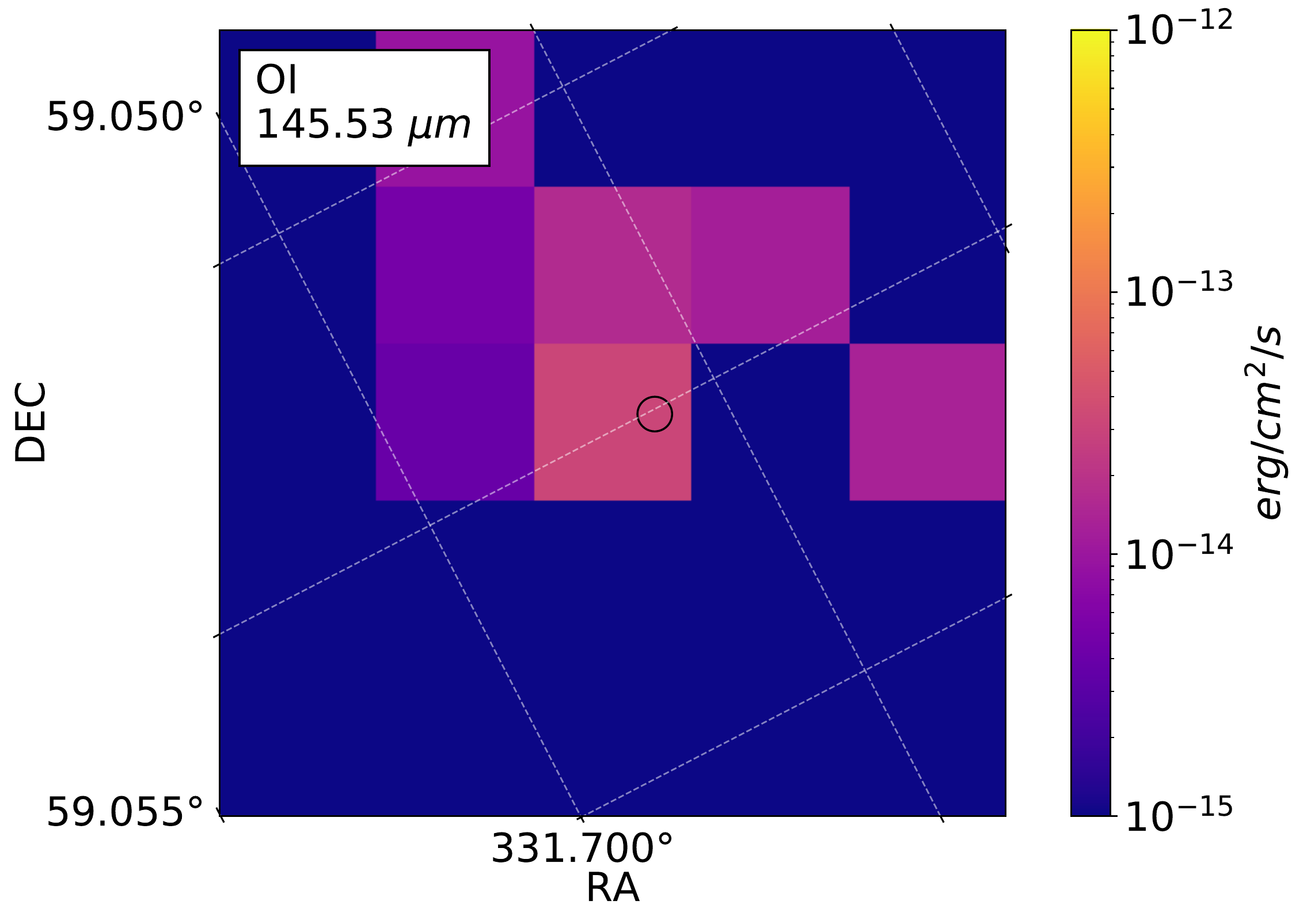}\hspace{-0.1cm}
\includegraphics[width=0.33\textwidth, trim={0cm 0 0cm 0}, clip]{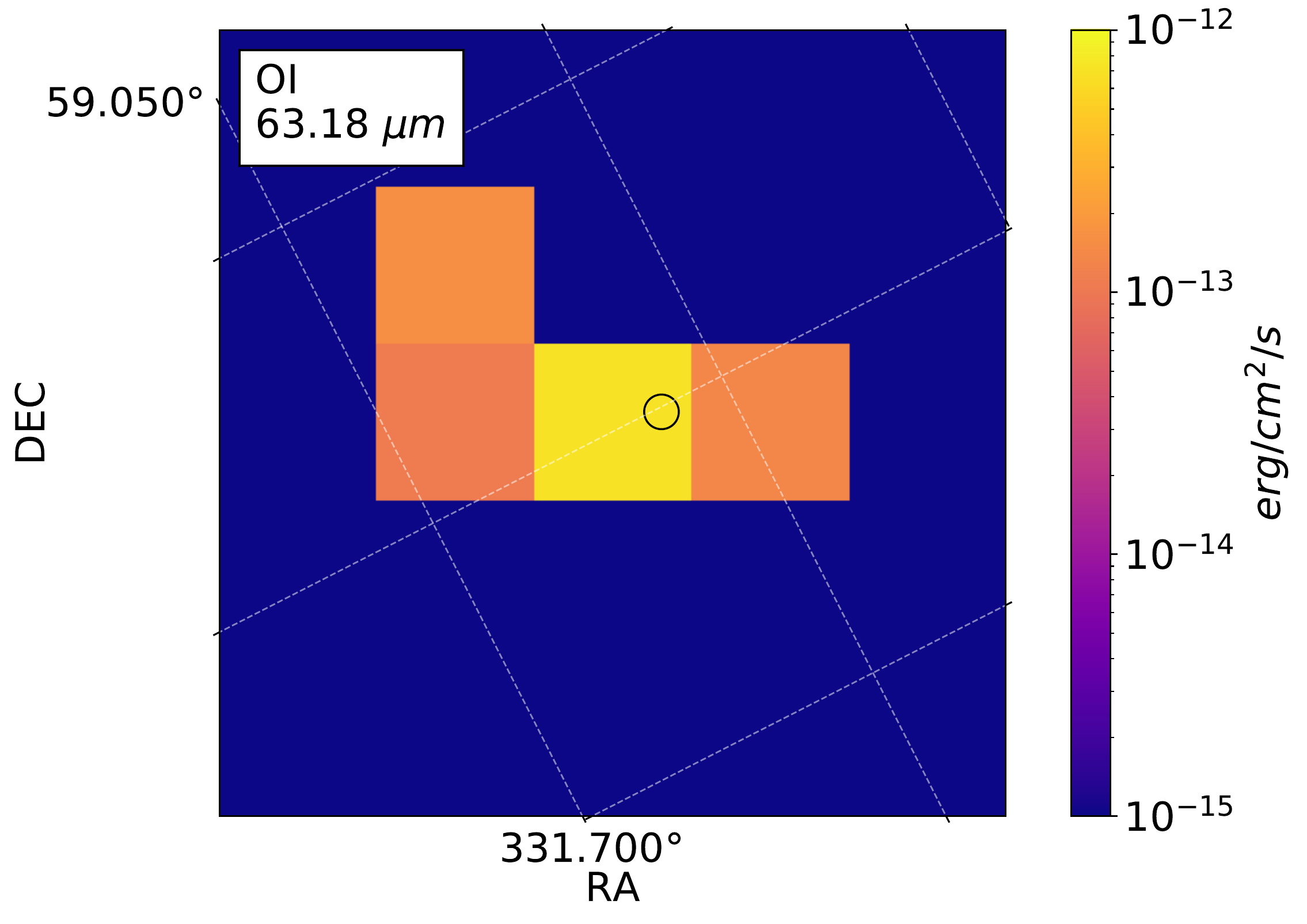}\hspace{-0.1cm}
    \caption{
                \footnotesize
                Line maps of PACS with visible lines for HH354 IRS. The maps show a specific bandwidth around emission lines of molecular or atomic transitions in the spectra coming from the integral field spectroscopy from PACS. The bandwidth is
                defined for each source by a fitting routine. A threshold cuts off noise below a specific value. For molecular lines, the species is mentioned in the upper left corner of the
                plot, together with the energy transition for CO and its isotopolog  and the wavelength for other molecules. For
                the atomic lines,   next to the isotope we give the frequency of the transition. The circle in each of the plots represents the coordinates of the optical FUor/EXor.
        }
\end{figure*}

\begin{figure*}
\includegraphics[width=0.33\textwidth, trim={0cm 0 0cm 0}, clip]{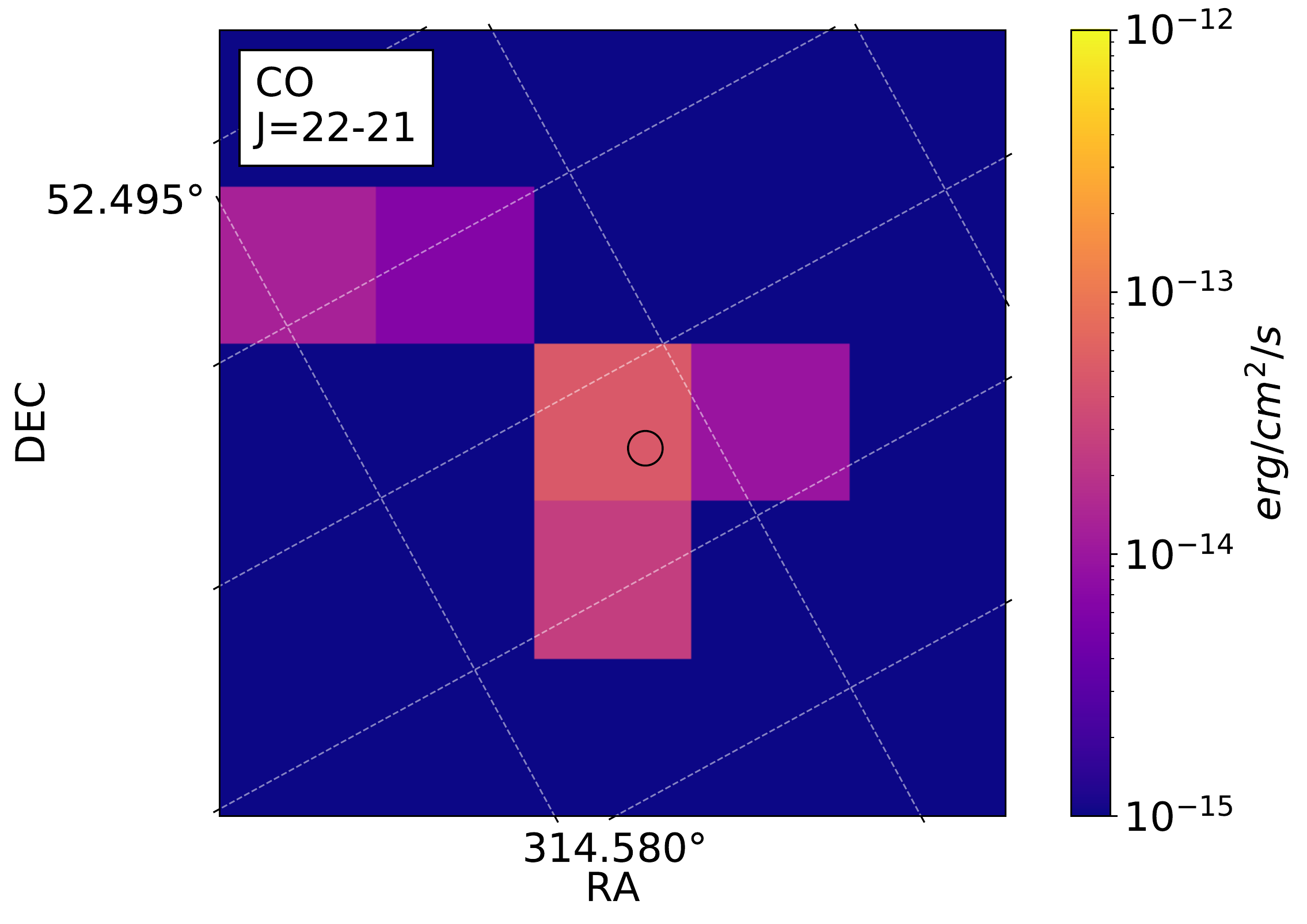}\hspace{-0.1cm}
\includegraphics[width=0.33\textwidth, trim={0cm 0 0cm 0}, clip]{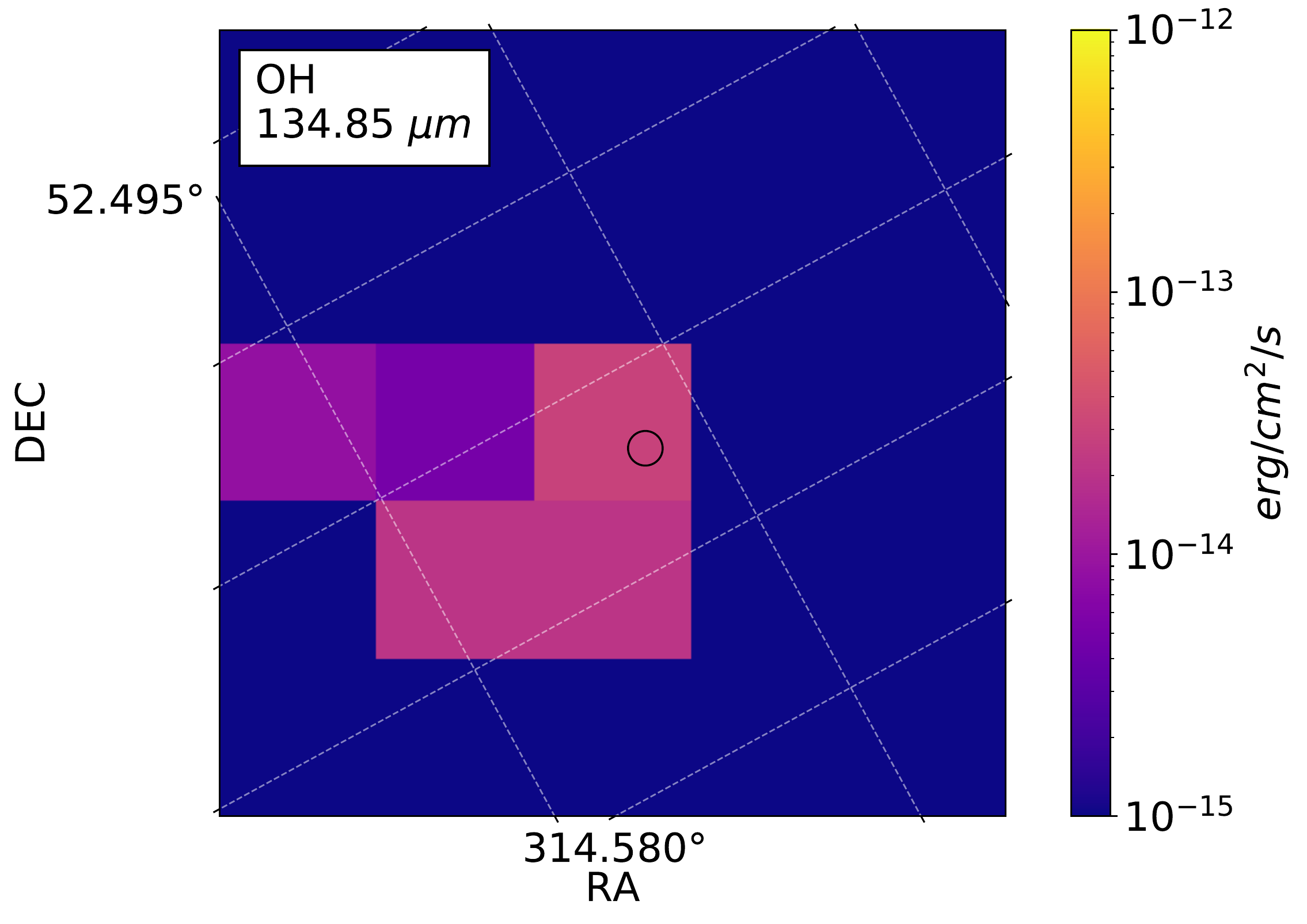}\hspace{-0.1cm}
\includegraphics[width=0.33\textwidth, trim={0cm 0 0cm 0}, clip]{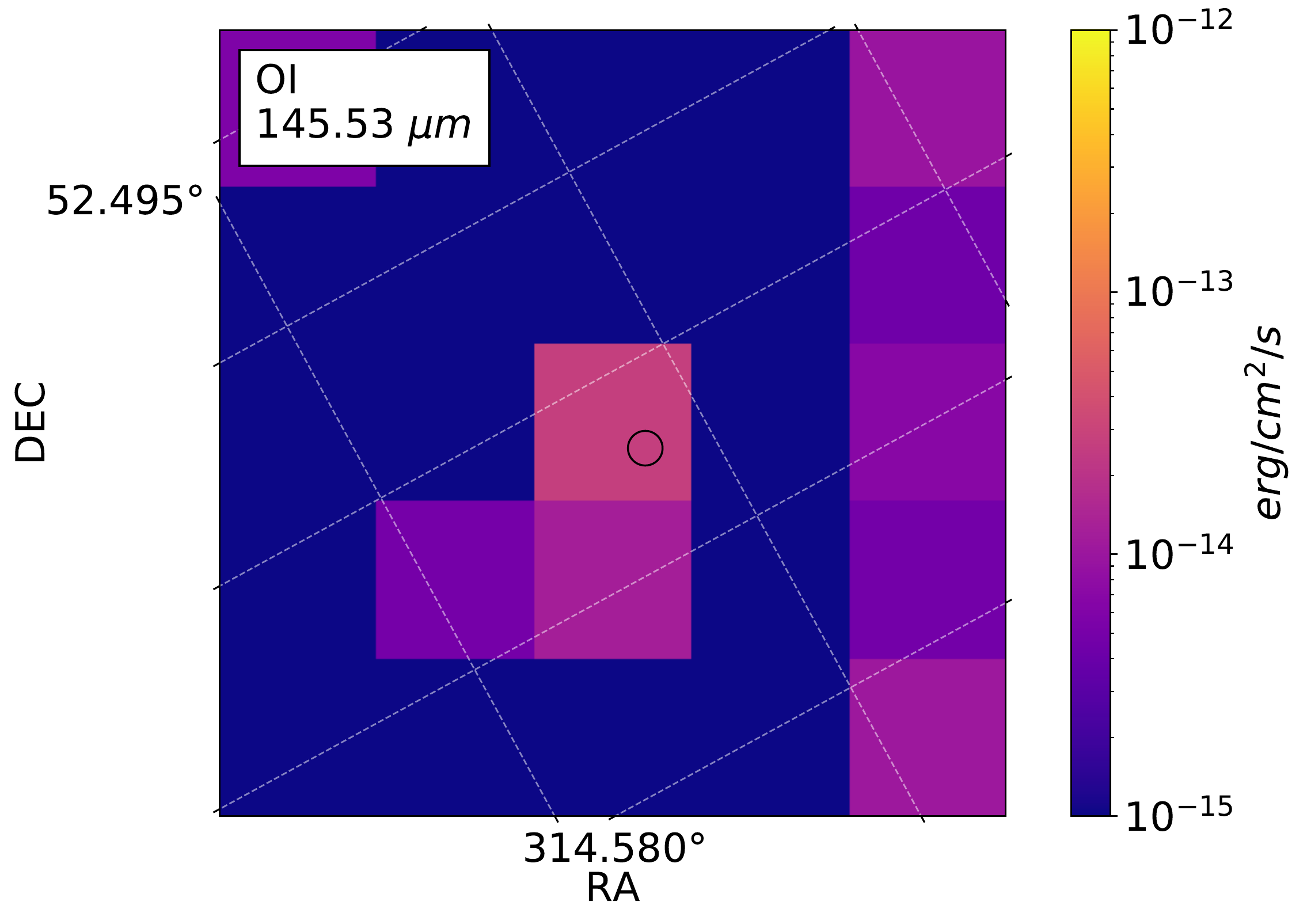}\hspace{-0.1cm}\\
\includegraphics[width=0.33\textwidth, trim={0cm 0 0cm 0}, clip]{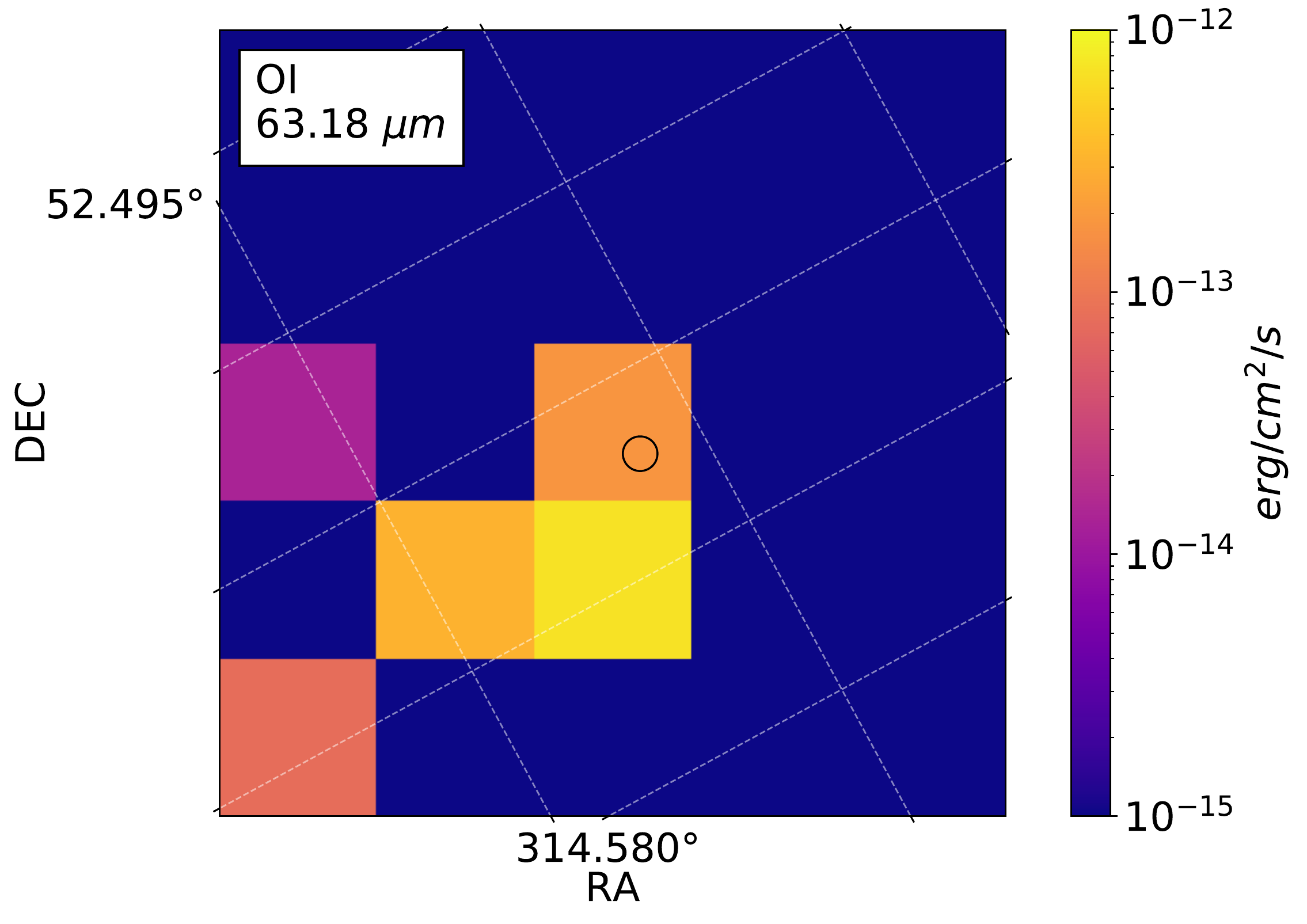}\hspace{-0.1cm}
    \caption{
                \footnotesize
                Line maps of PACS with visible lines for HH 381 IRS.
        }
\end{figure*}

\begin{figure*}
\includegraphics[width=0.33\textwidth, trim={0cm 0 0cm 0}, clip]{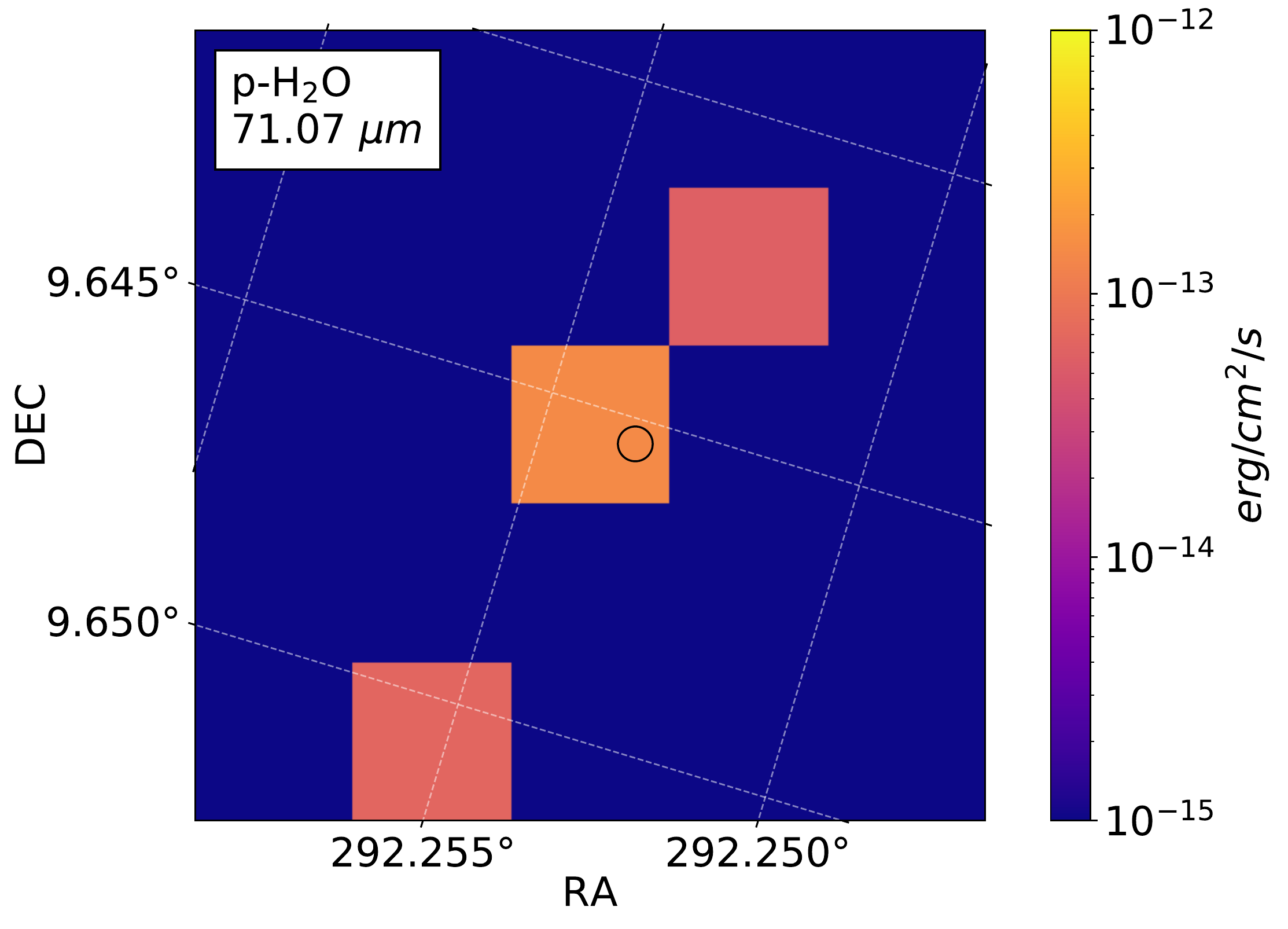}\hspace{-0.1cm}
    \caption{
                \footnotesize
                Line maps of PACS with visible lines for Parsamian 21.
        }
\end{figure*}

\begin{figure*}
\includegraphics[width=0.33\textwidth, trim={0cm 0 0cm 0}, clip]{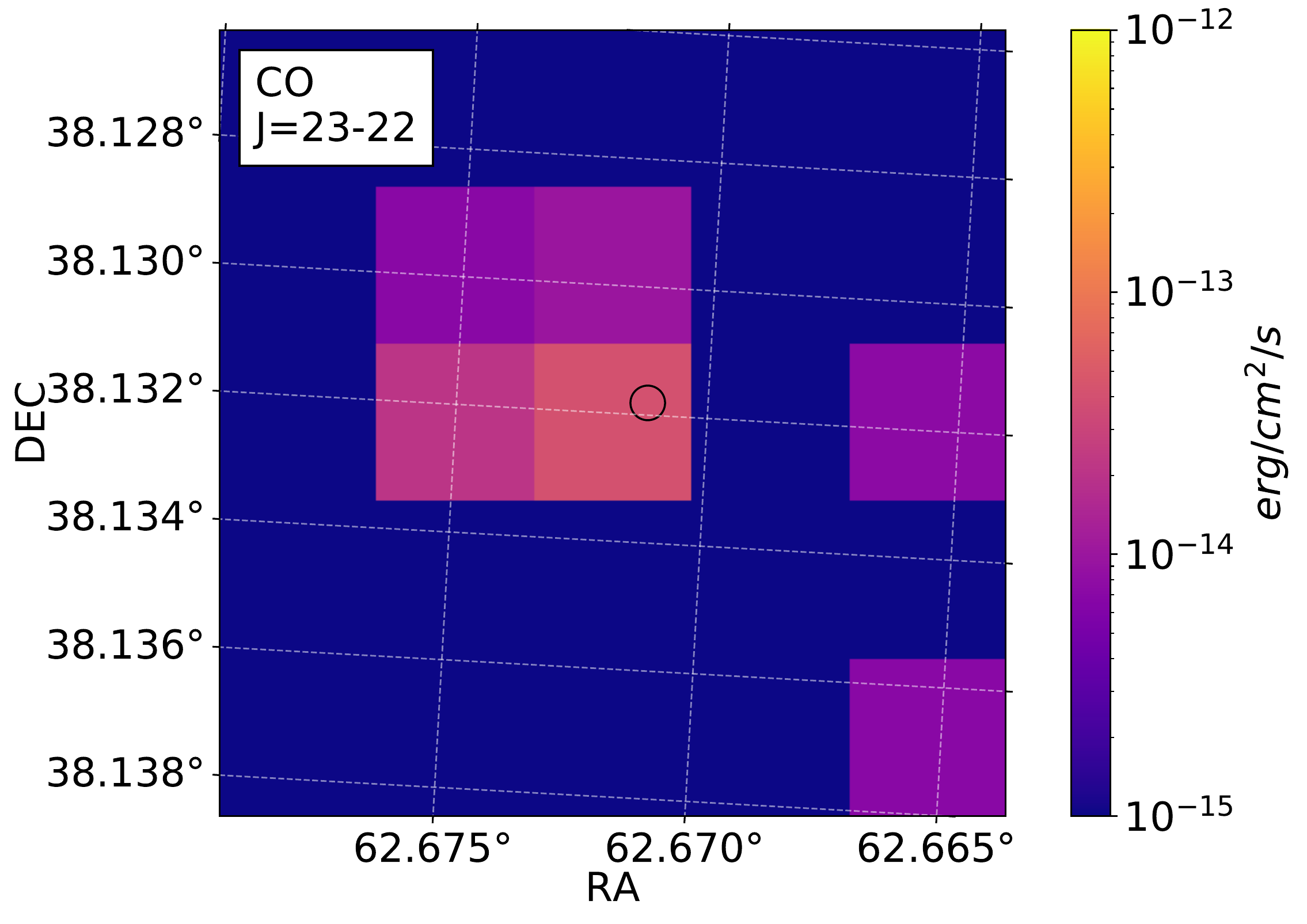}\hspace{-0.1cm}
\includegraphics[width=0.33\textwidth, trim={0cm 0 0cm 0}, clip]{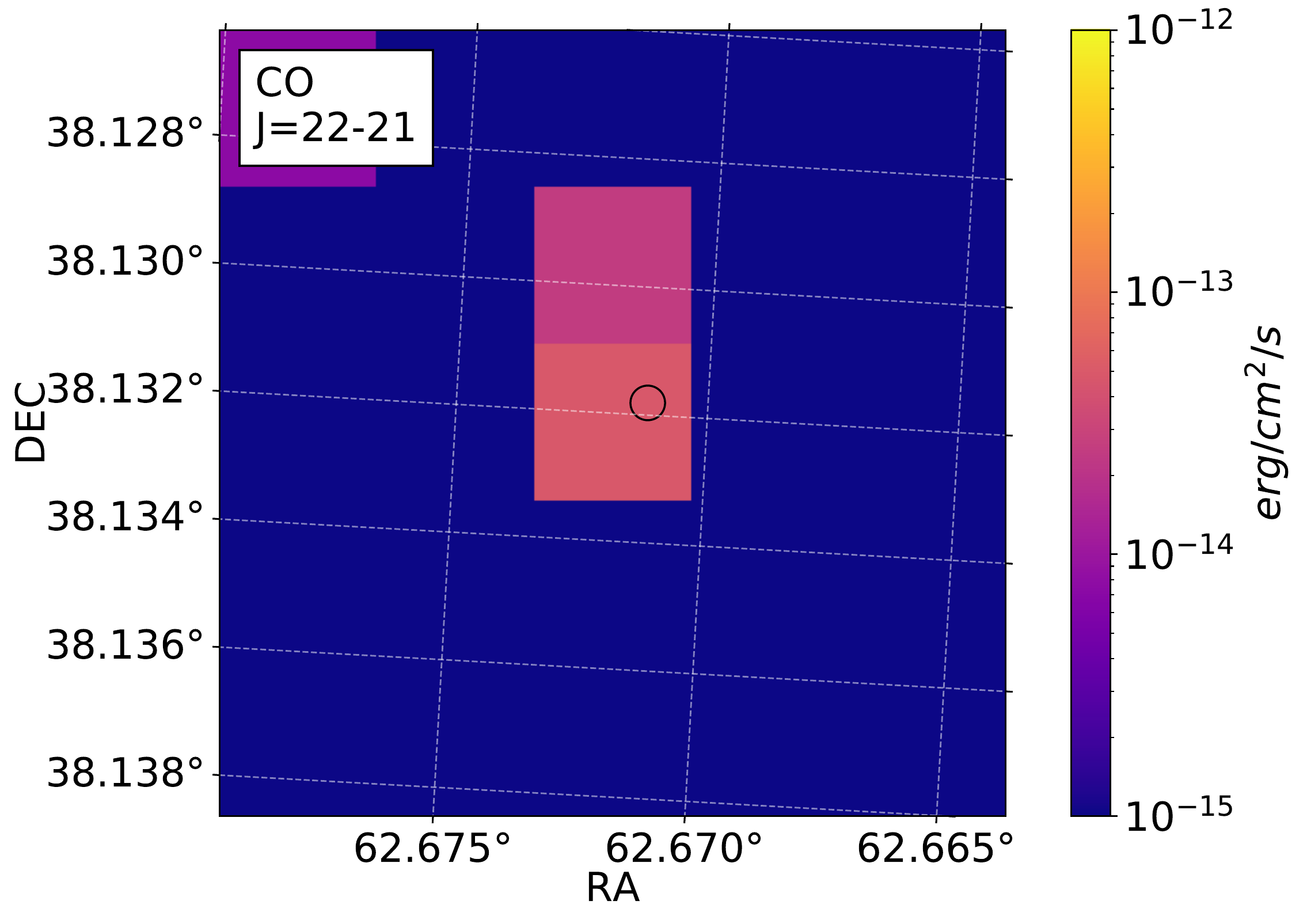}\hspace{-0.1cm}
\includegraphics[width=0.33\textwidth, trim={0cm 0 0cm 0}, clip]{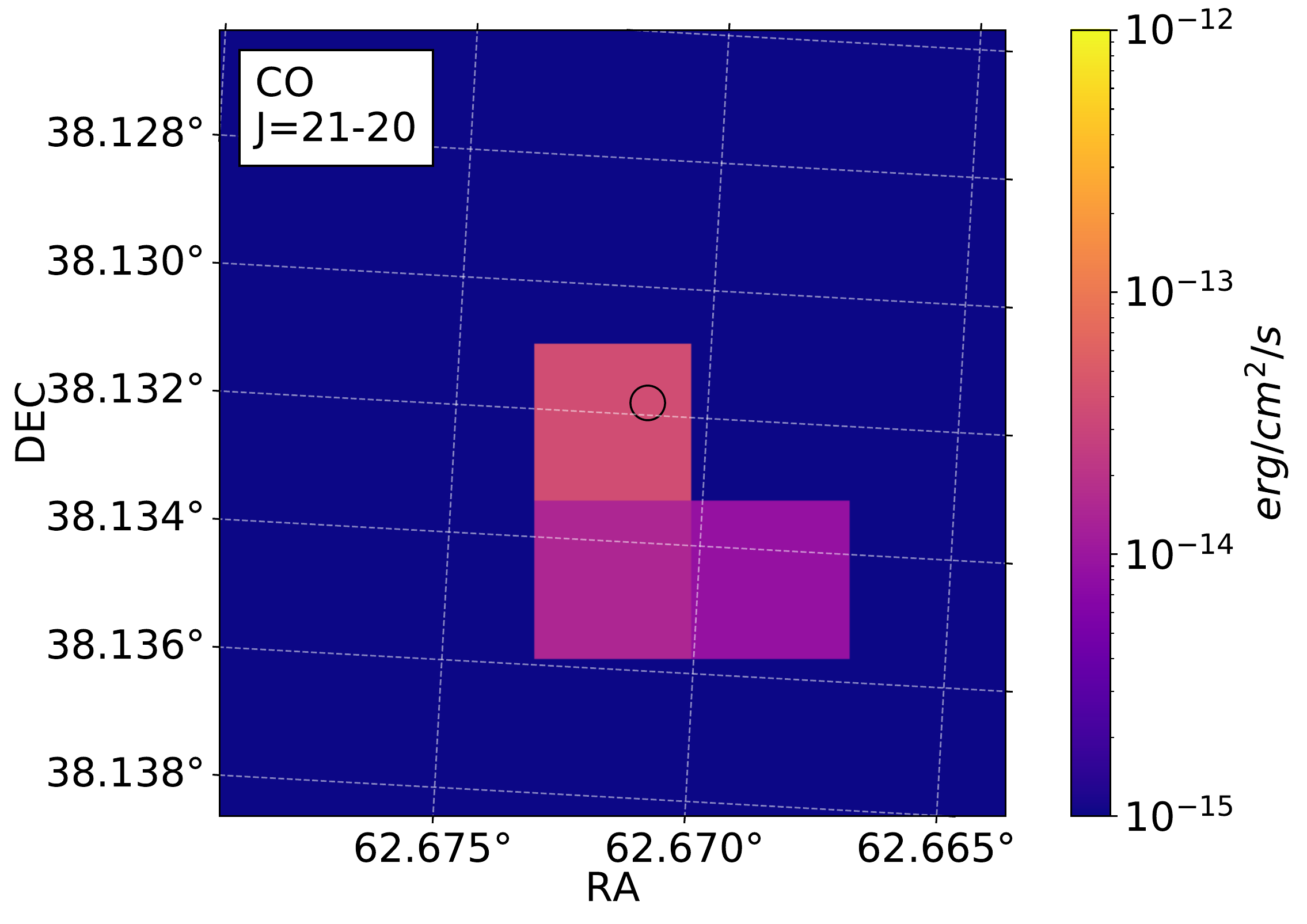}\hspace{-0.1cm}\\
\includegraphics[width=0.33\textwidth, trim={0cm 0 0cm 0}, clip]{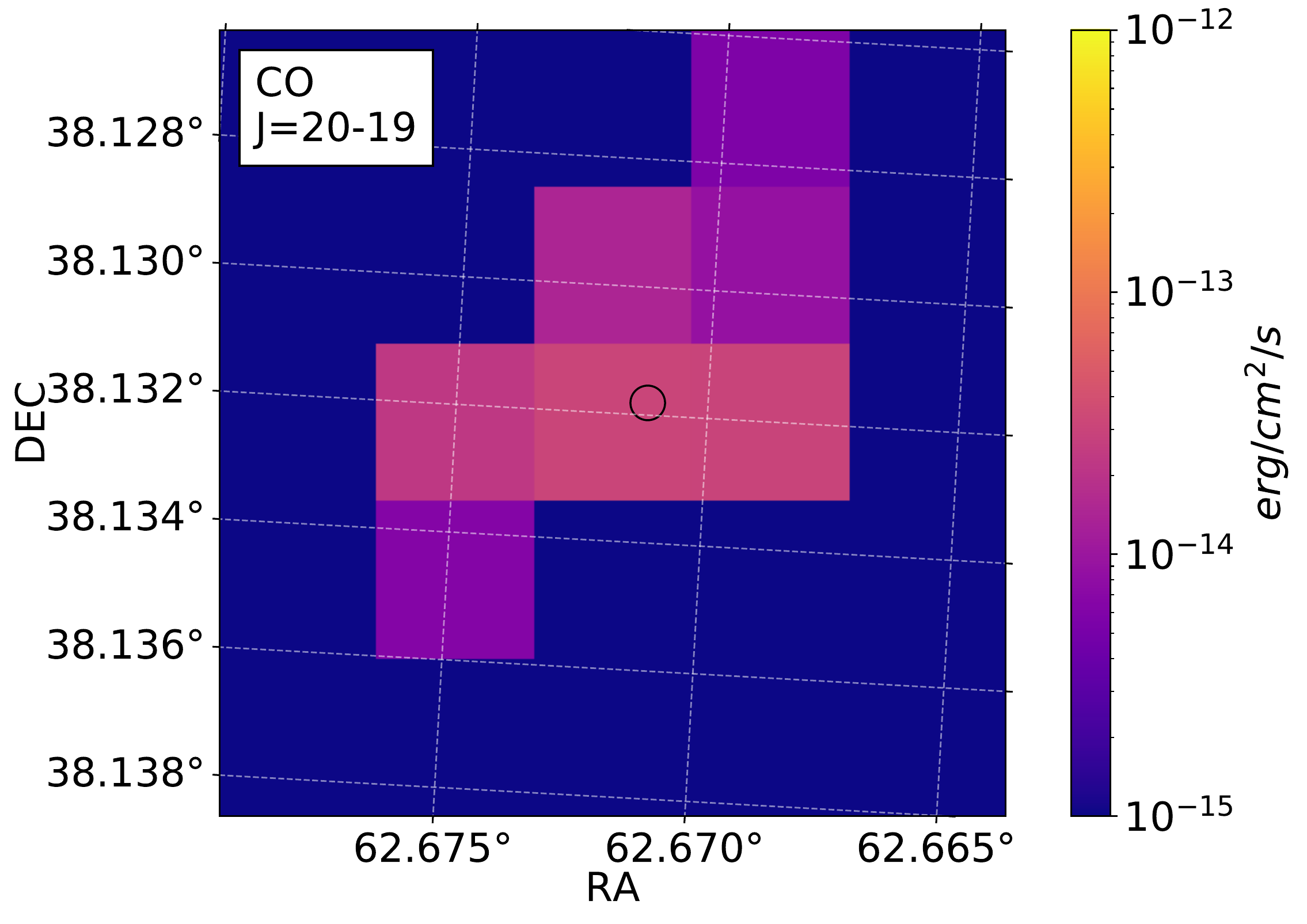}\hspace{-0.1cm}
\includegraphics[width=0.33\textwidth, trim={0cm 0 0cm 0}, clip]{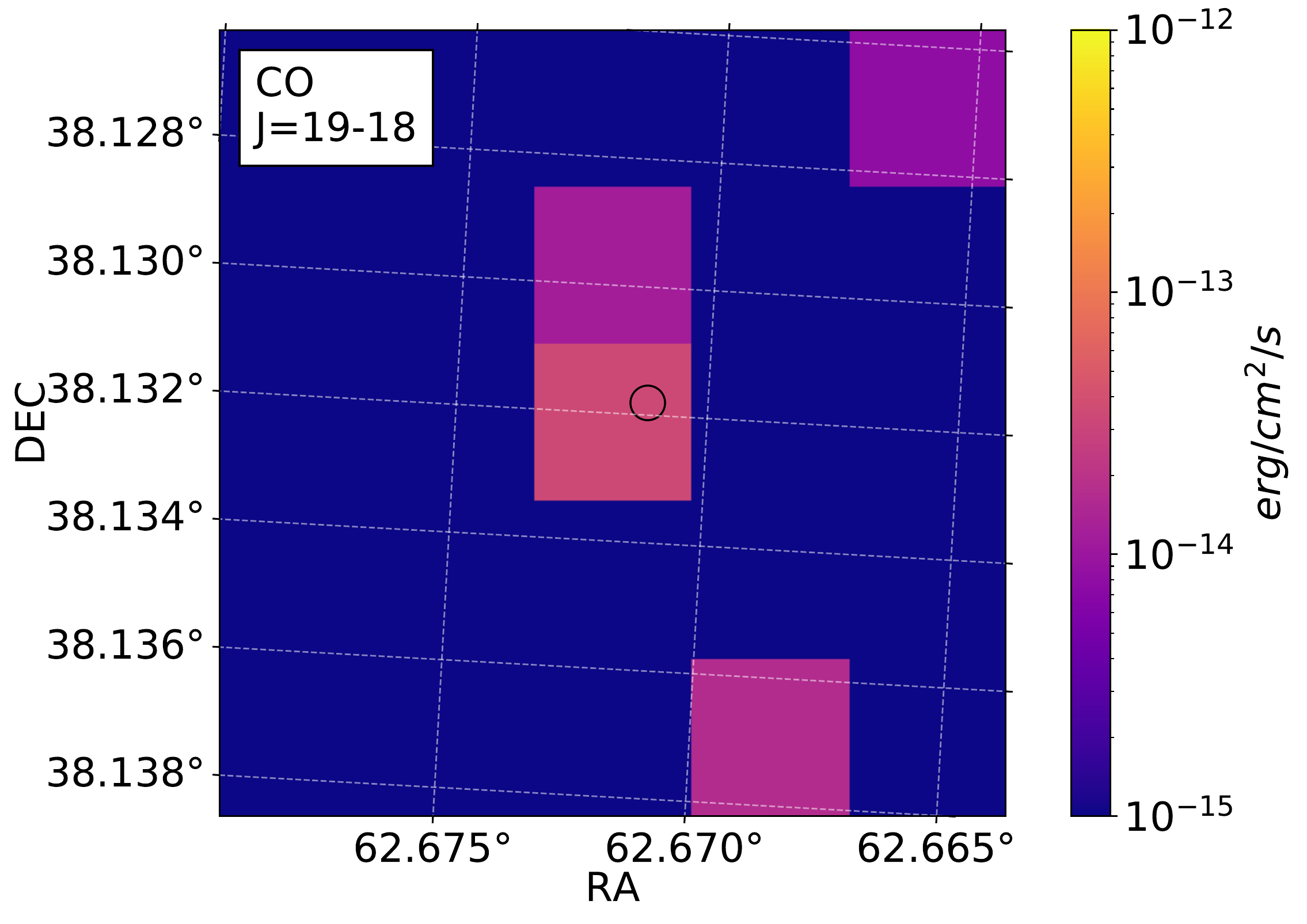}\hspace{-0.1cm}
\includegraphics[width=0.33\textwidth, trim={0cm 0 0cm 0}, clip]{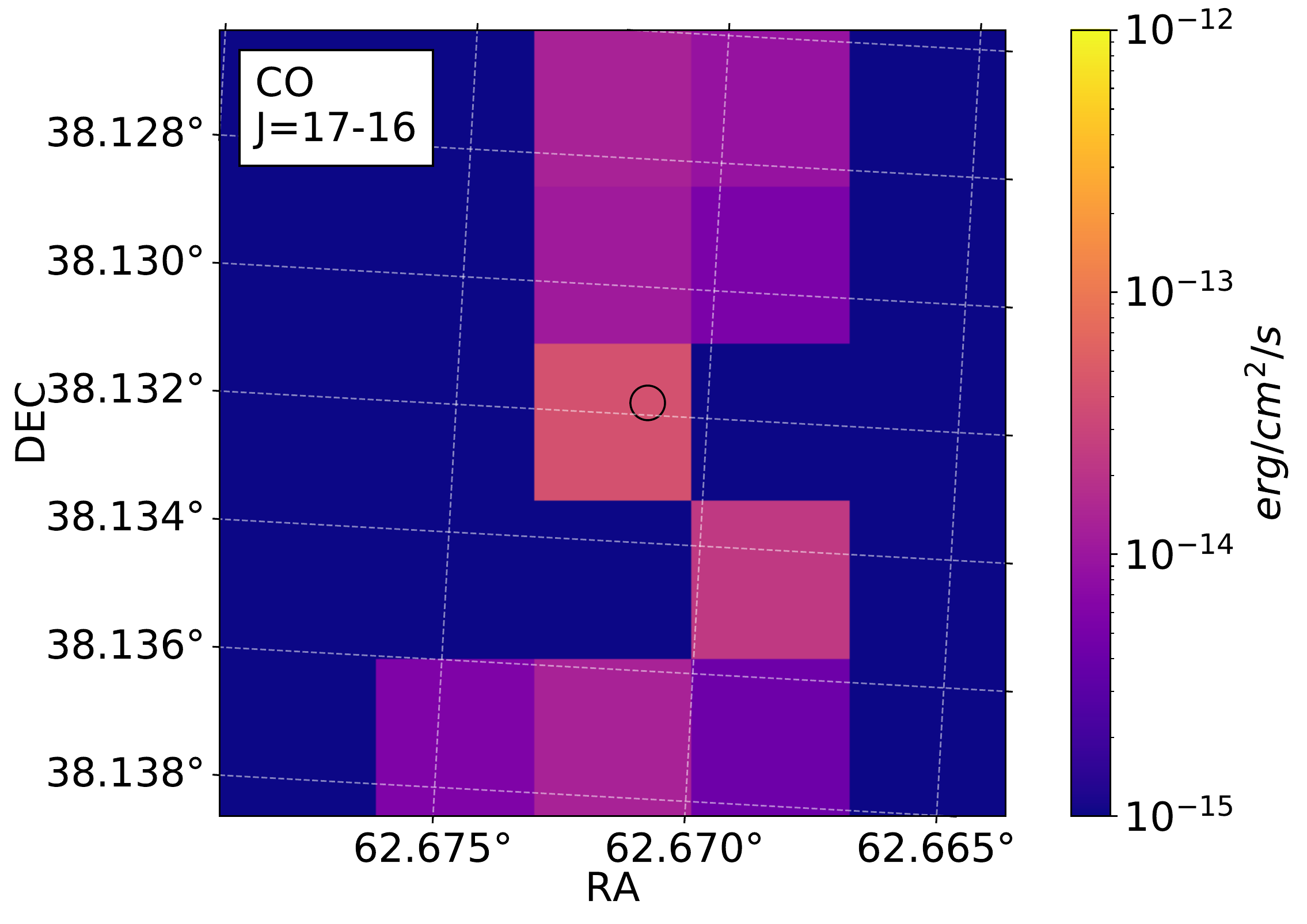}\hspace{-0.1cm}\\
\includegraphics[width=0.33\textwidth, trim={0cm 0 0cm 0}, clip]{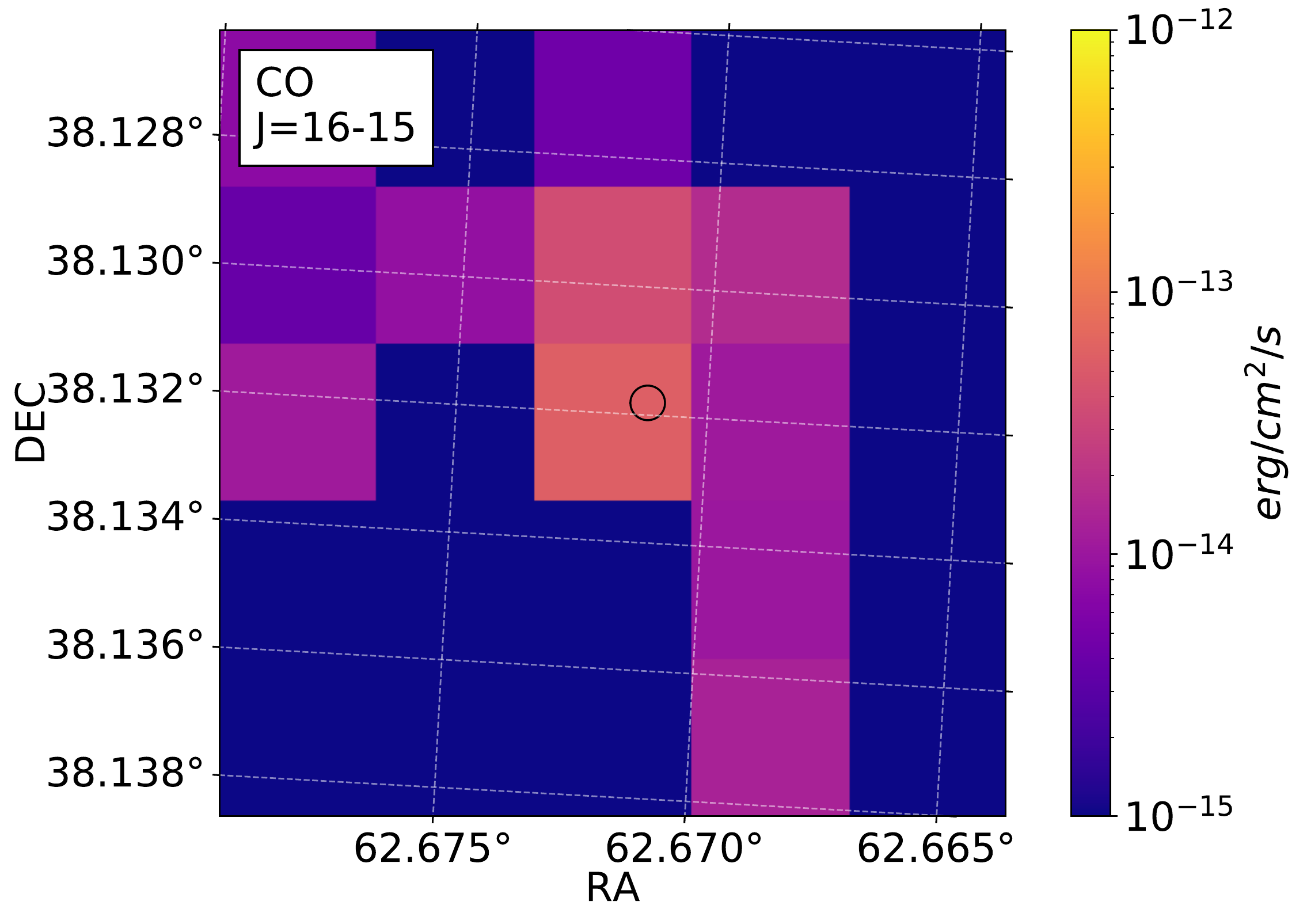}\hspace{-0.1cm}
\includegraphics[width=0.33\textwidth, trim={0cm 0 0cm 0}, clip]{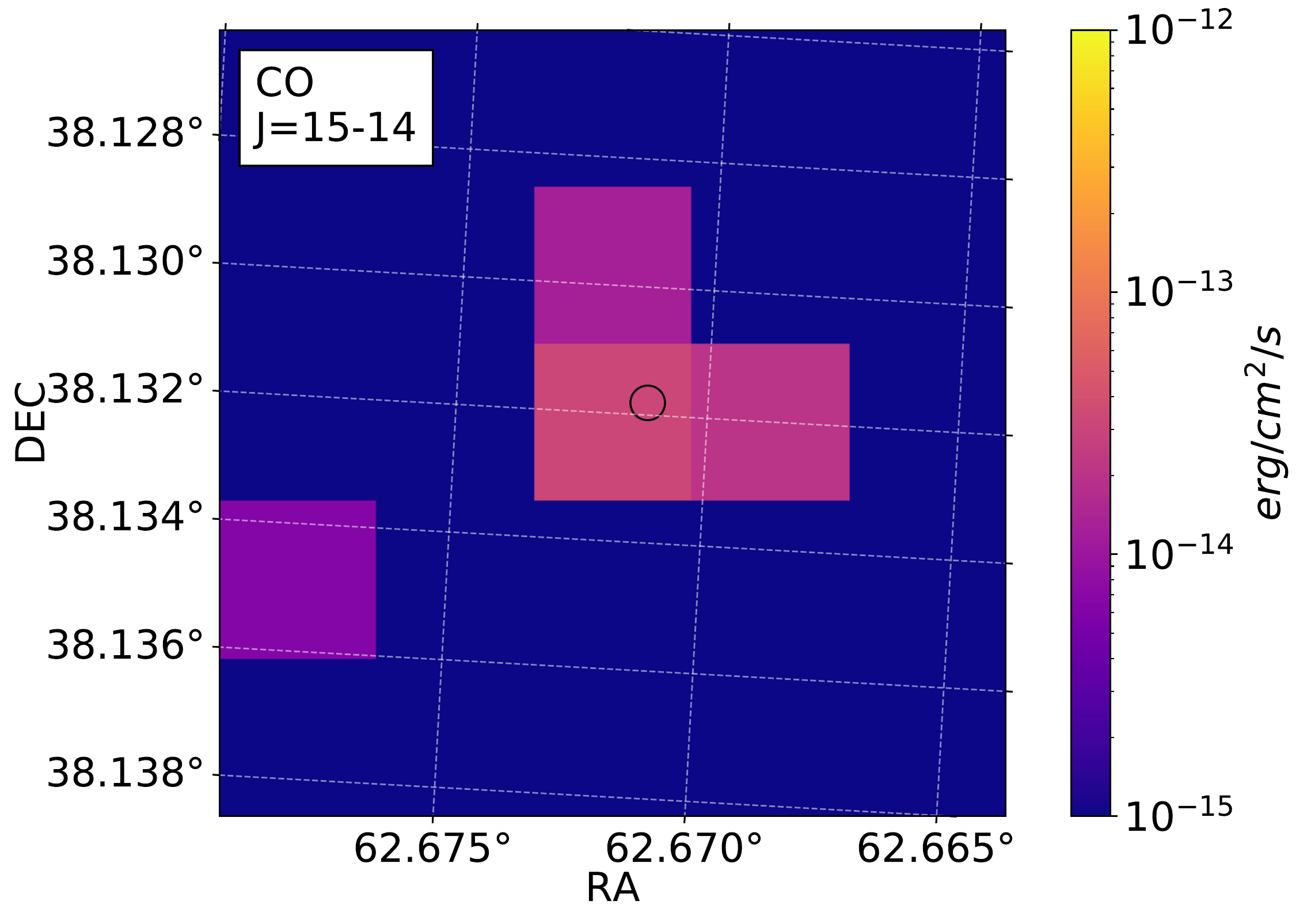}\hspace{-0.1cm}
\includegraphics[width=0.33\textwidth, trim={0cm 0 0cm 0}, clip]{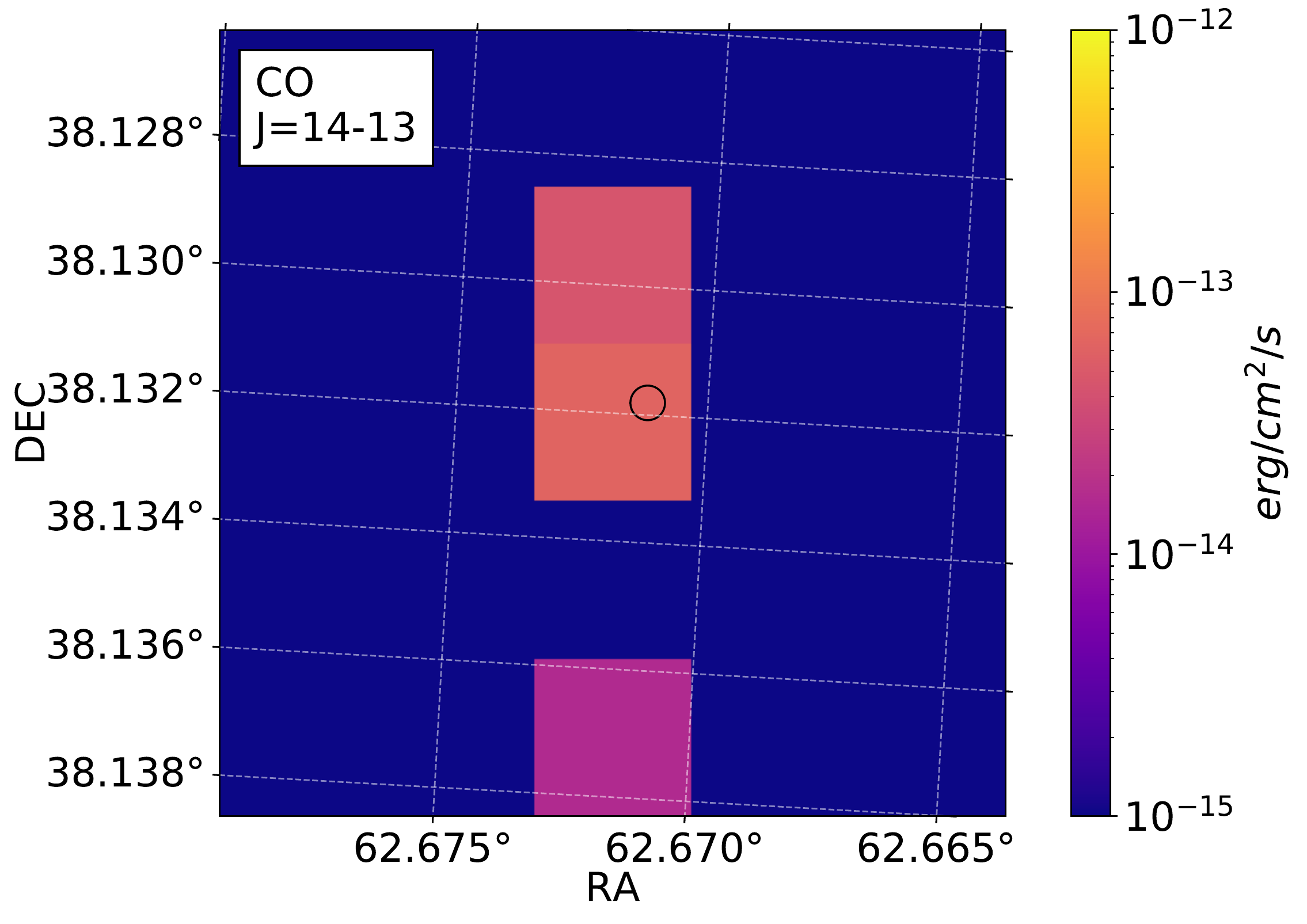}\hspace{-0.1cm}\\
\includegraphics[width=0.33\textwidth, trim={0cm 0 0cm 0}, clip]{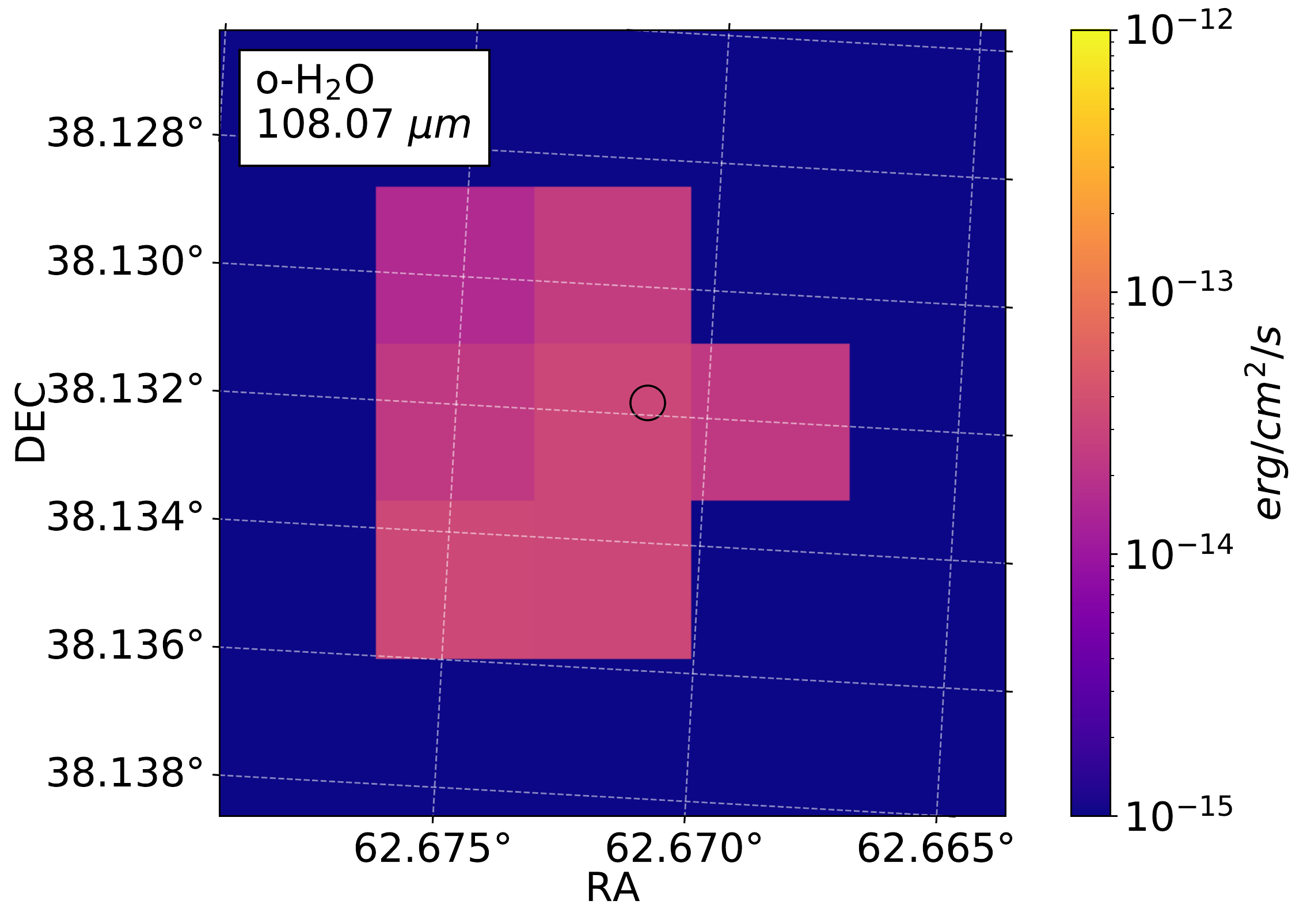}\hspace{-0.1cm}
\includegraphics[width=0.33\textwidth, trim={0cm 0 0cm 0}, clip]{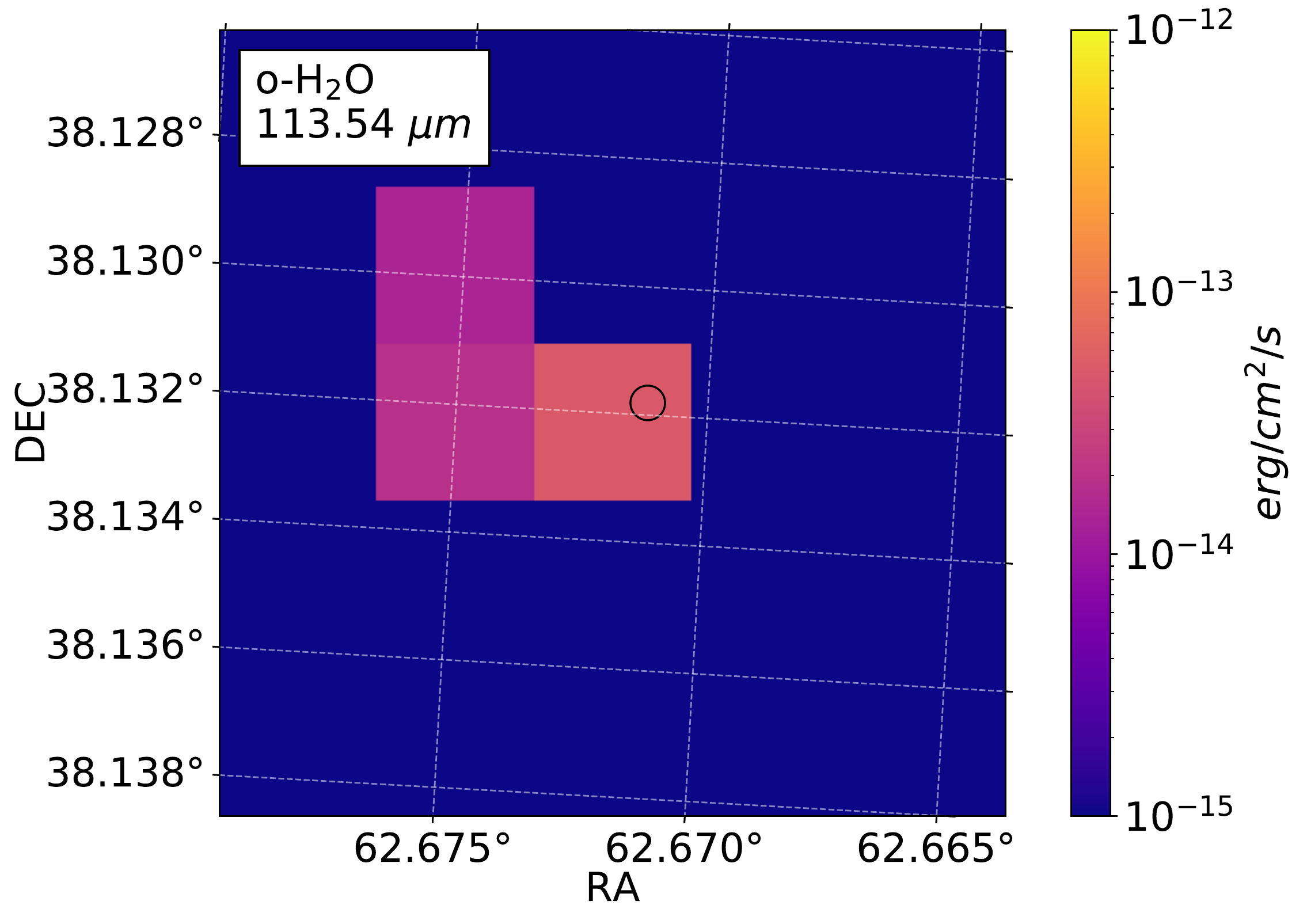}\hspace{-0.1cm}
\includegraphics[width=0.33\textwidth, trim={0cm 0 0cm 0}, clip]{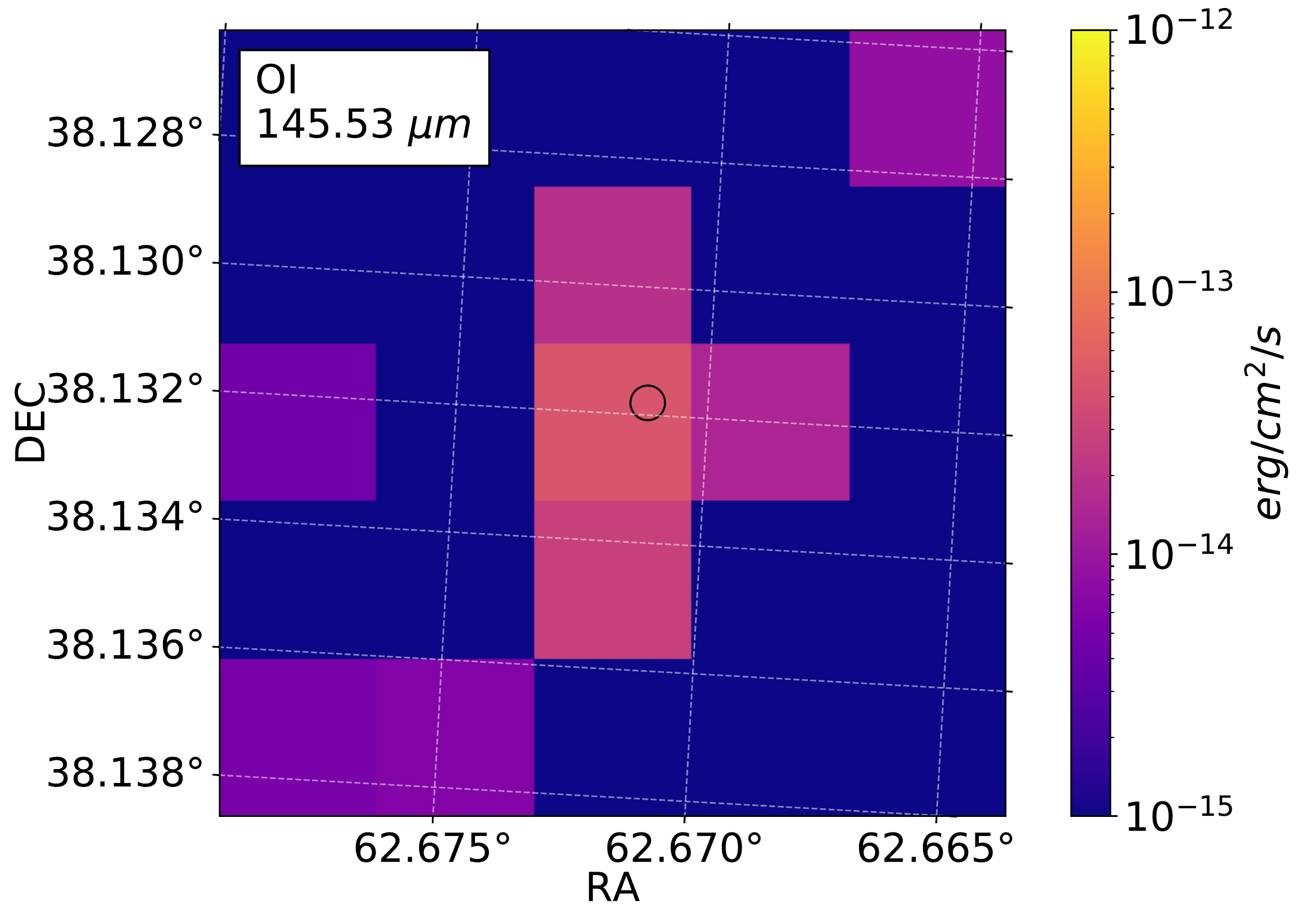}\hspace{-0.1cm}\\
\includegraphics[width=0.33\textwidth, trim={0cm 0 0cm 0}, clip]{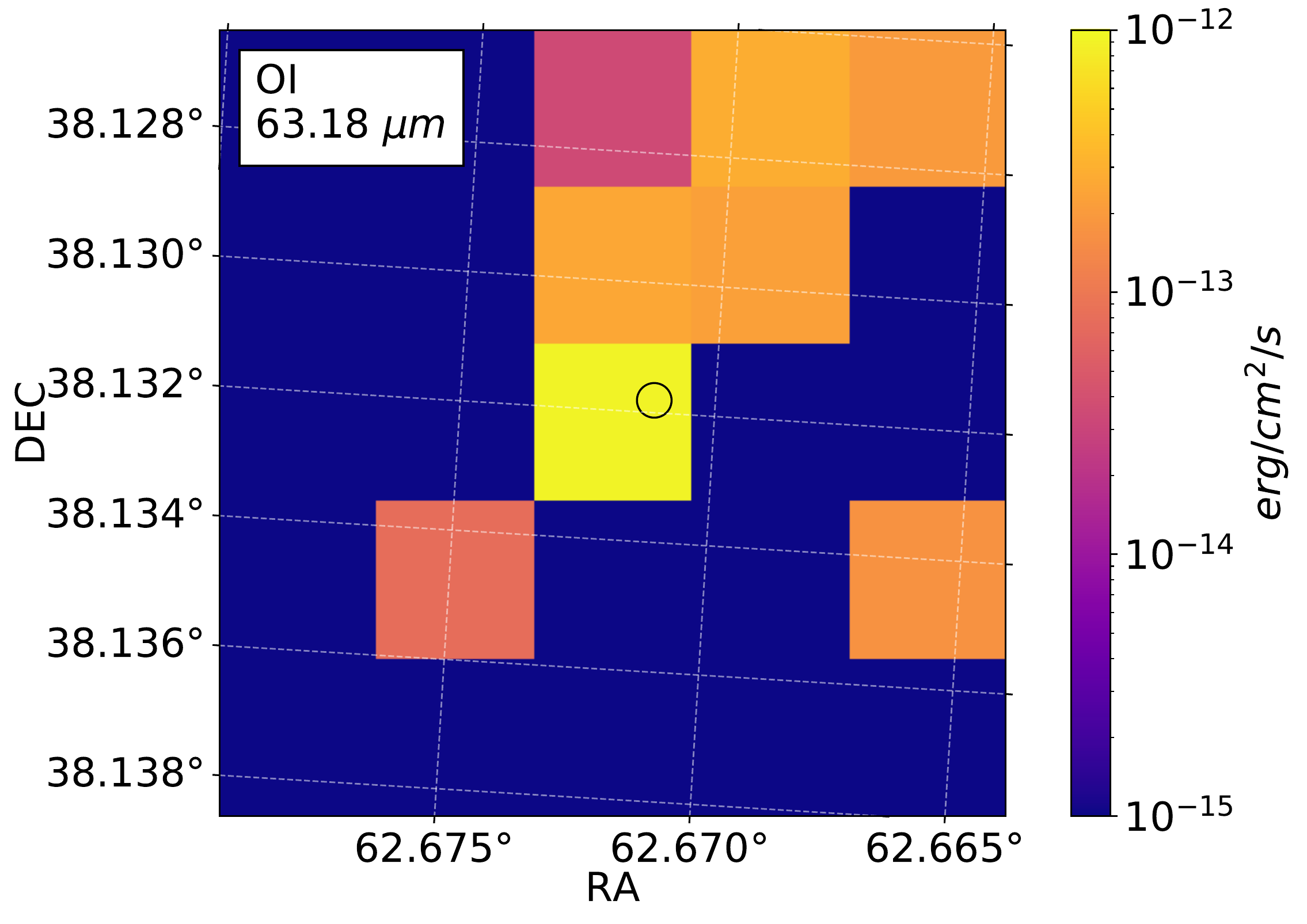}\hspace{-0.1cm}
\caption{
                \footnotesize
                Line maps of PACS with visible lines for PP 13 S.
        }
\end{figure*}

\begin{figure*}
\includegraphics[width=0.33\textwidth, trim={0cm 0 0cm 0}, clip]{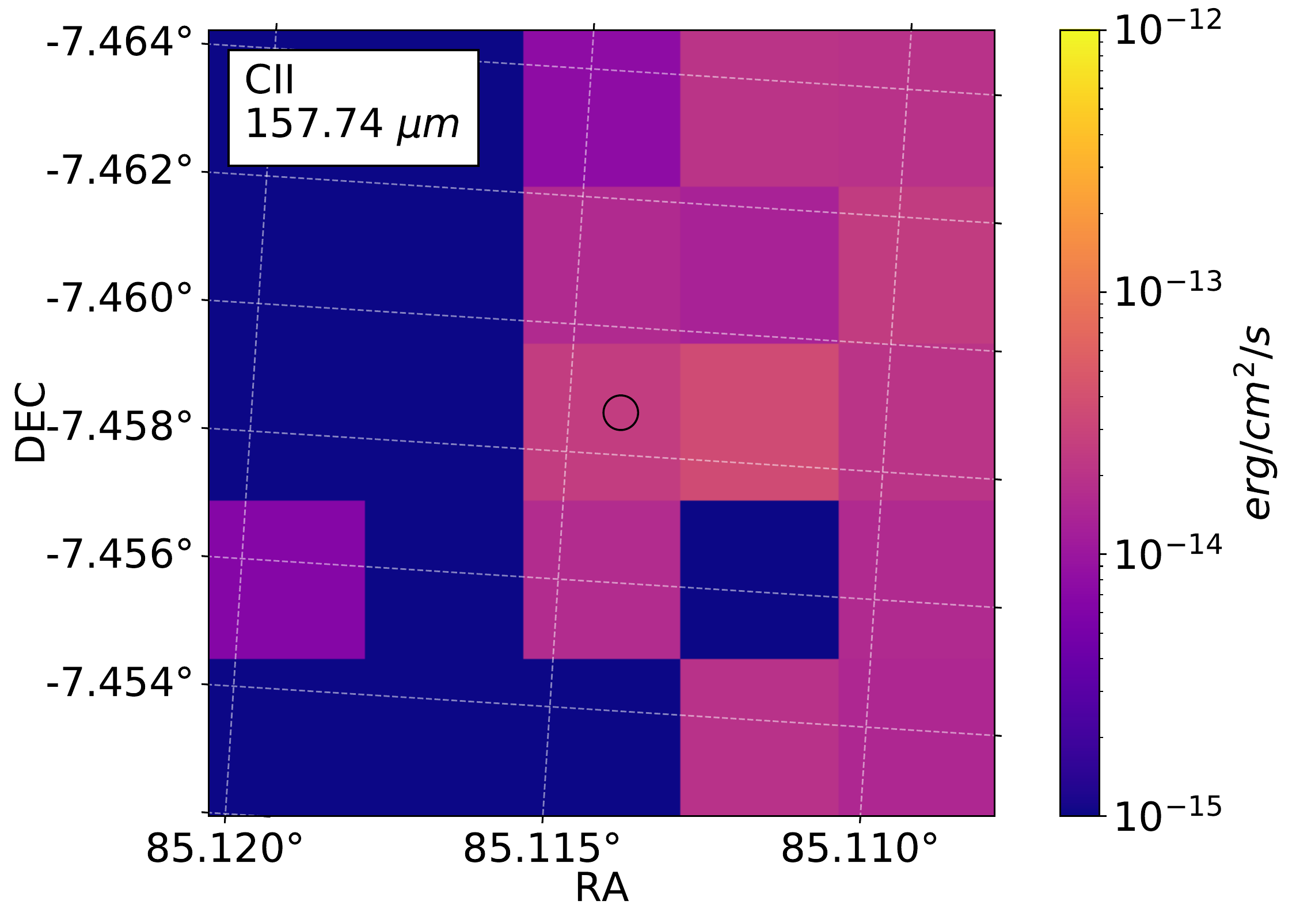}\hspace{-0.1cm}
\includegraphics[width=0.33\textwidth, trim={0cm 0 0cm 0}, clip]{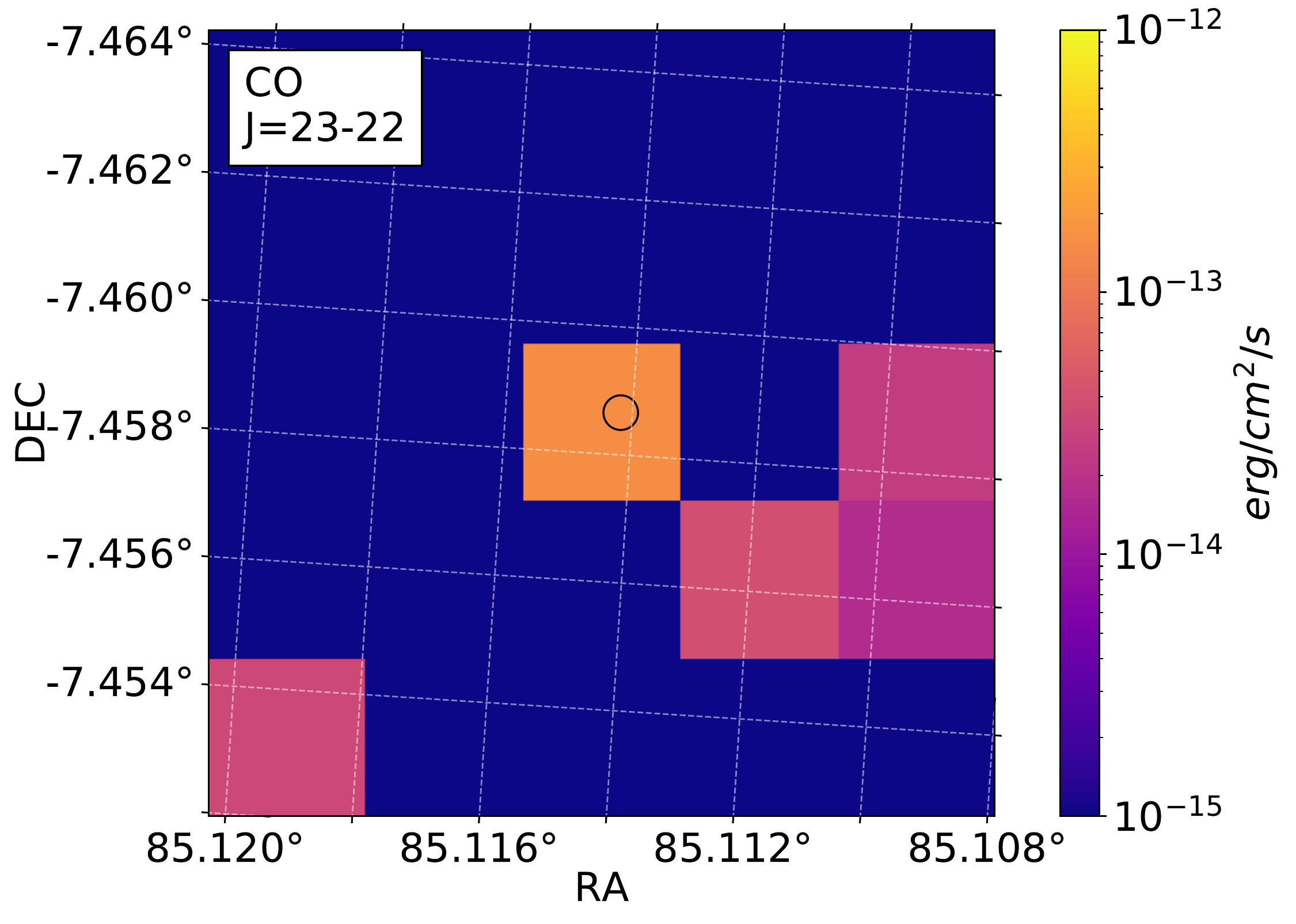}\hspace{-0.1cm}
\includegraphics[width=0.33\textwidth, trim={0cm 0 0cm 0}, clip]{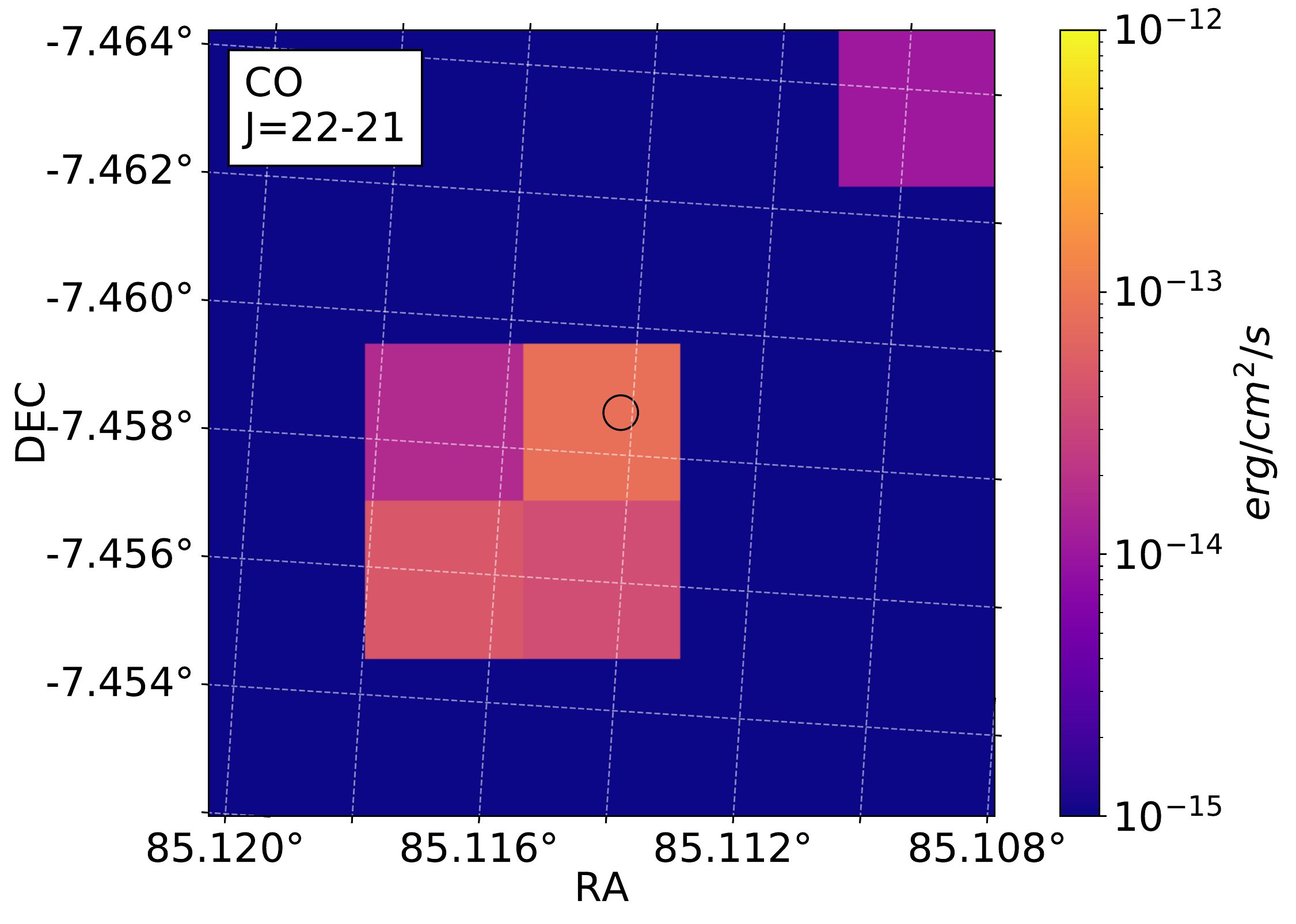}\hspace{-0.1cm}\\
\includegraphics[width=0.33\textwidth, trim={0cm 0 0cm 0}, clip]{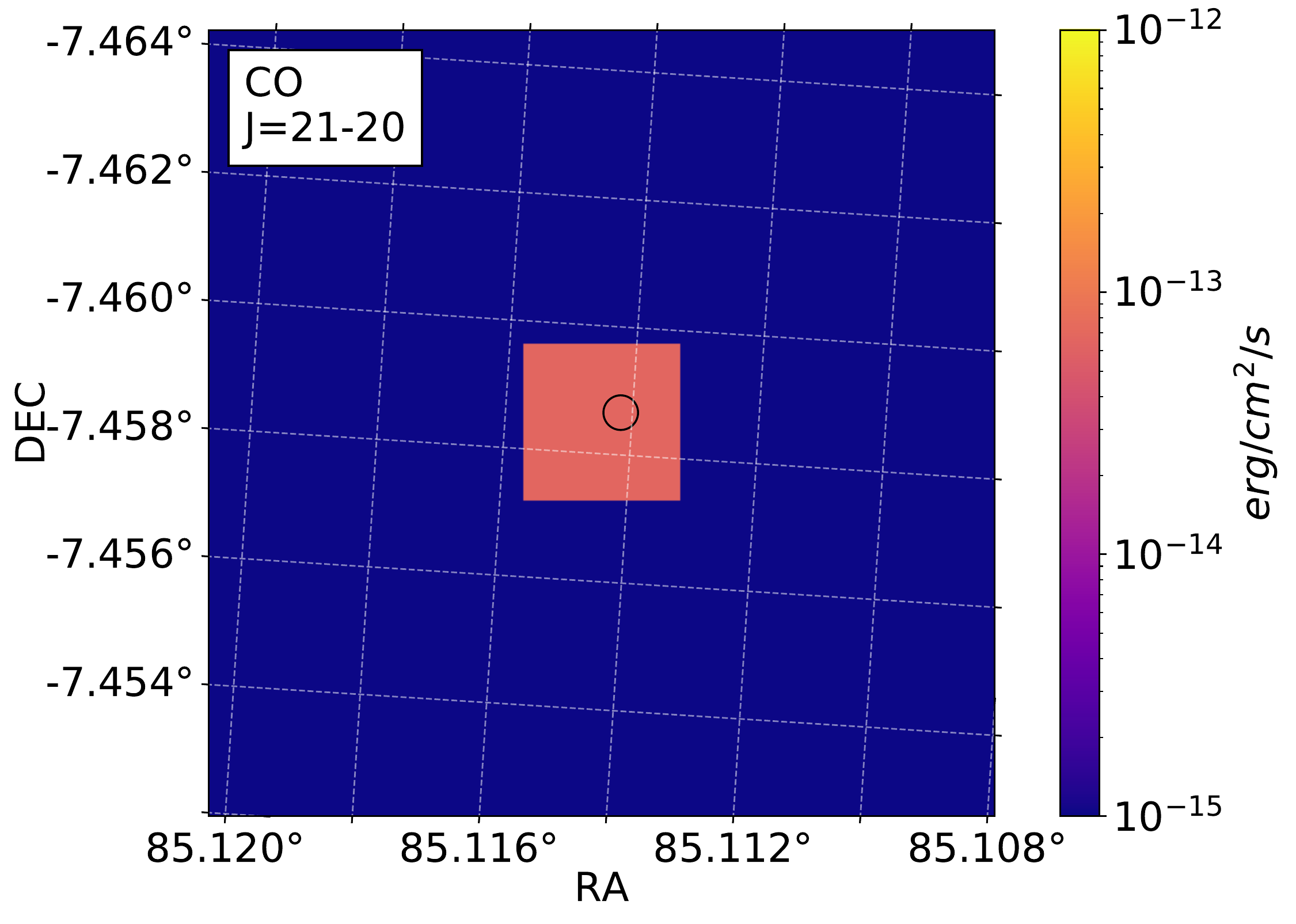}\hspace{-0.1cm}
\includegraphics[width=0.33\textwidth, trim={0cm 0 0cm 0}, clip]{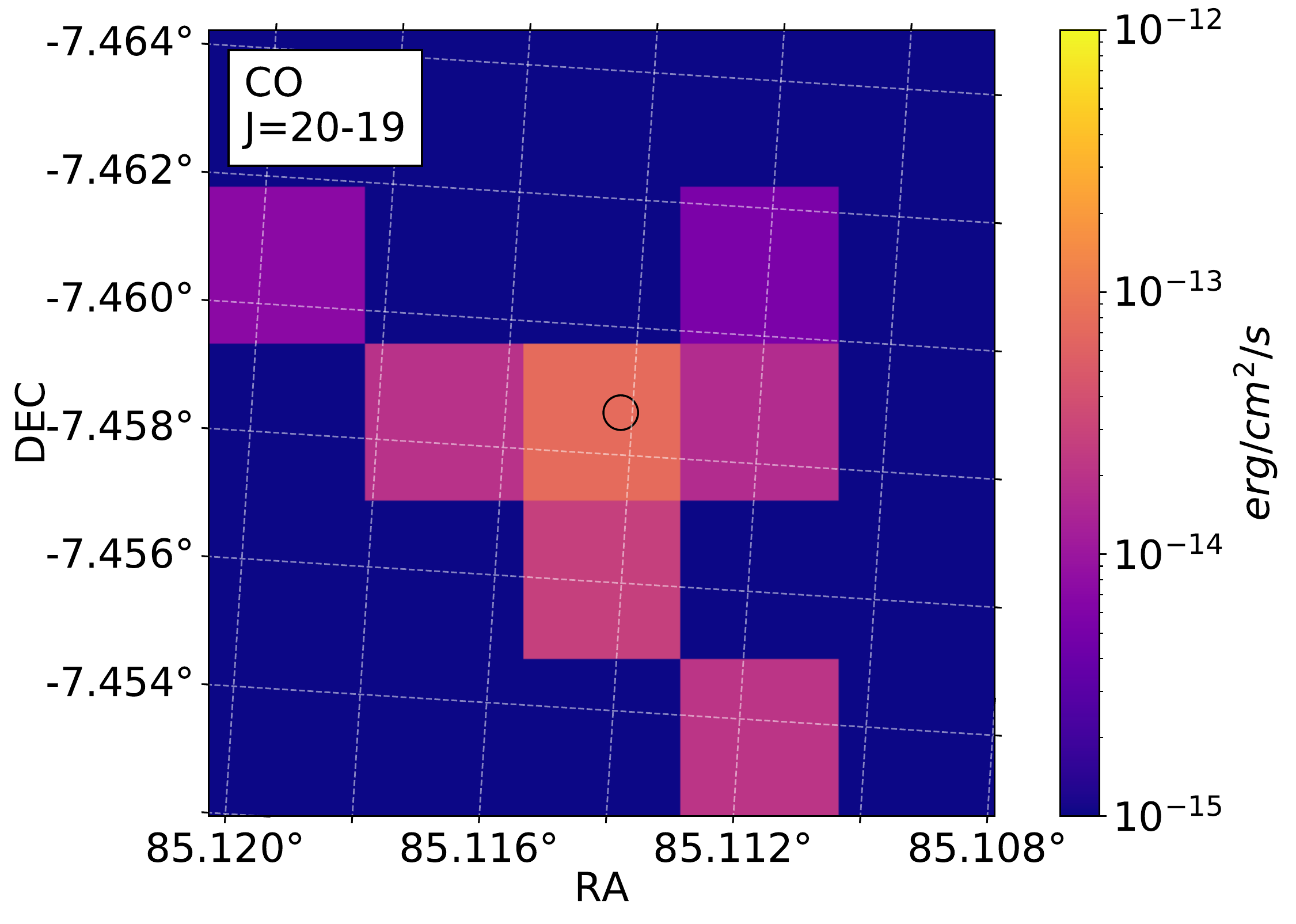}\hspace{-0.1cm}
\includegraphics[width=0.33\textwidth, trim={0cm 0 0cm 0}, clip]{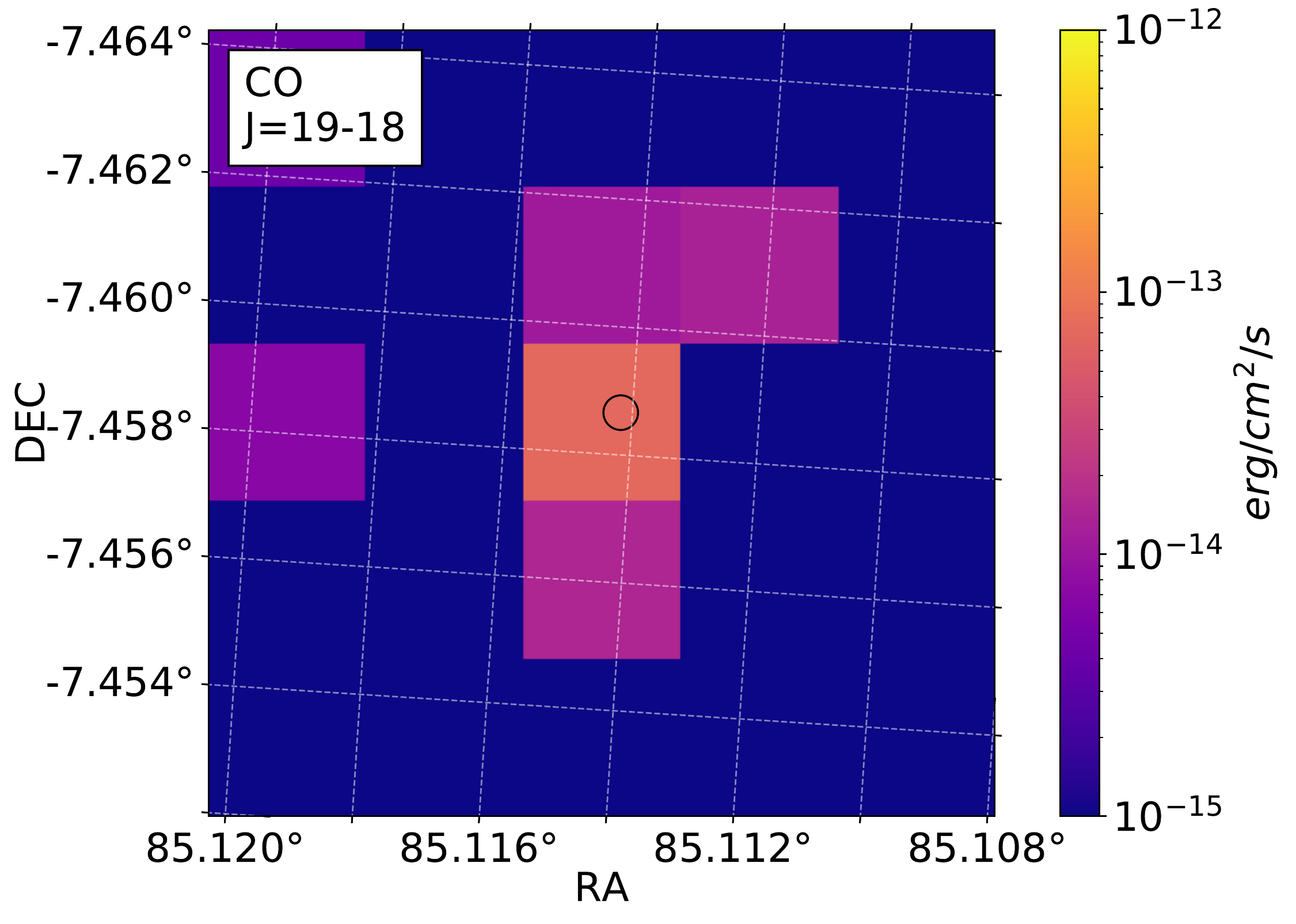}\hspace{-0.1cm}\\
\includegraphics[width=0.33\textwidth, trim={0cm 0 0cm 0}, clip]{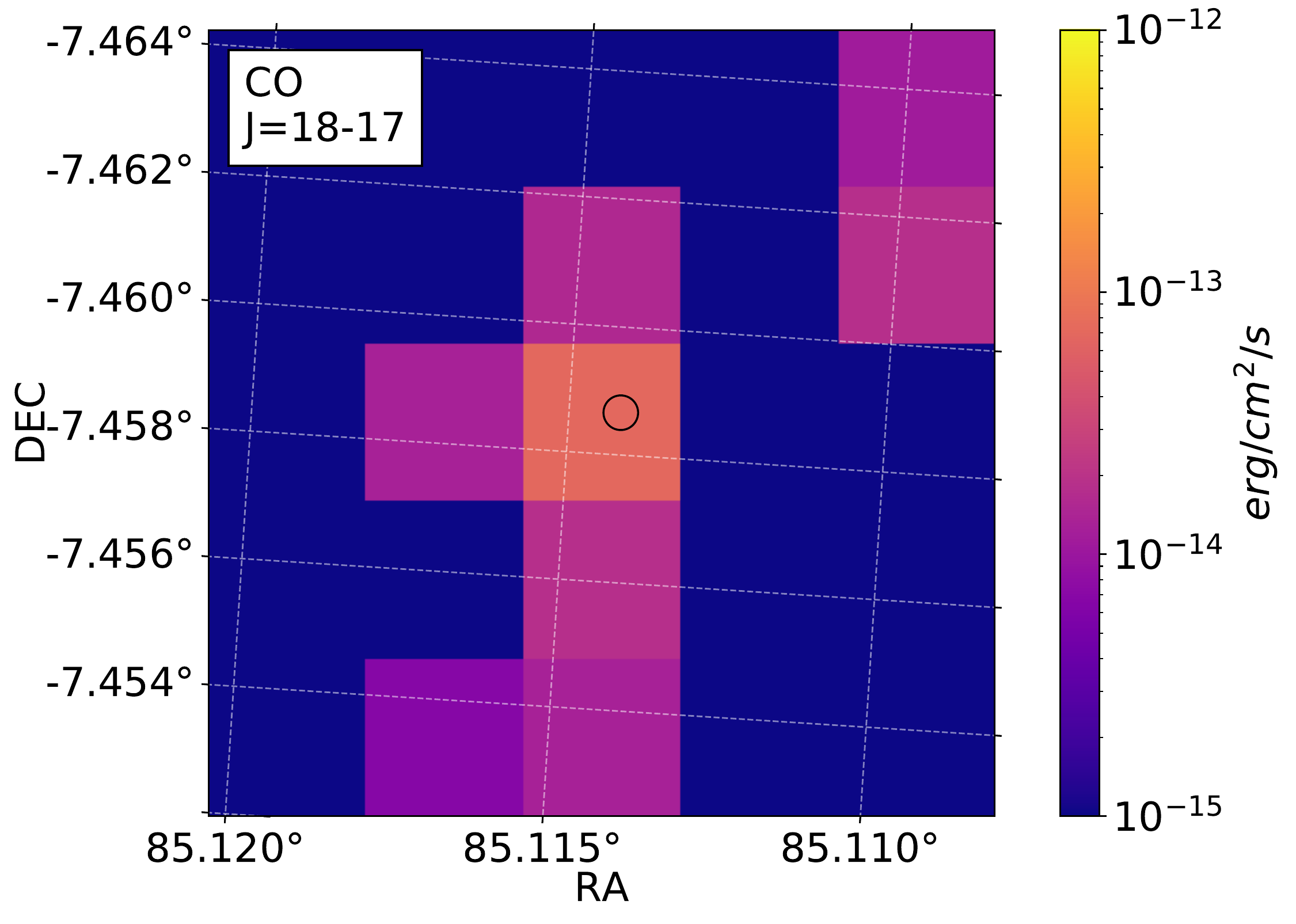}\hspace{-0.1cm}
\includegraphics[width=0.33\textwidth, trim={0cm 0 0cm 0}, clip]{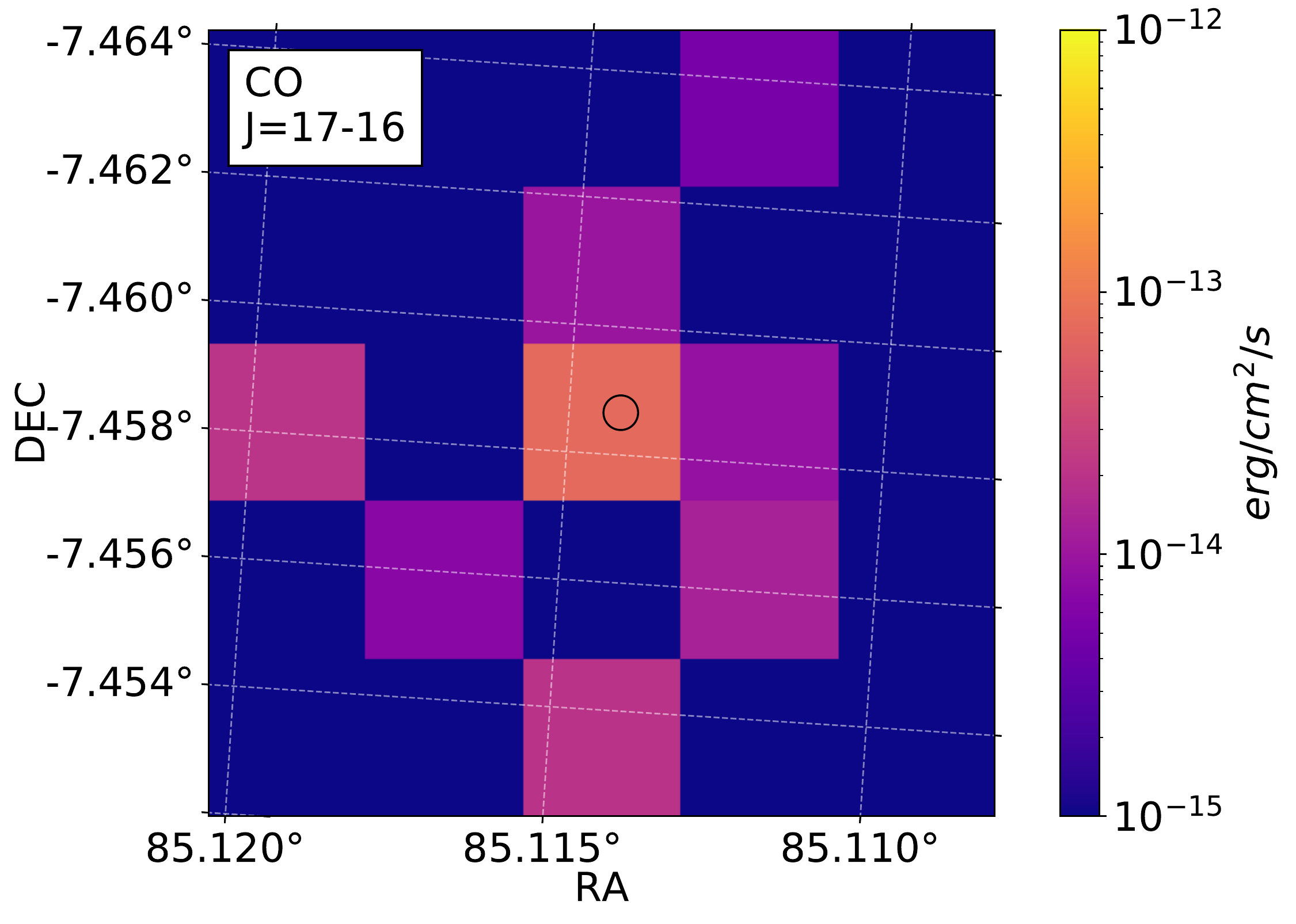}\hspace{-0.1cm}
\includegraphics[width=0.33\textwidth, trim={0cm 0 0cm 0}, clip]{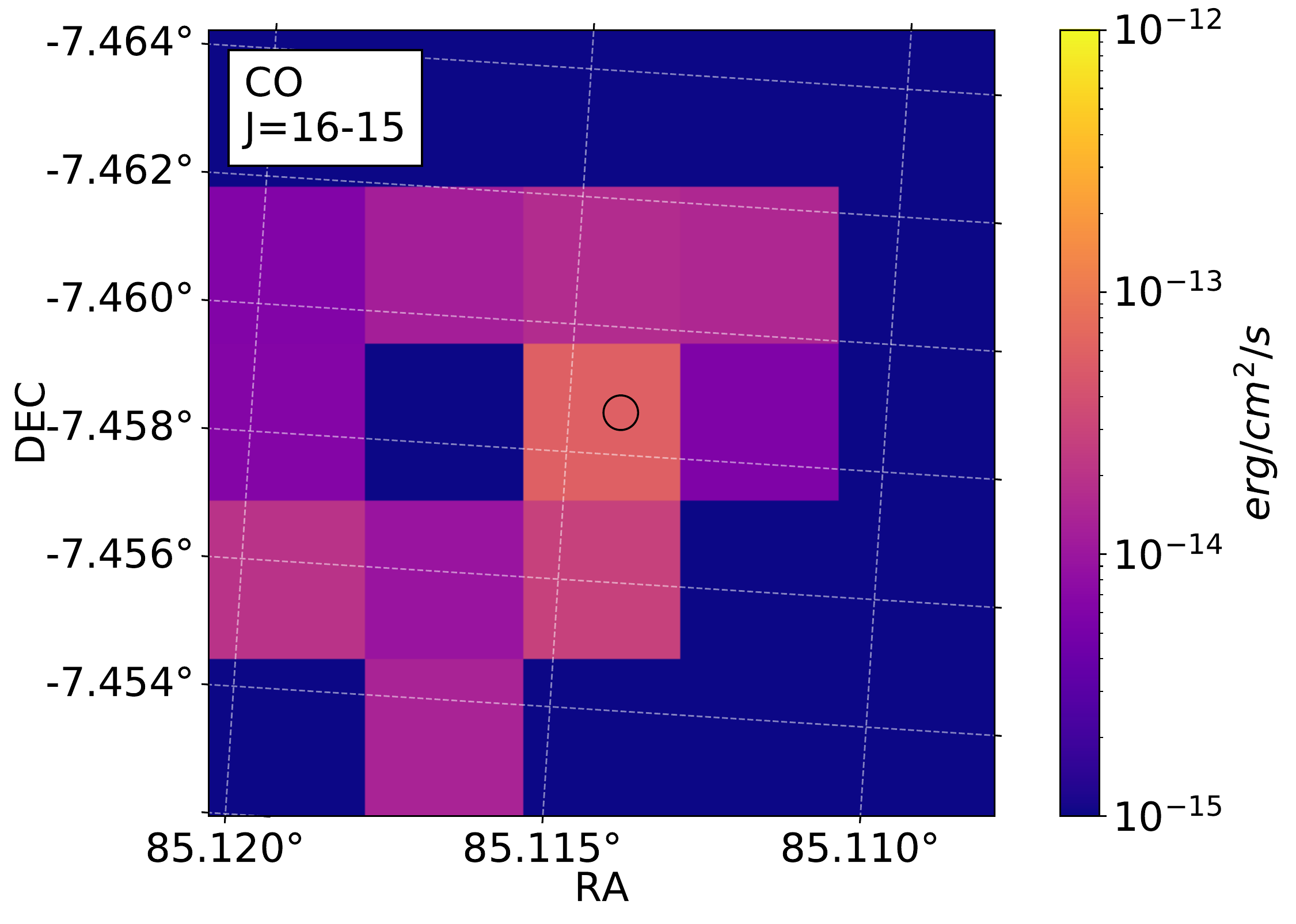}\hspace{-0.1cm}\\
\includegraphics[width=0.33\textwidth, trim={0cm 0 0cm 0}, clip]{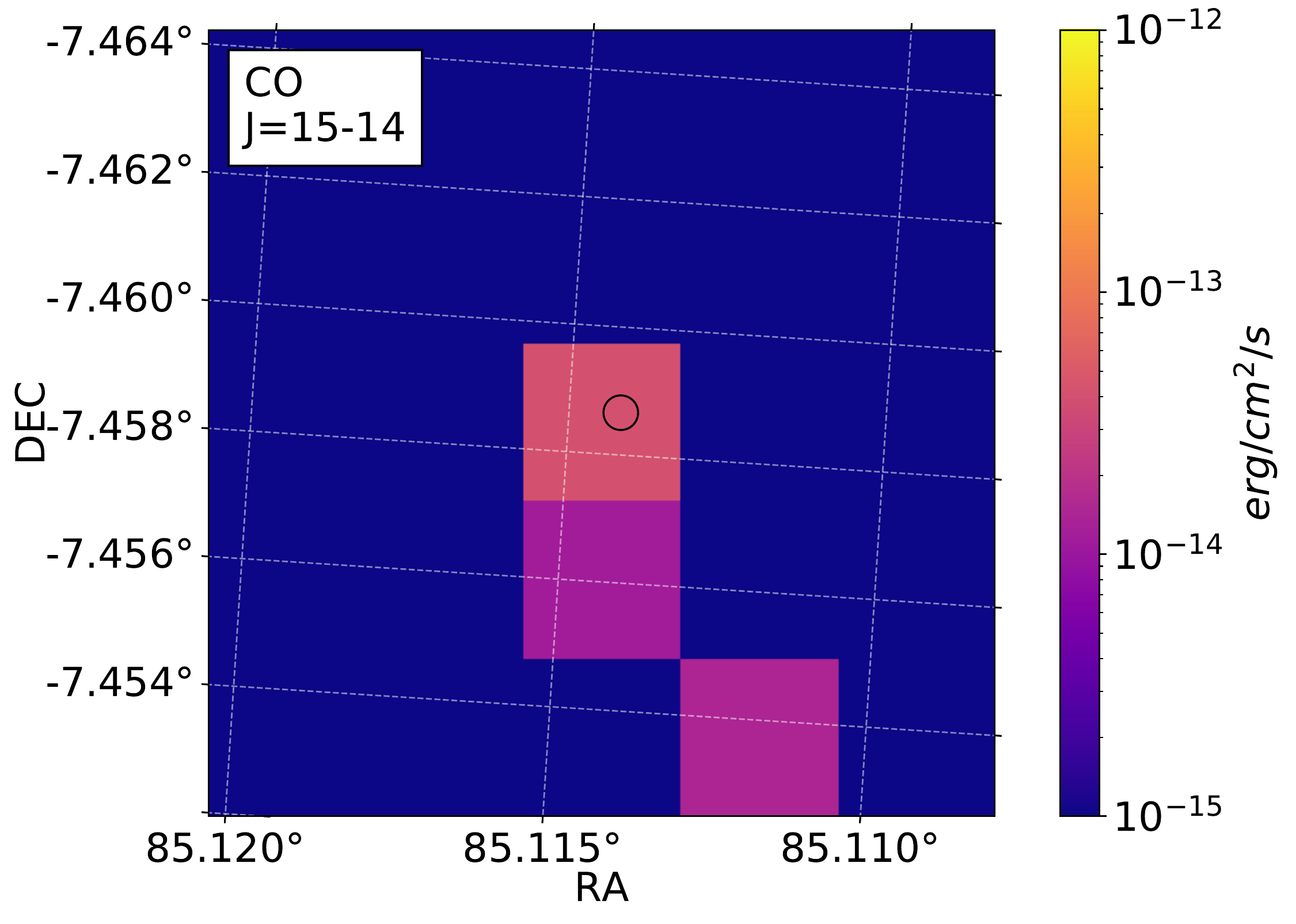}\hspace{-0.1cm}
\includegraphics[width=0.33\textwidth, trim={0cm 0 0cm 0}, clip]{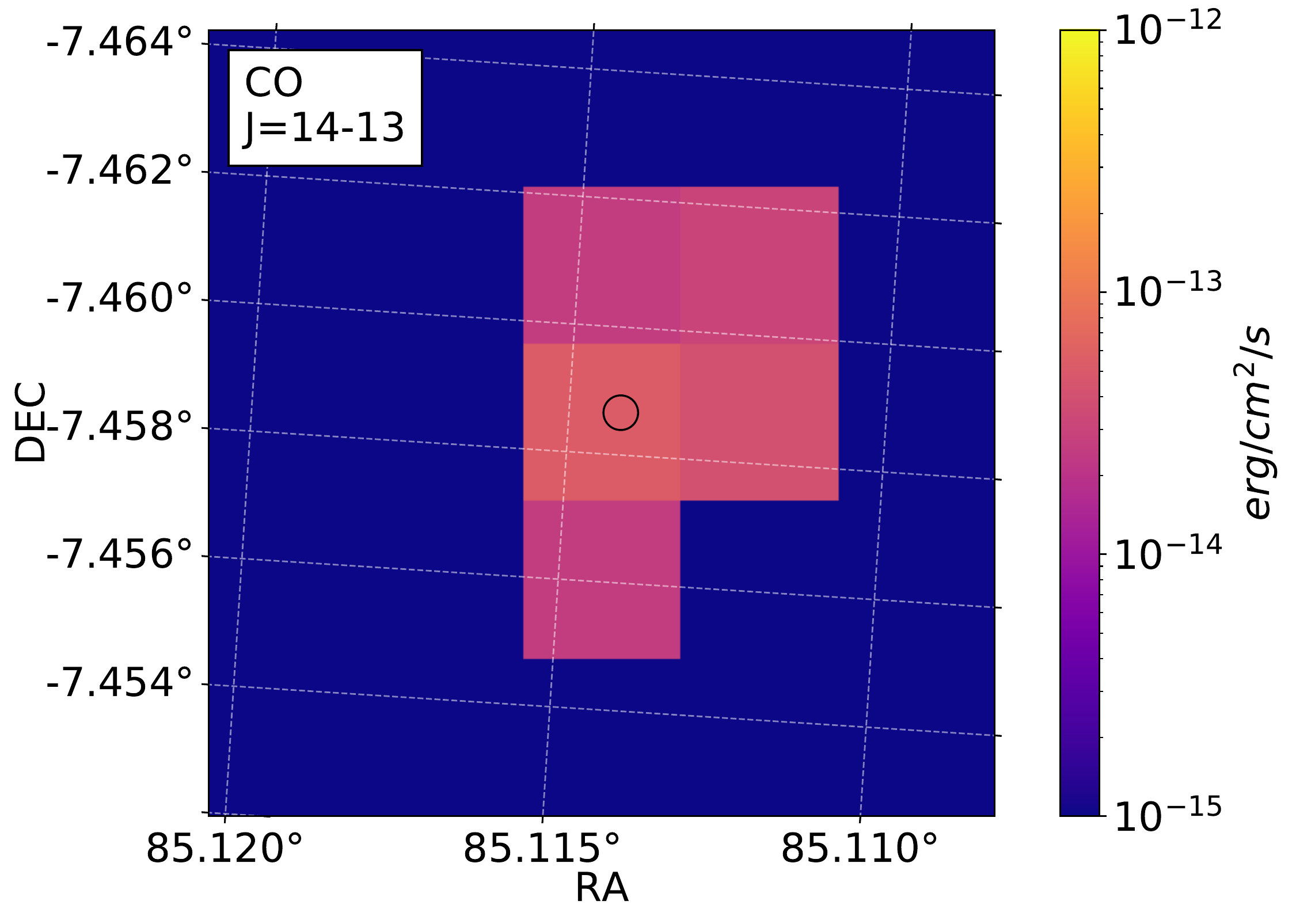}\hspace{-0.1cm}
\includegraphics[width=0.33\textwidth, trim={0cm 0 0cm 0}, clip]{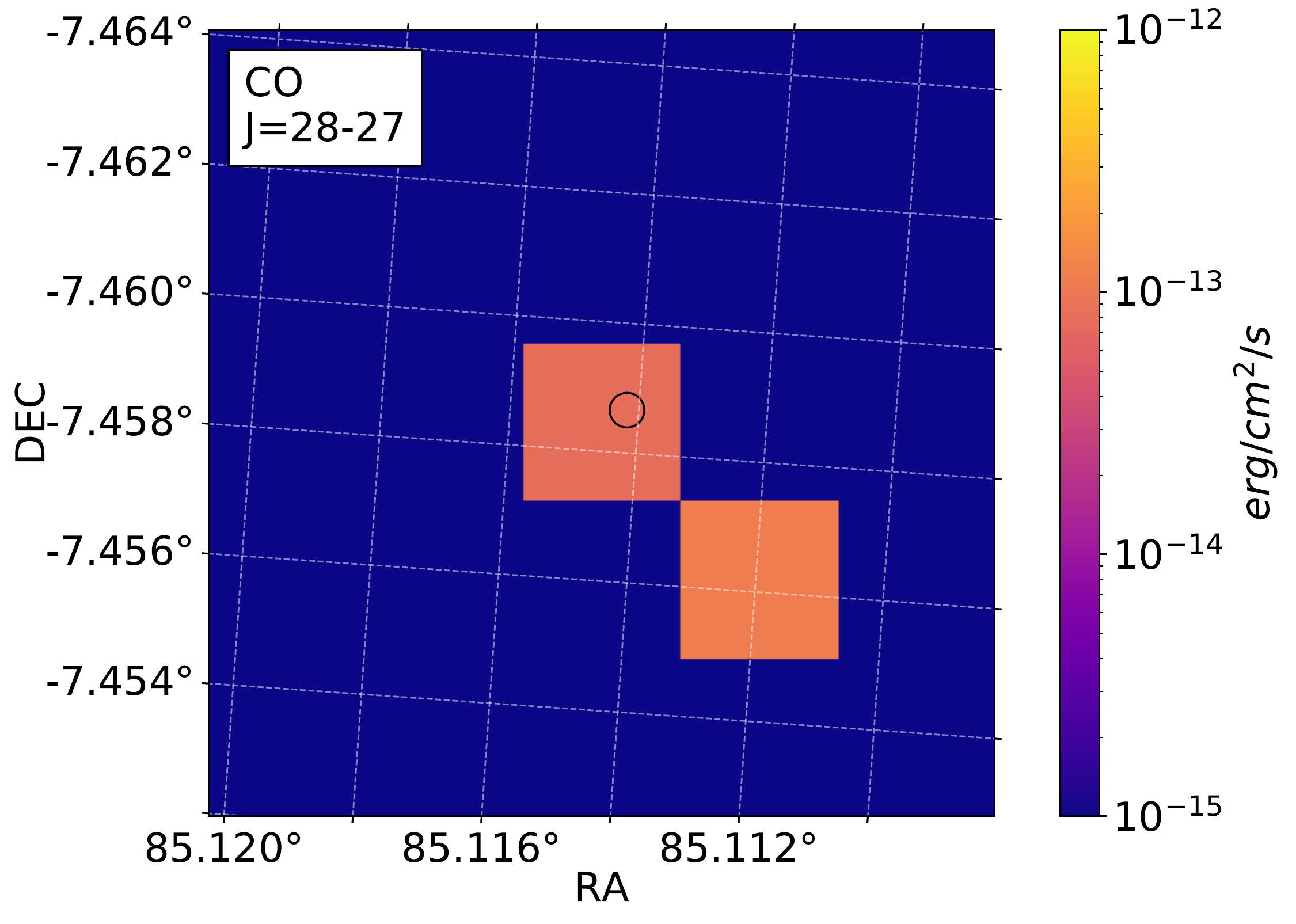}\hspace{-0.1cm}\\
\includegraphics[width=0.33\textwidth, trim={0cm 0 0cm 0}, clip]{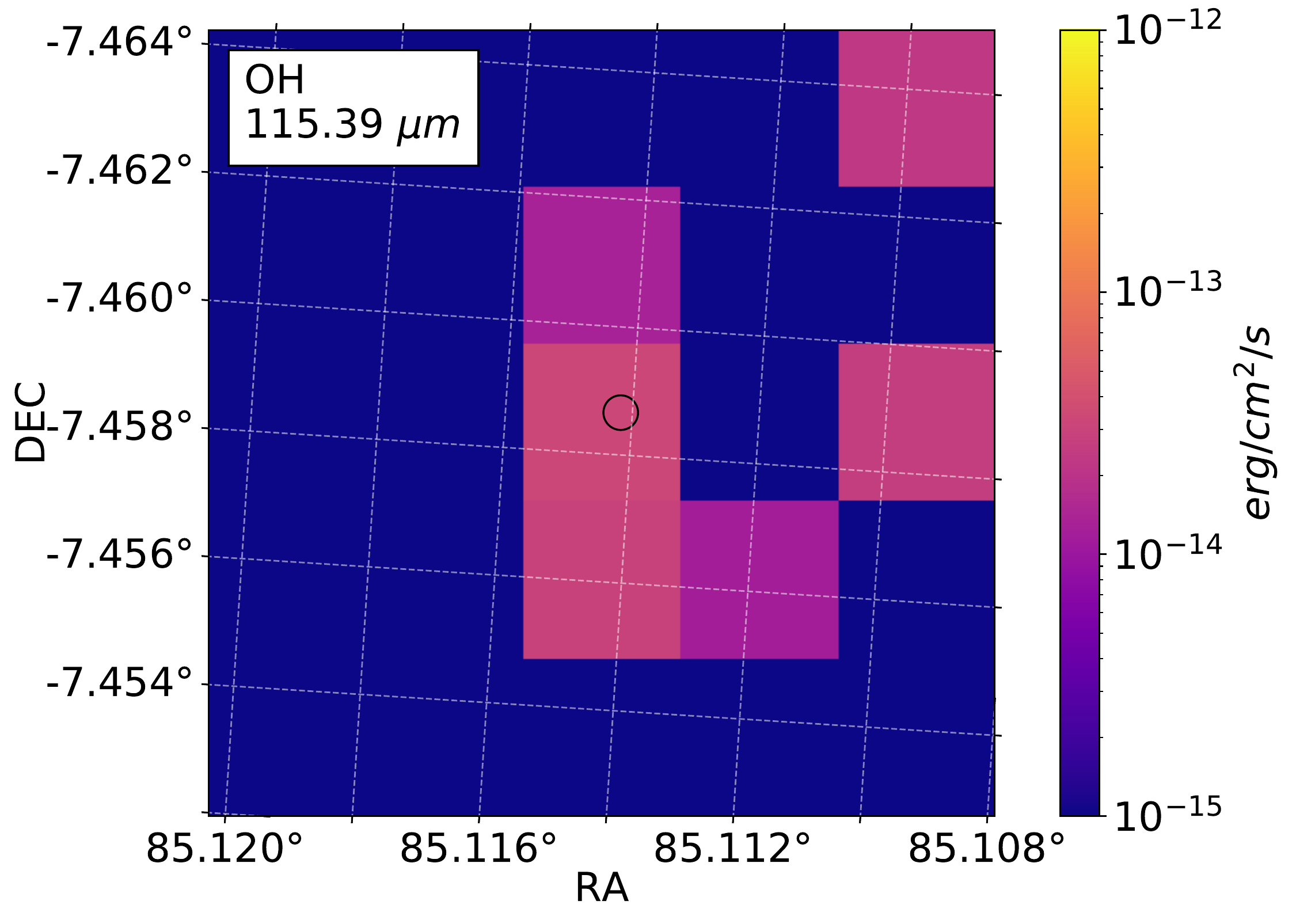}\hspace{-0.1cm}
\includegraphics[width=0.33\textwidth, trim={0cm 0 0cm 0}, clip]{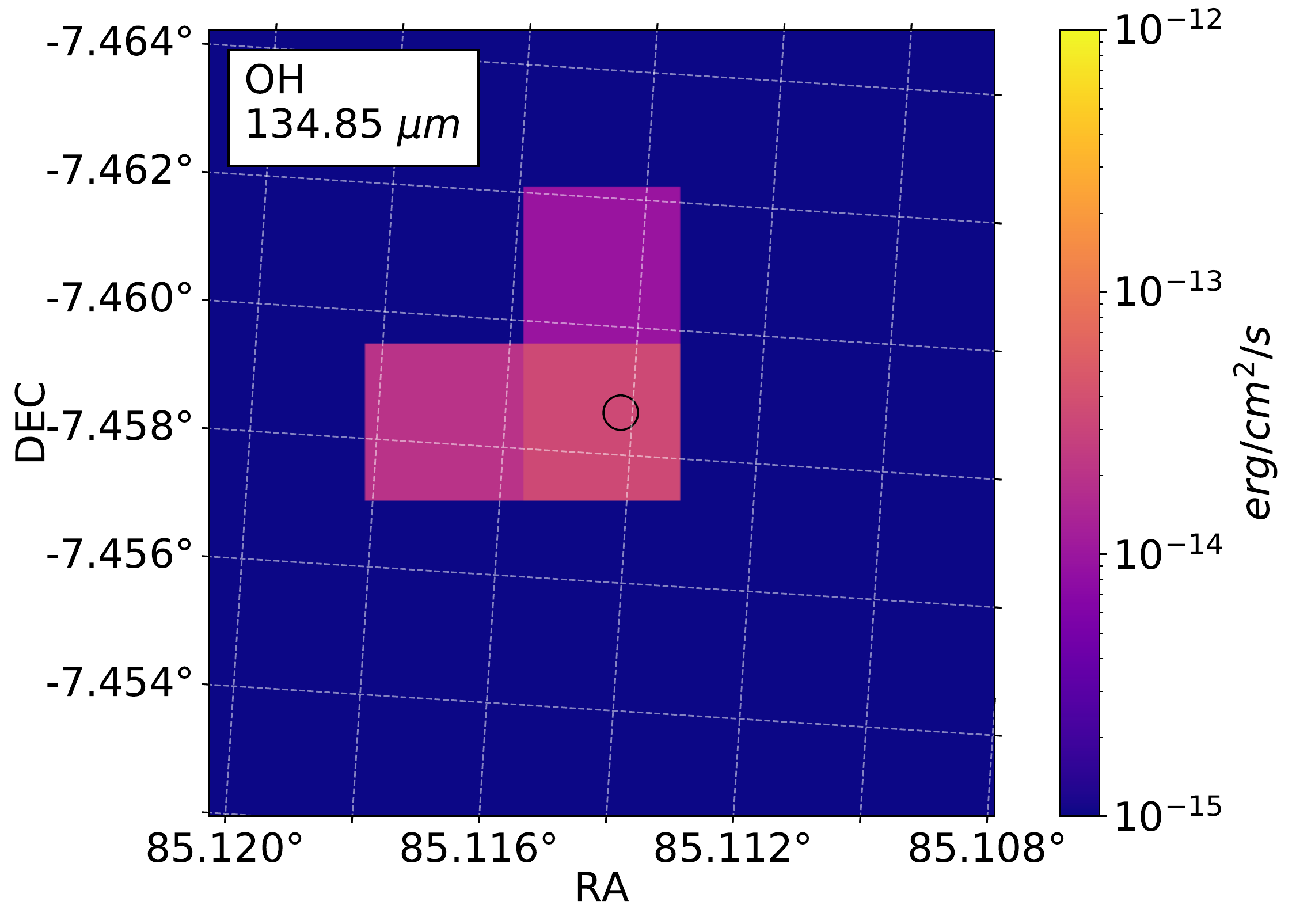}\hspace{-0.1cm}
\includegraphics[width=0.33\textwidth, trim={0cm 0 0cm 0}, clip]{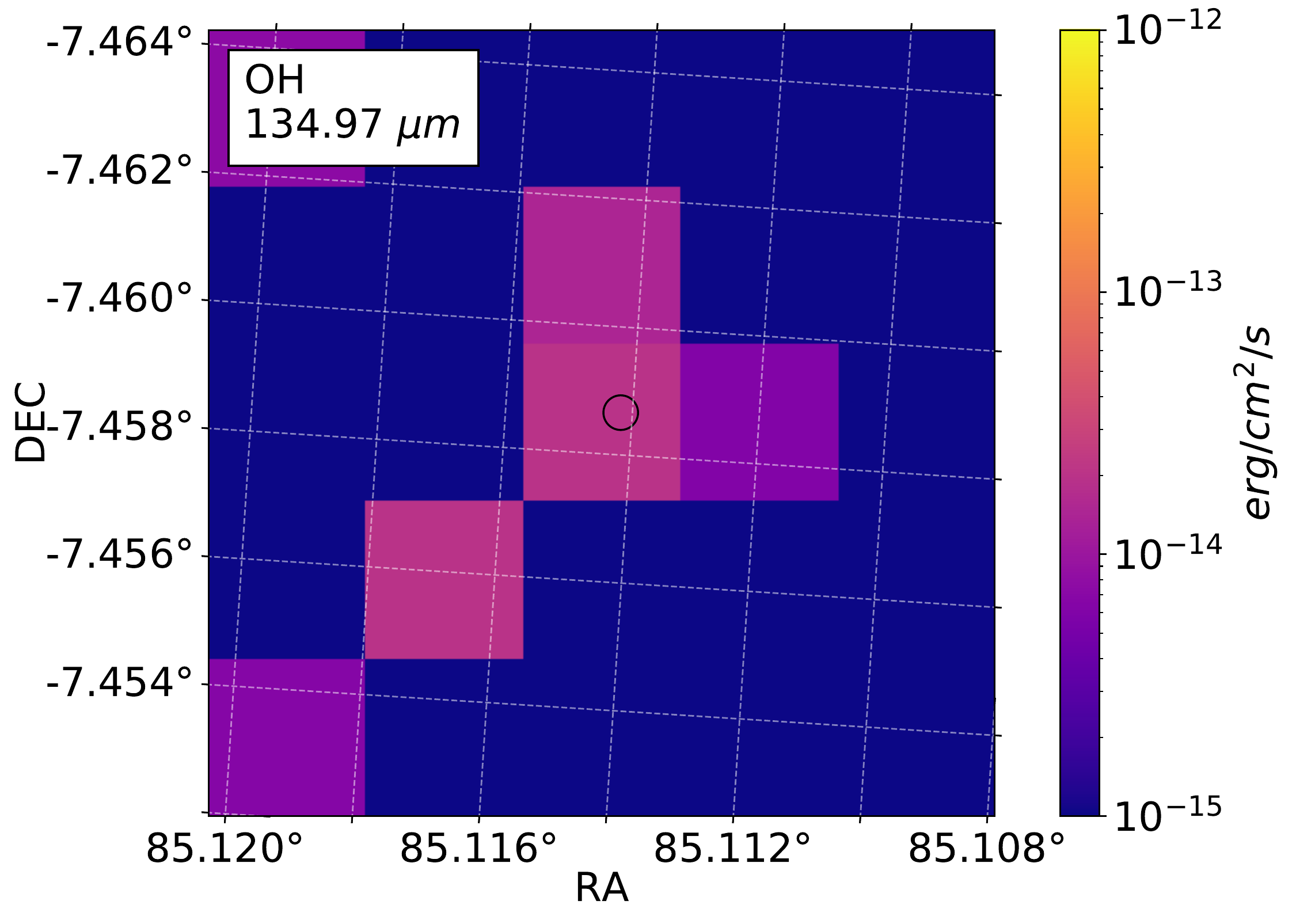}\hspace{-0.1cm}
\caption{
                \footnotesize
                Line maps of PACS with visible lines for Re 50 N IRS 1, part 1.
        }
\end{figure*}

\begin{figure*}
\includegraphics[width=0.33\textwidth, trim={0cm 0 0cm 0}, clip]{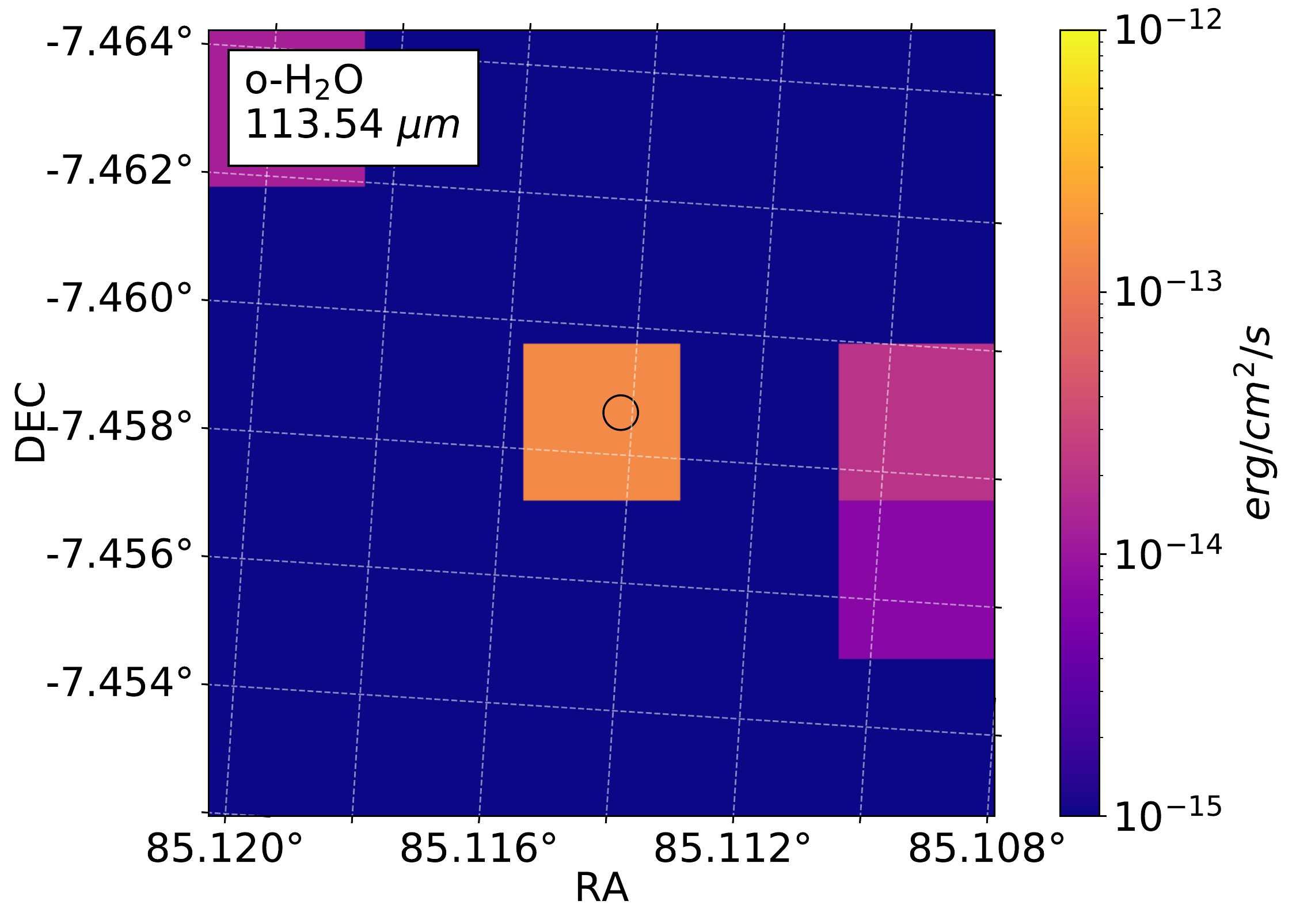}\hspace{-0.1cm}
\includegraphics[width=0.33\textwidth, trim={0cm 0 0cm 0}, clip]{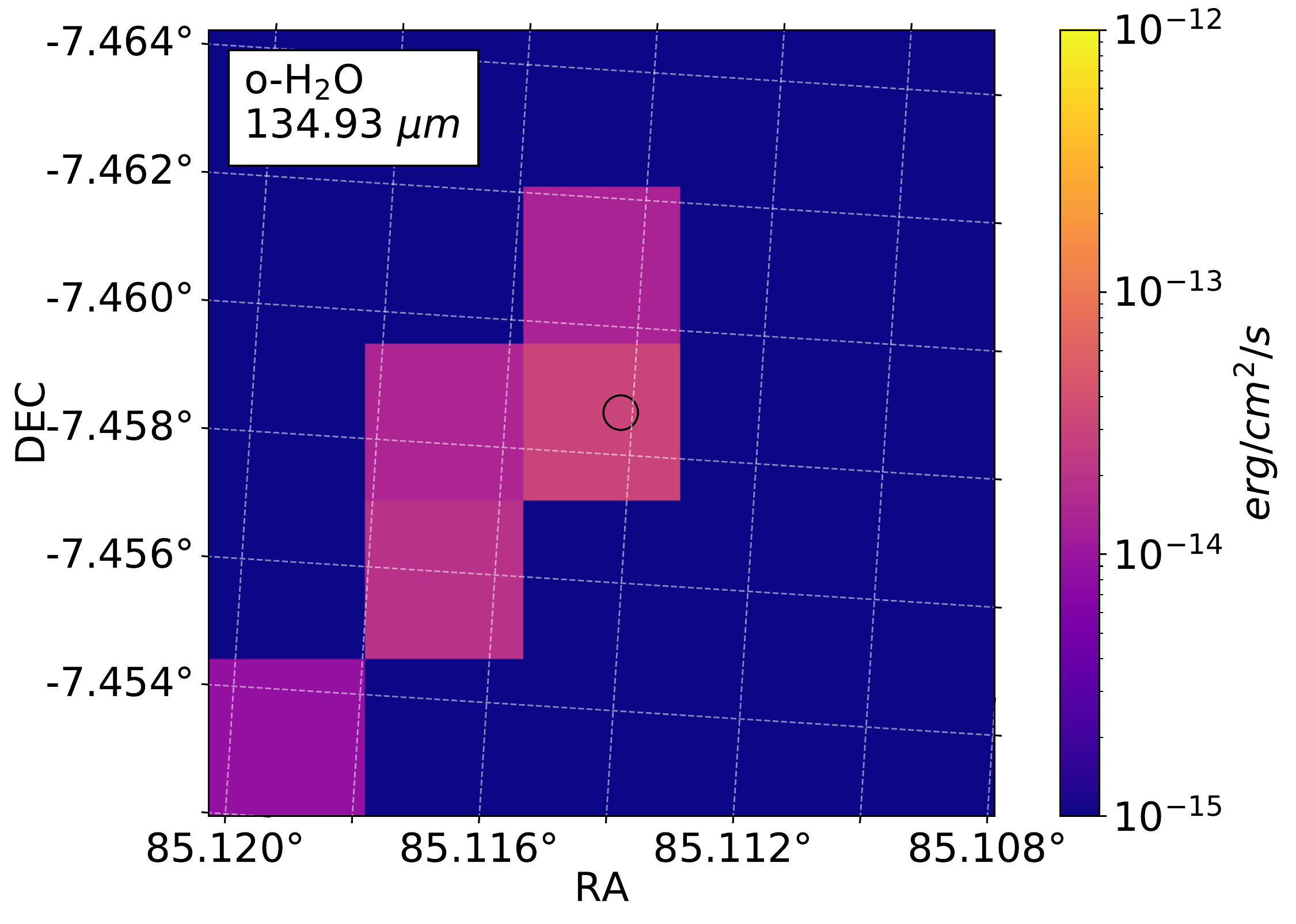}\hspace{-0.1cm}
\includegraphics[width=0.33\textwidth, trim={0cm 0 0cm 0}, clip]{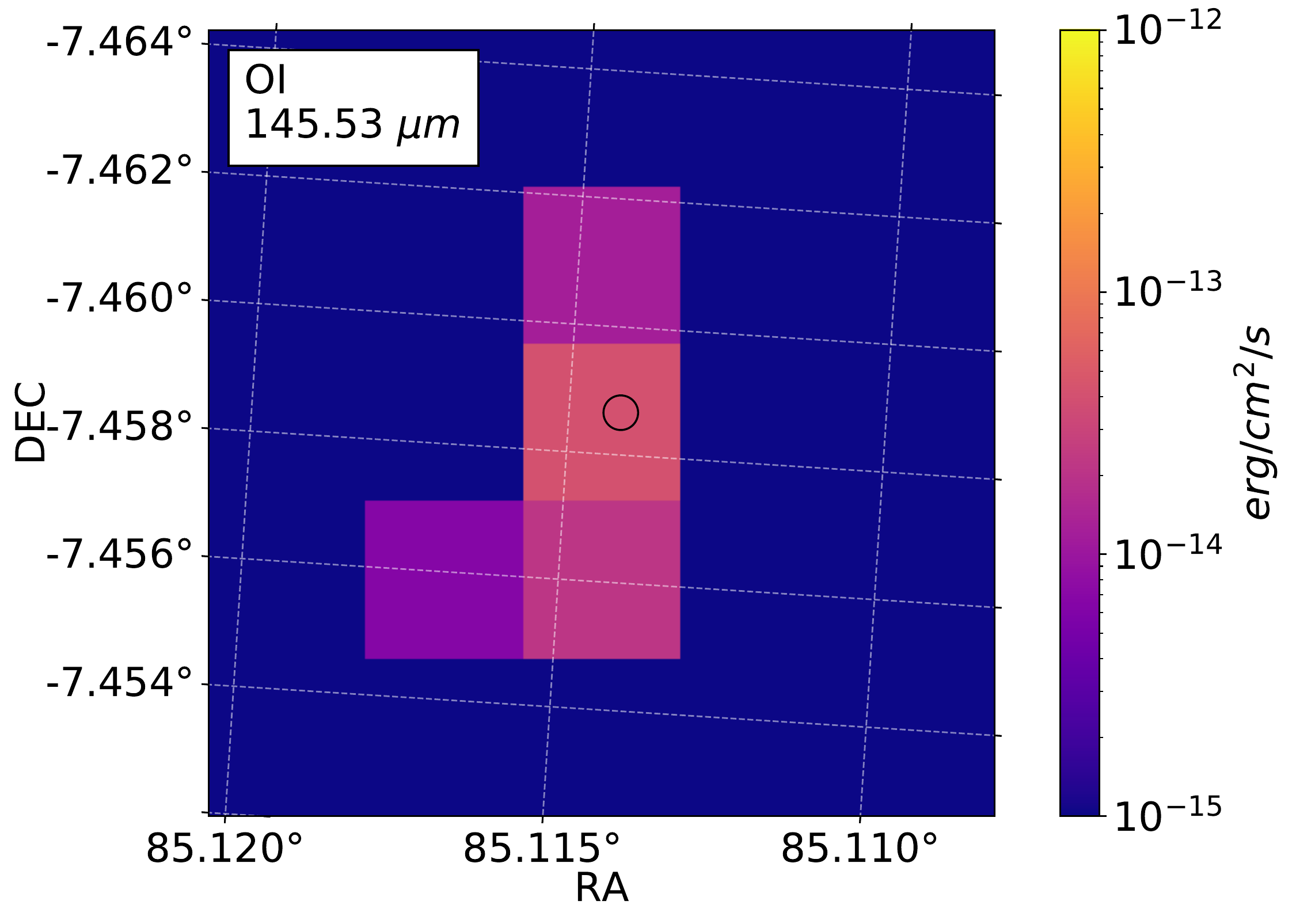}\hspace{-0.1cm}\\
\includegraphics[width=0.33\textwidth, trim={0cm 0 0cm 0}, clip]{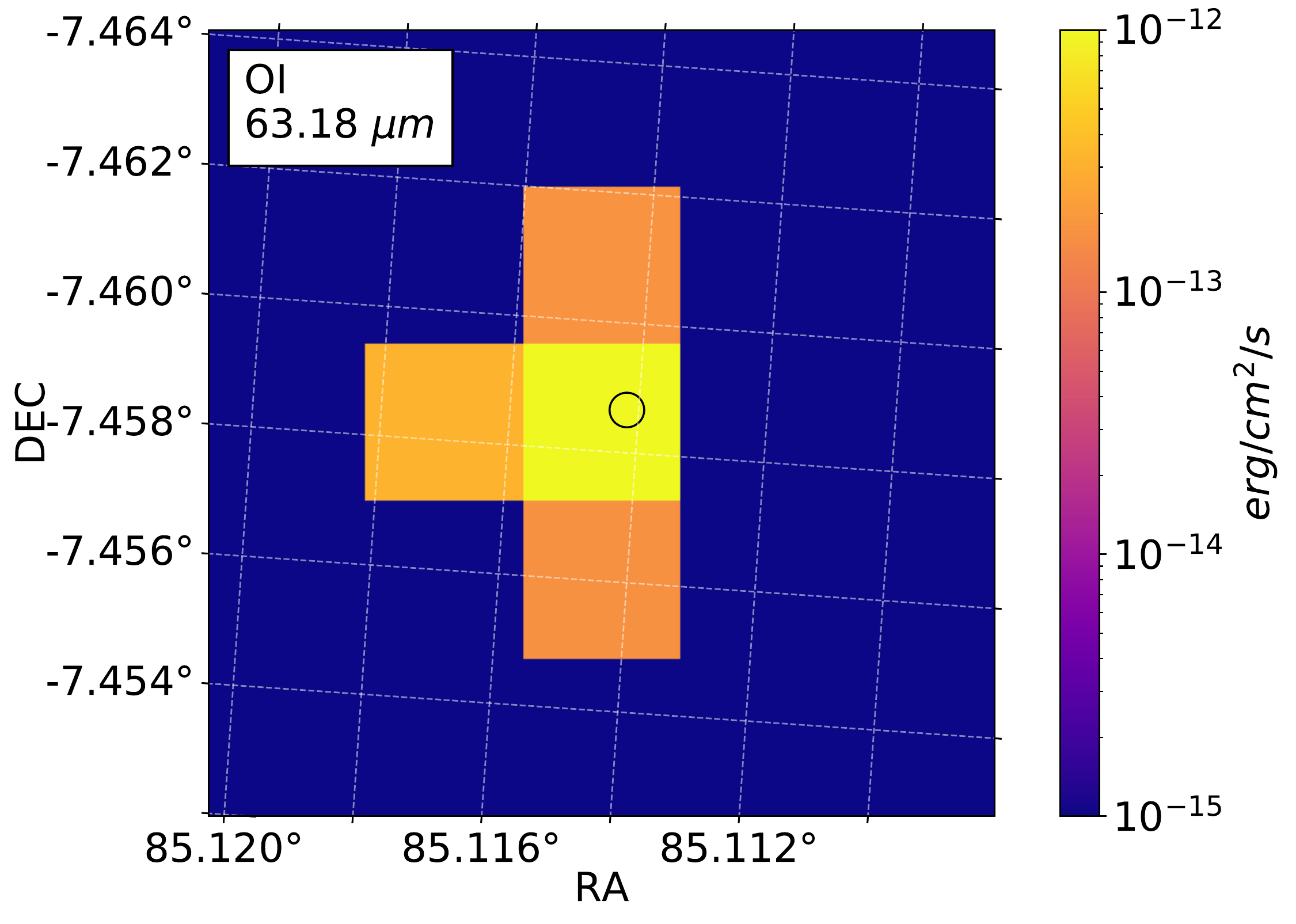}\hspace{-0.1cm}
\includegraphics[width=0.33\textwidth, trim={0cm 0 0cm 0}, clip]{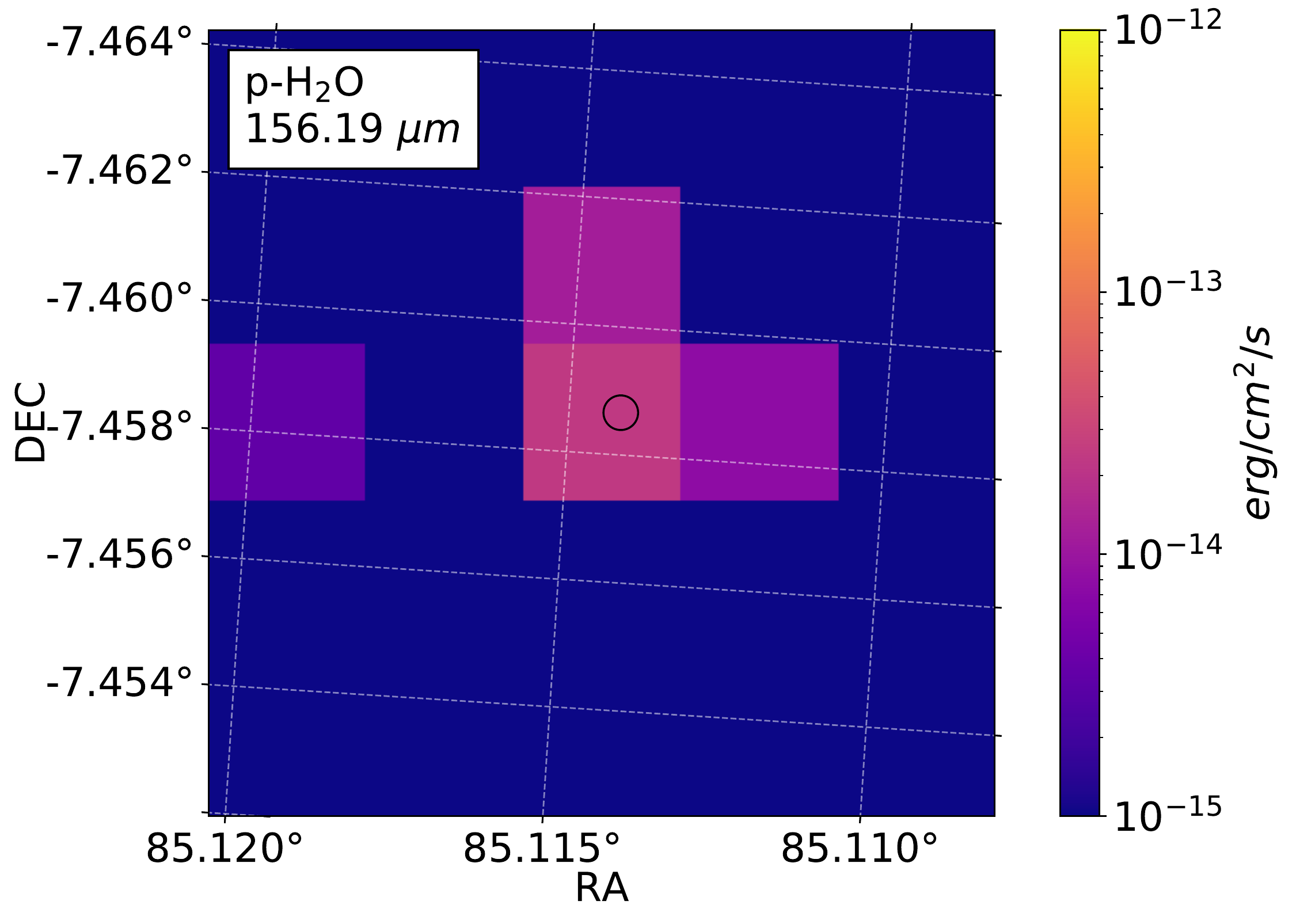}\hspace{-0.1cm}

    \caption{
                \footnotesize
                Line maps of PACS with visible lines for Re 50 N IRS 1, part 2.
        }
\end{figure*}

\begin{figure*}
\includegraphics[width=0.33\textwidth, trim={0cm 0 0cm 0}, clip]{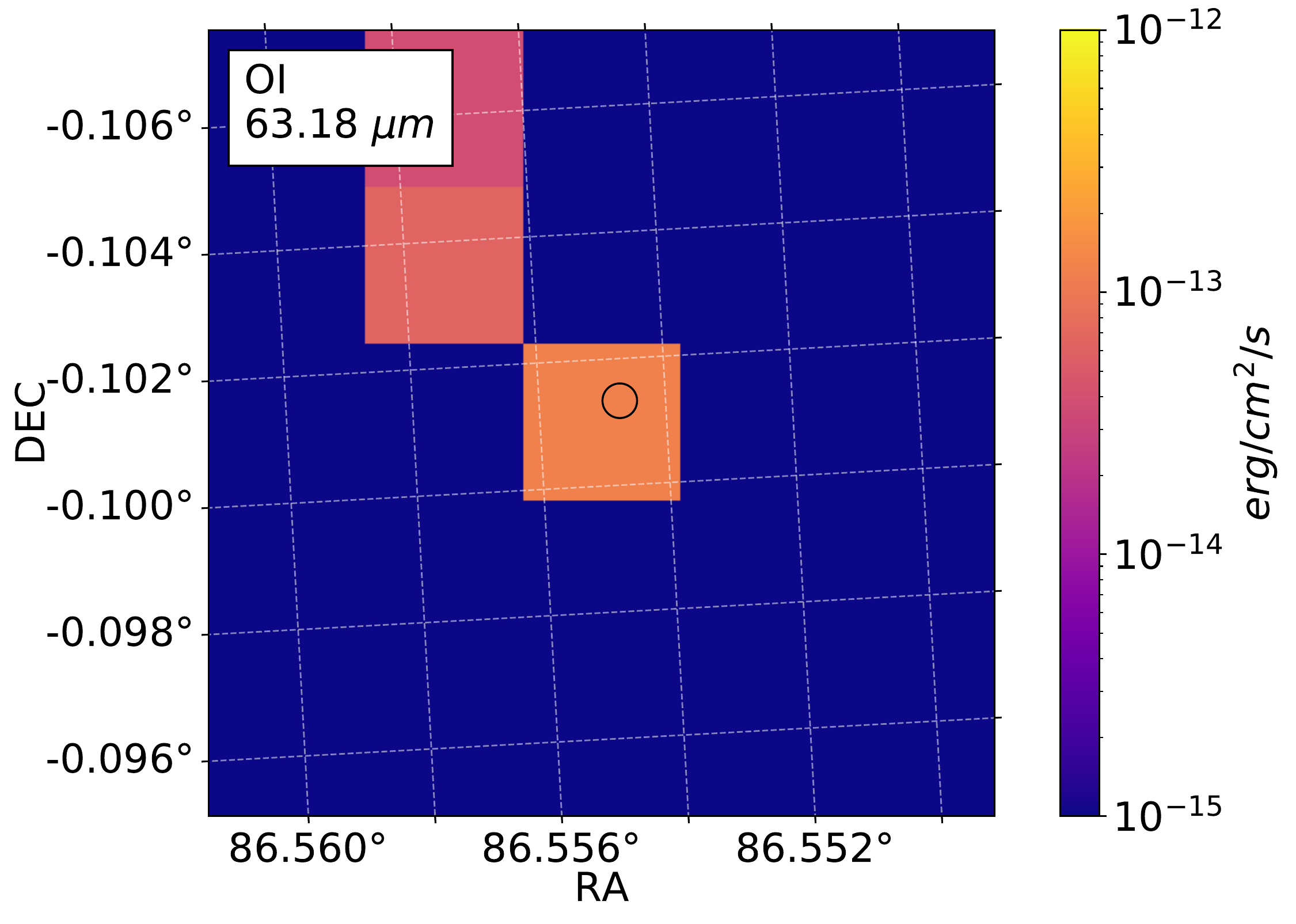}\hspace{-0.1cm}

    \caption{
                \footnotesize
                Line maps of PACS with visible lines for V1647 Ori.
        }
\end{figure*}

\begin{figure*}
\includegraphics[width=0.33\textwidth, trim={0cm 0 0cm 0}, clip]{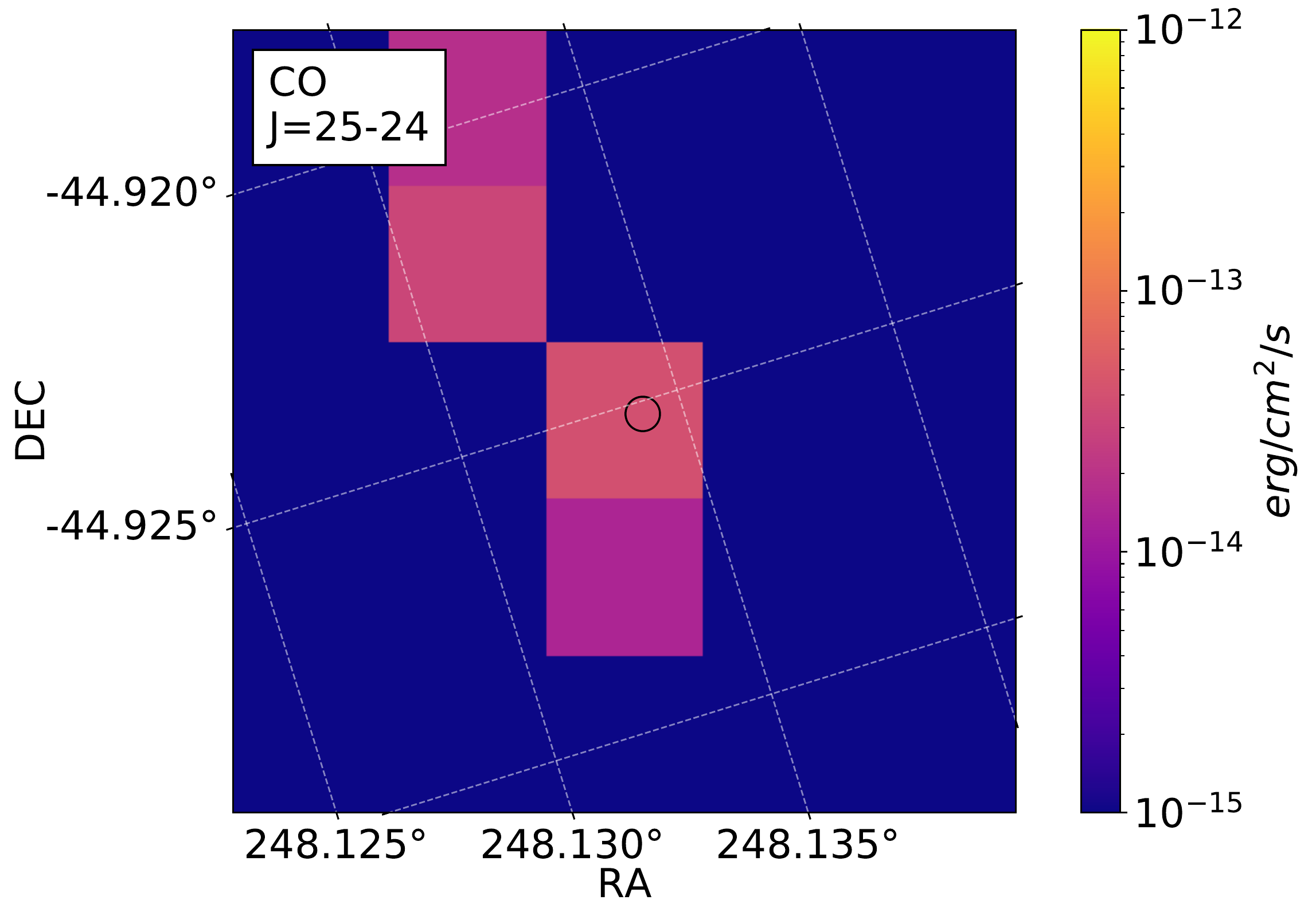}\hspace{-0.1cm}
\includegraphics[width=0.33\textwidth, trim={0cm 0 0cm 0}, clip]{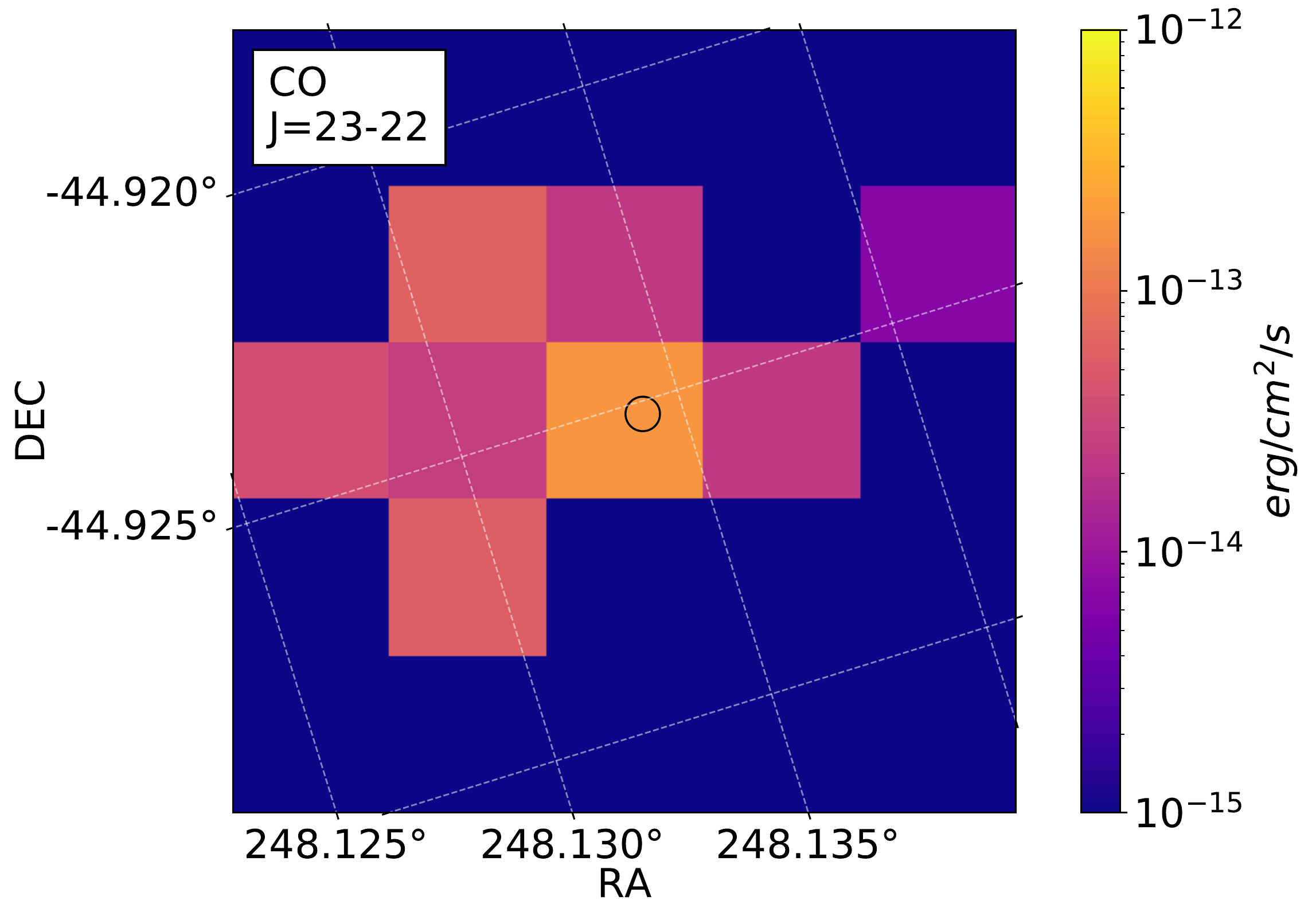}\hspace{-0.1cm}
\includegraphics[width=0.33\textwidth, trim={0cm 0 0cm 0}, clip]{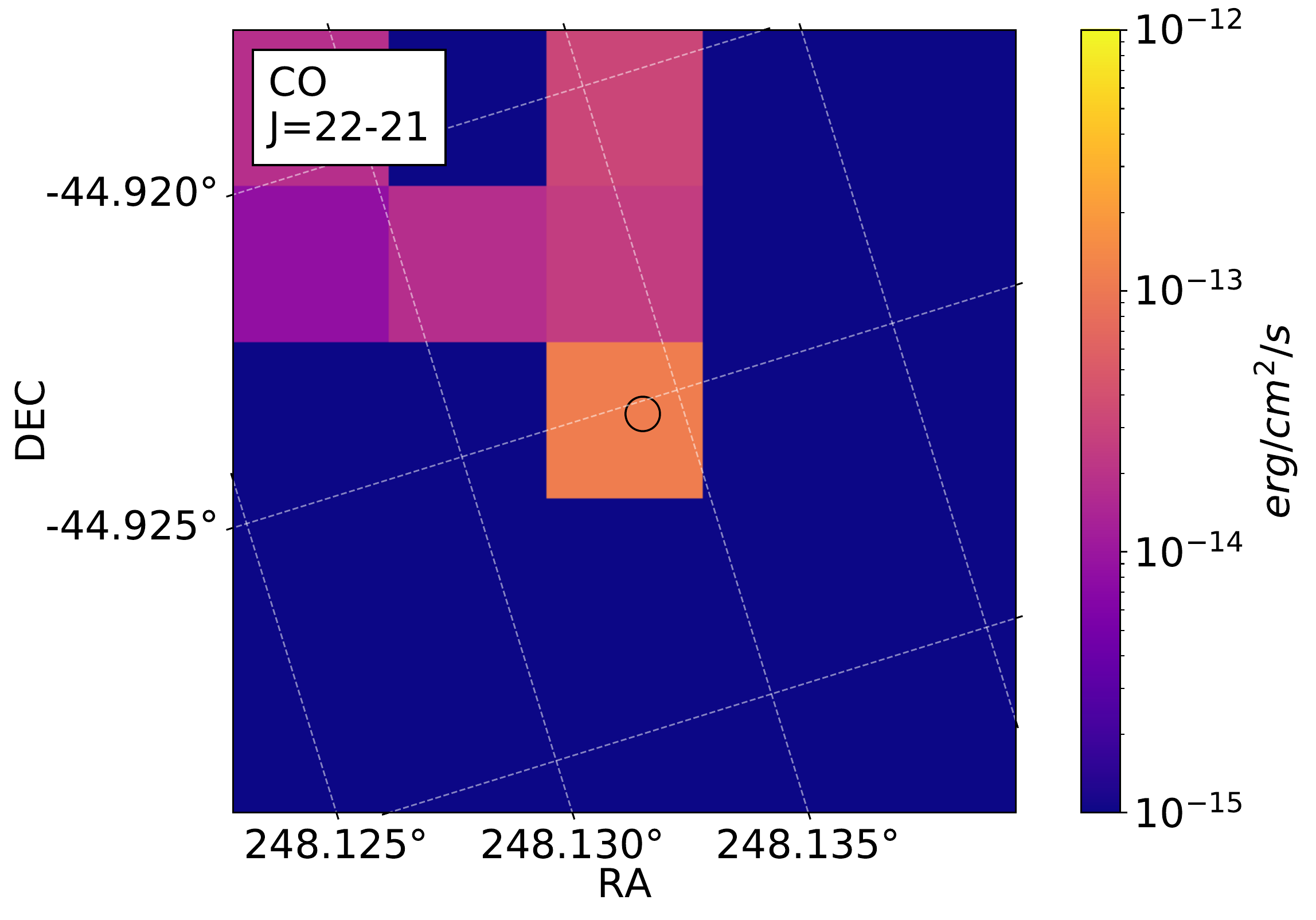}\hspace{-0.1cm}\\
\includegraphics[width=0.33\textwidth, trim={0cm 0 0cm 0}, clip]{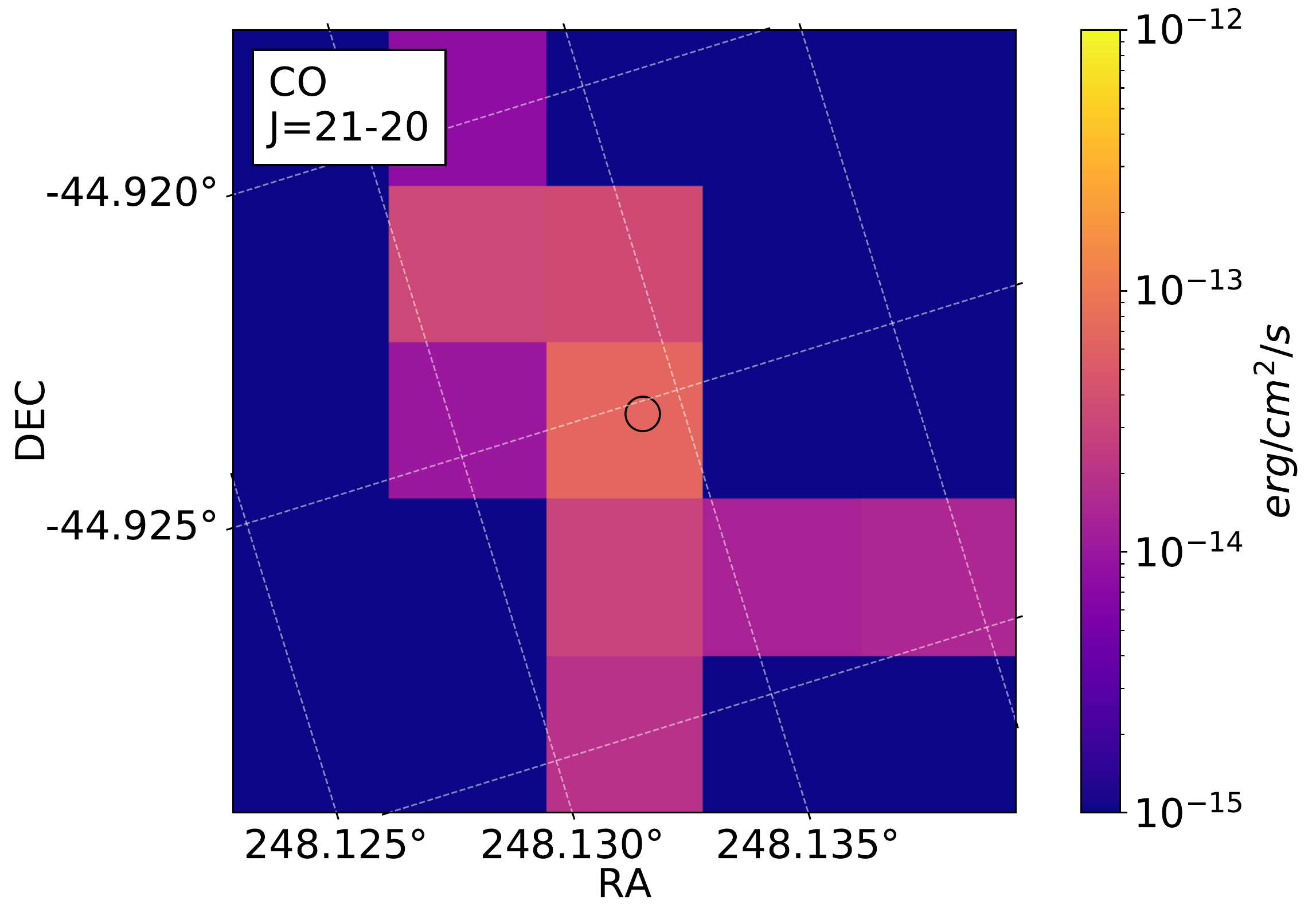}\hspace{-0.1cm}
\includegraphics[width=0.33\textwidth, trim={0cm 0 0cm 0}, clip]{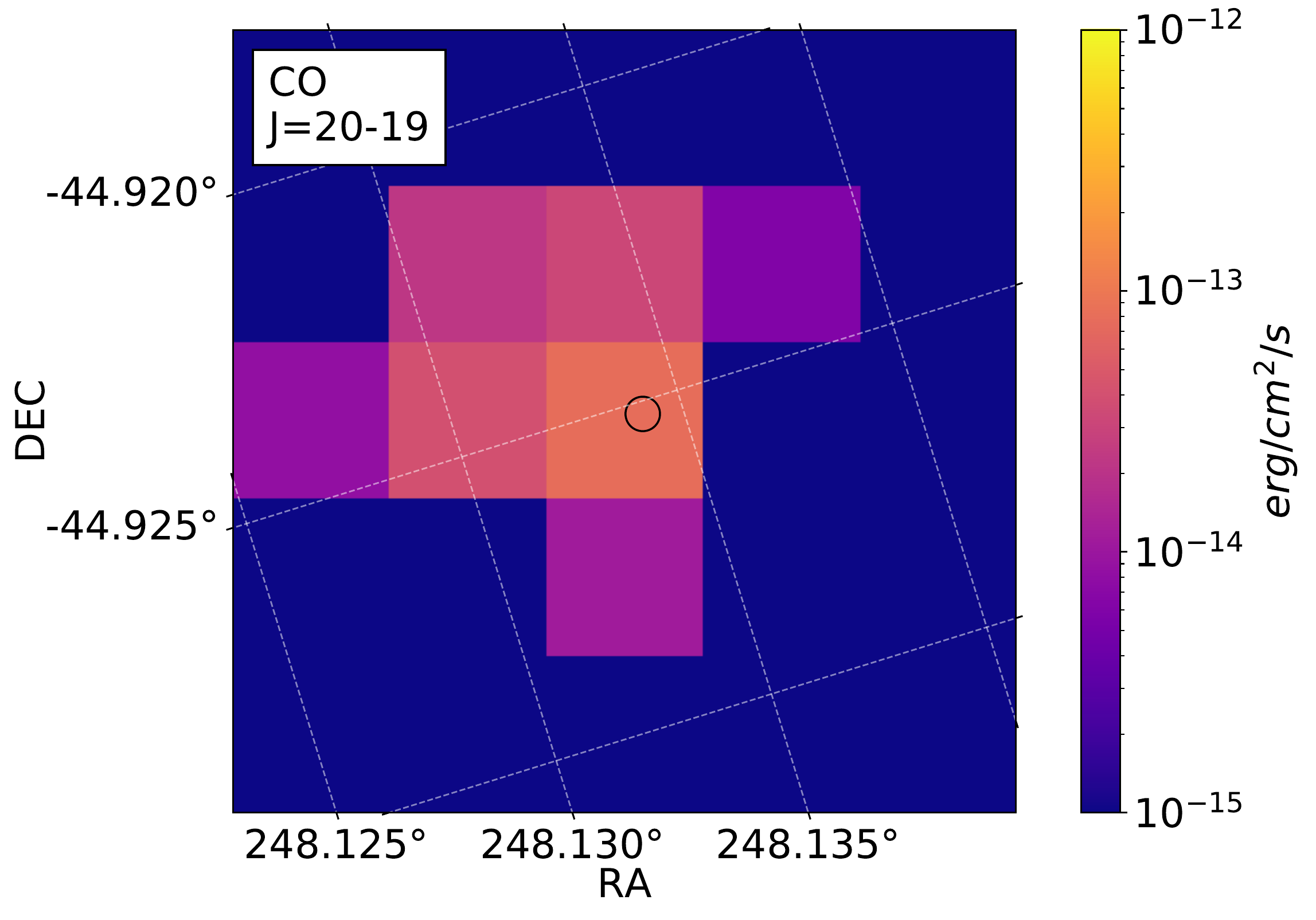}\hspace{-0.1cm}
\includegraphics[width=0.33\textwidth, trim={0cm 0 0cm 0}, clip]{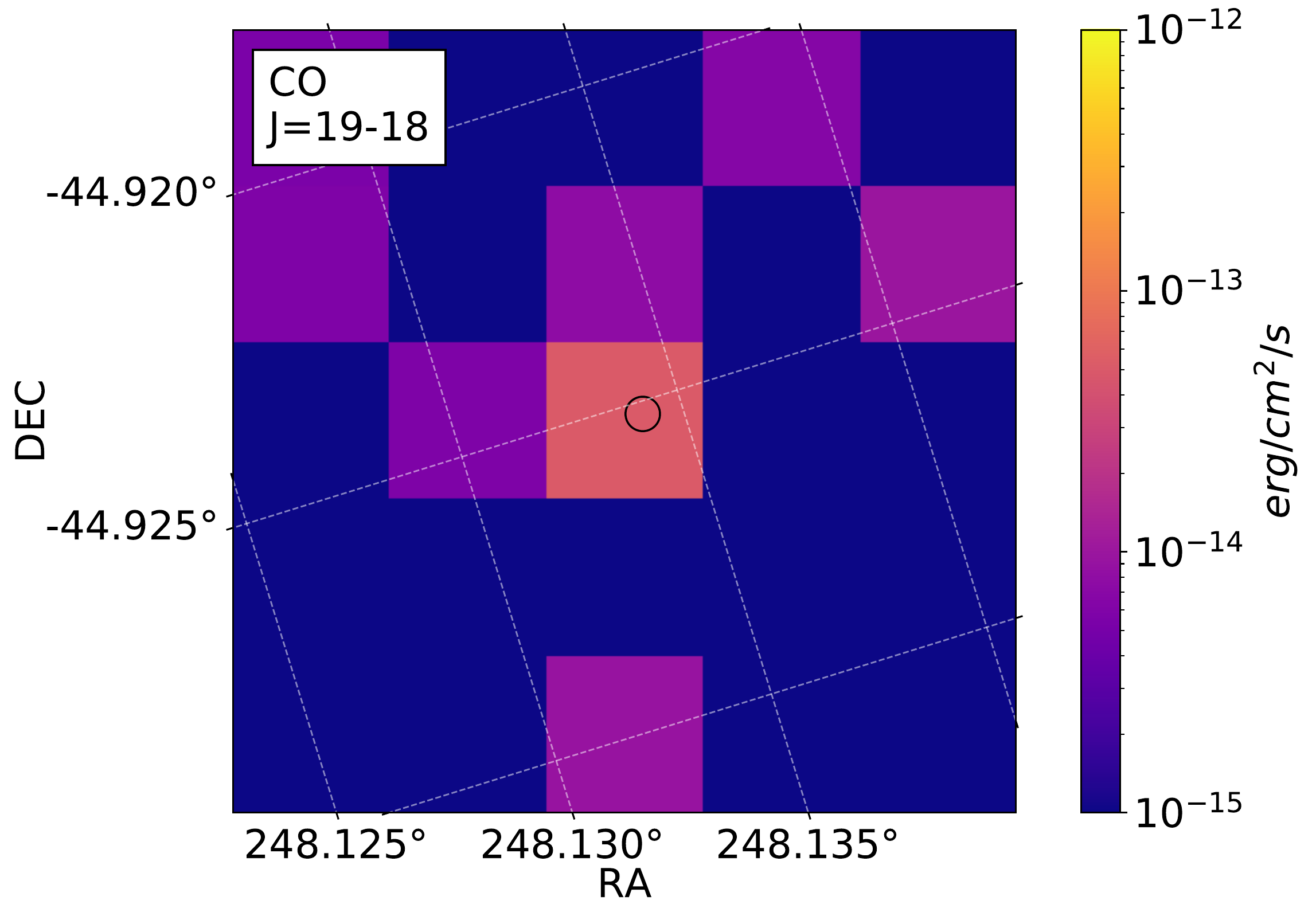}\hspace{-0.1cm}\\
\includegraphics[width=0.33\textwidth, trim={0cm 0 0cm 0}, clip]{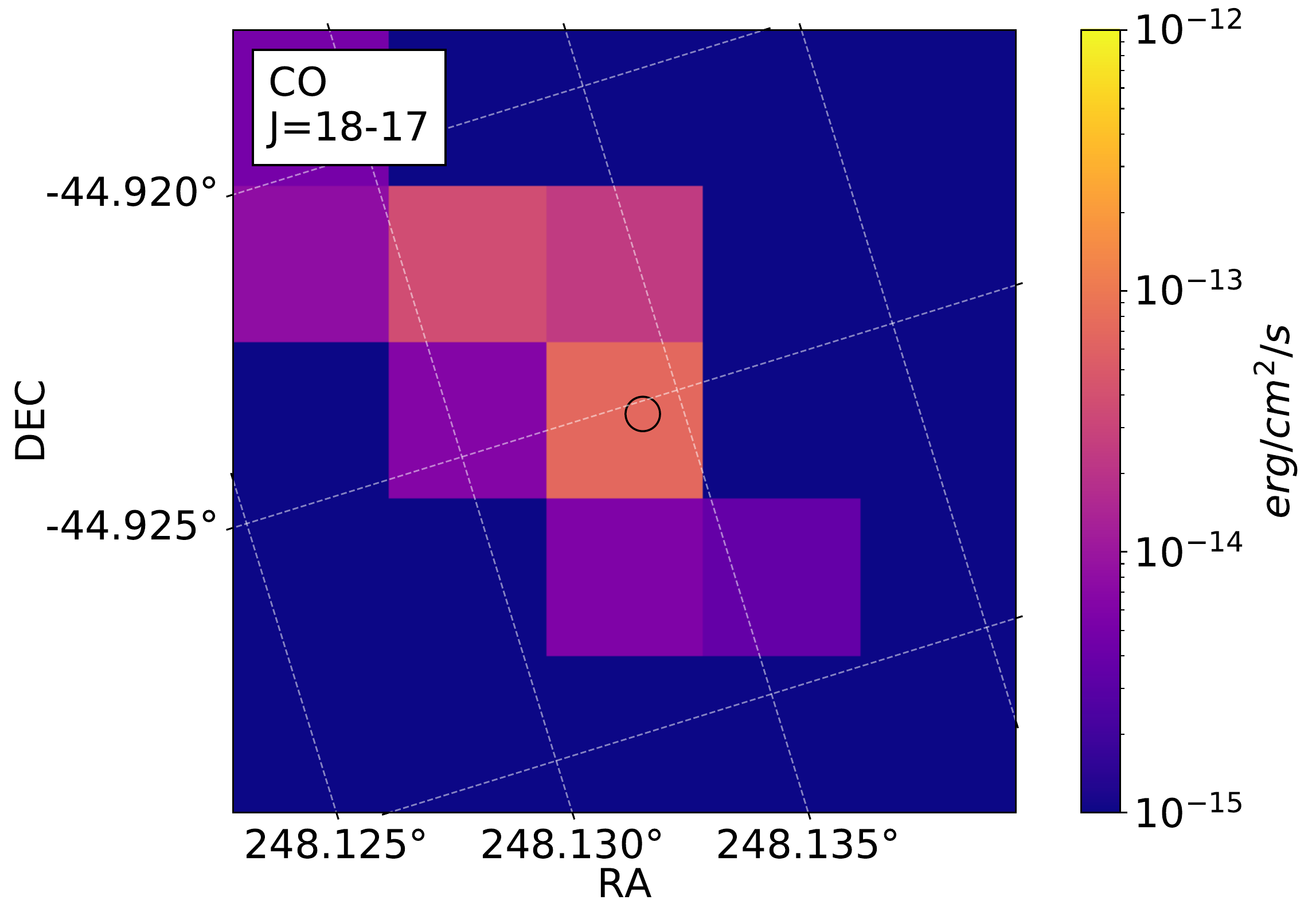}\hspace{-0.1cm}
\includegraphics[width=0.33\textwidth, trim={0cm 0 0cm 0}, clip]{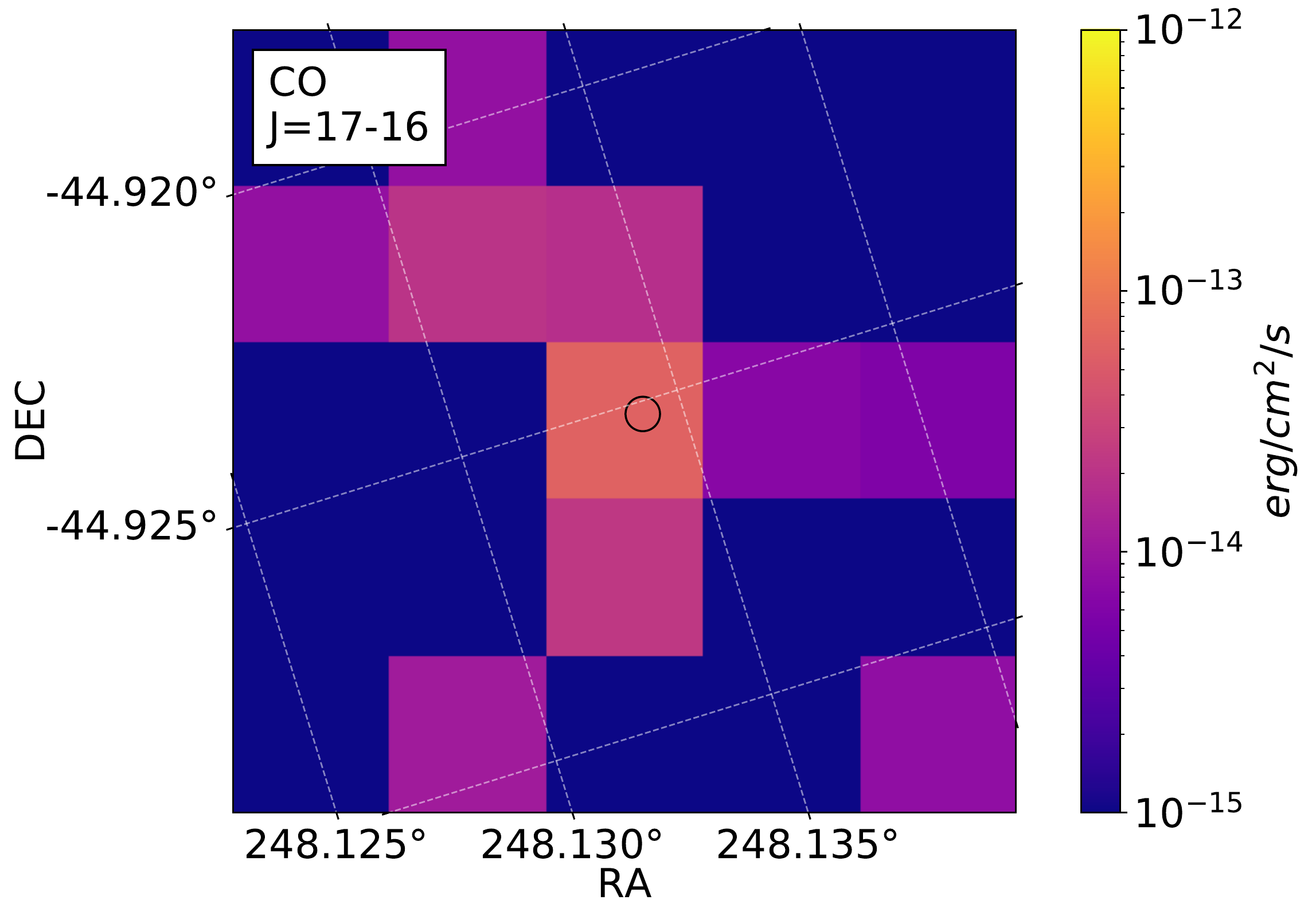}\hspace{-0.1cm}
\includegraphics[width=0.33\textwidth, trim={0cm 0 0cm 0}, clip]{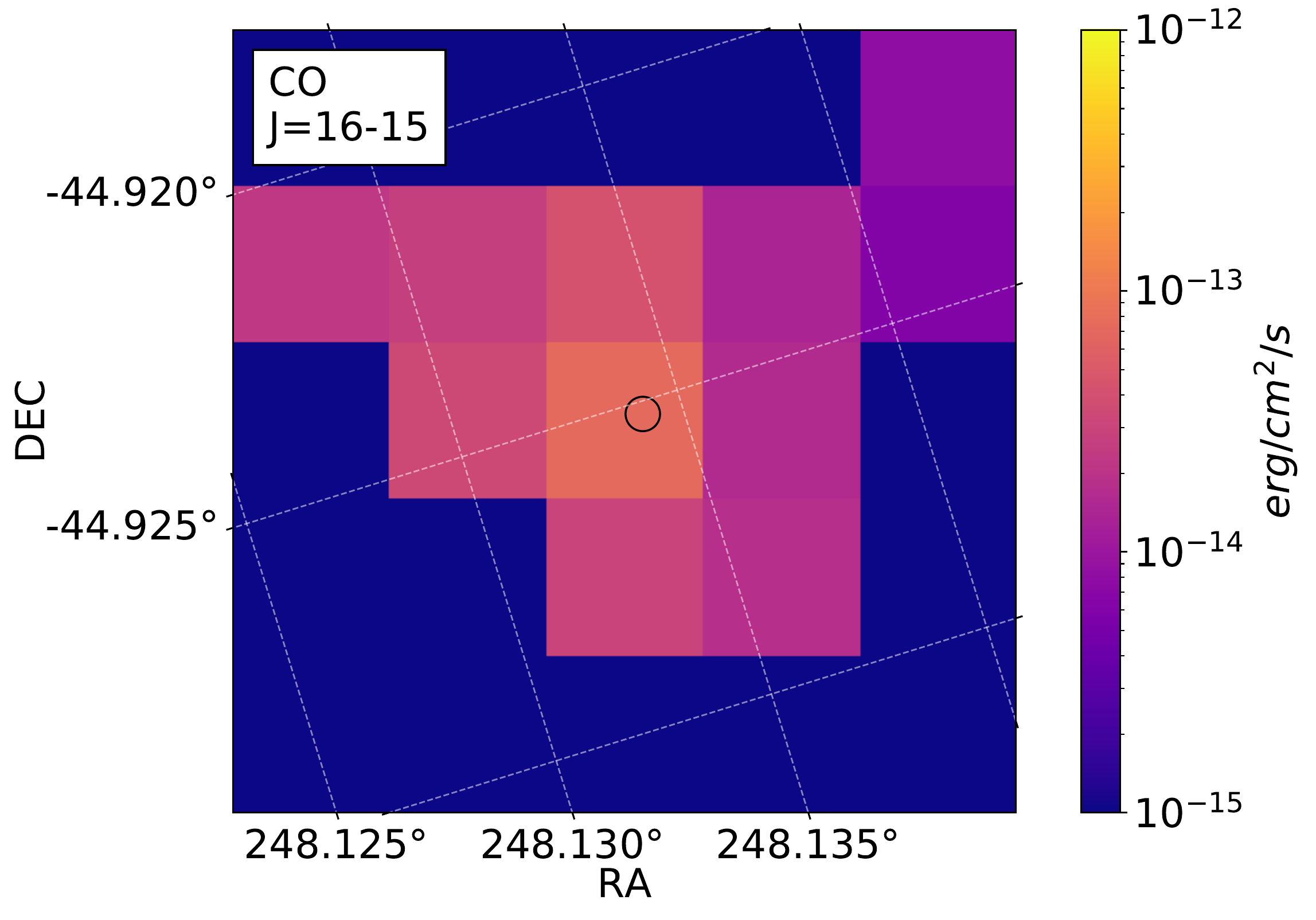}\hspace{-0.1cm}\\
\includegraphics[width=0.33\textwidth, trim={0cm 0 0cm 0}, clip]{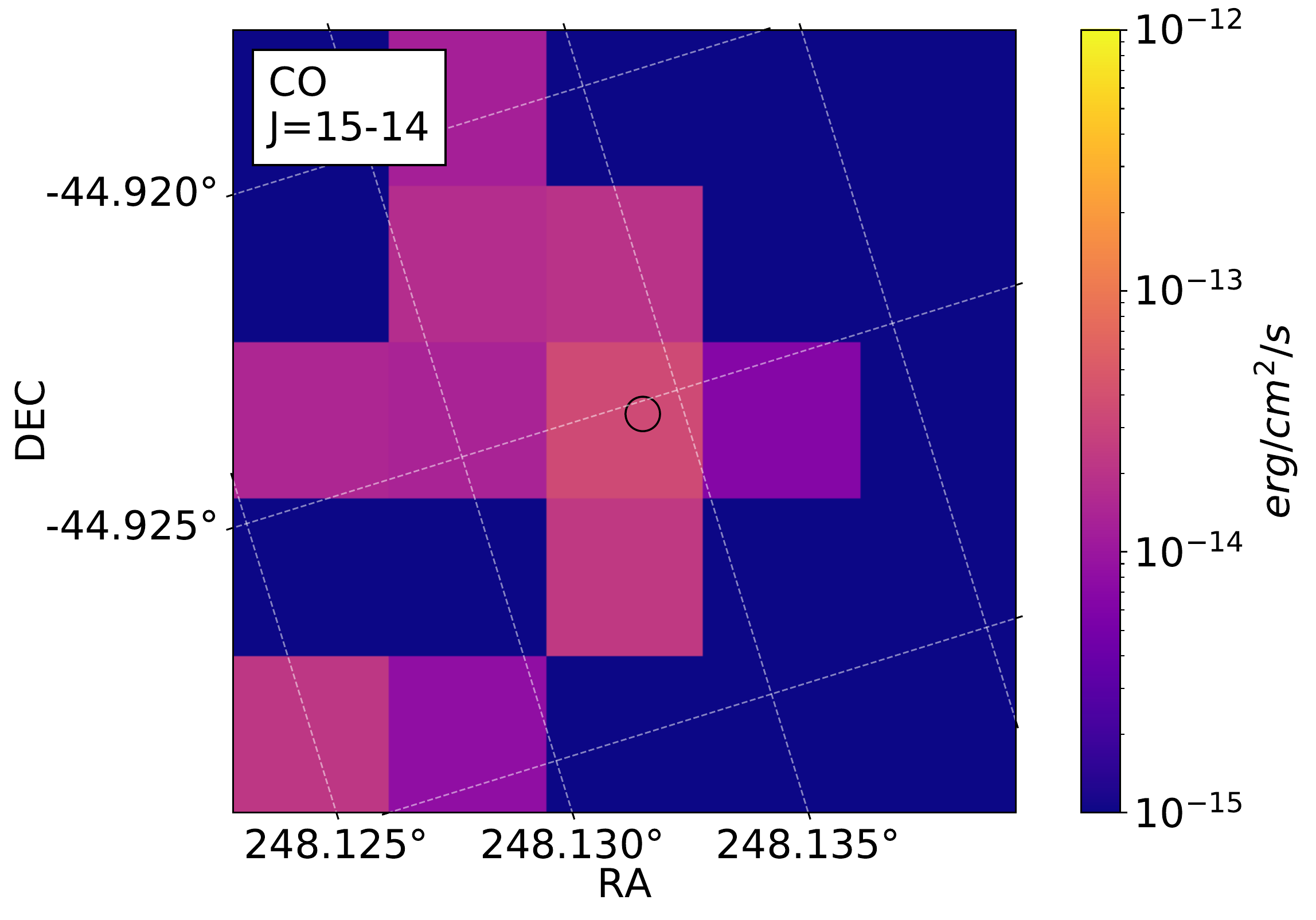}\hspace{-0.1cm}
\includegraphics[width=0.33\textwidth, trim={0cm 0 0cm 0}, clip]{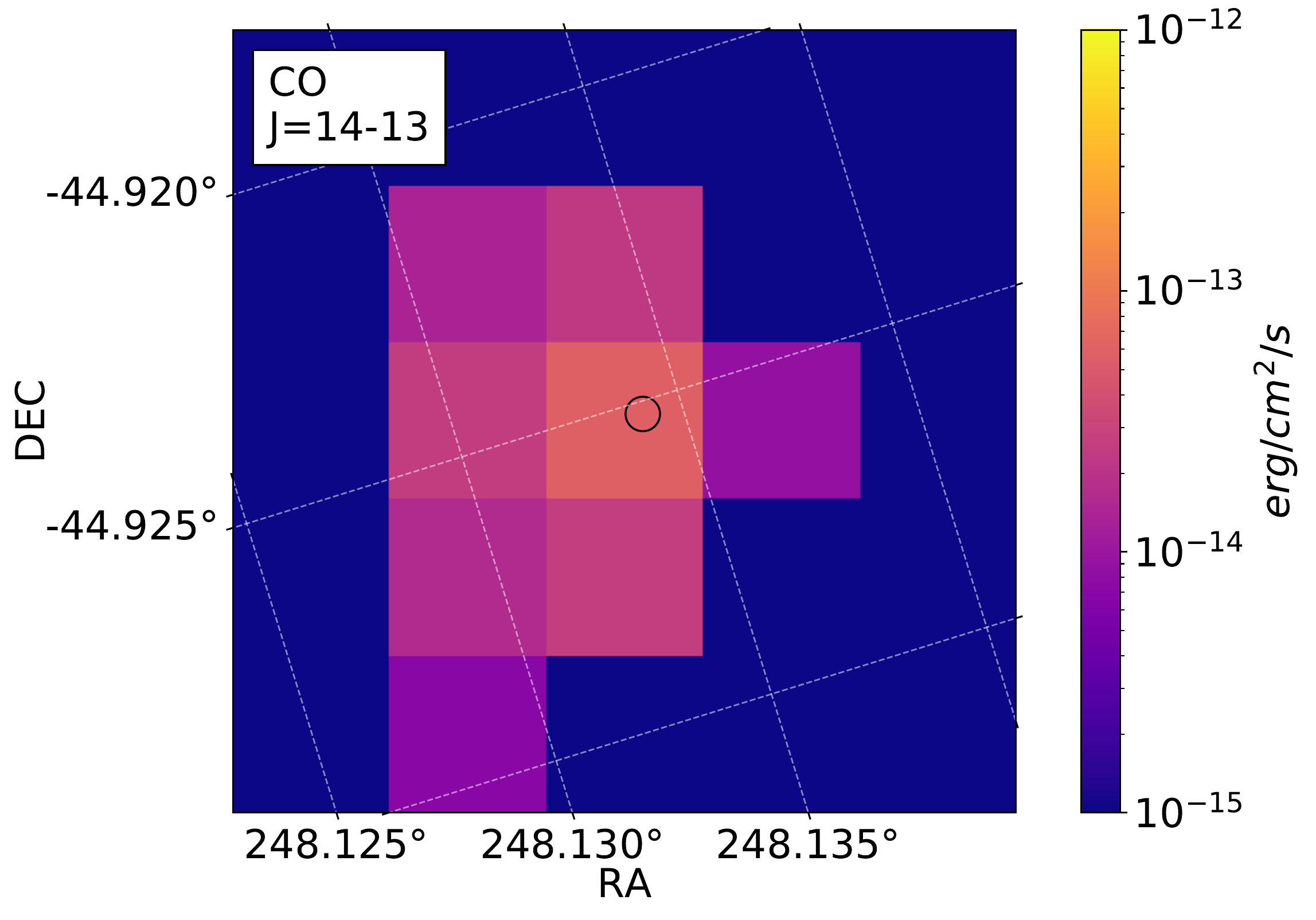}\hspace{-0.1cm}
\includegraphics[width=0.33\textwidth, trim={0cm 0 0cm 0}, clip]{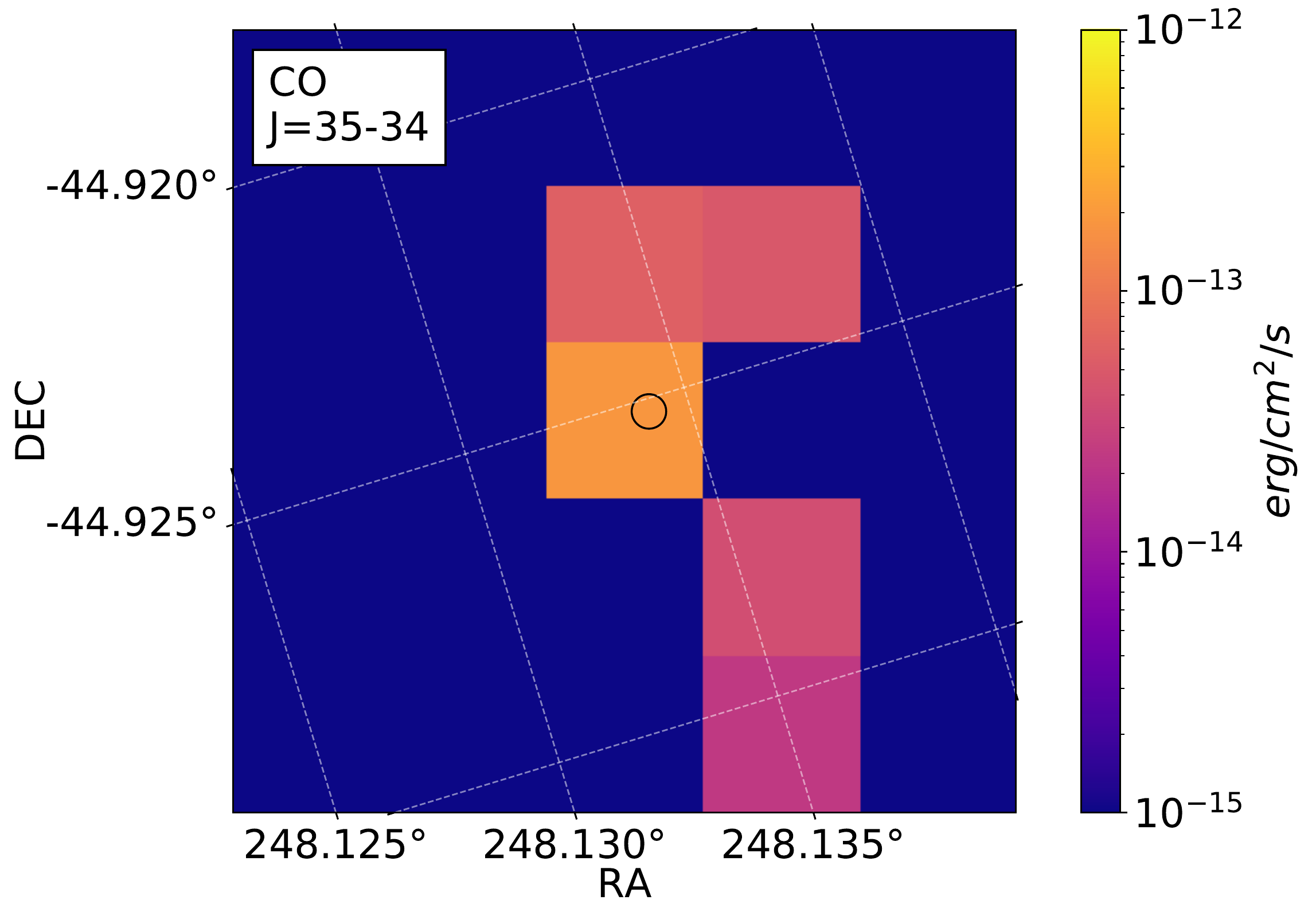}\hspace{-0.1cm}\\
\includegraphics[width=0.33\textwidth, trim={0cm 0 0cm 0}, clip]{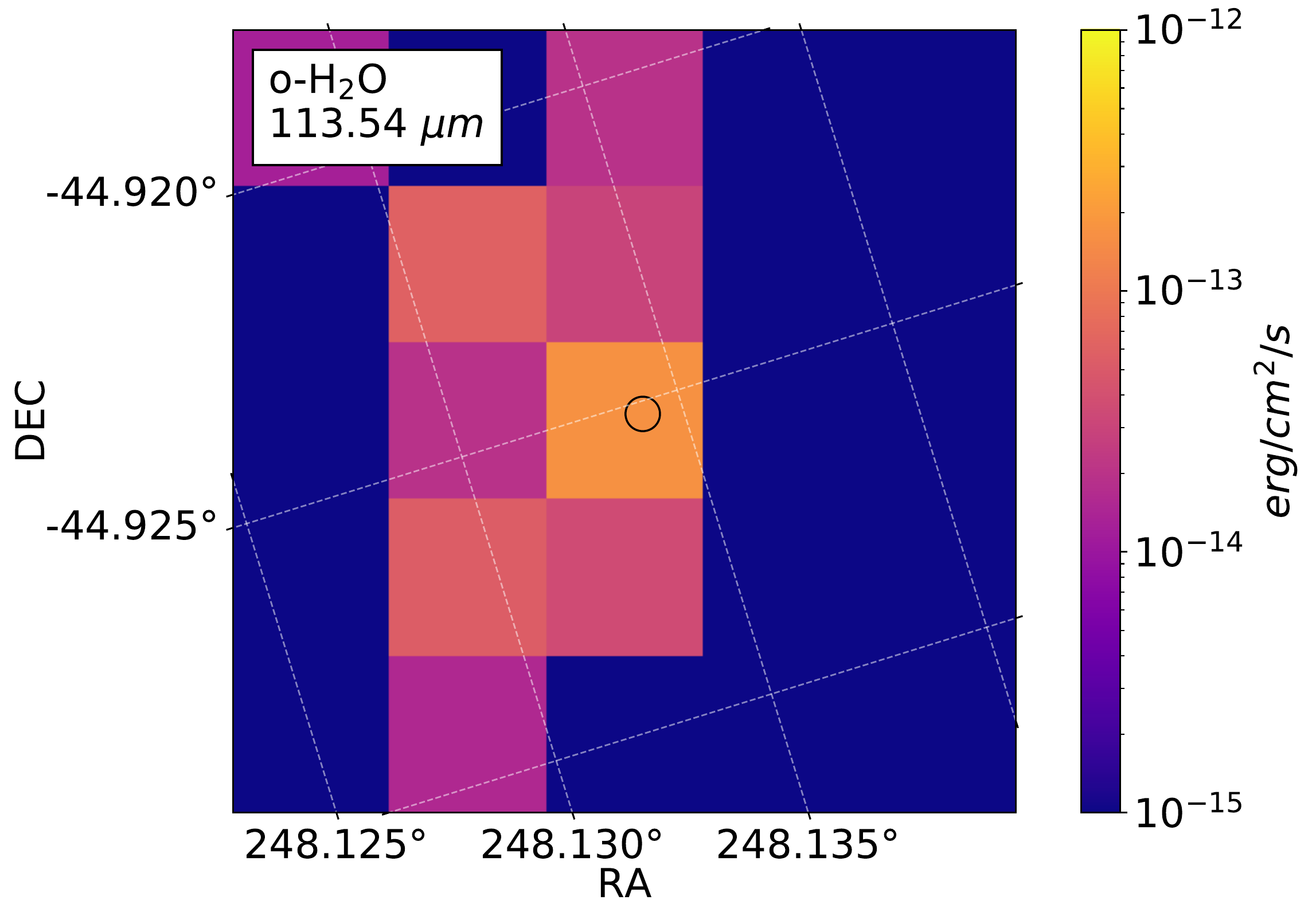}\hspace{-0.1cm}
\includegraphics[width=0.33\textwidth, trim={0cm 0 0cm 0}, clip]{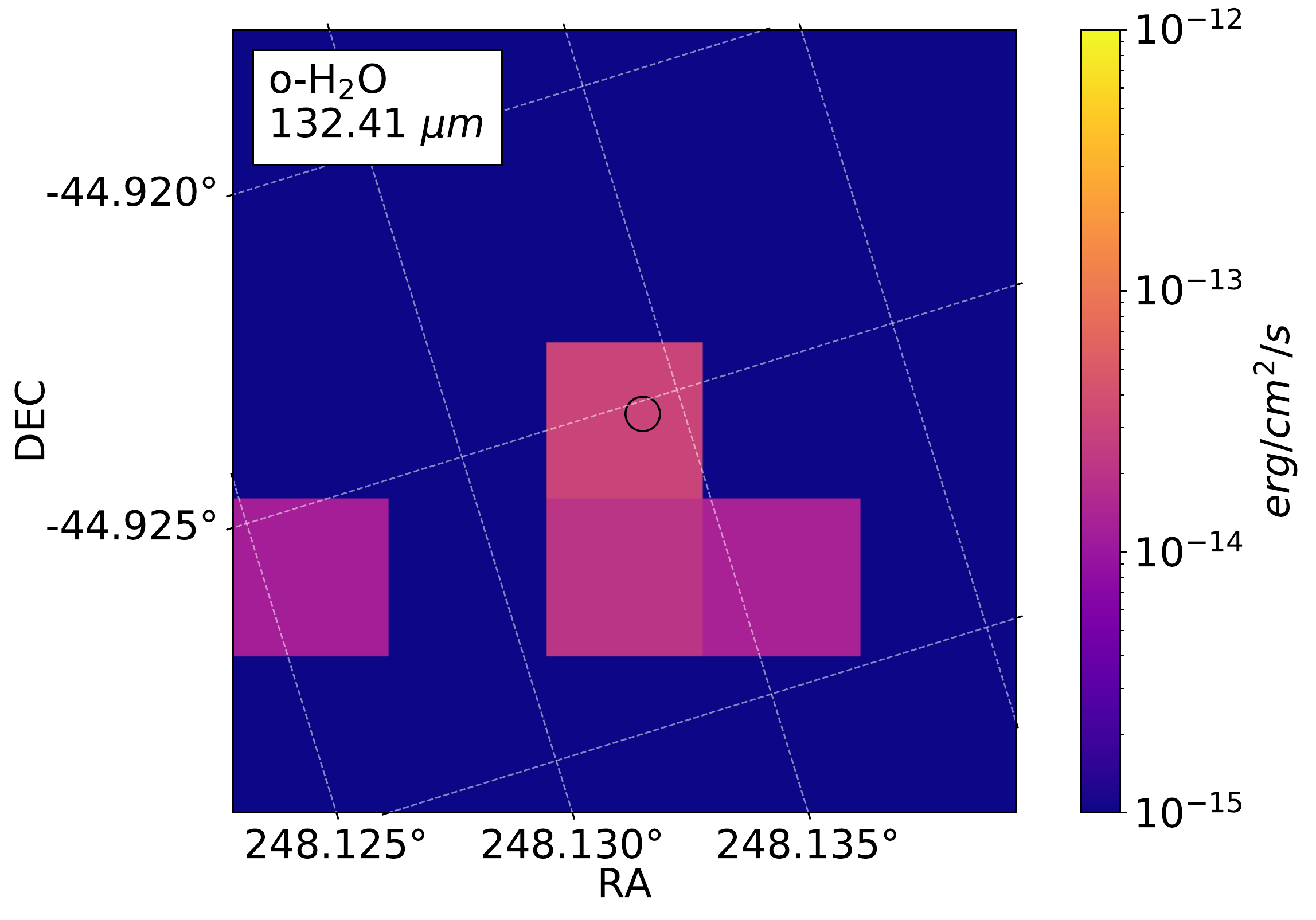}\hspace{-0.1cm}
\includegraphics[width=0.33\textwidth, trim={0cm 0 0cm 0}, clip]{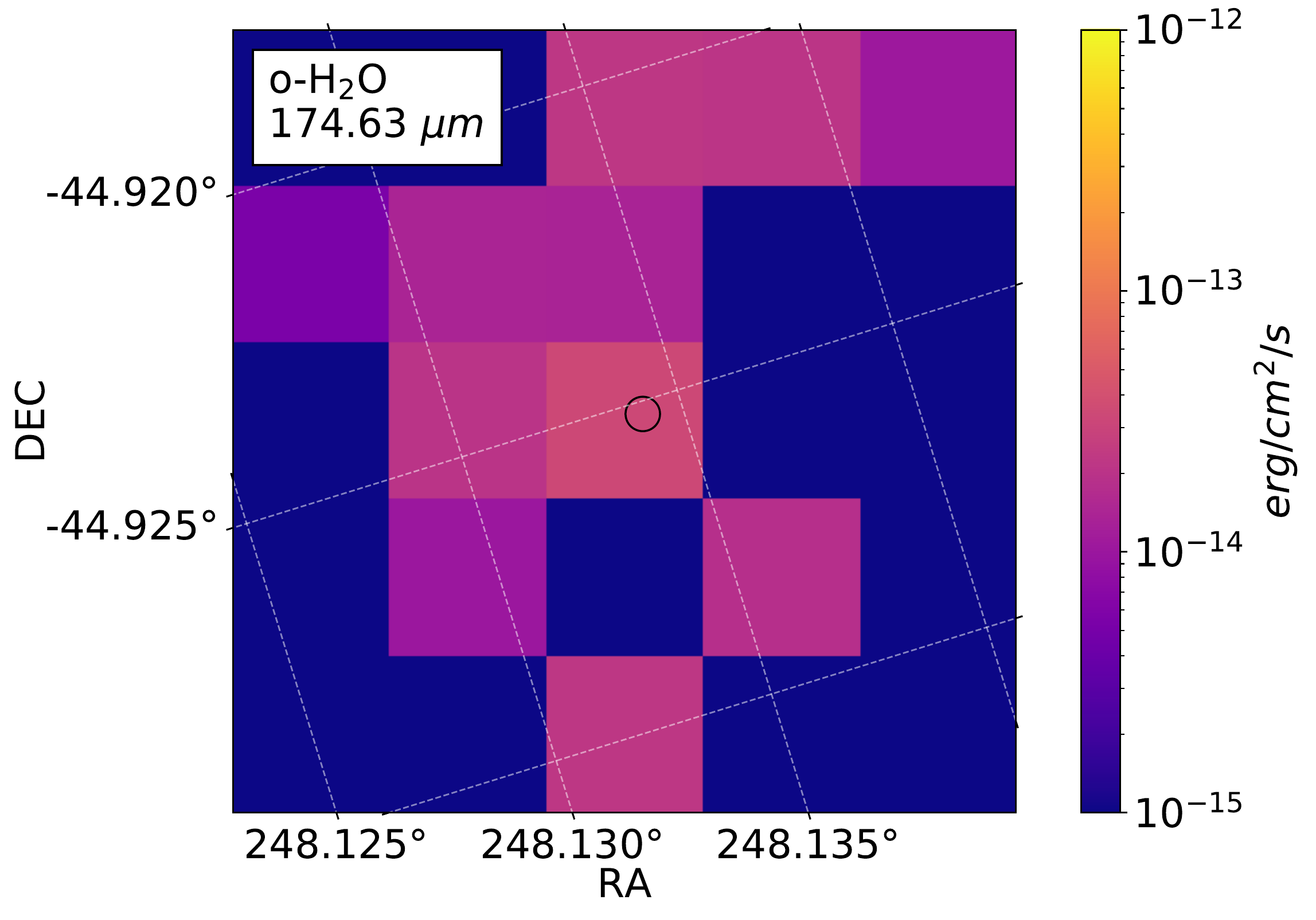}\hspace{-0.1cm}
\caption{
                \footnotesize
                Line maps of PACS with visible lines for V346 Nor, part 1.
        }
\end{figure*}

\begin{figure*}
\includegraphics[width=0.33\textwidth, trim={0cm 0 0cm 0}, clip]{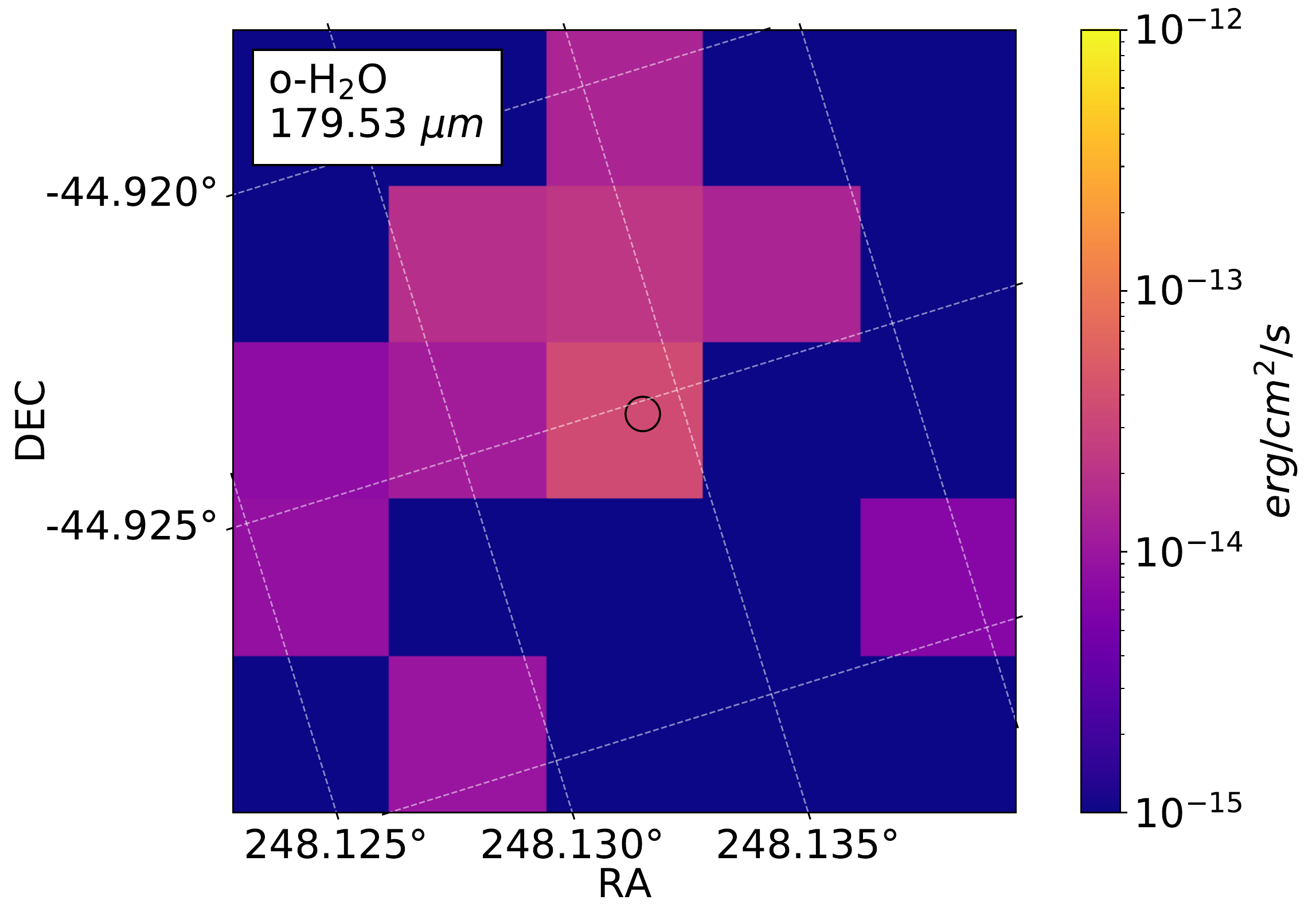}\hspace{-0.1cm}
\includegraphics[width=0.33\textwidth, trim={0cm 0 0cm 0}, clip]{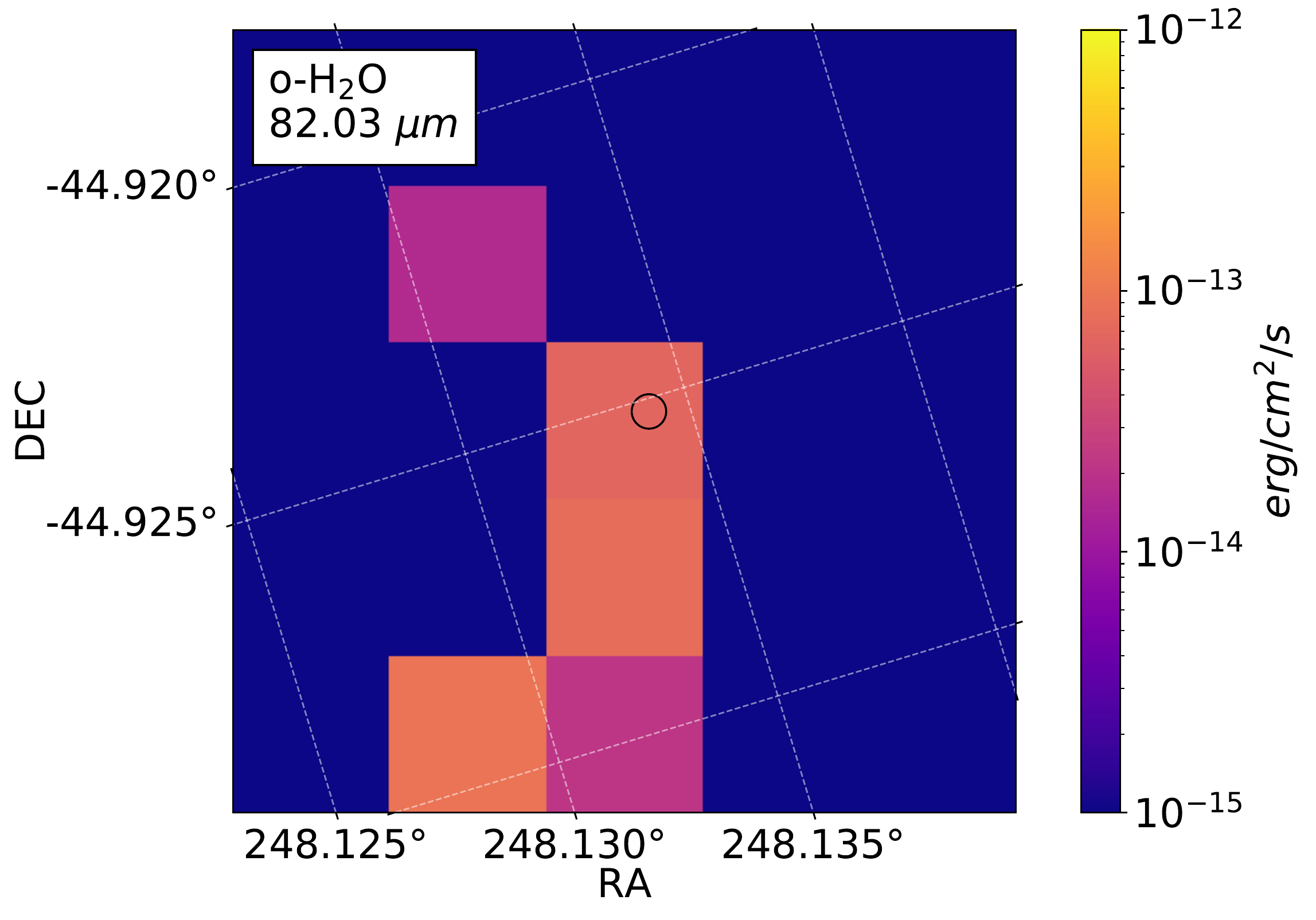}\hspace{-0.1cm}
\includegraphics[width=0.33\textwidth, trim={0cm 0 0cm 0}, clip]{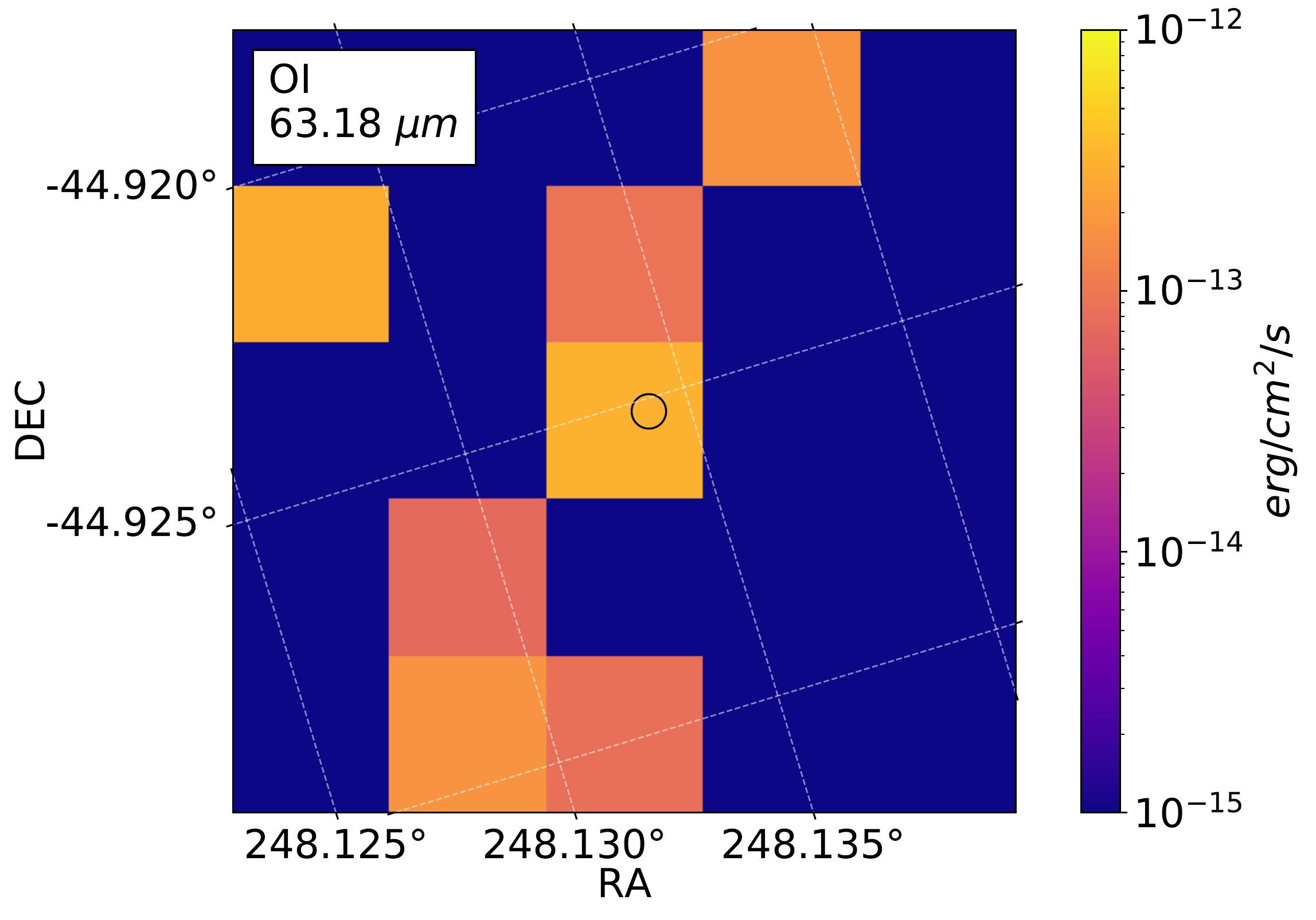}\hspace{-0.1cm}\\
\includegraphics[width=0.33\textwidth, trim={0cm 0 0cm 0}, clip]{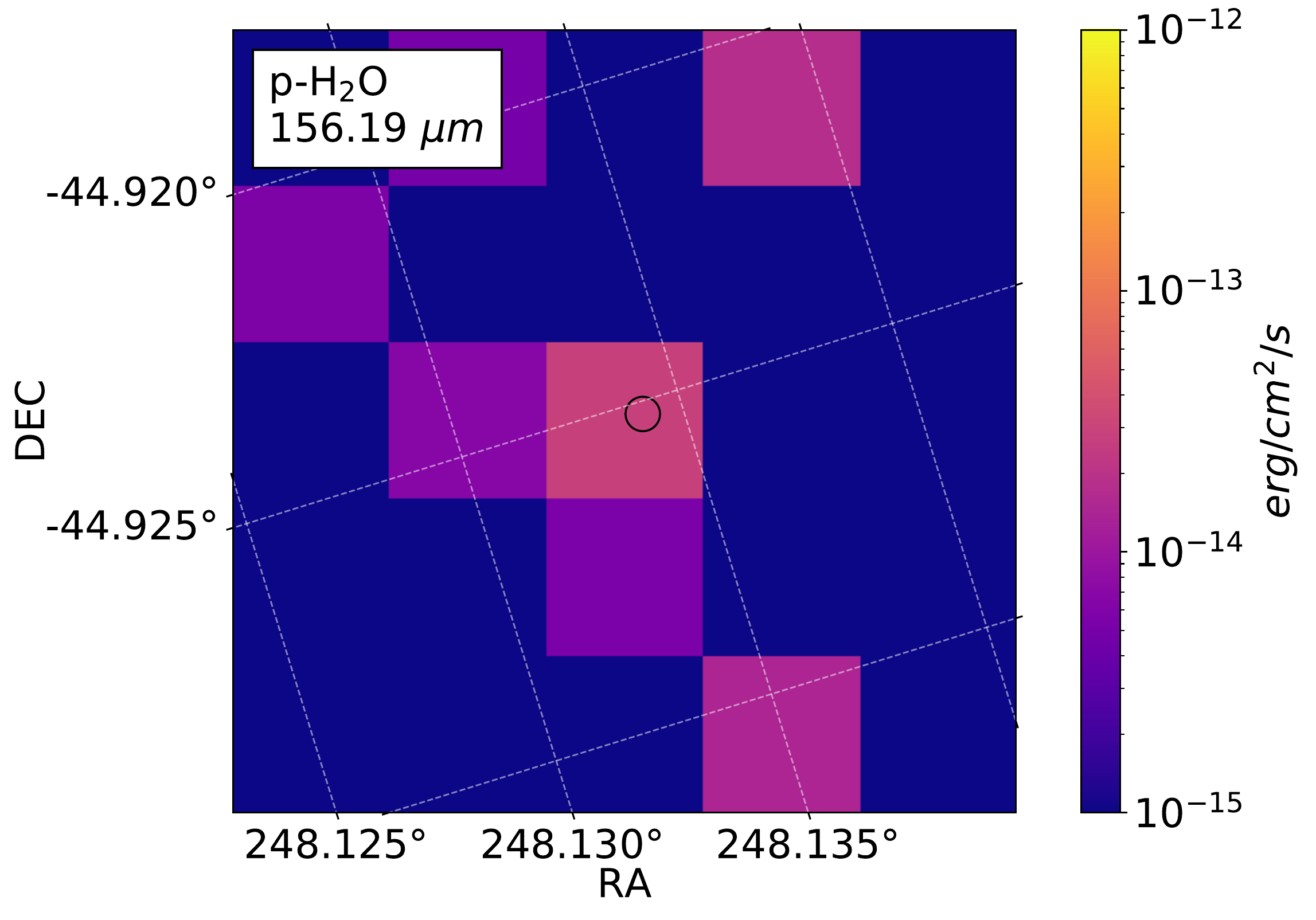}\hspace{-0.1cm}
    \caption{
                \footnotesize
                Line maps of PACS with visible lines for V346 Nor, part 2.
        }
\end{figure*}

\begin{figure*}
\includegraphics[width=0.33\textwidth, trim={0cm 0 0cm 0}, clip]{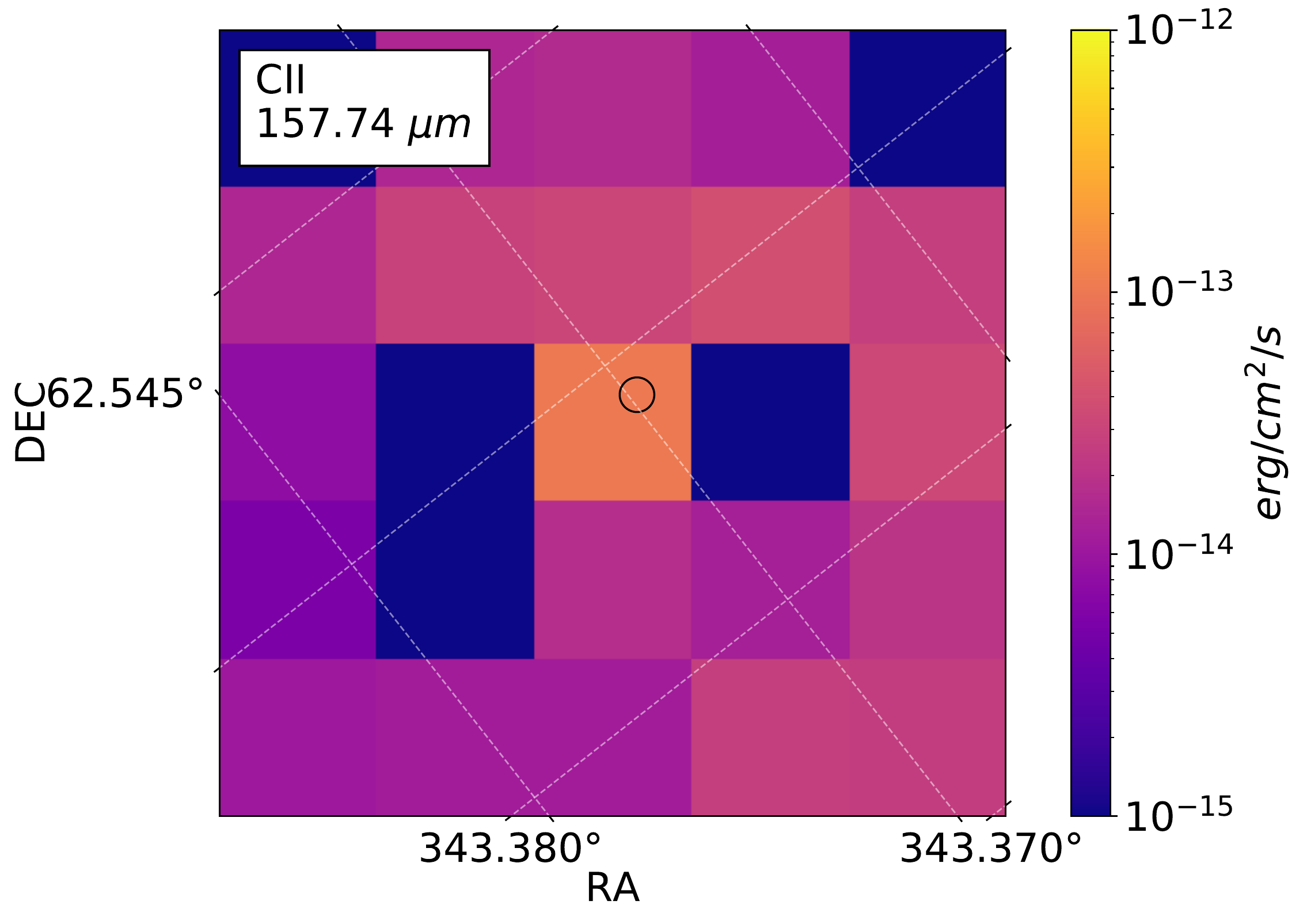}\hspace{-0.1cm}

    \caption{
                \footnotesize
                Line maps of PACS with visible lines for V733 Cep.
        }
\end{figure*}

\begin{figure*}
\includegraphics[width=0.33\textwidth, trim={0cm 0 0cm 0}, clip]{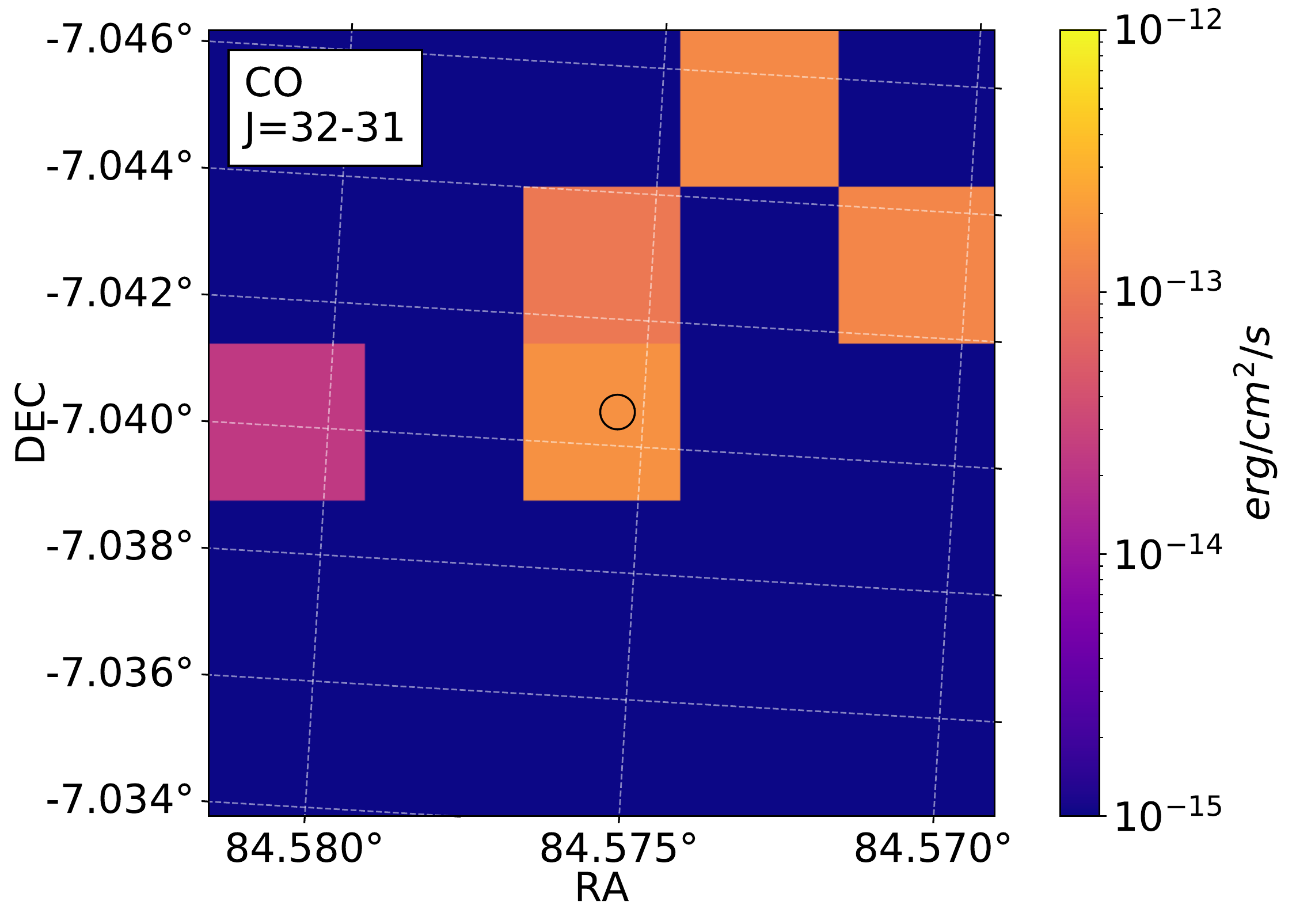}\hspace{-0.1cm}

    \caption{
                \footnotesize
                Line maps of PACS with visible lines for V883 Ori.
        }
\end{figure*}

\begin{figure*}
\includegraphics[width=0.33\textwidth, trim={0cm 0 0cm 0}, clip]{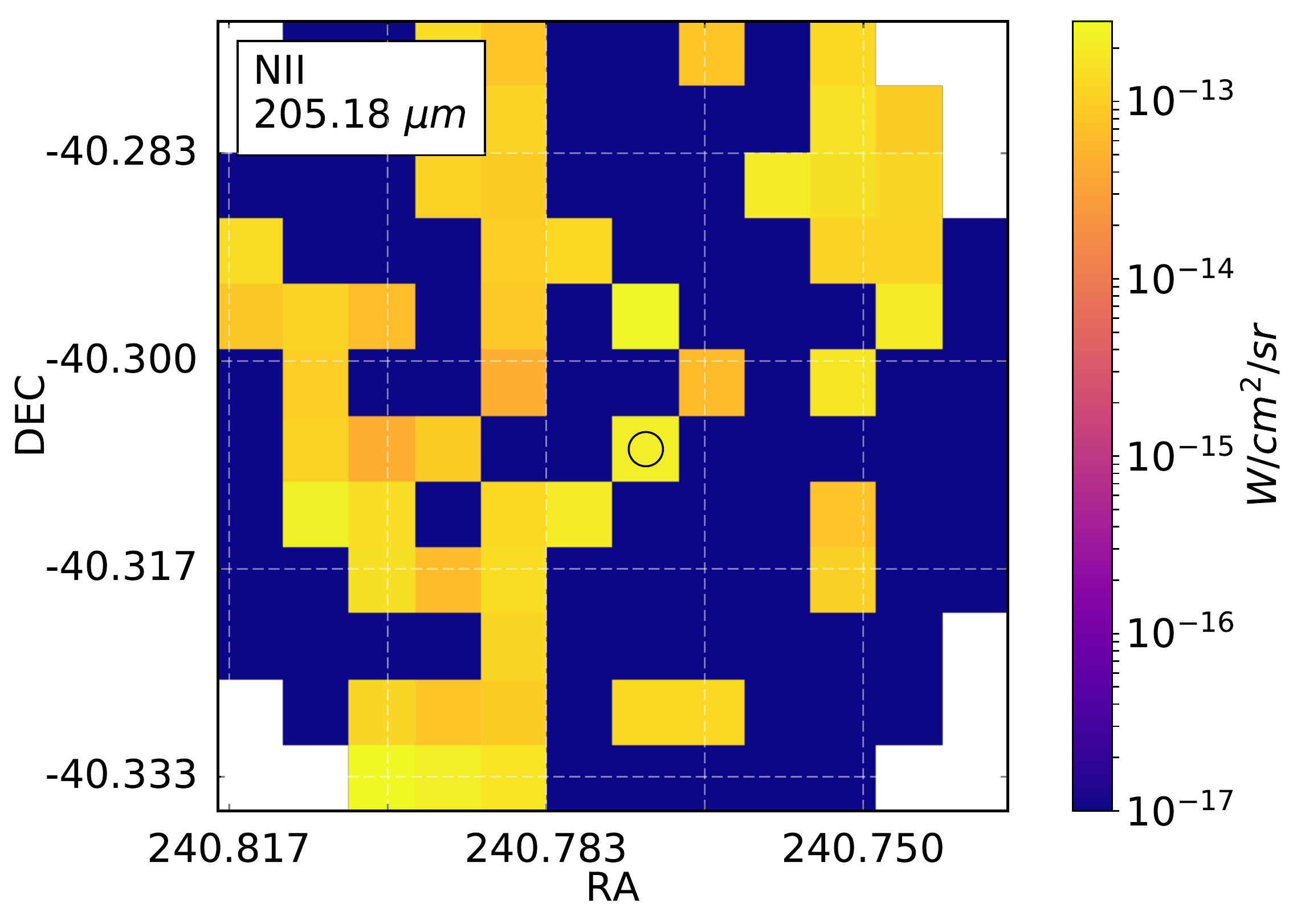}\hspace{-0.1cm}
\caption{
                \footnotesize
                Line maps of SPIRE with visible lines for EX Lup.
        }
\end{figure*}

\begin{figure*}
\includegraphics[width=0.33\textwidth, trim={0cm 0 0cm 0}, clip]{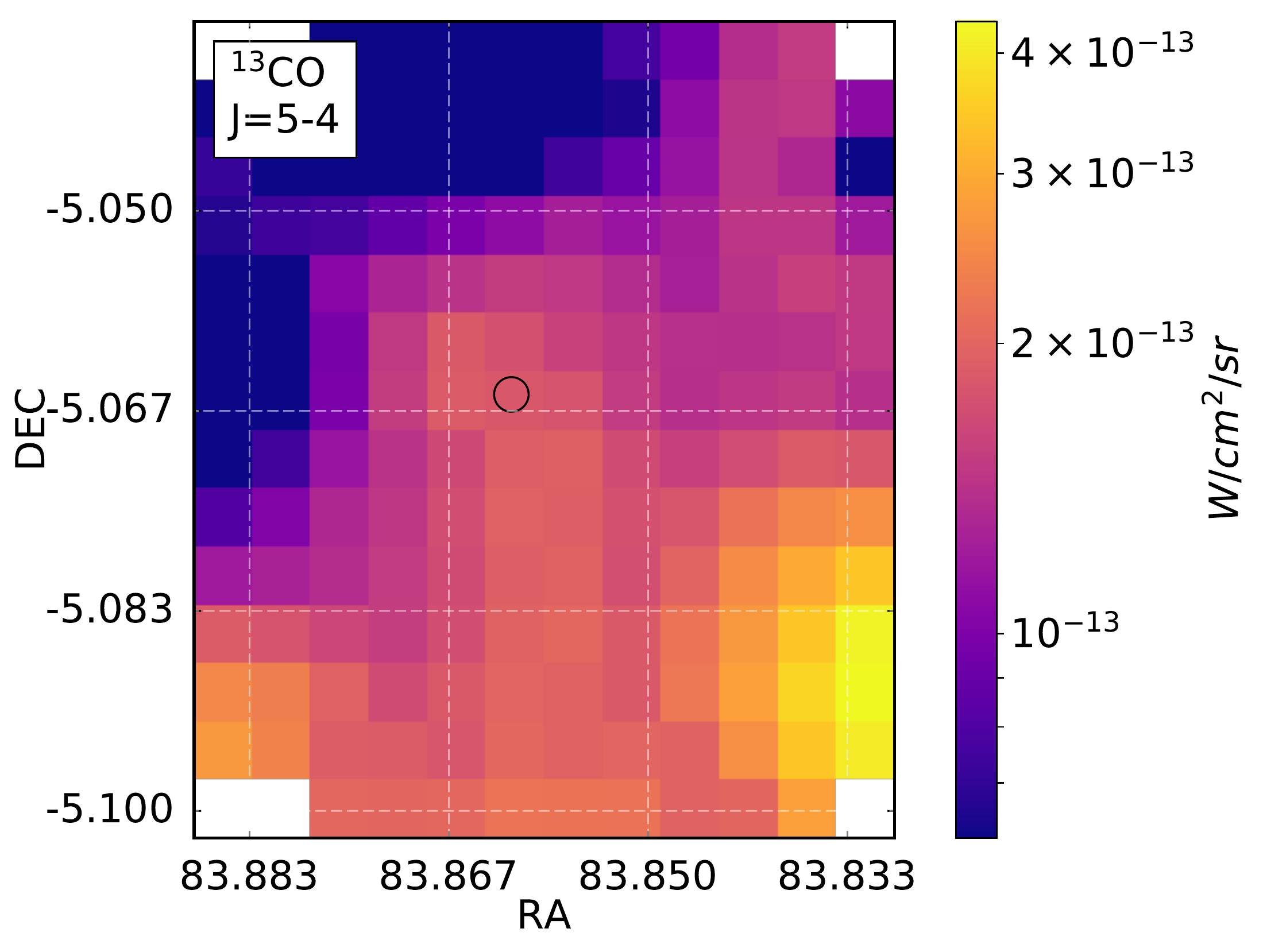}\hspace{-0.1cm}
\includegraphics[width=0.33\textwidth, trim={0cm 0 0cm 0}, clip]{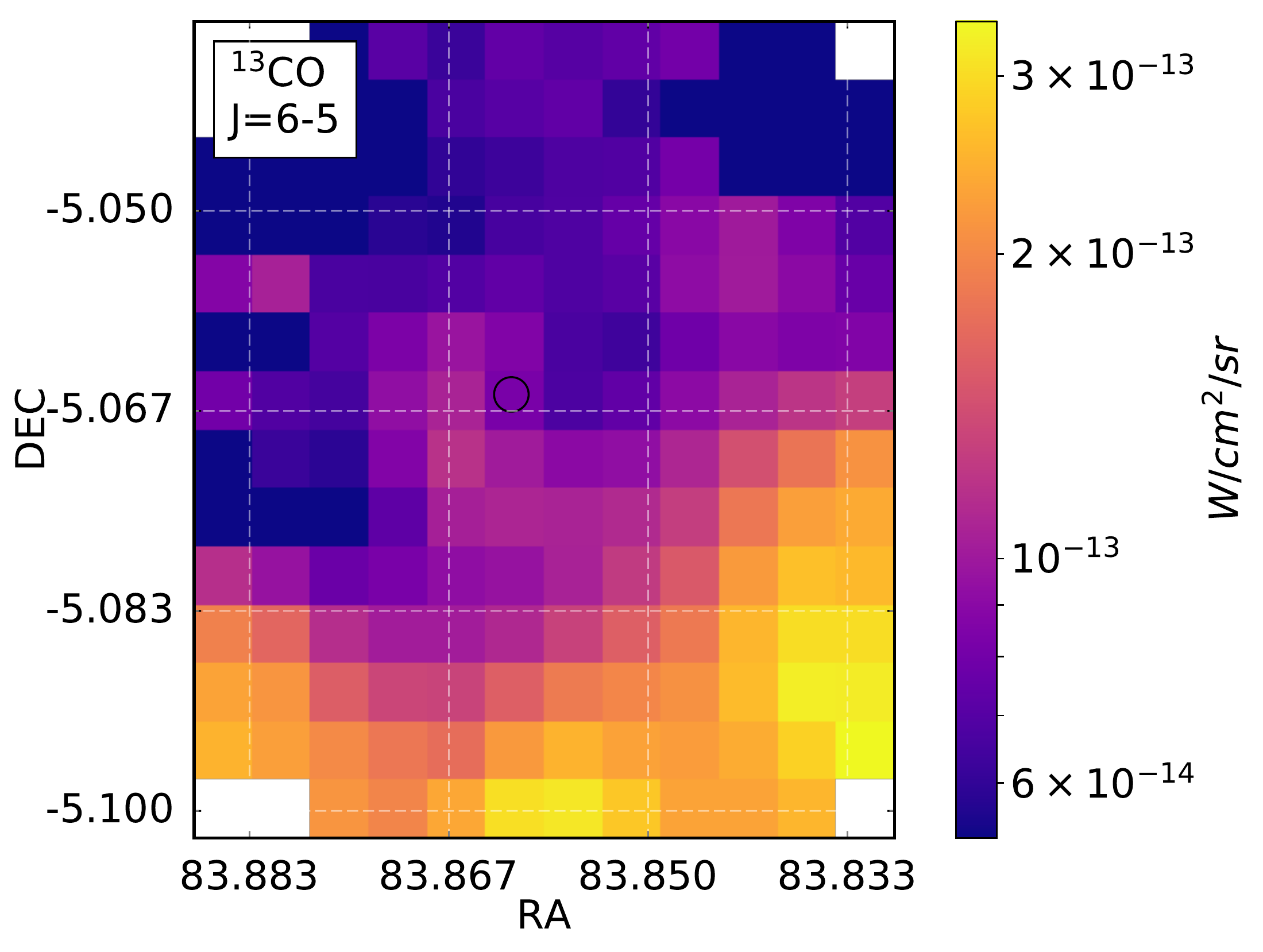}\hspace{-0.1cm}
\includegraphics[width=0.33\textwidth, trim={0cm 0 0cm 0}, clip]{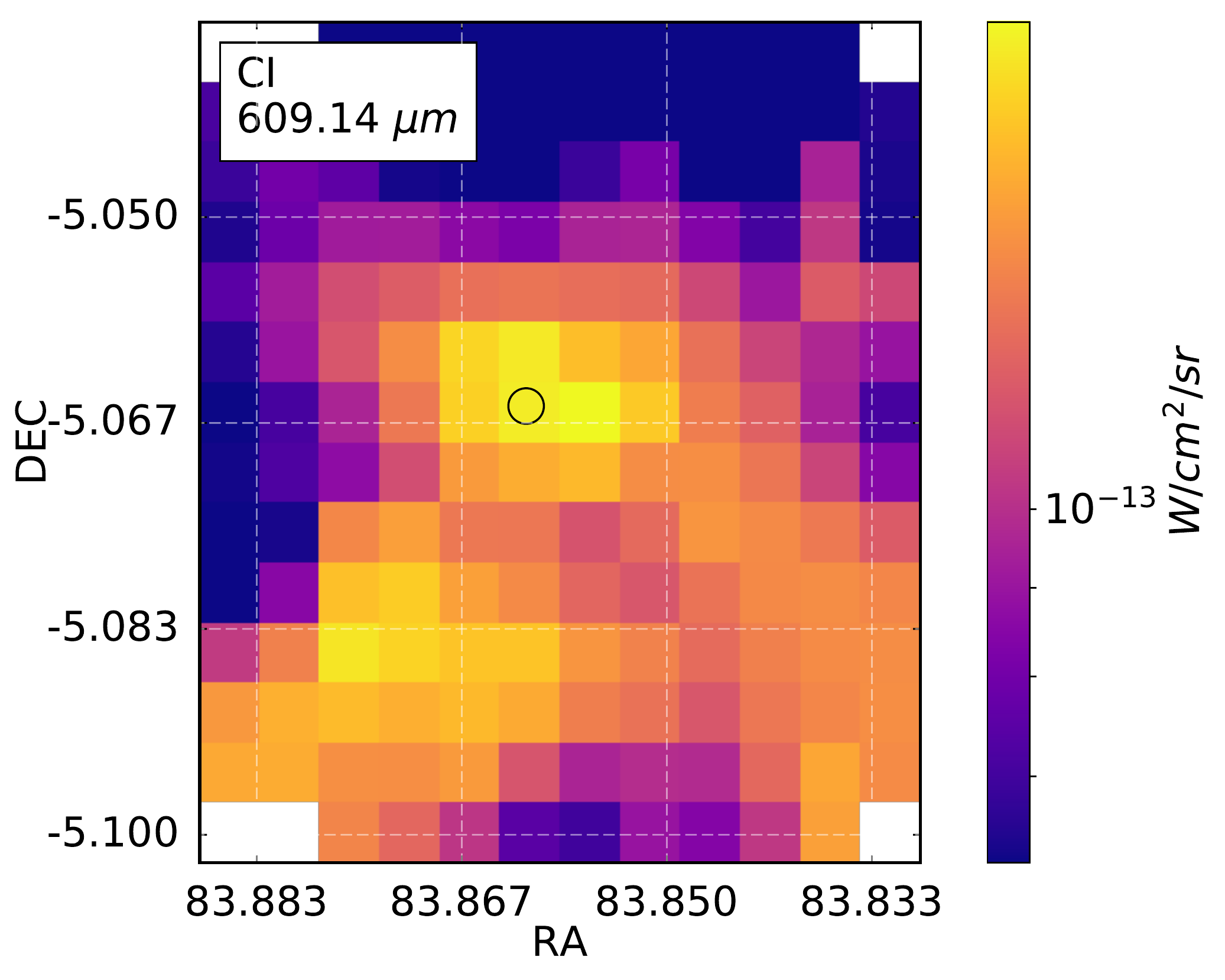}\hspace{-0.1cm}\\
\includegraphics[width=0.33\textwidth, trim={0cm 0 0cm 0}, clip]{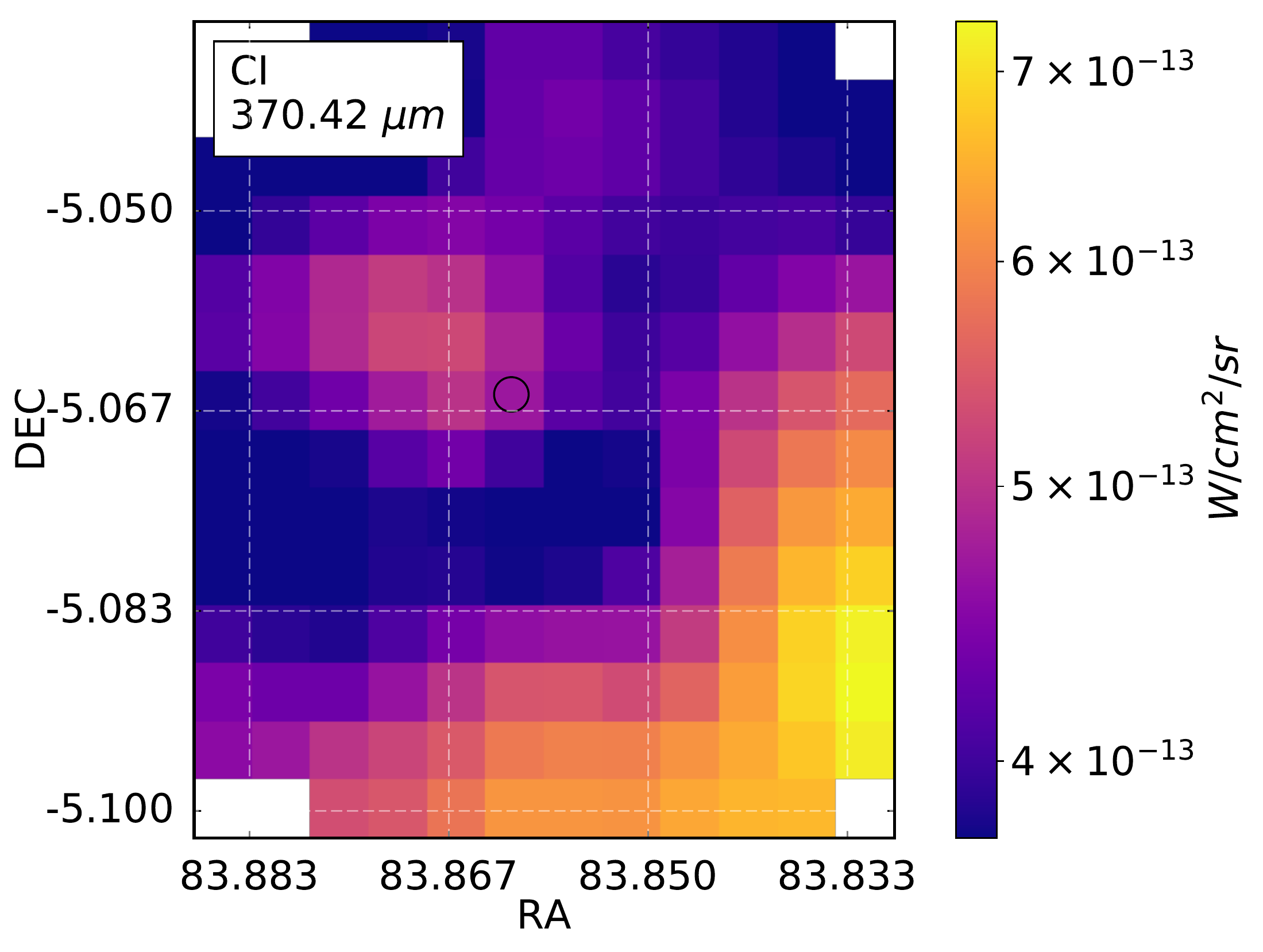}\hspace{-0.1cm}
\includegraphics[width=0.33\textwidth, trim={0cm 0 0cm 0}, clip]{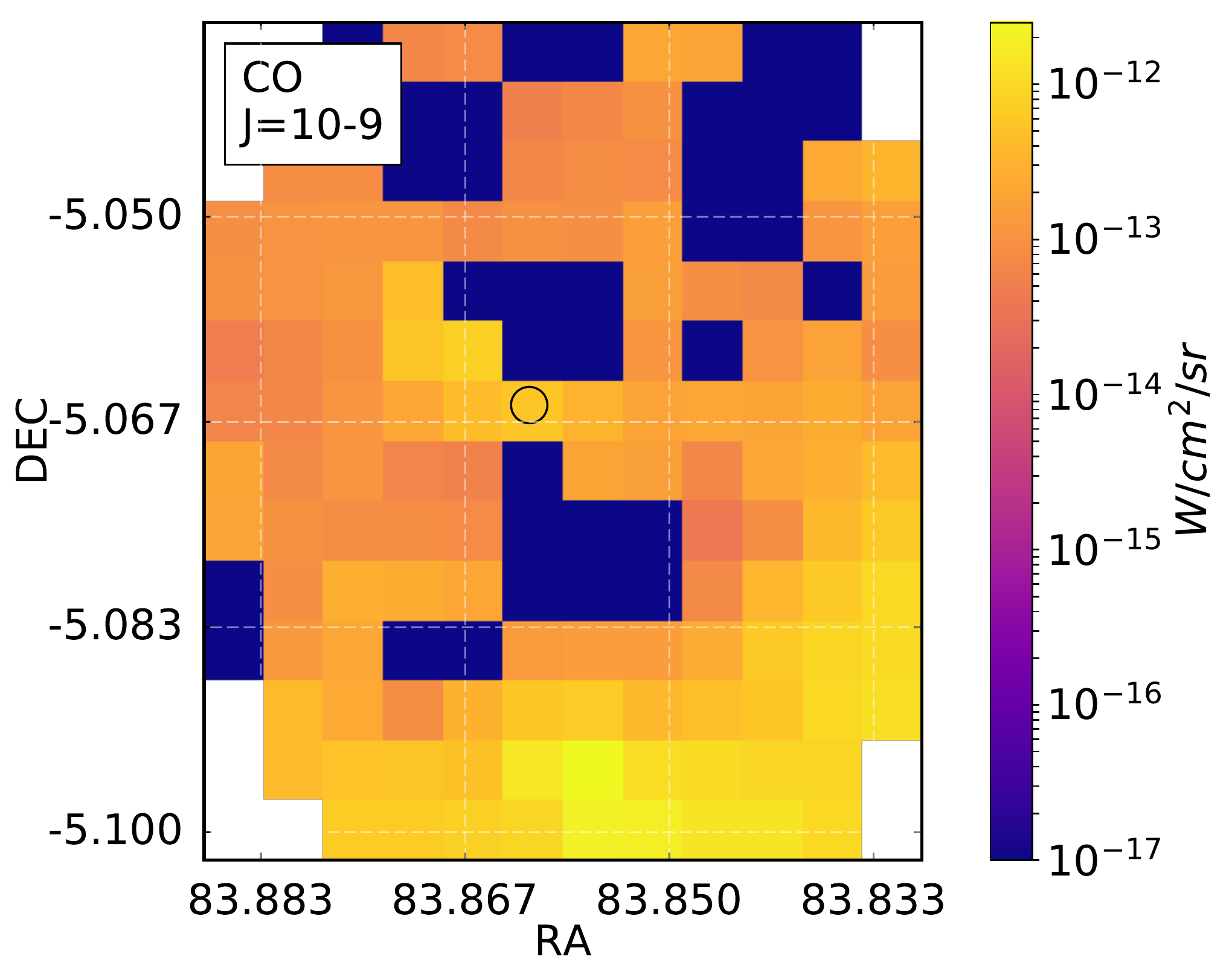}\hspace{-0.1cm}
\includegraphics[width=0.33\textwidth, trim={0cm 0 0cm 0}, clip]{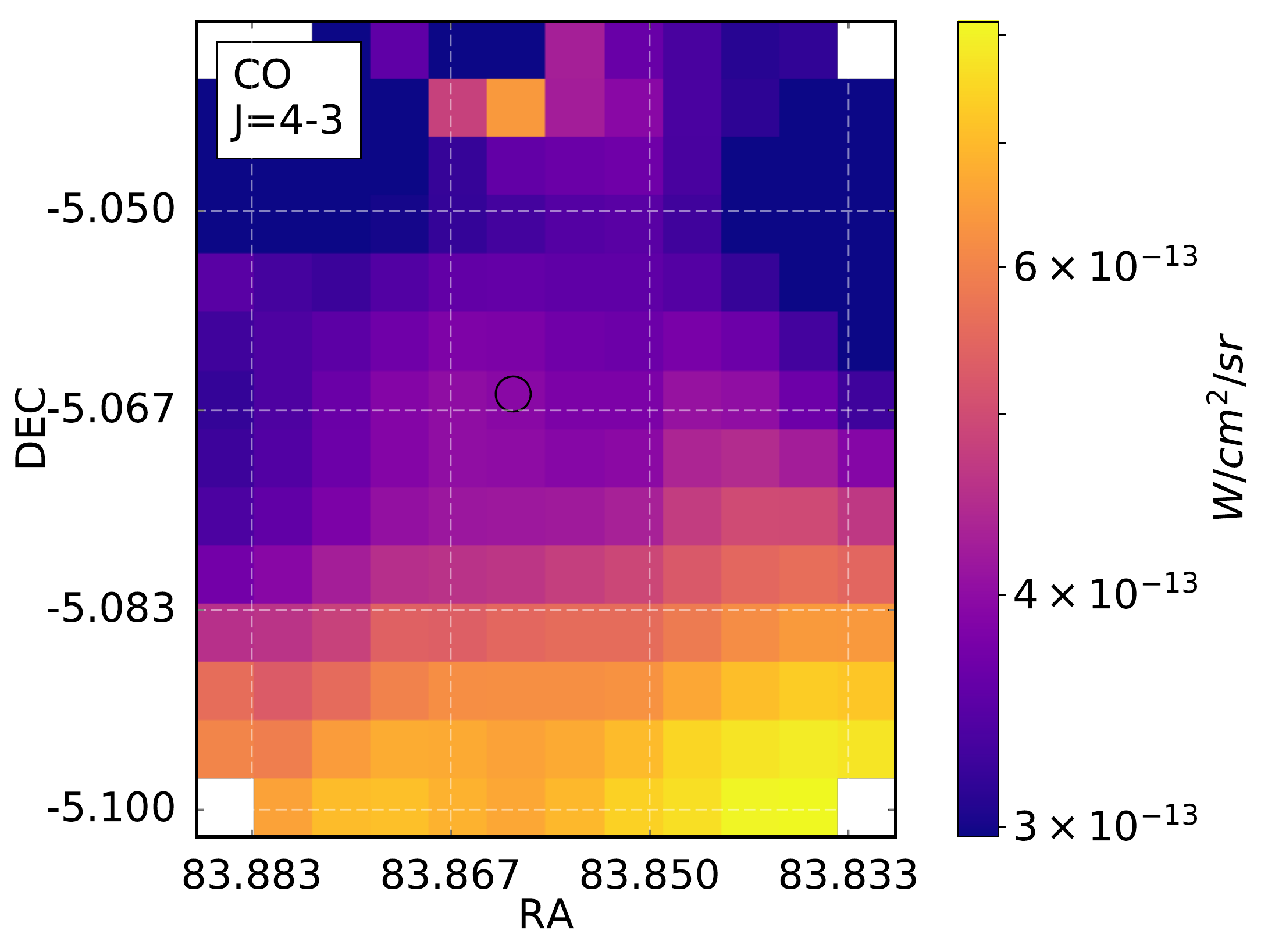}\hspace{-0.1cm}\\
\includegraphics[width=0.33\textwidth, trim={0cm 0 0cm 0}, clip]{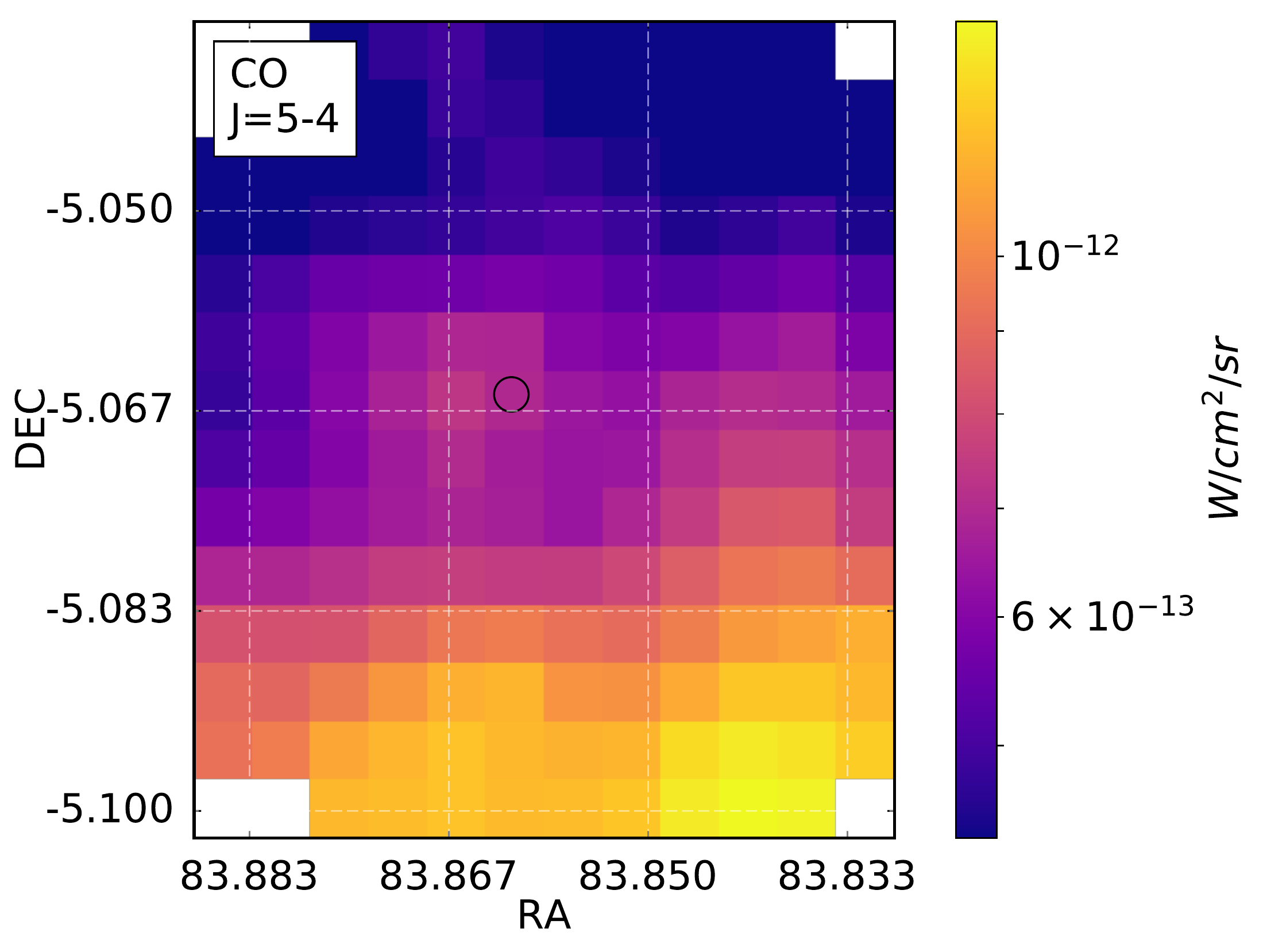}\hspace{-0.1cm}
\includegraphics[width=0.33\textwidth, trim={0cm 0 0cm 0}, clip]{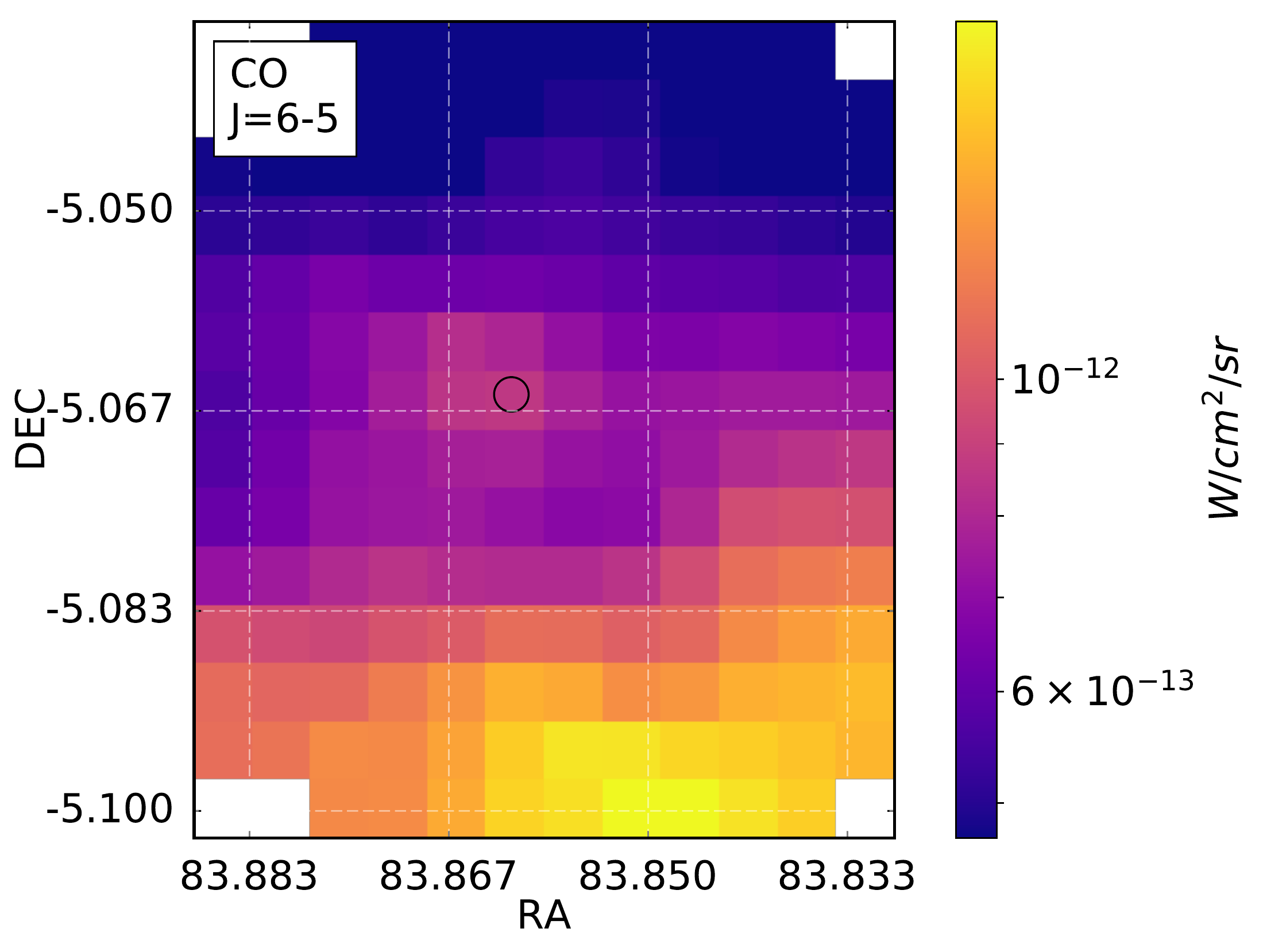}\hspace{-0.1cm}
\includegraphics[width=0.33\textwidth, trim={0cm 0 0cm 0}, clip]{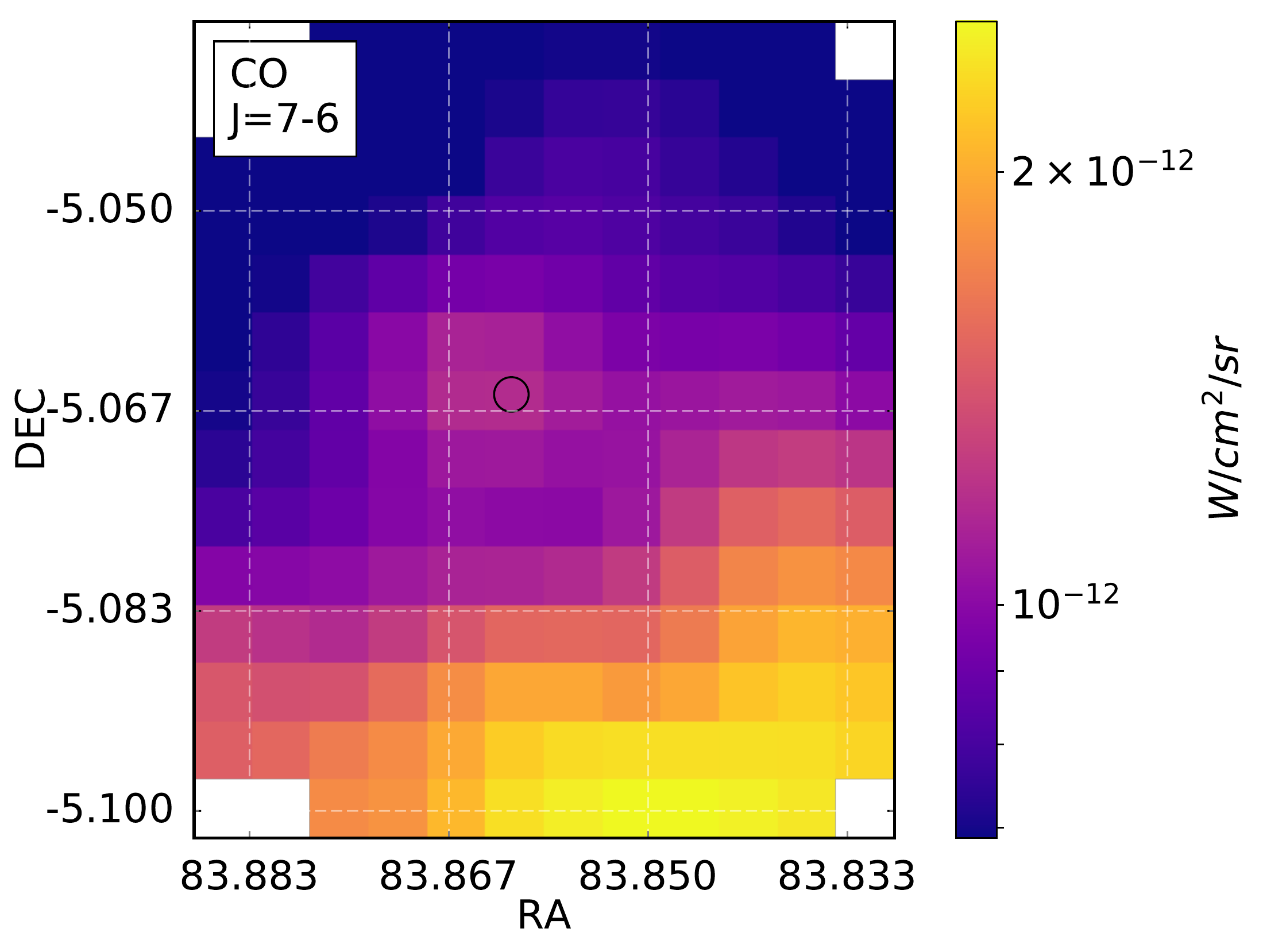}\hspace{-0.1cm}\\
\includegraphics[width=0.33\textwidth, trim={0cm 0 0cm 0}, clip]{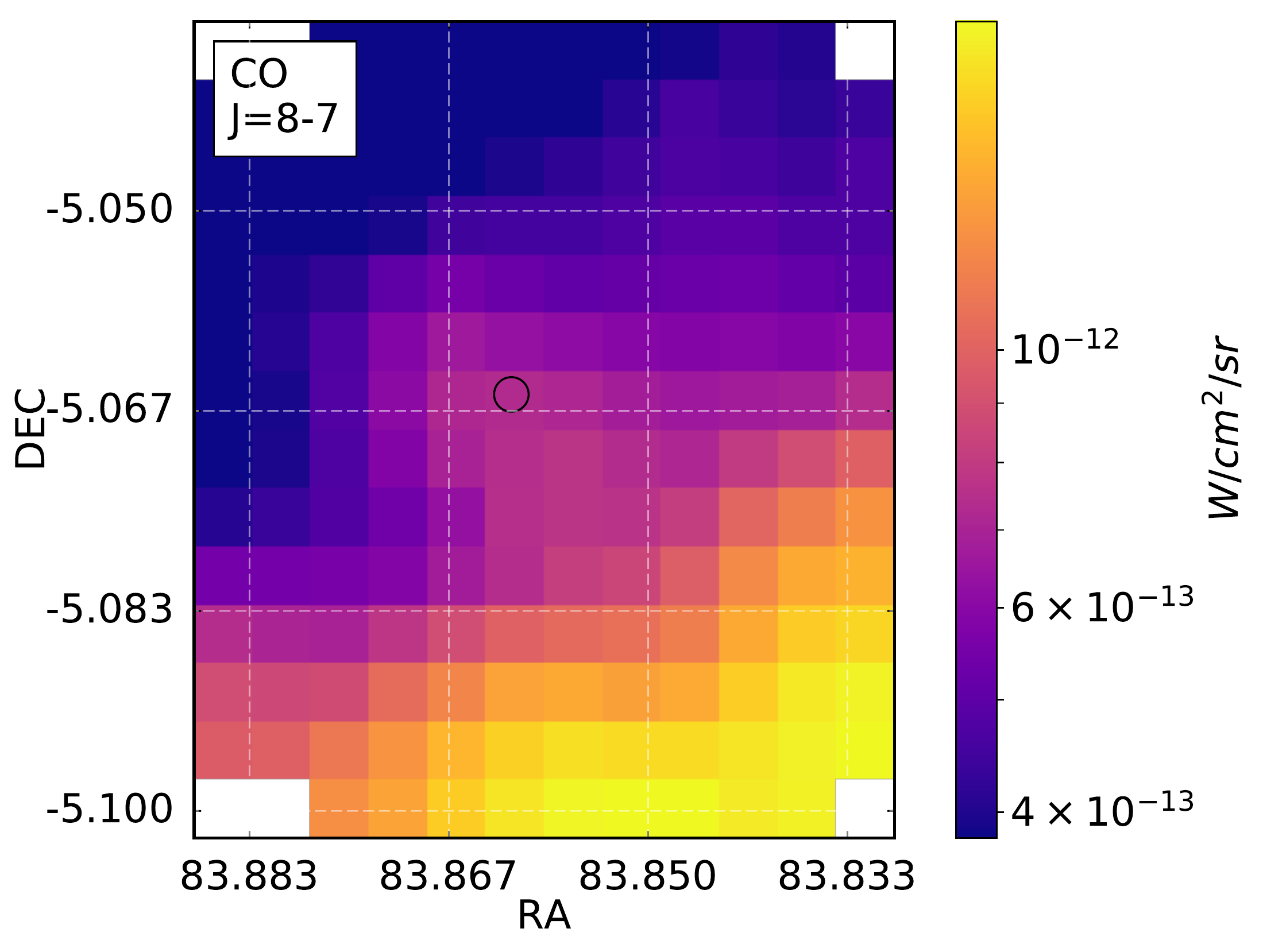}\hspace{-0.1cm}
\includegraphics[width=0.33\textwidth, trim={0cm 0 0cm 0}, clip]{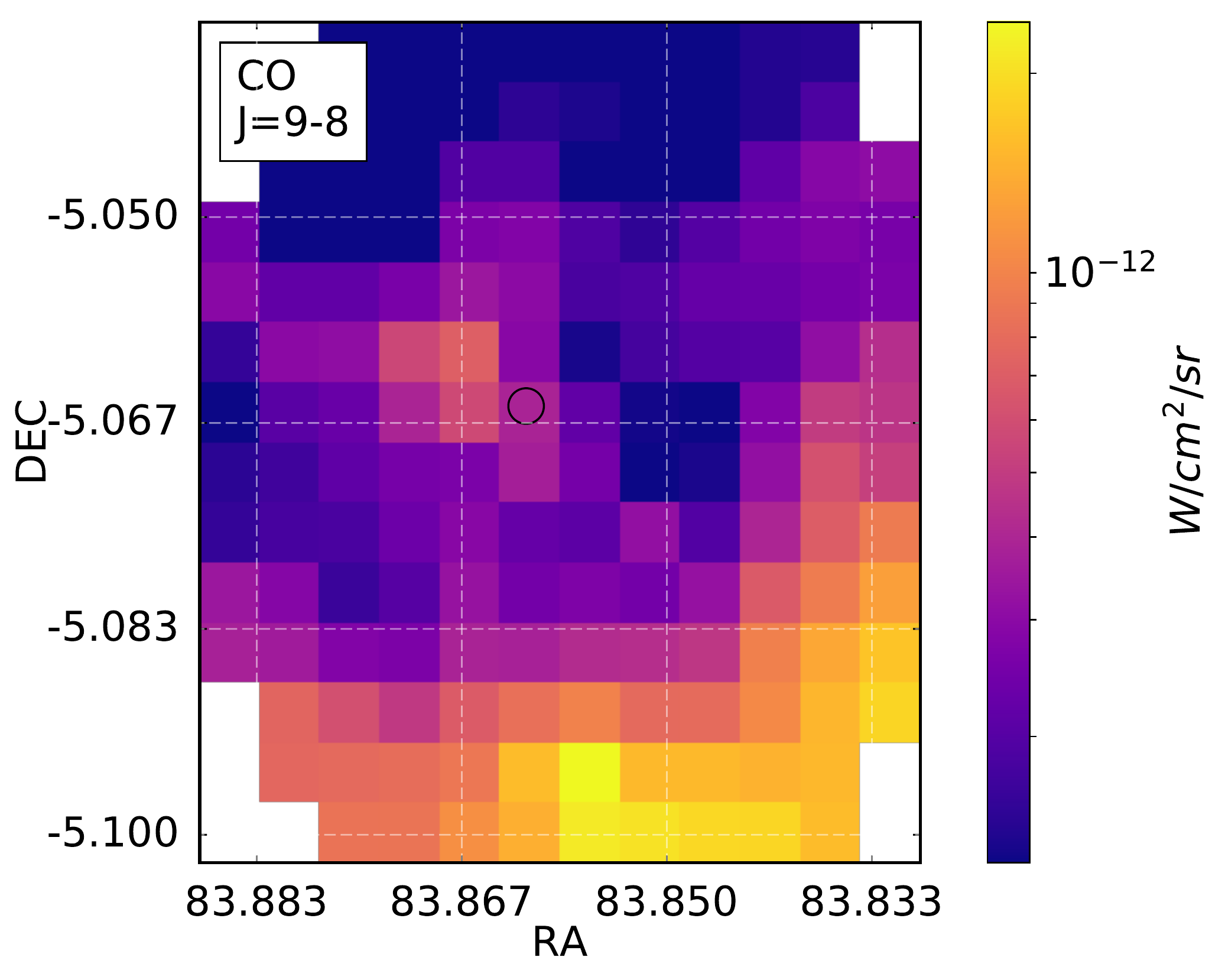}\hspace{-0.1cm}
\includegraphics[width=0.33\textwidth, trim={0cm 0 0cm 0}, clip]{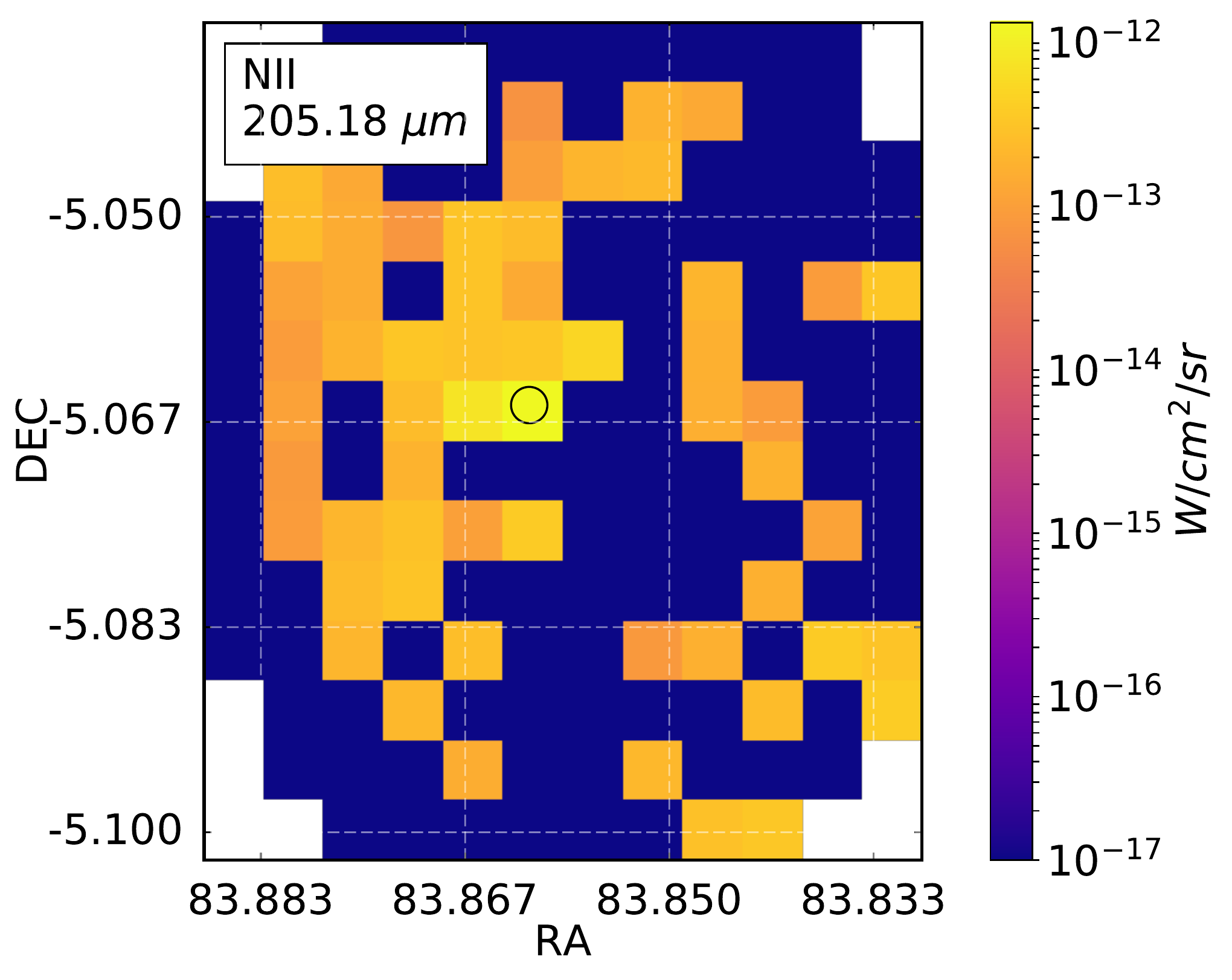}\hspace{-0.1cm}
\caption{
                \footnotesize
                Line maps of SPIRE with visible lines for Haro 5a IRS.
        }
\end{figure*}

\begin{figure*}
\includegraphics[width=0.33\textwidth, trim={0cm 0 0cm 0}, clip]{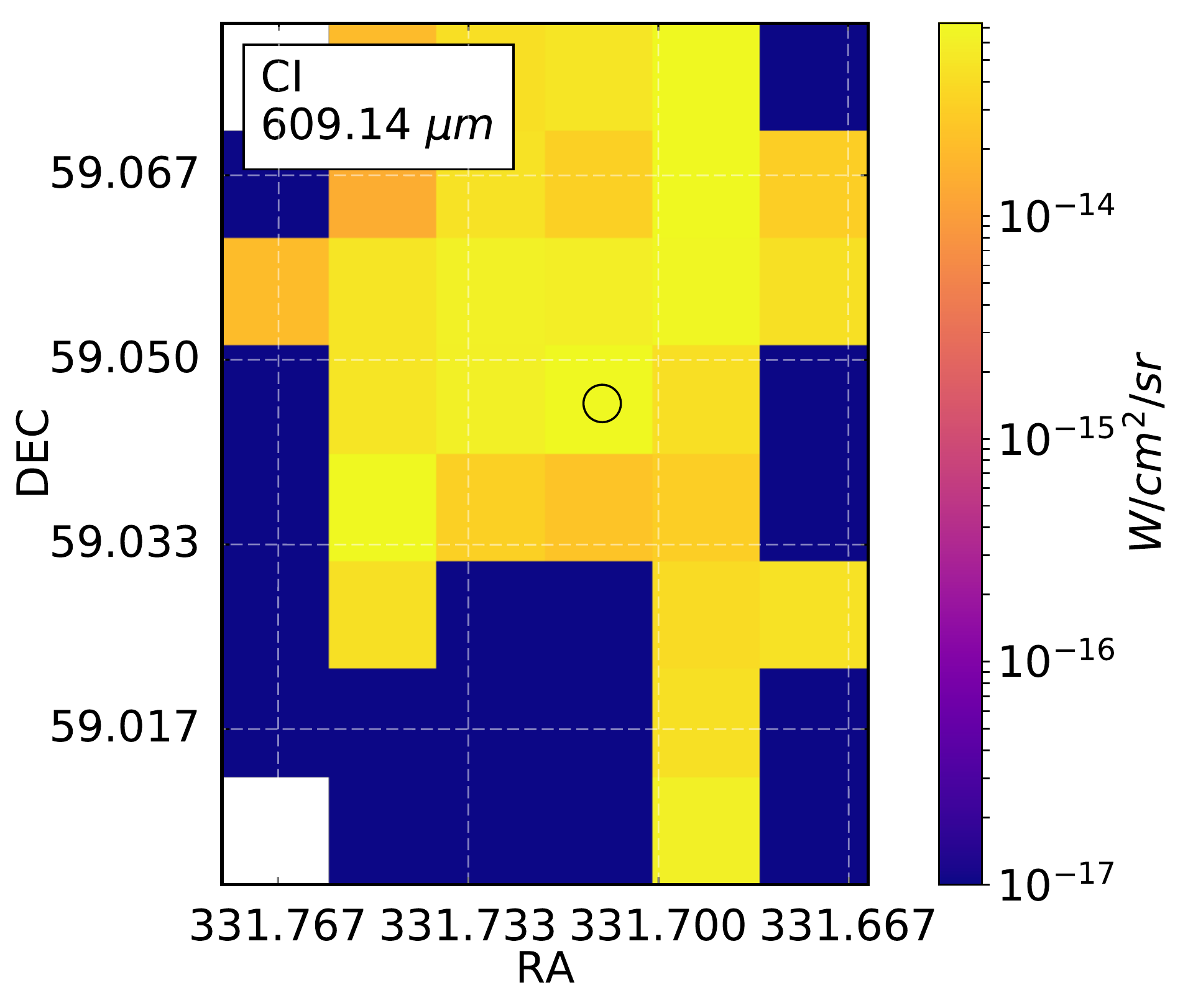}\hspace{-0.1cm}
\includegraphics[width=0.33\textwidth, trim={0cm 0 0cm 0}, clip]{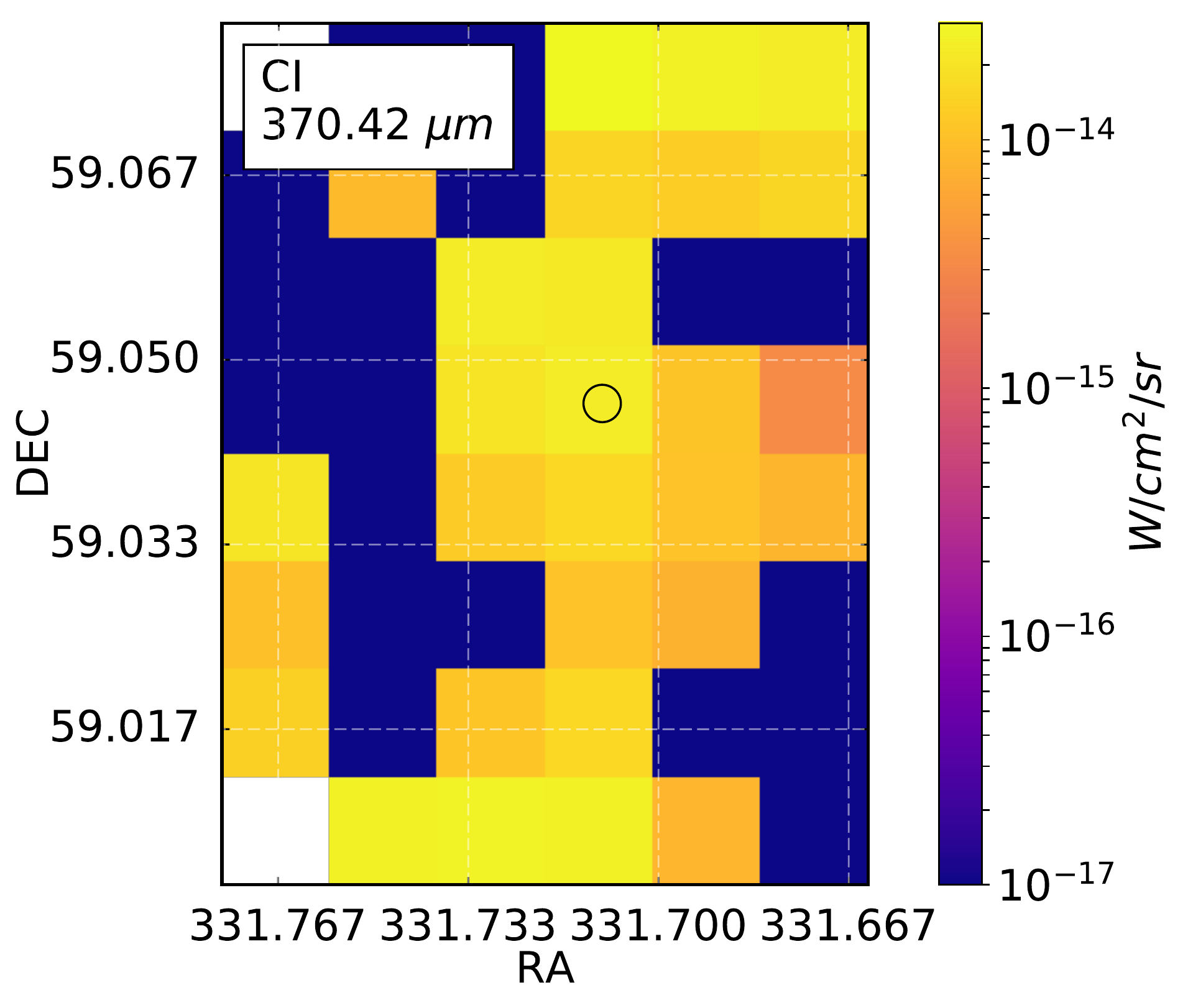}\hspace{-0.1cm}
\includegraphics[width=0.33\textwidth, trim={0cm 0 0cm 0}, clip]{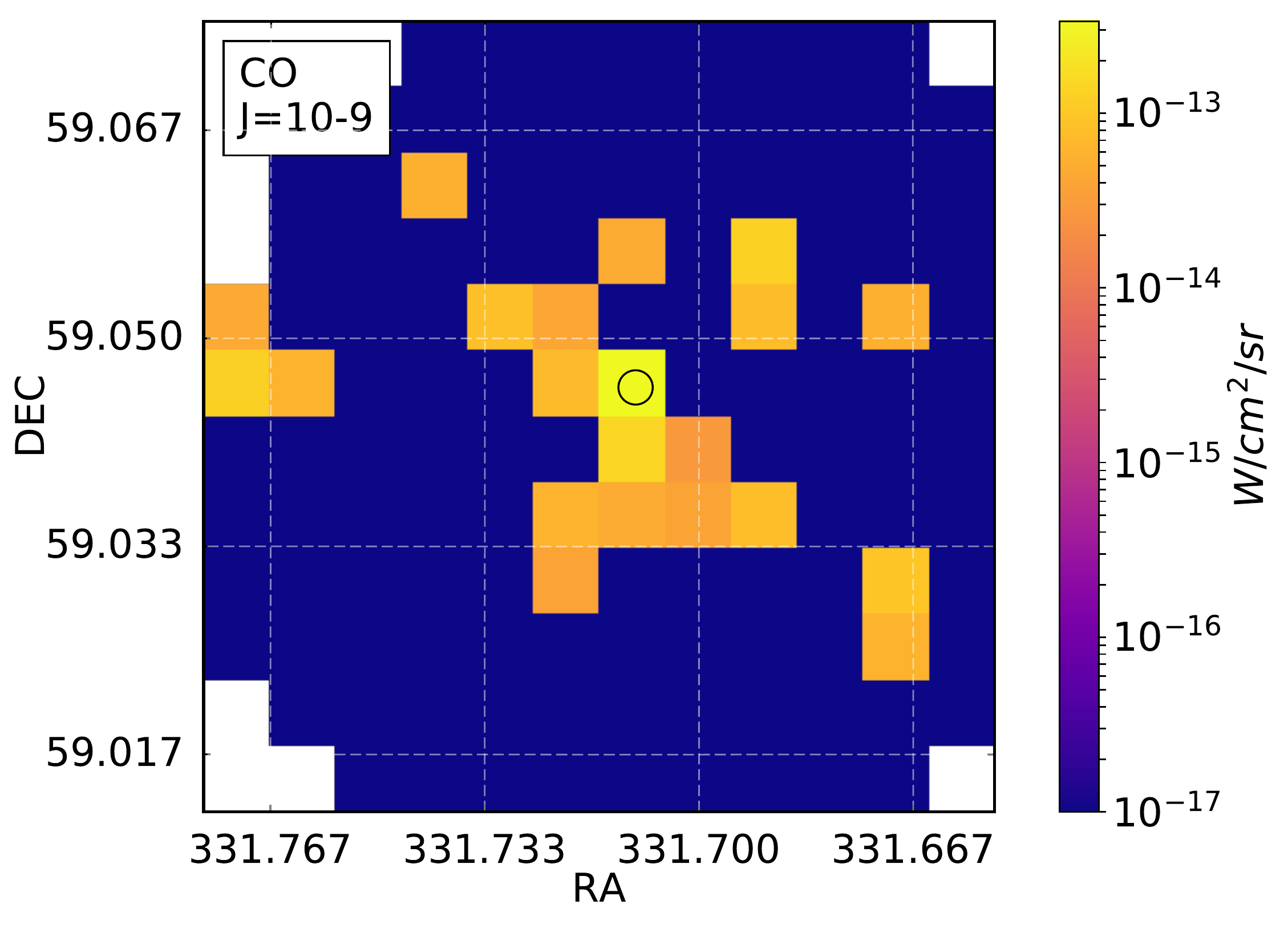}\hspace{-0.1cm}\\
\includegraphics[width=0.33\textwidth, trim={0cm 0 0cm 0}, clip]{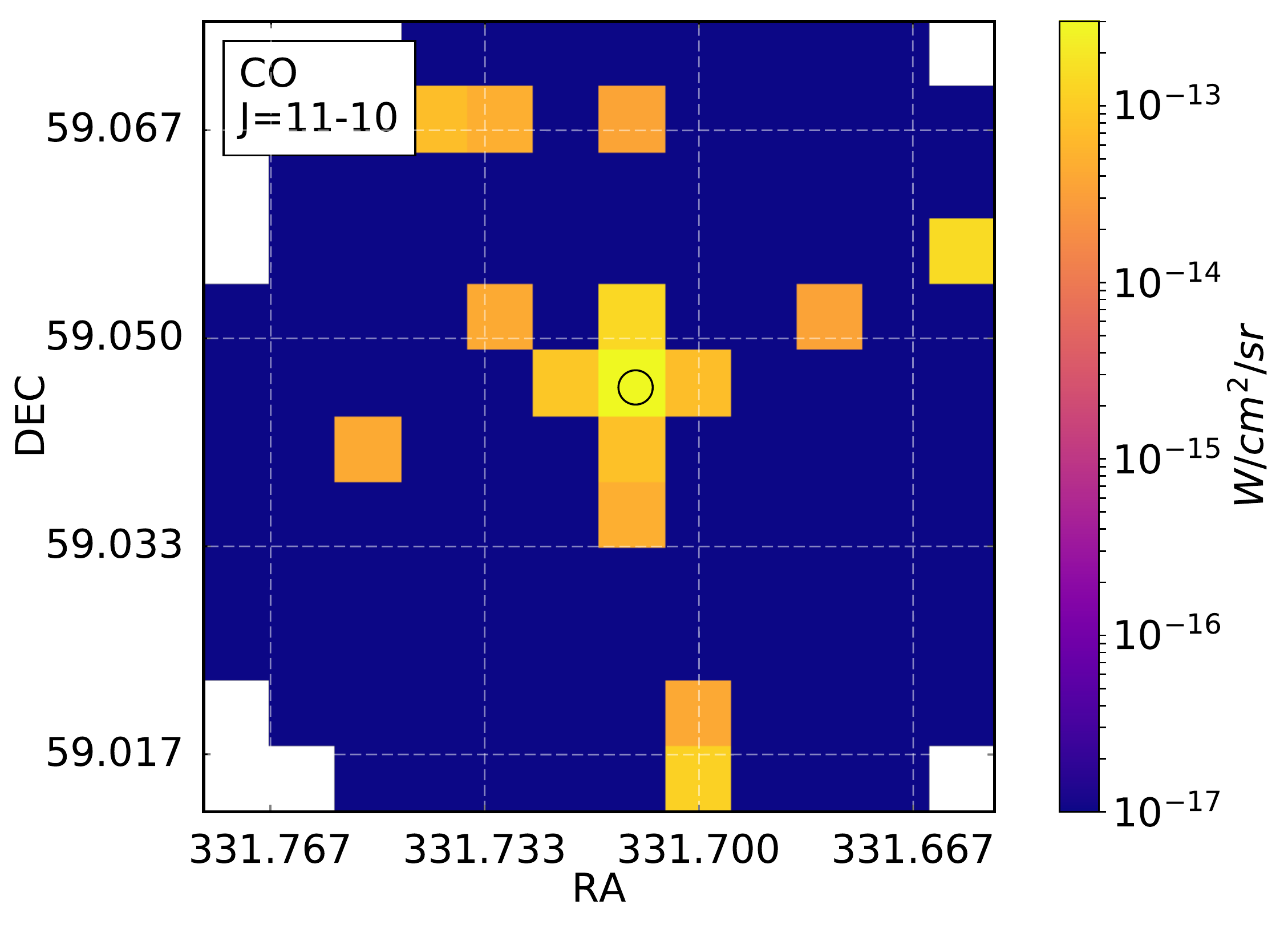}\hspace{-0.1cm}
\includegraphics[width=0.33\textwidth, trim={0cm 0 0cm 0}, clip]{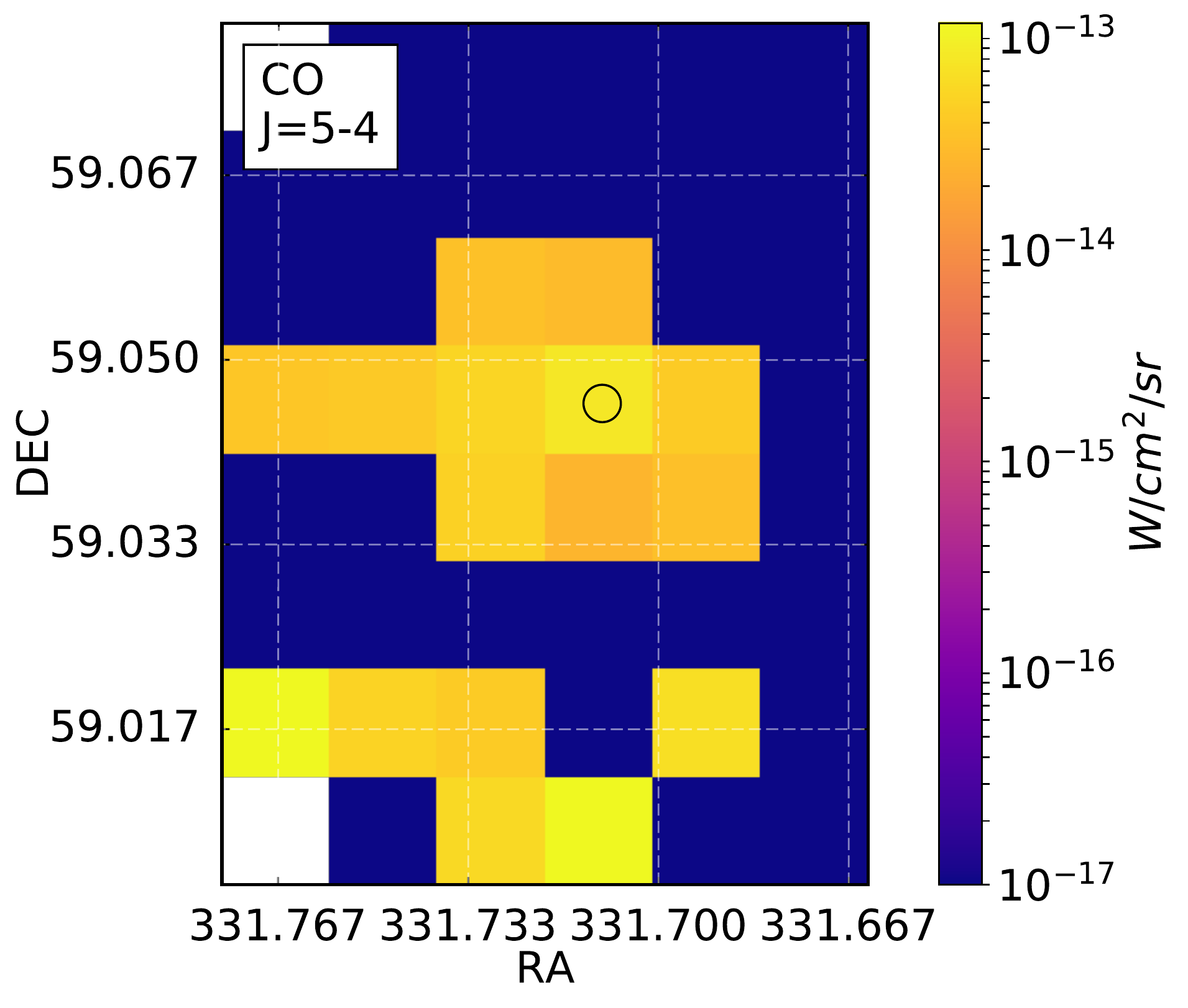}\hspace{-0.1cm}
\includegraphics[width=0.33\textwidth, trim={0cm 0 0cm 0}, clip]{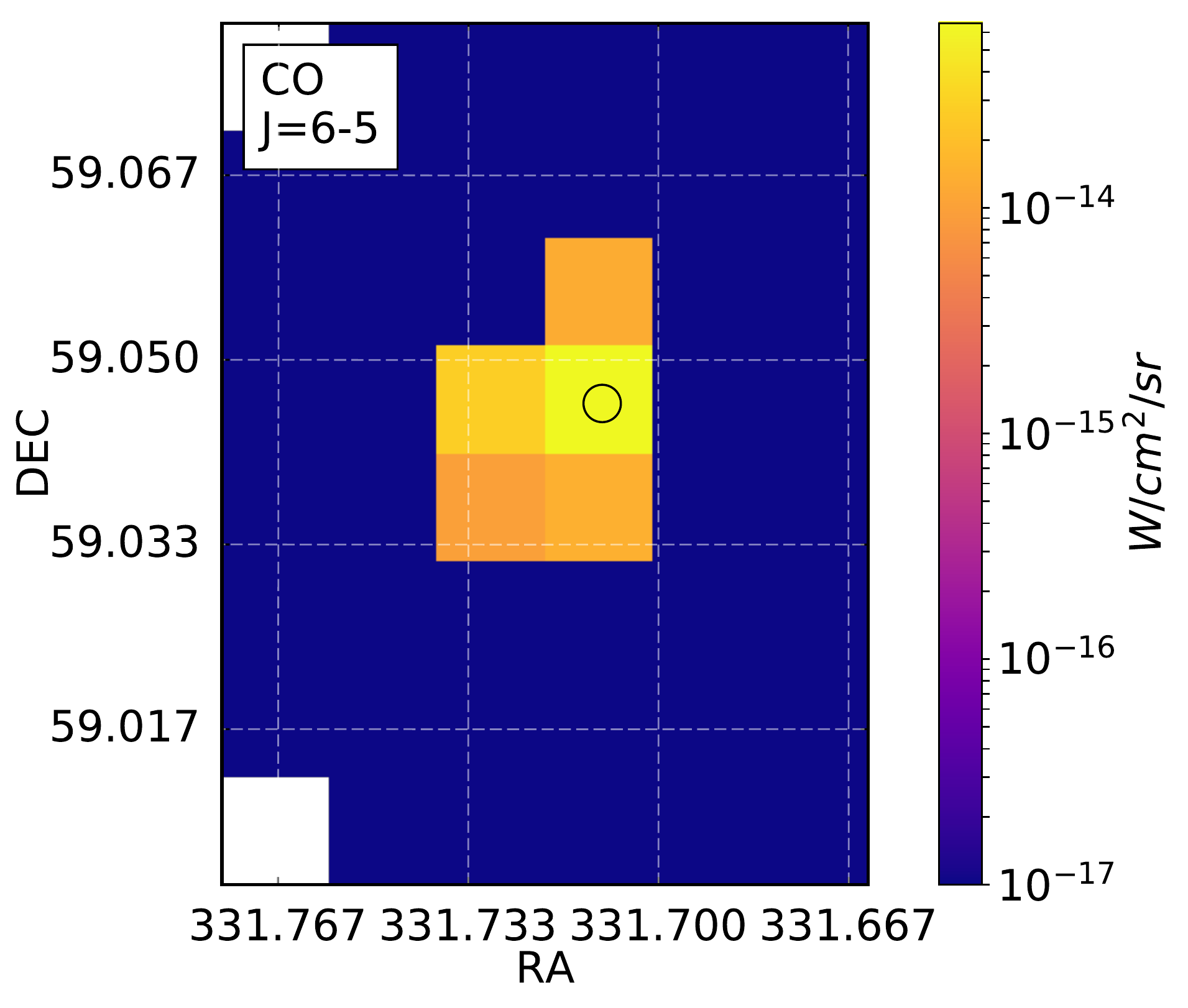}\hspace{-0.1cm}\\
\includegraphics[width=0.33\textwidth, trim={0cm 0 0cm 0}, clip]{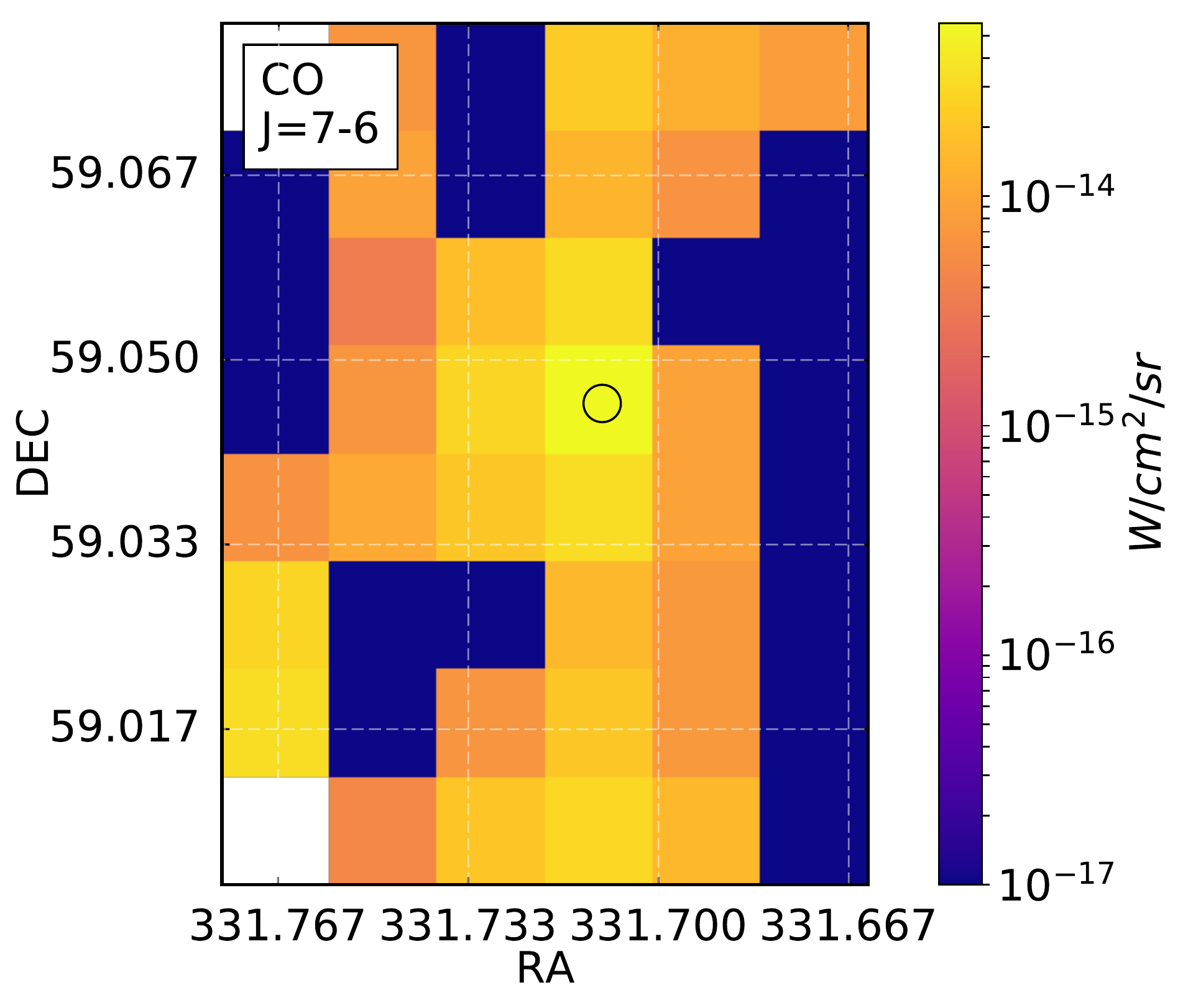}\hspace{-0.1cm}
\includegraphics[width=0.33\textwidth, trim={0cm 0 0cm 0}, clip]{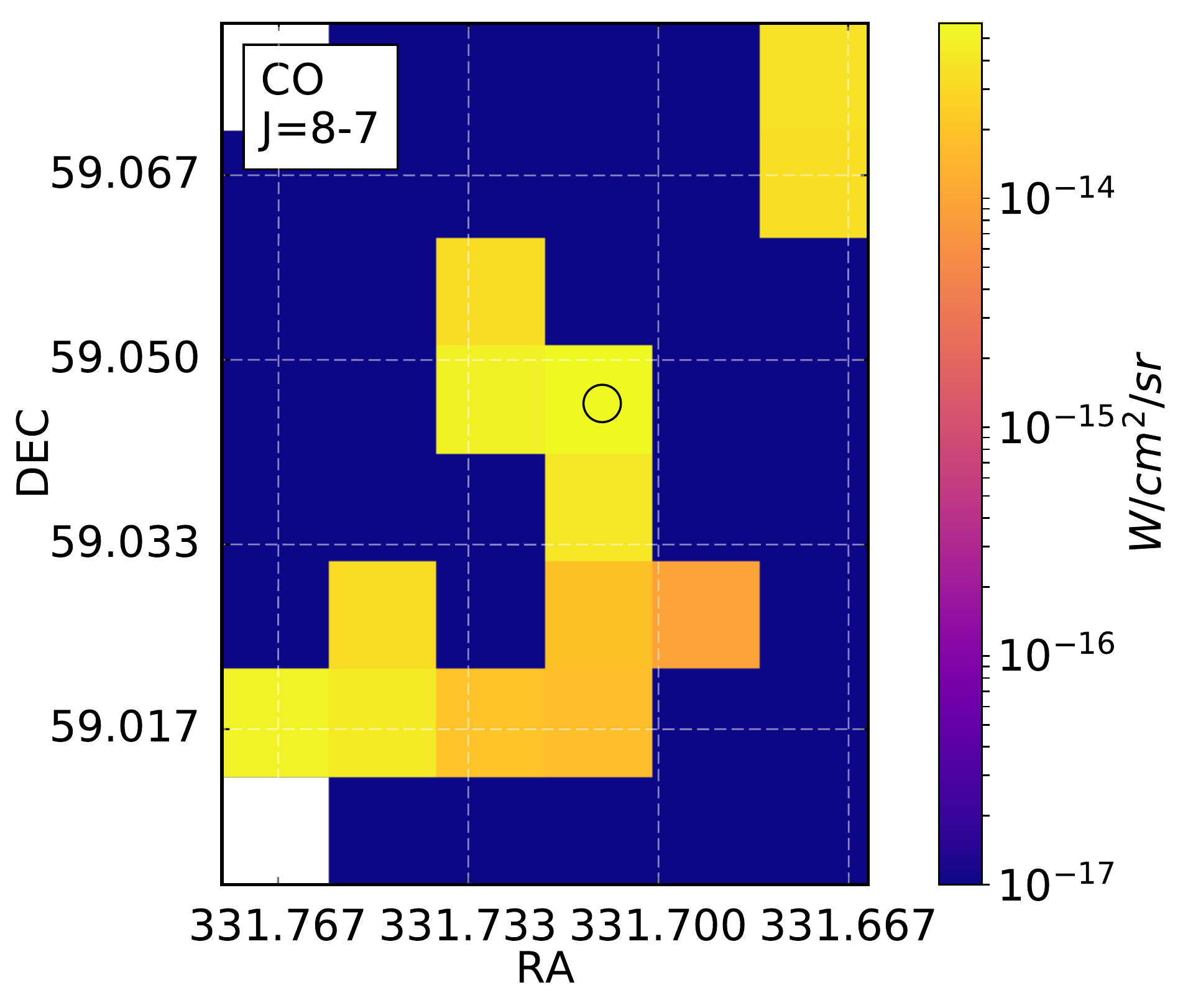}\hspace{-0.1cm}
\includegraphics[width=0.33\textwidth, trim={0cm 0 0cm 0}, clip]{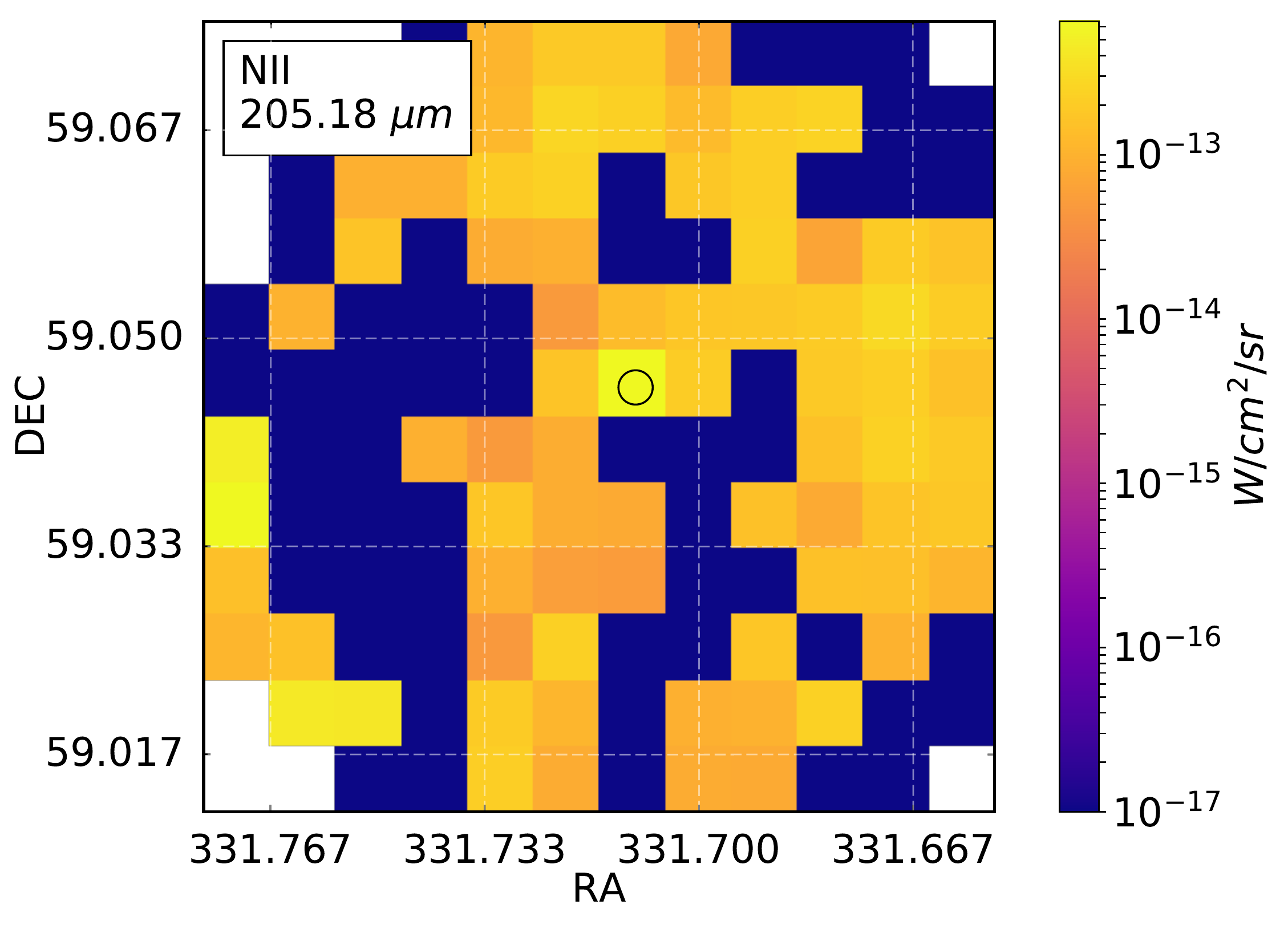}\hspace{-0.1cm}
\caption{
                \footnotesize
                Line maps of SPIRE with visible lines for HH 354 IRS.
        }
\end{figure*}

\begin{figure*}
\includegraphics[width=0.33\textwidth, trim={0cm 0 0cm 0}, clip]{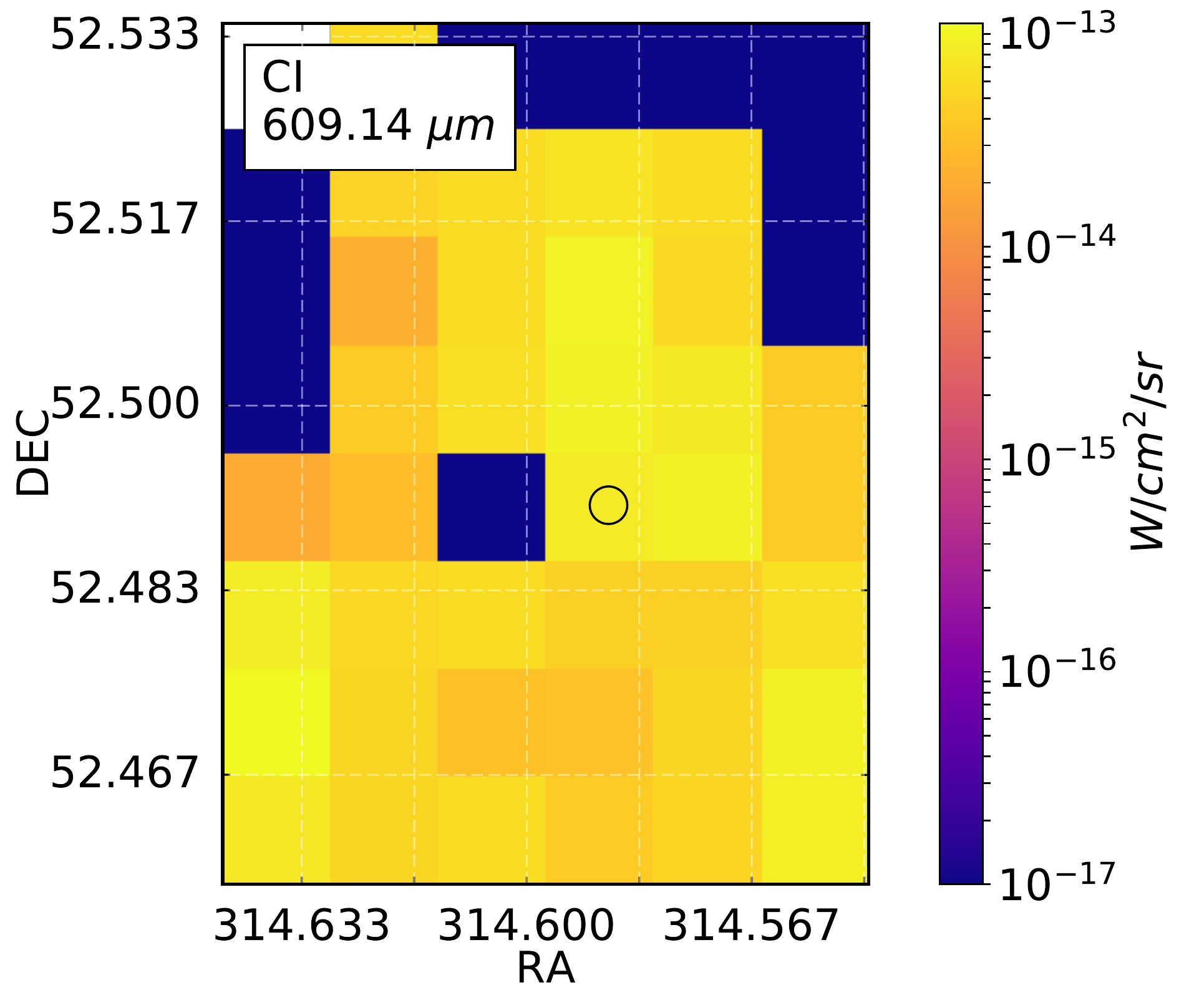}\hspace{-0.1cm}
\includegraphics[width=0.33\textwidth, trim={0cm 0 0cm 0}, clip]{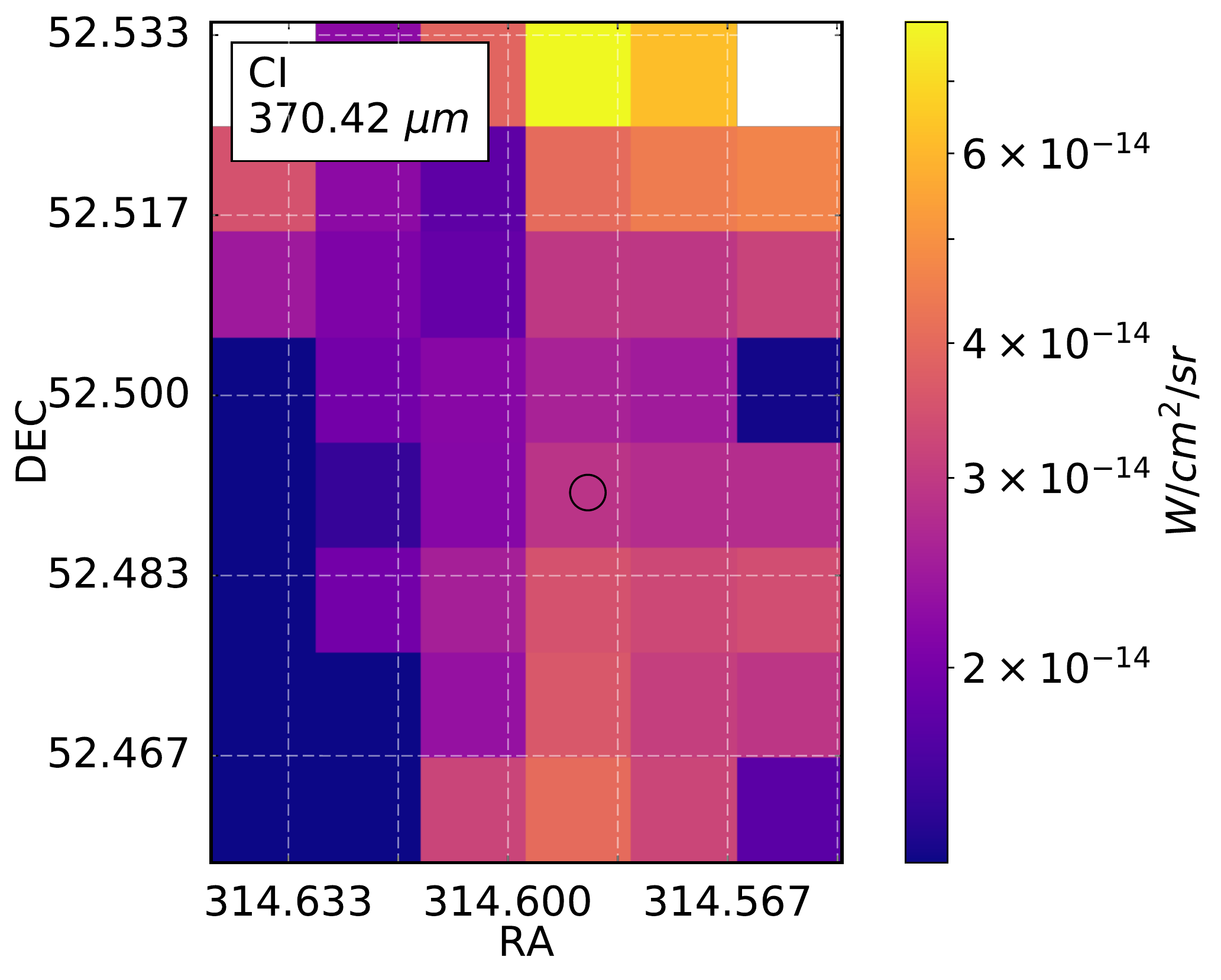}\hspace{-0.1cm}
\includegraphics[width=0.33\textwidth, trim={0cm 0 0cm 0}, clip]{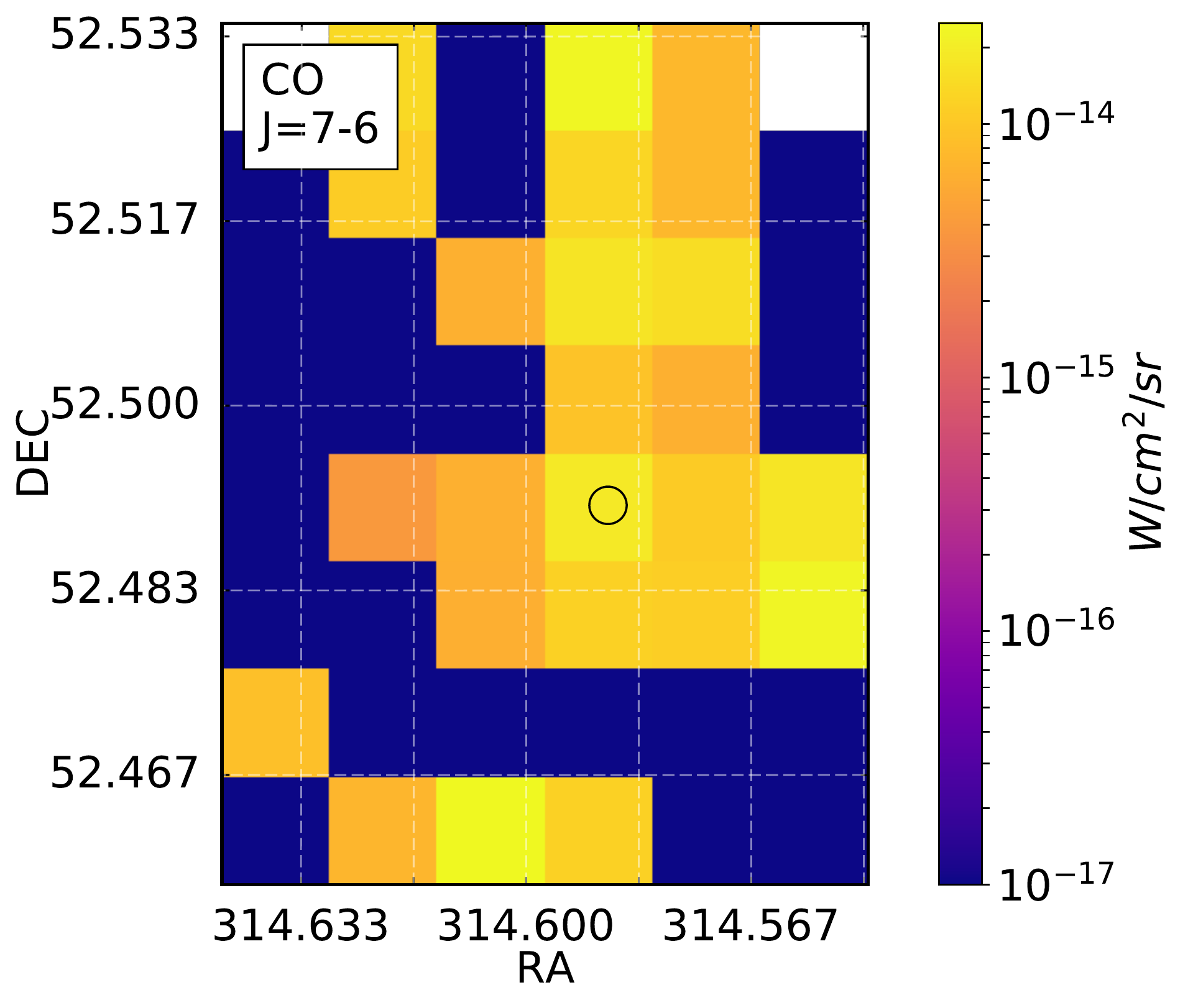}\hspace{-0.1cm}\\
\includegraphics[width=0.33\textwidth, trim={0cm 0 0cm 0}, clip]{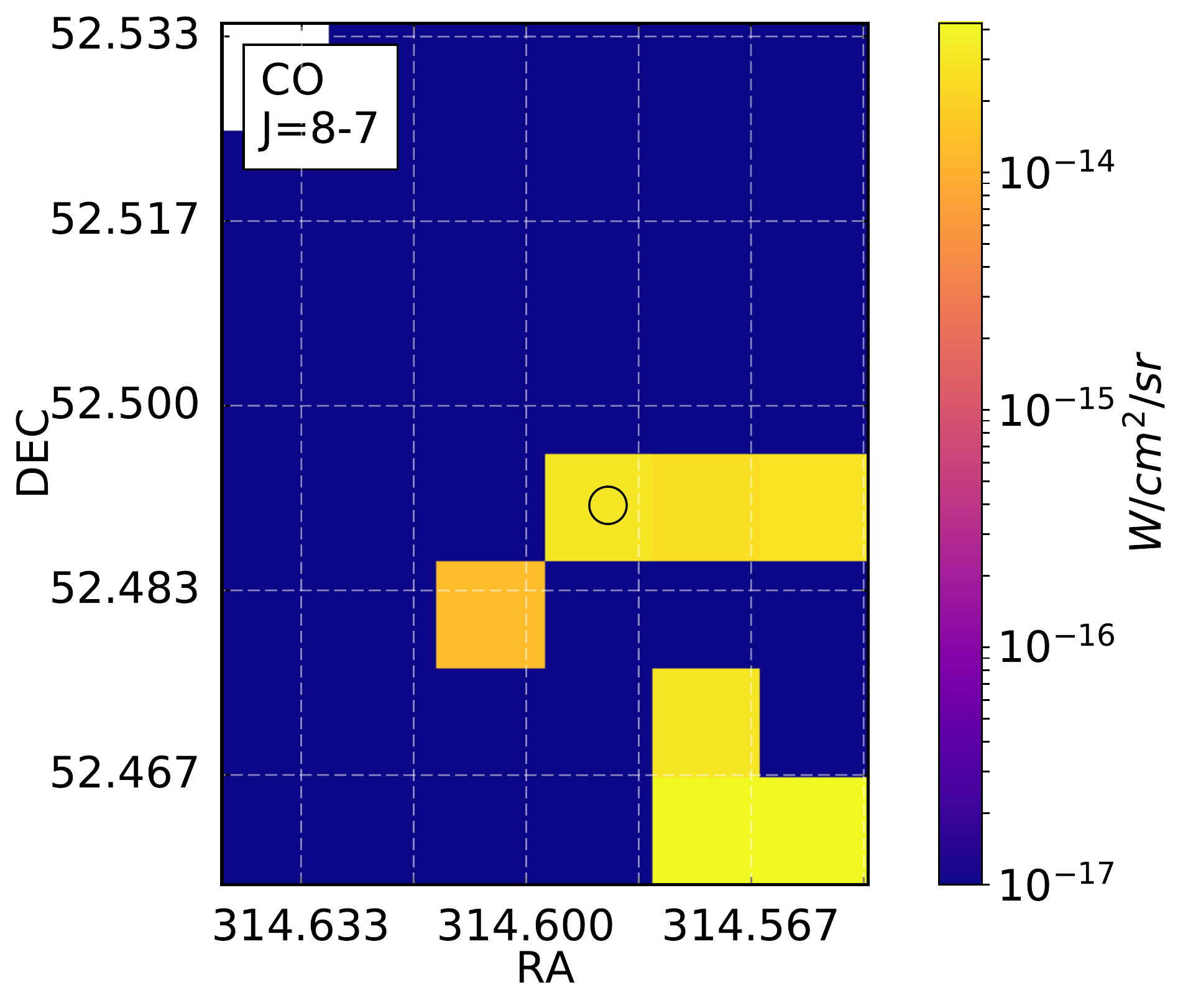}\hspace{-0.1cm}
\includegraphics[width=0.33\textwidth, trim={0cm 0 0cm 0}, clip]{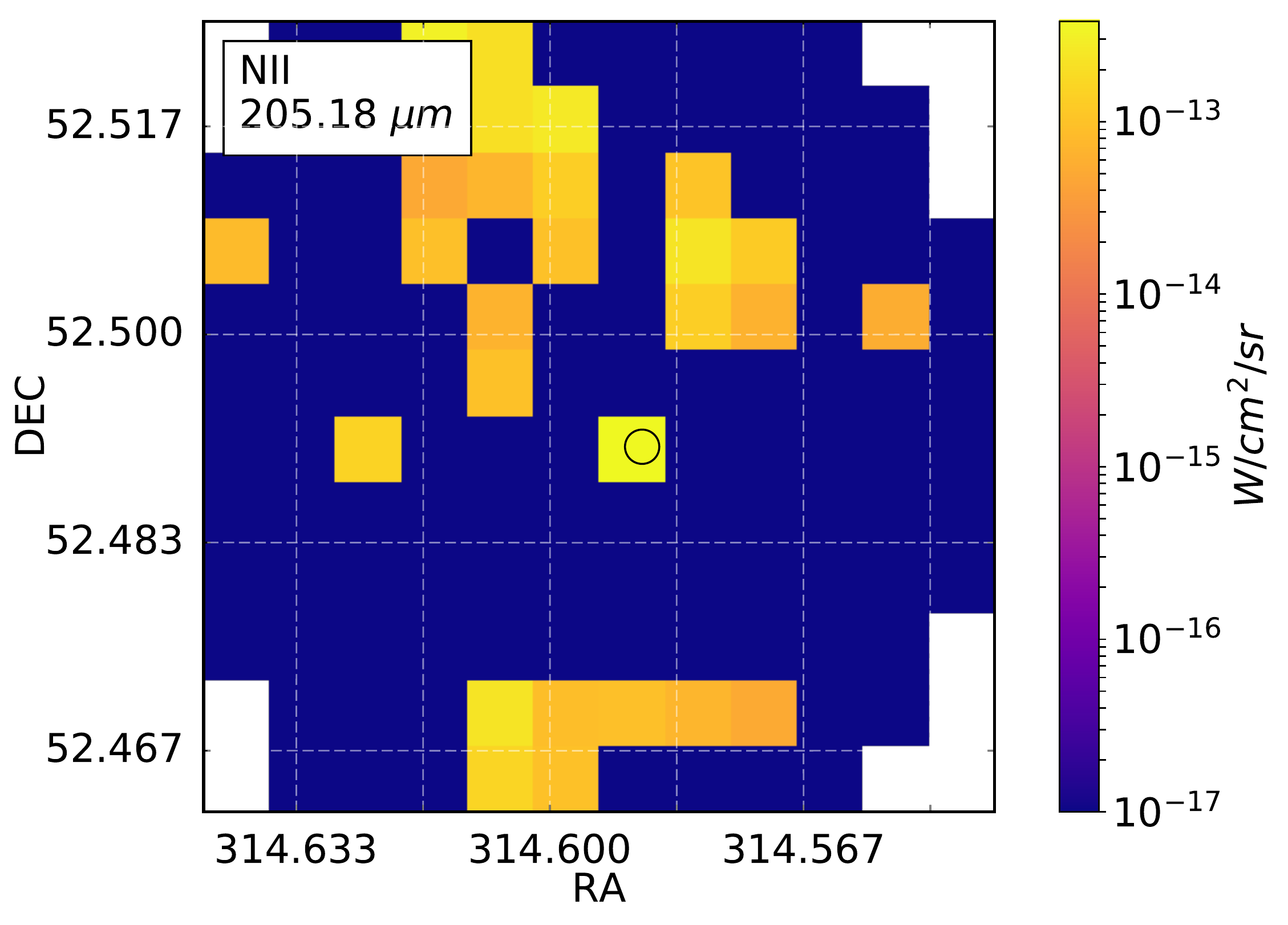}\hspace{-0.1cm}
\caption{
                \footnotesize
                Line maps of SPIRE with visible lines for HH 381 IRS.
        }
\end{figure*}

\begin{figure*}
\includegraphics[width=0.33\textwidth, trim={0cm 0 0cm 0}, clip]{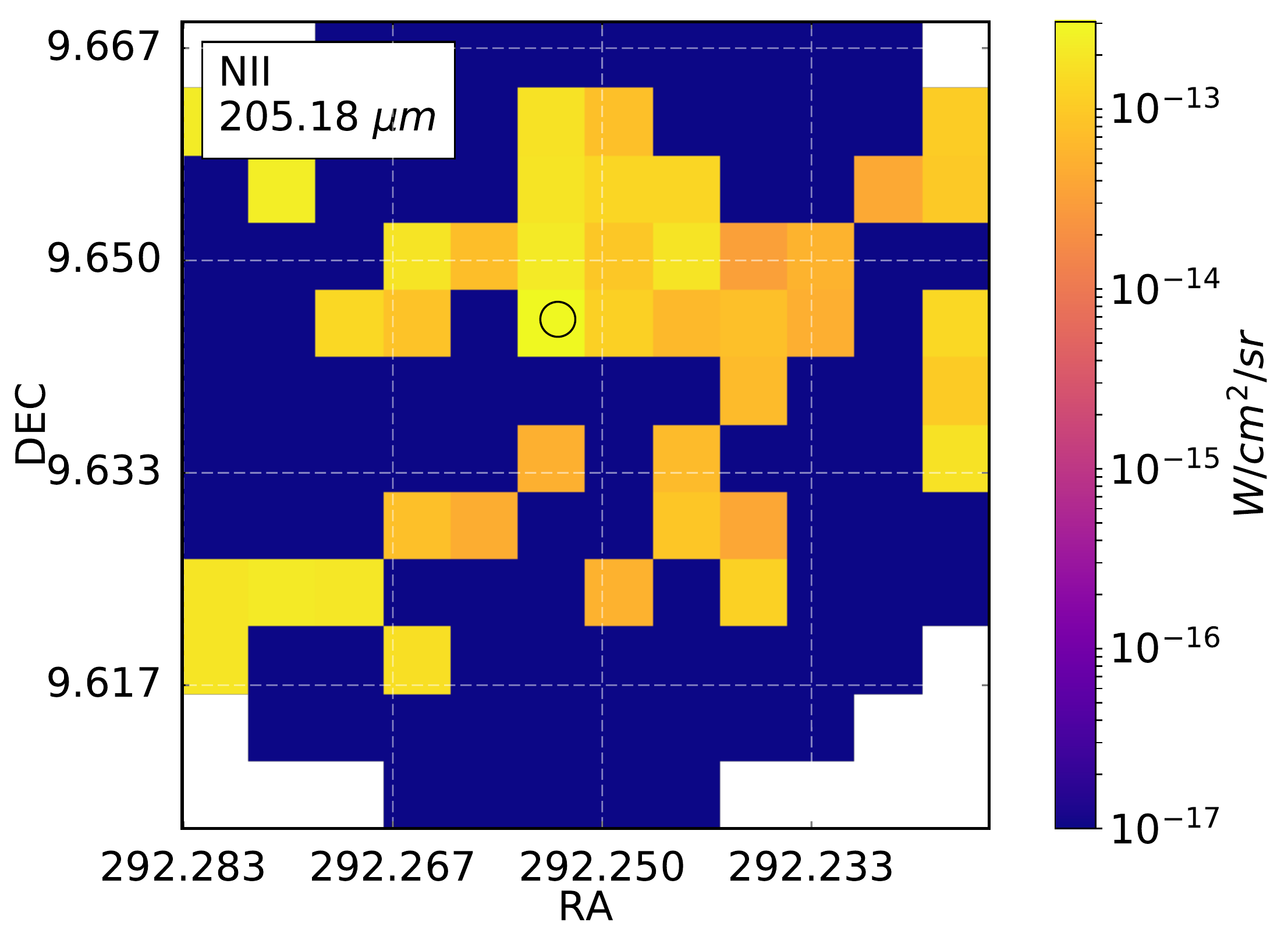}\hspace{-0.1cm}
\caption{
                \footnotesize
                Line maps of SPIRE with visible lines for Parsamian 21.
        }
\end{figure*}

\begin{figure*}
\includegraphics[width=0.33\textwidth, trim={0cm 0 0cm 0}, clip]{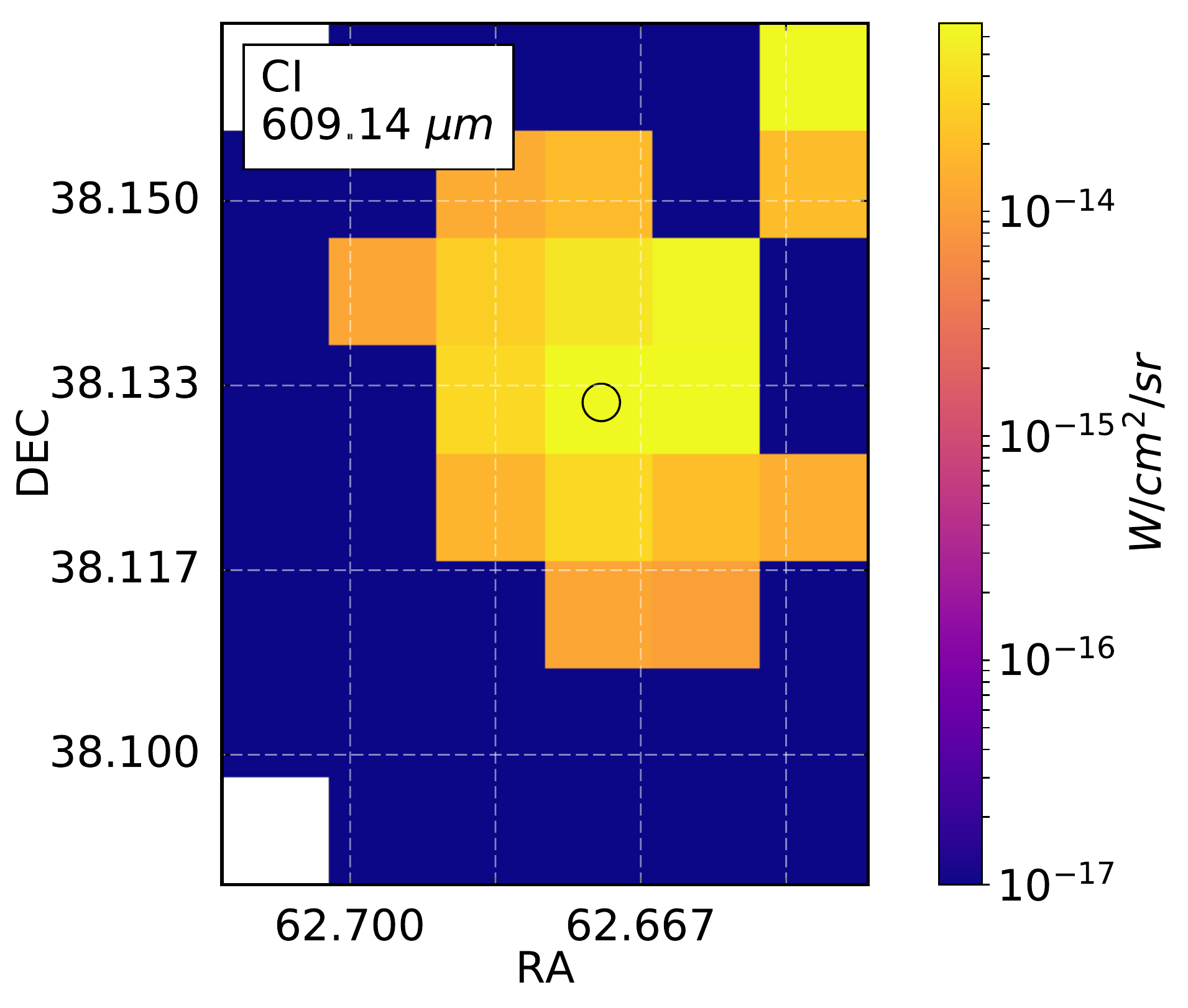}\hspace{-0.1cm}
\includegraphics[width=0.33\textwidth, trim={0cm 0 0cm 0}, clip]{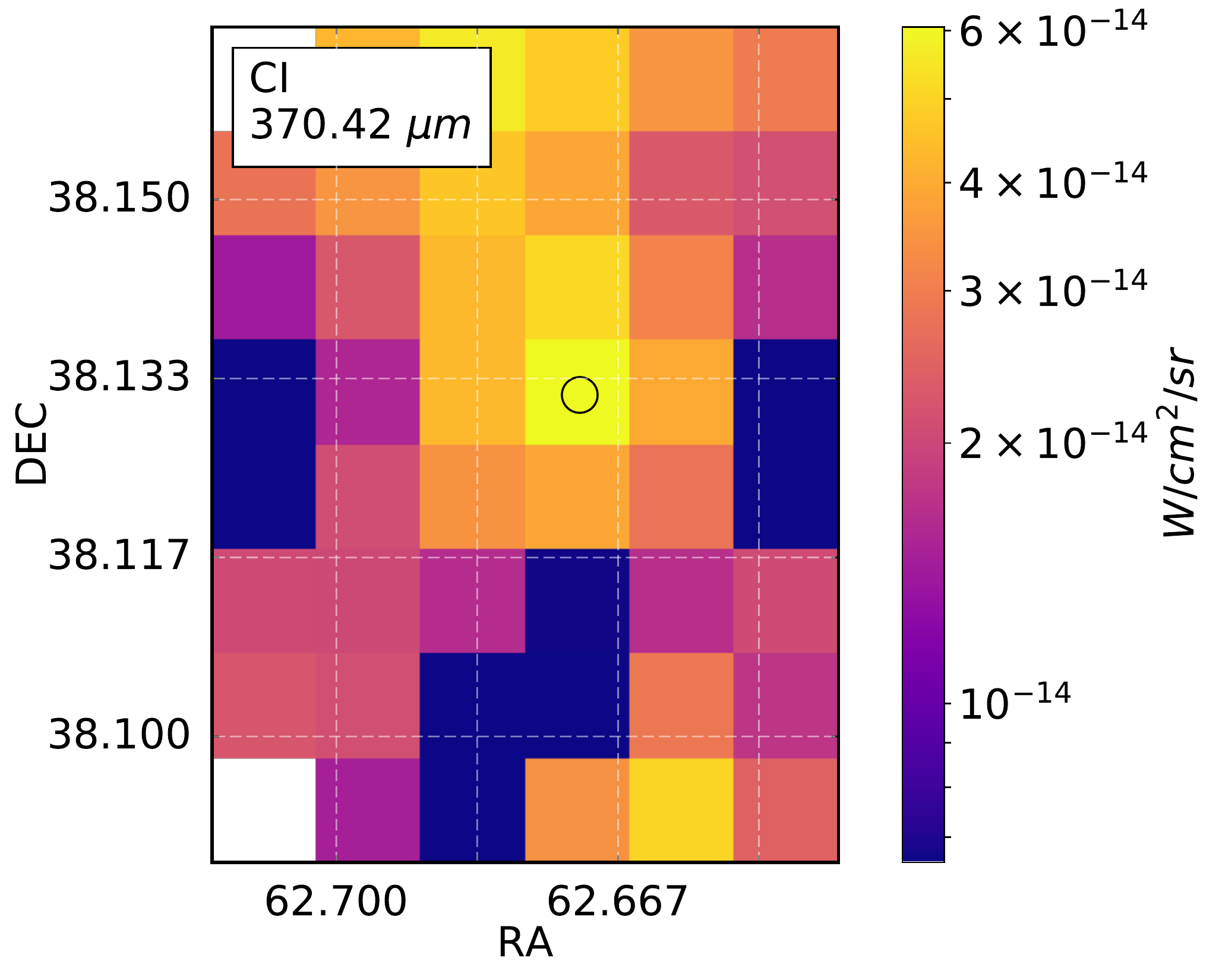}\hspace{-0.1cm}
\includegraphics[width=0.33\textwidth, trim={0cm 0 0cm 0}, clip]{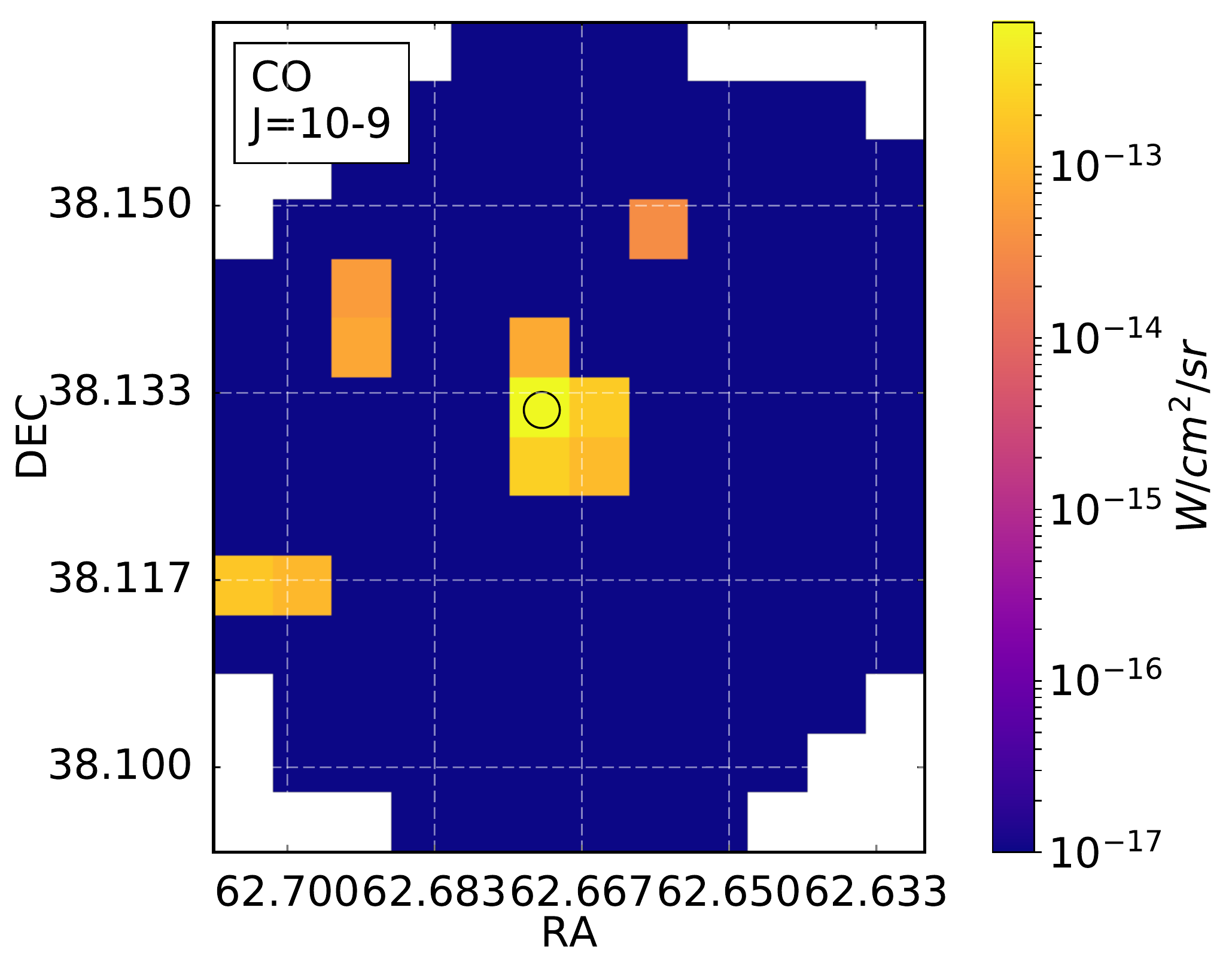}\hspace{-0.1cm}\\
\includegraphics[width=0.33\textwidth, trim={0cm 0 0cm 0}, clip]{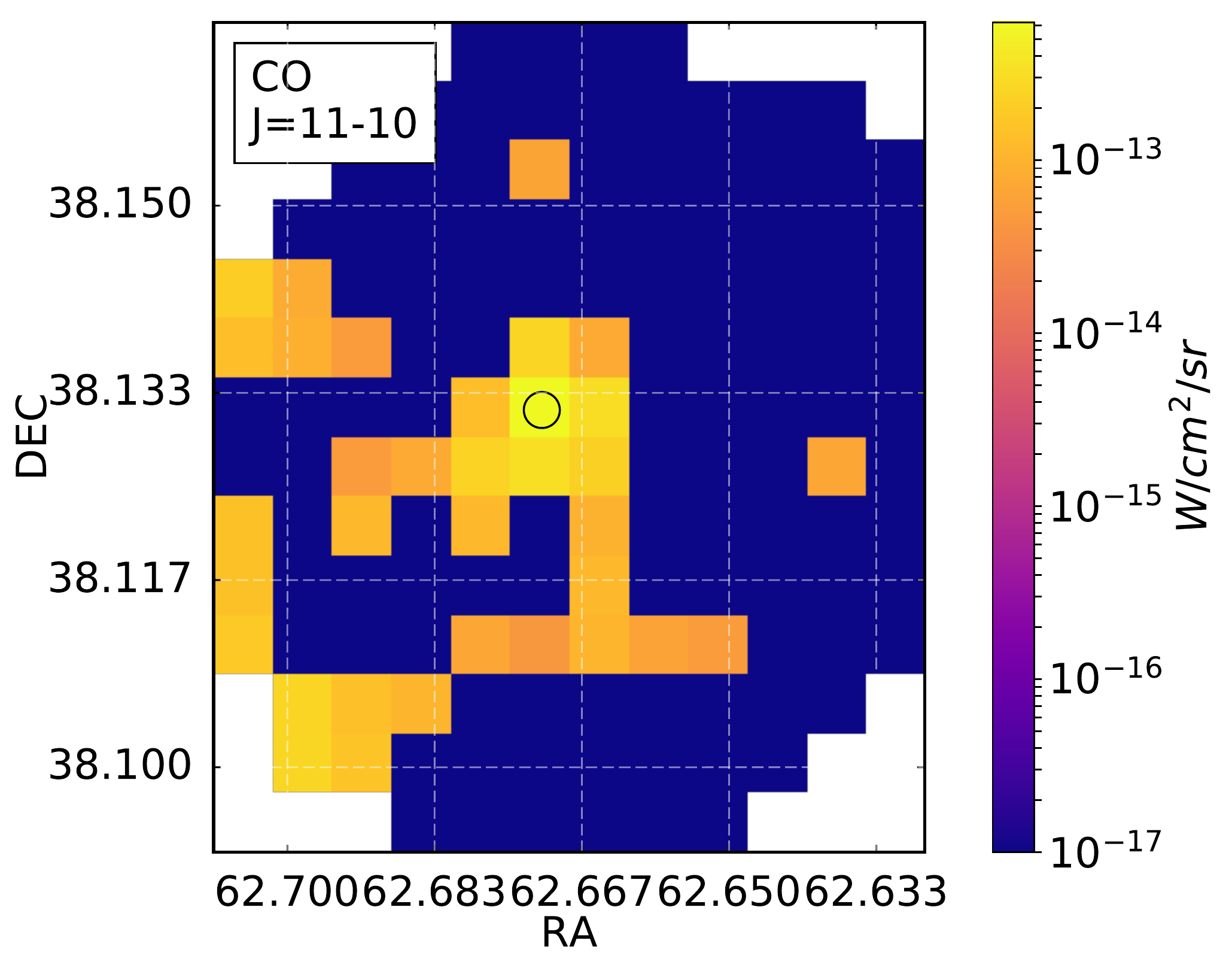}\hspace{-0.1cm}
\includegraphics[width=0.33\textwidth, trim={0cm 0 0cm 0}, clip]{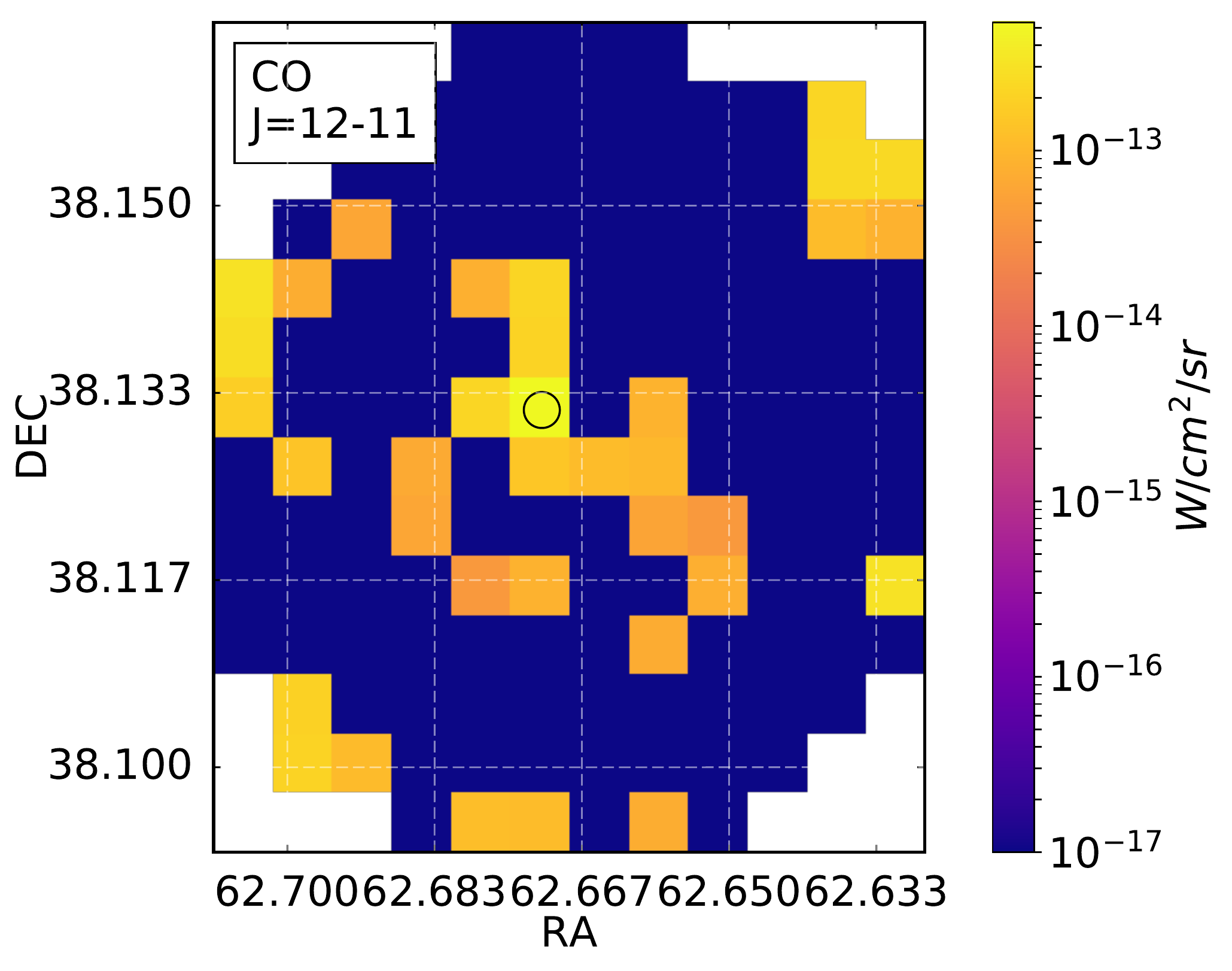}\hspace{-0.1cm}
\includegraphics[width=0.33\textwidth, trim={0cm 0 0cm 0}, clip]{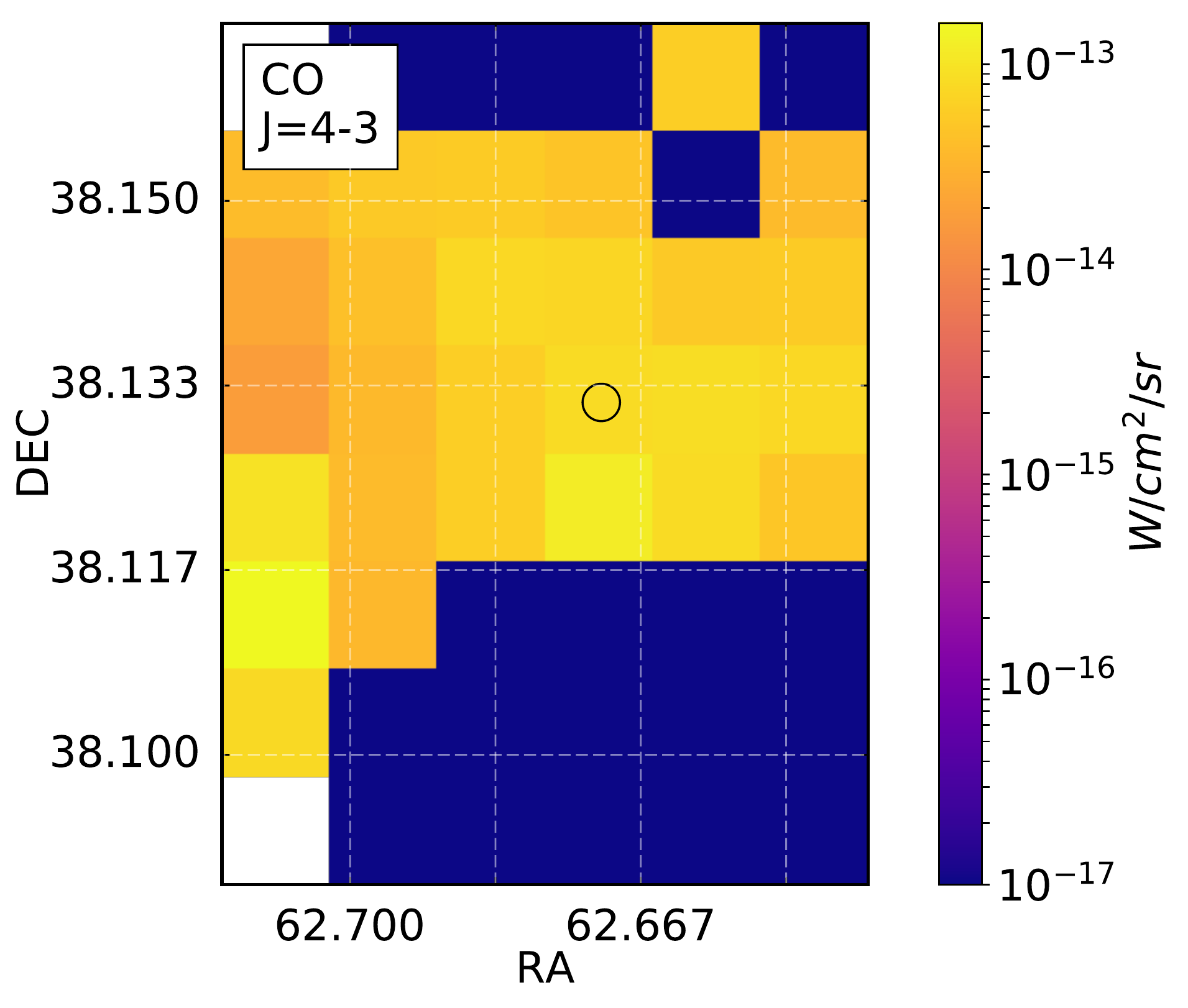}\hspace{-0.1cm}\\
\includegraphics[width=0.33\textwidth, trim={0cm 0 0cm 0}, clip]{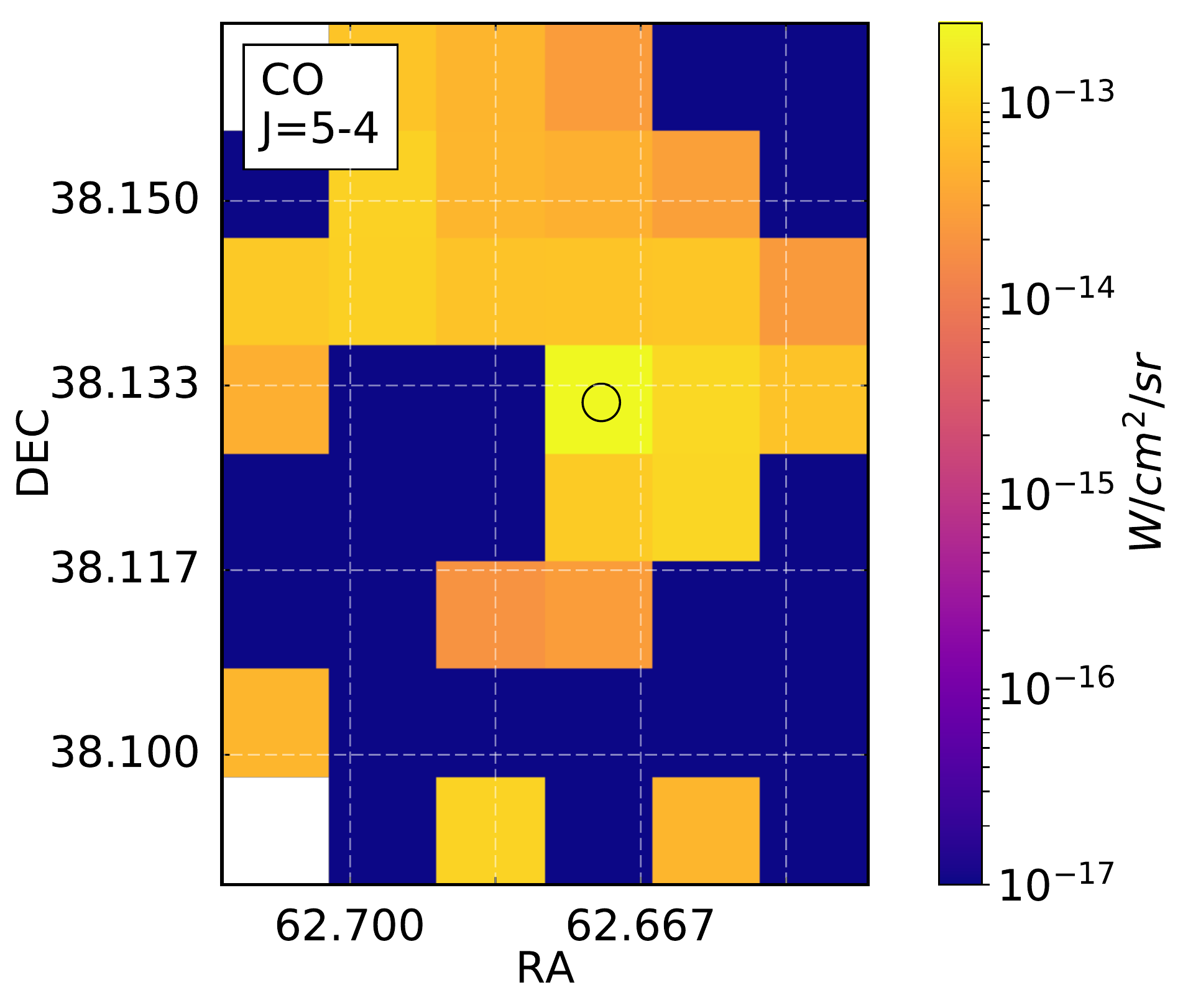}\hspace{-0.1cm}
\includegraphics[width=0.33\textwidth, trim={0cm 0 0cm 0}, clip]{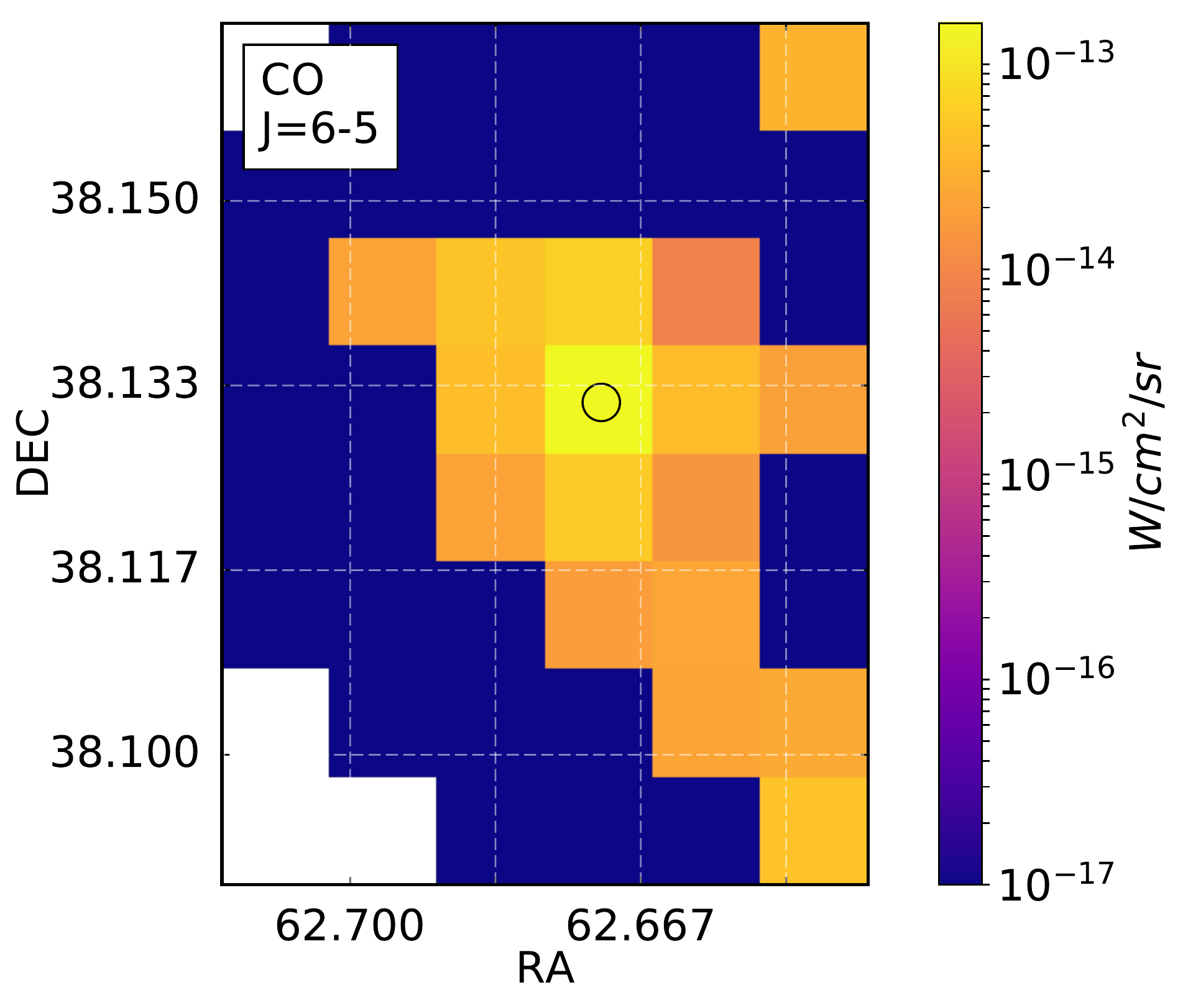}\hspace{-0.1cm}
\includegraphics[width=0.33\textwidth, trim={0cm 0 0cm 0}, clip]{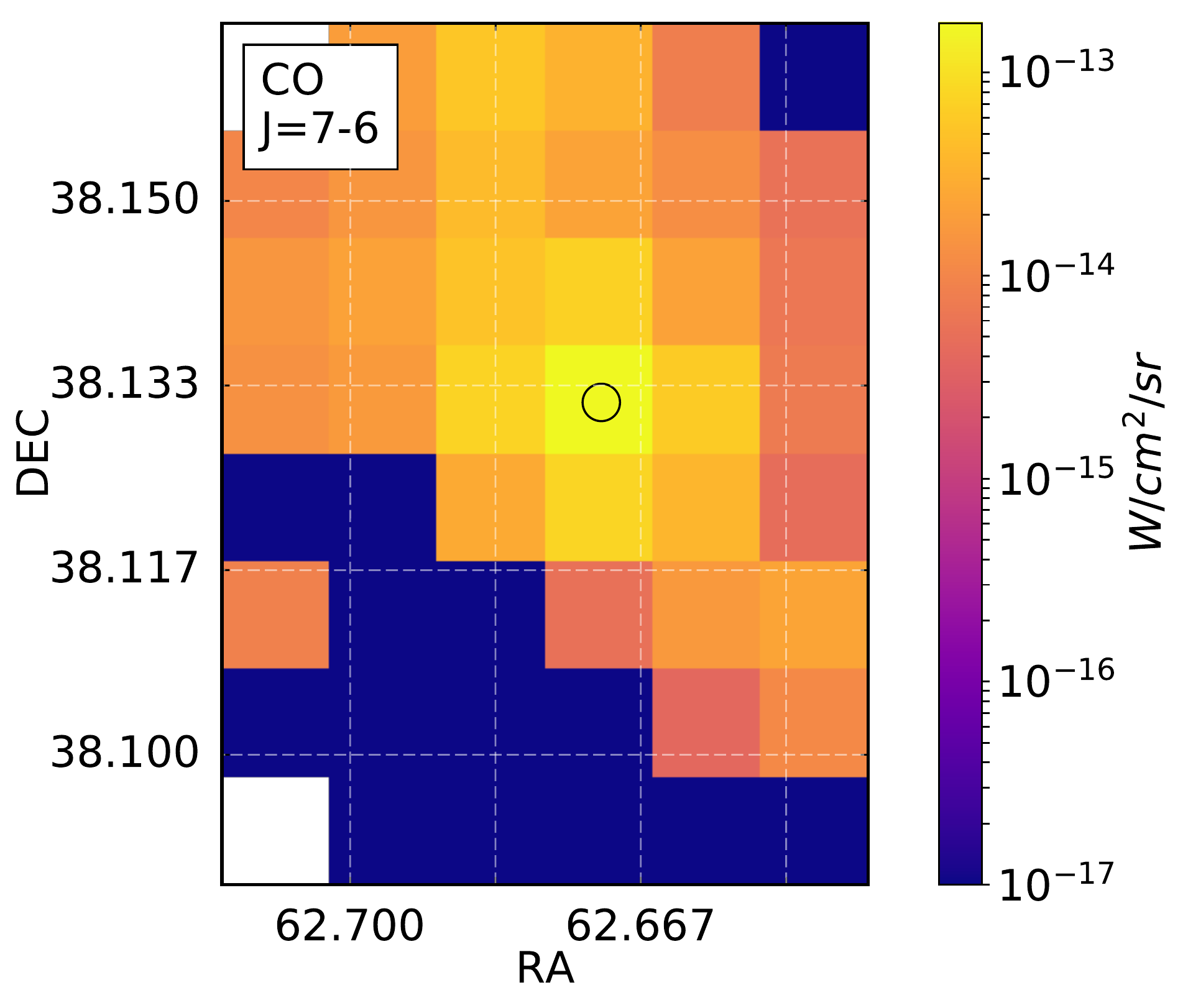}\hspace{-0.1cm}\\
\includegraphics[width=0.33\textwidth, trim={0cm 0 0cm 0}, clip]{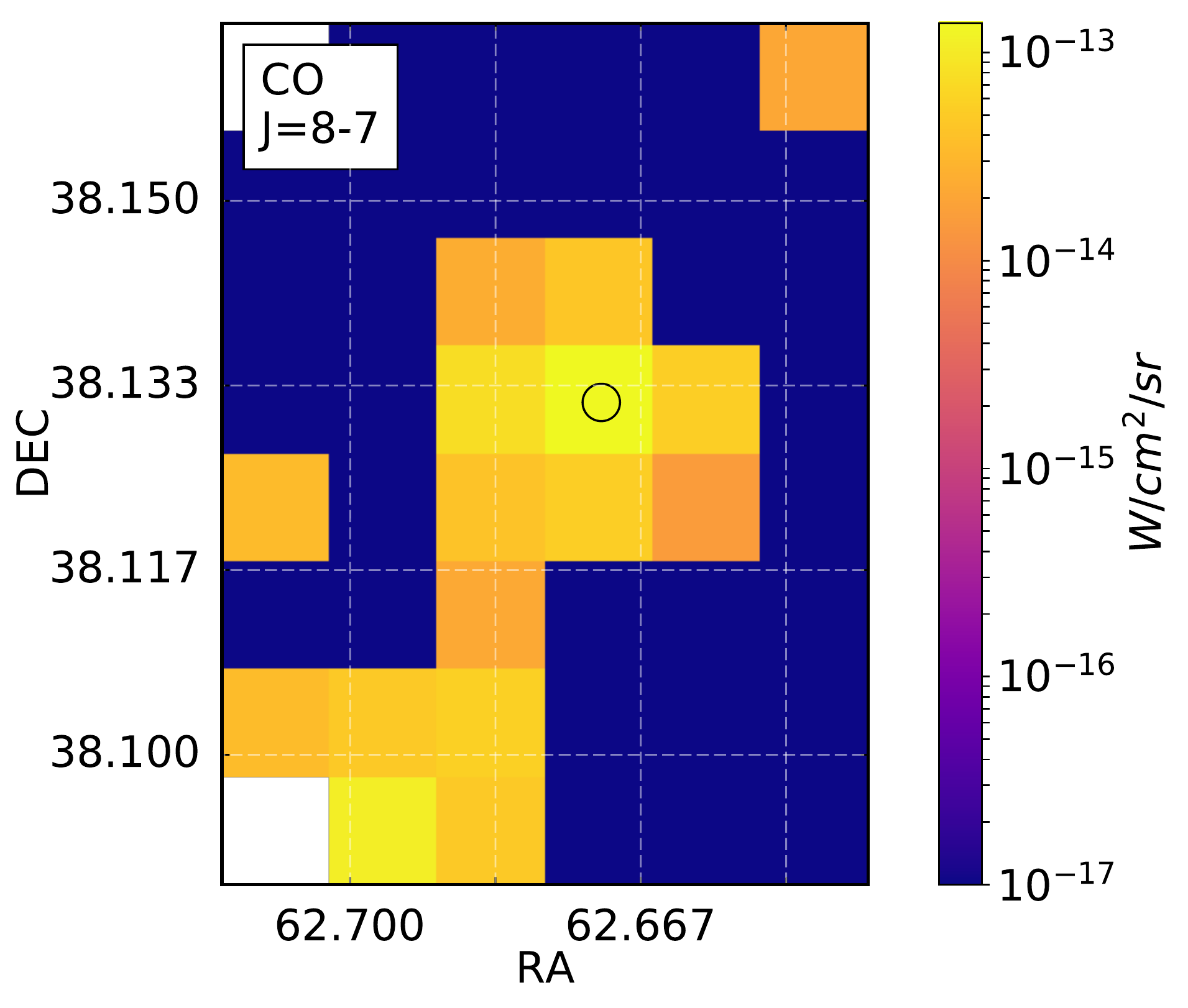}\hspace{-0.1cm}
\includegraphics[width=0.33\textwidth, trim={0cm 0 0cm 0}, clip]{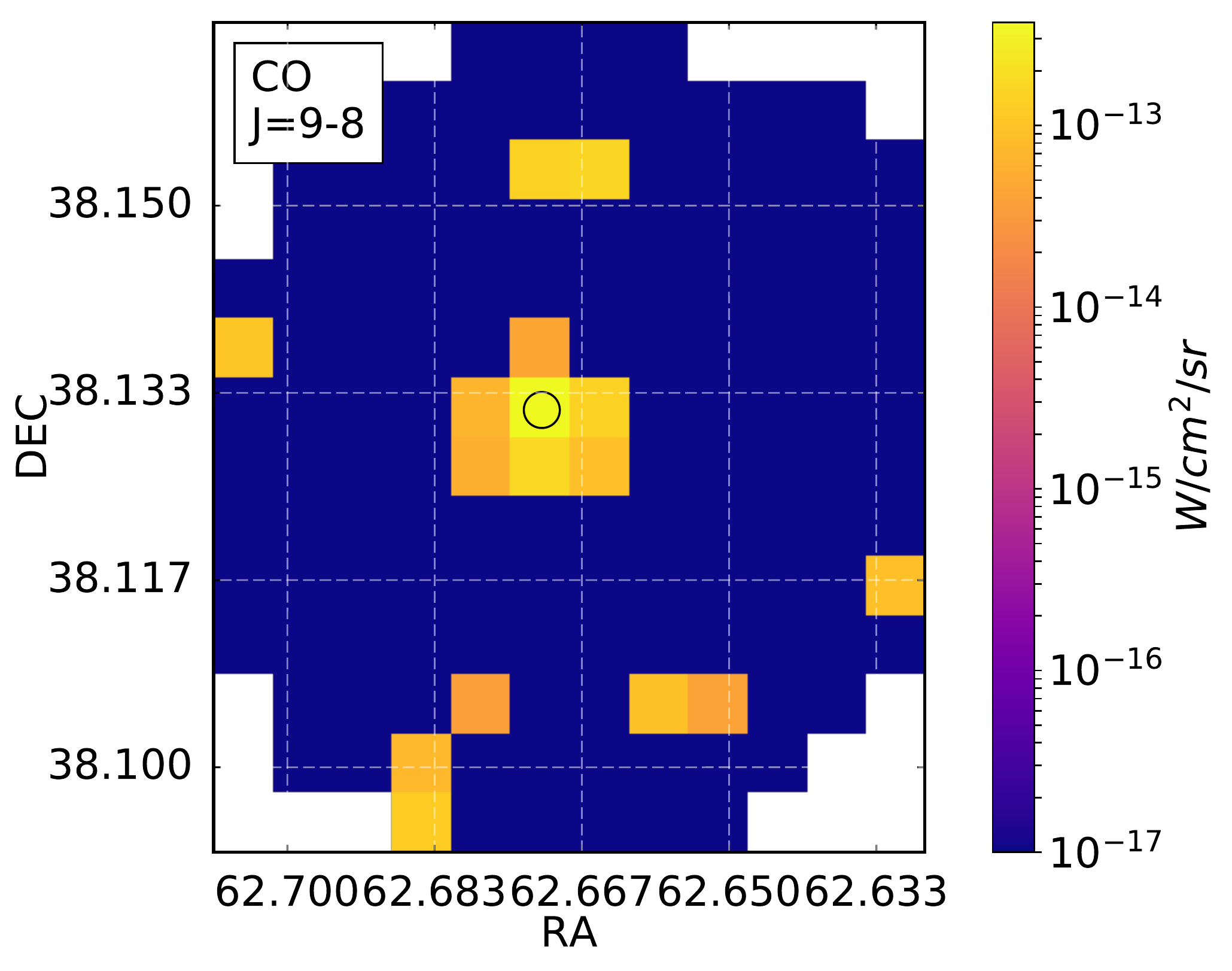}\hspace{-0.1cm}
\includegraphics[width=0.33\textwidth, trim={0cm 0 0cm 0}, clip]{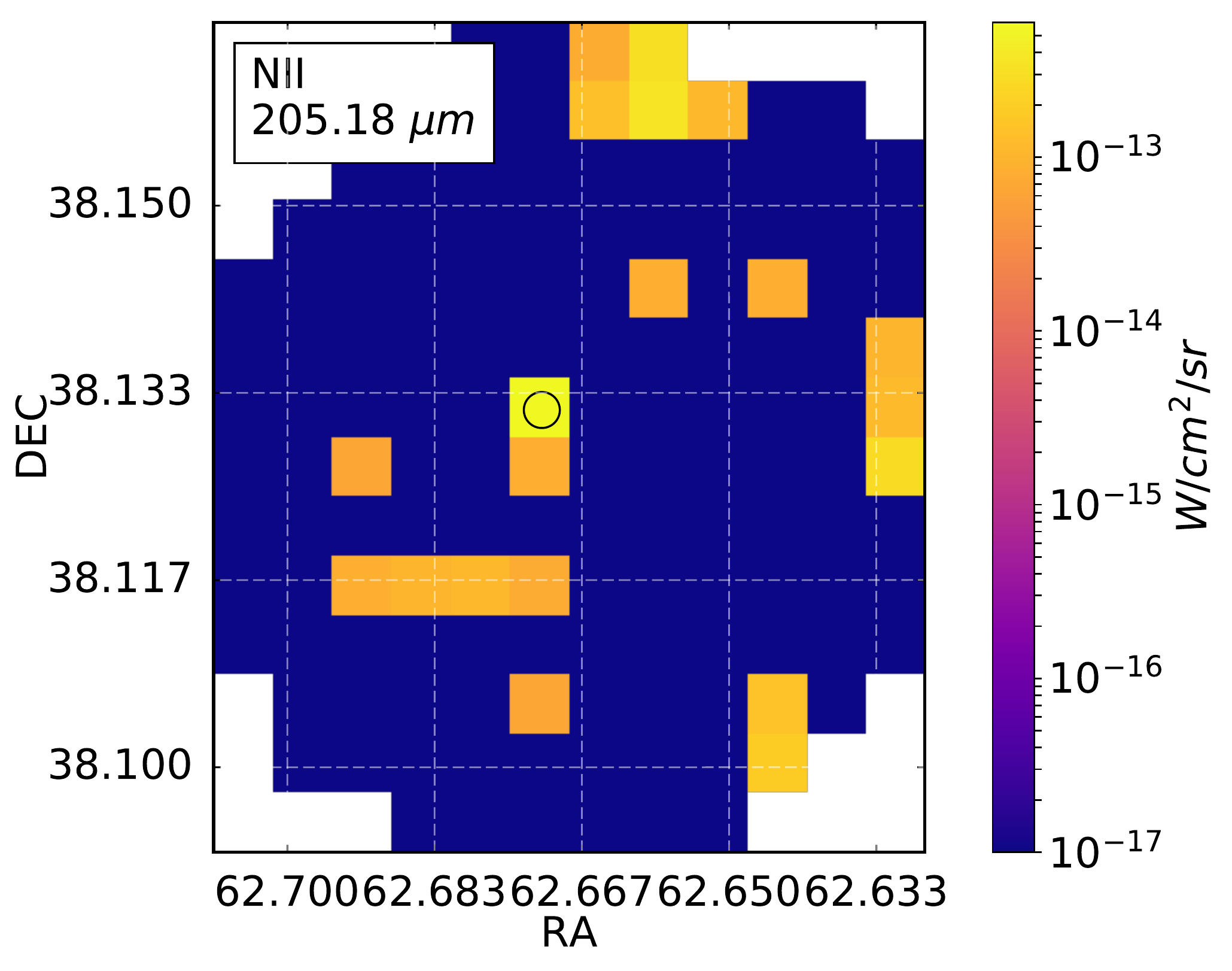}\hspace{-0.1cm}
\caption{
                \footnotesize
                Line maps of SPIRE with visible lines for PP 13 S.
        }
\end{figure*}

\begin{figure*}
\includegraphics[width=0.33\textwidth, trim={0cm 0 0cm 0}, clip]{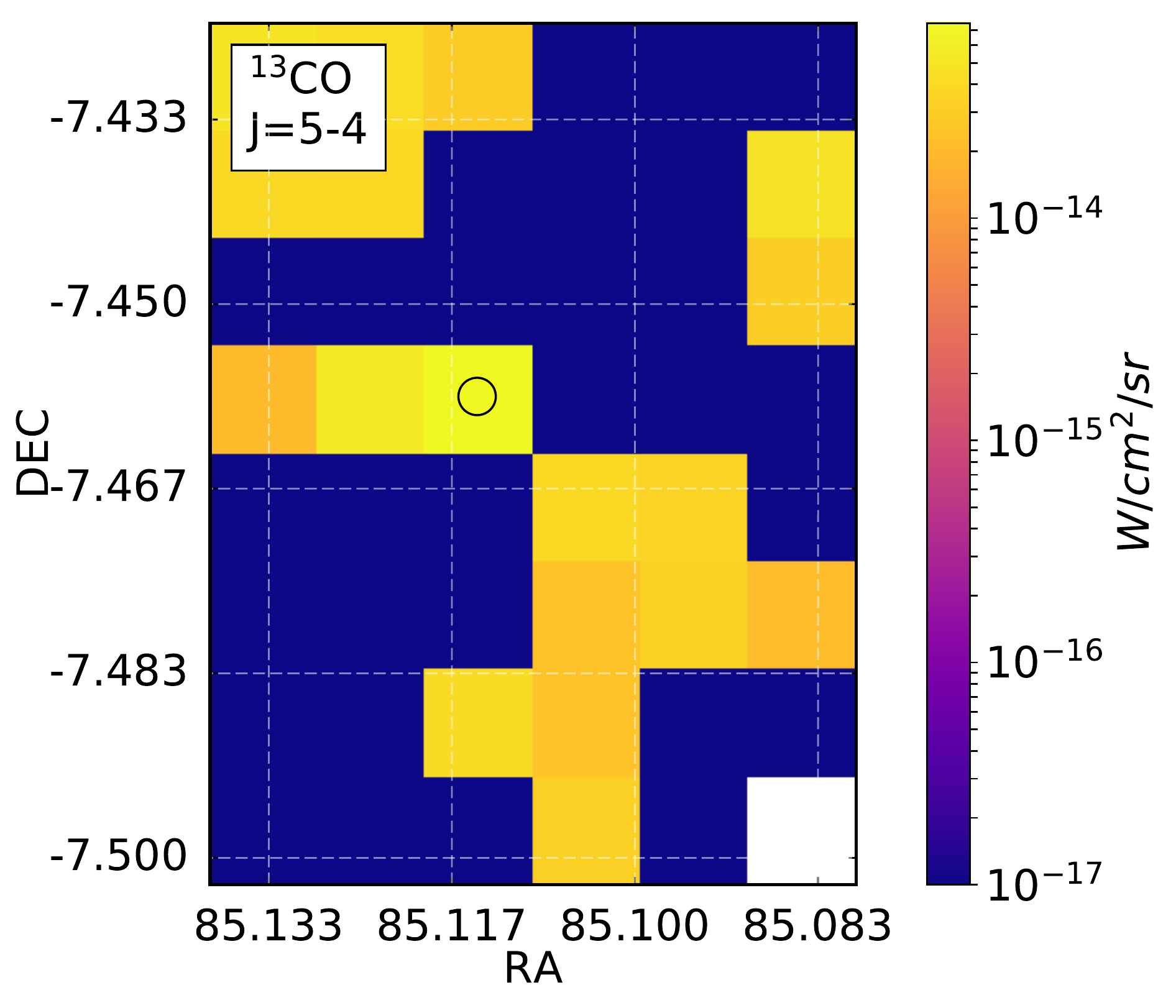}\hspace{-0.1cm}
\includegraphics[width=0.33\textwidth, trim={0cm 0 0cm 0}, clip]{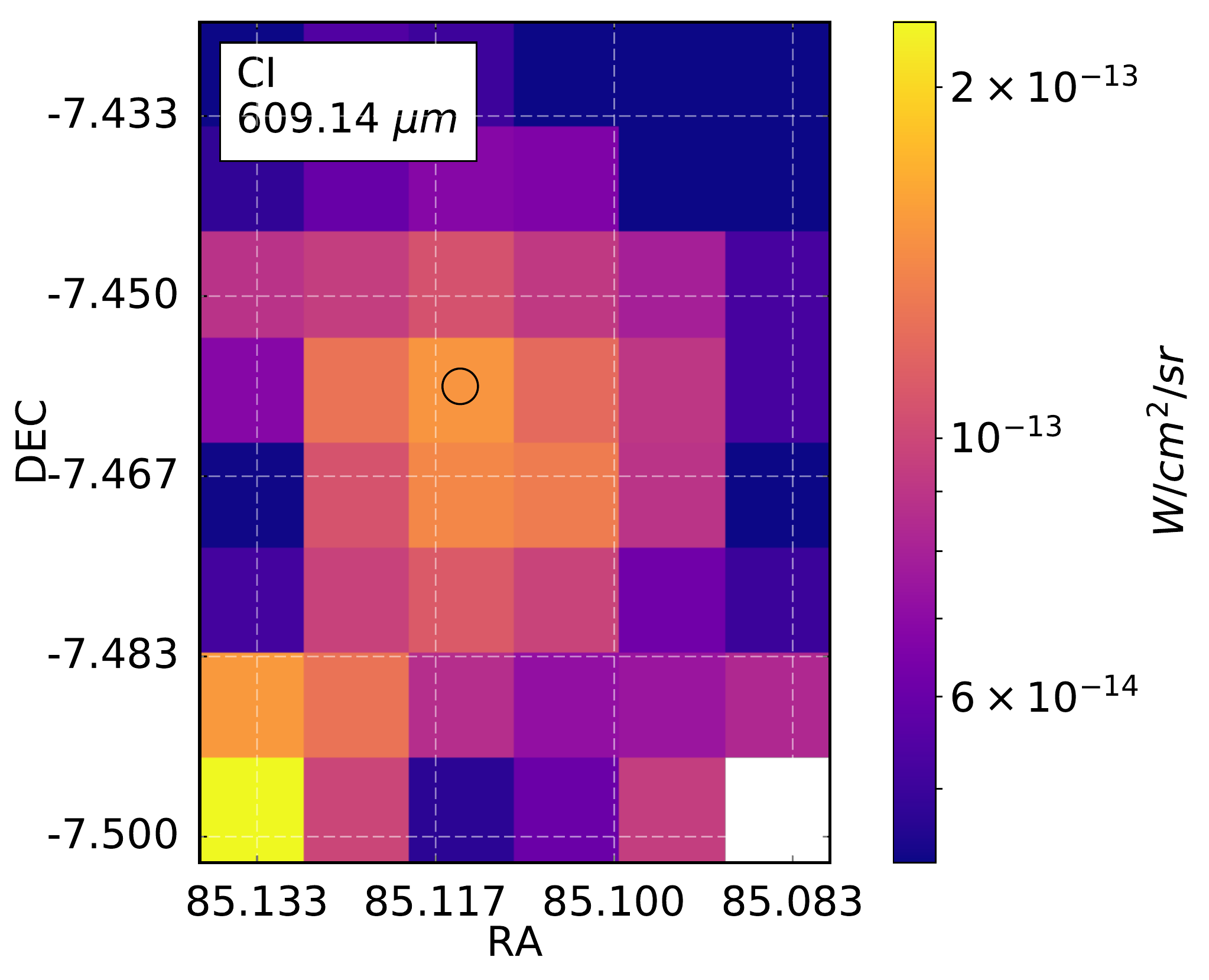}\hspace{-0.1cm}
\includegraphics[width=0.33\textwidth, trim={0cm 0 0cm 0}, clip]{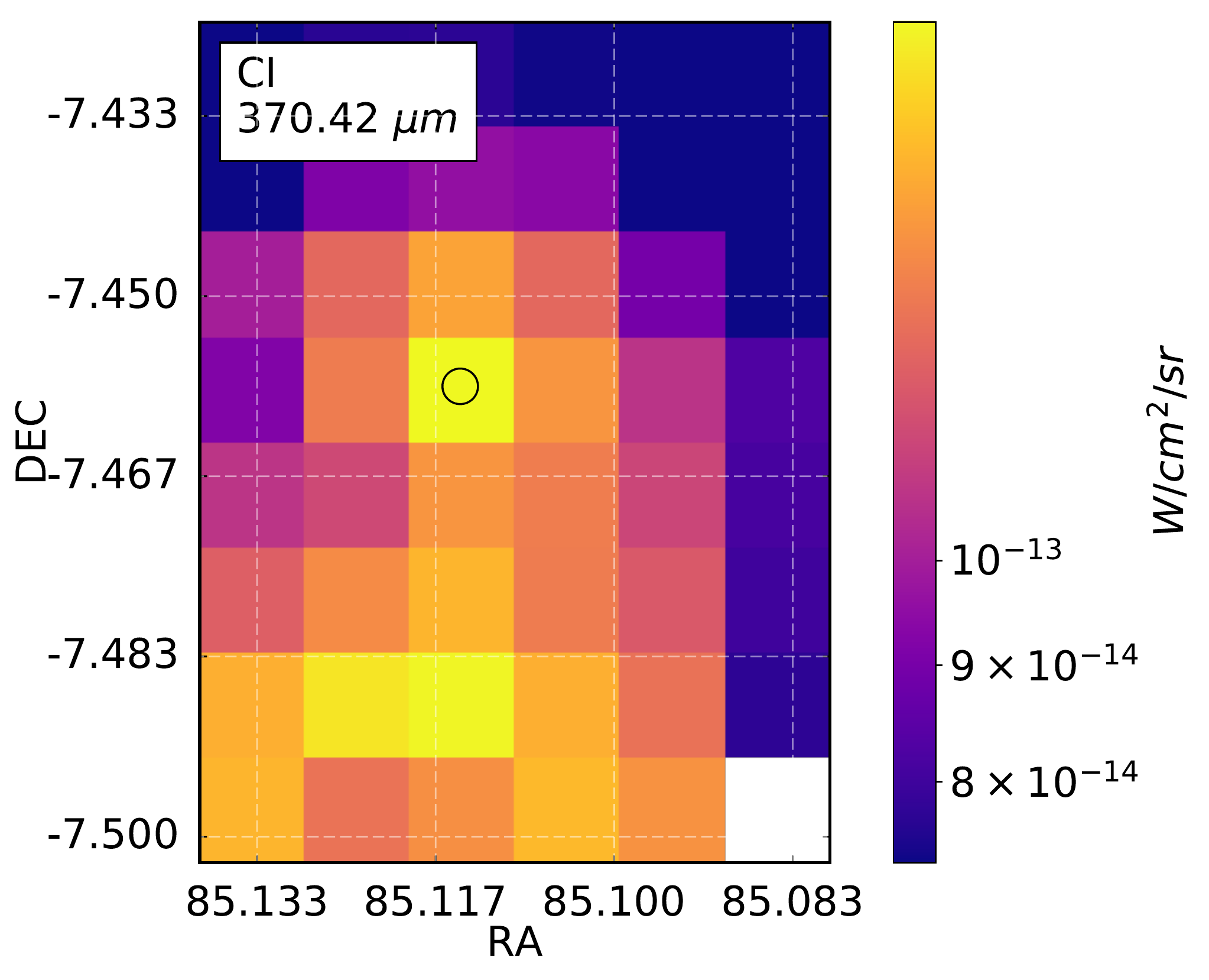}\hspace{-0.1cm}\\
\includegraphics[width=0.33\textwidth, trim={0cm 0 0cm 0}, clip]{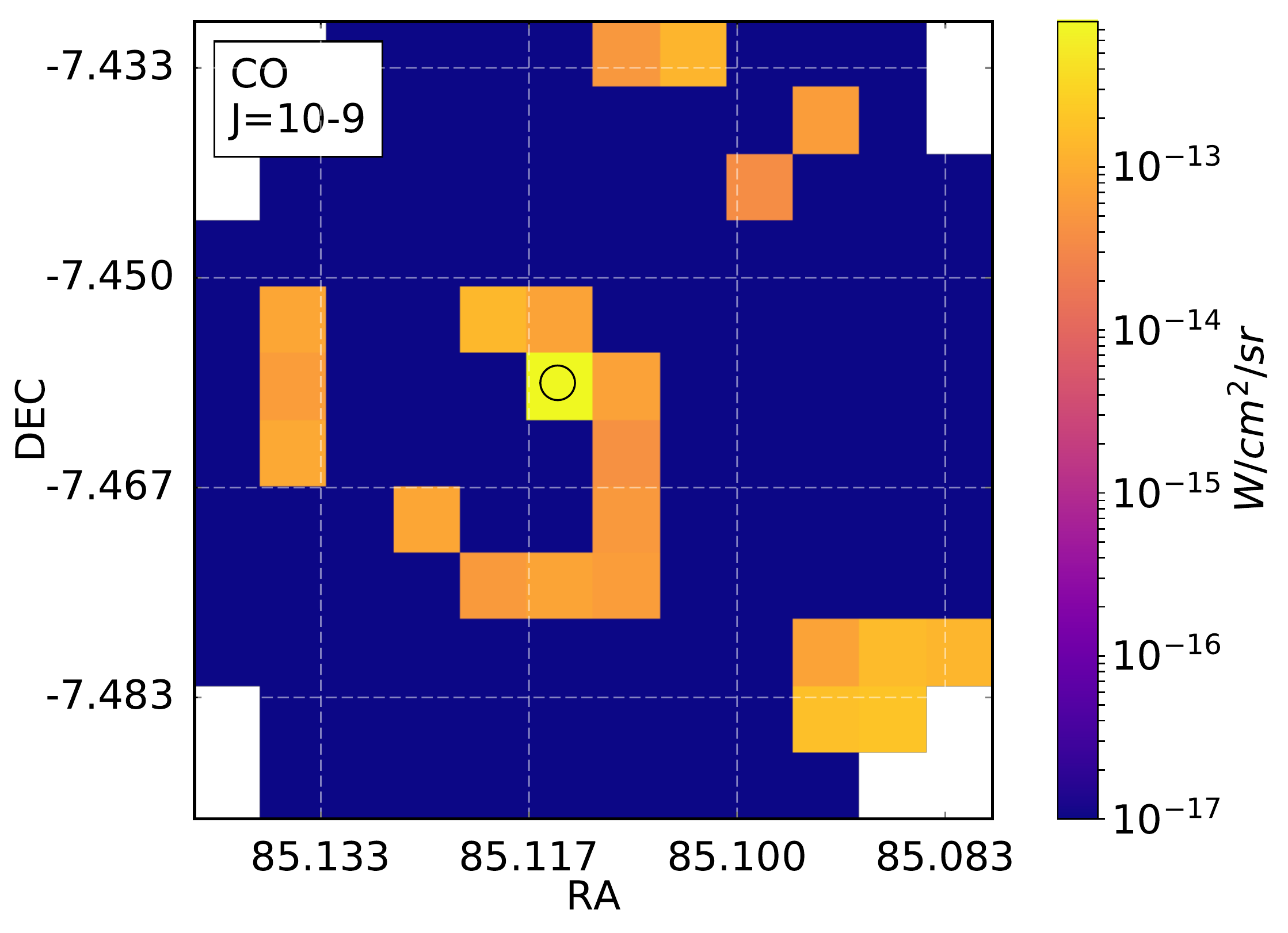}\hspace{-0.1cm}
\includegraphics[width=0.33\textwidth, trim={0cm 0 0cm 0}, clip]{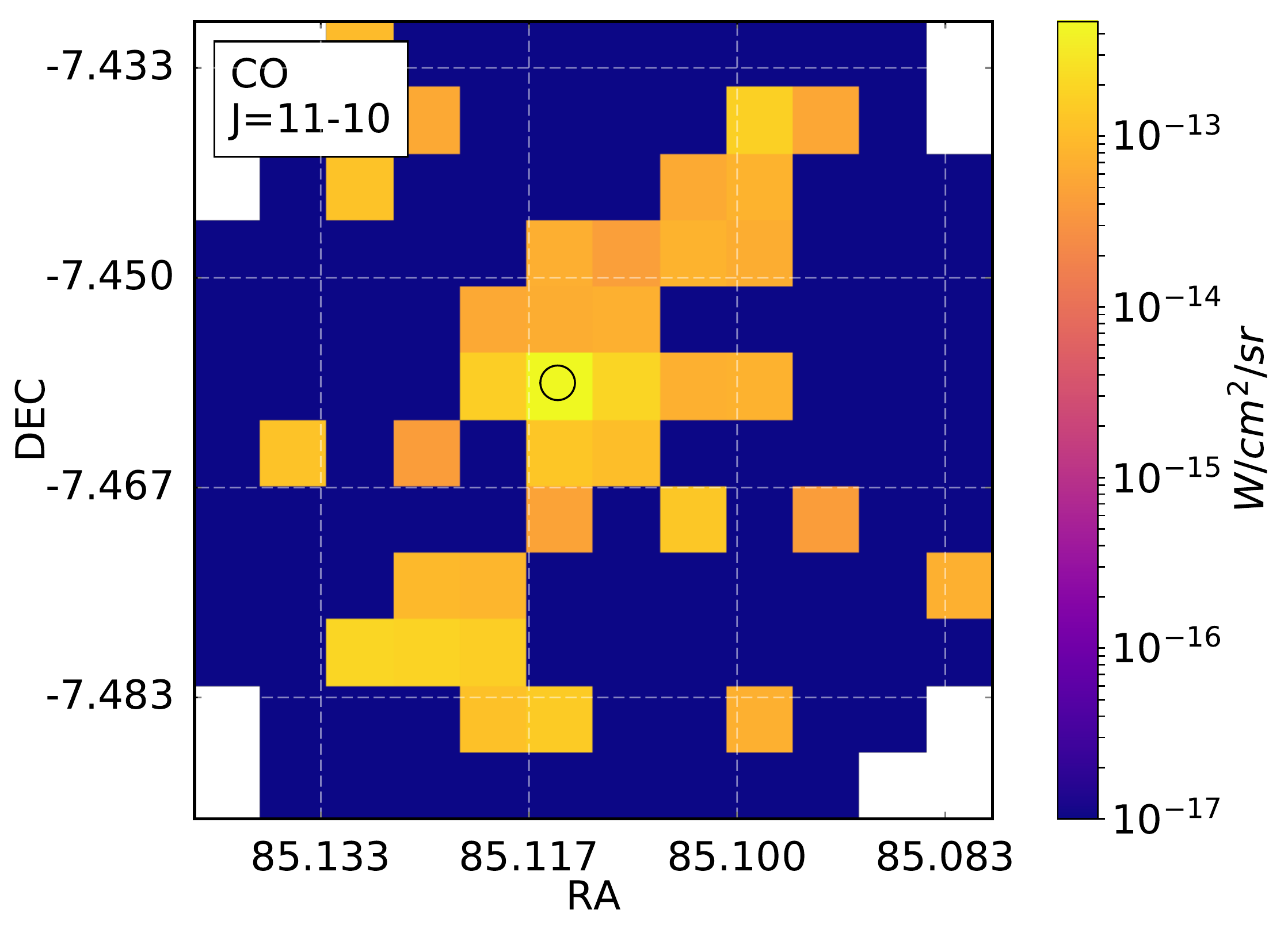}\hspace{-0.1cm}
\includegraphics[width=0.33\textwidth, trim={0cm 0 0cm 0}, clip]{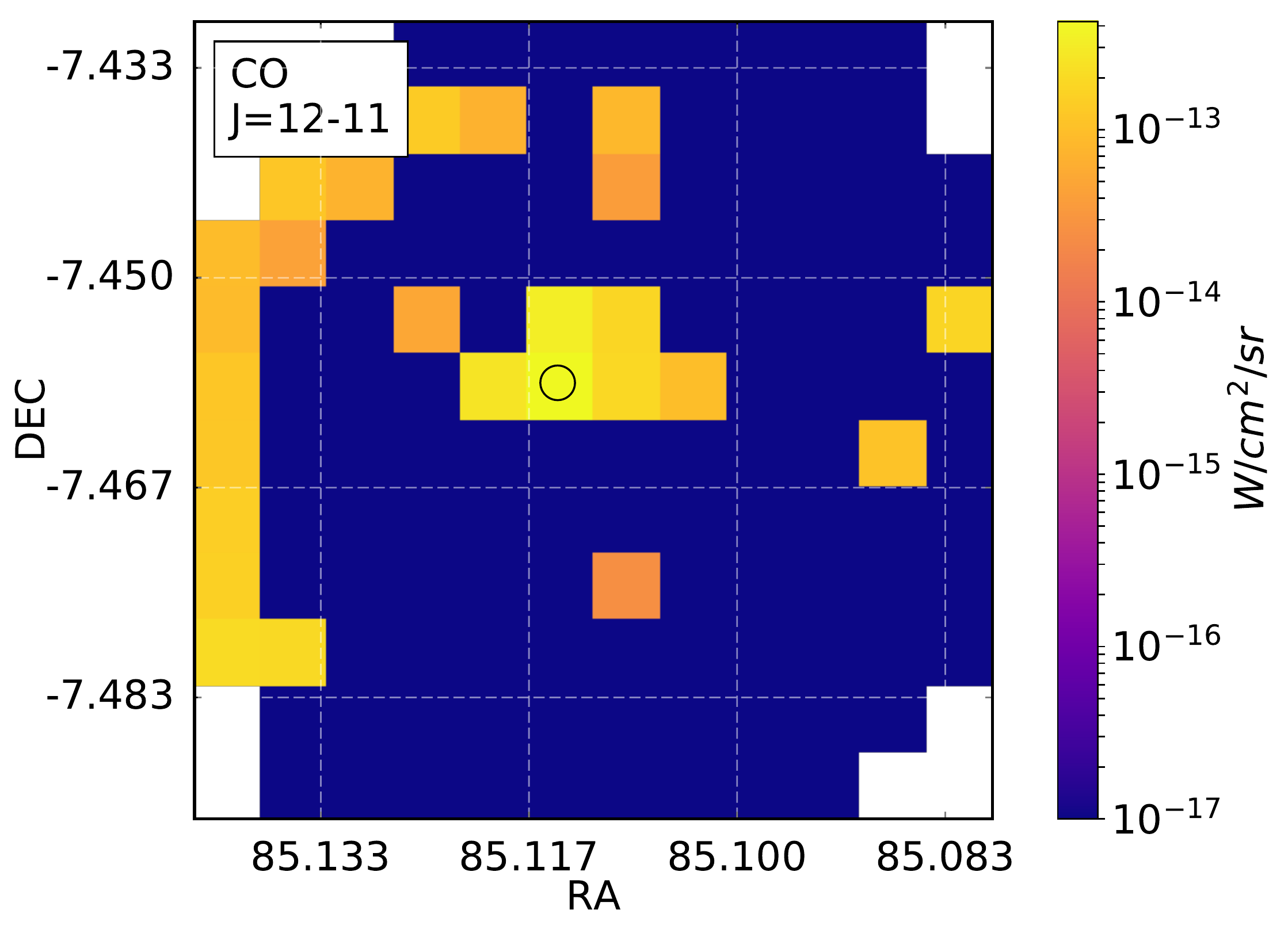}\hspace{-0.1cm}\\
\includegraphics[width=0.33\textwidth, trim={0cm 0 0cm 0}, clip]{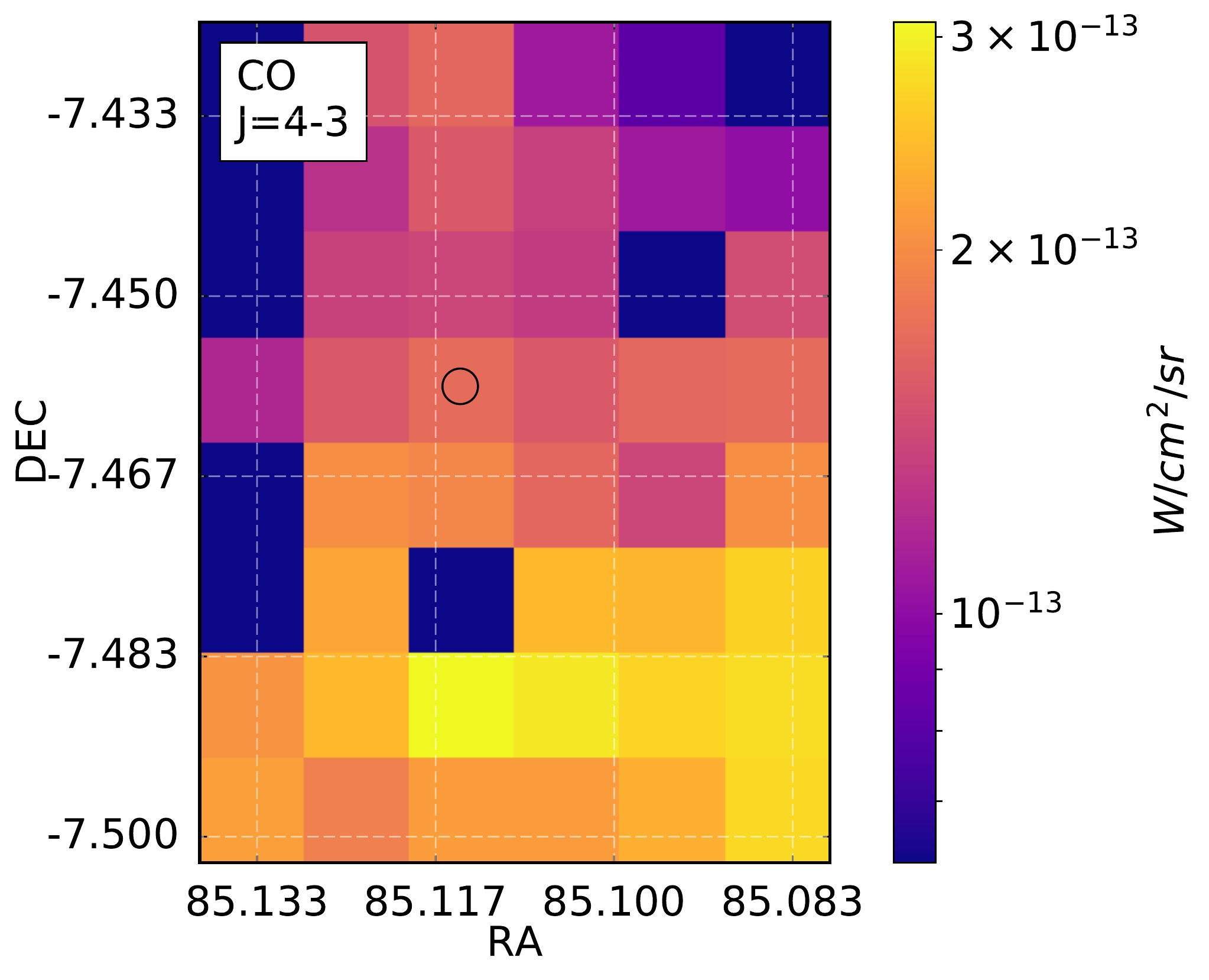}\hspace{-0.1cm}
\includegraphics[width=0.33\textwidth, trim={0cm 0 0cm 0}, clip]{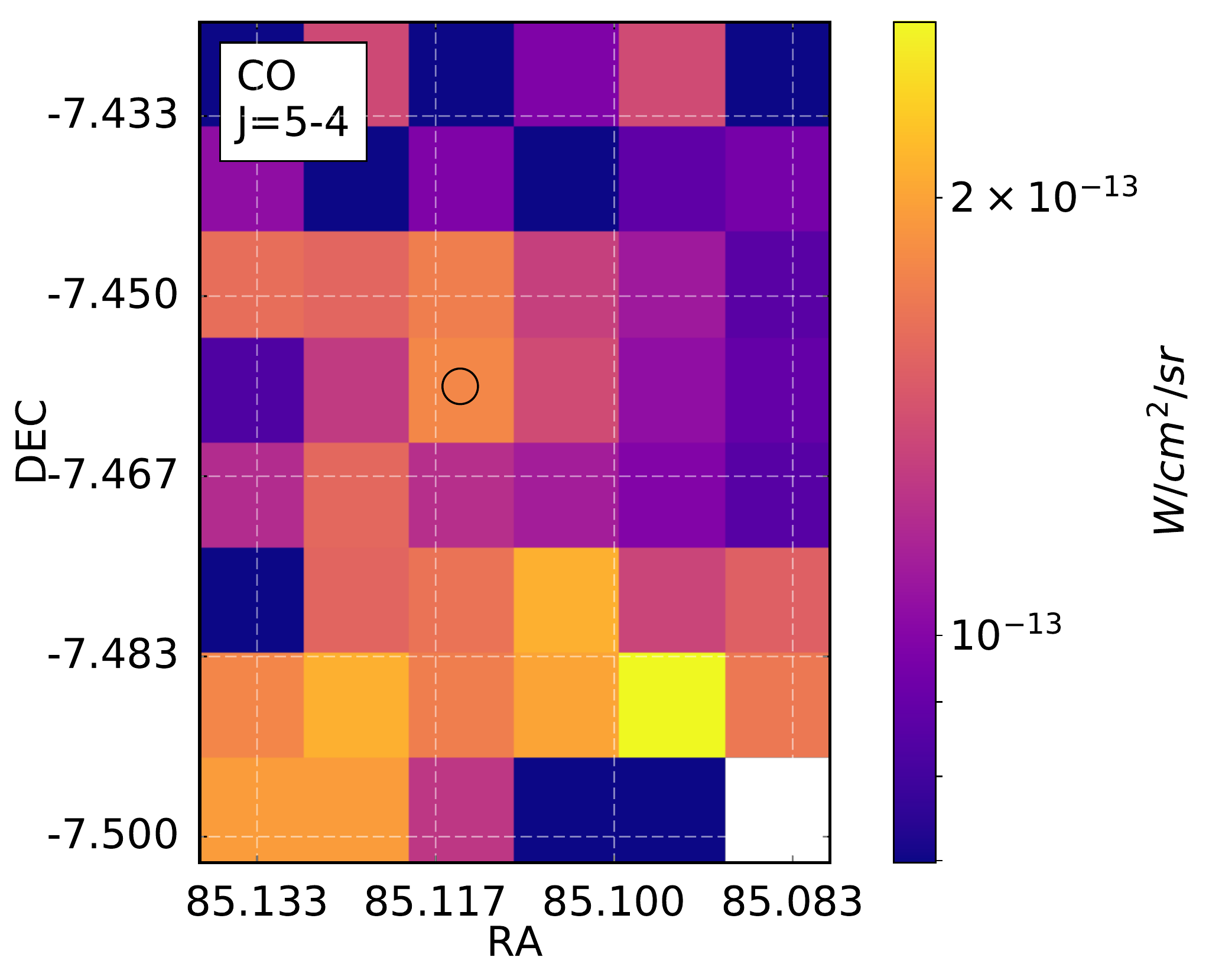}\hspace{-0.1cm}
\includegraphics[width=0.33\textwidth, trim={0cm 0 0cm 0}, clip]{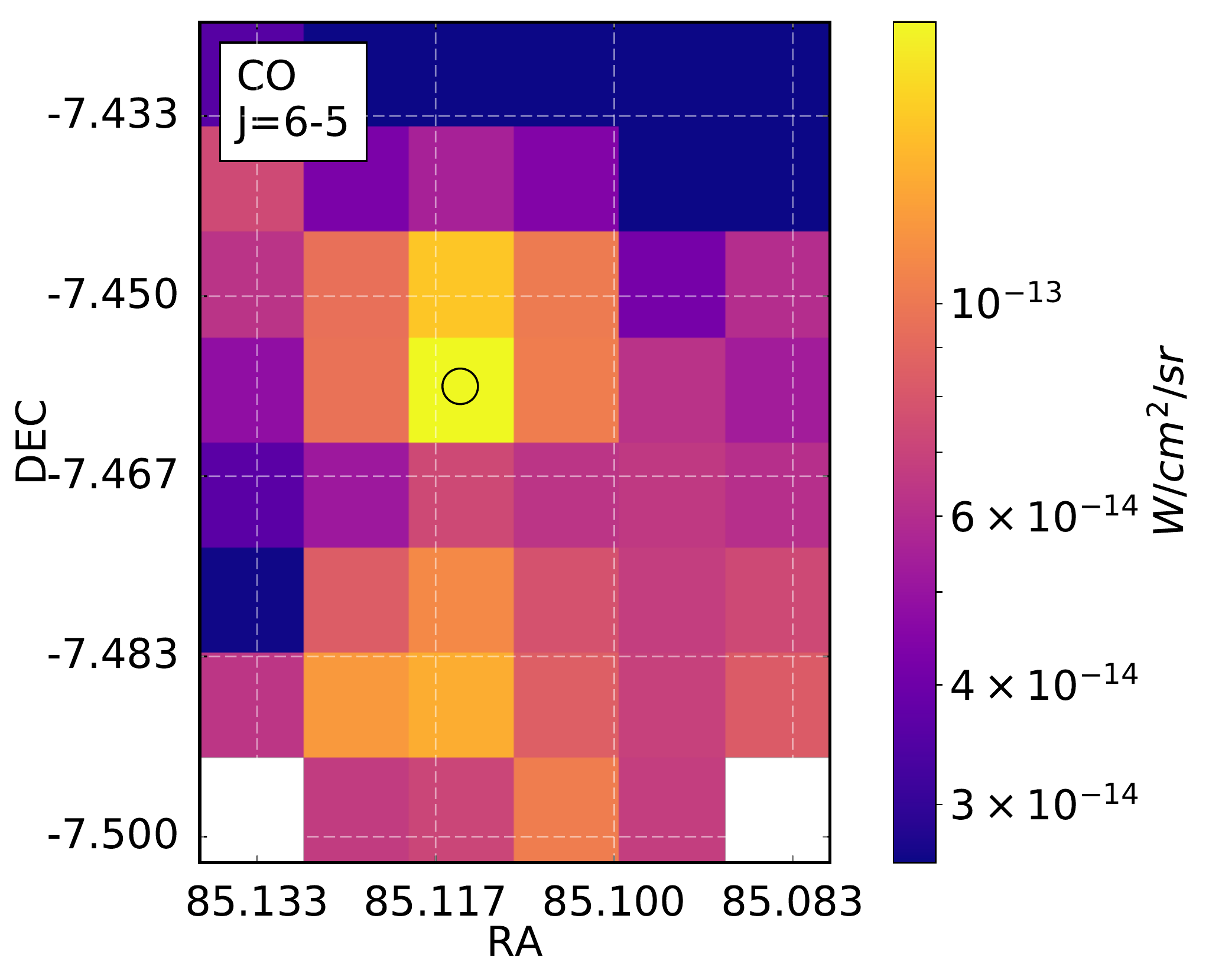}\hspace{-0.1cm}\\
\includegraphics[width=0.33\textwidth, trim={0cm 0 0cm 0}, clip]{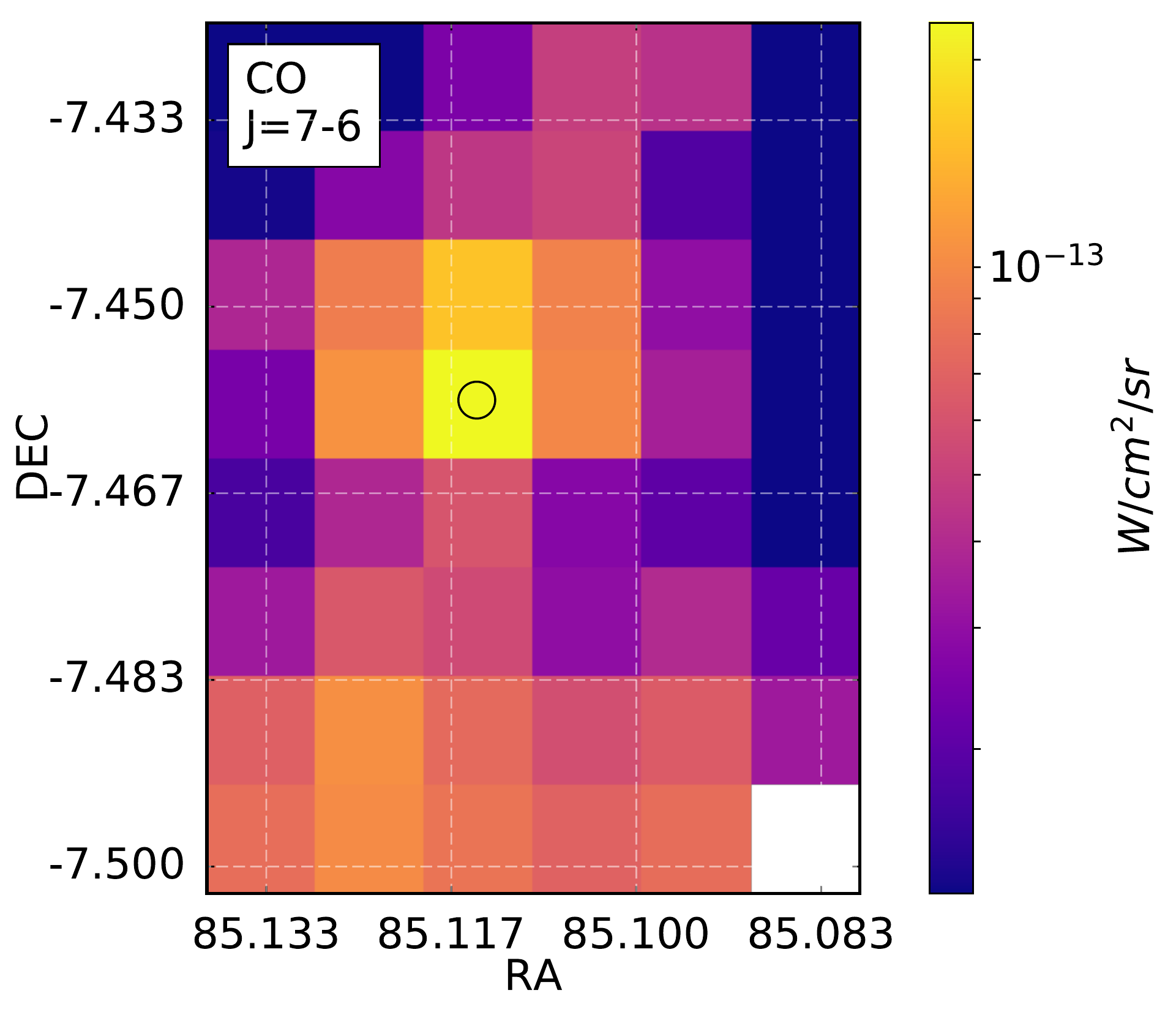}\hspace{-0.1cm}
\includegraphics[width=0.33\textwidth, trim={0cm 0 0cm 0}, clip]{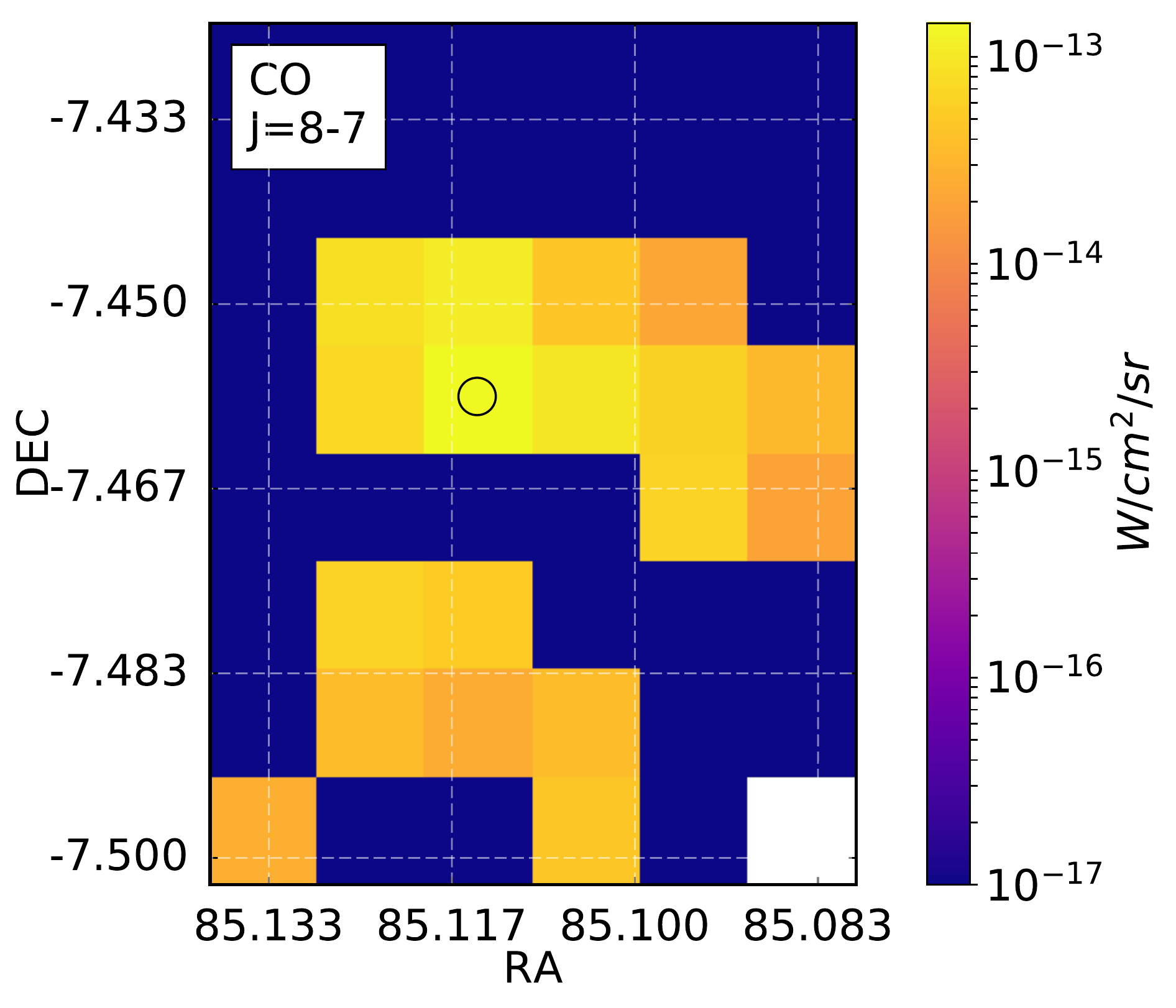}\hspace{-0.1cm}
\includegraphics[width=0.33\textwidth, trim={0cm 0 0cm 0}, clip]{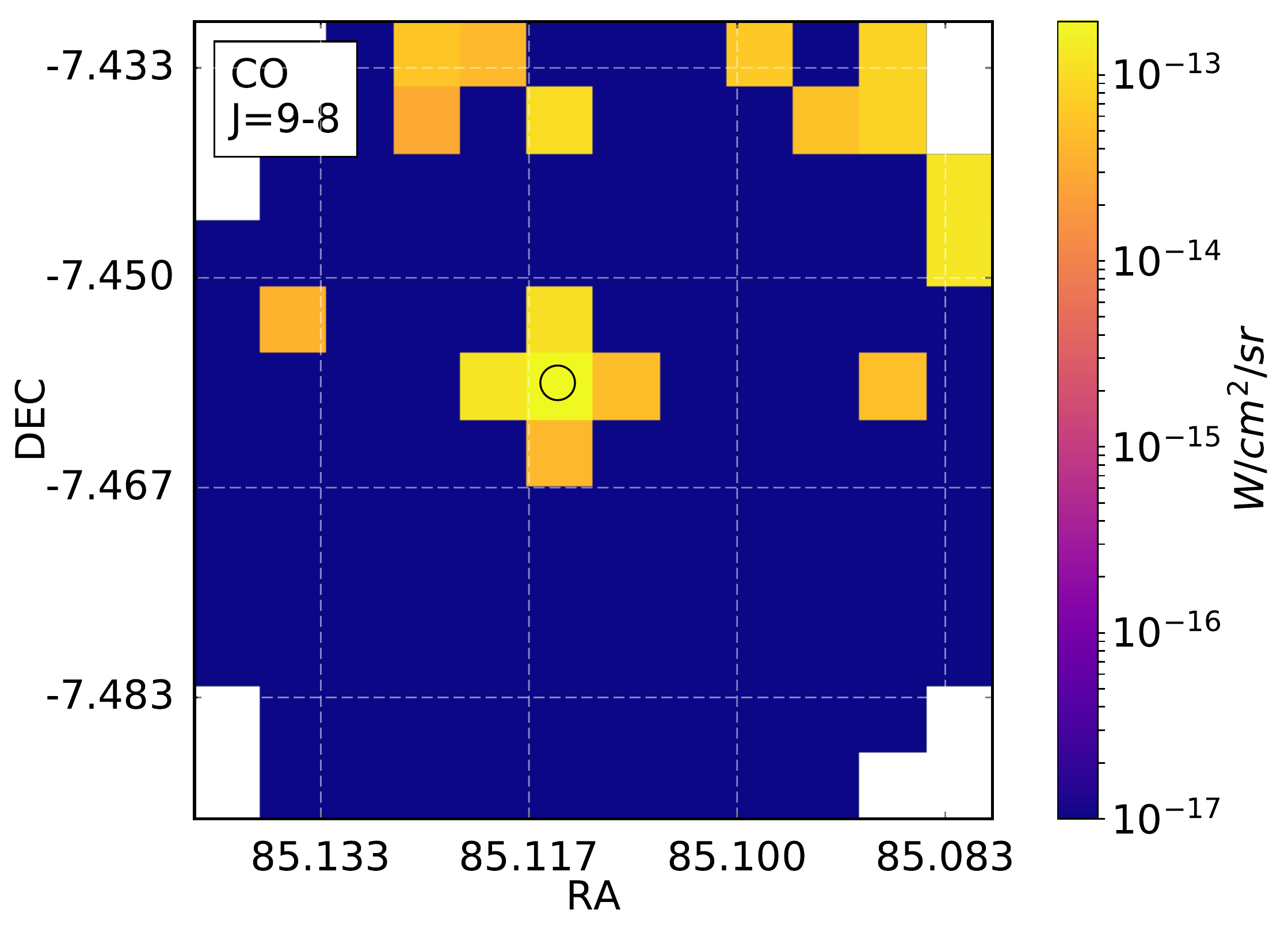}\hspace{-0.1cm}\\
\includegraphics[width=0.33\textwidth, trim={0cm 0 0cm 0}, clip]{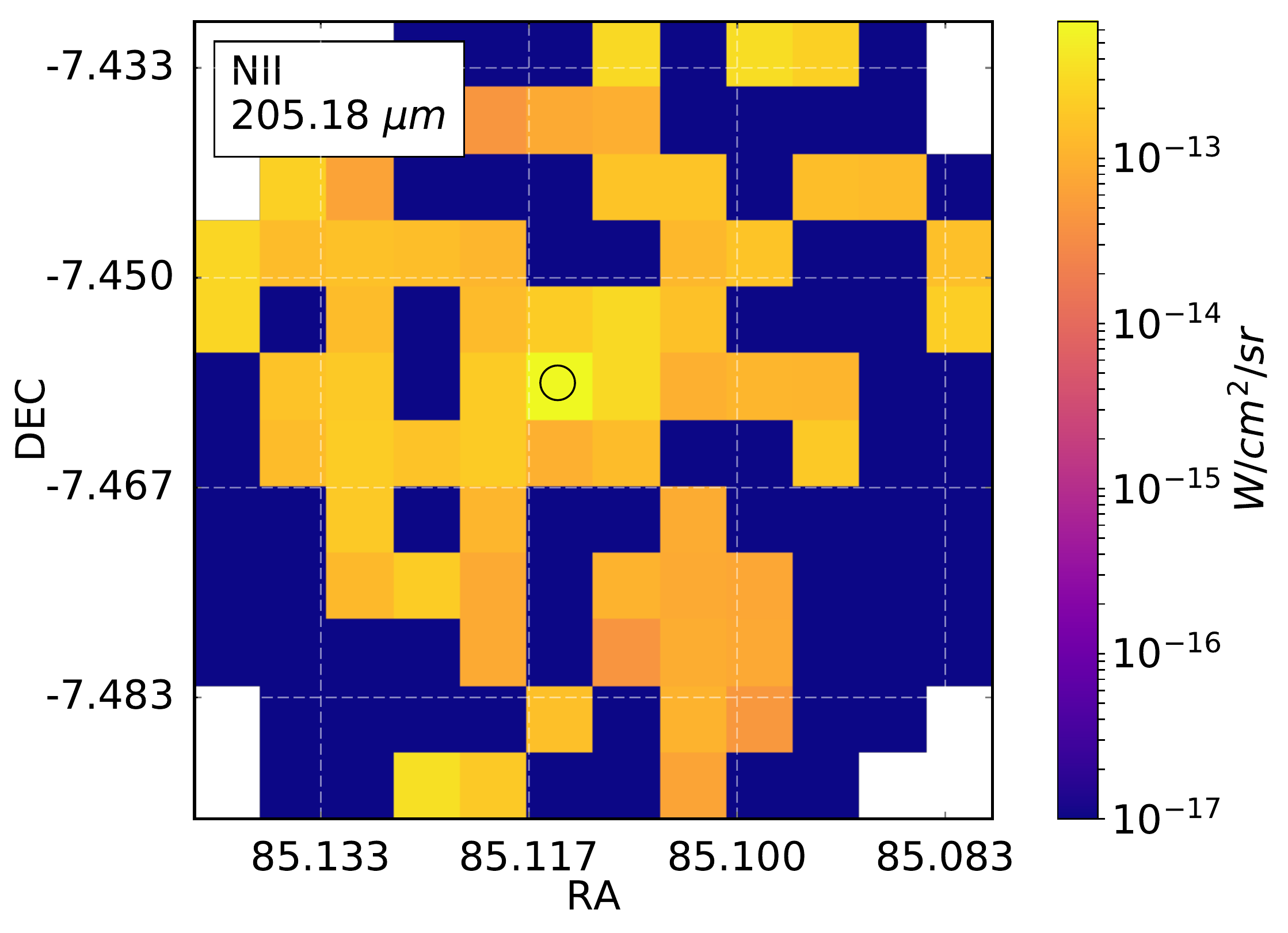}\hspace{-0.1cm}
\caption{
                \footnotesize
                Line maps of SPIRE with visible lines for Re 50 N IRS 1.
        }
\end{figure*}

\begin{figure*}
\includegraphics[width=0.33\textwidth, trim={0cm 0 0cm 0}, clip]{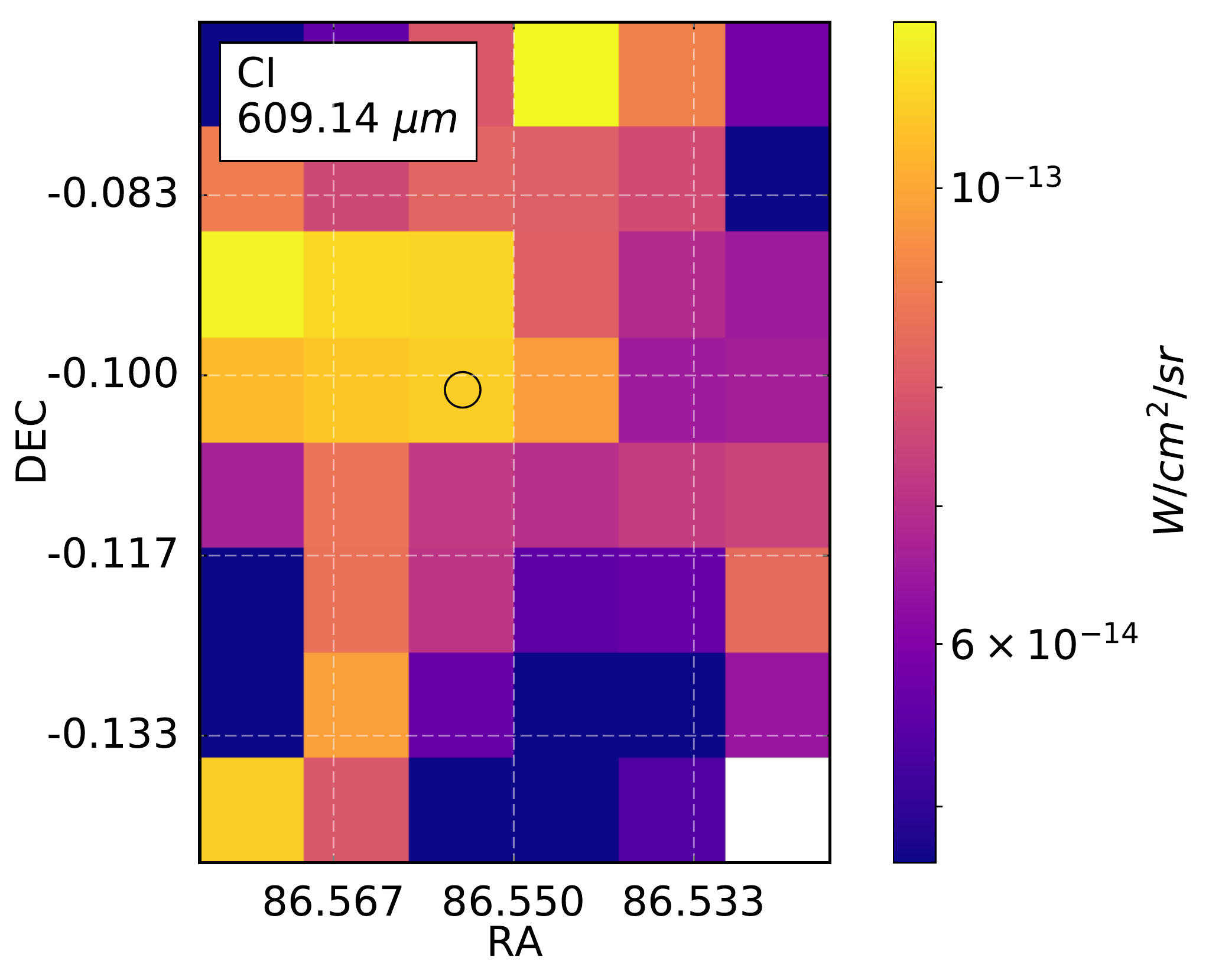}\hspace{-0.1cm}
\includegraphics[width=0.33\textwidth, trim={0cm 0 0cm 0}, clip]{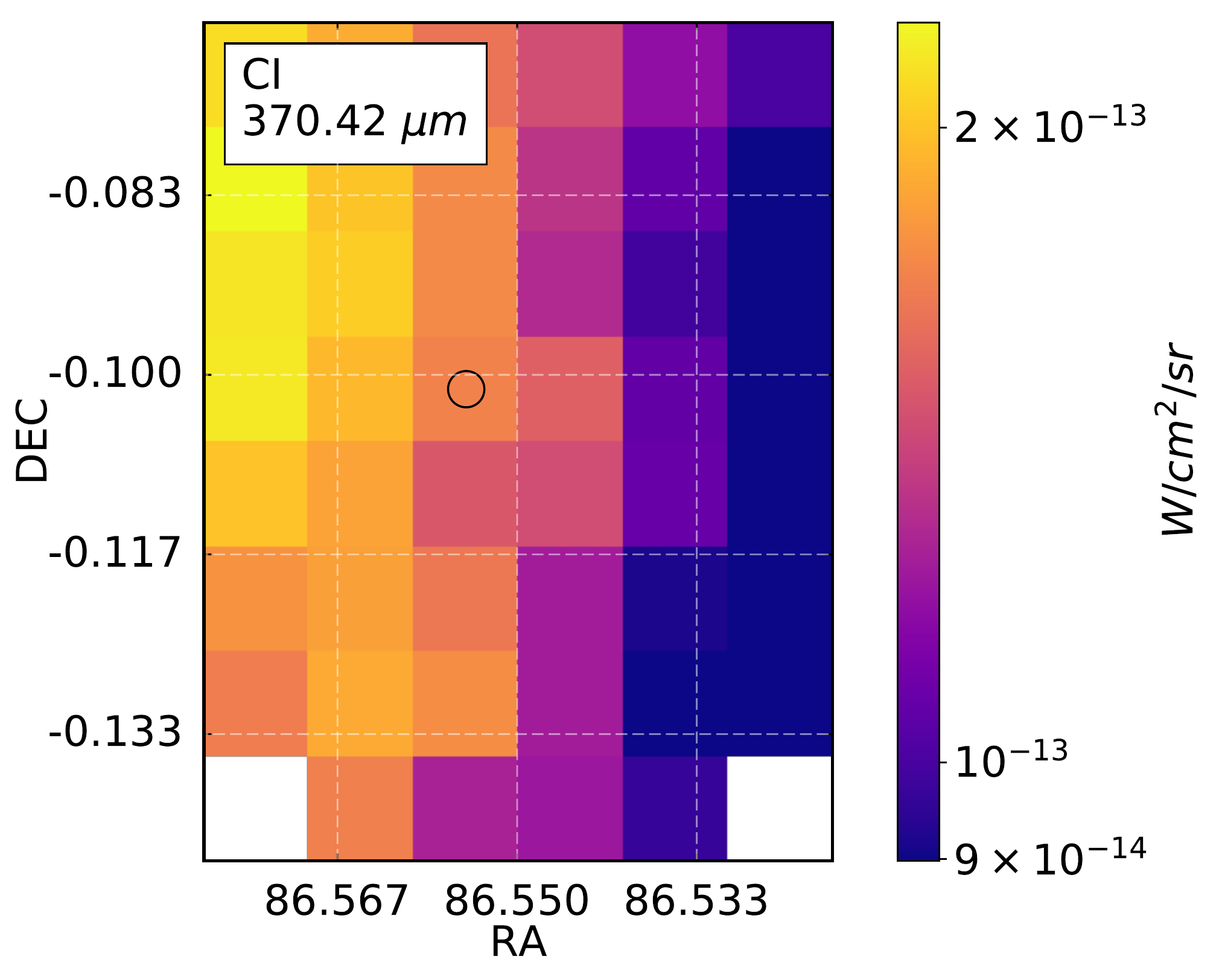}\hspace{-0.1cm}
\includegraphics[width=0.33\textwidth, trim={0cm 0 0cm 0}, clip]{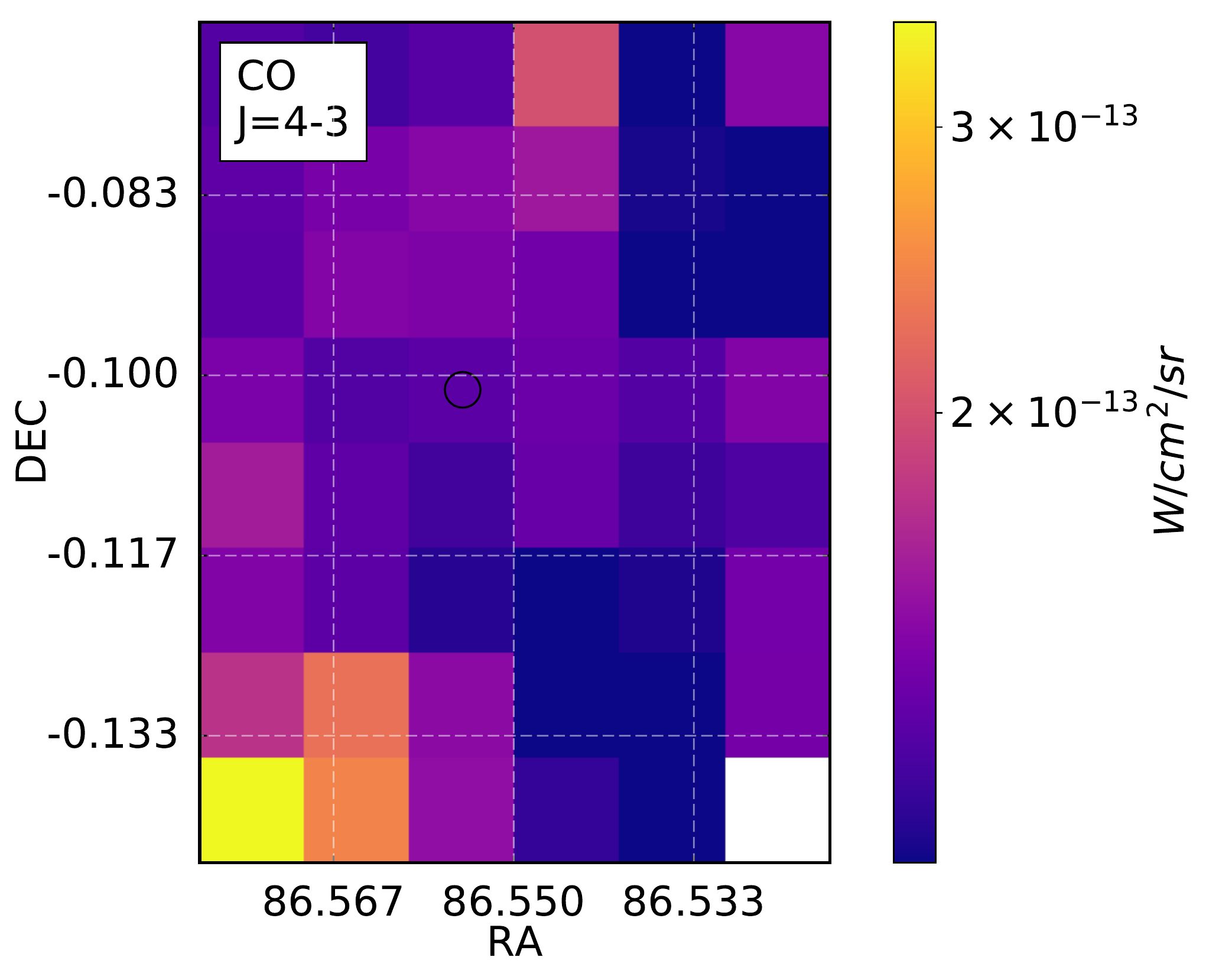}\hspace{-0.1cm}\\
\includegraphics[width=0.33\textwidth, trim={0cm 0 0cm 0}, clip]{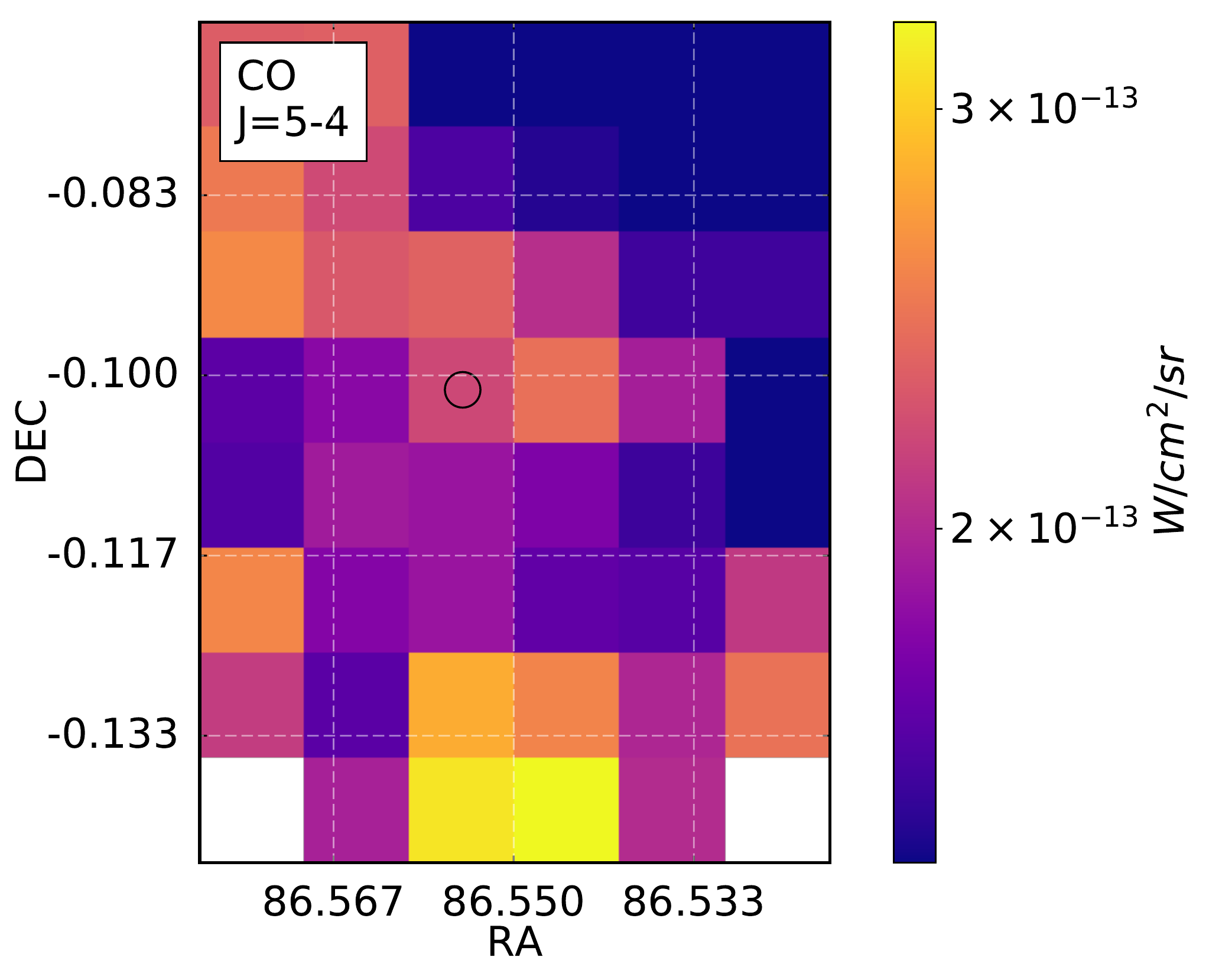}\hspace{-0.1cm}
\includegraphics[width=0.33\textwidth, trim={0cm 0 0cm 0}, clip]{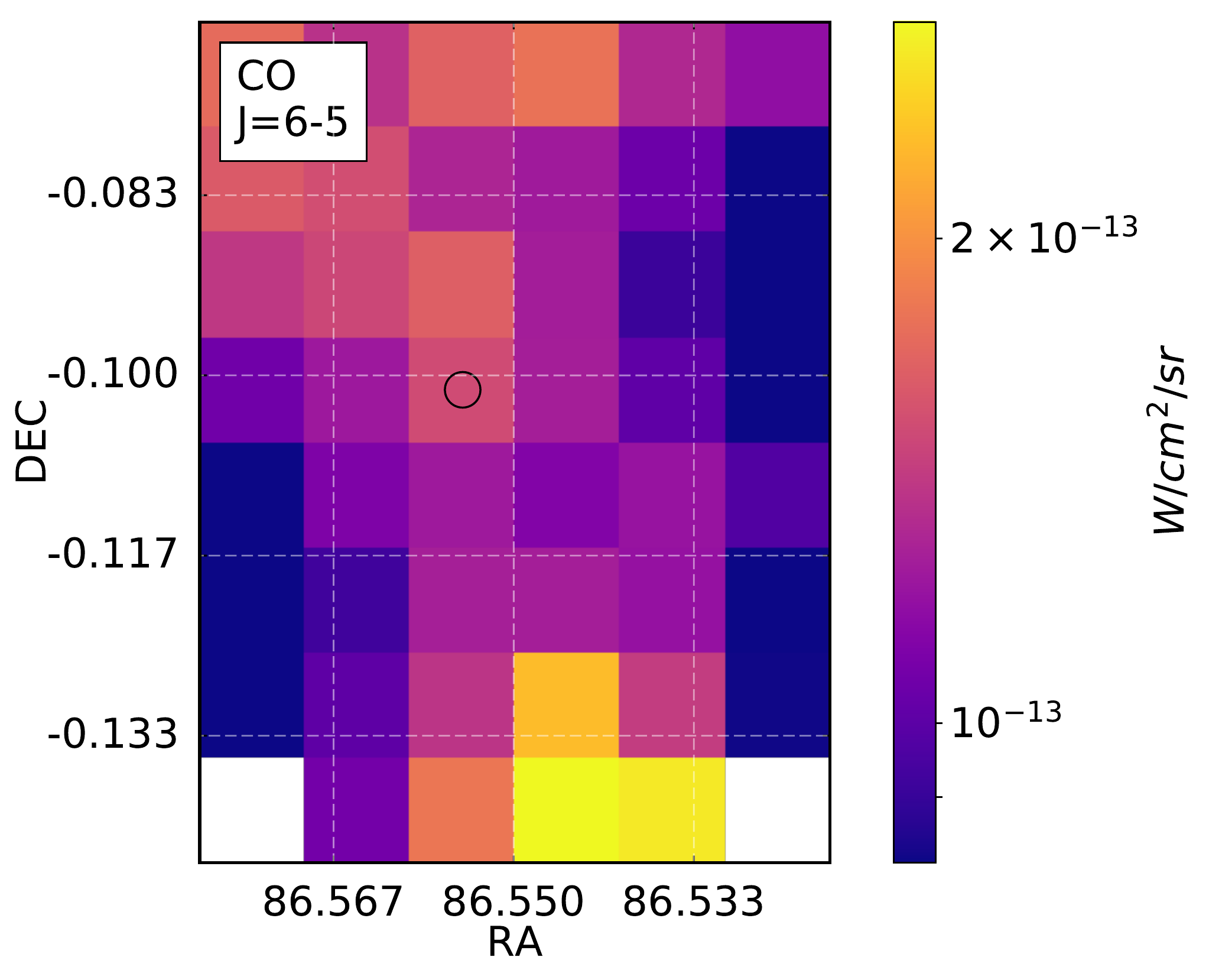}\hspace{-0.1cm}
\includegraphics[width=0.33\textwidth, trim={0cm 0 0cm 0}, clip]{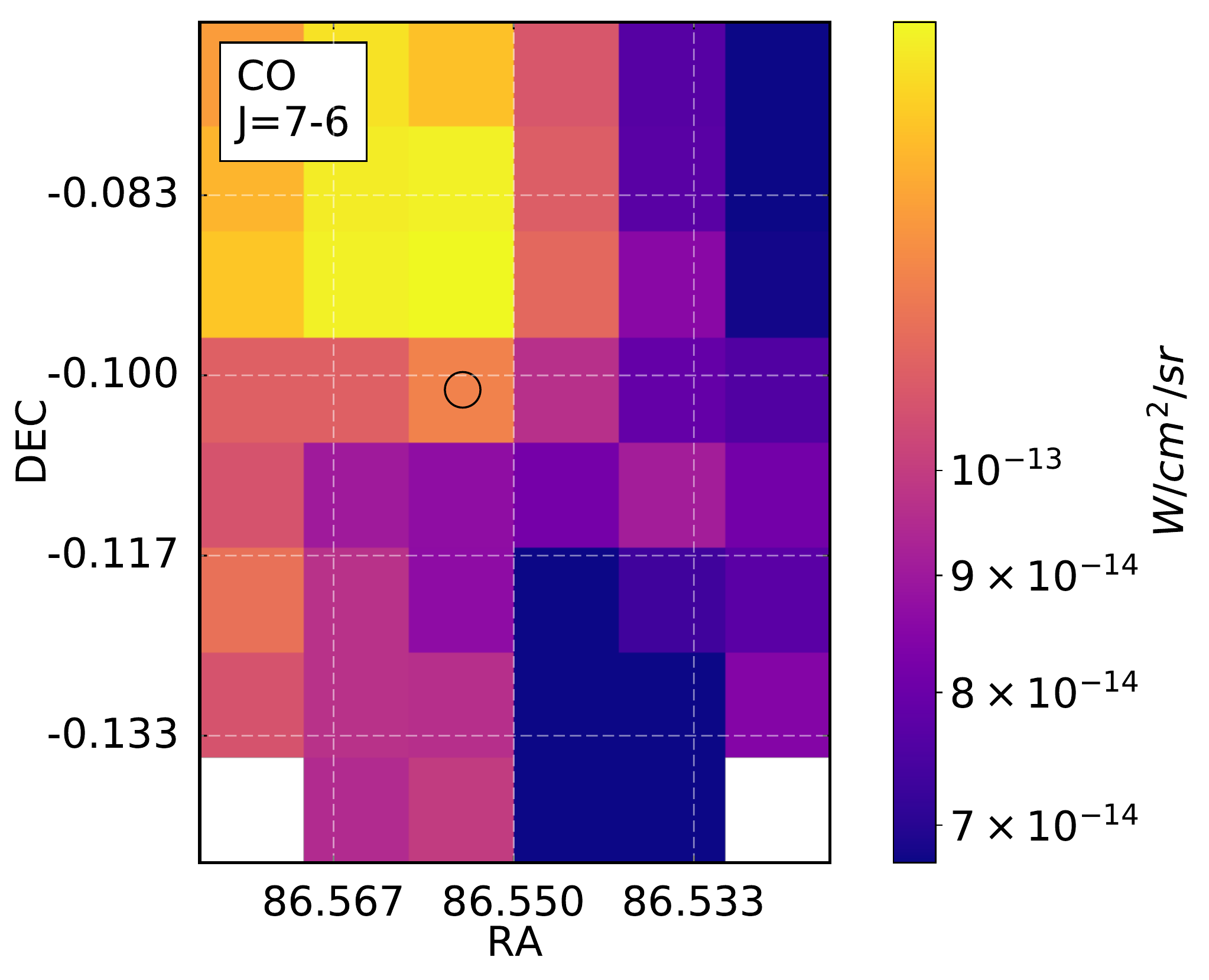}\hspace{-0.1cm}\\
\includegraphics[width=0.33\textwidth, trim={0cm 0 0cm 0}, clip]{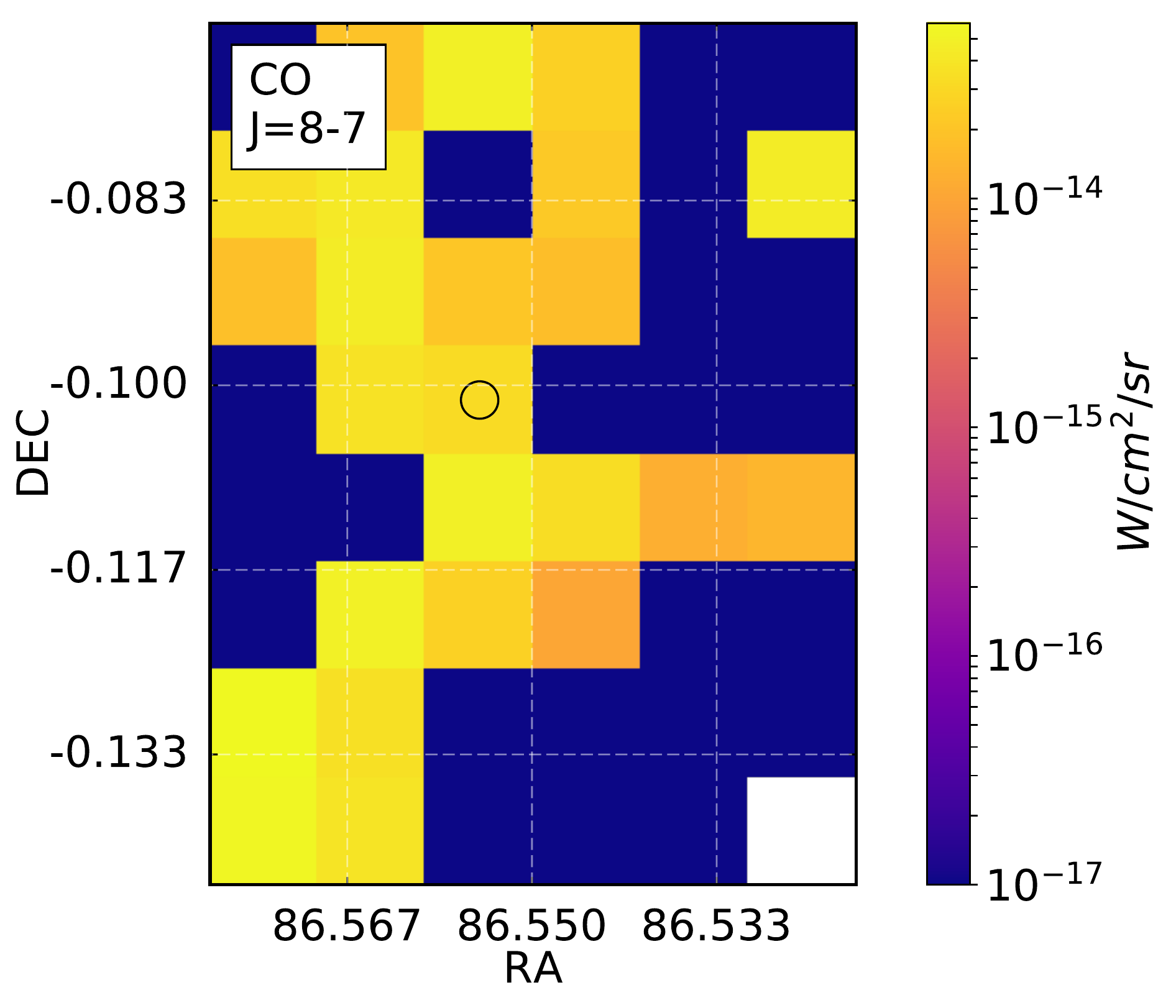}\hspace{-0.1cm}
\includegraphics[width=0.33\textwidth, trim={0cm 0 0cm 0}, clip]{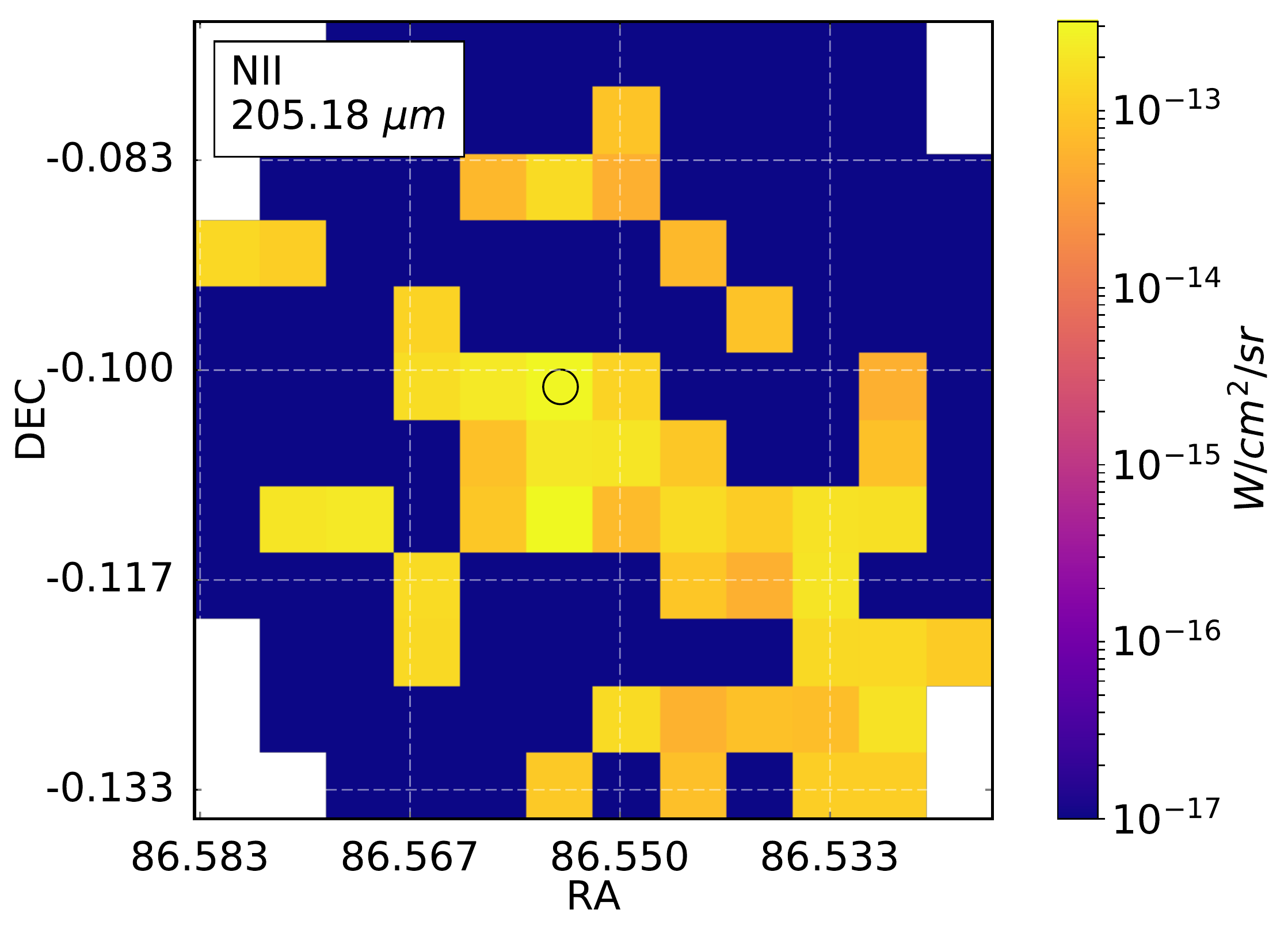}\hspace{-0.1cm}
\caption{
                \footnotesize
                Line maps of SPIRE with visible lines for V1647 Ori.
        }
\end{figure*}

\begin{figure*}
\includegraphics[width=0.33\textwidth, trim={0cm 0 0cm 0}, clip]{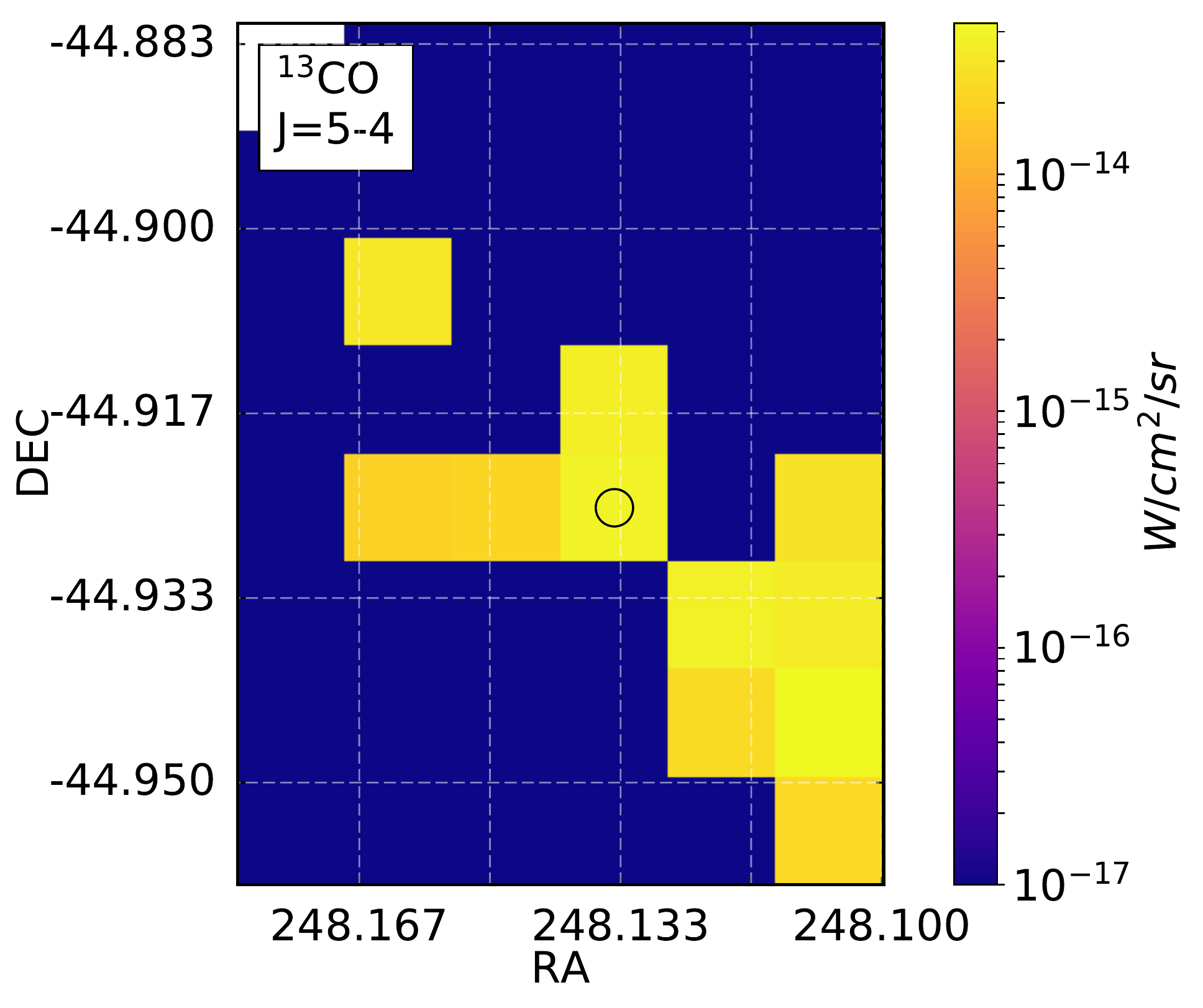}\hspace{-0.1cm}
\includegraphics[width=0.33\textwidth, trim={0cm 0 0cm 0}, clip]{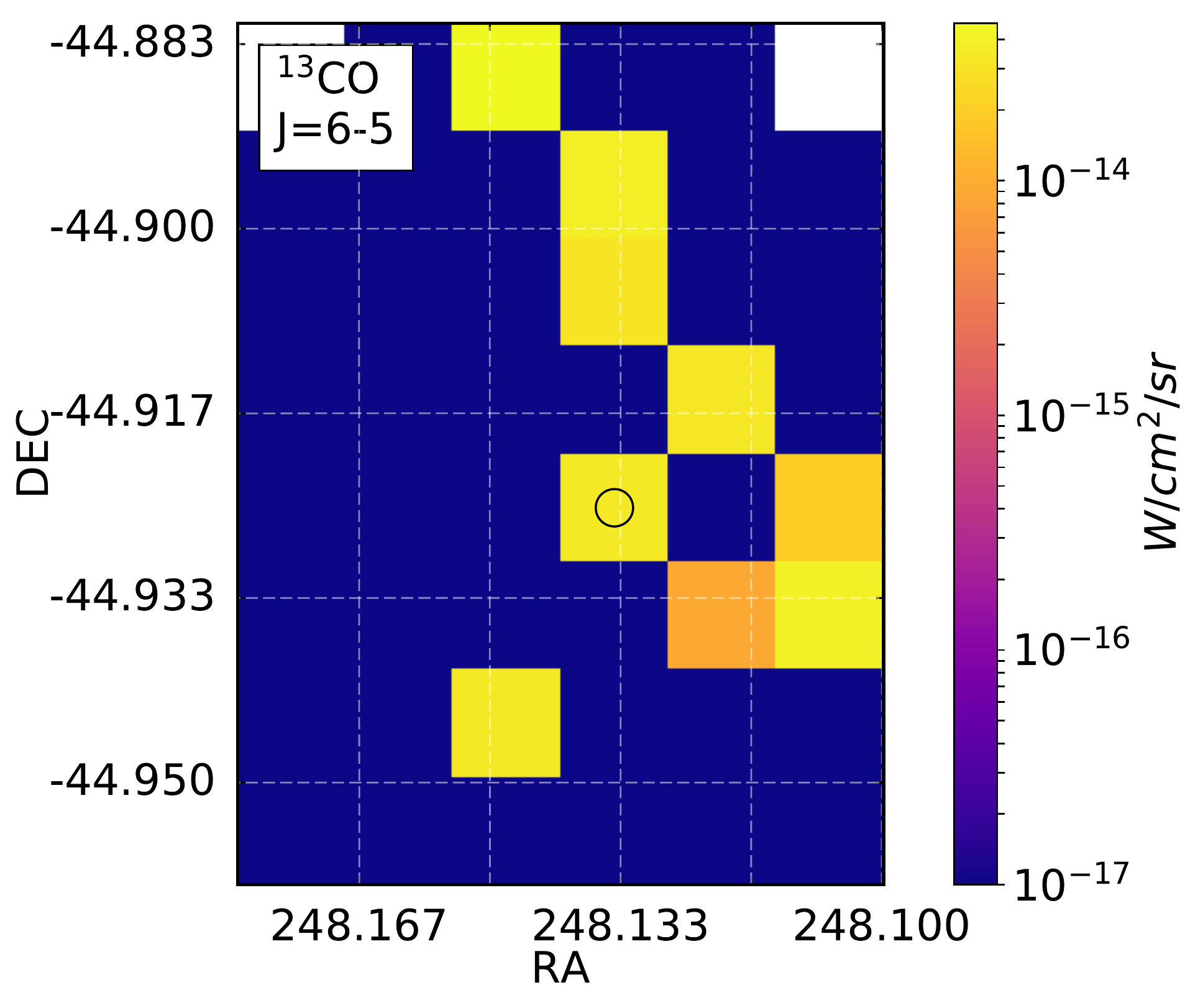}\hspace{-0.1cm}
\includegraphics[width=0.33\textwidth, trim={0cm 0 0cm 0}, clip]{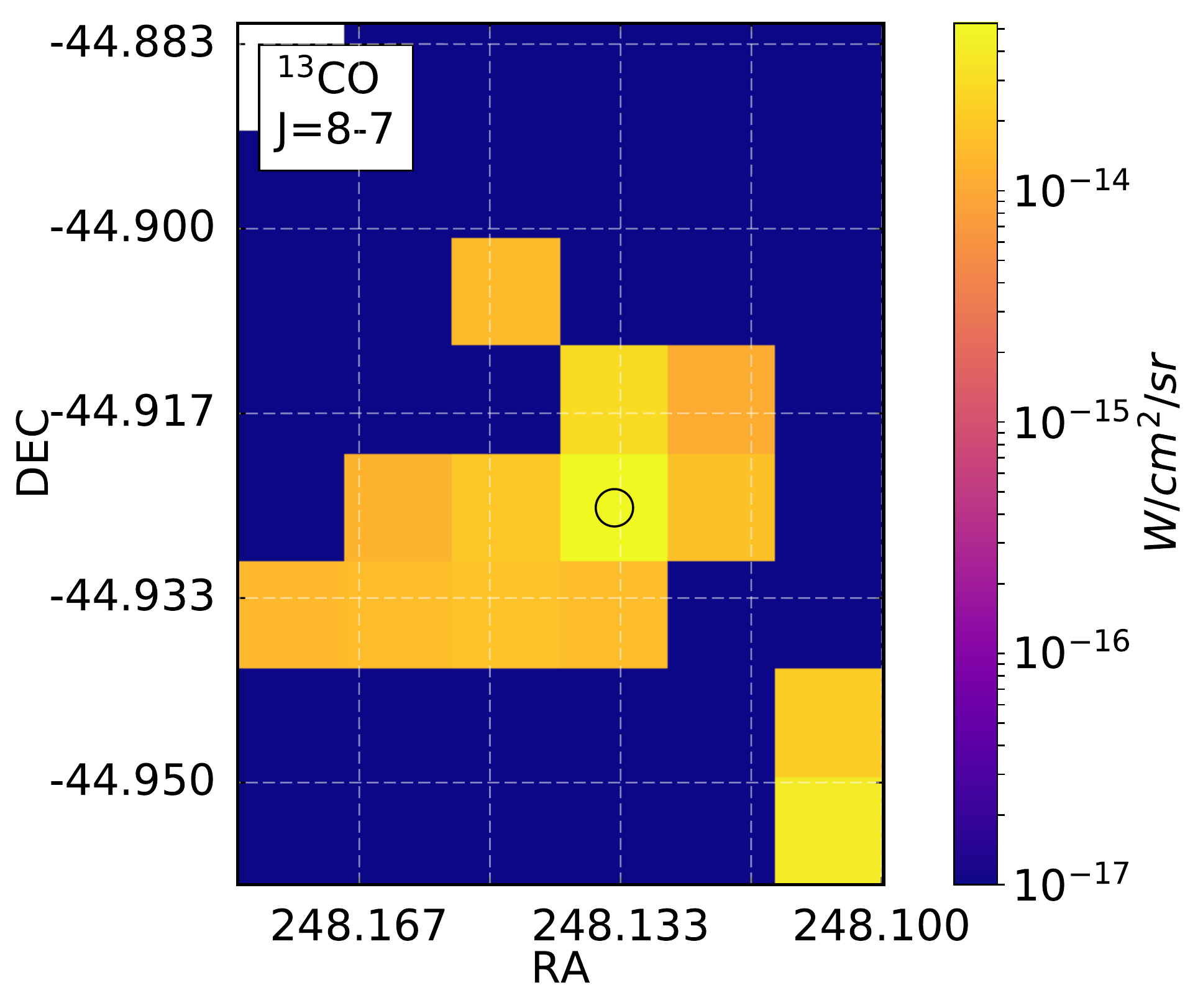}\hspace{-0.1cm}\\
\includegraphics[width=0.33\textwidth, trim={0cm 0 0cm 0}, clip]{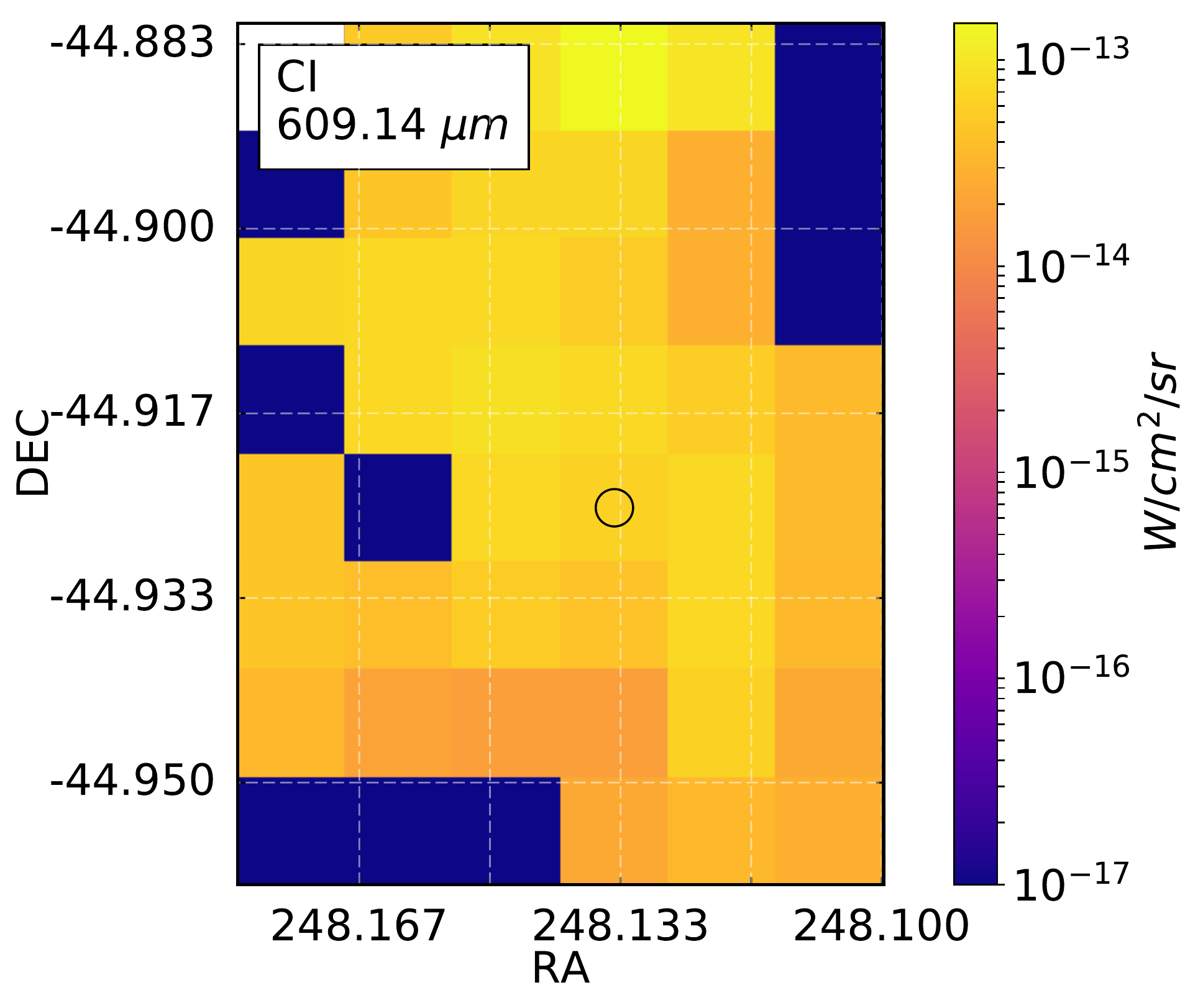}\hspace{-0.1cm}
\includegraphics[width=0.33\textwidth, trim={0cm 0 0cm 0}, clip]{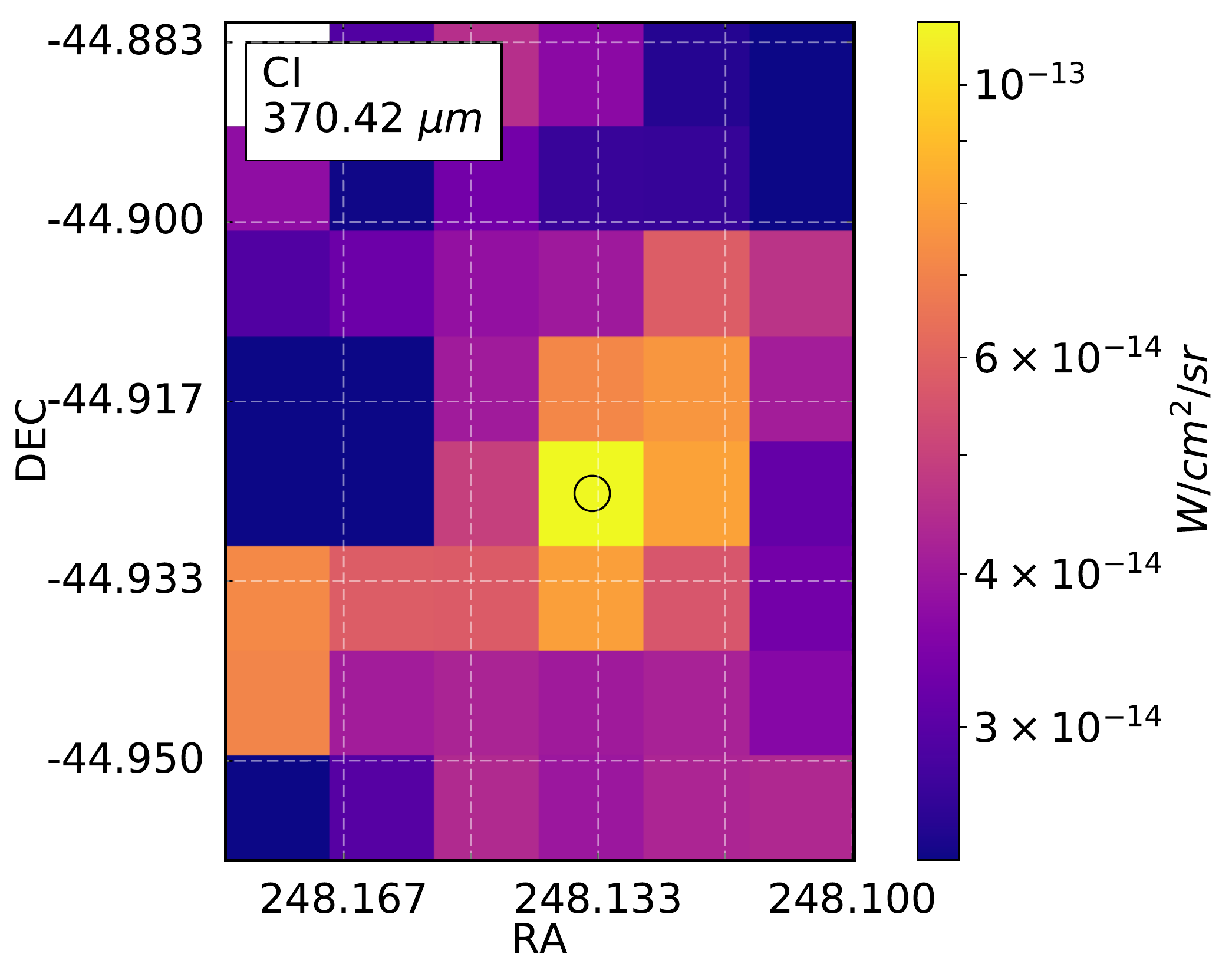}\hspace{-0.1cm}
\includegraphics[width=0.33\textwidth, trim={0cm 0 0cm 0}, clip]{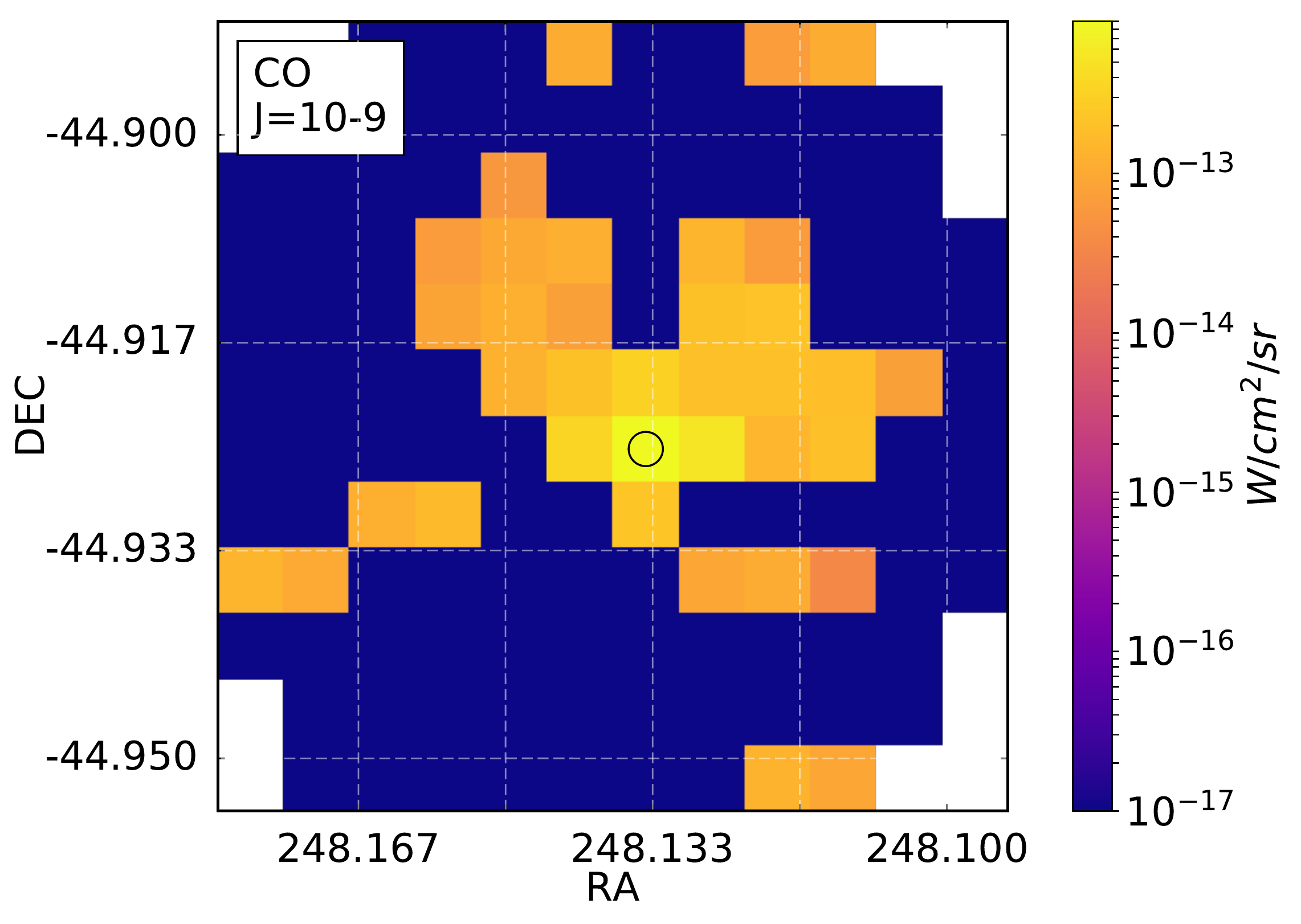}\hspace{-0.1cm}\\
\includegraphics[width=0.33\textwidth, trim={0cm 0 0cm 0}, clip]{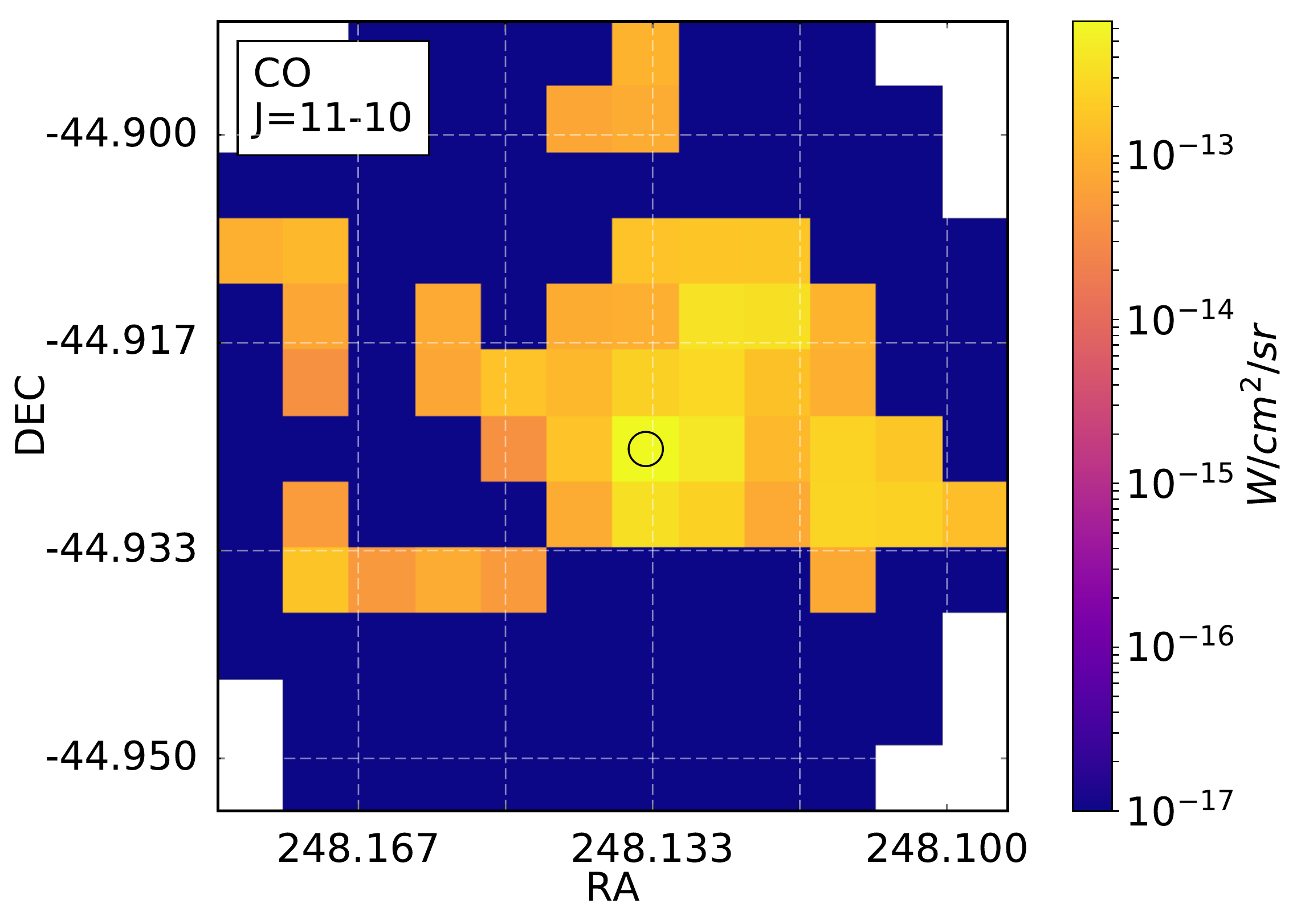}\hspace{-0.1cm}
\includegraphics[width=0.33\textwidth, trim={0cm 0 0cm 0}, clip]{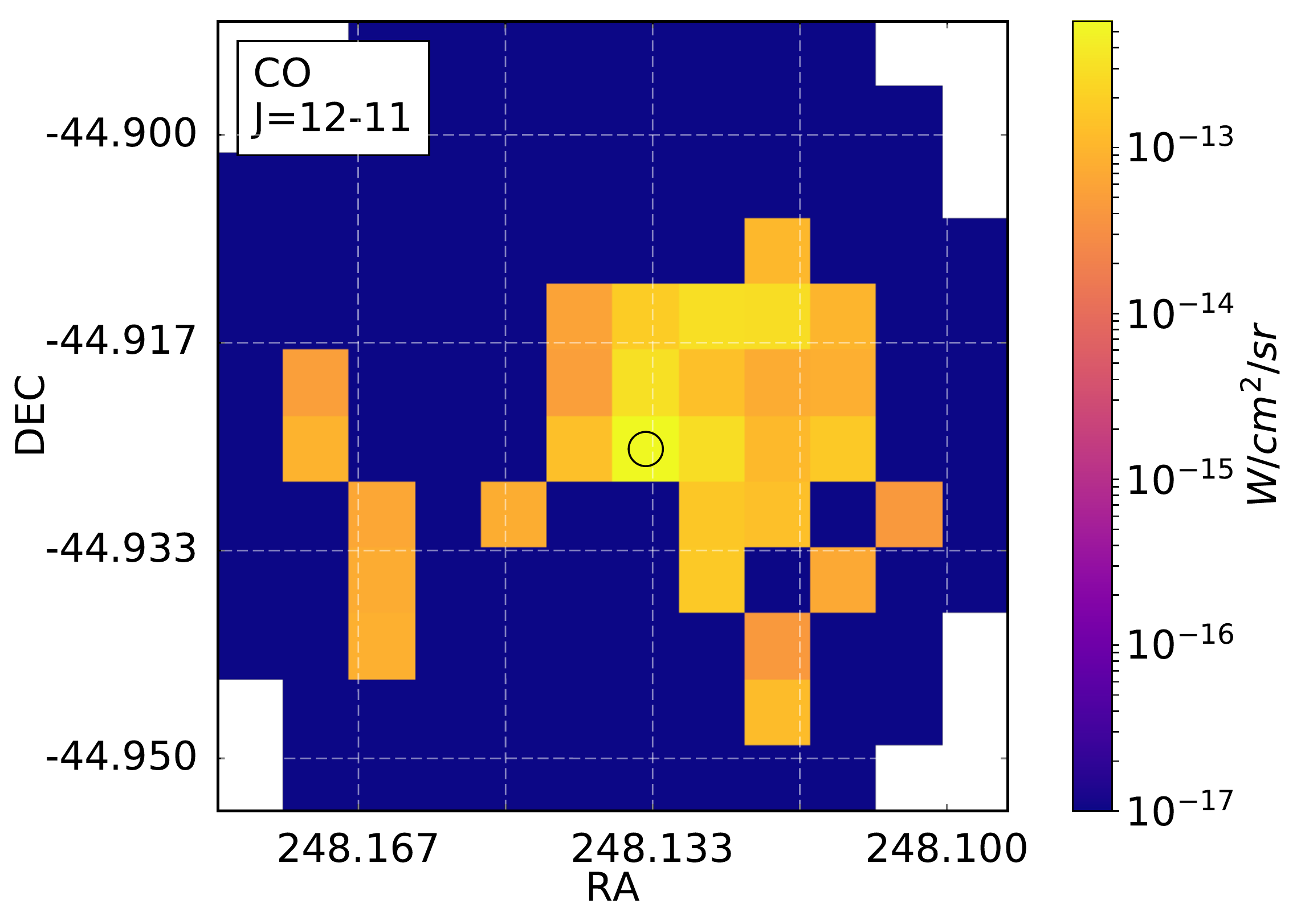}\hspace{-0.1cm}
\includegraphics[width=0.33\textwidth, trim={0cm 0 0cm 0}, clip]{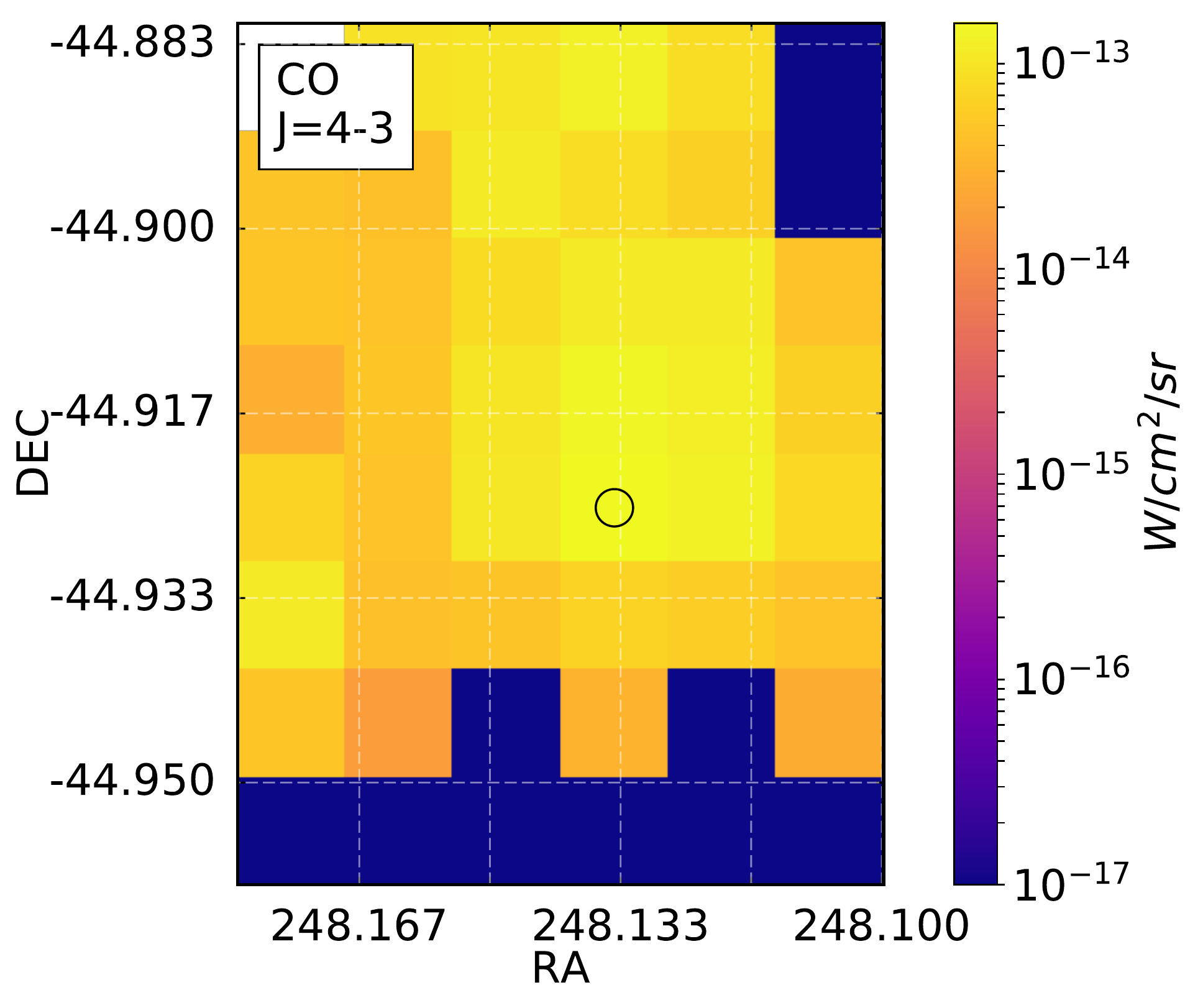}\hspace{-0.1cm}\\
\includegraphics[width=0.33\textwidth, trim={0cm 0 0cm 0}, clip]{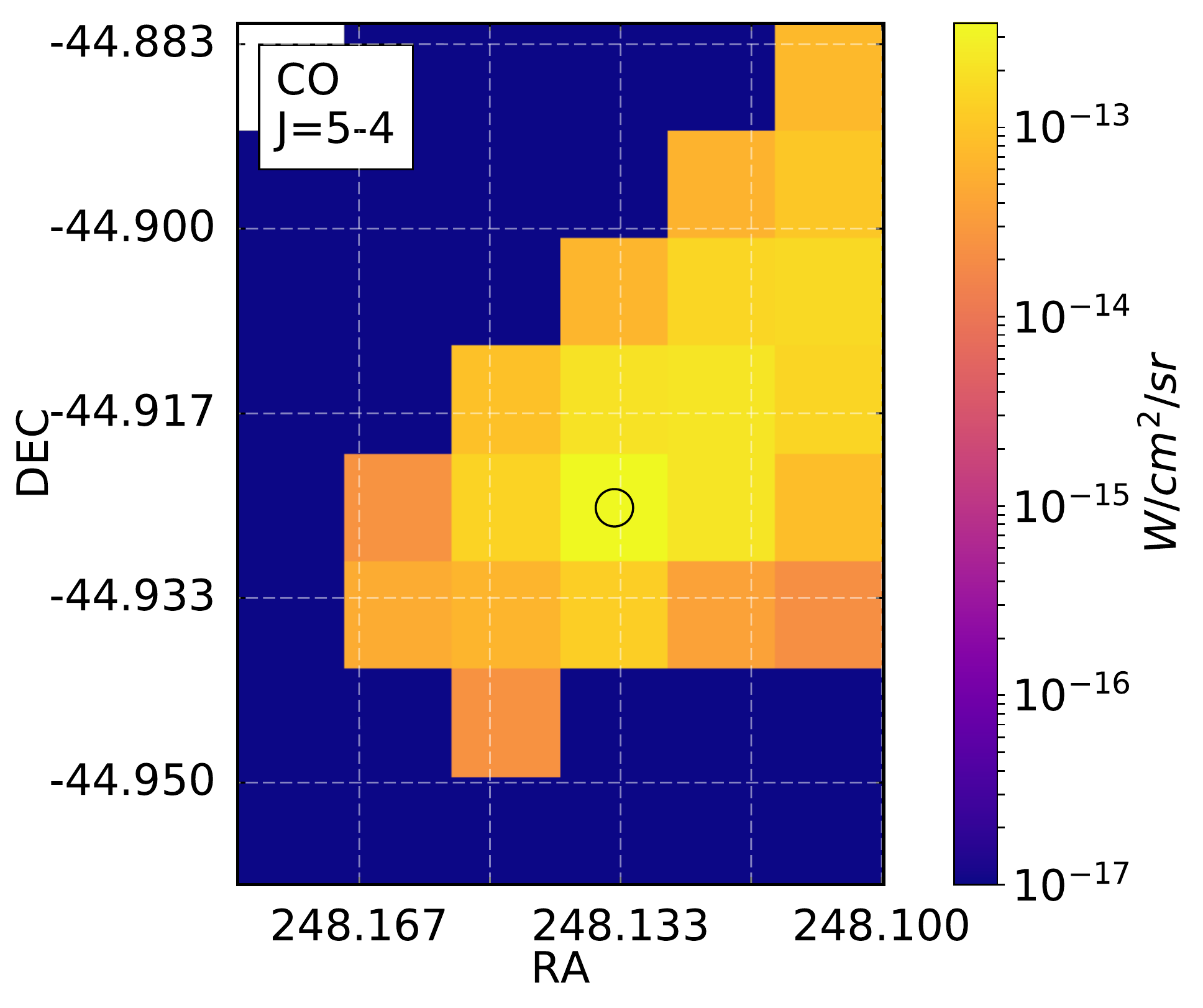}\hspace{-0.1cm}
\includegraphics[width=0.33\textwidth, trim={0cm 0 0cm 0}, clip]{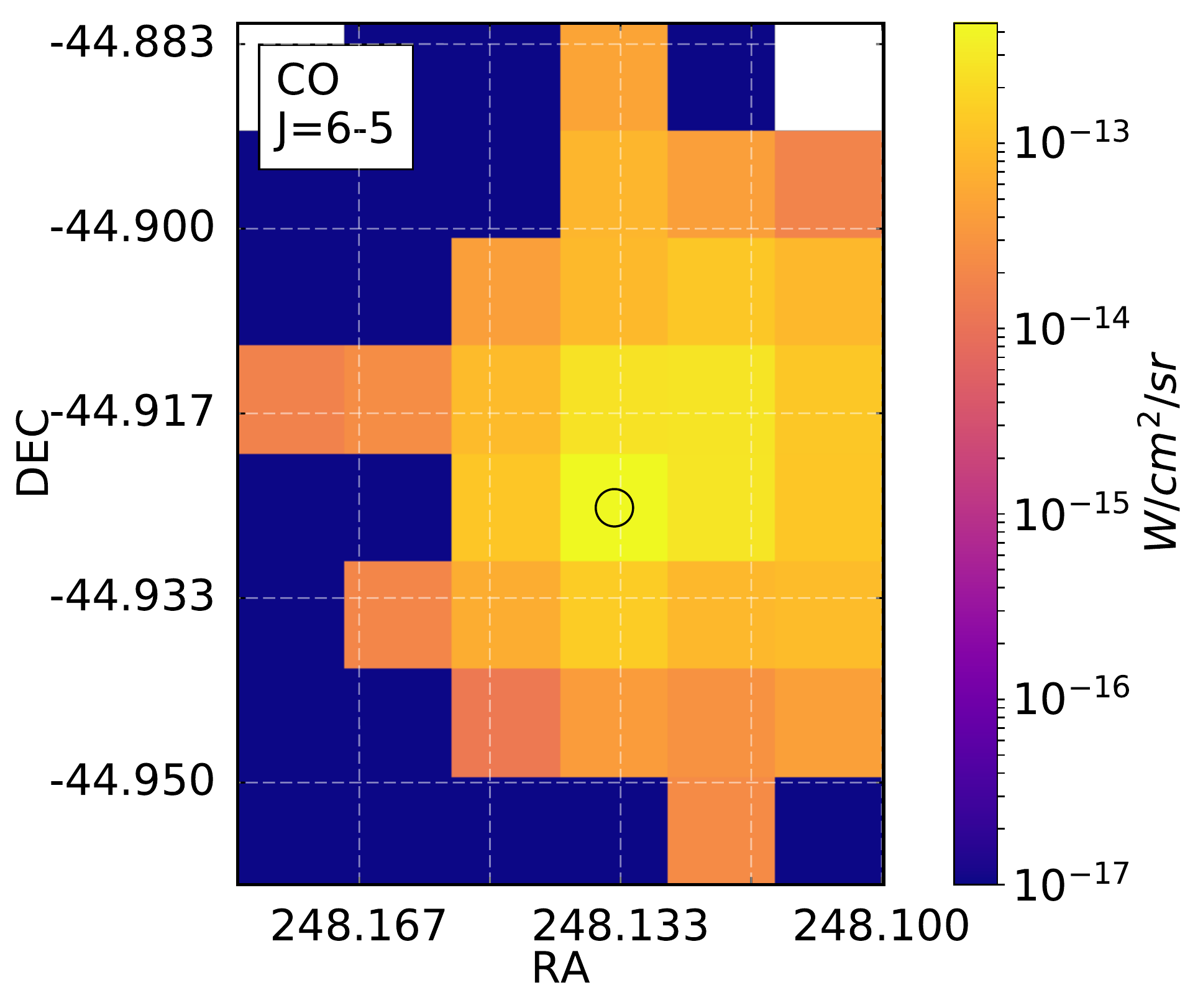}\hspace{-0.1cm}
\includegraphics[width=0.33\textwidth, trim={0cm 0 0cm 0}, clip]{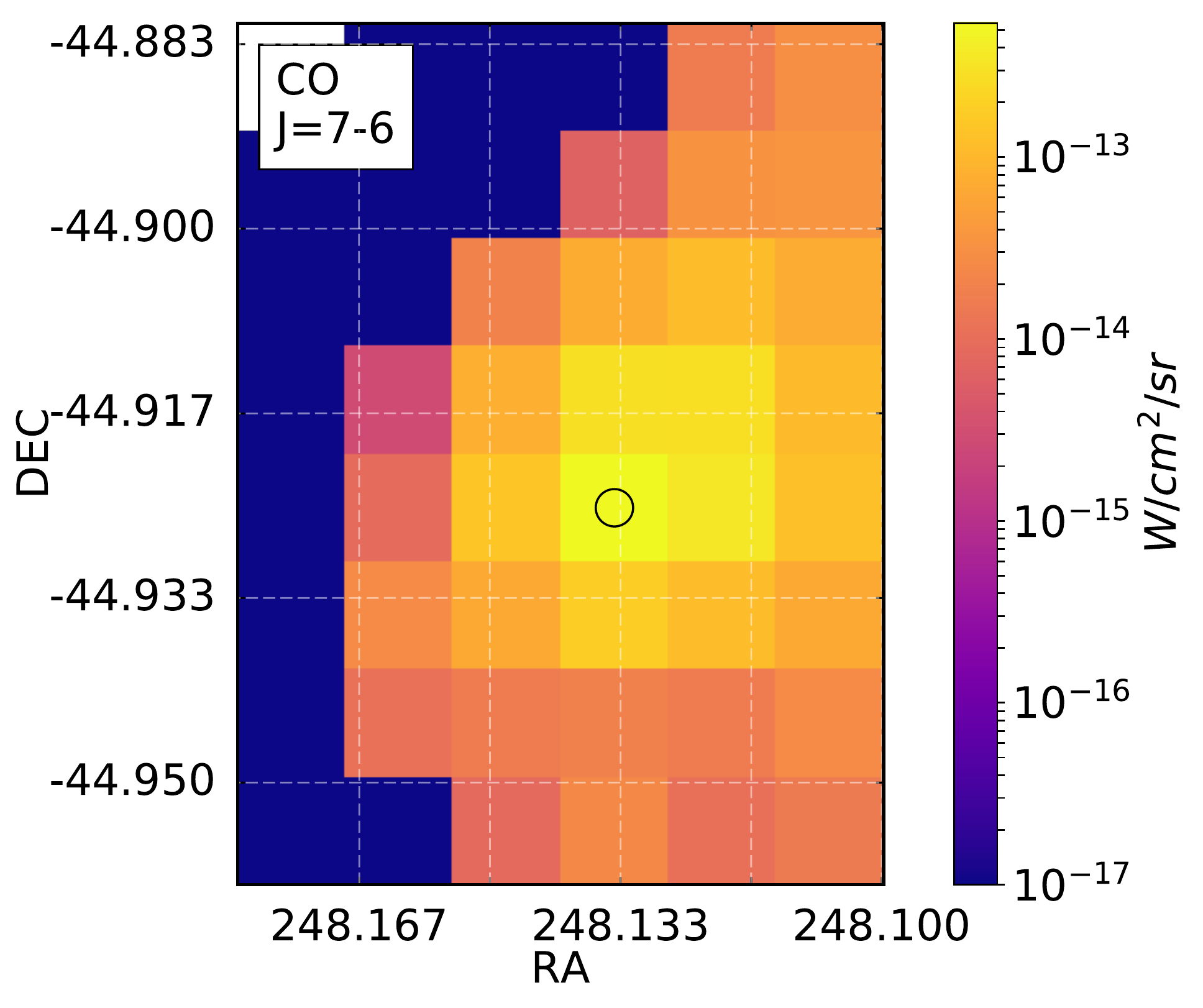}\hspace{-0.1cm}
\caption{
                \footnotesize
                Line maps of SPIRE with visible lines for V346 Nor, part 1.
        }
\end{figure*}

\begin{figure*}
\includegraphics[width=0.33\textwidth, trim={0cm 0 0cm 0}, clip]{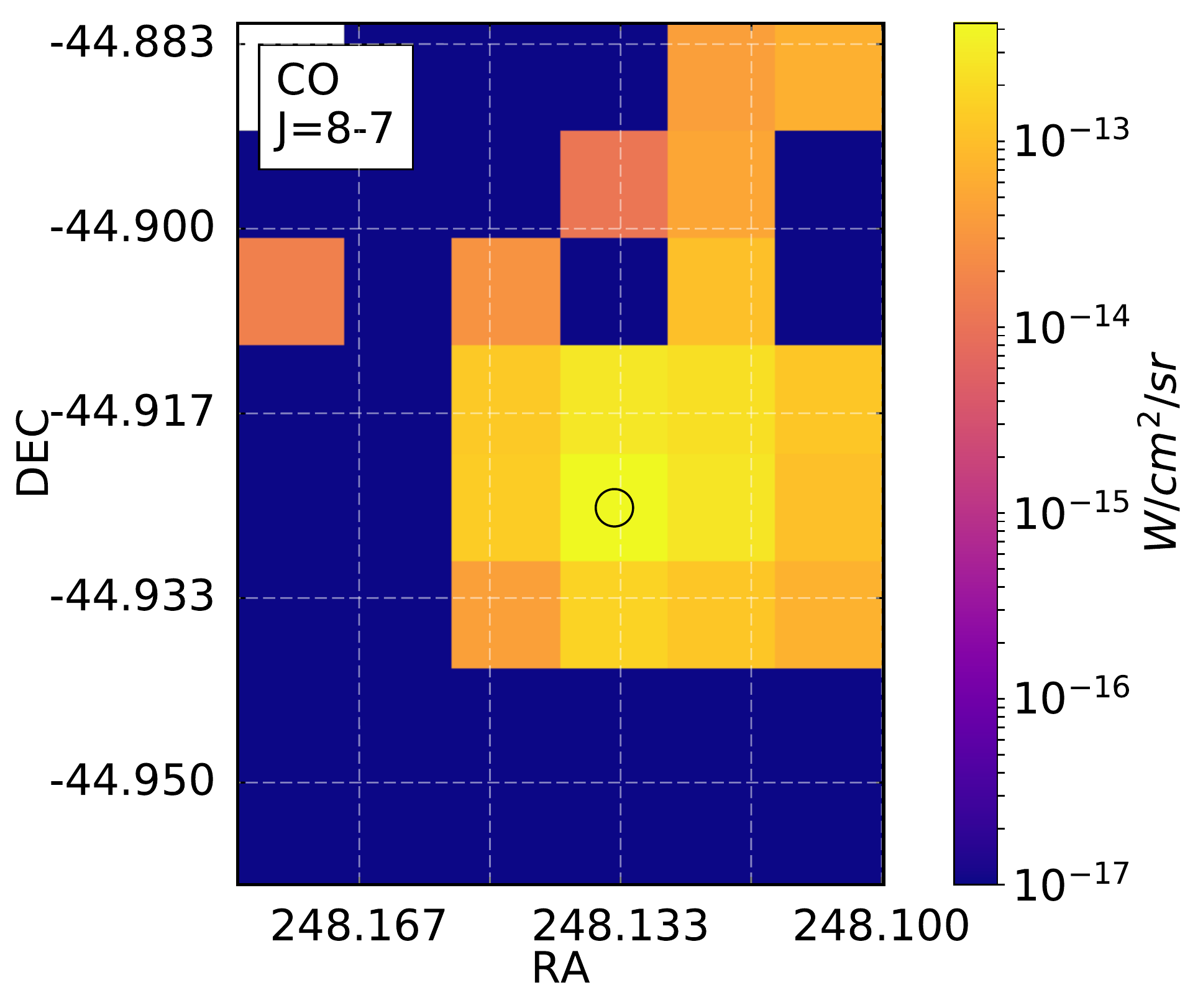}\hspace{-0.1cm}
\includegraphics[width=0.33\textwidth, trim={0cm 0 0cm 0}, clip]{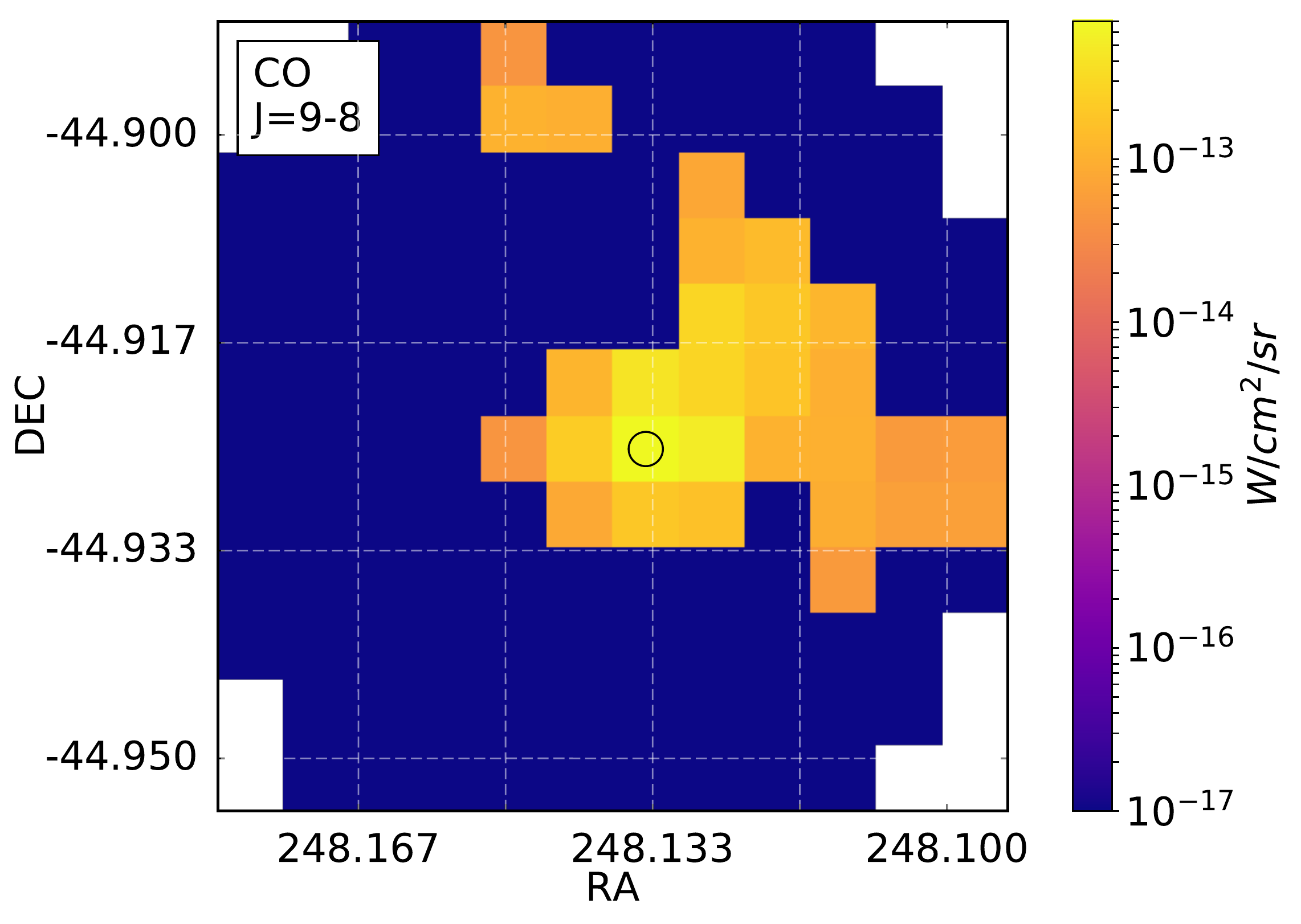}\hspace{-0.1cm}
\includegraphics[width=0.33\textwidth, trim={0cm 0 0cm 0}, clip]{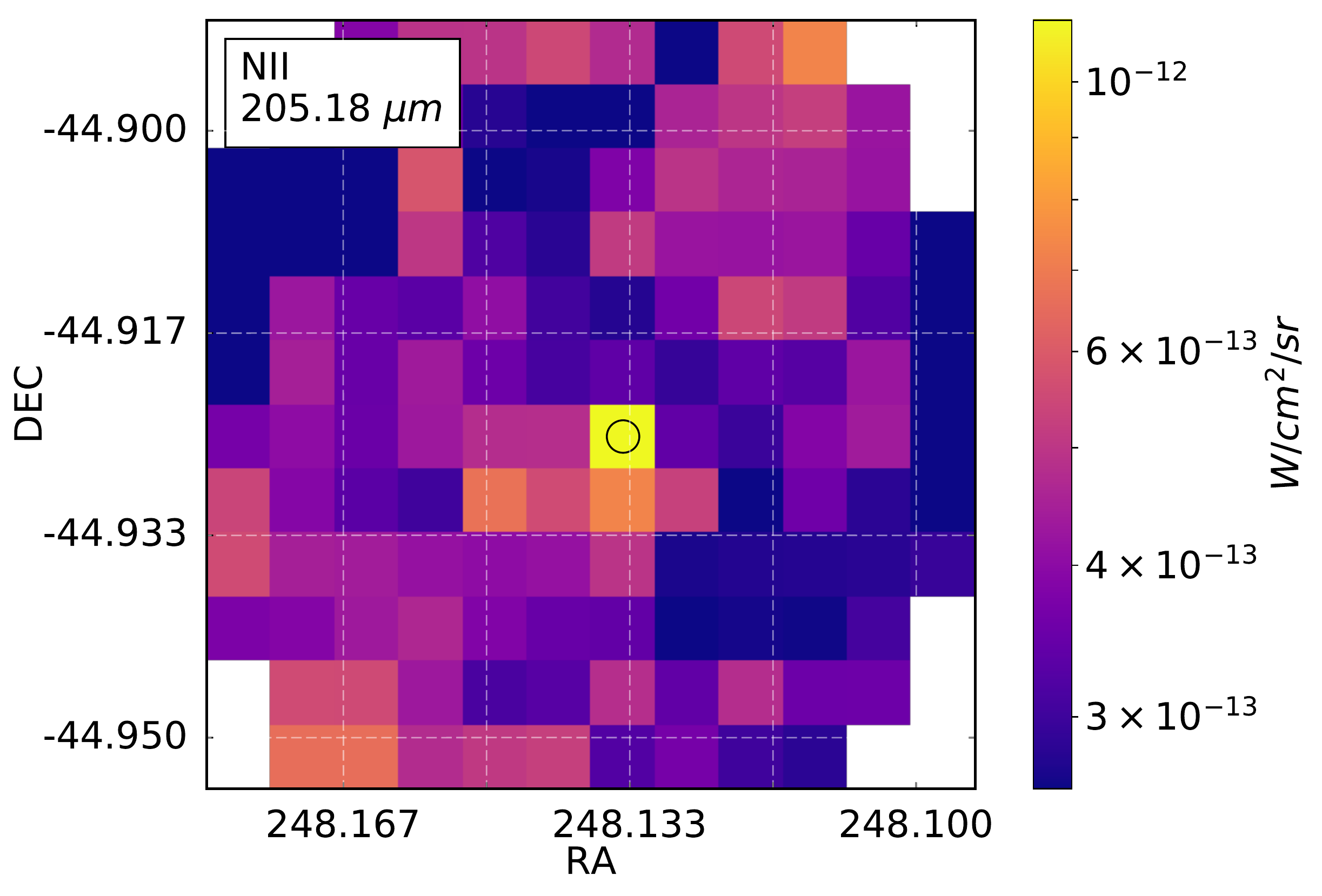}\hspace{-0.1cm}
\caption{
                \footnotesize
                Line maps of SPIRE with visible lines for V346 Nor, part 2.
        }
\end{figure*}

\begin{figure*}
\includegraphics[width=0.33\textwidth, trim={0cm 0 0cm 0}, clip]{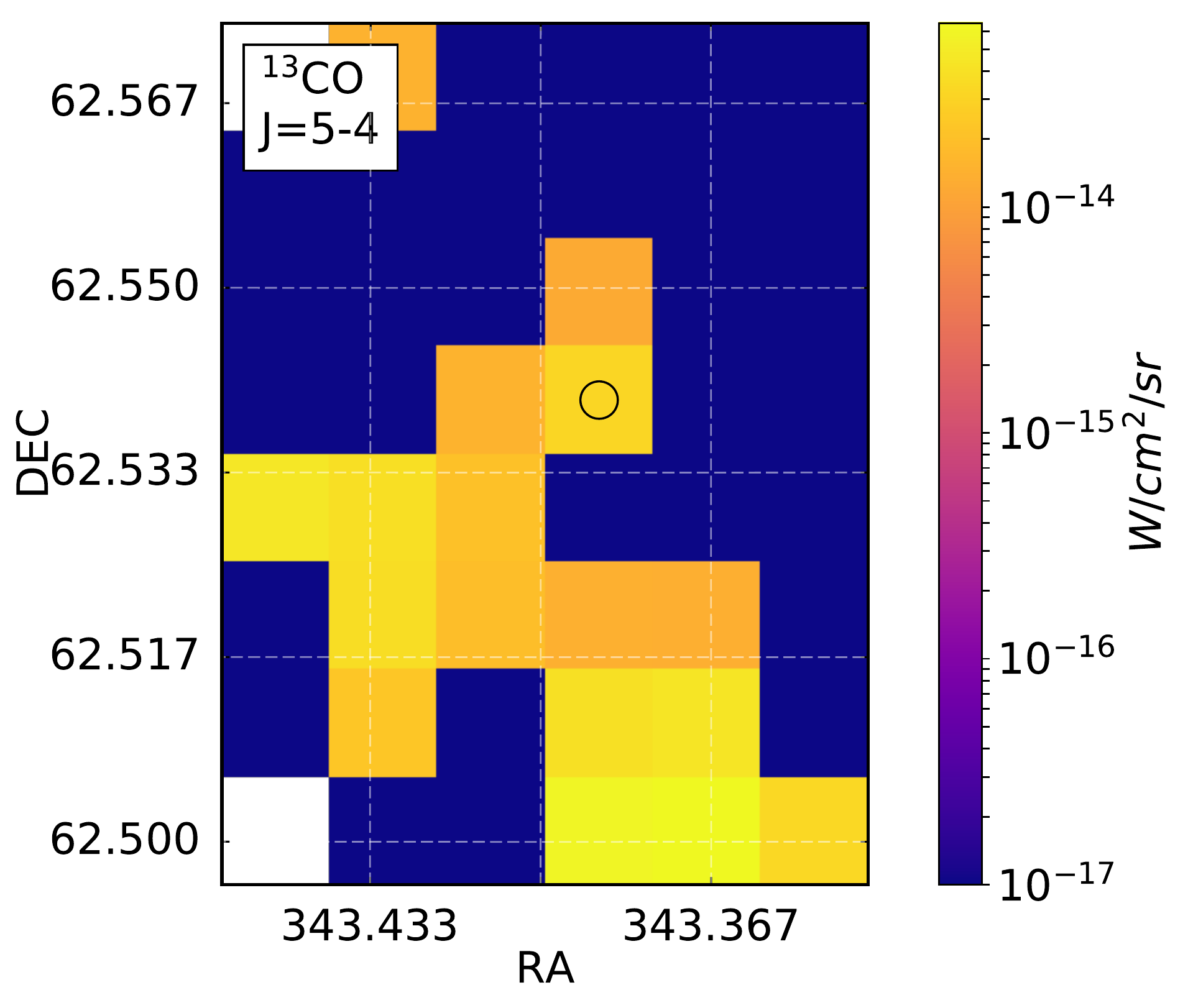}\hspace{-0.1cm}
\includegraphics[width=0.33\textwidth, trim={0cm 0 0cm 0}, clip]{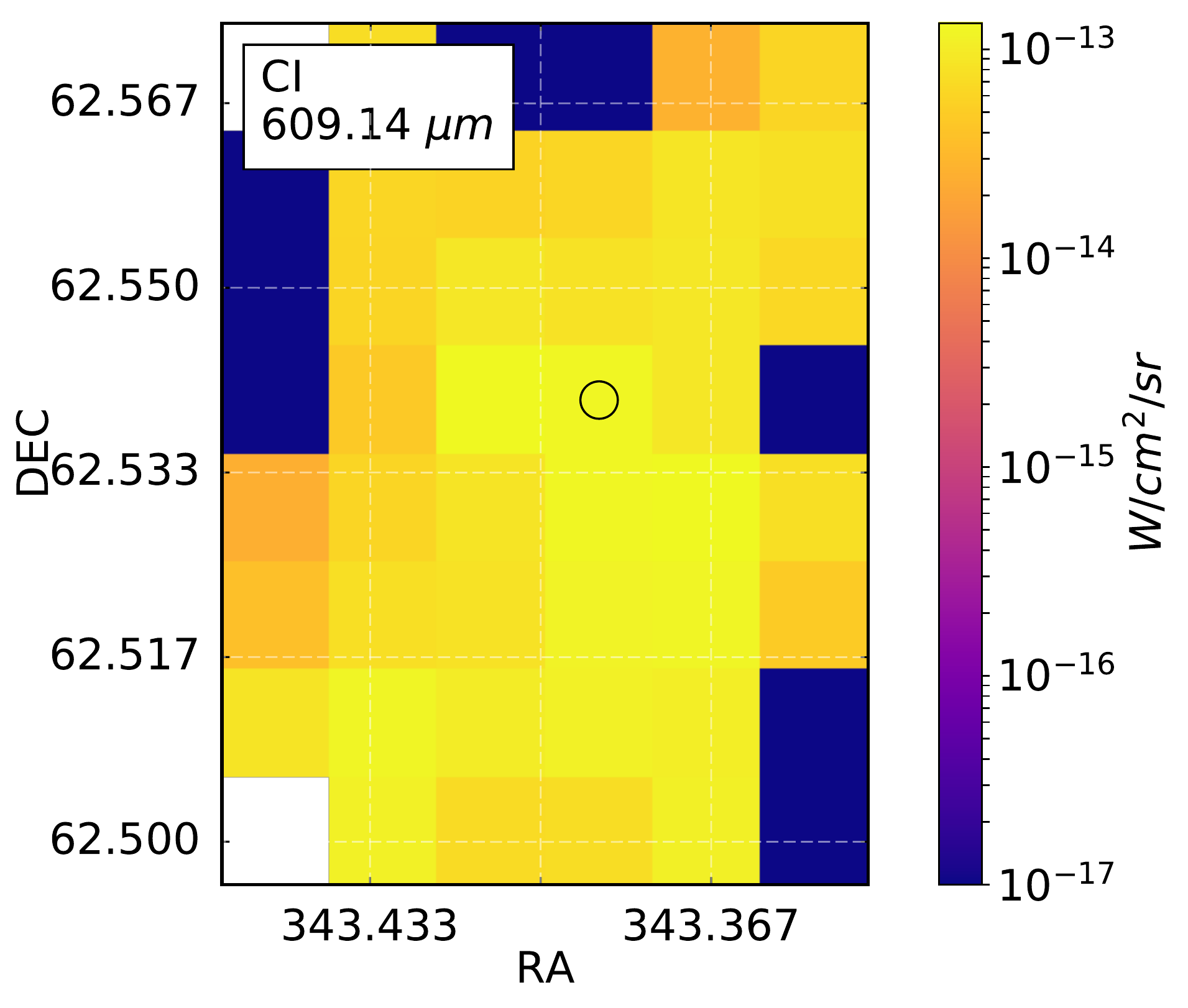}\hspace{-0.1cm}
\includegraphics[width=0.33\textwidth, trim={0cm 0 0cm 0}, clip]{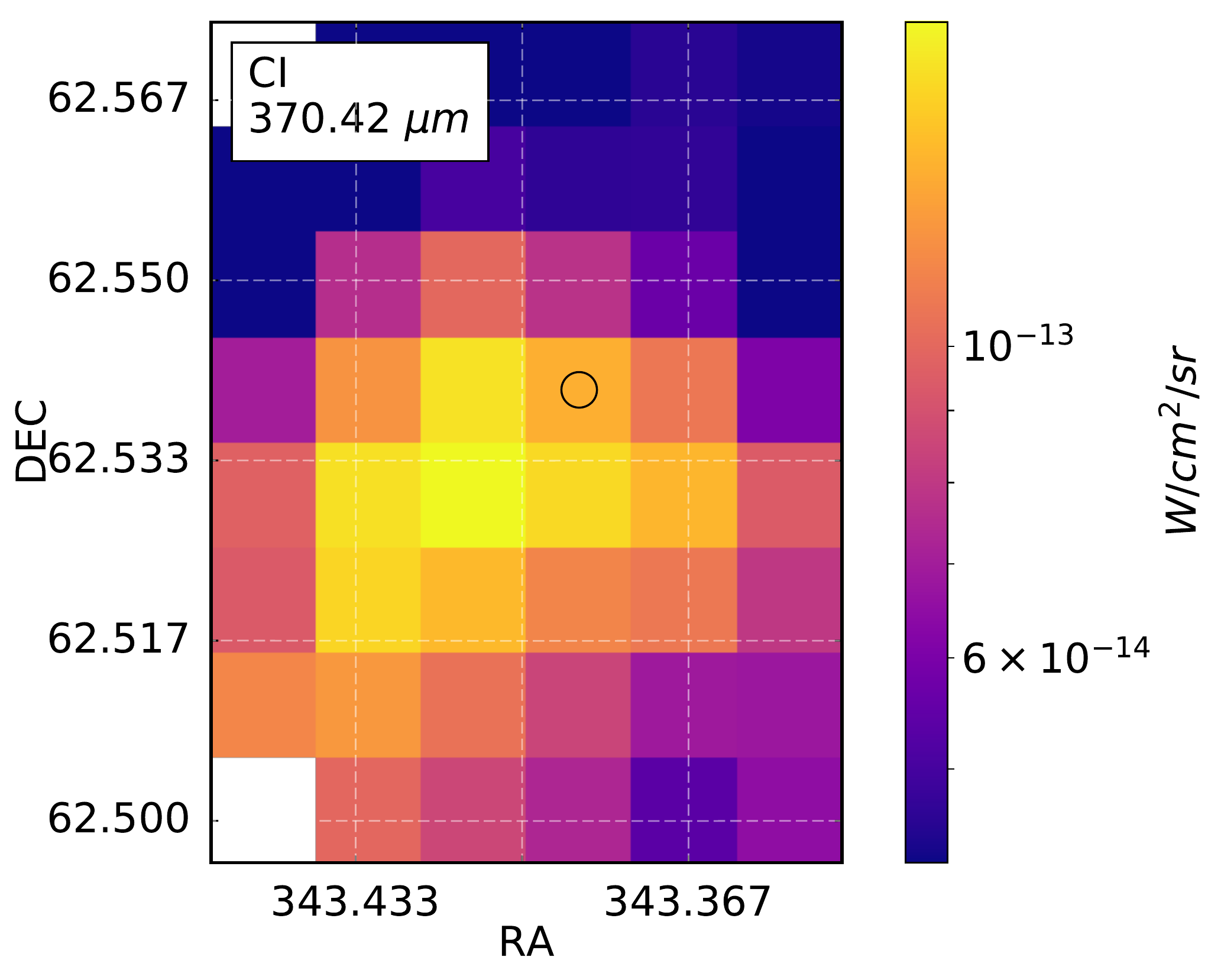}\hspace{-0.1cm}\\
\includegraphics[width=0.33\textwidth, trim={0cm 0 0cm 0}, clip]{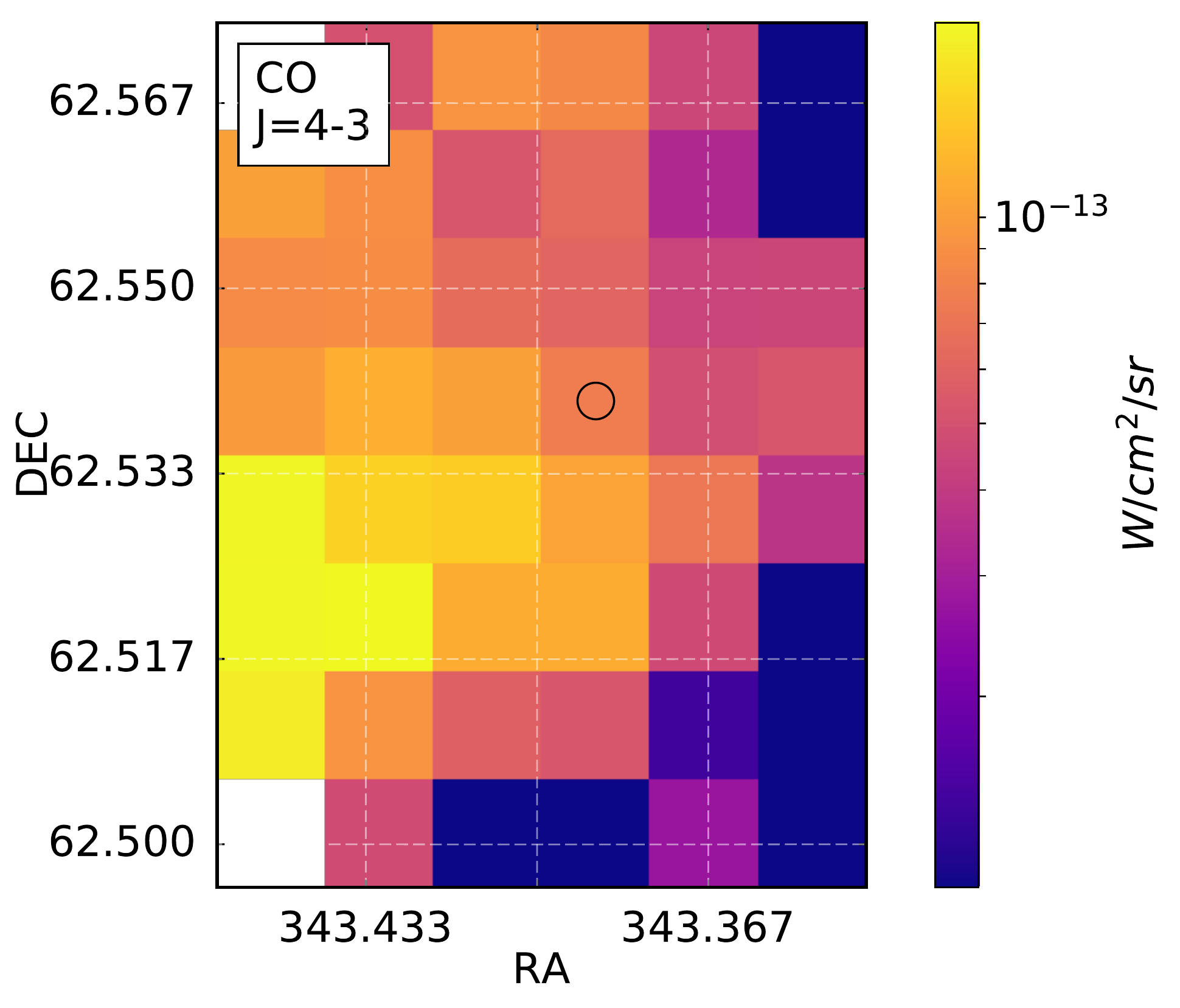}\hspace{-0.1cm}
\includegraphics[width=0.33\textwidth, trim={0cm 0 0cm 0}, clip]{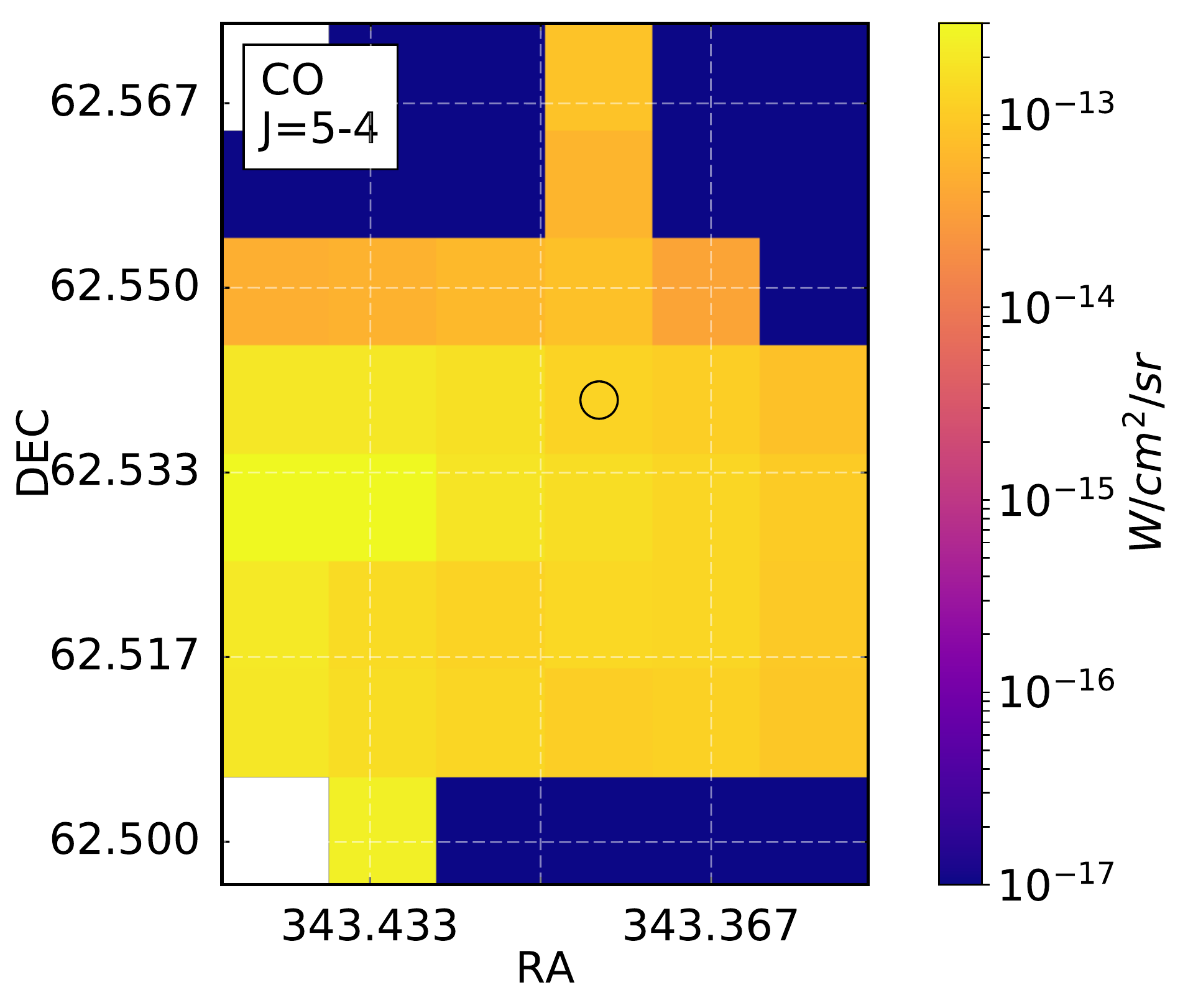}\hspace{-0.1cm}
\includegraphics[width=0.33\textwidth, trim={0cm 0 0cm 0}, clip]{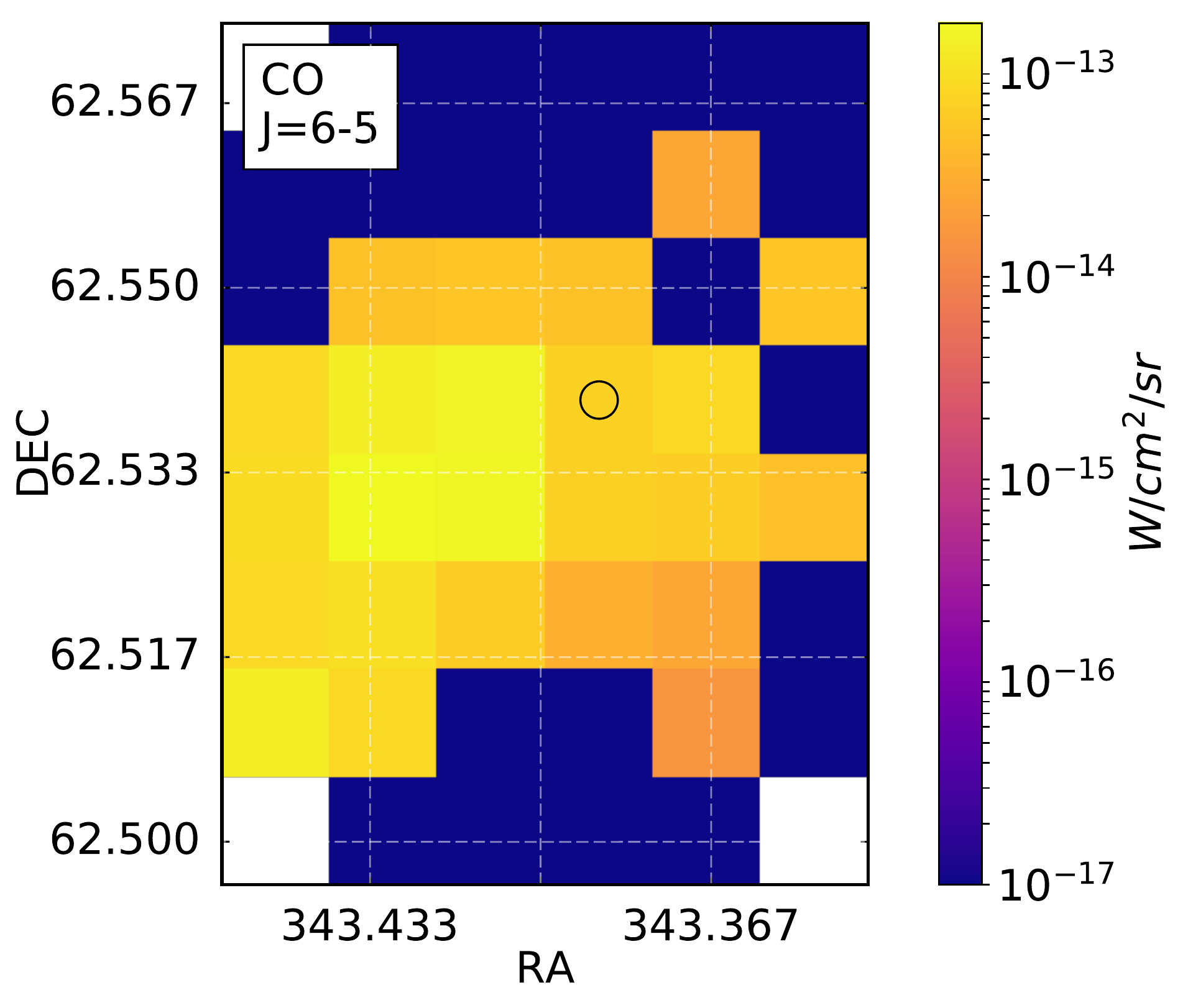}\hspace{-0.1cm}\\
\includegraphics[width=0.33\textwidth, trim={0cm 0 0cm 0}, clip]{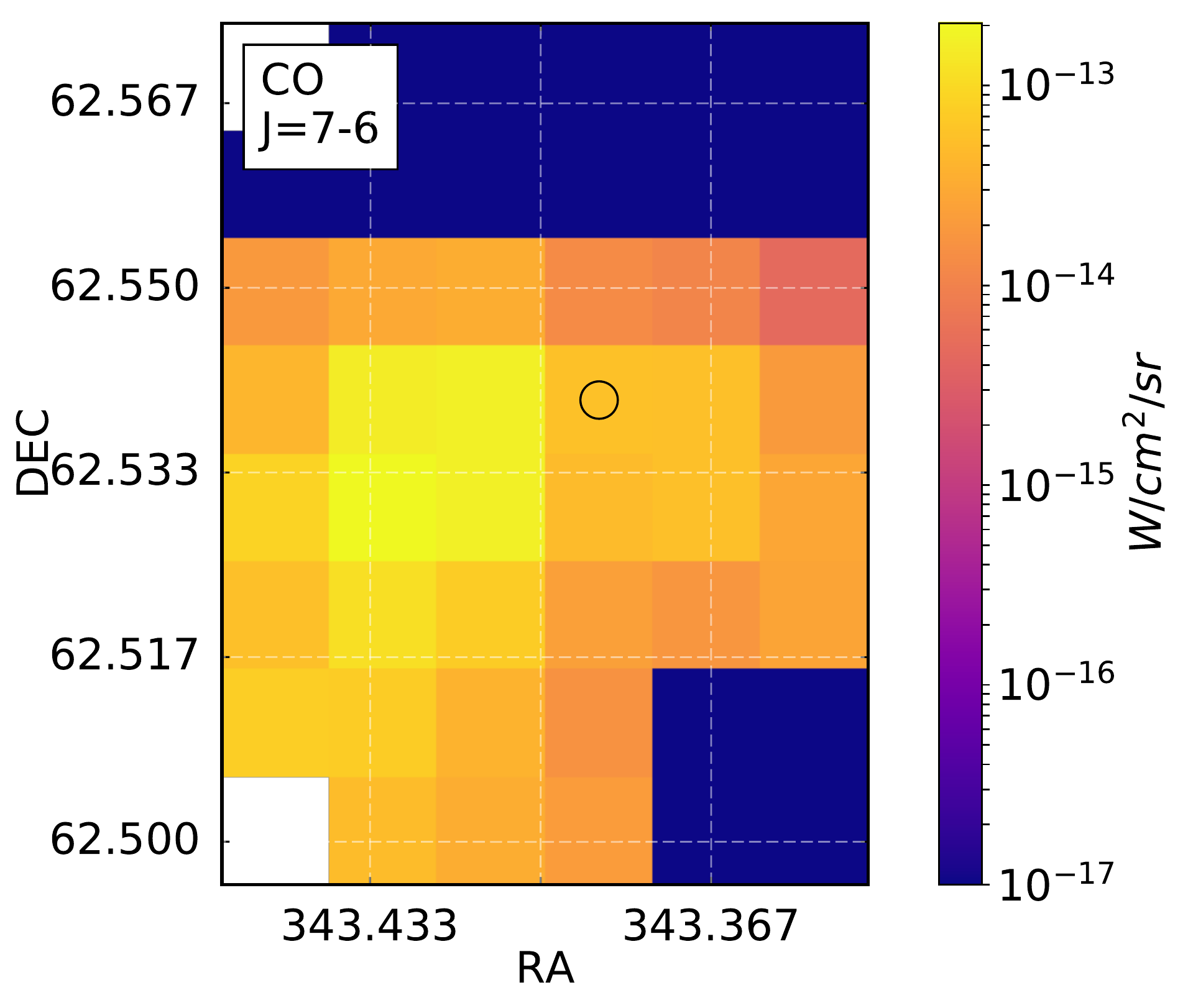}\hspace{-0.1cm}
\includegraphics[width=0.33\textwidth, trim={0cm 0 0cm 0}, clip]{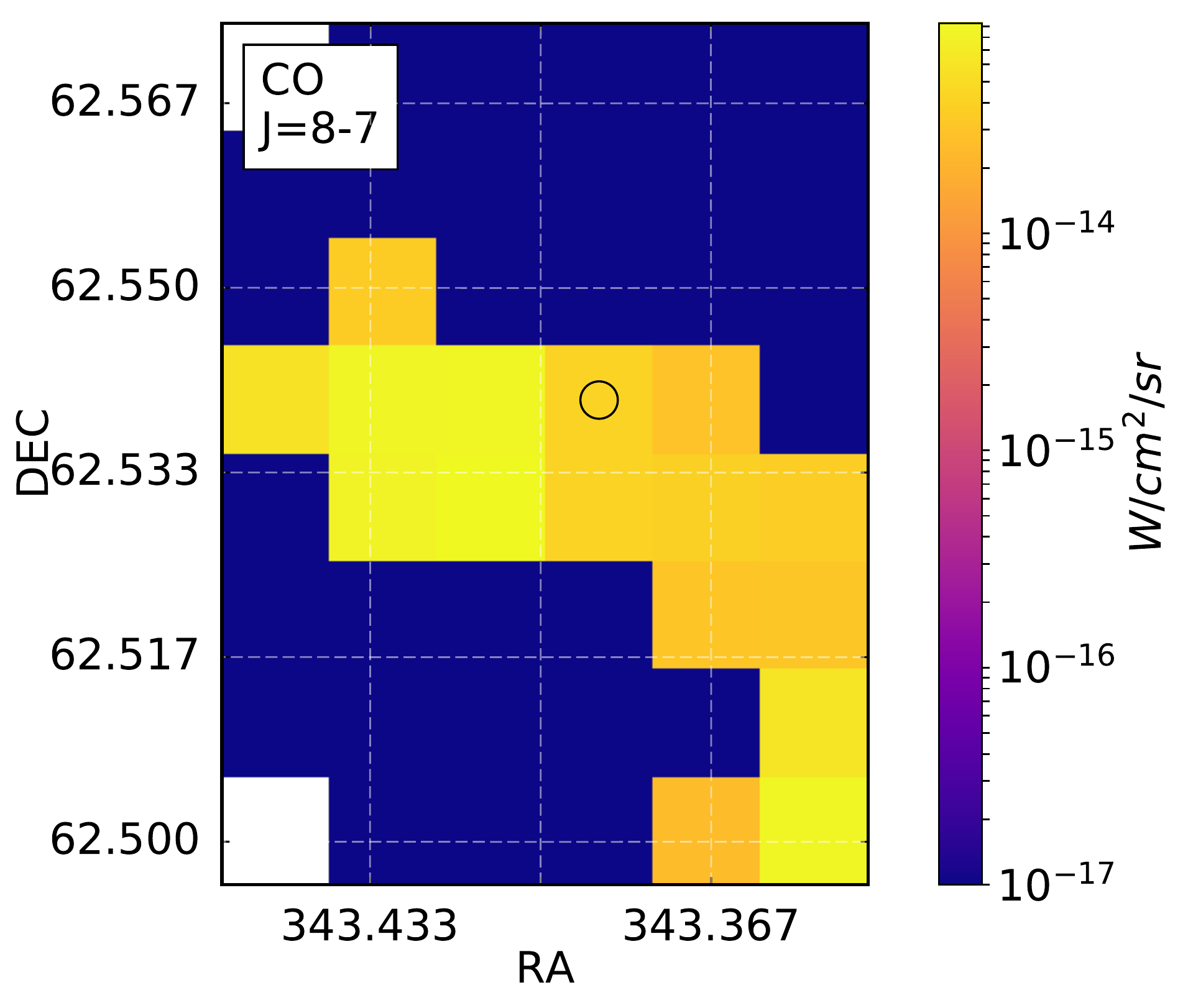}\hspace{-0.1cm}
\includegraphics[width=0.33\textwidth, trim={0cm 0 0cm 0}, clip]{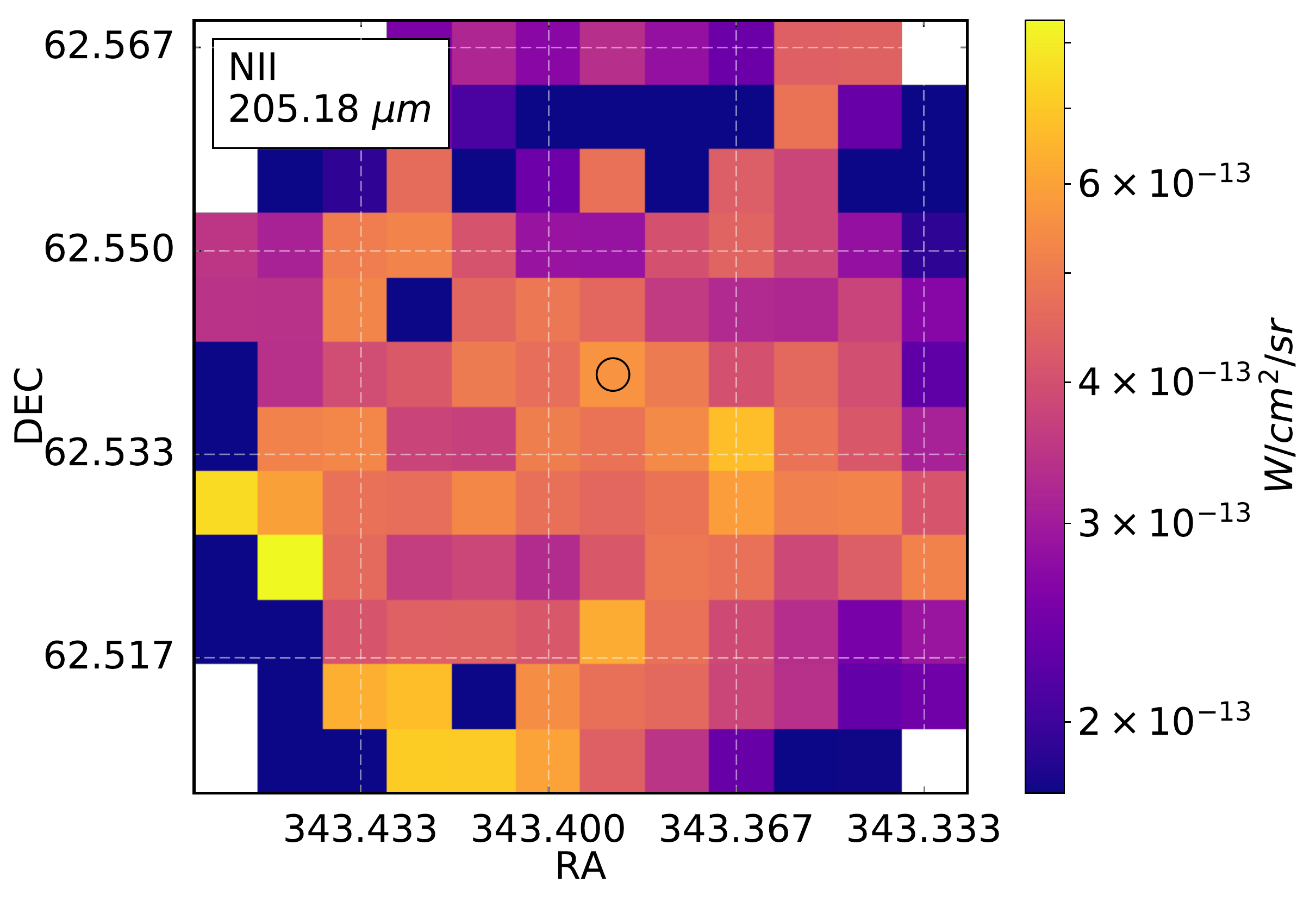}\hspace{-0.1cm}
\caption{
                \footnotesize
                Line maps of SPIRE with visible lines for V733 Cep.
        }
\end{figure*}

\begin{figure*}
\includegraphics[width=0.33\textwidth, trim={0cm 0 0cm 0}, clip]{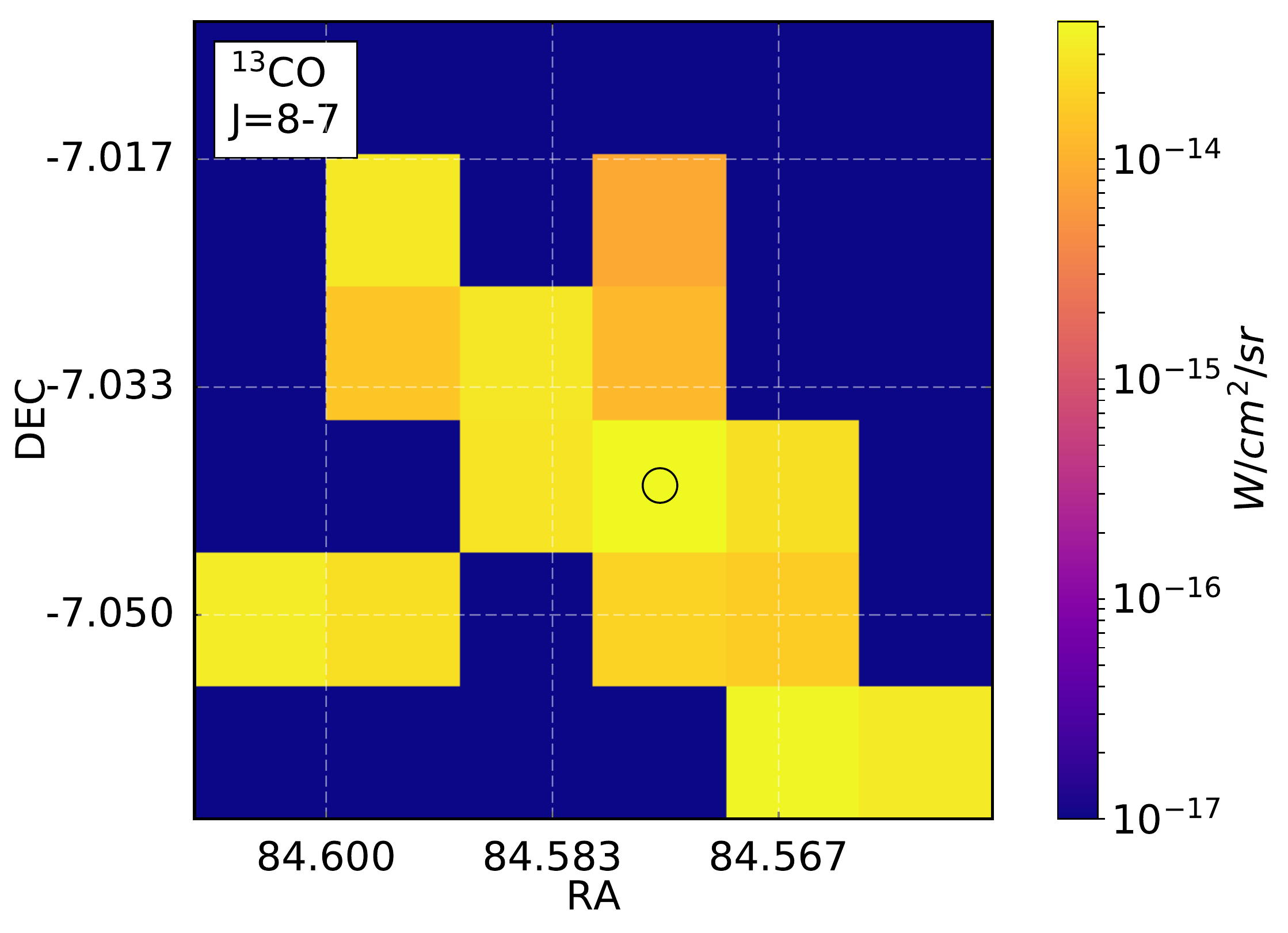}\hspace{-0.1cm}
\includegraphics[width=0.33\textwidth, trim={0cm 0 0cm 0}, clip]{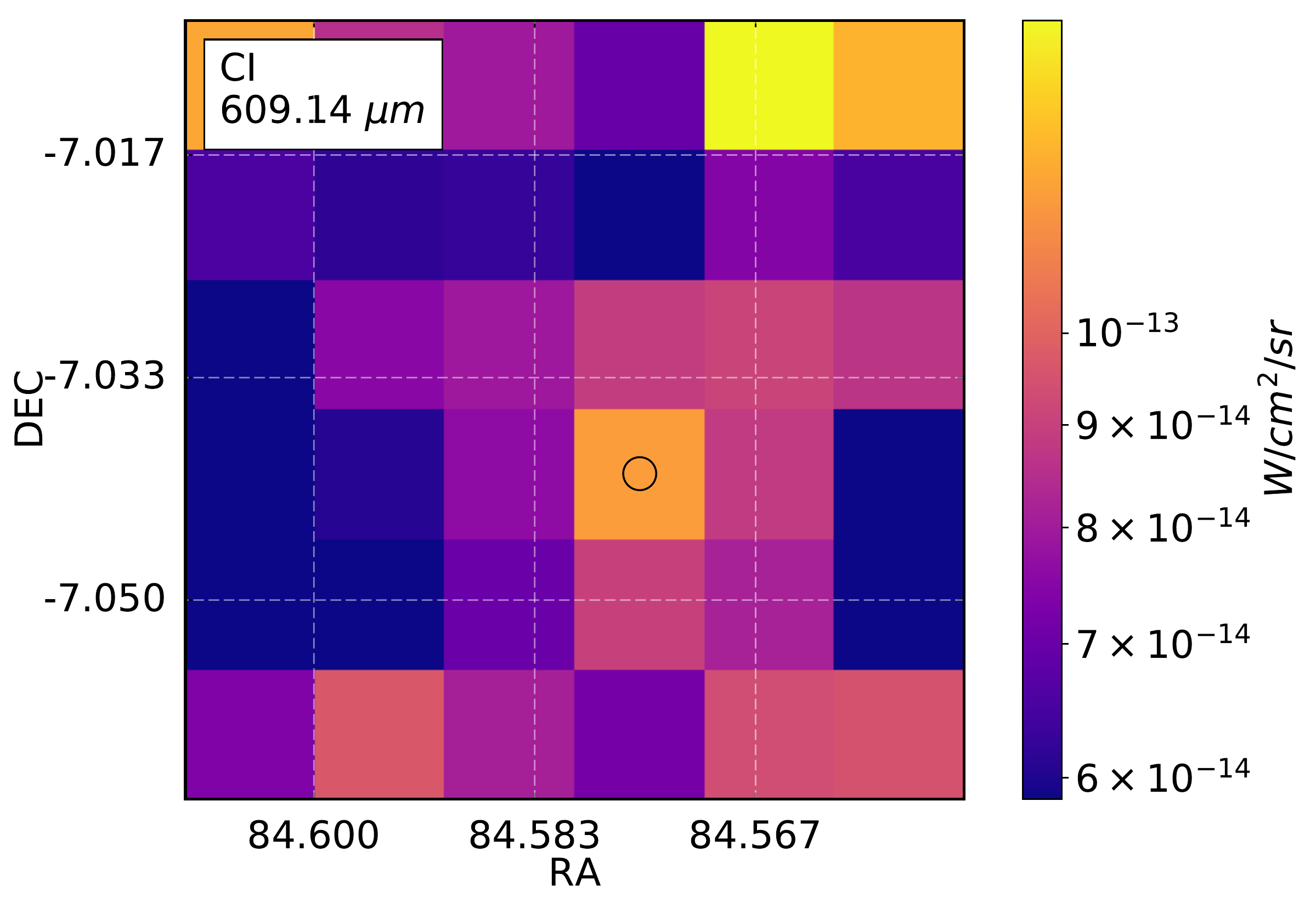}\hspace{-0.1cm}
\includegraphics[width=0.33\textwidth, trim={0cm 0 0cm 0}, clip]{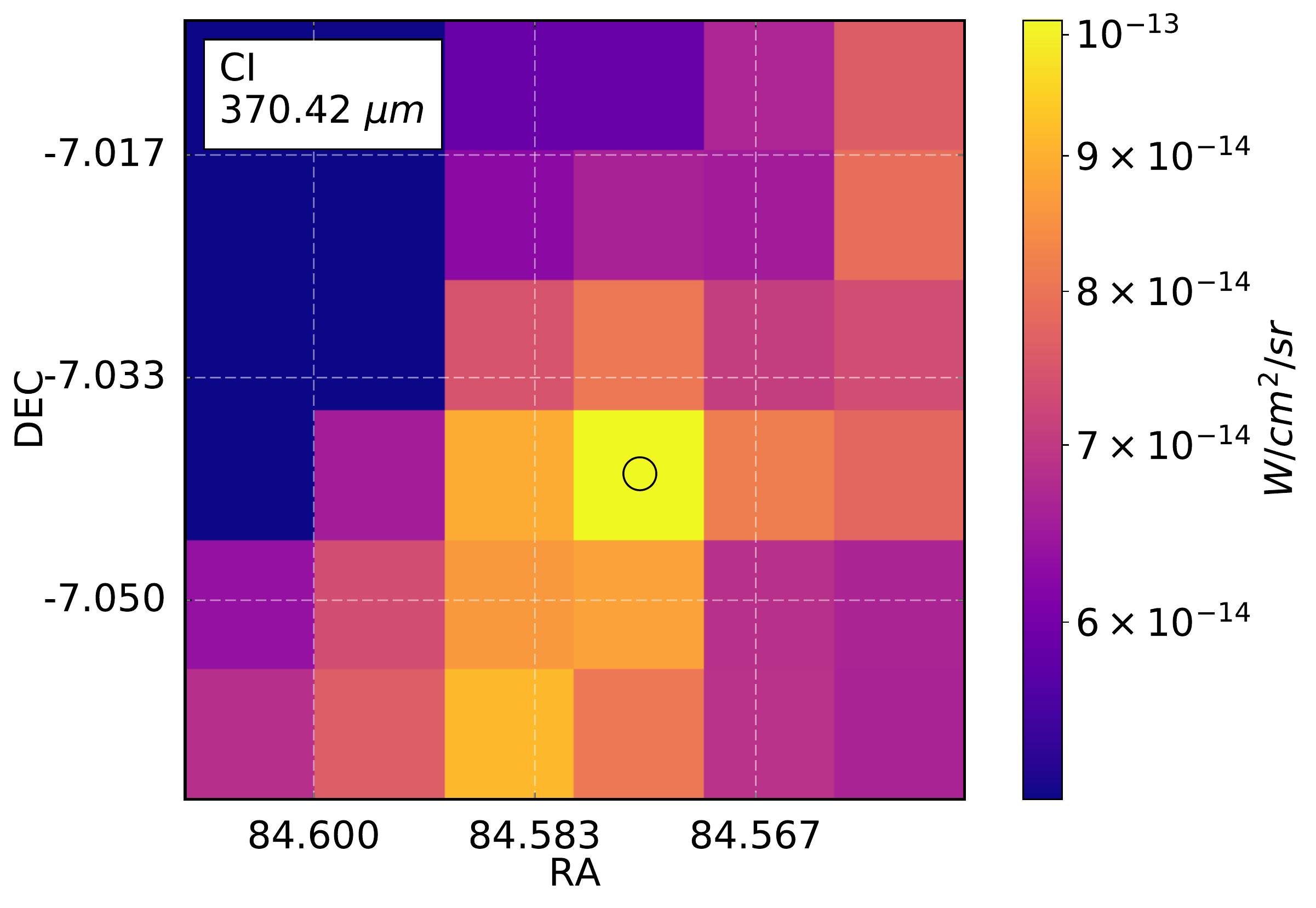}\hspace{-0.1cm}\\
\includegraphics[width=0.33\textwidth, trim={0cm 0 0cm 0}, clip]{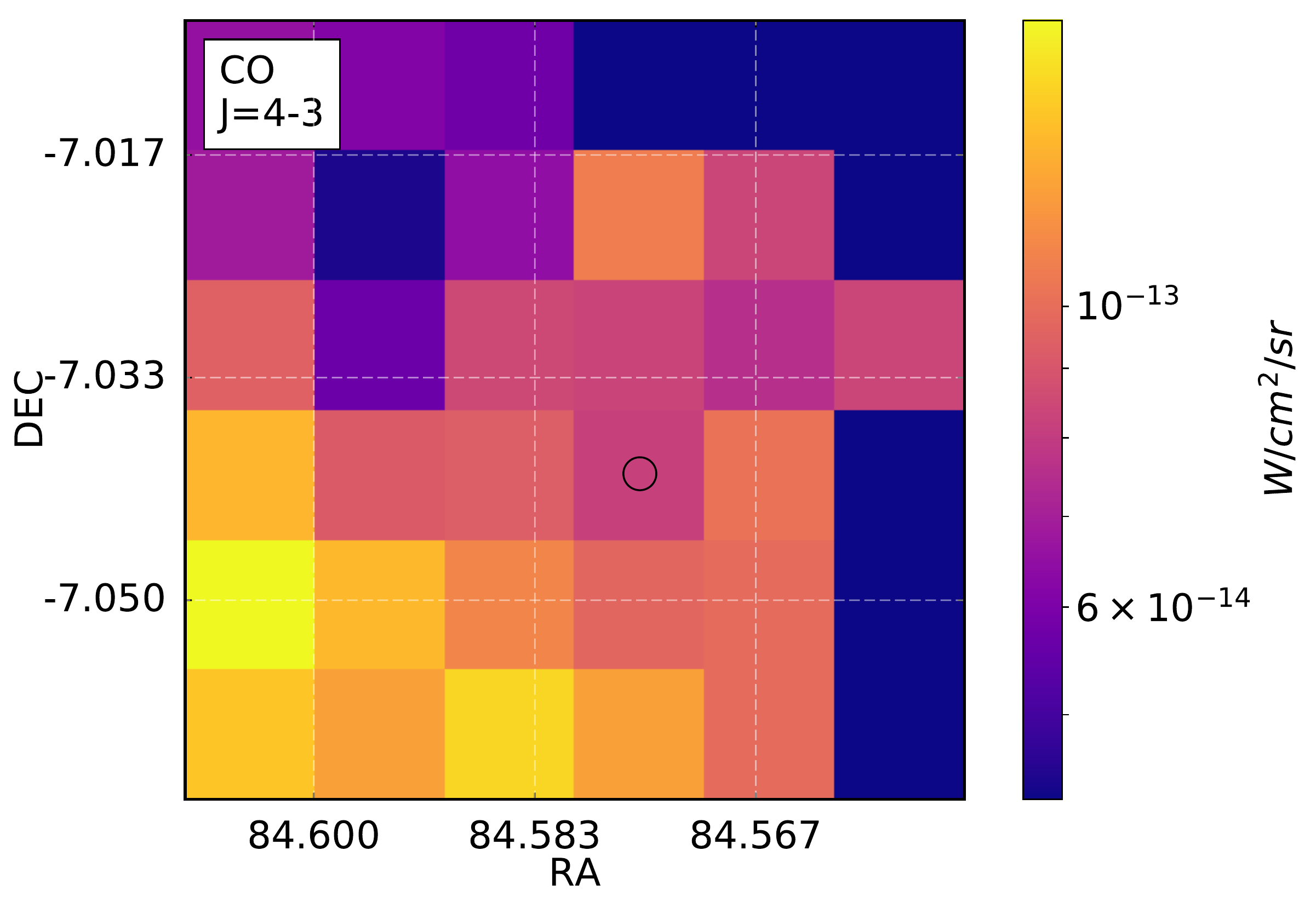}\hspace{-0.1cm}
\includegraphics[width=0.33\textwidth, trim={0cm 0 0cm 0}, clip]{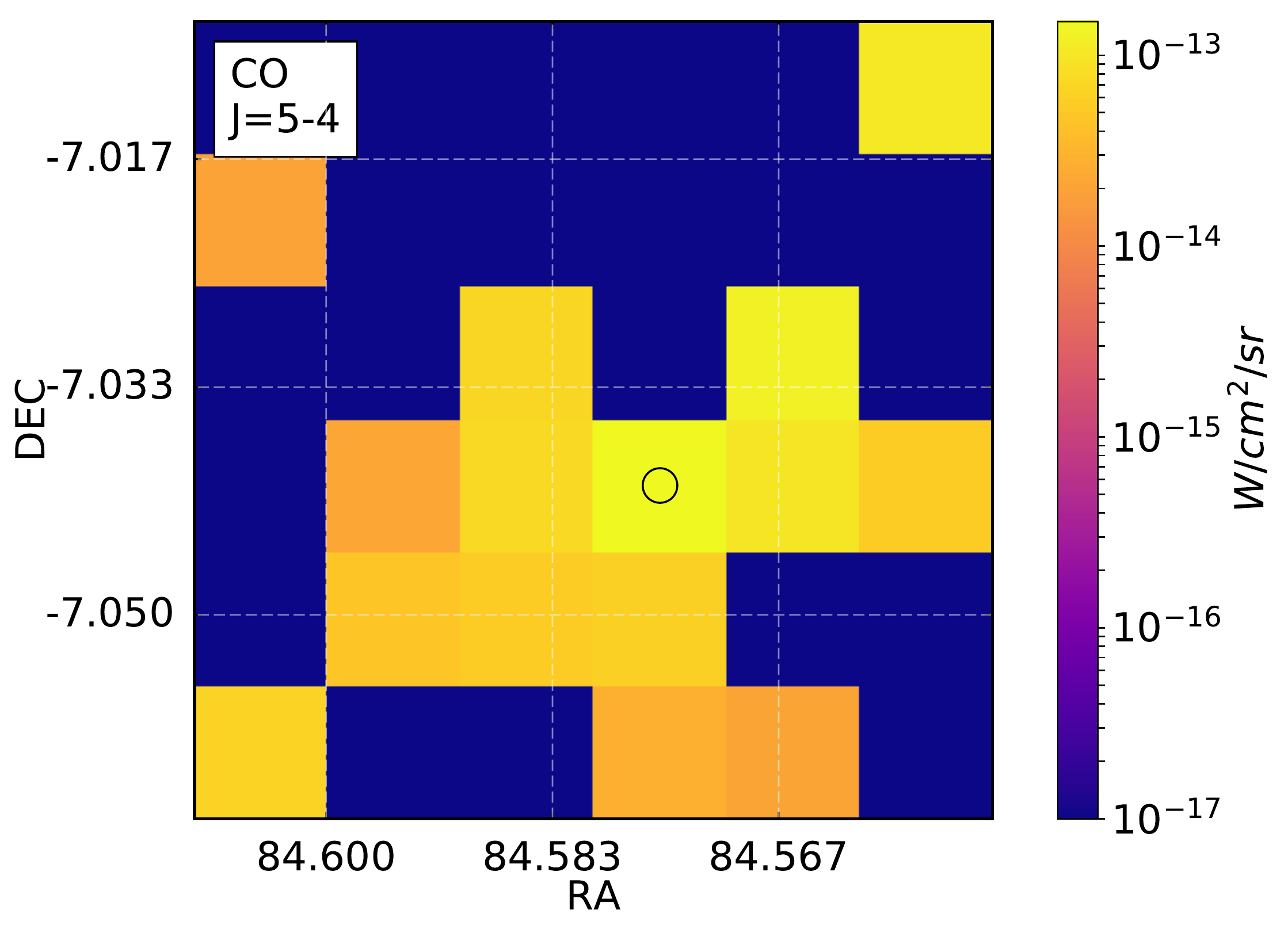}\hspace{-0.1cm}
\includegraphics[width=0.33\textwidth, trim={0cm 0 0cm 0}, clip]{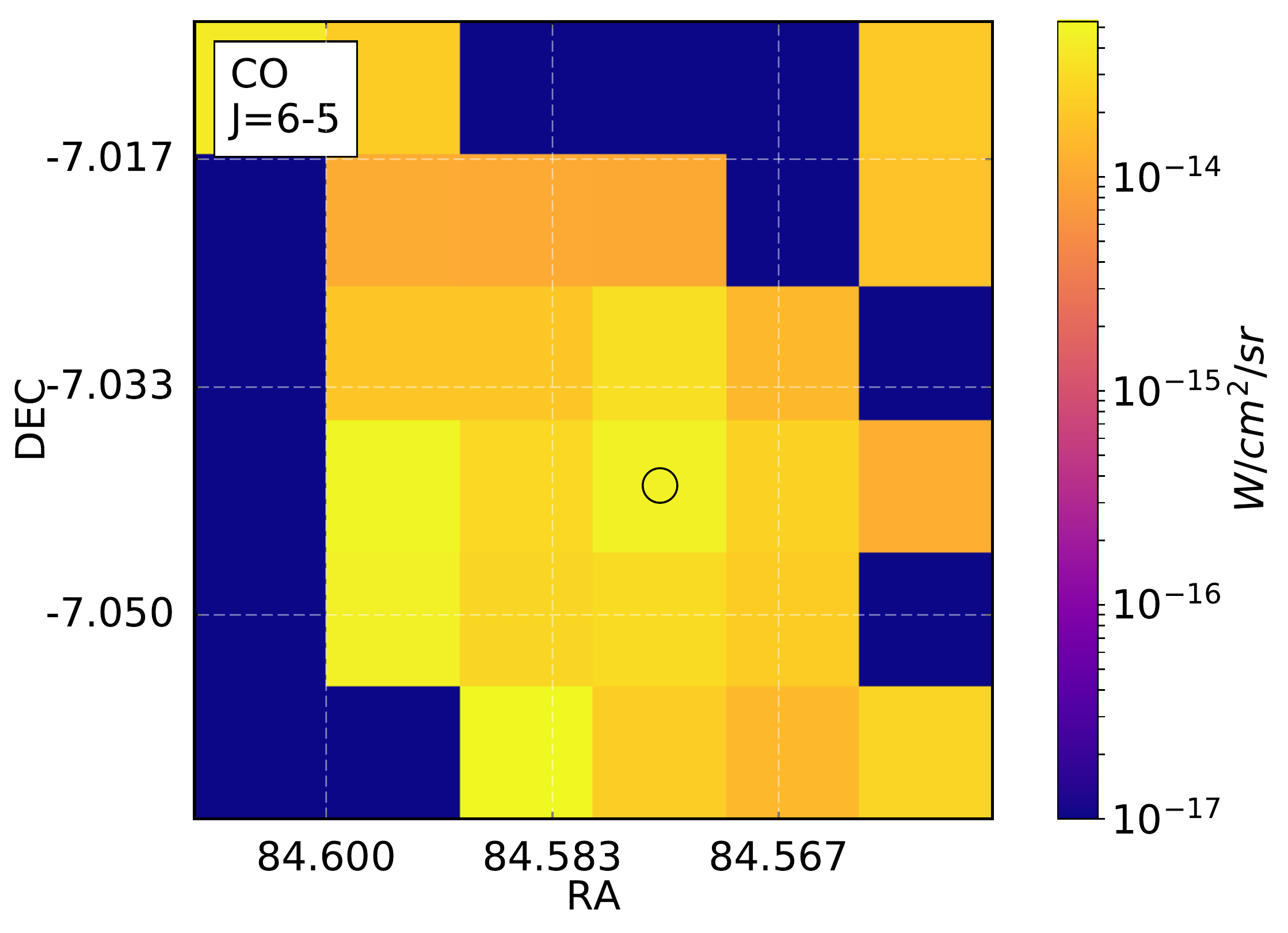}\hspace{-0.1cm}\\
\includegraphics[width=0.33\textwidth, trim={0cm 0 0cm 0}, clip]{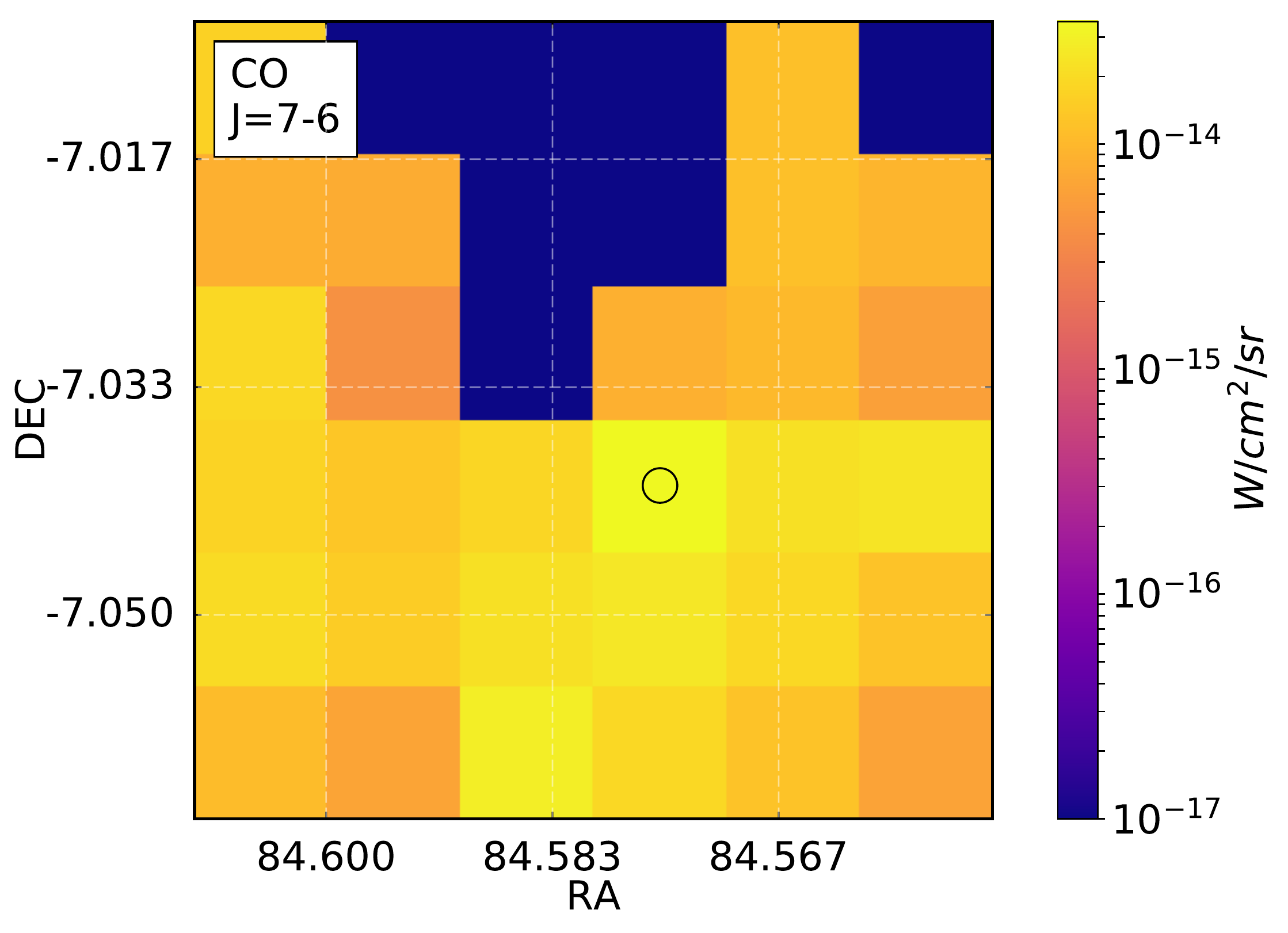}\hspace{-0.1cm}
\includegraphics[width=0.33\textwidth, trim={0cm 0 0cm 0}, clip]{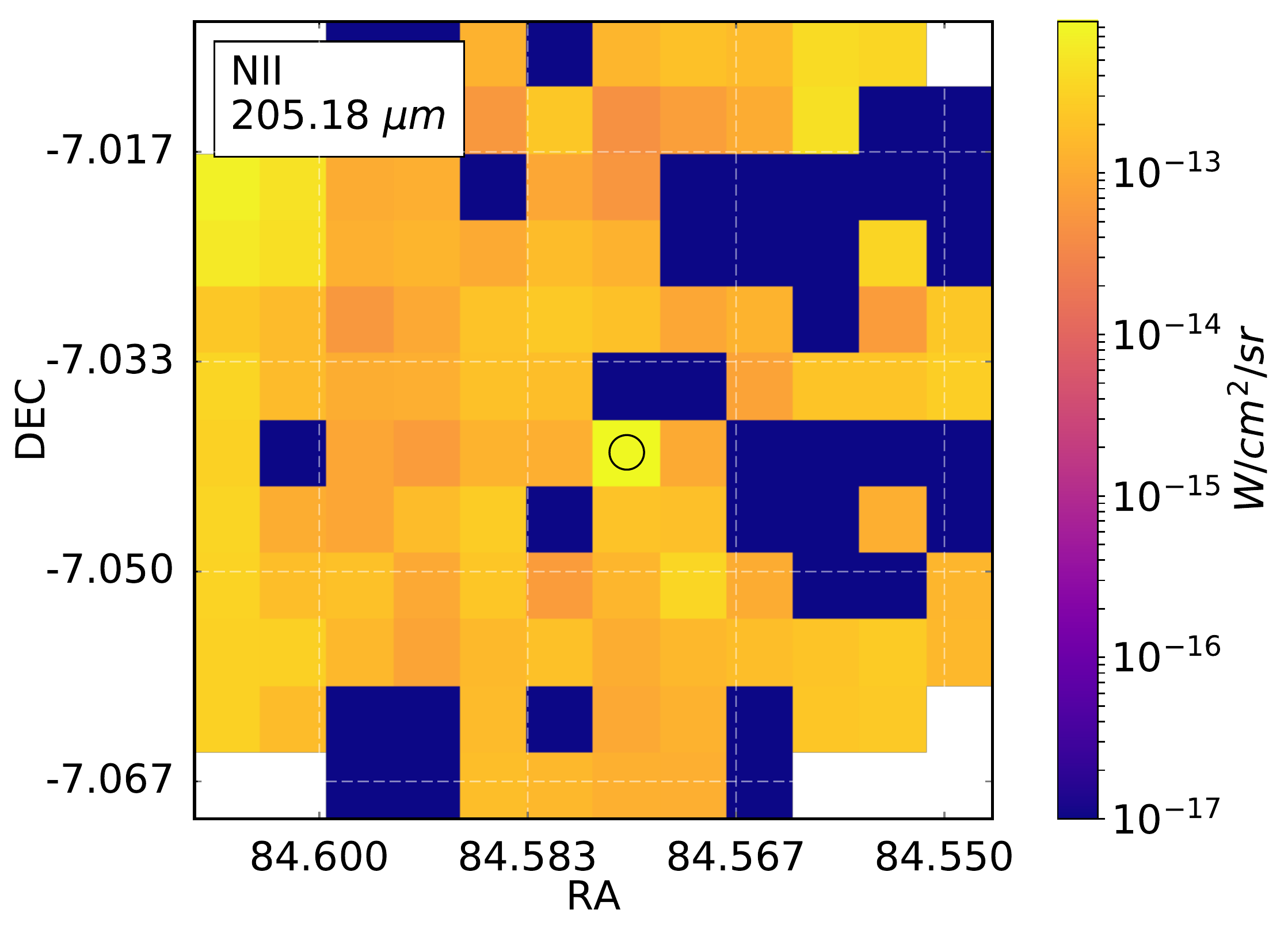}\hspace{-0.1cm}
\caption{
                \footnotesize
                Line maps of SPIRE with visible lines for V883 Ori.
        }
\end{figure*}
\clearpage

\section{Line flux tables}
    \label{line_flux_tables}
    \begin{table*}
\caption{Emission line flux densities}
\label{lineFes}
\centering
\begin{tabular}{lcc|cc|cc}
\hline\hline
& Molecule& &\multicolumn{2}{c|}{BRAN 76} & \multicolumn{2}{c}{EX Lup} \\
Wavelength &  or  & Transition & $F$ & $\Delta F$ & $F$ & $\Delta F$    \\
($\mu$m)&Atom&&\multicolumn{2}{|c|}{(W $m^{-2}$)} &\multicolumn{2}{c}{(W $m^{-2}$)} \\
\hline
272.204627849 & $^{13}$CO & 10-9 & -- & -- & <6.6e-18 & -- \\
302.414601259 & $^{13}$CO & 9-8 & -- & -- & <9.2e-18 & -- \\
340.181218118 & $^{13}$CO & 8-7 & -- & -- & <7.2e-18 & -- \\
388.743038433 & $^{13}$CO & 7-6 & -- & -- & <4.7e-18 & -- \\
453.497650043 & $^{13}$CO & 6-5 & -- & -- & <1.0e-17 & -- \\
544.160740632 & $^{13}$CO & 5-4 & -- & -- & <1.0e-17 & -- \\
230.349129089 & [\ion{C}{I}] & $^3P_{2}-^3P_{0}$ & -- & -- & <1.1e-17 & -- \\
370.415064475 & [\ion{C}{i}] & $^3P_{2}-^3P_{1}$ & -- & -- & <1.7e-18 & -- \\
609.135366673 & [\ion{C}{i}] & $^3P_{1}-^3P_{0}$ & -- & -- & <1.0e-17 & -- \\
157.7409 & [\ion{C}{ii}] & $^2P_{3/2}-^2P_{1/2}$ & <4.3e-17 & -- & 3.3e-17 & 1.2e-17 \\
69.074406 & CO & 38-37 & <1.6e-16 & -- & 2.22e-16 & 6.7e-17 \\
70.907239 & CO & 37-36 & <1.7e-16 & -- & 7.88e-16 & 7.4e-17 \\
72.842854 & CO & 36-35 & <1.4e-16 & -- & <1.7e-16 & -- \\
74.89006 & CO & 35-34 & <1.2e-16 & -- & <1.3e-16 & -- \\
77.058699 & CO & 34-33 & <1.1e-16 & -- & <1.3e-16 & -- \\
79.35981 & CO & 33-32 & <1.3e-16 & -- & <1.5e-16 & -- \\
81.80581 & CO & 32-31 & <1.3e-16 & -- & \textit{6.2e-17} & 5.2e-17 \\
84.410717 & CO & 31-30 & <1.4e-16 & -- & <1.8e-16 & -- \\
87.190421 & CO & 30-29 & <1.4e-16 & -- & <1.7e-16 & -- \\
90.163002 & CO & 29-28 & <1.4e-16 & -- & <1.6e-16 & -- \\
93.349123 & CO & 28-27 & <1.6e-16 & -- & <1.7e-16 & -- \\
104.44495 & CO & 25-24 & <7.6e-17 & -- & <7.1e-17 & -- \\
108.76281 & CO & 24-23 & 1.14e-16 & 2.3e-17 & <6.8e-17 & -- \\
113.4576 & CO & 23-22 & <5.5e-17 & -- & <5.7e-17 & -- \\
118.58072 & CO & 22-21 & \textit{1.7e-17} & 1.6e-17 & <4.4e-17 & -- \\
124.19335 & CO & 21-20 & \textit{2.9e-17} & 1.5e-17 & \textit{1.4e-17} & 1.3e-17 \\
130.36893 & CO & 20-19 & \textit{1.5e-17} & 1.3e-17 & \textit{1.9e-17} & 1.3e-17 \\
137.19633 & CO & 19-18 & 2.9e-17 & 1.2e-17 & <3.5e-17 & -- \\
144.78419 & CO & 18-17 & <4.8e-17 & -- & <3.5e-17 & -- \\
153.26671 & CO & 17-16 & <4.6e-17 & -- & 2.4e-17 & 1.1e-17 \\
162.81163 & CO & 16-15 & \textit{1.9e-17} & 1.5E-17 & \textit{1.2e-17} & 1.1e-17 \\
173.63143 & CO & 15-14 & <4.9e-17 & -- & <3.9e-17 & -- \\
185.9993 & CO & 14-13 & 8.5e-17 & 2.4e-17 & <5.7e-17 & -- \\
200.272477627 & CO & 13-12 & -- & -- & <1.4e-17 & -- \\
216.92729446 & CO & 12-11 & -- & -- & <8.8e-18 & -- \\
236.613281063 & CO & 11-10 & -- & -- & <9.3e-18 & -- \\
260.239806859 & CO & 10-9 & -- & -- & <7.8e-18 & -- \\
289.120332634 & CO & 9-8 & -- & -- & <8.4e-18 & -- \\
325.225163341 & CO & 8-7 & -- & -- & <6.6e-18 & -- \\
371.6503924 & CO & 7-6 & -- & -- & <1.7e-18 & -- \\
433.556224633 & CO & 6-5 & -- & -- & \textit{4.2e-18} & 2.1e-18 \\
520.231028647 & CO & 5-4 & -- & -- & <9.7e-18 & -- \\
650.251512464 & CO & 4-3 & -- & -- & <9.5e-18 & -- \\
205.17830087 & [\ion{N}{ii}] & $^3P_{1}-^3P_{0}$ & -- & -- & 1.88e-17 & 3.4e-18 \\
\hline
\end{tabular}
    \tablefoot{
                We only show lines that were detected in at least one of our sources. The extraction procedure for PACS and SPIRE is described in Sect. \ref{subsection:SEDs_and_photometry}.
                CO Lines below $3\sigma$ detection  used in the rotational diagrams are shown in italics.
        }
\end{table*}
    \begin{table*}
\centering
\begin{tabular}{lcc|cc|cc}
\hline\hline
& Molecule& &\multicolumn{2}{c|}{BRAN 76} & \multicolumn{2}{c}{EX Lup} \\
Wavelength &  or  & Transition & $F$ & $\Delta F$ & $F$ & $\Delta F$    \\
($\mu$m)&Atom&&\multicolumn{2}{|c|}{(W $m^{-2}$)} &\multicolumn{2}{c}{(W $m^{-2}$)} \\
\hline
65.1315 & OH & $^2P_{3/2}\ J=9/2+ \rightarrow J=7/2-$ & <2.0e-16 & -- & <1.8e-16 & -- \\
65.2788 & OH & $^2P_{3/2}\ J=9/2+ \rightarrow J=7/2-$ & <2.0e-16 & -- & <1.8e-16 & -- \\
71.17085 & OH & $^2P_{1/2}\ J=7/2+ \rightarrow J=5/2-$ & <2.0e-16 & -- & <2.5e-16 & -- \\
71.215869 & OH & $^2P_{1/2}\ J=7/2+ \rightarrow J=5/2-$ & 1.56e-16 & 6.8e-17 & 4.06e-16 & 8.6e-17 \\
79.118034 & OH & $^2P_{1/2}\ J=1/2+ \rightarrow J=3/2-$ & <1.3e-16 & -- & <1.5e-16 & -- \\
79.181726 & OH & $^2P_{1/2}\ J=1/2+ \rightarrow J=3/2-$ & <1.3e-16 & -- & <1.5e-16 & -- \\
84.419938 & OH & $^2P_{3/2}\ J=7/2+ \rightarrow J=5/2-$ & \multirow{ 2}{*}{<1.4e-16} & \multirow{ 2}{*}{--} &\multirow{ 2}{*}{<1.7e-16}&\multirow{ 2}{*}{--} \\
84.420391 & OH & $^2P_{3/2}\ J=7/2+ \rightarrow J=5/2-$ &  &  &  &  \\
84.596825 & OH & $^2P_{3/2}\ J=7/2+ \rightarrow J=5/2-$ & <1.4e-16 & -- & <1.7e-16 & -- \\
115.1541 & OH & $^2P_{1/2}\ J=5/2+ \rightarrow J=7/2-$ & 4.7e-17 & 1.7e-17 & <5.4e-17 & -- \\
115.389 & OH & $^2P_{1/2}\ J=5/2+ \rightarrow J=7/2-$ & <5.2e-17 & -- & <5.4e-17 & -- \\
119.23418 & OH & $^2P_{3/2}\ J=5/2+ \rightarrow J=3/2-$ & <4.9e-17 & -- & <4.5e-17 & -- \\
119.44167 & OH & $^2P_{3/2}\ J=5/2+ \rightarrow J=3/2-$ & <4.6e-17 & -- & <4.5e-17 & -- \\
134.8476 & OH & $^2P_{1/2}\ J=7/2+ \rightarrow J=9/2-$ & <3.7e-17 & -- & <3.5e-17 & -- \\
134.9696 & OH & $^2P_{1/2}\ J=7/2+ \rightarrow J=9/2-$ & <3.6e-17 & -- & <3.5e-17 & -- \\
163.12428 & OH & $^2P_{1/2}\ J=3/2+ \rightarrow J=1/2-$ & <4.5e-17 & -- & 2.5e-17 & 1.1e-17 \\
163.39718 & OH & $^2P_{1/2}\ J=3/2+ \rightarrow J=1/2-$ & <4.4e-17 & -- & <3.3e-17 & -- \\
63.183705 & [\ion{O}{i}] & $^3P_{1}-^3P_{2}$ & 3.87e-16 & 9.9e-17 & 4.74e-16 & 6.5e-17 \\
145.52544 & [\ion{O}{i}] & $^3P_{0}-^3P_{1}$ & 3.7e-17 & 1.5e-17 & <3.3e-17 & -- \\
58.6982 & o-H$_2$O & $4_{32}-3_{21}$ & <2.2e-16 & -- & <2.3e-16 & -- \\
66.4372 & o-H$_2$O & $3_{30}-2_{21}$ & <1.9e-16 & -- & <1.9e-16 & -- \\
67.2689 & o-H$_2$O & $3_{30}-3_{03}$ & <1.7e-16 & -- & <1.9e-16 & -- \\
71.946 & o-H$_2$O & $7_{07}-6_{16}$ & <2.2e-16 & -- & <2.7e-16 & -- \\
75.3804 & o-H$_2$O & $3_{21}-2_{12}$ & <1.1e-16 & -- & <1.2e-16 & -- \\
78.7414 & o-H$_2$O & $4_{23}-3_{12}$ & <1.3e-16 & -- & <1.5e-16 & -- \\
82.0304 & o-H$_2$O & $6_{16}-5_{05}$ & <1.3e-16 & -- & <1.6e-16 & -- \\
108.073 & o-H$_2$O & $2_{21}-1_{10}$ & <7.2e-17 & -- & <6.5e-17 & -- \\
113.5366 & o-H$_2$O & $4_{14}-3_{03}$ & <5.5e-17 & -- & <5.8e-17 & -- \\
116.7836 & o-H$_2$O & $7_{34}-6_{43}$ & <5.0e-17 & -- & <4.9e-17 & -- \\
121.7191 & o-H$_2$O & $4_{32}-4_{23}$ & <4.5e-17 & -- & <4.1e-17 & -- \\
132.407 & o-H$_2$O & $4_{23}-4_{14}$ & <3.7e-17 & -- & <3.4e-17 & -- \\
134.9346 & o-H$_2$O & $5_{14}-5_{05}$ & <3.7e-17 & -- & <3.5e-17 & -- \\
136.4944 & o-H$_2$O & $3_{30}-3_{21}$ & <3.6e-17 & -- & <3.4e-17 & -- \\
174.6264 & o-H$_2$O & $3_{03}-2_{12}$ & <5.1e-17 & -- & <4.1e-17 & -- \\
179.5265 & o-H$_2$O & $2_{12}-1_{01}$ & <5.2e-17 & -- & <4.1e-17 & -- \\
180.488 & o-H$_2$O & $2_{21}-2_{12}$ & <5.3e-17 & -- & <4.1e-17 & -- \\
56.3242 & p-H$_2$O & $4_{31}-3_{22}$ & <3.0e-16 & -- & 2.6e-16 & 1.1e-16 \\
57.6361 & p-H$_2$O & $4_{22}-3_{13}$ & 2.3e-16 & 8.3e-17 & <2.8e-16 & -- \\
61.8084 & p-H$_2$O & $4_{31}-4_{04}$ & <2.8e-16 & -- & <2.0e-16 & -- \\
63.457 & p-H$_2$O & $8_{08}-7_{17}$ & <2.9e-16 & -- & <1.9e-16 & -- \\
67.0886 & p-H$_2$O & $3_{31}-2_{20}$ & <1.7e-16 & -- & <1.9e-16 & -- \\
71.0662 & p-H$_2$O & $5_{24}-4_{13}$ & 1.16e-16 & 5.3e-17 & 4.35e-16 & 8.3e-17 \\
83.2831 & p-H$_2$O & $6_{06}-5_{15}$ & <1.4e-16 & -- & <1.7e-16 & -- \\
89.9878 & p-H$_2$O & $3_{22}-2_{11}$ & <1.4e-16 & -- & <1.6e-16 & -- \\
125.3534 & p-H$_2$O & $4_{04}-3_{13}$ & <4.0e-17 & -- & <4.0e-17 & -- \\
126.7126 & p-H$_2$O & $3_{31}-3_{22}$ & 3.1e-17 & 1.4e-17 & <4.1e-17 & -- \\
138.5272 & p-H$_2$O & $3_{13}-2_{02}$ & <3.5e-17 & -- & <3.4e-17 & -- \\
144.5181 & p-H$_2$O & $4_{13}-3_{22}$ & <4.7e-17 & -- & <3.5e-17 & -- \\
156.193 & p-H$_2$O & $3_{22}-3_{13}$ & <4.3e-17 & -- & <3.4e-17 & -- \\
158.309 & p-H$_2$O & $3_{31}-4_{04}$ & <4.3e-17 & -- & <3.4e-17 & -- \\
187.1104 & p-H$_2$O & $4_{13}-4_{04}$ & <7.9e-17 & -- & <5.5e-17 & -- \\
\hline
\end{tabular}
\end{table*}
    \begin{table*}
\centering
\begin{tabular}{lcc|cc|cc}
\hline\hline
& Molecule& &\multicolumn{2}{c|}{Haro 5a IRS} & \multicolumn{2}{c}{HH354 IRS} \\
Wavelength &  or  & Transition & $F$ & $\Delta F$ & $F$ & $\Delta F$    \\
($\mu$m)&Atom&&\multicolumn{2}{|c|}{(W $m^{-2}$)} &\multicolumn{2}{c}{(W $m^{-2}$)} \\
\hline
272.204627849 & $^{13}$CO & 10-9 & <2.9e-17 & -- & <1.5e-17 & -- \\
302.414601259 & $^{13}$CO & 9-8 & <1.7e-17 & -- & <2.2e-17 & -- \\
340.181218118 & $^{13}$CO & 8-7 & <9.5e-18 & -- & <9.2e-18 & -- \\
388.743038433 & $^{13}$CO & 7-6 & <4.6e-18 & -- & <5.0e-18 & -- \\
453.497650043 & $^{13}$CO & 6-5 & 7.0e-18 & 1.6e-18 & <8.5e-18 & -- \\
544.160740632 & $^{13}$CO & 5-4 & 1.56e-17 & 2.1e-18 & <1.0e-17 & -- \\
230.349129089 & [\ion{C}{i}] & $^3P_{2}-^3P_{0}$ & <3.2e-17 & -- & <2.5e-17 & -- \\
370.415064475 & [\ion{C}{i}] & $^3P_{2}-^3P_{1}$ & 3.977e-17 & 5.4e-19 & 6.26e-18 & 8.1e-19 \\
609.135366673 & [\ion{C}{i}] & $^3P_{1}-^3P_{0}$ & 1.57e-17 & 2.1e-18 & 1.96e-17 & 3.5e-18 \\
157.7409 & [\ion{C}{ii}] & $^2P_{3/2}-^2P_{1/2}$ & -- & -- & <3.9e-17 & -- \\
69.074406 & CO & 38-37 & -- & -- & <1.5e-16 & -- \\
70.907239 & CO & 37-36 & -- & -- & 2.82e-16 & 6.8e-17 \\
72.842854 & CO & 36-35 & -- & -- & <1.6e-16 & -- \\
74.89006 & CO & 35-34 & -- & -- & <1.3e-16 & -- \\
77.058699 & CO & 34-33 & -- & -- & <1.4e-16 & -- \\
79.35981 & CO & 33-32 & -- & -- & 1.89e-16 & 5.2e-17 \\
81.80581 & CO & 32-31 & -- & -- & \textit{9.2e-17} & 5.0e-17 \\
84.410717 & CO & 31-30 & -- & -- & <1.6e-16 & -- \\
87.190421 & CO & 30-29 & -- & -- & <1.7e-16 & -- \\
90.163002 & CO & 29-28 & -- & -- & <1.6e-16 & -- \\
93.349123 & CO & 28-27 & -- & -- & 3.43e-16 & 5.4e-17 \\
104.44495 & CO & 25-24 & -- & -- & <8.2e-17 & -- \\
108.76281 & CO & 24-23 & -- & -- & \textit{2.9e-17} & 2.4e-17 \\
113.4576 & CO & 23-22 & -- & -- & 6.5e-17 & 2.1e-17 \\
118.58072 & CO & 22-21 & -- & -- & 9.5e-17 & 1.7e-17 \\
124.19335 & CO & 21-20 & -- & -- & 4.2e-17 & 1.6e-17 \\
130.36893 & CO & 20-19 & -- & -- & 3.0e-17 & 1.4e-17 \\
137.19633 & CO & 19-18 & -- & -- & 6.5e-17 & 1.2e-17 \\
144.78419 & CO & 18-17 & -- & -- & 4.2e-17 & 1.3e-17 \\
153.26671 & CO & 17-16 & -- & -- & 5.0e-17 & 1.3e-17 \\
162.81163 & CO & 16-15 & -- & -- & \textit{2.3e-17} & 1.3e-17 \\
173.63143 & CO & 15-14 & -- & -- & 8.8e-17 & 1.5e-17 \\
185.9993 & CO & 14-13 & -- & -- & 9.6e-17 & 2.0e-17 \\
200.272477627 & CO & 13-12 & <9.5e-17 & -- & <5.7e-17 & -- \\
216.92729446 & CO & 12-11 & \textit{2.4e-17} & 1.4e-17 & 2.69e-17 & 9.5e-18 \\
236.613281063 & CO & 11-10 & \textit{2.0e-17} & 1.2e-17 & 2.56e-17 & 6.8e-18 \\
260.239806859 & CO & 10-9 & 4.83e-17 & 8.7e-18 & 2.86e-17 & 5.3e-18 \\
289.120332634 & CO & 9-8 & 3.31e-17 & 7.3e-18 & \textit{6.8e-18} & 3.6e-18 \\
325.225163341 & CO & 8-7 & 6.19e-17 & 3.3e-18 & 1.58e-17 & 3.0e-18 \\
371.6503924 & CO & 7-6 & 9.959e-17 & 5.4e-19 & 1.545e-17 & 8.1e-19 \\
433.556224633 & CO & 6-5 & 7.3e-17 & 1.5e-17 & 1.79e-17 & 3.8e-18 \\
520.231028647 & CO & 5-4 & 5.89e-17 & 1.7e-18 & 2.16e-17 & 3.4e-18 \\
650.251512464 & CO & 4-3 & 3.35e-17 & 1.8e-18 & <8.4e-18 & -- \\
205.17830087 & [\ion{N}{ii}] & $^3P_{1}-^3P_{0}$ & 1.13e-16 & 2.2e-17 & 5.5e-17 & 1.5e-17 \\

\hline
\end{tabular}
\end{table*}

    \begin{table*}
\centering
\begin{tabular}{lcc|cc|cc}
\hline\hline
& Molecule& &\multicolumn{2}{c|}{Haro 5a IRS} & \multicolumn{2}{c}{HH354 IRS} \\
Wavelength &  or  & Transition & $F$ & $\Delta F$ & $F$ & $\Delta F$    \\
($\mu$m)&Atom&&\multicolumn{2}{|c|}{(W $m^{-2}$)} &\multicolumn{2}{c}{(W $m^{-2}$)} \\
\hline
65.1315 & OH & $^2P_{3/2}\ J=9/2+ \rightarrow J=7/2-$ & -- & -- & <1.4e-16 & -- \\
65.2788 & OH & $^2P_{3/2}\ J=9/2+ \rightarrow J=7/2-$ & -- & -- & <1.4e-16 & -- \\
71.17085 & OH & $^2P_{1/2}\ J=7/2+ \rightarrow J=5/2-$ & -- & -- & <2.2e-16 & -- \\
71.215869 & OH & $^2P_{1/2}\ J=7/2+ \rightarrow J=5/2-$ & -- & -- & <2.3e-16 & -- \\
79.118034 & OH & $^2P_{1/2}\ J=1/2+ \rightarrow J=3/2-$ & -- & -- & <1.6e-16 & -- \\
79.181726 & OH & $^2P_{1/2}\ J=1/2+ \rightarrow J=3/2-$ & -- & -- & 1.89e-16 & 5.2e-17 \\
84.419938 & OH & $^2P_{3/2}\ J=7/2+ \rightarrow J=5/2-$ & \multirow{ 2}{*}{--} & \multirow{ 2}{*}{--} &\multirow{ 2}{*}{<1.6e-16}&\multirow{ 2}{*}{--} \\
84.420391 & OH & $^2P_{3/2}\ J=7/2+ \rightarrow J=5/2-$ &  &  &  &  \\
84.596825 & OH & $^2P_{3/2}\ J=7/2+ \rightarrow J=5/2-$ & -- & -- & <1.6e-16 & -- \\
115.1541 & OH & $^2P_{1/2}\ J=5/2+ \rightarrow J=7/2-$ & -- & -- & <5.8e-17 & -- \\
115.389 & OH & $^2P_{1/2}\ J=5/2+ \rightarrow J=7/2-$ & -- & -- & <6.0e-17 & -- \\
119.23418 & OH & $^2P_{3/2}\ J=5/2+ \rightarrow J=3/2-$ & -- & -- & <5.5e-17 & -- \\
119.44167 & OH & $^2P_{3/2}\ J=5/2+ \rightarrow J=3/2-$ & -- & -- & <5.6e-17 & -- \\
134.8476 & OH & $^2P_{1/2}\ J=7/2+ \rightarrow J=9/2-$ & -- & -- & <3.8e-17 & -- \\
134.9696 & OH & $^2P_{1/2}\ J=7/2+ \rightarrow J=9/2-$ & -- & -- & <3.9e-17 & -- \\
163.12428 & OH & $^2P_{1/2}\ J=3/2+ \rightarrow J=1/2-$ & -- & -- & 4.0e-17 & 1.4e-17 \\
163.39718 & OH & $^2P_{1/2}\ J=3/2+ \rightarrow J=1/2-$ & -- & -- & <4.2e-17 & -- \\
63.183705 & [\ion{O}{i}] & $^3P_{1}-^3P_{2}$ & -- & -- & 1.101e-15 & 5.2e-17 \\
145.52544 & [\ion{O}{i}] & $^3P_{0}-^3P_{1}$ & -- & -- & 6.9e-17 & 1.4e-17 \\
58.6982 & o-H$_2$O & $4_{32}-3_{21}$ & -- & -- & <1.9e-16 & -- \\
66.4372 & o-H$_2$O & $3_{30}-2_{21}$ & -- & -- & <1.5e-16 & -- \\
67.2689 & o-H$_2$O & $3_{30}-3_{03}$ & -- & -- & 1.74e-16 & 4.9e-17 \\
71.946 & o-H$_2$O & $7_{07}-6_{16}$ & -- & -- & <2.3e-16 & -- \\
75.3804 & o-H$_2$O & $3_{21}-2_{12}$ & -- & -- & <1.3e-16 & -- \\
78.7414 & o-H$_2$O & $4_{23}-3_{12}$ & -- & -- & <1.5e-16 & -- \\
82.0304 & o-H$_2$O & $6_{16}-5_{05}$ & -- & -- & <1.5e-16 & -- \\
108.073 & o-H$_2$O & $2_{21}-1_{10}$ & -- & -- & <7.5e-17 & -- \\
113.5366 & o-H$_2$O & $4_{14}-3_{03}$ & -- & -- & 6.0e-17 & 2.1e-17 \\
116.7836 & o-H$_2$O & $7_{34}-6_{43}$ & -- & -- & <5.6e-17 & -- \\
121.7191 & o-H$_2$O & $4_{32}-4_{23}$ & -- & -- & <4.6e-17 & -- \\
132.407 & o-H$_2$O & $4_{23}-4_{14}$ & -- & -- & 3.7e-17 & 1.5e-17 \\
134.9346 & o-H$_2$O & $5_{14}-5_{05}$ & -- & -- & <3.9e-17 & -- \\
136.4944 & o-H$_2$O & $3_{30}-3_{21}$ & -- & -- & <4.0e-17 & -- \\
174.6264 & o-H$_2$O & $3_{03}-2_{12}$ & -- & -- & <4.5e-17 & -- \\
179.5265 & o-H$_2$O & $2_{12}-1_{01}$ & -- & -- & <4.5e-17 & -- \\
180.488 & o-H$_2$O & $2_{21}-2_{12}$ & -- & -- & <4.3e-17 & -- \\
56.3242 & p-H$_2$O & $4_{31}-3_{22}$ & -- & -- & <2.6e-16 & -- \\
57.6361 & p-H$_2$O & $4_{22}-3_{13}$ & -- & -- & <2.2e-16 & -- \\
61.8084 & p-H$_2$O & $4_{31}-4_{04}$ & -- & -- & <2.0e-16 & -- \\
63.457 & p-H$_2$O & $8_{08}-7_{17}$ & -- & -- & <1.9e-16 & -- \\
67.0886 & p-H$_2$O & $3_{31}-2_{20}$ & -- & -- & <1.5e-16 & -- \\
71.0662 & p-H$_2$O & $5_{24}-4_{13}$ & -- & -- & <2.0e-16 & -- \\
83.2831 & p-H$_2$O & $6_{06}-5_{15}$ & -- & -- & <1.6e-16 & -- \\
89.9878 & p-H$_2$O & $3_{22}-2_{11}$ & -- & -- & <1.6e-16 & -- \\
125.3534 & p-H$_2$O & $4_{04}-3_{13}$ & -- & -- & <4.5e-17 & -- \\
126.7126 & p-H$_2$O & $3_{31}-3_{22}$ & -- & -- & <4.1e-17 & -- \\
138.5272 & p-H$_2$O & $3_{13}-2_{02}$ & -- & -- & <4.0e-17 & -- \\
144.5181 & p-H$_2$O & $4_{13}-3_{22}$ & -- & -- & <4.2e-17 & -- \\
156.193 & p-H$_2$O & $3_{22}-3_{13}$ & -- & -- & <4.0e-17 & -- \\
158.309 & p-H$_2$O & $3_{31}-4_{04}$ & -- & -- & <3.8e-17 & -- \\
187.1104 & p-H$_2$O & $4_{13}-4_{04}$ & -- & -- & <6.5e-17 & -- \\
\hline
\end{tabular}
\end{table*}
    \begin{table*}
\centering
\begin{tabular}{lcc|cc|cc}
\hline\hline
& Molecule& &\multicolumn{2}{c|}{HH381 IRS} & \multicolumn{2}{c}{Parsamian 21} \\
Wavelength &  or  & Transition & $F$ & $\Delta F$ & $F$ & $\Delta F$    \\
($\mu$m)&Atom&&\multicolumn{2}{|c|}{(W $m^{-2}$)} &\multicolumn{2}{c}{(W $m^{-2}$)} \\
\hline
272.204627849 & $^{13}$CO & 10-9 & <1.3e-17 & -- & <9.4e-18 & -- \\
302.414601259 & $^{13}$CO & 9-8 & <1.3e-17 & -- & <7.2e-18 & -- \\
340.181218118 & $^{13}$CO & 8-7 & <1.1e-17 & -- & <5.3e-18 & -- \\
388.743038433 & $^{13}$CO & 7-6 & <6.7e-18 & -- & <4.2e-18 & -- \\
453.497650043 & $^{13}$CO & 6-5 & <1.1e-17 & -- & <7.6e-18 & -- \\
544.160740632 & $^{13}$CO & 5-4 & <1.1e-17 & -- & <5.2e-18 & -- \\
230.349129089 & [\ion{C}{i}] & $^3P_{2}-^3P_{0}$ & 1.76e-17 & 6.0e-18 & <9.3e-18 & -- \\
370.415064475 & [\ion{C}{i}] & $^3P_{2}-^3P_{1}$ & 7.81e-18 & 8.9e-19 & <1.8e-18 & -- \\
609.135366673 & [\ion{C}{i}] & $^3P_{1}-^3P_{0}$ & 2.21e-17 & 4.1e-18 & <5.9e-18 & -- \\
157.7409 & [\ion{C}{ii}] & $^2P_{3/2}-^2P_{1/2}$ & 4.3e-17 & 1.2e-17 & <3.3e-17 & -- \\
69.074406 & CO & 38-37 & <1.6e-16 & -- & <1.4e-16 & -- \\
70.907239 & CO & 37-36 & 1.48e-16 & 6.9e-17 & <1.9e-16 & -- \\
72.842854 & CO & 36-35 & <1.7e-16 & -- & \textit{8.0e-17} & 4.6e-17 \\
74.89006 & CO & 35-34 & 1.95e-16 & 4.8e-17 & <1.1e-16 & -- \\
77.058699 & CO & 34-33 & <1.4e-16 & -- & <1.1e-16 & -- \\
79.35981 & CO & 33-32 & <1.5e-16 & -- & <1.2e-16 & -- \\
81.80581 & CO & 32-31 & <1.4e-16 & -- & <1.1e-16 & -- \\
84.410717 & CO & 31-30 & <1.6e-16 & -- & <1.4e-16 & -- \\
87.190421 & CO & 30-29 & <1.7e-16 & -- & <1.4e-16 & -- \\
90.163002 & CO & 29-28 & <1.6e-16 & -- & <1.3e-16 & -- \\
93.349123 & CO & 28-27 & \textit{5.9e-17} & 5.5e-17 & <1.4e-16 & -- \\
104.44495 & CO & 25-24 & 9.9e-17 & 3.0e-17 & <7.4e-17 & -- \\
108.76281 & CO & 24-23 & \textit{3.3e-17} & 2.5e-17 & <6.4e-17 & -- \\
113.4576 & CO & 23-22 & <6.3e-17 & -- & 4.9e-17 & 1.8e-17 \\
118.58072 & CO & 22-21 & 9.1e-17 & 1.9e-17 & 5.3e-17 & 1.6e-17 \\
124.19335 & CO & 21-20 & \textit{2.3e-17} & 1.5e-17 & \textit{2.2e-17} & 1.4e-17 \\
130.36893 & CO & 20-19 & 4.6e-17 & 1.4e-17 & 3.5e-17 & 1.2e-17 \\
137.19633 & CO & 19-18 & <4.1e-17 & -- & <3.4e-17 & -- \\
144.78419 & CO & 18-17 & \textit{1.2e-17} & 1.2e-17 & <3.5e-17 & -- \\
153.26671 & CO & 17-16 & \textit{1.9e-17} & 1.2e-17 & <3.4e-17 & -- \\
162.81163 & CO & 16-15 & \textit{1.3e-17} & 1.2e-17 & \textit{1.6e-17} & 1.2e-17 \\
173.63143 & CO & 15-14 & <4.0e-17 & -- & <3.5e-17 & -- \\
185.9993 & CO & 14-13 & \textit{2.4e-17} & 1.9e-17 & <5.3e-17 & -- \\
200.272477627 & CO & 13-12 & <3.1e-17 & -- & <1.6e-17 & -- \\
216.92729446 & CO & 12-11 & <1.7e-17 & -- & <9.8e-18 & -- \\
236.613281063 & CO & 11-10 & \textit{8.0e-18} & 6.5e-18 & <1.2e-17 & -- \\
260.239806859 & CO & 10-9 & <1.3e-17 & -- & <1.1e-17 & -- \\
289.120332634 & CO & 9-8 & <8.3e-18 & -- & <7.4e-18 & -- \\
325.225163341 & CO & 8-7 & 8.2e-18 & 2.6e-18 & <6.6e-18 & -- \\
371.6503924 & CO & 7-6 & 5.07e-18 & 8.9e-19 & 1.6e-18 & 6.2e-19 \\
433.556224633 & CO & 6-5 & <7.5e-18 & -- & <6.5e-18 & -- \\
520.231028647 & CO & 5-4 & <1.4e-17 & -- & 5.3e-18 & 2.2e-18 \\
650.251512464 & CO & 4-3 & <1.6e-17 & -- & <7.5e-18 & -- \\
205.17830087 & [\ion{N}{ii}] & $^3P_{1}-^3P_{0}$ & 3.24e-17 & 8.0e-18 & 2.6e-17 & 4.6e-18 \\

\hline
\end{tabular}
\end{table*}

    \begin{table*}
\centering
\begin{tabular}{lcc|cc|cc}
\hline\hline
& Molecule& &\multicolumn{2}{c|}{HH381 IRS} & \multicolumn{2}{c}{Parsamian 21} \\
Wavelength &  or  & Transition & $F$ & $\Delta F$ & $F$ & $\Delta F$    \\
($\mu$m)&Atom&&\multicolumn{2}{|c|}{(W $m^{-2}$)} &\multicolumn{2}{c}{(W $m^{-2}$)} \\
\hline
65.1315 & OH & $^2P_{3/2}\ J=9/2+ \rightarrow J=7/2-$ & <1.5e-16 & -- & <1.3e-16 & -- \\
65.2788 & OH & $^2P_{3/2}\ J=9/2+ \rightarrow J=7/2-$ & <1.6e-16 & -- & <1.3e-16 & -- \\
71.17085 & OH & $^2P_{1/2}\ J=7/2+ \rightarrow J=5/2-$ & 3.44e-16 & 9.0e-17 & <2.1e-16 & -- \\
71.215869 & OH & $^2P_{1/2}\ J=7/2+ \rightarrow J=5/2-$ & 2.31e-16 & 9.1e-17 & <2.1e-16 & -- \\
79.118034 & OH & $^2P_{1/2}\ J=1/2+ \rightarrow J=3/2-$ & <1.5e-16 & -- & 2.29e-16 & 4.0e-17 \\
79.181726 & OH & $^2P_{1/2}\ J=1/2+ \rightarrow J=3/2-$ & <1.5e-16 & -- & 1.92e-16 & 4.0e-17 \\
84.419938 & OH & $^2P_{3/2}\ J=7/2+ \rightarrow J=5/2-$ & \multirow{ 2}{*}{<1.6e-16} & \multirow{ 2}{*}{--} &\multirow{ 2}{*}{<1.4e-16}&\multirow{ 2}{*}{--} \\
84.420391 & OH & $^2P_{3/2}\ J=7/2+ \rightarrow J=5/2-$ &  &  &  &  \\
84.596825 & OH & $^2P_{3/2}\ J=7/2+ \rightarrow J=5/2-$ & <1.6e-16 & -- & <1.4e-16 & -- \\
115.1541 & OH & $^2P_{1/2}\ J=5/2+ \rightarrow J=7/2-$ & <6.1e-17 & -- & <5.0e-17 & -- \\
115.389 & OH & $^2P_{1/2}\ J=5/2+ \rightarrow J=7/2-$ & 5.1e-17 & 2.1e-17 & <5.2e-17 & -- \\
119.23418 & OH & $^2P_{3/2}\ J=5/2+ \rightarrow J=3/2-$ & <5.4e-17 & -- & 3.7e-17 & 1.6e-17 \\
119.44167 & OH & $^2P_{3/2}\ J=5/2+ \rightarrow J=3/2-$ & <5.1e-17 & -- & <4.8e-17 & -- \\
134.8476 & OH & $^2P_{1/2}\ J=7/2+ \rightarrow J=9/2-$ & 7.6e-17 & 1.3e-17 & <3.6e-17 & -- \\
134.9696 & OH & $^2P_{1/2}\ J=7/2+ \rightarrow J=9/2-$ & <3.9e-17 & -- & <3.7e-17 & -- \\
163.12428 & OH & $^2P_{1/2}\ J=3/2+ \rightarrow J=1/2-$ & <3.7e-17 & -- & <3.5e-17 & -- \\
163.39718 & OH & $^2P_{1/2}\ J=3/2+ \rightarrow J=1/2-$ & <3.8e-17 & -- & <3.6e-17 & -- \\
63.183705 & [\ion{O}{i}] & $^3P_{1}-^3P_{2}$ & 1.178e-15 & 5.7e-17 & <1.5e-16 & -- \\
145.52544 & [\ion{O}{i}] & $^3P_{0}-^3P_{1}$ & 4.3e-17 & 1.1e-17 & <3.6e-17 & -- \\
58.6982 & o-H$_2$O & $4_{32}-3_{21}$ & <1.9e-16 & -- & <1.6e-16 & -- \\
66.4372 & o-H$_2$O & $3_{30}-2_{21}$ & <1.6e-16 & -- & <1.3e-16 & -- \\
67.2689 & o-H$_2$O & $3_{30}-3_{03}$ & <1.6e-16 & -- & <1.3e-16 & -- \\
71.946 & o-H$_2$O & $7_{07}-6_{16}$ & 2.67e-16 & 8.9e-17 & <2.1e-16 & -- \\
75.3804 & o-H$_2$O & $3_{21}-2_{12}$ & <1.4e-16 & -- & 1.31e-16 & 3.7e-17 \\
78.7414 & o-H$_2$O & $4_{23}-3_{12}$ & <1.5e-16 & -- & <1.2e-16 & -- \\
82.0304 & o-H$_2$O & $6_{16}-5_{05}$ & <1.4e-16 & -- & <1.2e-16 & -- \\
108.073 & o-H$_2$O & $2_{21}-1_{10}$ & <7.8e-17 & -- & <6.9e-17 & -- \\
113.5366 & o-H$_2$O & $4_{14}-3_{03}$ & <6.2e-17 & -- & 4.2e-17 & 1.7e-17 \\
116.7836 & o-H$_2$O & $7_{34}-6_{43}$ & <5.7e-17 & -- & <4.9e-17 & -- \\
121.7191 & o-H$_2$O & $4_{32}-4_{23}$ & <4.7e-17 & -- & <4.3e-17 & -- \\
132.407 & o-H$_2$O & $4_{23}-4_{14}$ & <4.2e-17 & -- & <3.7e-17 & -- \\
134.9346 & o-H$_2$O & $5_{14}-5_{05}$ & <3.9e-17 & -- & <3.7e-17 & -- \\
136.4944 & o-H$_2$O & $3_{30}-3_{21}$ & <3.8e-17 & -- & <3.4e-17 & -- \\
174.6264 & o-H$_2$O & $3_{03}-2_{12}$ & <4.0e-17 & -- & <3.6e-17 & -- \\
179.5265 & o-H$_2$O & $2_{12}-1_{01}$ & <4.7e-17 & -- & <4.3e-17 & -- \\
180.488 & o-H$_2$O & $2_{21}-2_{12}$ & <4.7e-17 & -- & <4.4e-17 & -- \\
56.3242 & p-H$_2$O & $4_{31}-3_{22}$ & 2.75e-16 & 9.0e-17 & 4.57e-16 & 7.5e-17 \\
57.6361 & p-H$_2$O & $4_{22}-3_{13}$ & 1.83e-16 & 7.8e-17 & <1.9e-16 & -- \\
61.8084 & p-H$_2$O & $4_{31}-4_{04}$ & <1.8e-16 & -- & <1.4e-16 & -- \\
63.457 & p-H$_2$O & $8_{08}-7_{17}$ & <1.8e-16 & -- & <1.4e-16 & -- \\
67.0886 & p-H$_2$O & $3_{31}-2_{20}$ & <1.5e-16 & -- & <1.3e-16 & -- \\
71.0662 & p-H$_2$O & $5_{24}-4_{13}$ & <2.1e-16 & -- & 2.05e-16 & 6.6e-17 \\
83.2831 & p-H$_2$O & $6_{06}-5_{15}$ & <1.5e-16 & -- & <1.3e-16 & -- \\
89.9878 & p-H$_2$O & $3_{22}-2_{11}$ & <1.6e-16 & -- & <1.3e-16 & -- \\
125.3534 & p-H$_2$O & $4_{04}-3_{13}$ & <4.3e-17 & -- & <4.2e-17 & -- \\
126.7126 & p-H$_2$O & $3_{31}-3_{22}$ & <4.3e-17 & -- & <3.9e-17 & -- \\
138.5272 & p-H$_2$O & $3_{13}-2_{02}$ & <3.8e-17 & -- & <3.5e-17 & -- \\
144.5181 & p-H$_2$O & $4_{13}-3_{22}$ & <3.7e-17 & -- & <3.6e-17 & -- \\
156.193 & p-H$_2$O & $3_{22}-3_{13}$ & <3.6e-17 & -- & <3.3e-17 & -- \\
158.309 & p-H$_2$O & $3_{31}-4_{04}$ & <3.7e-17 & -- & 4.8e-17 & 1.0e-17 \\
187.1104 & p-H$_2$O & $4_{13}-4_{04}$ & <5.7e-17 & -- & 6.7e-17 & 1.9e-17 \\
\hline
\end{tabular}
\end{table*}
    \begin{table*}
\centering
\begin{tabular}{lcc|cc|cc}
\hline\hline
& Molecule& &\multicolumn{2}{c|}{PP13 S} & \multicolumn{2}{c}{Re 50 N IRS 1 / HBC494} \\
Wavelength &  or  & Transition & $F$ & $\Delta F$ & $F$ & $\Delta F$    \\
($\mu$m)&Atom&&\multicolumn{2}{|c|}{(W $m^{-2}$)} &\multicolumn{2}{c}{(W $m^{-2}$)} \\
\hline
272.204627849 & $^{13}$CO & 10-9 & <1.4e-17 & -- & <1.4e-17 & -- \\
302.414601259 & $^{13}$CO & 9-8 & <1.3e-17 & -- & <1.4e-17 & -- \\
340.181218118 & $^{13}$CO & 8-7 & <1.2e-17 & -- & <1.5e-17 & -- \\
388.743038433 & $^{13}$CO & 7-6 & <5.4e-18 & -- & <7.3e-18 & -- \\
453.497650043 & $^{13}$CO & 6-5 & <7.4e-18 & -- & <8.6e-18 & -- \\
544.160740632 & $^{13}$CO & 5-4 & <9.1e-18 & -- & 2.05e-17 & 3.2e-18 \\
230.349129089 & [\ion{C}{i}] & $^3P_{2}-^3P_{0}$ & <3.1e-17 & -- & <2.6e-17 & -- \\
370.415064475 & [\ion{C}{i}] & $^3P_{2}-^3P_{1}$ & 1.645e-17 & 8.6e-19 & 4.68e-17 & 7.7e-19 \\
609.135366673 & [\ion{C}{i}] & $^3P_{1}-^3P_{0}$ & 1.87e-17 & 2.8e-18 & 4.11e-17 & 2.8e-18 \\
157.7409 & [\ion{C}{ii}] & $^2P_{3/2}-^2P_{1/2}$ & 2.6e-17 & 1.3e-17 & 1.07e-16 & 1.3e-17 \\
69.074406 & CO & 38-37 & <1.5e-16 & -- & <1.6e-16 & -- \\
70.907239 & CO & 37-36 & 4.14e-16 & 6.1e-17 & 4.24e-16 & 7.2e-17 \\
72.842854 & CO & 36-35 & <1.4e-16 & -- & 1.57e-16 & 5.6e-17 \\
74.89006 & CO & 35-34 & <1.2e-16 & -- & <1.5e-16 & -- \\
77.058699 & CO & 34-33 & 9.0e-17 & 3.8e-17 & \textit{6.6e-17} & 4.6e-17 \\
79.35981 & CO & 33-32 & <1.4e-16 & -- & <1.6e-16 & -- \\
81.80581 & CO & 32-31 & <1.3e-16 & -- & \textit{5.1e-17} & 5.0e-17 \\
84.410717 & CO & 31-30 & 1.26e-16 & 5.0e-17 & 1.19e-16 & 5.8e-17 \\
87.190421 & CO & 30-29 & <1.4e-16 & -- & <1.7e-16 & -- \\
90.163002 & CO & 29-28 & <1.5e-16 & -- & \textit{9.3e-17} & 5.5e-17 \\
93.349123 & CO & 28-27 & <1.6e-16 & -- & 1.89e-16 & 6.0e-17 \\
104.44495 & CO & 25-24 & <8.2e-17 & -- & 8.5e-17 & 3.1e-17 \\
108.76281 & CO & 24-23 & <7.0e-17 & -- & \textit{4.7e-17} & 2.7e-17 \\
113.4576 & CO & 23-22 & 7.9e-17 & 2.0e-17 & 1.93e-16 & 2.2e-17 \\
118.58072 & CO & 22-21 & 7.2e-17 & 1.8e-17 & 1.83e-16 & 2.1e-17 \\
124.19335 & CO & 21-20 & 6.0e-17 & 1.5e-17 & 6.6e-17 & 1.6e-17 \\
130.36893 & CO & 20-19 & 1.12e-16 & 1.4e-17 & 1.43e-16 & 1.5e-17 \\
137.19633 & CO & 19-18 & 4.6e-17 & 1.3e-17 & 1.09e-16 & 1.3e-17 \\
144.78419 & CO & 18-17 & 3.1e-17 & 1.4e-17 & 1.16e-16 & 1.5e-17 \\
153.26671 & CO & 17-16 & 8.0e-17 & 1.2e-17 & 1.11e-16 & 1.3e-17 \\
162.81163 & CO & 16-15 & 1.4e-16 & 1.2e-17 & 1.47e-16 & 1.4e-17 \\
173.63143 & CO & 15-14 & 6.5e-17 & 1.4e-17 & 5.3e-17 & 1.5e-17 \\
185.9993 & CO & 14-13 & 1.07e-16 & 1.8e-17 & 1.73e-16 & 2.0e-17 \\
200.272477627 & CO & 13-12 & \textit{2.3e-17} & 1.8e-17 & 3.8e-17 & 1.7e-17 \\
216.92729446 & CO & 12-11 & 4.58e-17 & 1.0e-17 & 3.6e-17 & 8.9e-18 \\
236.613281063 & CO & 11-10 & 5.33e-17 & 6.7e-18 & 3.99e-17 & 6.2e-18 \\
260.239806859 & CO & 10-9 & 5.91e-17 & 5.3e-18 & 6.65e-17 & 4.4e-18 \\
289.120332634 & CO & 9-8 & 3.14e-17 & 3.7e-18 & 1.65e-17 & 4.9e-18 \\
325.225163341 & CO & 8-7 & 3.77e-17 & 4.7e-18 & 3.96e-17 & 5.0e-18 \\
371.6503924 & CO & 7-6 & 4.763e-17 & 8.6e-19 & 6.144e-17 & 7.7e-19 \\
433.556224633 & CO & 6-5 & 4.31e-17 & 8.2e-18 & 5.4e-17 & 1.1e-17 \\
520.231028647 & CO & 5-4 & 6.98e-17 & 3.3e-18 & 4.89e-17 & 3.3e-18 \\
650.251512464 & CO & 4-3 & 2.2e-17 & 2.3e-18 & 4.61e-17 & 3.3e-18 \\
205.17830087 & [\ion{N}{ii}] & $^3P_{1}-^3P_{0}$ & 5.1e-17 & 1.1e-17 & 5.73e-17 & 9.1e-18 \\

\hline
\end{tabular}
\end{table*}

    \begin{table*}
\centering
\begin{tabular}{lcc|cc|cc}
\hline\hline
& Molecule& &\multicolumn{2}{c|}{PP13 S} & \multicolumn{2}{c}{Re 50 N IRS 1 / HBC494} \\
Wavelength &  or  & Transition & $F$ & $\Delta F$ & $F$ & $\Delta F$    \\
($\mu$m)&Atom&&\multicolumn{2}{|c|}{(W $m^{-2}$)} &\multicolumn{2}{c}{(W $m^{-2}$)} \\
\hline
65.1315 & OH & $^2P_{3/2}\ J=9/2+ \rightarrow J=7/2-$ & <1.4e-16 & -- & <1.5e-16 & -- \\
65.2788 & OH & $^2P_{3/2}\ J=9/2+ \rightarrow J=7/2-$ & 1.49e-16 & 4.5e-17 & 1.48e-16 & 5.1e-17 \\
71.17085 & OH & $^2P_{1/2}\ J=7/2+ \rightarrow J=5/2-$ & <2.2e-16 & -- & <2.3e-16 & -- \\
71.215869 & OH & $^2P_{1/2}\ J=7/2+ \rightarrow J=5/2-$ & <2.3e-16 & -- & <2.3e-16 & -- \\
79.118034 & OH & $^2P_{1/2}\ J=1/2+ \rightarrow J=3/2-$ & 2.21e-16 & 4.4e-17 & 1.66e-16 & 5.1e-17 \\
79.181726 & OH & $^2P_{1/2}\ J=1/2+ \rightarrow J=3/2-$ & <1.3e-16 & -- & <1.5e-16 & -- \\
84.419938 & OH & $^2P_{3/2}\ J=7/2+ \rightarrow J=5/2-$ & \multirow{ 2}{*}{1.25e-16} & \multirow{ 2}{*}{5.0e-17} &\multirow{ 2}{*}{<1.7e-16}&\multirow{ 2}{*}{--} \\
84.420391 & OH & $^2P_{3/2}\ J=7/2+ \rightarrow J=5/2-$ &  &  &  &  \\
84.596825 & OH & $^2P_{3/2}\ J=7/2+ \rightarrow J=5/2-$ & <1.5e-16 & -- & <1.7e-16 & -- \\
115.1541 & OH & $^2P_{1/2}\ J=5/2+ \rightarrow J=7/2-$ & <5.6e-17 & -- & <6.6e-17 & -- \\
115.389 & OH & $^2P_{1/2}\ J=5/2+ \rightarrow J=7/2-$ & <5.5e-17 & -- & 8.6e-17 & 2.2e-17 \\
119.23418 & OH & $^2P_{3/2}\ J=5/2+ \rightarrow J=3/2-$ & <5.0e-17 & -- & <6.5e-17 & -- \\
119.44167 & OH & $^2P_{3/2}\ J=5/2+ \rightarrow J=3/2-$ & <5.3e-17 & -- & <6.5e-17 & -- \\
134.8476 & OH & $^2P_{1/2}\ J=7/2+ \rightarrow J=9/2-$ & <4.1e-17 & -- & 6.3e-17 & 1.5e-17 \\
134.9696 & OH & $^2P_{1/2}\ J=7/2+ \rightarrow J=9/2-$ & <4.1e-17 & -- & 6.0e-17 & 1.5e-17 \\
163.12428 & OH & $^2P_{1/2}\ J=3/2+ \rightarrow J=1/2-$ & <4.9e-17 & -- & <5.4e-17 & -- \\
163.39718 & OH & $^2P_{1/2}\ J=3/2+ \rightarrow J=1/2-$ & <4.8e-17 & -- & <5.3e-17 & -- \\
63.183705 & [\ion{O}{i}] & $^3P_{1}-^3P_{2}$ & 1.48e-15 & 4.9e-17 & 1.648e-15 & 5.3e-17 \\
145.52544 & [\ion{O}{i}] & $^3P_{0}-^3P_{1}$ & 1.07e-16 & 1.4e-17 & 8.1e-17 & 1.8e-17 \\
58.6982 & o-H$_2$O & $4_{32}-3_{21}$ & <1.7e-16 & -- & <1.8e-16 & -- \\
66.4372 & o-H$_2$O & $3_{30}-2_{21}$ & <1.4e-16 & -- & <1.6e-16 & -- \\
67.2689 & o-H$_2$O & $3_{30}-3_{03}$ & <1.4e-16 & -- & <1.6e-16 & -- \\
71.946 & o-H$_2$O & $7_{07}-6_{16}$ & <2.2e-16 & -- & <2.3e-16 & -- \\
75.3804 & o-H$_2$O & $3_{21}-2_{12}$ & <1.1e-16 & -- & <1.4e-16 & -- \\
78.7414 & o-H$_2$O & $4_{23}-3_{12}$ & <1.3e-16 & -- & <1.6e-16 & -- \\
82.0304 & o-H$_2$O & $6_{16}-5_{05}$ & <1.3e-16 & -- & <1.5e-16 & -- \\
108.073 & o-H$_2$O & $2_{21}-1_{10}$ & 1.85e-16 & 2.3e-17 & 5.6e-17 & 2.8e-17 \\
113.5366 & o-H$_2$O & $4_{14}-3_{03}$ & 8.2e-17 & 2.0e-17 & 1.45e-16 & 2.2e-17 \\
116.7836 & o-H$_2$O & $7_{34}-6_{43}$ & <5.4e-17 & -- & <6.6e-17 & -- \\
121.7191 & o-H$_2$O & $4_{32}-4_{23}$ & <4.8e-17 & -- & <4.8e-17 & -- \\
132.407 & o-H$_2$O & $4_{23}-4_{14}$ & <4.0e-17 & -- & <4.2e-17 & -- \\
134.9346 & o-H$_2$O & $5_{14}-5_{05}$ & <4.1e-17 & -- & 7.8e-17 & 1.5e-17 \\
136.4944 & o-H$_2$O & $3_{30}-3_{21}$ & <4.2e-17 & -- & <5.3e-17 & -- \\
174.6264 & o-H$_2$O & $3_{03}-2_{12}$ & <4.5e-17 & -- & <5.3e-17 & -- \\
179.5265 & o-H$_2$O & $2_{12}-1_{01}$ & 4.0e-17 & 1.5e-17 & <4.7e-17 & -- \\
180.488 & o-H$_2$O & $2_{21}-2_{12}$ & <5.0e-17 & -- & <5.0e-17 & -- \\
56.3242 & p-H$_2$O & $4_{31}-3_{22}$ & <2.3e-16 & -- & 6.77e-16 & 8.8e-17 \\
57.6361 & p-H$_2$O & $4_{22}-3_{13}$ & <2.0e-16 & -- & <2.2e-16 & -- \\
61.8084 & p-H$_2$O & $4_{31}-4_{04}$ & <1.9e-16 & -- & <2.3e-16 & -- \\
63.457 & p-H$_2$O & $8_{08}-7_{17}$ & <1.9e-16 & -- & <2.2e-16 & -- \\
67.0886 & p-H$_2$O & $3_{31}-2_{20}$ & <1.4e-16 & -- & <1.6e-16 & -- \\
71.0662 & p-H$_2$O & $5_{24}-4_{13}$ & <1.9e-16 & -- & <2.1e-16 & -- \\
83.2831 & p-H$_2$O & $6_{06}-5_{15}$ & <1.4e-16 & -- & <1.6e-16 & -- \\
89.9878 & p-H$_2$O & $3_{22}-2_{11}$ & <1.5e-16 & -- & <1.7e-16 & -- \\
125.3534 & p-H$_2$O & $4_{04}-3_{13}$ & <4.5e-17 & -- & 3.4e-17 & 1.7e-17 \\
126.7126 & p-H$_2$O & $3_{31}-3_{22}$ & <3.9e-17 & -- & <4.3e-17 & -- \\
138.5272 & p-H$_2$O & $3_{13}-2_{02}$ & <4.2e-17 & -- & 3.3e-17 & 1.6e-17 \\
144.5181 & p-H$_2$O & $4_{13}-3_{22}$ & <4.5e-17 & -- & <5.7e-17 & -- \\
156.193 & p-H$_2$O & $3_{22}-3_{13}$ & <3.8e-17 & -- & 4.3e-17 & 1.3e-17 \\
158.309 & p-H$_2$O & $3_{31}-4_{04}$ & <3.8e-17 & -- & <4.4e-17 & -- \\
187.1104 & p-H$_2$O & $4_{13}-4_{04}$ & 4.8e-17 & 2.1e-17 & <7.0e-17 & -- \\
\hline
\end{tabular}
\end{table*}
    \begin{table*}
\centering
\begin{tabular}{lcc|cc|cc}
\hline\hline
& Molecule& &\multicolumn{2}{c|}{V346 Nor} & \multicolumn{2}{c}{V733 Cep} \\
Wavelength &  or  & Transition & $F$ & $\Delta F$ & $F$ & $\Delta F$    \\
($\mu$m)&Atom&&\multicolumn{2}{|c|}{(W $m^{-2}$)} &\multicolumn{2}{c}{(W $m^{-2}$)} \\
\hline
272.204627849 & $^{13}$CO & 10-9 & <1.6e-17 & -- & <1.3e-17 & -- \\
302.414601259 & $^{13}$CO & 9-8 & <1.6e-17 & -- & <1.1e-17 & -- \\
340.181218118 & $^{13}$CO & 8-7 & 1.44e-17 & 4.2e-18 & <6.8e-18 & -- \\
388.743038433 & $^{13}$CO & 7-6 & <5.9e-18 & -- & <4.4e-18 & -- \\
453.497650043 & $^{13}$CO & 6-5 & 9.5e-18 & 2.8e-18 & <8.1e-18 & -- \\
544.160740632 & $^{13}$CO & 5-4 & 1.07e-17 & 3.2e-18 & 8.9e-18 & 2.8e-18 \\
230.349129089 & [\ion{C}{i}] & $^3P_{2}-^3P_{0}$ & <3.5e-17 & -- & <8.1e-18 & -- \\
370.415064475 & [\ion{C}{i}] & $^3P_{2}-^3P_{1}$ & 3.058e-17 & 7.1e-19 & 3.618e-17 & 8.3e-19 \\
609.135366673 & [\ion{C}{i}] & $^3P_{1}-^3P_{0}$ & 1.69e-17 & 3.4e-18 & 3.39e-17 & 2.4e-18 \\
157.7409 & [\ion{C}{ii}] & $^2P_{3/2}-^2P_{1/2}$ & <3.8e-17 & -- & 2.32e-16 & 3.0e-17 \\
69.074406 & CO & 38-37 & \textit{5.2e-17} & 5.0e-17 & <1.4e-16 & -- \\
70.907239 & CO & 37-36 & 7.31e-16 & 6.5e-17 & <1.8e-16 & -- \\
72.842854 & CO & 36-35 & <1.6e-16 & -- & <1.4e-16 & -- \\
74.89006 & CO & 35-34 & 3.31e-16 & 3.8e-17 & <1.1e-16 & -- \\
77.058699 & CO & 34-33 & <1.2e-16 & -- & \textit{5.7e-17} & 3.7e-17 \\
79.35981 & CO & 33-32 & 1.44e-16 & 4.6e-17 & 1.73e-16 & 4.3e-17 \\
81.80581 & CO & 32-31 & \textit{8.9e-17} & 4.7e-17 & <1.2e-16 & -- \\
84.410717 & CO & 31-30 & <1.5e-16 & -- & <1.4e-16 & -- \\
87.190421 & CO & 30-29 & <1.6e-16 & -- & <1.5e-16 & -- \\
90.163002 & CO & 29-28 & <1.4e-16 & -- & <1.3e-16 & -- \\
93.349123 & CO & 28-27 & \textit{8.1e-17} & 5.1e-17 & <1.3e-16 & -- \\
104.44495 & CO & 25-24 & 8.5e-17 & 2.7e-17 & <7.7e-17 & -- \\
108.76281 & CO & 24-23 & 5.7e-17 & 2.3e-17 & <6.5e-17 & -- \\
113.4576 & CO & 23-22 & 3.71e-16 & 2.1e-17 & <5.3e-17 & -- \\
118.58072 & CO & 22-21 & 1.54e-16 & 2.7e-17 & \textit{1.8e-17} & 1.6e-17 \\
124.19335 & CO & 21-20 & 1.88e-16 & 1.5e-17 & <3.8e-17 & -- \\
130.36893 & CO & 20-19 & 1.9e-16 & 1.3e-17 & <3.6e-17 & -- \\
137.19633 & CO & 19-18 & 6.4e-17 & 1.4e-17 & <3.3e-17 & -- \\
144.78419 & CO & 18-17 & 1.47e-16 & 1.4e-17 & <1.4e-16 & -- \\
153.26671 & CO & 17-16 & 1.28e-16 & 1.3e-17 & <9.8e-17 & -- \\
162.81163 & CO & 16-15 & 2.52e-16 & 1.4e-17 & \textit{3.6e-17} & 3.4e-17 \\
173.63143 & CO & 15-14 & 1.15e-16 & 1.5e-17 & <1.3e-16 & -- \\
185.9993 & CO & 14-13 & 1.71e-16 & 2.0e-17 & <1.6e-16 & -- \\
200.272477627 & CO & 13-12 & 5.3e-17 & 1.8e-17 & <1.7e-17 & -- \\
216.92729446 & CO & 12-11 & 4.9e-17 & 1.1e-17 & 1.09e-17 & 4.6e-18 \\
236.613281063 & CO & 11-10 & 5.64e-17 & 6.6e-18 & <1.0e-17 & -- \\
260.239806859 & CO & 10-9 & 7.66e-17 & 5.5e-18 & <1.1e-17 & -- \\
289.120332634 & CO & 9-8 & 5.96e-17 & 4.4e-18 & <1.1e-17 & -- \\
325.225163341 & CO & 8-7 & 1.173e-16 & 4.9e-18 & 1.11e-17 & 2.7e-18 \\
371.6503924 & CO & 7-6 & 1.483e-16 & 7.1e-19 & 1.533e-17 & 8.3e-19 \\
433.556224633 & CO & 6-5 & 1.21e-16 & 2.4e-17 & 2.01e-17 & 3.6e-18 \\
520.231028647 & CO & 5-4 & 9.66e-17 & 2.9e-18 & 3.26e-17 & 2.1e-18 \\
650.251512464 & CO & 4-3 & 4.29e-17 & 4.3e-18 & 2.07e-17 & 2.0e-18 \\
205.17830087 & [\ion{N}{ii}] & $^3P_{1}-^3P_{0}$ & 9.5e-17 & 1.4e-17 & 4.78e-17 & 4.9e-18 \\

\hline
\end{tabular}
\end{table*}

    \begin{table*}
\centering
\begin{tabular}{lcc|cc|cc}
\hline\hline
& Molecule& &\multicolumn{2}{c|}{V346 Nor} & \multicolumn{2}{c}{V733 Cep} \\
Wavelength &  or  & Transition & $F$ & $\Delta F$ & $F$ & $\Delta F$    \\
($\mu$m)&Atom&&\multicolumn{2}{|c|}{(W $m^{-2}$)} &\multicolumn{2}{c}{(W $m^{-2}$)} \\
\hline
65.1315 & OH & $^2P_{3/2}\ J=9/2+ \rightarrow J=7/2-$ & <1.4e-16 & -- & <1.3e-16 & -- \\
65.2788 & OH & $^2P_{3/2}\ J=9/2+ \rightarrow J=7/2-$ & <1.4e-16 & -- & <1.3e-16 & -- \\
71.17085 & OH & $^2P_{1/2}\ J=7/2+ \rightarrow J=5/2-$ & <2.0e-16 & -- & <2.0e-16 & -- \\
71.215869 & OH & $^2P_{1/2}\ J=7/2+ \rightarrow J=5/2-$ & <2.1e-16 & -- & <2.0e-16 & -- \\
79.118034 & OH & $^2P_{1/2}\ J=1/2+ \rightarrow J=3/2-$ & <1.4e-16 & -- & <1.3e-16 & -- \\
79.181726 & OH & $^2P_{1/2}\ J=1/2+ \rightarrow J=3/2-$ & <1.4e-16 & -- & <1.3e-16 & -- \\
84.419938 & OH & $^2P_{3/2}\ J=7/2+ \rightarrow J=5/2-$ & <1.5e-16 & -- & <1.4e-16 & -- \\
84.419938 & OH & $^2P_{3/2}\ J=7/2+ \rightarrow J=5/2-$ & \multirow{ 2}{*}{<1.5e-16} & \multirow{ 2}{*}{--} &\multirow{ 2}{*}{<1.4e-16}&\multirow{ 2}{*}{--} \\
84.420391 & OH & $^2P_{3/2}\ J=7/2+ \rightarrow J=5/2-$ &  &  &  &  \\
84.596825 & OH & $^2P_{3/2}\ J=7/2+ \rightarrow J=5/2-$ & <1.5e-16 & -- & <1.5e-16 & -- \\
115.1541 & OH & $^2P_{1/2}\ J=5/2+ \rightarrow J=7/2-$ & <5.7e-17 & -- & <4.7e-17 & -- \\
115.389 & OH & $^2P_{1/2}\ J=5/2+ \rightarrow J=7/2-$ & <5.9e-17 & -- & <4.9e-17 & -- \\
119.23418 & OH & $^2P_{3/2}\ J=5/2+ \rightarrow J=3/2-$ & <7.5e-17 & -- & 3.8e-17 & 1.6e-17 \\
119.44167 & OH & $^2P_{3/2}\ J=5/2+ \rightarrow J=3/2-$ & <7.4e-17 & -- & <4.8e-17 & -- \\
134.8476 & OH & $^2P_{1/2}\ J=7/2+ \rightarrow J=9/2-$ & <4.2e-17 & -- & <3.5e-17 & -- \\
134.9696 & OH & $^2P_{1/2}\ J=7/2+ \rightarrow J=9/2-$ & <4.2e-17 & -- & <3.5e-17 & -- \\
163.12428 & OH & $^2P_{1/2}\ J=3/2+ \rightarrow J=1/2-$ & <5.5e-17 & -- & <1.1e-16 & -- \\
163.39718 & OH & $^2P_{1/2}\ J=3/2+ \rightarrow J=1/2-$ & <5.5e-17 & -- & <1.1e-16 & -- \\
63.183705 & [\ion{O}{i}] & $^3P_{0}-^3P_{1}$ & 4.73e-16 & 5.1e-17 & <1.5e-16 & -- \\
145.52544 & [\ion{O}{i}] & $^3P_{1}-^3P_{2}$ & 5.3e-17 & 1.8e-17 & 1.41e-16 & 4.3e-17 \\
58.6982 & o-H$_2$O & $4_{32}-3_{21}$ & <1.7e-16 & -- & <1.6e-16 & -- \\
66.4372 & o-H$_2$O & $3_{30}-2_{21}$ & <1.4e-16 & -- & <1.3e-16 & -- \\
67.2689 & o-H$_2$O & $3_{30}-3_{03}$ & <1.4e-16 & -- & <1.3e-16 & -- \\
71.946 & o-H$_2$O & $7_{07}-6_{16}$ & <2.2e-16 & -- & <1.5e-15 & -- \\
75.3804 & o-H$_2$O & $3_{21}-2_{12}$ & <1.1e-16 & -- & <1.0e-16 & -- \\
78.7414 & o-H$_2$O & $4_{23}-3_{12}$ & <1.4e-16 & -- & <1.2e-16 & -- \\
82.0304 & o-H$_2$O & $6_{16}-5_{05}$ & 1.61e-16 & 4.7e-17 & <1.2e-16 & -- \\
108.073 & o-H$_2$O & $2_{21}-1_{10}$ & <7.5e-17 & -- & <6.7e-17 & -- \\
113.5366 & o-H$_2$O & $4_{14}-3_{03}$ & 3.66e-16 & 2.1e-17 & <5.2e-17 & -- \\
116.7836 & o-H$_2$O & $7_{34}-6_{43}$ & 4.7e-17 & 1.9e-17 & <4.7e-17 & -- \\
121.7191 & o-H$_2$O & $4_{32}-4_{23}$ & <4.7e-17 & -- & <4.4e-17 & -- \\
132.407 & o-H$_2$O & $4_{23}-4_{14}$ & 6.4e-17 & 1.3e-17 & <3.6e-17 & -- \\
134.9346 & o-H$_2$O & $5_{14}-5_{05}$ & 3.9e-17 & 1.4e-17 & <3.5e-17 & -- \\
136.4944 & o-H$_2$O & $3_{30}-3_{21}$ & <4.7e-17 & -- & <3.4e-17 & -- \\
174.6264 & o-H$_2$O & $3_{03}-2_{12}$ & 1.08e-16 & 1.7e-17 & <1.3e-16 & -- \\
179.5265 & o-H$_2$O & $2_{12}-1_{01}$ & 1.02e-16 & 1.5e-17 & <1.1e-16 & -- \\
180.488 & o-H$_2$O & $2_{21}-2_{12}$ & <5.1e-17 & -- & <1.2e-16 & -- \\
56.3242 & p-H$_2$O & $4_{31}-3_{22}$ & <2.4e-16 & -- & 3.5e-16 & 7.7e-17 \\
57.6361 & p-H$_2$O & $4_{22}-3_{13}$ & <2.0e-16 & -- & 1.58e-16 & 6.5e-17 \\
61.8084 & p-H$_2$O & $4_{31}-4_{04}$ & <1.8e-16 & -- & <1.5e-16 & -- \\
63.457 & p-H$_2$O & $8_{08}-7_{17}$ & <1.7e-16 & -- & <1.5e-16 & -- \\
67.0886 & p-H$_2$O & $3_{31}-2_{20}$ & <1.4e-16 & -- & <1.3e-16 & -- \\
71.0662 & p-H$_2$O & $5_{24}-4_{13}$ & <2.0e-16 & -- & <1.9e-16 & -- \\
83.2831 & p-H$_2$O & $6_{06}-5_{15}$ & <1.4e-16 & -- & <1.3e-16 & -- \\
89.9878 & p-H$_2$O & $3_{22}-2_{11}$ & <1.4e-16 & -- & <1.3e-16 & -- \\
125.3534 & p-H$_2$O & $4_{04}-3_{13}$ & <4.9e-17 & -- & <3.5e-17 & -- \\
126.7126 & p-H$_2$O & $3_{31}-3_{22}$ & <4.6e-17 & -- & <3.5e-17 & -- \\
138.5272 & p-H$_2$O & $3_{13}-2_{02}$ & 7.2e-17 & 1.5e-17 & <3.4e-17 & -- \\
144.5181 & p-H$_2$O & $4_{13}-3_{22}$ & <5.3e-17 & -- & <1.4e-16 & -- \\
156.193 & p-H$_2$O & $3_{22}-3_{13}$ & 4.0e-17 & 1.2e-17 & <9.8e-17 & -- \\
158.309 & p-H$_2$O & $3_{31}-4_{04}$ & <4.2e-17 & -- & <1.2e-16 & -- \\
187.1104 & p-H$_2$O & $4_{13}-4_{04}$ & <7.4e-17 & -- & <1.7e-16 & -- \\
\hline
\end{tabular}
\end{table*}
    \begin{table*}
\centering
\begin{tabular}{lcc|cc|cc}
\hline\hline
& Molecule& &\multicolumn{2}{c|}{V883 Ori} & \multicolumn{2}{c}{V1647 Ori} \\
Wavelength &  or  & Transition & $F$ & $\Delta F$ & $F$ & $\Delta F$    \\
($\mu$m)&Atom&&\multicolumn{2}{|c|}{(W $m^{-2}$)} &\multicolumn{2}{c}{(W $m^{-2}$)} \\
\hline
272.204627849 & $^{13}$CO & 10-9 & <1.4e-17 & -- & <1.0e-17 & -- \\
302.414601259 & $^{13}$CO & 9-8 & <1.5e-17 & -- & <1.2e-17 & -- \\
340.181218118 & $^{13}$CO & 8-7 & 1.15e-17 & 3.4e-18 & <6.8e-18 & -- \\
388.743038433 & $^{13}$CO & 7-6 & <4.7e-18 & -- & <4.5e-18 & -- \\
453.497650043 & $^{13}$CO & 6-5 & <6.8e-18 & -- & <9.1e-18 & -- \\
544.160740632 & $^{13}$CO & 5-4 & <7.8e-18 & -- & <1.1e-17 & -- \\
230.349129089 & [\ion{C}{i}] & $^3P_{2}-^3P_{0}$ & <2.2e-17 & -- & <1.1e-17 & -- \\
370.415064475 & [\ion{C}{i}] & $^3P_{2}-^3P_{1}$ & 2.751e-17 & 9.3e-19 & 4.614e-17 & 6.6e-19 \\
609.135366673 & [\ion{C}{i}] & $^3P_{1}-^3P_{0}$ & 3.19e-17 & 3.3e-18 & 2.98e-17 & 3.8e-18 \\
157.7409 & [\ion{C}{ii}] & $^2P_{3/2}-^2P_{1/2}$ & 6.5e-17 & 1.2e-17 & <3.6e-17 & -- \\
69.074406 & CO & 38-37 & <1.5e-16 & -- & <1.5e-16 & -- \\
70.907239 & CO & 37-36 & \textit{8.3e-17} & 6.7e-17 & <1.8e-16 & -- \\
72.842854 & CO & 36-35 & <1.5e-16 & -- & <1.4e-16 & -- \\
74.89006 & CO & 35-34 & <1.2e-16 & -- & <1.2e-16 & -- \\
77.058699 & CO & 34-33 & <1.3e-16 & -- & <1.2e-16 & -- \\
79.35981 & CO & 33-32 & 1.08e-16 & 4.9e-17 & <1.3e-16 & -- \\
81.80581 & CO & 32-31 & 2.66e-16 & 4.7e-17 & <1.3e-16 & -- \\
84.410717 & CO & 31-30 & <1.5e-16 & -- & <1.5e-16 & -- \\
87.190421 & CO & 30-29 & <1.5e-16 & -- & \textit{6.0e-17} & 4.9e-17 \\
90.163002 & CO & 29-28 & <1.4e-16 & -- & <1.4e-16 & -- \\
93.349123 & CO & 28-27 & <1.6e-16 & -- & <1.5e-16 & -- \\
104.44495 & CO & 25-24 & <8.5e-17 & -- & 8.4e-17 & 2.7e-17 \\
108.76281 & CO & 24-23 & <8.1e-17 & -- & <7.0e-17 & -- \\
113.4576 & CO & 23-22 & <6.4e-17 & -- & <6.1e-17 & -- \\
118.58072 & CO & 22-21 & 5.6e-17 & 2.3e-17 & \textit{2.8e-17} & 1.6e-17 \\
124.19335 & CO & 21-20 & 4.7e-17 & 1.6e-17 & <4.4e-17 & -- \\
130.36893 & CO & 20-19 & \textit{2.3e-17} & 1.3e-17 & \textit{1.5e-17} & 1.3e-17 \\
137.19633 & CO & 19-18 & \textit{1.6e-17} & 1.4e-17 & 4.4e-17 & 1.2e-17 \\
144.78419 & CO & 18-17 & <3.9e-17 & -- & 2.6e-17 & 1.1e-17 \\
153.26671 & CO & 17-16 & <3.7e-17 & -- & 2.8e-17 & 1.2e-17 \\
162.81163 & CO & 16-15 & \textit{1.3e-17} & 1.3e-17 & <3.7e-17 & -- \\
173.63143 & CO & 15-14 & \textit{1.6e-17} & 1.3e-17 & <3.9e-17 & -- \\
185.9993 & CO & 14-13 & \textit{2.8e-17} & 1.9e-17 & \textit{1.9e-17} & 1.8e-17 \\
200.272477627 & CO & 13-12 & <6.5e-17 & -- & <2.5e-17 & -- \\
216.92729446 & CO & 12-11 & <2.7e-17 & -- & <1.1e-17 & -- \\
236.613281063 & CO & 11-10 & <2.3e-17 & -- & 8.1e-18 & 3.6e-18 \\
260.239806859 & CO & 10-9 & <1.4e-17 & -- & <1.1e-17 & -- \\
289.120332634 & CO & 9-8 & <1.1e-17 & -- & <9.4e-18 & -- \\
325.225163341 & CO & 8-7 & <9.5e-18 & -- & 8.7e-18 & 2.4e-18 \\
371.6503924 & CO & 7-6 & 9.56e-18 & 9.3e-19 & 3.304e-17 & 6.6e-19 \\
433.556224633 & CO & 6-5 & 1.26e-17 & 2.9e-18 & 4.15e-17 & 8.5e-18 \\
520.231028647 & CO & 5-4 & 4.09e-17 & 3.3e-18 & 5.93e-17 & 2.9e-18 \\
650.251512464 & CO & 4-3 & 2.22e-17 & 3.5e-18 & 3.49e-17 & 2.8e-18 \\
205.17830087 & [\ion{N}{ii}] & $^3P_{1}-^3P_{0}$ & 7.4e-17 & 1.5e-17 & 2.5e-17 & 6.0e-18 \\

\hline
\end{tabular}
\end{table*}

    \begin{table*}
\centering
\begin{tabular}{lcc|cc|cc}
\hline\hline
& Molecule& &\multicolumn{2}{c|}{V883 Ori} & \multicolumn{2}{c}{V1647 Ori} \\
Wavelength &  or  & Transition & $F$ & $\Delta F$ & $F$ & $\Delta F$    \\
($\mu$m)&Atom&&\multicolumn{2}{|c|}{(W $m^{-2}$)} &\multicolumn{2}{c}{(W $m^{-2}$)} \\
\hline
65.1315 & OH & $^2P_{3/2}\ J=9/2+ \rightarrow J=7/2-$ & <1.4e-16 & -- & <1.4e-16 & -- \\
65.2788 & OH & $^2P_{3/2}\ J=9/2+ \rightarrow J=7/2-$ & <1.4e-16 & -- & <1.3e-16 & -- \\
71.17085 & OH & $^2P_{1/2}\ J=7/2+ \rightarrow J=5/2-$ & 2.43e-16 & 6.6e-17 & 1.63e-16 & 7.1e-17 \\
71.215869 & OH & $^2P_{1/2}\ J=7/2+ \rightarrow J=5/2-$ & <2.0e-16 & -- & <2.2e-16 & -- \\
79.118034 & OH & $^2P_{1/2}\ J=1/2+ \rightarrow J=3/2-$ & <1.5e-16 & -- & <1.3e-16 & -- \\
79.181726 & OH & $^2P_{1/2}\ J=1/2+ \rightarrow J=3/2-$ & <1.4e-16 & -- & <1.3e-16 & -- \\
84.419938 & OH & $^2P_{3/2}\ J=7/2+ \rightarrow J=5/2-$ & <1.5e-16 & -- & <1.5e-16 & -- \\
84.419938 & OH & $^2P_{3/2}\ J=7/2+ \rightarrow J=5/2-$ & \multirow{ 2}{*}{<1.5e-16} & \multirow{ 2}{*}{--} &\multirow{ 2}{*}{<1.5e-16}&\multirow{ 2}{*}{--} \\
84.420391 & OH & $^2P_{3/2}\ J=7/2+ \rightarrow J=5/2-$ &  &  &  &  \\
84.596825 & OH & $^2P_{3/2}\ J=7/2+ \rightarrow J=5/2-$ & <1.5e-16 & -- & <1.5e-16 & -- \\
115.1541 & OH & $^2P_{1/2}\ J=5/2+ \rightarrow J=7/2-$ & <5.8e-17 & -- & <5.5e-17 & -- \\
115.389 & OH & $^2P_{1/2}\ J=5/2+ \rightarrow J=7/2-$ & <6.0e-17 & -- & <5.4e-17 & -- \\
119.23418 & OH & $^2P_{3/2}\ J=5/2+ \rightarrow J=3/2-$ & <6.1e-17 & -- & <4.9e-17 & -- \\
119.44167 & OH & $^2P_{3/2}\ J=5/2+ \rightarrow J=3/2-$ & <6.1e-17 & -- & <4.8e-17 & -- \\
134.8476 & OH & $^2P_{1/2}\ J=7/2+ \rightarrow J=9/2-$ & <4.2e-17 & -- & <3.6e-17 & -- \\
134.9696 & OH & $^2P_{1/2}\ J=7/2+ \rightarrow J=9/2-$ & <4.1e-17 & -- & <3.6e-17 & -- \\
163.12428 & OH & $^2P_{1/2}\ J=3/2+ \rightarrow J=1/2-$ & <3.9e-17 & -- & <3.7e-17 & -- \\
163.39718 & OH & $^2P_{1/2}\ J=3/2+ \rightarrow J=1/2-$ & <4.0e-17 & -- & <3.6e-17 & -- \\
63.183705 & [\ion{O}{i}] & $^3P_{0}-^3P_{1}$ & <1.5e-16 & -- & 1.85e-16 & 5.0e-17 \\
145.52544 & [\ion{O}{i}] & $^3P_{1}-^3P_{2}$ & 2.8e-17 & 1.3e-17 & 3.1e-17 & 1.2e-17 \\
58.6982 & o-H$_2$O & $4_{32}-3_{21}$ & <1.8e-16 & -- & <1.7e-16 & -- \\
66.4372 & o-H$_2$O & $3_{30}-2_{21}$ & <1.5e-16 & -- & <1.3e-16 & -- \\
67.2689 & o-H$_2$O & $3_{30}-3_{03}$ & <1.4e-16 & -- & <1.4e-16 & -- \\
71.946 & o-H$_2$O & $7_{07}-6_{16}$ & <2.1e-16 & -- & <2.1e-16 & -- \\
75.3804 & o-H$_2$O & $3_{21}-2_{12}$ & <1.2e-16 & -- & <1.2e-16 & -- \\
78.7414 & o-H$_2$O & $4_{23}-3_{12}$ & <1.5e-16 & -- & <1.3e-16 & -- \\
82.0304 & o-H$_2$O & $6_{16}-5_{05}$ & <1.4e-16 & -- & <1.3e-16 & -- \\
108.073 & o-H$_2$O & $2_{21}-1_{10}$ & <8.0e-17 & -- & <7.2e-17 & -- \\
113.5366 & o-H$_2$O & $4_{14}-3_{03}$ & <6.5e-17 & -- & <6.1e-17 & -- \\
116.7836 & o-H$_2$O & $7_{34}-6_{43}$ & <6.3e-17 & -- & <5.4e-17 & -- \\
121.7191 & o-H$_2$O & $4_{32}-4_{23}$ & <4.9e-17 & -- & <4.4e-17 & -- \\
132.407 & o-H$_2$O & $4_{23}-4_{14}$ & <4.0e-17 & -- & <3.8e-17 & -- \\
134.9346 & o-H$_2$O & $5_{14}-5_{05}$ & 3.0e-17 & 1.4e-17 & <3.7e-17 & -- \\
136.4944 & o-H$_2$O & $3_{30}-3_{21}$ & <4.0e-17 & -- & <3.6e-17 & -- \\
174.6264 & o-H$_2$O & $3_{03}-2_{12}$ & <4.3e-17 & -- & <4.0e-17 & -- \\
179.5265 & o-H$_2$O & $2_{12}-1_{01}$ & <4.6e-17 & -- & <4.4e-17 & -- \\
180.488 & o-H$_2$O & $2_{21}-2_{12}$ & <5.3e-17 & -- & 3.0e-17 & 1.5e-17 \\
56.3242 & p-H$_2$O & $4_{31}-3_{22}$ & <2.5e-16 & -- & <2.4e-16 & -- \\
57.6361 & p-H$_2$O & $4_{22}-3_{13}$ & <2.2e-16 & -- & <2.0e-16 & -- \\
61.8084 & p-H$_2$O & $4_{31}-4_{04}$ & <1.5e-16 & -- & <1.5e-16 & -- \\
63.457 & p-H$_2$O & $8_{08}-7_{17}$ & <1.5e-16 & -- & <1.5e-16 & -- \\
67.0886 & p-H$_2$O & $3_{31}-2_{20}$ & <1.4e-16 & -- & <1.4e-16 & -- \\
71.0662 & p-H$_2$O & $5_{24}-4_{13}$ & <2.0e-16 & -- & <1.9e-16 & -- \\
83.2831 & p-H$_2$O & $6_{06}-5_{15}$ & <1.5e-16 & -- & <1.4e-16 & -- \\
89.9878 & p-H$_2$O & $3_{22}-2_{11}$ & <1.4e-16 & -- & <1.4e-16 & -- \\
125.3534 & p-H$_2$O & $4_{04}-3_{13}$ & <4.5e-17 & -- & <4.0e-17 & -- \\
126.7126 & p-H$_2$O & $3_{31}-3_{22}$ & <4.1e-17 & -- & <3.8e-17 & -- \\
138.5272 & p-H$_2$O & $3_{13}-2_{02}$ & <3.9e-17 & -- & <3.7e-17 & -- \\
144.5181 & p-H$_2$O & $4_{13}-3_{22}$ & <4.1e-17 & -- & <3.5e-17 & -- \\
156.193 & p-H$_2$O & $3_{22}-3_{13}$ & <3.9e-17 & -- & <3.6e-17 & -- \\
158.309 & p-H$_2$O & $3_{31}-4_{04}$ & <3.9e-17 & -- & <3.8e-17 & -- \\
187.1104 & p-H$_2$O & $4_{13}-4_{04}$ & <6.4e-17 & -- & <5.7e-17 & -- \\
\hline
\end{tabular}
\end{table*}

\end{appendix}

\end{document}